\pdfoutput=1
\documentclass[twoside,12pt]{article}
\usepackage{graphicx}
\usepackage{latexsym}
\begin{document}

\title{Primordial backgrounds of relic gravitons}
\author{Massimo Giovannini\footnote{Electronic address: massimo.giovannini@cern.ch}\\
\\
{\small Department of Physics, CERN, 1211 Geneva 23, Switzerland }
\\
{\small INFN, Section of Milan-Bicocca, 20126 Milan, Italy}}
\date{}
\maketitle

\begin{abstract} 
The diffuse backgrounds of relic gravitons with frequencies ranging between the aHz band and the GHz region encode the ultimate information on the primeval evolution of the plasma and on the underlying theory of gravity well before the electroweak epoch. While the temperature and polarization anisotropies of the microwave background radiation probe the low-frequency tail of the graviton spectra, during the next score year the pulsar timing arrays and the wide-band interferometers (both terrestrial and hopefully space-borne) will explore a much larger frequency window encompassing the nHz domain and the audio band. The salient theoretical aspects of the relic gravitons are reviewed in a cross-disciplinary perspective touching upon various unsettled questions of particle physics, cosmology and astrophysics.
\end{abstract}
\begin{center}
CERN-TH-2019-166
\end{center}
\newpage
\tableofcontents

\newpage

\renewcommand{\theequation}{1.\arabic{equation}}
\setcounter{equation}{0}
\section{The cosmic spectrum of relic gravitons}
\label{sec1}
The LIGO/Virgo collaboration recently published the three-detector observations 
of gravitational waves from black-hole coalescence \cite{LIGO1}, the evidence of gravitational waves 
from neutron star inspiral \cite{LIGO2}, and the observation of a $50$-solar-mass binary black hole 
coalescence for a redshift $z =0.2$ \cite{LIGO3}.  It is fair to say that the recent discoveries 
of wide-band interferometers \cite{LIGO1,LIGO2,LIGO3} are fostering an unprecedented development of gravitational wave astronomy. The goal of this article is instead to scrutinize  the theory of the relic gravitons, i.e. {\em gravitons produced during the earliest stages of the evolution of the cosmic plasma}. The energies and the frequencies of these massless particles are suppressed by the cosmic expansion so that their corresponding wavelengths are redshifted and may even be comparable with the diameter of the current Hubble patch. In many respects the cosmic gravitons bear analogy to the diffuse electromagnetic backgrounds with the notable difference that the microwave photons observed today last scattered free electrons and protons at a typical redshift $z_{dec} \simeq 1090$, while the relic gravitons come from a stage when the Hubble rate was nearly Planckian.  Even if specific attention will be paid to the hoped sensitivities of the forthcoming detectors, the main emphasis of this article will be on the theoretical aspects and on the uncertainties of the potential signals. 
\begin{table}[!ht]
\begin{center}
\caption{Typical frequencies, energies and wavelengths of the electromagnetic spectrum.}
\vskip 0.4 cm
\begin{tabular}{||c|c|c|c||}
\hline
\hline
\rule{0pt}{4ex}  band & frequency [Hz] & energy [eV] & wavelength [m]\\
\hline
\hline
$\gamma$-rays & $\mathrm{ZHz} = 10^{21}  \,\mathrm{Hz}$ & $4\,\mathrm{MeV} = 4\times10^{6}\, \mathrm{eV}$  & $3\times10^{-13} \, \mathrm{m}$ \\
soft $x$-rays & $\mathrm{EHz} = 10^{18} \,\mathrm{Hz}$ & $4 \,\mathrm{keV} = 4\times 10^{3}\, \mathrm{eV}$  & $3\times 10^{-10} \, \mathrm{m}$ \\
ultraviolet & $\mathrm{PHz} = 10^{15}\, \mathrm{Hz}$ & $4\,\mathrm{eV}$  & $3\times 10^{-7} \, \mathrm{m}$ \\
infrared & $\mathrm{THz} = 10^{12} \,\mathrm{Hz}$ & $4\,\mathrm{meV}= 4\times 10^{-3} \, \mathrm{eV}$  & $3\times 10^{-4} \, \mathrm{m}$ \\
microwaves & $\mathrm{GHz} = 10^{9} \, \mathrm{Hz}$ & $4 \mu\mathrm{eV} =4\times 10^{-6} \, \mathrm{eV}$  & $ 0.3\,\mathrm{m}$ \\
radio waves (HF) & $\mathrm{MHz} = 10^{6} \, \mathrm{Hz}$ & $4 \mathrm{neV} = 4\times 10^{-9} \, \mathrm{eV}$  & $300 \, \mathrm{m} $ \\
radio waves (LF) & $\mathrm{kHz} =  10^{3} \, \mathrm{Hz}$ & $4\,\mathrm{peV} = 4\times10^{-12} \, \mathrm{eV}$  & $ 3\times 10^{5} \, \mathrm{m}$ \\
low frequency waves & Hz&    $4\, \mathrm{feV} = 4\times 10^{-15} \, \mathrm{eV}$  & $3\times 10^{8} \,\mathrm{m}$ \\
\hline
\hline
\end{tabular}
\label{SEC1TABLE1}
\end{center}
\end{table}
\subsection{Typical frequencies of the relic gravitons} 
In a preliminary perspective it is useful to compare the gravitational and the electromagnetic spectra even 
if the origins of the two emissions are very different. The typical frequencies, wavelengths and energies  
 of the electromagnetic spectra are illustrated in Tab. \ref{SEC1TABLE1}  from the ultra low-frequency region\footnote{Standard prefixes will be used throughout to simplify the discussions: for instance $1\, \mathrm{GHz} = 10^{-9}$ Hz, $1\, \mathrm{feV} = 10^{-15} \, \mathrm{eV}$ 
and so on and so forth. } (of the order of the Hz) up to the $\gamma$-ray band (of the order of few ZHz). The gravitational spectra of Tab. \ref{SEC1TABLE2} extend instead between few aHz and, approximately, $100$ GHz. The range of frequencies roughly spans $20$ orders of magnitude in the case of the photons and $30$ orders of magnitude in the case of the gravitons. 
While the various frequency bands introduced in Tab. \ref{SEC1TABLE2} have their own peculiar 
features, the spectrum of the relic gravitons can be divided in three major domains that will be described in this introductory section and more extensively discussed in the remaining part of the article. 

In the electromagnetic case we talk about electromagnetic waves when discussing radio waves
but already for wavelengths of the order of the mm the particle description is preferable so that 
it is customary to discuss, for instance, the energy density and the concentration of the microwave background photons.  For the same reasons we use the wording $x$-ray 
photons or $\gamma$-ray photons for the high-frequency branch of the electromagnetic spectrum. 
Along the same perspective the random gravitational waves that are potentially measurable at the present time were really produced as gravitons in the early universe: their corresponding wavelengths, initially very short,  are subsequently stretched by the expansion of the universe and may now be observed as long-wavelength modes.  To the complement the previous remark 
we must add that even gravitons in a gravitational-wave burst are phase-coherent while photons in electromagnetic signals are usually phase-incoherent. This arises from the fact that each graviton is generated from the same bulk motion of matter or of spacetime curvature, while each photon is normally generated by different, independent events involving atoms, ions or electrons. As we shall specifically argue throughout this article the relic gravitons are similar, in many respects, to the laser lights or to the so-called squeezed lights\footnote{According to the quantum theory of optical coherence 
the laser light is made by coherent states of the electromagnetic field. Coherent states minimize the indetermination relations.
In the last $30$ years more general states of the radiation field became technologically feasible and are commonly 
referred to as {\em squeezed states}. The squeezed states span the same surface of coherent states 
in the phase space but the indetermination in one of the two canonically conjugate operators can undershoot the quantum noise while the other exceeds the variance of the quantum noise. Relic gravitons are produced 
as squeezed states of the gravitational field.  Moreover the squeezed states of the radiation 
field are customarily injected in wide-band interferometers with the aim of increasing their sensitivity 
for the gravitational wave detection.} and we could even use the phase coherence of relic gravitons to enhance their detectability. 
\begin{table}[!ht]
\begin{center}
\caption{Typical frequencies, energies and wavelengths of the gravitational spectrum.}
\vskip 0.4 cm
\begin{tabular}{||c|c|c|c||}
\hline
\hline
\rule{0pt}{4ex}  band  & frequency [Hz] & energy [eV] & wavelength [m]\\
\hline
\hline
ultra high-frequency & $\mathrm{GHz} = 10^{9}  \,\mathrm{Hz}$ & $4 \, \mu\mathrm{eV} =4\times 10^{-6} \, \mathrm{eV}$  & $0.3 \,\mathrm{m}$ \\
high-frequency  & $\mathrm{MHz} = 10^{6} \, \mathrm{Hz}$ & $4\,\mathrm{neV} = 4\times 10^{-9} \, \mathrm{eV}$  & $300 \, \mathrm{m} $ \\
audio (HF)  & $\mathrm{kHz} = 10^{3} \, \mathrm{Hz}$ & $4\,\mathrm{peV} = 4\times 10^{-12} \, \mathrm{eV}$  & $3\times 10^{5} \, \mathrm{m}$ \\
audio (LF) & $\mathrm{Hz}$ &  $4\,\mathrm{feV} = 4\times 10^{-15} \, \mathrm{eV}$  & $3\times 10^{8} \,\mathrm{m}$ \\
intermediate (HF)  & $\mathrm{mHz} = 10^{-3} \,  \mathrm{Hz}$ &  $4 \, \mathrm{aeV} = 4\times 10^{-18} \, \mathrm{eV}$  & $3\times10^{11} \,\mathrm{m}$ \\
intermediate (LF)  & $\mu\mathrm{Hz} = 10^{-6} \,  \mathrm{Hz}$ &  $4\, \mathrm{zeV} = 4\times 10^{-21} \, \mathrm{eV}$  & $3\times 10^{14} \,\mathrm{m}$ \\
low-frequency (HF) & $\mathrm{nHz} = 10^{-9} \,  \mathrm{Hz}$ &  $4\,\mathrm{yeV} = 4\times 10^{-24} \, \mathrm{eV}$  & $3\times 10^{17} \,\mathrm{m}$ \\
low-frequency (LF) & $\mathrm{pHz} = 10^{-12} \,  \mathrm{Hz}$ &  $ 4\times 10^{-30} \, 
\mathrm{eV}$  & $3\times 10^{20} \,\mathrm{m}$ \\
ultra low-frequency (HF) & $\mathrm{fHz} = 10^{-15} \,  \mathrm{Hz}$ &  $ 4\times10^{-33} \, 
\mathrm{eV}$  & $3\times 10^{23} \,\mathrm{m}$ \\
ultra low-frequency (LF) & $\mathrm{aHz} = 10^{-18} \,  \mathrm{Hz}$ &  $ 4\times10^{-36} \, \mathrm{eV}$  & $3 \times 10^{26} \,\mathrm{m}$ \\
\hline
\hline
\end{tabular}
\label{SEC1TABLE2}
\end{center}
\end{table}

\subsubsection{Low-frequencies}
 The low-frequency band ranges approximately between few aHz and the fHz region. Recalling that 
$1\, \mathrm{Mpc} = 3.085 \times 10^{22} \mathrm{m}$, the frequencies ${\mathcal O}(\mathrm{aHz})$ correspond to 
wavelengths that are comparable with the Hubble radius
$H_{0}^{-1} = 4282.7 \, (h_{0}/0.7)^{-1} \, \mathrm{Mpc}$, where $h_{0}$ will denote throughout the 
value of the Hubble rate in units of $100 \, (\mathrm{km}/\mathrm{Mpc})\, \mathrm{Hz}$.
The  low-frequency gravitons contribute to the temperature and the polarization anisotropies of the Cosmic Microwave Background  
(CMB in what follows). The initial conditions of the Einstein-Boltzmann hierarchy (ultimately determining the CMB anisotropies) are customarily assigned prior to matter-radiation equality and at a pivot wavenumber\footnote{The initial conditions of the temperature and polarization anisotropies are assigned where the (statistical) errors are small and while the choice of $k_{p}$ is largely conventional it is preferable to use the same $k_{p}$ for the scalar and for the tensor power spectra.  This approach for the normalization of the temperature and polarization 
anisotropies differs from the earlier strategies preferentially employing even lower 
multipoles (i.e. between $2$ and $20$) \cite{CMB3}.} 
 \cite{RT1,RT2,RT3,RT4}
 \begin{equation}
 k_{p} \,\,=\,\,0.002 \,\,\mathrm{Mpc}^{-1},
 \label{FF0a}
 \end{equation}
whose associated pivot frequency is ${\mathcal O}(3)$ aHz, i.e. 
\begin{equation}
\nu_{p} \,\,= \,\,\frac{k_{p}}{2\pi} \,\,=\,\, 3.092 \biggl(\frac{k_{p}}{0.002 \, \mathrm{Mpc}^{-1}} \biggr) \times 10^{-18} \mathrm{Hz} = 3.092 \,\,  \biggl(\frac{k_{p}}{0.002 \, \mathrm{Mpc}^{-1}} \biggr) \,\, \mathrm{aHz}.
\label{FF1}
\end{equation}
 The frequency related with the dominance of dark energy is of the same order 
of Eq. (\ref{FF1}) and it depends on $\Omega_{M0}$ and on $\Omega_{\Lambda}$ (i.e. the critical fractions 
of matter and dark energy):
\begin{equation}
 \nu_{\Lambda} \,\,= \,\, \frac{k_{\Lambda}}{2\pi} =1.638 \biggl(\frac{h_{0}}{0.719}\biggr) \biggl(\frac{\Omega_{M0}}{0.258}\biggr)^{1/3} \biggl(\frac{\Omega_{\Lambda}}{0.742}\biggr)^{-1/3} \,\, \mathrm{aHz}.
\label{FF2}
\end{equation}
The equality wavenumber is
given by 
\begin{equation}
k_{eq}  \,\,= \,\, 0.0732\,[h_{0}^2 \Omega_{R0}/(4.15\times 10^{-5})]^{-1/2}\,\,  h_{0}^2 \Omega_{M0} \,\, \mathrm{Mpc}^{-1},
\label{FF2a}
\end{equation}
and  it corresponds to a comoving wavelength comparable with the Hubble radius at the time of matter-radiation equality.
The frequency $\nu_{eq}$ associated with $k_{eq}$ always exceeds $\nu_{p}$ and it is:
\begin{equation}
\nu_{eq}  \,\,= \,\,  \frac{k_{eq}}{2\pi} = 15.97 \biggl(\frac{h_{0}^2 \Omega_{M0}}{0.1411}\biggr) \biggl(\frac{h_{0}^2 \Omega_{R0}}{4.15 \times 10^{-5}}\biggr)^{-1/2}\,\, \mathrm{aHz},
\label{FF3}
\end{equation}
where $\Omega_{R0}$ denotes the critical fraction of relativistic species in the framework of the 
concordance paradigm. The low-frequency gravitons contribute both to the 
temperature and to the polarization inhomogeneities of the CMB. After the seminal discovery 
of the microwave background temperature by Penzias and Wilson \cite{CMB1} the COBE satellite
not only measured with great accuracy the CMB spectrum and its distortions \cite{CMB2,CMB4,CMB5},
but also the novel temperature anisotropies \cite{CMB3} that have been subsequently complemented by 
the polarization inhomogeneities, originally observed by the WMAP experiment 
\cite{WMAP1,WMAP1a,WMAP1b,WMAP2,WMAP2a}. These important measurements and their descendants 
did not allow so far to disentangle the contribution of the (scalar) curvature inhomogeneities from 
the effect of the relic gravitons.  This is why both satellite missions and ground-based observations 
(like the ones of the partially successful Bicep2 collaboration \cite{B1}) 
should either measure the $B$-mode polarization caused by the low-frequency gravitons or 
substantially improve some of the current upper limits \cite{RT3,RT4}. 

\subsubsection{Intermediate frequencies}
The intermediate frequency band ranges approximately between the pHz and the few Hz.  
In this region the gravitons directly affect  the big-bang nucleosynthesis (BBN) occurring when 
the approximate temperature of the plasma was  ${\mathcal O}(0.1)$ MeV and ultimately accounting 
for the abundances of the light nuclear elements. The gravitons with larger frequencies 
may also impact BBN but their potential relevance depends, in this case, on the properties of their spectral slope. The Hubble rate at the BBN time defines a typical frequency whose comoving value is given by\footnote{The present value of the 
scale factor will be normalized to $1$ (i.e. $a_{0} =1$); consequently {\em comoving and physical frequencies 
coincide at the present time}.}:
\begin{equation}
\nu_{bbn} \,\,= \,\, 2.3\times 10^{-2} \biggl(\frac{g_{\rho}}{10.75}\biggr)^{1/4} \biggl(\frac{T_{bbn}}{\,\,\mathrm{MeV}}\biggr) 
\biggl(\frac{h_{0}^2 \Omega_{R0}}{4.15 \times 10^{-5}}\biggr)^{1/4}\,\,\mathrm{nHz},
\label{FF4}
\end{equation}
where $g_{\rho}$ denotes the effective number of relativistic degrees of freedom entering the total energy 
density of the plasma and $T_{bbn}$ is the putative BBN temperature. 
The expansion rate at the time of the formation of the light elements depends on the total energy density 
of the gravitons with frequencies larger than the $0.01$ nHz.  In the intermediate frequency band the spectral energy density 
of the relic gravitons is also constrained by  the millisecond pulsar timing observations that
 set an upper bound on the energy density of the relic gravitons for a typical
frequency 
\begin{equation}
\nu_{pulsar} \,\,= \,\,{\mathcal O}(10)\,\,\mathrm{nHz}, 
\label{FF4a}
\end{equation}
roughly corresponding to the inverse of the observation time
along which the pulsars timing has been monitored \cite{PUL1,PUL2}. Between a fraction of the mHz and few Hz  will be eventually located the operating window of space-borne interferometers such as LISA
 (Laser Interferometer Space Antenna) \cite{LISA,LISAa}, BBO (Big Bang Observer) \cite{BBO}, and DECIGO (Deci-hertz Interferometer Gravitational Wave Observatory) \cite{DECIGO1}.   In the $\mu$Hz range falls the so-called 
 electroweak frequency:
 \begin{equation}
\nu_{ew}  \,\,= \,\, 7.05  \biggl(\frac{g_{\rho}}{106.75}\biggr)^{1/4} \biggl(\frac{T_{ew}}{173\, \mathrm{GeV}} \biggr) \, \mu\mathrm{Hz},
\label{FF5}
\end{equation}
which is ${\mathcal O}(6)\, \mu\mathrm{Hz}$ if the electroweak phase transition takes place around $150$ GeV.  The reference value of the temperature appearing in Eq. (\ref{FF5}) 
corresponds to the approximate top-quark mass $m_{t} \simeq 173$ GeV while an estimate of $g_{\rho}$ 
follows by considering  all the degrees of freedom of the standard model 
$SU_{L}(2)\otimes U_{Y}(1) \otimes SU_{c}(3)$ in local thermal equilibrium slightly above  $m_{t}$. 
Given the value of the Higgs mass and of the $W$ mass \cite{PT0a,PT0b} the electroweak phase transition will not be strongly first-order and will not lead to a sizable burst of gravitational radiation at least in the context of the standard model of particle interactions where a cross-over is expected. 
Even in this case, however, the electroweak epoch might be the source of a secondary backgrounds of gravitational radiation as a consequence of the dynamical evolution of the plasma.

\subsubsection{High-frequencies}
The high-frequency region  ranges between few Hz and the GHz. Within this broad frequency domain 
is located the so-called audio band (i.e. between few Hz and $10$ kHz) where wide-band interferometers are now operating. 
The advanced LIGO/Virgo projects \cite{LV1,LV2} will be complemented by further interferometers with different designs (but always operational in the audio band) namely the Japanese Kagra (Kamioka Gravitational Wave Detector) \cite{kagra1,kagra2} (i.e. the prosecution of the Tama-300 experiment \cite{TAMA}) and probably the Indian LIGO \cite{LIGOindia}. Among the 
wide-band detectors we should also mention the GEO-600 experiment \cite{GEO1} (which is 
now included in the LIGO/Virgo consortium \cite{GEO2}) and the futuristic Einstein telescope \cite{ET1}. 
 The black-body spectrum associated with the relic gravitons peaks in the region of $100$ GHz corresponding to a temperature 
 ${\mathcal O}(0.9)$ K (i.e. slightly smaller than the one of photons \cite{CMB2,CMB4,CMB5}).  The early variation of the space-time curvature also produces a stochastic backgrounds of relic gravitons, with a signal that is roughly $10$ orders of magnitude below the one of thermal gravitons. In the context of conventional inflationary models the maximal frequency of the relic graviton background is given by:
\begin{equation}
\nu_{max}  \,\,= \,\, 1.9\times 10^{2} \biggl(\frac{\epsilon}{0.001}\biggr)^{1/4} 
\biggl(\frac{{\mathcal A}_{{\mathcal R}}}{2.41\times 10^{-9}}\biggr)^{1/4} 
\biggl(\frac{h_{0}^2 \Omega_{R0}}{4.15 \times 10^{-5}}\biggr)^{1/4}  \,\, \mathrm{MHz},
\label{ONEb}
\end{equation}
where ${\mathcal A}_{{\mathcal R}}$ denotes the amplitude 
of the power spectrum of curvature inhomogeneities at the wavenumber $k_{p}$ (see Eq. (\ref{FF1}) and discussion therein); $\epsilon$ denotes the slow-roll parameter characterizing the evolution of standard inflationary scenarios. The conventional 
inflationary scenarios are not the only potential source of high-frequency gravitational waves and 
the maximal frequency of the spectrum might exceed  ${\mathcal O}(200)$ MHz. 
In the high-frequency region there could be detection strategies based on electro-mechanical detectors where the electromagnetic field and the field of elastic deformations are simultaneously present and are both affected by the high-frequency gravitational wave \cite{brag1,cav1}.  

\subsection{The concordance paradigm}
All in all the concordance paradigm\footnote{The concordance paradigm is often referred to as $\Lambda$CDM where $\Lambda$ stands for the dark energy component and CDM refers to the cold dark matter component.} is just a useful compromise between the available data, the standard cosmological model and the number of ascertainable parameters. In the minimal version 
of the popular $\Lambda$CDM scenario {\em the dark energy does not fluctuate, the three 
neutrinos species are strictly massless and the presence of the relic gravitons can be safely ignored}.

Even if the first observational evidence of the large-scale polarization of the CMB 
has been  obtained by the DASI (Degree Angular Scale Interferometer) 
experiment \cite{LCDM1}, the turning point shaping the present form of 
the $\Lambda$CDM scenario came with the evidence 
of the cross-correlation between the temperature and the 
polarization anisotropies \cite{WMAP1,WMAP1a,WMAP1b,WMAP2,WMAP2a} obtained by the WMAP collaboration. 
The position of the first Doppler peak (of the temperature autocorrelations) and  the first anti-correlation 
peak (of the cross-correlation between temperature and polarization) implied that the source of large-scale inhomogeneities 
accounting for the CMB anisotropies had to be {\em adiabatic and Gaussian fluctuations of the spatial 
curvature} \cite{LCDM4a,LCDM4b,LCDM4c,LCDM5}. This important observation, confirmed by the subsequent 
data releases\footnote{There have been 5 different releases of the WMAP  data corresponding to one, 
 three, five, seven and nine years of integrated observations. The various releases led to compatible (but slightly different) determinations of the pivotal parameters of the $\Lambda$CDM paradigm. A similar comment holds for the three releases of the Planck collaboration. This point, together with other features of the $\Lambda$CDM paradigm is illustrated in Tab. \ref{SEC1TABLE3}.} of the WMAP experiment (see e.g. \cite{RT1,RT2}) and by the 
Planck collaboration \cite{LCDM6,LCDM7,LCDM8}, justifies and motivates the current formulation of the 
concordance paradigm where the dominant source of 
large-scale inhomogeneity are the adiabatic curvature
perturbations. 
\begin{table}[!ht]
\begin{center}
\caption{The six pivotal parameters of $\Lambda$CDM scenario. The best fits of the WMAP5, 
WMAP7 and WMAP9  data alone are compared with the Planck fiducial set of parameters.}
\vskip 0.4 cm
\begin{tabular}{||l|c|c|c|c|c|c||}
\hline
\hline
\rule{0pt}{4ex} Data & WMAP5  & WMAP7  & WMAP9 & PLANCK 2015\\
\hline
$\Omega_{b0}$& $0.0441\pm 0.0030$ & $ 0.0449\pm 0.0028$& $0.0463 \pm 0.0024$ &$0.0486 \pm 0.0010$ \\
$\Omega_{c0}$& $0.214\pm 0.027$& $ 0.222\pm0.026$& $0.233\pm0.023 $& $0.2589 \pm 0.0057$\\
$\Omega_{\Lambda}$& $0.742\pm 0.030 $ &$ 0.734\pm 0.029$&$0.721\pm0.025$&$0.6911\pm 0.0062$\\
$H_{0}$ & $71.9^{+2.6}_{-2.7} $ & $71.0\pm 2.5$&$ 70.0\pm 2.2 $ &$67.74 \pm 0.46$\\
$n_{s}$ & $0.963^{+0.014}_{-0.015}$ & $0.963\pm 0.014$&$ 0.972\pm 0.013$&$0.9667\pm 0.0040$\\
$\epsilon_{re}$ & $0.087\pm 0.017$ & $0.088\pm 0.015$&$0.089\pm0.014$&$0.066 \pm0.012$\\
$r_{T}$ & $< 0.43$& $< 0.36$ & $< 0.34$& $<0.1$\\
\hline
\end{tabular}
\label{SEC1TABLE3}
\end{center}
\end{table}
The $\Lambda$CDM paradigm is specified by assigning six 
 pivotal parameters that are customarily selected as follows:
{\it (i)} the present critical fraction of baryonic matter  [i.e. $\Omega_{b0}= \rho_{b0}/\rho_{crit}= {\mathcal O}(0.048)$];  
{\it (ii)} the present critical fraction of CDM particles, [i.e. $\Omega_{c0}= \rho_{c0}/\rho_{crit}={\mathcal O}(0.26)$];
{\it (iii)} the present critical fraction of dark energy, [i.e. $\Omega_{\Lambda}= \rho_{\Lambda}/\rho_{crit}={\mathcal O}(0.7)$];
{\it (iv)} the indetermination on the Hubble rate [i.e.  $h_{0}= {\mathcal O}(0.7)$]; {\it (v)} the spectral index of scalar inhomogeneities [i.e. $n_{s}= {\mathcal O}(0.96)$];
{\it (vi)} the optical depth at reionization [i.e.  $\epsilon_{re} ={\mathcal O}(0.06)$]. 
The parameters of the $\Lambda$CDM paradigm can be inferred either by considering 
a single class of data (e.g. microwave background observations) or by 
requiring the consistency of the scenario with the three observational data sets represented
by the temperature and polarization anisotropies of the microwave background,
 by the extended galaxy surveys (see e.g. \cite{LCDM9,LCDM10}) and by the supernova observations (see e.g. \cite{LCDM11,LCDM12}). Various terrestrial observations of the temperature and polarization anisotropies have been reported so far (see e.g. \cite{LCDM13,LCDM14,LCDM15,LCDM17,LCDM17a,LCDM19}). In Tab. \ref{SEC1TABLE3} the $\Lambda$CDM parameters derived 
from the last three data releases of the WMAP experiment are compared with the Planck 
fiducial set of parameters obtained from the 2015 release.
Even if the relic gravitons and the neutrino masses are absent from the conventional 
formulation of the $\Lambda$CDM paradigm, various bounds have been obtained through 
the years. In the last line of Tab. \ref{SEC1TABLE3}, for reference,  the limits on the 
tensor to scalar ratio $r_{T}$ have been illustrated for each corresponding data release. In the case 
$r_{T} =0$ the gravitons would be totally absent. Even if the bound $r_{T} < 0.1$ 
seems already constraining, there have been recent attempts 
to make the bound even more stringent.  So for instance in Ref. \cite{RT3} the combination 
of different data sets implies $r_{T} <0.07$ while in Ref. \cite{RT4} values $r_{T} <0.064$ are quoted. 
These bounds are not based on a direct assessment of the $B$-mode polarization (i.e. the $BB$ 
correlations) but rather on the effect of the tensor modes of the geometry on the remaining temperature and polarization 
anisotropies\footnote{The tensor modes of the geometry directly affect the temperature 
anisotropies ($TT$ correlations in the jargon), the measured polarization 
anisotropies (i.e. the $EE$ correlations) and their cross correlation (i.e. the $TE$ correlations).}.
The $BB$ correlations are particularly relevant, in this context, since in the concordance paradigm 
they must be strictly zero when $r_{T}=0$ (barring for the $B$-mode polarization induced by the lensing of the primary anisotropies). In what follows we shall use a fiducial set of parameters following from the 2018 
Planck release \cite{RT4,RT5} where
\begin{equation}
(h_{0}^2 \Omega_{b0}, \, h_{0}^2 \Omega_{c0},\, \Omega_{\Lambda}, \, h_{0},\, n_{s}\, \epsilon_{re}) = (0.02237, \, 0.1200,\, 0.6847, \, 0.67,\, 0.9649,\, 0.0544).
\label{PARSET}
\end{equation}
For an introduction on the relevant aspects 
of CMB physics there exist various extended monographs \cite{book1,book2,book3}. Besides 
the last line which is directly relevant for the present ends, Tab. \ref{SEC1TABLE3} gives an interesting 
account of the trend of the data in the context of minimal $\Lambda$CDM paradigm supplemented by the 
tensor modes.
 
\subsection{Cosmic photons versus cosmic gravitons}
\begin{figure}[!ht]
\centering
\includegraphics[width=0.8\textwidth]{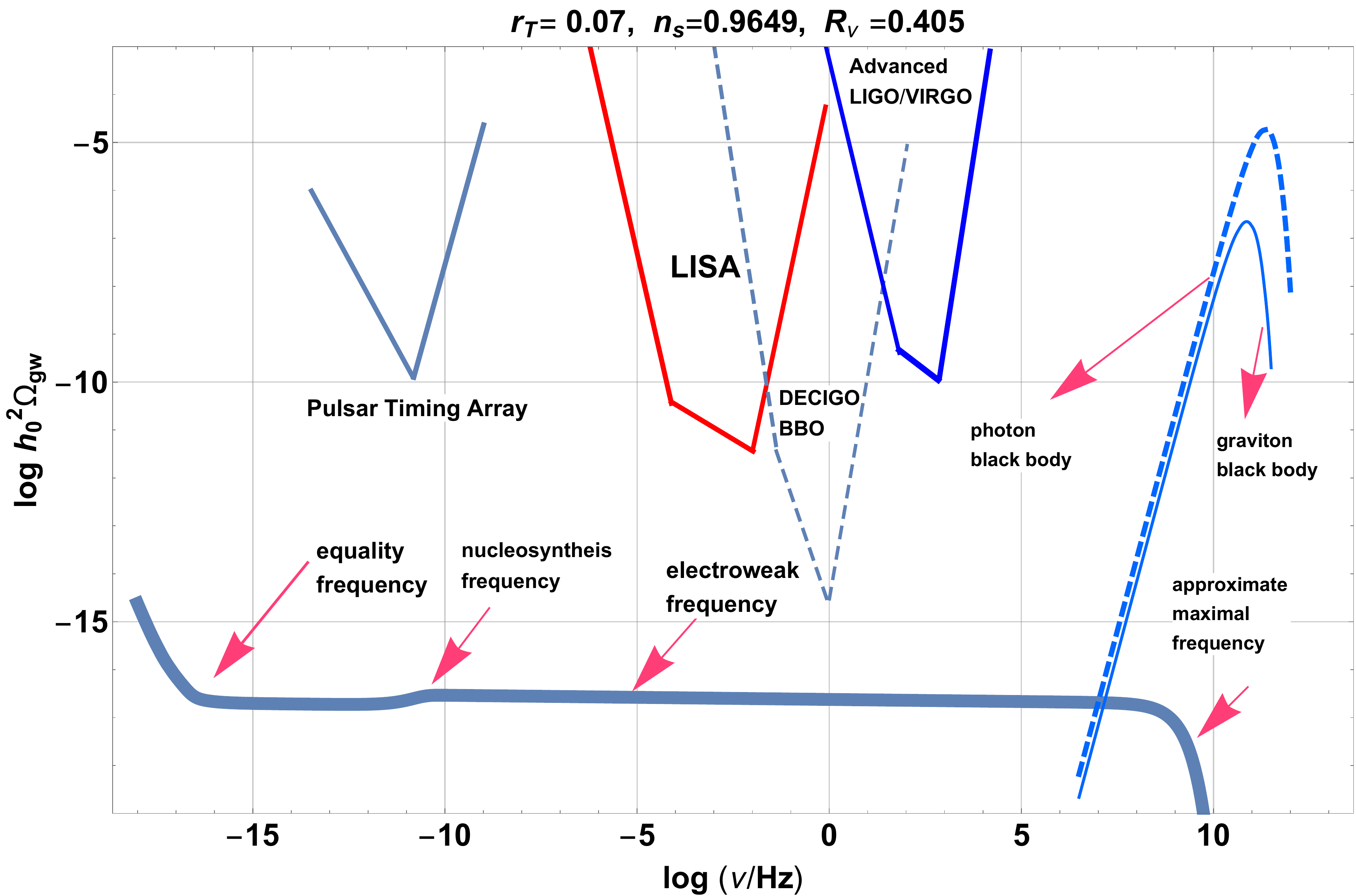}
\caption[a]{The relic gravitons and the CMB photons are illustrated in critical units. It is useful to compare this figure with Fig. \ref{SEC1FIG2} where the CMB photons and the thermal gravitons are illustrated together with the other diffuse electromagnetic emissions. Common logarithms are employed on both axes. The two curves appearing in the rightmost 
part of the plot denote the photon black-body (dashed line) and the graviton black-body (full thin line). }
\label{SEC1FIG1} 
\end{figure}

\begin{figure}[!ht]
\centering
\includegraphics[width=0.8\textwidth]{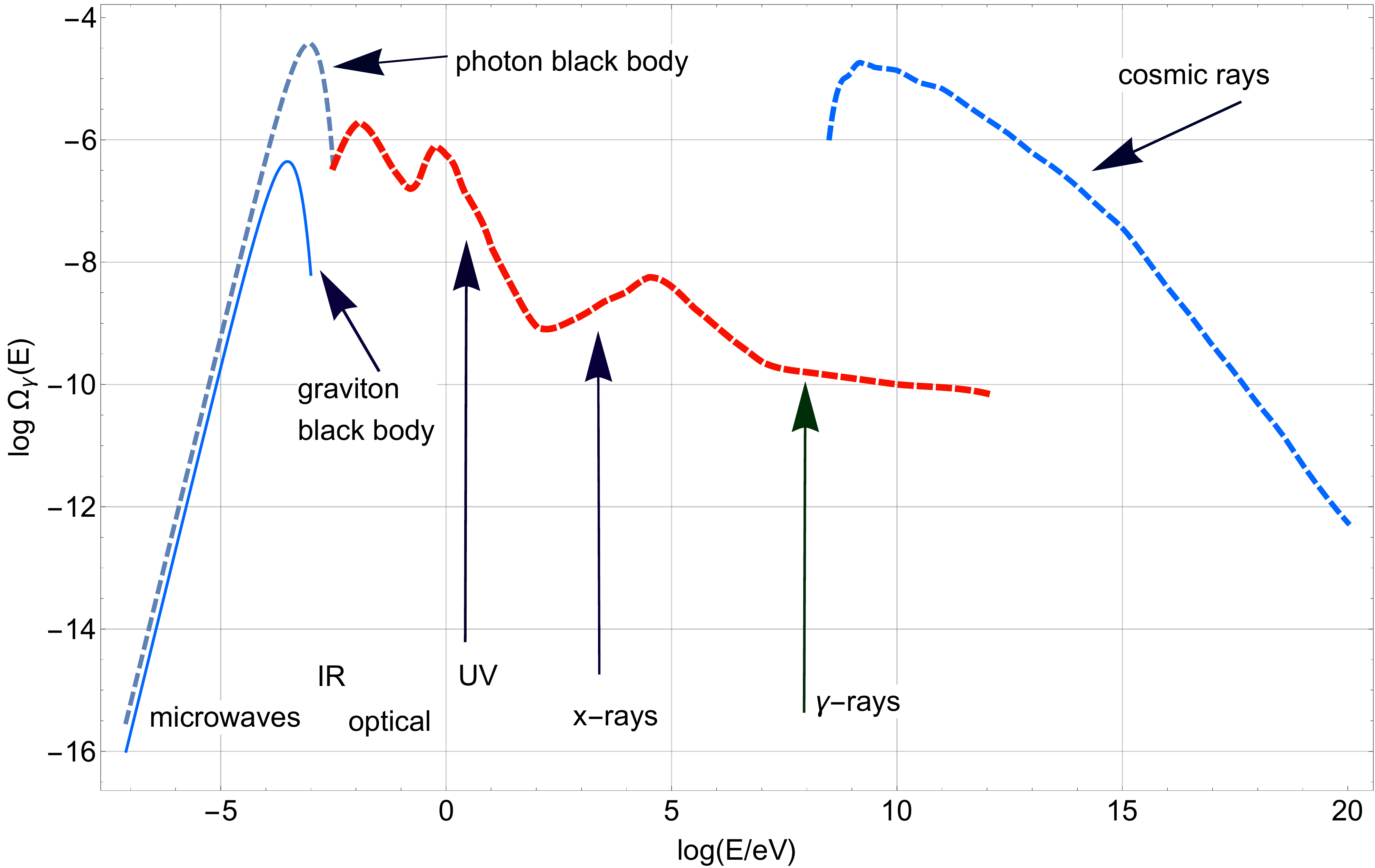}
\caption[a]{The  (extragalactic) electromagnetic emissions are compared with the thermal gravitons and the cosmic rays. On the vertical axis the logarithm (to base $10$)  of the emitted energy density is reported in units of $\rho_{crit}$. The cosmic ray spectrum is also included for comparison and in the same units used to describe the electromagnetic contribution. Unlike Fig. \ref{SEC1FIG1} we are plotting here $\Omega_{\gamma}(E)$ (and not $h_{0}^2 \Omega_{\gamma}(E)$). The value selected for $h_{0}$ is $0.67$.}
\label{SEC1FIG2} 
\end{figure}

The relative magnitude of the gravitational and electromagnetic signals can be appreciated 
by comparing the energy density of the diffuse electromagnetic emissions of cosmological origin with the spectral energy density of the relic gravitons. In Fig. \ref{SEC1FIG1} the energy densities of the  CMB photons and of cosmic gravitons are compared on the same scale. In Fig.  \ref{SEC1FIG2} the thermal gravitons are also compared with the other diffuse electromagnetic emissions of cosmological origin, including the microwave photons. The cartoons of Figs.  \ref{SEC1FIG1} and \ref{SEC1FIG2} concisely illustrate the orders of magnitude involved in the problem, the typical frequencies and the approximate sensitivities of the major planned detectors which will be discussed in more depth later on.
 The temperature of the graviton black-body depends on the number of relativistic  degrees 
of freedom of the plasma.  In the hypothesis the number of spin degrees of freedom 
is the one counted in the standard model of particle interactions the temperature of the 
black body would be $0.9$ K to be compared with $2.725$ K which is the result for the photons.

Since the characteristic energy scale provided by the $\Lambda$CDM paradigm is the 
critical energy density, it is instructive to measure the energy density of cosmic
gravitons and of cosmic photons in units of $\rho_{crit}$. The energy density of CMB 
photons per logarithmic interval of (angular) frequency is:
\begin{equation}
h_{0}^2 \Omega_{\gamma} (\nu, \tau_{0}) = \frac{h_{0}^2}{\rho_{crit}} \frac{d\rho_{\gamma}}{d\ln{k}}= \frac{15}{\pi^4} h_{0}^2 \Omega_{\gamma 0}\, F(k/T_{\gamma\,0}), \qquad 
h_{0}^2 \Omega_{\gamma 0} = 2.47 \times 10^{-5},
\label{MAGN1}
\end{equation}
where $F(x)=  x^{4}/(e^{x} -1)$ and  $x = k/T_{\gamma 0}$; note that
$\omega = k= 2 \pi \nu$ in the natural units is employed throughout the present discussion (for more details see also Tab. \ref{SEC1TABLE4} and comments thereafter). While $h_{0}^2 \Omega_{\gamma}(\nu, \tau_{0})$ 
does depend on the frequency, $h_{0}^2 \Omega_{\gamma 0}$ is frequency-independent and it defines 
the constant normalization of the spectral energy density. Note that $F(x)$ is maximized\footnote{Similarly the maximum of the brightness of the microwave background spectrum follows from the extremum of $B(x) = x^{3}/(e^{x} -1)$ corresponding  to $x_{max}^{(b)} = 2.821$. While the maximal {\em frequency} of the spectral energy density, thanks to the value of  $x_{max}^{(\rho)}$, is ${\mathcal O}(226.6)$ GHz [see Eq. (\ref{MAGN3})] the maximum of the brightness, thanks to the value of $x_{max}^{(b)}$, is ${\mathcal O}(160.3)$ GHz.} for $x_{max}^{(\rho)} = 3.920$ and this value also determines the maximum 
of the spectral energy density. In the rightmost part of Fig. \ref{SEC1FIG1} the spectrum of the microwave photons given of Eq. (\ref{MAGN1}) is illustrated with a dashed curve; the black-body spectrum of thermal gravitons in critical units
\begin{equation}
h_{0}^2 \Omega_{gw} (\nu, \tau_{0}) = \frac{h_{0}^2}{\rho_{crit}} \frac{d\rho_{gw}}{d\ln{k}}= \frac{15}{\pi^4} \,\,h_{0}^2 \Omega_{\gamma 0}\,\, \biggl(\frac{T_{g\, 0}}{T_{\gamma\, 0}}\biggr)^4 \,  F(k/T_{g\,0}),
\label{MAGN2}
\end{equation}
 is instead reported with a thin line.
Assuming an adiabatic evolution throughout the early stages of the primordial plasma the graviton temperature is related to the photon temperature as 
\begin{equation}
T_{g}(t_{0}) = T_{\gamma\, 0} \biggl(\frac{g_{s}(t_{0})}{g_{s}(t_{i})}\biggr)^{1/3}= 0.9 \biggl(\frac{T_{\gamma\, 0}}{2.725 \, \mathrm{K}}\biggr) \, \biggl(\frac{g_{s}(t_{0})}{3.91}\biggr)^{1/3} \biggl(\frac{g_{s}(t_{i})}{106.73}\biggr)^{-1/3}\, \, \mathrm{K}. 
\label{MAGN2a}
\end{equation}
According to Eq. (\ref{MAGN2a}) the temperature of the gravitons is $0.9$ K for $g_{s}(t_{i}) =106.75$ (when all the species of the standard model of particle interactions are in thermal equilibrium at $t_{i}$) and $g_{s}(t_{0}) = 3.91$ in the case of three massless neutrino species. If the pivotal model of particle interactions is different the total number of relativistic degrees of freedom can only increase so that  
 $T_{g\,0} \leq 0.9$ K. This means, in particular, that $T_{g\, 0} < T_{\gamma\, 0}$  and the graviton black-body is suppressed in comparison with the photon black-body by a factor  $(T_{g\, 0}/T_{\gamma\, 0})^4$.  This 
 aspect is emphasized both in Fig. \ref{SEC1FIG1} and in Fig. \ref{SEC1FIG2}: the two rightmost curves of Fig. \ref{SEC1FIG1} 
 denote the photon black-body (dashed line) and the graviton black-body (full thin line). The same portion of the spectrum 
 is illustrated in the leftmost region of Fig. \ref{SEC1FIG2} where the graviton black-body (always with the full thin line)
 undershoots the photon black-body. In Fig. \ref{SEC1FIG2}, on the horizontal axis, we report the energy while 
 in Fig. \ref{SEC1FIG1} we plot the comoving frequency: this is the reason why the position of the black-bodies is inverted.
 
Before plunging into the description of Figs. \ref{SEC1FIG1} and \ref{SEC1FIG2} we note that, according to Wien's law, 
the maximal frequencies for photons and gravitons are therefore given by:
\begin{equation}
\nu_{max\, \gamma}  \,\,= \,\, 226.6 \biggl(\frac{T_{\gamma\, 0}}{2.725\, \mathrm{K}}\biggr) \, \mathrm{GHz}, \qquad 
\nu_{max\, g}  \,\,= \,\, 73.50 \biggl(\frac{T_{g\, 0}}{0.9 \, \mathrm{K}}\biggr) \, \mathrm{GHz},
\label{MAGN3}
\end{equation}
where, as in Eq. \ref{MAGN2a}, we used the most accurate determination of the photon temperature $T_{\gamma 0} = (2.72548 \pm 0.00057) \, \mathrm{K}$ provided by the COBE satellite mission\footnote{It is remarkable that the COBE data 
still provide the most accurate determination of the CMB temperature and of the potential black-body distortions 
associated with it.} \cite{CMB5}. When evaluated at their respective maxima the energy densities of the cosmic photons and gravitons are\footnote{It is relevant to mention that Eq. (\ref{MAGN4})
does not depend on the specific value of $h_{0}$: since $h_{0}^2$ appears in the critical energy density $h_{0}^2 \Omega_{gw} (\nu , \tau_{0})$ {\rm is independent on the particular value of $h_{0}$}. }:
\begin{equation}
h_{0}^2 \Omega_{\gamma} (\nu_{max\, \gamma} , \tau_{0}) = 1.818 \times 10^{-5}  \biggl(\frac{T_{\gamma\, 0}}{2.725\, \mathrm{K}}\biggr)^{4},\qquad 
h_{0}^2 \Omega_{gw} (\nu_{max\, g} , \tau_{0}) = 2.161 \times 10^{-7}  \biggl(\frac{T_{g\, 0}}{0.9\, \mathrm{K}}\biggr)^{4}.
\label{MAGN4}
\end{equation}
To obtain Eqs. (\ref{MAGN3}) and (\ref{MAGN4}) the easiest procedure is to observe that $F[x_{max}^{(\rho)}]= 4.779$
for $x_{max}^{(\rho)} = 3.920$. The frequencies of the maxima given in Eq. (\ref{MAGN3}) are 
comparable (within one order of magnitude) but the frequency of the gravitons 
is always smaller than the frequency of the photons. For this reason we also 
have that while for photons $h_{0}^2 \Omega_{\gamma} (\nu , \tau_{0}) = {\mathcal O}(10^{-5})$, for 
gravitons $h_{0}^2 \Omega_{gw} (\nu , \tau_{0}) = {\mathcal O}(10^{-7})< h_{0}^2 \Omega_{\gamma} (\nu , \tau_{0})$. 
In Fig. \ref{SEC1FIG1}  the relic graviton backgrounds are illustrated in the case $r_{T}= 0.07$ and for the fiducial set of parameters given in Eq. (\ref{PARSET}).  In the same plot the equality, the BBN and the electroweak frequencies (defined in Eqs. (\ref{FF3}), (\ref{FF4}) and (\ref{FF5}) respectively) are illustrated with the arrows.  In the conventional lore the flat branch of the spectrum comes from the quantum fluctuations of the gravitational field 
amplified during a stage of inflationary expansion. The low-frequency branch (i.e. for $\nu < \nu_{eq}$) 
is associated with the wavelengths that have been further amplified all along the dust-dominated phase.
 The concordance spectrum of the relic gravitons illustrated in Fig. \ref{SEC1FIG1}  is roughly captured by the 
following semi-analytic parametrization:
\begin{equation}
h_{0}^2 \Omega_{gw}(\nu,\tau_{0}) = {\mathcal N}_{\rho}  \,T^2_{\rho}(\nu/\nu_{eq})\, r_{T}\, \biggl(\frac{\nu}{\nu_{\mathrm{p}}}\biggr)^{n_{T}} e^{- 2 \beta \frac{\nu}{\nu_{max}}}, \qquad
{\mathcal N}_{\rho} = 4.165 \times 10^{-15} \biggl(\frac{h_{0}^2 \Omega_{R0}}{4.15\times 10^{-5}}\biggr).
\label{MAGN5}
\end{equation}
The normalization is controlled by the tensor-to-scalar ratio $r_{T}$ and the transfer function of the energy density $T_{\rho}(\nu/\nu_{eq})$ (thoroughly discussed in section \ref{sec4}) 
accounts for the transition to the dust-dominated epoch. On a general ground, recalling the definition of Eq. (\ref{FF2}), 
 $T^2_{\rho} \to 1$ for $\nu \gg \nu_{eq}$ while
$T^2_{\rho} \to (\nu/\nu_{eq})^{-2}$ for $\nu \ll \nu_{eq}$.  
The parameter $\beta= {\mathcal O}(1)$ appearing in Eq. (\ref{MAGN5}) depends upon the width of the transition between the inflationary phase and the subsequent radiation dominated phase. The parametrization of Eq. (\ref{MAGN5}) neglects various corrections (which are however considered in Fig.  \ref{SEC1FIG1}) such as 
the damping effect associated with the neutrinos, the evolution of the number of relativistic species 
and the contribution of dark energy. All in all we can say that the 
spectral energy density corresponding to thermal gravitons (and thermal photons) 
exceeds the quantum contribution by $10$ or even  $12$ orders of magnitude.
However while the thermal photons and the thermal gravitons are peaked in the 
region of $100$ GHz the quantum spectrum may extend over a much wider 
range of frequencies.

In Fig. \ref{SEC1FIG1} we also illustrated, in a preliminary perspective, the approximate sensitivities 
of the pulsar timing array at the approximate frequency of Eq. (\ref{FF4a}); this range roughly corresponds 
to the inverse of the observation time along which the pulsars timing has been monitored \cite{PUL1,PUL2}.
A more thorough discussion of this theme can be found in section \ref{sec6}.
In the central part of Fig. \ref{SEC1FIG1} we also indicated {\em the minimal detectable spectral energy density 
of diffuse backgrounds of relic gravitons}  in the case of the advanced LIGO/Virgo detectors \cite{LV1,LV2} 
and for the hoped space-borne interferometers \cite{LISA,LISAa}, BBO (Big Bang Observer) \cite{BBO}, and DECIGO (Deci-hertz Interferometer Gravitational Wave Observatory) \cite{DECIGO1}.  The shapes 
of Fig. \ref{SEC1FIG1} are purely illustrative since the minimal detectable spectral energy density 
will ultimately depend also on the slope of $h_{0}^2 \Omega_{gw}$ in the operating window 
of the various instruments. While a more complete 
discussion of the interplay between the diffuse backgrounds of gravitational radiation and the 
wide-band interferometers will be presented section \ref{sec8},  the approximate 
sensitivities of Fig. \ref{SEC1FIG1} shall be assumed throughout the whole discussion.

Figure \ref{SEC1FIG2} finally illustrates the comparison between the relic gravitons and the 
diffuse electromagnetic emissions. While in Fig. \ref{SEC1FIG1} the common logarithm of the frequency (expressed in Hz) 
is reported on the horizontal axis, in Fig. \ref{SEC1FIG2} the various emissions are
illustrated as a function of the common logarithm of the energy (expressed in eV). 
Barring for the different units used on the horizontal axis, the leftmost portion of Fig. \ref{SEC1FIG2} 
coincides, in practice, with the rightmost part of Fig. \ref{SEC1FIG1}. 
In Fig. \ref{SEC1FIG2} $\Omega_{\gamma}(E)$ is maximal for typical 
energies ${\mathcal O}(10^{-3})\,\mathrm{eV}$ corresponding 
to wavelengths  ${\mathcal O}(\mathrm{mm})$. In the optical and 
ultraviolet range of wavelengths the energy density drops almost by two orders of magnitude.
In the $x$-rays (i.e. for wavelengths $10^{-6} \mathrm{mm}<\lambda < 10^{-9} \mathrm{mm}$) 
the energy density of the emitted radiation drops of more than three  orders 
of magnitude in comparison with the maximum. The $x$-ray range 
corresponds to photon energies $E > \mathrm{keV}$.
In the $\gamma$-rays (i.e. $10^{-9} \mathrm{mm} < \lambda< 10^{-12} \mathrm{mm}$) 
the spectral amplitude is roughly $5$ orders of 
magnitude smaller than in the case of the mm maximum.
The range of $\gamma$-rays occurs for photon energies 
$E > \mathrm{GeV}$.  While the CMB represents $0.93$ of the 
extragalactic emission, the infra-red and visible part give, respectively, 
$0.05$ and $0.02$. The $x$-ray and $\gamma$-ray 
branches contribute, respectively, by $2.5\times 10^{-4}$ and $2.5\times 10^{-5}$. 
The CMB is therefore the 93 \% of the total extragalactic emission.
In Fig. \ref{SEC1FIG2} the approximate form of the spectrum 
of the cosmic rays  is also reported, for comparison. This inclusion is somehow arbitrary since the cosmic rays 
of moderate energy are known to come from within the galaxy. It is however useful to report 
also the energy spectrum of cosmic rays and to compare it, in the same units, with the energy spectrum 
of CMB photons\footnote{I am indebted with the late G. Cocconi for suggesting the comparison among the various (extra-galactic) electromagnetic emissions and the thermal gravitons in the critical units employed in Fig. \ref{SEC1FIG2}.}. The energy density of the cosmic rays is, roughly,  of the same order of  the energy density of the CMB. 
For energies smaller than $10^{15}$ eV the rate is approximately of 
one particle per $\mathrm{m^2}$ and per second.
For energies larger than $10^{15}$ eV the rate is approximately of 
one particle per $\mathrm{m^2}$ and per year.  The difference in these 
two rates corresponds to a slightly different spectral behaviour 
of the cosmic ray spectrum, the so-called knee.
Finally, for energies larger than $10^{18}$ eV, the rate of the so-called 
ultra-high-energy cosmic rays is even smaller and of the 
order of one particle per $\mathrm{km^2}$ and per 
year\footnote{ The sudden drop in the flux corresponds to another small change in the spectral behaviour, 
the so-called ankle. In the parametrization chosen in Fig. \ref{SEC1FIG2} the cosmic ray spectrum 
does not decrease as $E^{-3}$ but rather as $E^{-2}$. The rationale for this difference stems from the fact that we plot the energy density of cosmic rays per logarithmic interval of $E$. }.

\subsection{Relic gravitons and large-scale inhomogeneities}
\subsubsection{Quantum origin of cosmological inhomogeneities}
Prior to the formulation of conventional inflationary models, A. D. Sakharov  suggested that the complicated patterns observed 
 in the galaxy distributions could come from the zero-point fluctuations of quantum fields in
curved backgrounds \cite{SAK1}. This visionary conjecture also implied that the production of quanta with opposite momenta from 
the vacuum could be viewed, in classical terms, as a generation of standing waves (occasionally referred to as 
Sakharov oscillations). The emergence of standing waves has been also independently discussed in the classic paper of Peebles 
and Yu \cite{SAK2} (see also \cite{SAK3}) on the original formulation of the adiabatic paradigm 
which is today at the heart of the concordance scenario. One of the first thorough studies on the production of fields of various spin in isotropic 
backgrounds was conducted by Parker in a well known series of papers \cite{PAR1,PAR2,PAR3,PAR4}
but these pivotal analyses did not include the spin-2 fields  since, for a long period, 
the evolution of the tensor modes of the geometry was believed to be invariant under the 
Weyl rescaling of the four-dimensional metric exactly as it happens in the case of the free electromagnetic fields and chiral fermions in general relativity.  Weak gravitational waves in an isotropic universe have been studied  by Lifshitz \cite{HIS1} in one of the first investigations on the evolution of inhomogeneities in cosmological backgrounds.

\subsubsection{Weyl invariance and relic gravitons}
According to the standard lore gravitational waves could not be amplified in (four-dimensional)
isotropic  background geometries as a consequence of their presumed Weyl invariance.
However, in the presence of highly anisotropic background evolutions
(as it happens, for instance, in Bianchi-type models \cite{ryan}) gravitational waves 
could experience some amplification. It is useful to mention, at this point, that
in those years anisotropic backgrounds also played a central role in the analyses of 
Lifshitz, Khalatnikov \cite{HIS1a,HIS1aa,HIS1b,HIS1bb} and Belinskii \cite{HIS1c}. 
In an attempt towards the construction of the general cosmological solution to the Einstein equation,
 these authors  analyzed the dynamical approach to the cosmological singularity in terms of an expansion
 in spatial gradients of the geometry \cite{HIS1a,HIS1aa,HIS1b,HIS1bb}.
It was demonstrated, in particular, that if all matter components have standard equations of state,
 a contracting universe is unstable against perturbative deviations from isotropy so that an initially imposed anisotropy may increase and bring the metric in a Kasner-like form. It was later argued \cite{HIS1c,HIS1d} that the asymptotic evolution of the metric near the singularity is described by a mixmaster- type behaviour \cite{HIS1e,HIS1f}.  Grishchuk was the first one to appreciate, contrary to the general wisdom, that weak gravitational waves could indeed be amplified  in a non-stationary isotropic universe to a greater degree 
 than indicated by the adiabatic law \cite{HIS5,HIS5a}. For this reason this process is sometimes 
 referred to as {\it super-adiabatic amplification}. Today we know that super-adiabatic amplification takes place 
 both in inflationary models as well as in other cosmological scenarios based on totally different premises:  
 the production of relic gravitons is a necessary consequence of any scenario where the space-time curvature evolves at early 
times. Since the waves obtained by super-adiabatic amplification are statistically independent  the relic gravitons indeed constitute a diffuse background of isotropic (but inhomogeneous) random fields.  If the evolution of the background deviates from the radiation-dominated case, gravitational waves can be amplified to various degrees and are described by the tensor analog of the Sakharov oscillations \cite{SAK1,SAK2,SAK3}.
\subsubsection{Inflation, concordance paradigm and beyond}
The pioneering analyses of Refs. \cite{HIS5,HIS5a}  suggested  various cosmological implications \cite{HIS5b,HIS6}. Ford and Parker \cite{HIS7,HIS8} (see also \cite{HIS9}) analyzed for the first time the effective action 
of relic gravitons.  Starobinsky \cite{HIS10} discussed the scale-invariance of the spectral energy density of the gravitons coming 
from a de Sitter stage of expansion. Shortly after Lukash investigated the production of phonons  
of an irrotational relativistic plasma evolving in an isotropic universe \cite{HIS10a} by using the method of the effective action of Ref. \cite{HIS7}.  The first versions of the conventional inflationary scenarios 
\cite{HIS10b,HIS10c,HIS10d,HIS10e} stimulated a more accurate estimate of the spectra of relic gravitational waves 
especially at low-frequencies \cite{HIS11}. Almost at the same time the effects of primordially produced gravitons 
upon the CMB anisotropies were estimated \cite{HIS12,HIS12a,HIS12aa}. 
The constraints on inflationary cosmologies became implicitly constraints on the produced gravitons 
in the low frequency branch \cite{HIS13,HIS14} and they are now customarily 
expressed as bounds on the tensor to scalar ratio $r_{T}$ (see also Tab. \ref{SEC1TABLE3}). 
The description of the relic gravitons with quantum optical techniques pioneered in Ref. \cite{HIS15} is today 
essential for a correct discussion of the correlation properties of the relic gravitons and for other 
applications \cite{HIS15b}. As long as the conventional inflationary phase is almost suddenly 
followed by the radiation epoch the spectral energy density of relic gravity waves 
is always red\footnote{Denoting by $\delta$ the spectral slope 
[i.e.  $h_{0}^2 \Omega_{gw}(\nu, \tau_{0}) \propto (\nu/\nu_{*})^{\delta}$] the graviton spectrum of the concordance 
scenario is typically {\em red} (i.e. slightly decreasing with $\delta \leq 0$). When $0 < \delta <1$ we shall talk about  {\em blue} slopes; when $\delta> 1$ the slopes are said to be {\em violet}. } \cite{HIS16,HIS16aa} but various other possibilities have been suggested through the years. In particular violet spectra may arise whenever the curvature scale increases 
for a portion of the evolution of the universe \cite{HIS16a} and, in this case, 
the low-frequency constraints on the graviton spectrum become less relevant.
The same kind of evolution may arise in scalar-tensor theories of gravity 
inspired by the low-energy string effective action \cite{HIS18}. 
Gravitational wave spectra from generalized theories of gravity have been 
also analyzed in Refs. \cite{HIS19,HIS20} (in the case of scalar-tensor theories)
and in Ref. \cite{HIS20a} (in the case higher-curvature gravity theories). 
 It was then argued that a large class of string 
cosmology backgrounds may lead to a violet spectrum of relic gravitons 
strongly tilted towards high frequencies \cite{HIS20b}. Violet spectra may also arise 
in the context of ekpyrotic and cyclic models \cite{HIS20c,HIS20d}. The increase of the 
space-time curvature during a certain portion of the cosmological evolution is 
a sufficient condition for an increasing spectral energy density. This condition 
is however not necessary since the blue slopes may also arise when the post-inflationary 
phase is dominated by effective sources whose equation of state is stiffer than radiation \cite{HIS23,HIS24a,HIS24b,HIS24}. 
In different frequency domains the spectra of the relic gravitons can be used not only to infer the early evolution of the space-time curvature, but also to test the evolution of the plasma for temperatures 
in the MeV range, i.e. prior to the formation of light nuclei. 
In the forthcoming sections the results swiftly recalled in this quasi-historic introduction 
will be often mentioned even if the attention will be mostly focussed on the themes 
that emerged {\em after} the formulation of the concordance scenario. There is however 
no doubt that the original approaches mentioned here are instrumental in understanding 
the most recent developments.
\begin{table}
\begin{center}
\caption{Definition of some of the most frequent symbols.}
\vskip 0.5truecm
\begin{tabular}{| c | l | c | }
\hline
\, Symbol \quad  & 
\quad\qquad\qquad Definition 
      \\ \hline
$ M_{P} = 1.22 \times 10^{19} \, \mathrm{GeV}$ & Planck mass
       \\ \hline 
$ \overline{M}_{P} = M_{P}/\sqrt{ 8 \pi}$ & reduced Planck mass
        \\ \hline
$\ell_{P} = \sqrt{8 \pi G} = \sqrt{8 \pi}/M_{P} = 1/\overline{M}_{P}$ & Planck length 
         \\ \hline
$\nu, \quad k$ & comoving frequency ($\nu$) and wavenumber ($k$)
         \\ \hline
$\nu_{p},\quad k_{p}$ &  pivot frequency and pivot wavenumber
        \\ \hline   
$\nu_{eq}$ & equality frequency
        \\ \hline
$\nu_{bbn}$, $\nu_{ew}$ & nucleosynthesis and electroweak frequencies
       \\ \hline
$\ln{}$, $\log{}$ & Neperian (i.e. natural) and common logarithms
       \\ \hline      
$\tau,\,\, t$ & conformal ($\tau$) and cosmic ($t$) time coordinates      
       \\ \hline   
$^{\cdot}= \partial/\partial t$ & derivation with respect to the cosmic time coordinate $t$
       \\ \hline
$^{\prime} = \partial/\partial \tau$  & derivation with respect to the conformal time coordinate $\tau$
        \\ \hline
$T_{\mu\mu},\,\,\, {\mathcal F}_{\mu\nu},\,\,\,{\mathcal L}_{\mu\nu},\,\,\,{\mathcal B}_{\mu\nu}$ & energy-momentum tensors
       \\ \hline
$G_{\mu\nu}, \qquad g_{\mu\nu}$ & metric tensors
       \\ \hline
$\eta_{\mu\nu}$ & Minkowski metric tensor with signature mostly minus
       \\ \hline
$R_{\alpha\mu\beta\nu},\,\,\, R_{\mu\nu} = g^{\alpha\beta}R_{\alpha\mu\beta\nu} $ & Riemann and Ricci tensors
       \\ \hline
$ \nabla_{\alpha} A_{\mu} = A_{\mu\, ; \alpha}$, $\nabla_{\beta} \nabla_{\alpha} A_{\mu}  
=  A_{\mu\, ; \alpha\, \beta}$& covariant derivatives 
  \\ \hline
$a(\tau)$ & scale factor of a conformally flat metric
      \\ \hline 
$\partial_{i} = \partial/\partial x^{i} = \nabla_{i}, \quad \nabla^2 = \delta^{ij}\nabla_{i} \nabla_{j}$ & spatial gradients and Laplacian in a conformally flat metric
      \\ \hline
   $g^{\alpha\beta} \nabla_{\alpha} \nabla_{\beta}= \Box$ & covariant d'Alembertian   
      \\ \hline          
${\mathcal H} = a^{\prime}/ a=  a H, \quad H = \dot{a}/a$ & expansion rates 
      \\ \hline
$\rho_{X}, \qquad p_{X}$ & energy density and pressure of a given species $X$       
        \\ \hline
$\rho_{crit} = 3 H^2 \,\overline{M}_{P}^2$ & critical energy density
       \\ \hline
$\Omega_{X} = \rho_{X}/\rho_{crit}$ & critical fraction of a given species $X$
       \\ \hline
$\Omega_{X\, 0}$ & present critical fraction of a given species $X$       
        \\ \hline
$n_{X}$ & concentration of a given species $X$   
         \\ \hline
$H_{0}$ & present value of the Hubble rate
         \\ \hline
 $h_{ij}$ & amplitude  of the tensor modes of the geometry
         \\ \hline
 $\phi$, $\psi$, $E$ and $B$ & amplitudes of the scalar modes of the geometry
         \\ \hline 
 ${\mathcal R}$ & amplitude of the gauge-invariant curvature inhomogeneities         
         \\ \hline
  $W_{i}$ and $Q_{i}$ & amplitudes of the vector modes of the geometry      
         \\ \hline
 $ P_{Y}(k,\tau)$ & power spectrum of a given random field $Y$
       \\ \hline
${\mathcal A}_{{\mathcal R}}$ & amplitude of the power spectrum of curvature inhomogeneities
        \\ \hline
${\mathcal A}_{T}$ & amplitude of the power spectrum of tensor inhomogeneities
        \\ \hline
 $r_{T}$ & tensor to scalar ratio at the pivot frequency 
        \\ \hline
$ P_{T}(k,\tau)$ & tensor power spectrum 
          \\ \hline
$n_{s}$ and $n_{T}$ & scalar and tensor spectral index          
         \\ \hline
$h_{c}(\nu)$ &  chirp  amplitude 
         \\ \hline
$ S_{h}(\nu)$  & spectral amplitude     
        \\ \hline
$ \Omega_{gw}(\nu,\tau)$ &  spectral  energy density of the relic gravitons        
       \\ \hline
$\hat{a}_{\vec{k},\,\alpha}^{\dagger},\,\,\hat{a}_{\vec{k},\,\alpha}$ & 
creation and annihilation operators for a given polarization
\\ \hline
$f_{k}(\tau),\,\,g_{k}(\tau)$ & mode functions
\\ \hline
$F_{k}(\tau),\,\,G_{k}(\tau)$ & rescaled mode functions
\\ \hline
\end{tabular}
\label{SEC1TABLE4}
\end{center}
\end{table}
\subsection{Notations, units and summary} 
The natural system of units $\hbar= c= k_{B}= 1$ (where $k_{B}$ is the Boltzmann constant) will be implicitly  used in most of the estimates. In this system  $\hbar c= 197.327 
\, \mathrm{MeV}\, \mathrm{fm}$ equals $1$. The relation between Kelvin degrees and eV 
is notoriously given by $\mathrm{K} = 8.617\times 10^{-5} \,\mathrm{eV}$. Moreover, since 
$c= 2.99792\times 10^{10} \,\,\mathrm{cm}/\mathrm{sec}$, in the natural system of units $1 \, \mathrm{GeV} = 1.5507\times10^{24}\,\, \mathrm{Hz}$. The {\em standard metric prefixes will be used to denote the multiples (or the fractions) of a given unit}. When needed the astronomical length-scales will be often expressed in parsecs and their multiples:  (e.g. $1\, \mathrm{kpc} = 3.085\times 10^{21} \mathrm{cm} $).  The {\em natural gravitational units} where $\overline{M}_{P} =1$ will also be occasionally employed. The signature of the metric will be taken to be mostly minus; in the case of the Minkowski 
metric this choice implies $\eta_{\mu\nu} = {\mathrm{diag}}(+1, \, -1, \, -1\, -1)$. The Riemann tensor is defined as $R^{\beta}_{\,\,\mu\alpha\nu} = \partial_{\alpha} \Gamma^{\,\,\,\,\,\,\beta}_{\mu\nu} - \partial_{\nu}  \Gamma^{\,\,\,\,\,\,\beta}_{\alpha\mu} +  \Gamma^{\,\,\,\,\,\,\gamma}_{\mu\nu} \, \Gamma^{\,\,\,\,\,\,\beta}_{\gamma\alpha} - \Gamma^{\,\,\,\,\,\,\gamma}_{\mu\alpha} \, \Gamma^{\,\,\,\,\,\,\beta}_{\gamma\nu}$. The Ricci tensor is defined from the 
contraction of the first and third indices of the Riemann tensor, i.e. $R_{\mu\nu} = R^{\alpha}_{\,\,\,\mu\alpha\nu}$. Even if the notations will
be progressively introduced, the definitions of some of the most used symbols appearing in the script are collected Tab. \ref{SEC1TABLE4}. The layout of the present article is in short the following.  Section \ref{sec2} is devoted to the evolution of the tensor modes in curved backgrounds. The effective actions 
and the energy densities of gravitational waves in cosmological backgrounds are analyzed in section \ref{sec3} where the mutual relations among the various parametrizations of the cosmic graviton spectra are also scrutinized. Section \ref{sec4} discusses the spectra of the relic gravitons in the case of the minimal concordance scenario.  In  section \ref{sec5}   the low-frequency band of the spectrum is presented by addressing the interplay between the stochastic backgrounds of relic gravitons and CMB physics. In section \ref{sec6}  the intermediate frequency band will be analyzed with particular attention to the various phenomenological bounds. Section \ref{sec7} is devoted to the high-frequency domain where wide-band detectors are now operating. Explicit references to the detection strategies and to the sensitivities of the various instruments have been made 
throughout the article and,  to avoid excessive digressions, section \ref{sec8} is focussed on the
interplay between the diffuse backgrounds of relic gravitons and the current (or planned) detectors of gravitational radiation
with particular attention to the recent bounds provided by wide-band interferometers. 
The concluding considerations and some perspectives for the future are finally collected in section \ref{sec9}.

\newpage 

\renewcommand{\theequation}{2.\arabic{equation}}
\setcounter{equation}{0}
\section{The tensor modes of the geometry}
\label{sec2}
In four-dimensional curved backgrounds the relativistic fluctuations of the geometry are described by a symmetric rank-two tensor containing overall $10$ independent components. This covariant approach has been successfully applied in the pioneering studies of relic gravitons  \cite{HIS1,HIS1d,HIS5,HIS5a}. In cosmological space-times the $10$ independent components of the perturbed metric can be preliminarily separated in scalar, vector and tensor modes depending on their transformation properties with respect to the three-dimensional rotations. In this non-covariant strategy the two degrees of freedom associated with the tensor modes are automatically invariant under infinitesimal coordinate transformations while the scalar and vector fluctuations of the metric are gauge-dependent. The scalar and vector inhomogeneities  admit however 
various gauge-invariant descriptions \cite{bard1} (see also \cite{book1,book2,book3} for some topical monographs).  The same kind of non-covariant decomposition simplifies the analysis of the $6$ polarizations of gravitational waves in non-Einsteinian theories of gravity (e.g. scalar-tensor theories or scalar-vector-tensor theories) that will be briefly addressed at the end of this section.

\subsection{The tensor modes in flat space-time}
When the Minkowski space-time is slightly curved because of the presence of a tensor fluctuation, the metric $g_{\mu\nu}$ can be written as the sum the flat Minkowski metric (i.e. $\eta_{\mu\nu}$) supplemented by the corresponding tensor fluctuation $h_{\mu\nu}$, i.e.
\begin{equation}
g_{\mu\nu}(x) = \eta_{\mu\nu} + h_{\mu\nu}(x), \qquad g^{\mu\nu}(x) = \eta^{\mu\nu} - h^{\mu\nu}(x),\qquad  |h_{\mu\nu}(x)| \ll 1,
\label{MM1}
\end{equation}
where $x =( \vec{x},\, \tau)$ denotes the space-time point which will be omitted in the subsequent equations if not strictly necessary.
Equation (\ref{MM1}) can be inserted into the definition of the Christoffel connections so that, to first-order in $h_{\mu\nu}$, 
we have that 
\begin{equation}
\delta^{(1)} \Gamma^{\alpha}_{\mu\nu} = \frac{1}{2}\biggl( - \partial^{\alpha} h_{\mu\nu} + \partial_{\nu} h^{\alpha}_{\mu} + \partial_{\mu} h^{\alpha}_{\nu}\biggr).
\label{CC}
\end{equation}
Within the same strategy, the linearized expressions of the Ricci tensor and of the 
Ricci scalar are:
\begin{equation}
\delta^{(1)} R_{\mu\nu} = \frac{1}{2} \biggl( - \Box h_{\mu\nu} + \partial_{\nu} \partial_{\alpha} h^{\alpha}_{\mu} + \partial_{\mu} \partial_{\alpha} h^{\alpha}_{\nu} 
- \partial_{\mu}\partial_{\nu} h\, \biggr),\qquad 
 \delta^{(1)} R = - \Box h + \partial_{\alpha}\partial_{\beta} h^{\alpha\beta},
 \label{CC2CC3}
 \end{equation}
 where $ h = \eta^{\alpha\beta} h_{\alpha\beta} = h_{\alpha}^{\,\,\alpha}$
 and the symbol $\Box$ denotes the standard d'Alembertian. 
Since the background values of the Ricci tensor and of the Ricci scalar vanish in flat space-time, 
from Eqs. (\ref{CC2CC3}) the linearized Einstein equations simply read
\begin{equation}
\Box \psi_{\mu\nu} + \eta_{\mu\nu} \,\partial_{\alpha} \partial_{\beta} \psi^{\alpha\beta} - \partial_{\alpha} \partial_{\nu} \psi^{\alpha}_{\,\,\,\mu} - 
\partial_{\alpha} \partial_{\mu}\psi^{\alpha}_{\,\,\,\,\nu} = - 2 \ell_{P}^2 T_{\mu\nu},
\label{FT}
\end{equation}
where $h_{\mu\nu} =\psi_{\mu\nu} + \eta_{\mu\nu}\,h/2$ and $T_{\mu\nu}$ is  the 
energy-momentum tensor of the sources. For infinitesimal coordinate transformations of the form $x^{\mu} \to \widetilde{\,x\,}^{\mu} = x^{\mu} + \epsilon^{\mu}$ the field $h_{\mu\nu}$ changes as:
\begin{equation}
h_{\mu\nu} \to \widetilde{\,h\,}_{\mu\nu} = h_{\mu\nu} - \partial_{\mu} \epsilon_{\nu} - \partial_{\nu} \epsilon_{\mu}.
\label{GT}
\end{equation}
Now, if $\partial_{\alpha} \psi^{\alpha\beta} =0$ it is immediate to conclude 
from Eq. (\ref{FT}) that the field equations simplify as $\Box \psi_{\mu\nu} = - 2 \ell_{P}^2 T_{\mu\nu}$. Conversely, 
whenever $\partial_{\alpha} \psi^{\alpha\beta} \neq 0$ it is always possible to change the coordinate system 
as in Eq. (\ref{GT}) by selecting a new gauge where $\widetilde{\partial_{\alpha}\,\psi\,}^{\alpha\beta} =0$. While $h_{\mu\nu}$ 
follows Eq. (\ref{GT}), for an infinitesimal coordinate transformation we have 
that $\partial_{\nu} \psi^{\,\,\,\nu}_{\mu}$ changes instead as:
\begin{equation}
\partial_{\nu} \psi^{\,\,\,\nu}_{\mu} \to \widetilde{\partial_{\nu}\,\psi\,}^{\,\,\,\nu}_{\mu}= \partial_{\nu} \psi^{\,\,\,\nu}_{\mu}  -\Box \epsilon_{\mu}.
\label{HT}
\end{equation}
Equation (\ref{HT}) implies that if $\partial_{\nu} \psi^{\,\,\,\nu}_{\mu}\neq 0$ in a coordinate system we can always find a different gauge where $\widetilde{\partial_{\nu} \psi\,}^{\,\,\,\nu}_{\mu}=0$ by requiring $\Box \epsilon_{\mu} = \partial_{\nu} \psi^{\,\,\,\nu}_{\mu} $.
The condition $\partial_{\alpha} \psi^{\alpha\beta} =0$ does not univocally select the coordinate system since we can always perform a new transformation for which $\epsilon^{\mu}$ satisfies $\Box \epsilon^{\mu}=0$. However, as long as  $\Box \epsilon^{\mu}=0$ the gauge condition $\partial_{\alpha} \psi^{\alpha\beta}=0$ is preserved by any further gauge transformation\footnote{Note that the so-called de Donder gauge requires  $g^{\mu\nu} \Gamma_{\mu\nu}^{\alpha}=0$; by perturbing this condition  we recover the gauge 
condition $\partial_{\alpha} \psi^{\alpha\beta}=0$ thanks to the first-order form of the Christoffel connections given in Eq. (\ref{CC}).}.
 
\subsection{The tensor modes in curved space-times}
In four-dimensional curved space-times the metric fluctuations can be represented as:
\begin{equation}
g_{\mu\nu}(x) = \overline{g}_{\mu\nu}(x) + \delta^{(1)} g_{\mu\nu}(x),\qquad \delta^{(1)}g_{\mu\nu}(x) = \overline{h}_{\mu\nu}(x),\qquad   |\overline{h}_{\mu\nu}(x)| \ll 1,
\label{CUR0a}
\end{equation}
where $\overline{g}_{\mu\nu}(x)$ is the background metric and  $\delta^{(1)}$ denotes the (first-order) covariant  fluctuation of the corresponding tensor. As before we shall omit the dependence on the space-time point unless strictly necessary. To avoid possible notational conflicts with the non-covariant decomposition of the metric fluctuations discussed later on, the covariant perturbation has been defined with a bar (i.e. $\overline{h}_{\mu\nu}$).  The fluctuations of the inverse metric and of $\sqrt{-g}$ are:
\begin{eqnarray}
\delta^{(1)} \,g^{\mu\nu} &=&  - \overline{h}^{\,\,\mu\nu}, \qquad \delta^{(2)} \,g^{\mu\nu}  = \overline{h}^{\,\,\nu\alpha} \,\, \overline{h}_{\alpha}^{\,\,\,\,\mu},\qquad \delta^{(1)}\sqrt{- g} = \frac{1}{2} \sqrt{- \overline{g}} \,\,\overline{h},
\label{CUR0b}\\
\delta^{(2)}\sqrt{- g} &=&
\frac{\sqrt{- \overline{g}}}{8} \biggl[ \overline{h}^2  - 2 \,\overline{g}^{\mu\alpha} \,\overline{g}^{\nu\beta} \,\overline{h}_{\alpha\nu} \, \overline{h}_{\beta\mu}\biggr],\qquad \overline{h} = \overline{g}^{\alpha\beta} \, \overline{h}_{\alpha\beta},
\label{CUR0c}
\end{eqnarray}
where $\delta^{(2)}$ denotes the second-order  
fluctuations of the corresponding quantity. Using  Eqs. (\ref{CUR0b}) and (\ref{CUR0c}) 
the first- and second-order expressions of the Christoffel connections are easily obtained so that the Ricci tensor can be expressed, order by order, as: 
\begin{eqnarray}
\overline{R}_{\mu\nu} &=& \partial_{\alpha} \overline{\Gamma}_{\mu\nu}^{\,\,\,\,\,\,\alpha} - \partial_{\nu} \overline{\Gamma}_{\mu\alpha}^{\,\,\,\,\,\,\alpha} + \overline{\Gamma}_{\mu\nu}^{\,\,\,\,\,\,\alpha} \, \overline{\Gamma}_{\alpha\beta}^{\,\,\,\,\,\,\beta} - 
\overline{\Gamma}_{\nu\alpha}^{\,\,\,\,\,\,\beta} \, \overline{\Gamma}_{\beta\mu}^{\,\,\,\,\,\,\alpha},
\label{CUR0d1}\\
\delta^{(1)} R_{\mu\nu} &=& \frac{1}{2} \biggl( - \overline{h}_{;\mu\,;\nu} - \overline{h}_{\mu\nu\,;\alpha}^{\,\,\,\,\,;\alpha} + \overline{h}_{\alpha\mu\,;\nu}^{\,\,\,\,\,\,;\alpha} + \overline{h}_{\alpha\nu;\, \mu}^{\,\,\,\,\,;\alpha}\biggr),
\label{CUR0d}\\
\delta^{(2)} R_{\mu\nu} &=& \frac{1}{2} \biggl[ \frac{1}{2} \overline{h}_{\alpha\beta\,;\mu}\,\, \overline{h}^{\alpha\beta}_{;\nu} + \overline{h}^{\alpha\beta}\biggl(  \overline{h}_{\alpha\beta\,;\mu;\nu} + 
\overline{h}_{\mu\nu\, ; \alpha;\beta} - \overline{h}_{\alpha\mu\,;\nu;\beta} - \overline{h}_{\alpha\nu\,;\mu;\beta}\biggr) 
\nonumber\\
&+& \overline{h}_{\nu}^{\alpha\,;\beta}\biggl( \overline{h}_{\alpha\mu;\beta} - \overline{h}_{\beta\mu;\alpha}\biggr) - \biggl(\overline{h}^{\,\,\alpha\beta}_{;\beta} - \frac{1}{2} \overline{h}^{\,\,;\alpha}\biggr)\biggl(\overline{h}_{\alpha\mu;\nu} + \overline{h}_{\alpha\nu;\mu} - \overline{h}_{\mu\nu;\alpha}\biggr) \biggr],
\label{CUR0e}
\end{eqnarray}
where $\overline{\Gamma}_{\alpha\beta}^{\,\,\,\,\,\,\gamma}$ denotes the background values of the Christoffel connections
and the covariant derivatives (expressed through the usual shorthand notation) are defined in terms of the background 
metric and of the background connections. Note that when $\overline{g}_{\mu\nu} = \eta_{\mu\nu}$ and $\overline{h}_{\mu\nu} = h_{\mu\nu}$ the results 
of Minkowski space-time are easily recovered. From Eqs. (\ref{CUR0a}) and (\ref{CUR0d}) the perturbed Einstein's equations are:
\begin{equation}
\delta^{(1)} R_{\mu\nu} = \ell_{P}^2 \biggl[ \delta^{(1)} T_{\mu\nu} - \frac{\delta^{(1)}T}{2} \overline{g}_{\mu\nu} - \frac{\overline{T}}{2} \, \overline{h}_{\mu\nu} \biggr],
\label{CUR0f}
\end{equation}
where $T_{\mu\nu}$ denotes a covariantly conserved but otherwise generic energy-momentum 
tensor\footnote{As usual  $T = T_{\mu}^{\,\,\,\, \mu}$ denotes the trace and
$\overline{T} = \overline{T}_{\mu}^{\,\,\,\, \mu}$ is the corresponding background value.}. 
 Inserting Eq. (\ref{CUR0d})  into Eq. (\ref{CUR0f})  we obtain
\begin{equation}
\overline{\nabla}_{\alpha} \overline{\nabla}^{\alpha} \, \overline{h}_{\mu\nu} + \overline{\nabla}_{\mu} \overline{\nabla}_{\nu} \overline{h} - 
\overline{g}^{\alpha\beta} \biggl[ \overline{\nabla}_{\alpha} \overline{\nabla}_{\nu} \overline{h}_{\mu\beta}  + \overline{\nabla}_{\alpha} \overline{\nabla}_{\mu} \overline{h}_{\nu\beta} \biggr] =- 2 \ell_{P}^2 \biggl[ \delta^{(1)} T_{\mu\nu} - \frac{\delta^{(1)}T}{2} \overline{g}_{\mu\nu} - \frac{\overline{T}}{2} \overline{h}_{\mu\nu} \biggr].
\label{CUR0g}
\end{equation}
Equation (\ref{CUR0g}) can be further simplified by flipping the order of the covariant derivatives appearing inside the square brackets thanks to the well known identity valid for
 Riemannian and pseudo-Riemannian manifolds:
\begin{equation}
 \overline{h}_{\mu\nu;\alpha\,\, ; \beta} - \overline{h}_{\mu\nu\,\, ;\beta\,\, ; \alpha} = 
h_{\mu\lambda} \, \, \overline{R}^{\lambda}_{\,\,\,\,\,\,\nu\alpha\beta} + 
h_{\nu\lambda} \, \, \overline{R}^{\lambda}_{\,\,\,\,\,\,\mu\alpha\beta}.
\label{CUR30a}
\end{equation}
When the sources are absent Eq. (\ref{CUR0g}) becomes:
 \begin{equation}
 \overline{\nabla}_{\alpha} \overline{\nabla}^{\alpha}\psi_{\mu\nu} - 2 \psi^{\beta}_{\,\,\,\gamma} \,\,\overline{R}^{\gamma}_{\,\,\,\,\,\,\mu\nu\beta} =0, \qquad \psi_{\mu\nu} = \overline{h}_{\mu\nu} - \frac{\overline{h}}{2} \,\overline{g}_{\mu\nu},
 \label{CUR30b}
 \end{equation}
where  $\psi_{\mu\nu}$ must satisfy $\overline{\nabla}_{\mu} \psi^{\mu\nu} =0$. This gauge condition remains unaltered  for an infinitesimal coordinate transformation of the type $x^{\mu} \to \widetilde{\,x\,}^{\mu} = x^{\mu} + \epsilon^{\mu}$ where $\epsilon_{\mu}$ satisfies the equation
$ \epsilon_{\mu;\alpha}^{\,\,;\alpha} + \epsilon^{\alpha} \overline{R}_{\alpha\mu} =0$; the latter condition reduces to $\epsilon_{\mu;\alpha}^{;\alpha} =0$ whenever $\overline{R}_{\alpha\mu}=0$. A perfect, relativistic and irrotational fluid in a generic curved background 
(not necessarily cosmological) is characterized by the energy-momentum tensor
\begin{equation} 
T_{\mu}^{\,\,\,\nu} = ( p_{t} + \rho_{t}) u_{\mu} \, u^{\nu} - p_{t} \delta_{\mu}^{\nu},
\label{CUR1}
\end{equation}
where $u^{\mu}$ is the four-velocity while $\rho_{t}$ and $p_{t}$ denote the (total) 
energy density and pressure. As originally discussed by Lifshitz, the gravitational waves in a generic fluid background are associated with the symmetric perturbations that satisfy the following pair of conditions  \cite{HIS1a,HIS1b,HIS1c,HIS5}
\begin{equation}
u^{\mu} \overline{h}_{\mu\nu} =0, \qquad \overline{\nabla}_{\mu} \overline{h}^{\mu\nu}=0.
\label{CUR2}
\end{equation}
If the two conditions of Eq. (\ref{CUR2}) are simultaneously imposed,
 the covariant derivative of the fluctuation projected along $u_{\nu}$ obeys the following 
 identity:
\begin{equation}
\overline{\nabla}_{\mu} ( u_{\nu} \overline{h}^{\mu\nu}) =  (\overline{\nabla}_{\mu} u_{\nu}) \, \overline{h}^{\mu\nu} +(\overline{\nabla}_{\mu} \overline{h}^{\mu\nu} ) u_{\nu} =0.
\label{CUR2a}
\end{equation}
Taking into account the conditions (\ref{CUR2}), the left hand side of Eq. (\ref{CUR2a}) vanishes so that, after a trivial rearrangement 
of the indices we have 
\begin{equation}
u_{\nu} \overline{\nabla}_{\mu} \overline{h}^{\,\,\mu\nu} = - \overline{h}^{\,\,\alpha\beta} \, B_{\alpha\beta}, \qquad B_{\alpha\beta} =\overline{\nabla}_{\beta} u_{\alpha} \equiv u_{\alpha\,;\beta}.
\label{CUR3} 
\end{equation}
The tensor $B_{\alpha\beta}$ can be covariantly decomposed  in terms of the total expansion 
$\theta = \overline{\nabla}_{\alpha} u^{\alpha}$ supplemented by the 
shear tensor $\sigma_{\alpha\beta}$ and by the 
vorticity tensor $\omega_{\alpha\beta}$; the result is:
\begin{equation}
B_{\alpha\beta} = \dot{u}_{\alpha} u_{\beta} + \sigma_{\alpha\beta} + \omega_{\alpha\beta} + \frac{\theta}{3} {\mathcal P}_{\alpha\beta},
\label{CUR3a}
\end{equation} 
where $\dot{u}_{\alpha}  = u^{\gamma} \nabla_{\gamma} u_{\alpha}$ and ${\mathcal P}_{\alpha\beta} = (\overline{g}_{\alpha\beta} - u_{\alpha} u_{\beta})$. Inserting Eq. (\ref{CUR3a}) into Eq. (\ref{CUR3}) the condition $u_{\nu} \overline{\nabla}_{\mu} \overline{h}^{\,\,\mu\nu}$ simplifies since the vorticity, being anti-symmetric (i.e.  $\omega_{\alpha\beta} = - \omega_{\beta\alpha}$) does not contribute:
\begin{equation}
u_{\nu} \overline{\nabla}_{\mu} \overline{h}^{\,\,\mu\nu} = - \overline{h}^{\alpha\beta}\, \sigma_{\alpha\beta} - \frac{\theta}{3} \overline{h}.
\label{CUR3b}
\end{equation}
The remaining terms not appearing in Eq. (\ref{CUR3b}) have been eliminated thanks to the first condition of Eq. (\ref{CUR2}).
The meaning of Eq. (\ref{CUR3b}) becomes more transparent by considering a conformally flat background geometry where $\sigma_{\alpha\beta} =0$; in this case  Eq. (\ref{CUR3b}) reads
\begin{equation}
u_{\nu} \overline{\nabla}_{\mu} \overline{h}^{\,\,\mu\nu} = - \frac{\theta}{3} \,\,\overline{h}, \qquad \theta \geq 0.
\label{CUR3c}
\end{equation}
There are then two distinct possibilities: ({\em a}) if $\theta \neq 0$ the validity of Eq. (\ref{CUR3}) implies, according to Eq. (\ref{CUR3c}), the traceless condition $\overline{h} = \overline{h}_{\mu}^{\,\,\,\mu} =0$; ({\em b})
if $\theta =0$ Eqs. (\ref{CUR3c}) and (\ref{CUR3}) are identical so that the requirement $\overline{h} = \overline{h}_{\mu}^{\,\,\,\mu} =0$ {\em must be separately imposed and does not follow from Eq. (\ref{CUR2})}. 
With these specifications, in the case of a perfect relativistic fluid Eqs. (\ref{CUR0g}) and (\ref{CUR2}) lead to the following  evolution equation of the tensor modes of the geometry:
\begin{equation}
 \overline{\nabla}_{\alpha} \overline{\nabla}^{\alpha} \overline{h}_{\mu\nu} - 2 \, \overline{h}^{\,\,\beta\gamma} \,\,\overline{R}_{\gamma\mu\nu\beta} - \overline{h}_{\mu}^{\,\,\,\gamma} \, \overline{R}_{\gamma\nu} - \overline{h}_{\nu}^{\,\,\,\gamma} \,\overline{R}_{\gamma\mu} = \ell_{P}^2 ( \rho_{t} - p_{t}) \overline{h}_{\mu\nu}.
 \label{CUR4}
 \end{equation}
The evolution equations of the background associated with Eq. (\ref{CUR1}) can be written as:
\begin{equation}
\overline{R}_{\mu\nu} = \ell_{P}^2\biggl[ (\rho_{t} + p_{t}) u_{\mu} \, u_{\nu} + \frac{p_{t} - \rho_{t}}{2} \overline{g}_{\mu\nu} \biggr].
\label{CUR4a}
\end{equation} 
Equation (\ref{CUR4a}) together with the first condition of Eq. (\ref{CUR2}) implies that the evolution equation describing the  propagation of the gravitational waves in a fluid background is:
\begin{equation}
\overline{\nabla}_{\alpha} \,\overline{\nabla}^{\alpha} \,\overline{h}_{\mu\nu} - 2 \overline{h}^{\beta}_{\,\,\,\gamma} \overline{R}^{\gamma}_{\,\,\,\,\,\mu\nu\beta} =0,
 \label{CUR5}
 \end{equation}
and it coincides with the result obtained in the vacuum case (see Eq. (\ref{CUR30b})). 
Equation (\ref{CUR5}) shows that the gravitational waves decouple from the background sources to first-order in the tensor amplitude. 

\subsection{The tensor modes in cosmological space-times}
In four space-time dimensions the assumption of homogeneity and isotropy implies 
that the independent components of the background geometry can be reduced 
from $10$ to $4$ (having taken into account the $3$ spatial rotations and the
$3$ spatial translations). The most general form 
of a line element which is invariant under spatial rotations and spatial translations 
is:
\begin{equation}
ds^2 =  e^{\nu} dt^2 - e^{\lambda} dr^2 - e^{\mu}( r^2 d\vartheta^2 + r^2 \sin^2{\vartheta} d\varphi^2)
- e^{\sigma} d r d t.
\label{genFRW}
\end{equation}
From Eq. (\ref{genFRW}) we still have the possibility of choosing a gauge so that the 
metric can be ultimately reduced to its canonical Friedmann-Robertson-Walker 
(FRW) form\footnote{The series of transformations bringing the line element (\ref{genFRW}) in the form Eq. (\ref{FRW})  can be followed in the 
classic book of Tolman \cite{tolman} and in a more recent monograph \cite{hobson}. }:
\begin{equation}
ds^2 = \overline{g}_{\mu\nu} dx^{\mu} dx^{\nu} = dt^2 - a^2(t) \biggl[ \frac{dr^2}{1 - k r^2} + r^2 \biggl(d\vartheta^2 +
\sin^2{\vartheta} d\varphi^2\biggr)\biggr],
\label{FRW}
\end{equation}
where $\overline{g}_{\mu\nu}$ is the metric tensor of the FRW geometry, $a(t)$ is the (dimensionless) scale factor and $k$ is the (dimensionful) spatial curvature. In the parametrization of Eq. (\ref{FRW}), $k=0$ corresponds to a spatially flat 
Universe; if  $k>0$ the Universe is spatially closed and, finally,   $k<0$ corresponds to a spatially open Universe. In what follows we shall often switch from the cosmic time coordinate $t$ to the conformal time 
parametrization $\tau$ where the line element of Eq. (\ref{FRW}) takes the form:
\begin{equation}
ds^2 = 
a^2(\tau)\biggl\{ d\tau^2 - \biggl[ \frac{dr^2}{1 - k r^2} + r^2 \biggl(d\vartheta^2 +
\sin^{2}\vartheta d\varphi^2\biggr)\biggr]\biggr\}.
\label{FRW2}
\end{equation}
The background Einstein's equations of Eq. (\ref{CUR4a}) in the metric (\ref{FRW}) become the standard Friedmann-Lema\^itre equations:
\begin{equation}
 {\mathcal H}^2 = \frac{\ell_{P}^2}{3} a^2 \rho_{t} - k,\qquad  2 ({\mathcal H}^2 - {\mathcal H}' )= \ell_{P}^2 a^2 (\rho_{t} + p_{t}) - 2 k,
\label{FL2C}
\end{equation}
where the prime denotes a derivation with respect to the conformal 
time coordinate $\tau$ and ${\mathcal H} = a'/a$ (see also Tab. \ref{SEC1TABLE4} and notations therein). 
In connection with Eq. (\ref{FL2C}) it is useful to recall that the standard Hubble rate can be expressed as $
H = {\mathcal H}/a$, while the (cosmic) time derivative of $H$ is $\dot{H} = ({\mathcal H}' - {\mathcal H}^2)/a^2$.
Note that Eq. (\ref{FL2C}) follows from Eq. (\ref{CUR4a}) 
since the $4$-velocity of a fundamental observer in comoving coordinates 
is $u^{\mu}= (1, \, 0,\,0,\,0)$. By combining the two equations 
of Eq. (\ref{FL2C}) it is immediate to obtain  that $\rho_{t}' + 3 {\mathcal H}(\rho_{t} + p_{t})=0$
as it follows from the covariant conservation of the total energy-momentum tensor of the fluid sources. In the case of a FRW space-time Eq. (\ref{CUR5}) becomes:
\begin{equation}
h_{i}^{\,\,j\,\prime\prime}+ 2 {\mathcal H} h_{i}^{\,\,j \,\,\prime} - \widehat{\nabla}^2 h_{i}^{\,\, j}  + 2\, k \,h_{i}^{\,\, j} =0, \qquad \widehat{\nabla}^2 = \gamma^{ij} \widehat{\nabla}_{i} \, \widehat{\nabla}_{j} ,
\label{CUR7}
\end{equation}
where $\widehat{\nabla}_{i}$ now indicates the covariant derivative with respect to the metric $\gamma_{ij}$ and it coincides with the standard partial derivative (i.e. $\widehat{\nabla}_{i} = \partial_{i}$) in the case $k=0$ (i.e.when the intrinsic 
curvature vanishes and $\gamma_{ij}  = \delta_{ij}$ ). For the sake of simplicity and for later convenience Eq. (\ref{CUR7}) 
has been written directly in terms of the rescaled tensor amplitude $h_{ij}$ : 
\begin{equation}
\overline{h}_{ij} = - a^2(\tau) \, h_{ij},  \qquad \widehat{\nabla}_{i} h^{ij} =0 = h_{i}^{i}.
\label{CUR7a}
\end{equation}
Equation (\ref{CUR7a}) explains why  $\overline{h}_{\mu\nu}$ has been formally distinguished 
from $h_{i j}$: the tensor amplitude $h_{i j}$ (appearing in the non-covariant decomposition 
of cosmological perturbations) does not coincide with the spatial components of 
$\overline{h}_{\mu\nu}$ but it is proportional to it via a time-dependent redefinition. 
Depending on the value of $k$ the explicit form of Eq. (\ref{CUR7}) 
is different. In fact recalling that $\widehat{\nabla}^2 h_{i}^{\,\,j} = \nabla^2 h_{i}^{j} + 3 \,k\,h_{i}^{\,\,j}$, Eq. (\ref{CUR7}) can also be written as:
\begin{equation}
\mu_{i}^{\,\,j\,\prime\prime} - \nabla^2 \, \mu_{i}^{\,\,j } - k \,\mu_{i}^{\,\, j}  - \frac{a^{\prime\prime}}{a} \mu_{i}^{\,\, j}  =0, \qquad 
\mu_{i}^{\,\,j} = a \, \,h_{i}^{\,\,j}.
\label{CUR7ab}
\end{equation}
 While in the cases of open or flat universes the spectrum of the three-dimensional Laplacian 
 of Eq. (\ref{CUR7ab}) is continuous, for closed FRW backgrounds it takes discrete values. 
 In the context of the concordance paradigm 
 the spatial curvature is today subleading in comparison with the extrinsic (i.e. Hubble) 
 curvature. It is however possible to conceive both open and closed
 models of inflationary dynamics  \cite{open0,open1,open2,closed1} and, in these cases, 
 the role of the spatial curvature in Eq. (\ref{CUR7ab}) cannot be neglected at early times.

\subsection{Relativistic cosmological fluctuations}
According to the results of the covariant treatment (see e.g. Eqs. (\ref{CUR7a}) and (\ref{CUR7ab})) the 
tensor modes of the geometry in a spatially flat cosmological background are 
ultimately described by a (transverse and traceless) rank-two tensor in three dimensions. This observation will now be taken as the starting point of a  non-covariant description where the scalar, vector and tensor modes are separated at the beginning of the derivation (and {\em not at the end} as in the covariant case discussed above). In terms of the conformal time parametrization of Eq. (\ref{FRW2}) the relativistic fluctuations of the metric $\delta^{(1)} g_{\mu\nu}(\vec{x}, \tau)$ shall then be decomposed as the sum three different contributions
\begin{equation}
\delta^{(1)} g_{\mu\nu}(\vec{x}, \tau) = \delta_{s}^{(1)} g_{\mu\nu}(\vec{x}, \tau)
+  \delta_{v}^{(1)} g_{\mu\nu}(\vec{x}, \tau) +
\delta_{t}^{(1)} g_{\mu\nu}(\vec{x}, \tau),
\end{equation}
where the subscripts $s$, $v$ and $t$ denote the scalar, vector and  tensor
inhomogeneities respectively; as in the covariant case, for the sake of conciseness,
 the explicit dependence upon the space-time points shall be omitted.  Furthermore, for consistency with the 
parameters of the concordance paradigm the attention shall be mainly focussed on the case of conformally flat geometries\footnote{Recently the Planck collaboration suggested some possible evidence of closed universe \cite{RT5}. At the moment the 
analyses presented by the collaboration only suggest a mild indication, not a 
compelling conclusion.}  where the background metric is $\overline{g}_{\mu\nu}(\tau) = a^2(\tau) \eta_{\mu\nu}$; some of the results discussed below can be however translated 
to closed and open universes by replacing the partial derivatives with the appropriate 
spatial derivatives defined in terms of the three-dimensional manifold (i.e. $\partial_{i} \to \widehat{\nabla}_{i}$). In the non-covariant approach the $10$ independent components of the perturbed metric $\delta^{(1)} g_{\mu\nu}$ can be separated into $4$ scalars, $2$ solenoidal vectors in three-dimensions and 
$1$ (traceless and solenoidal) rank-two tensor \cite{bard1}: 
\begin{eqnarray}
&& \delta^{(1)} g_{00} = 2 a^2 \phi,\qquad \delta^{(1)} g_{0i} = - a^2 \partial_{i} B - a^2 Q_{i},
\label{COS1}\\
&& \delta^{(1)} g_{ij} = 2 a^2 ( \psi \delta_{ij} - \partial_{i} \partial_{j}E) 
- a^2 h_{ij} + a^2 ( \partial_{i} W_{j} + \partial_{j} W_{i}),
\label{COS2}
\end{eqnarray}
where $\phi$, $\psi$, $B$ and $E$ parametrize the scalar fluctuations; $Q_{i}$ and $W_{i}$ account  for the vector inhomogeneities and are both solenoidal (i.e. $\partial_{i} Q^{i} = \partial_{i} W^{i} =0$); $h_{ij}$, as already established 
in Eq. (\ref{CUR7}), is solenoidal and traceless (i.e. $h_{i}^{i} = \partial_{i}h^{i}_{j} =0$). The decomposition expressed by Eqs. (\ref{COS1})--(\ref{COS2})  is typical of the Bardeen formalism \cite{bard1} and for infinitesimal coordinate transformations of the type $x^{\mu} \to \widetilde{\,x\,}^{\mu} = x^{\mu} + \epsilon^{\mu}$ the fluctuations of the metric change according to the Lie derivative in the direction $\epsilon^{\mu}$:
\begin{equation}
\delta^{(1)} {g}_{\mu\nu} \to \widetilde{\,\delta^{(1)}g\,}_{\mu\nu} = \delta^{(1)}g_{\mu\nu} - \overline{\nabla}_{\mu} \epsilon_{\nu} - \overline{\nabla}_{\nu} \epsilon_{\mu}.
\label{COS6}
\end{equation}
Since $\epsilon_{\mu}$ can be written as $\epsilon_{\mu} = a^2(\tau)(\epsilon_{0}, - \epsilon_{i})$, the three gauge parameters $\epsilon_{i}$ 
will affect both the scalars and the vectors but not the tensors. Indeed $\epsilon_{i}$ can be separated as $\epsilon_{i} = \partial_{i} \epsilon + \zeta_{i}$
where, again, $\zeta^{i}$ is solenoidal (i.e.  $ \partial_{i} \zeta^{i} =0$). The gauge transformations involving $\epsilon_{0}$ and $ \epsilon$ preserve the scalar nature of the fluctuations while the transformations parametrized by $\zeta_{i}$ 
preserve the vector nature of the fluctuation. Thanks to Eq. (\ref{COS6}) the original fluctuations of Eq. (\ref{COS1}) 
and (\ref{COS2}) become, in the new frame,
\begin{eqnarray}
&& \phi \to \widetilde{\,\phi\,} = \phi - {\mathcal H} \epsilon_0 - \epsilon_{0}' ,\qquad \psi \to \widetilde{\,\psi\,} = \psi + {\mathcal H} \epsilon_{0},
\label{phipsi}\\
&& B \to \widetilde{\,B\,} = B +\epsilon_{0} - \epsilon',\qquad E \to \widetilde{\,E\,} = E - \epsilon.
\label{EB}
\end{eqnarray}
Under a coordinate transformation preserving the 
vector nature of the fluctuation the rotational modes of the geometry transform instead as:
\begin{equation}
Q_{i} \to \widetilde{\,Q\,}_{i} = Q_{i} - \zeta_{i}', \qquad W_{i} \to \widetilde{\,W\,}_{i}= W_{i} + \zeta_{i},\qquad \partial_{i} \, \zeta^{i} =0.
\label{W}
\end{equation}
The tensor fluctuations, in the parametrization of Eqs. (\ref{COS2}) are automatically invariant under infinitesimal diffeomorphisms, i.e. $\widetilde{\,h\,}_{ij} = h_{ij}$. The gauge-invariance 
of the tensor amplitude $h_{ij}$ implies, for instance, the gauge-invariance of the fluctuations of the Ricci tensor
\begin{equation}
 \delta^{(1)}_{t} R_{i j} = \frac{1}{2}\biggl[ h_{i j}^{\prime\prime}+ 2 {\cal H} 
h_{ij}' +  2 ( {\cal H}' + 2 {\cal H}^2) h_{i j} - \nabla^2 h_{ij} \biggr],
\label{driccit1gen}
\end{equation}
so that the tensor fluctuations of the Einstein equations will lead to the gauge-invariant evolution 
of $h_{ij}$ without any supplementary manipulation:
\begin{equation}
\delta_{t}^{(1)} R_{ij} = -\frac{a^2}{2}\, \overline{R} \,h_{i\,j} + \ell_{P}^2 \,\delta_{t}^{(1)} T_{i\,j},
\label{perttensAA}
\end{equation}
where $\overline{R} = - 6 a^{\prime\prime}/a^3$ is the background Ricci scalar and $\delta_{t}^{(1)} T_{i\,j} = a^2 p_{t} \,h_{i\,j}$. Inserting Eq. (\ref{driccit1gen}) into Eq. (\ref{perttensAA}) we obtain:
\begin{equation}
h_{ij}^{\prime\prime} + 2 {\mathcal H} h_{i j}^{\prime} - \nabla^2 h_{ij} + h_{ij} \biggl( 4 {\mathcal H}^{\prime} + 2 
{\mathcal H}^2 + \ell_{P}^2 \, p_{t} a^2 \biggr) =0.
\label{perttensBB}
\end{equation}
But the last term at the left hand side of Eq. (\ref{perttensBB}) vanishes since, by combining the two equations of Eq. (\ref{FL2C}) we obtain exactly $4 {\mathcal H}^{\prime} + 2 {\mathcal H}^2 + \ell_{P}^2 \, p_{t} a^2 =0$. Consequently the rescaled amplitude 
$\mu_{ij} = a \, h_{ij}$ obeys 
\begin{equation}
\mu_{ij}^{\prime\prime} - \frac{a^{\prime\prime}}{a} \mu_{ij}^{\prime} - \nabla^2 \mu_{ij} =0,
\label{NM1}
\end{equation}
which coincides with Eq. (\ref{CUR7ab}) in the flat case when $k=0$.
Note that if we perturb the Einstein equations with mixed indices the evolution 
of the tensor perturbations is automatically on shell and we do not need to use any background relation 
to simplify the obtained result:
\begin{equation} 
\delta_{t}^{(1)} R_{i}^{\,\,\,j} = \delta_{t}^{(1)}R_{i \, k} \, \overline{g}^{k\,\, j} + \overline{R}_{i \, k} \,\frac{h^{i\,k}}{a^2}
= - \frac{1}{2 a^2} \biggl( h_{i}^{\,\,j\,\,\,\prime\prime} + 2 {\mathcal H} h_{i}^{\,\,j\prime} - \nabla^2 h_{i}^{\,\,j} \biggr) =0,
\label{perttensCC}
\end{equation}
where we recalled that the tensor fluctuations with doubly contravariant indices are $\delta^{(1)}_{t} g^{ij} = h^{ij}/a^2$. It is finally useful to remark that the tensor modes $h_{i\,j}$ actually coincide, up to a trivial rescaling, with the tensor normal modes of the system (see the discussions of the  effective actions in section \ref{sec3}). 

\subsection{Gauge-invariant normal modes}
Unlike the tensor modes, the scalar and the vector fluctuations of the geometry defined 
in Eqs. (\ref{phipsi}), (\ref{EB}) and (\ref{W}) are not  
gauge-invariant. In the scalar case the perturbed components of the Ricci tensor can be expressed as:
\begin{eqnarray}
 \delta^{(1)}_{s} R_{00} &=& \nabla^2\biggl[ \phi + (B- E^{\,\prime})^{\,\prime}+ {\mathcal H}( B - E^{\,\prime})\biggr]
+ 3 \biggl[ \psi^{\,\prime\prime} + {\mathcal H}( \phi^{\,\prime} + \psi^{\,\prime}) \biggr],
\nonumber\\
 \delta_{s}^{(1)} R_{0 i} &=& \partial_{i} \biggl[( {\mathcal H}^{\prime} + 2 {\mathcal H}^2) B + 
2 ( \psi^{\prime} + {\mathcal H} \phi)\biggr],
\nonumber\\
 \delta^{(1)}_{s} R_{i j} &=& - \delta_{i j}\biggl\{ \biggl[\psi^{\,\prime\prime} + 2 ( {\mathcal H}^{\prime} + 2 {\mathcal H}^2) 
(\psi + \phi)  + {\mathcal H} ( \phi^{\prime} + 5 \psi^{\prime}) - \nabla^2 \psi \biggr]+ {\mathcal H} 
\nabla^2 ( B - E^{\prime}) \biggl\}
\nonumber\\
&+&\partial_{i} \partial_{j}\biggl[ (E^{\,\prime} - B)^{\prime} + 2 ( {\mathcal H}^{\prime} + 2 {\mathcal H}^2 )E + 
2 {\mathcal H} (E^{\prime} -B) + ( \psi - \phi) \biggr],
\label{riccigenfl}
\end{eqnarray}
and are not gauge invariant as a consequence of the transformations of the 
individual fluctuations given in Eqs. (\ref{phipsi})--(\ref{EB}). For the same 
reason Eq. (\ref{W}) implies that, in the vector case, the fluctuations of the Ricci tensor 
are not immediately gauge-invariant: 
\begin{eqnarray}
 \delta^{(1)}_{v} R_{0 i} &=& \biggl({\mathcal H}' + 2 {\cal H}^2 \biggr) Q_{i} 
 - \frac{1}{2} \nabla^2 Q_{i}  - \frac{1}{2} \nabla^2 W_{i}',
\nonumber\\
\delta^{(1)}_{v} R_{i j} &=& 
- \frac{1}{2} \biggl[ \biggl( \partial_{i} Q_{j} + \partial_{j} Q_{i}\biggr)^{\,\prime} + 2 {\mathcal H}   ( \partial_{i} Q_{j} + \partial_{j} Q_{i}\biggr)\biggr]- \frac{1}{2} \biggl(\partial_{i}W_{j} + \partial_{j}W_{i}\biggr) ( 2 {\mathcal H}' + 4 {\mathcal H}^2 )
\nonumber\\
&-& \frac{1}{2}\biggl(\partial_{i}W_{j} + \partial_{j}W_{i}\biggr)^{\, \prime\prime} 
-{\mathcal H}  \biggl(\partial_{i}W_{j}+ \partial_{j}W_{i}\biggr)^{\, \prime},
\label{VRICCIgen}
\end{eqnarray}
as it happens, on the contrary, in the tensor case (see e.g. Eq. (\ref{driccit1gen})).
Consequently, in the scalar and vector case the strategy is either to work in a specific gauge 
or to define an appropriate gauge-invariant variable \cite{bard1}. 
The gauge choices for the vector modes are limited since there are only two solenoidal
vectors in the perturbed form of the metric given in Eqs. (\ref{COS1}) and (\ref{COS2}).
In the scalar case there are instead $4$ scalars and larger number of potentially viable choices for the 
coordinate system. If $B$ and $E$ are set to zero the gauge freedom is 
completely fixed and this choice pins down the conformally Newtonian gauge \cite{bard1,bertschingerma} where 
the longitudinal fluctuations of the metric read, in Fourier space,
\begin{equation}
\delta_{s}^{(1)}\, g_{00}(k,\tau) = 2 a^2 \,\phi(\vec{k},\tau),\qquad \delta_{s}^{(1)} g_{ij} = 2 a^2 \psi(\vec{k},\tau) \delta_{ij},
\label{STR4}
\end{equation}  
and $\vec{k}$ denotes the comoving three-momentum.
By instead setting $\phi$ and $B$ to zero we recover the standard choice of the synchronous 
coordinate system \cite{syn1,syn2} (i.e. $\phi_{S} = B_{S} =0$) where the metric fluctuations are expressed, in the conventional notation, as
\cite{bertschingerma}:
\begin{equation}
\delta_{s}^{(1)} g_{i j}(k,\tau) = 
a^2(\tau)\biggl[ \hat{k}_{i} \hat{k}_{j}\, h(\vec{k},\tau) + 6\, \xi(\vec{k},\tau)\biggl(\hat{k}_{i} \hat{k}_{j} - \frac{1}{3} \delta_{ij}\biggr)\biggr],
\label{SYN2}
\end{equation}
and  $\hat{k}_{i} = k_{i}/|\vec{k}|$; note that $h$ in Eq. (\ref{SYN2}) has nothing to do with the tensor fluctuations but it is related with the trace of the perturbed metric in the scalar case. In the parametrization of Eqs. (\ref{COS1})--(\ref{COS2})  the synchronous fluctuations of the metric are given by $\delta_{s} g_{ij}(k,\tau) = 2 a^2 (\psi_{S} \delta_{ij} + k_{i} k_{j} E_{S})$ implying that 
$\psi_{S} = - \xi$ and $E_{S} = (h + 6 \xi)/(2 k^2)$. Finally a third convenient choice is the off-diagonal (or uniform curvature) gauge demanding that $\psi_{U}= E_{U} = 0$ in Eqs. (\ref{COS1})--(\ref{COS2}) \cite{hw1,hw2,hw2a,NoM1}. While the choice of a specific gauge is dictated by various practical considerations, there obviously exist appropriate gauge-invariant combinations that are invariant under infinitesimal coordinate transformations.  For instance the gauge-invariant generalizations of the longitudinal fluctuations of the metric are given by the Bardeen potentials \cite{bard1} $\Phi = \phi + ( B - E^{\prime})^{\prime} + {\mathcal H}( B- E^{\prime})$ and $\Psi=\psi - {\mathcal H} ( B - E^{\prime})$ \cite{bard1}; the evolution of $\Phi$ and $\Psi$ coincides with  the equations of the longitudinal gauge where  $\Psi \equiv \psi$ and $\Phi \equiv \phi$.  Since the different gauge-invariant combinations do not have a particular physical relevance, it is better to focus, in analogy with the tensor case, on those gauge-invariant combinations that coincide with the normal modes of the system. The scalar normal modes are instead associated with the curvature perturbations on the hypersurfaces that are comoving (i.e. the four-velocity vanishes) and also orthogonal (i.e. $B=0$ in Eq. (\ref{COS1}) so that the scalar fluctuations of the off-diagonal entries of the metric vanish). The curvature perturbations are conventionally denoted by ${\mathcal R}$ and in the comoving orthogonal gauge they correspond to the scalar fluctuation of the spatial curvature, i.e.  $\delta_{s} ^{(3)}R = - (4/a^2) \nabla^2 {\mathcal R}$ \cite{lythpert,KSa}. When the background is  dominated by an irrotational relativistic fluid the evolution of ${\mathcal R}$ is:
\begin{equation}
{\mathcal R}^{\prime\prime} + 2 \frac{z_{t}^{\prime}}{z_{t}} {\mathcal R}^{\prime} - c_{st}^2 \nabla^2 {\mathcal R}=0, \qquad z_{t} = \frac{a^2 \sqrt{\rho_{t} + p_{t}}}{{\mathcal H} \, c_{st}},
\label{NM2}
\end{equation}
where $c_{st}^2 = p_{t}^{\prime}/\rho_{t}^{\prime}$; note that $\rho_{t}$ and $p_{t}$ enter directly the background equations (\ref{FL2C}). The variable of Eqs. (\ref{NM2}) and  has been discussed for the first time by Lukash \cite{lukash} (see also \cite{strokov,luk2}) while analyzing the
quantum excitations of an irrotational and relativistic fluid.  The canonical normal mode identified in Ref. \cite{lukash}  is invariant under infinitesimal coordinate transformations. 
If the background is  dominated by a single scalar field $\varphi$ the analog 
of Eq. (\ref{NM2}) is:
\begin{equation}
{\mathcal R}^{\prime\prime} + 2 \frac{z_{\varphi}^{\prime}}{z_{\varphi}} {\mathcal R}^{\prime} - \nabla^2 {\mathcal R}=0, \qquad
 z_{\varphi}= \frac{a \varphi^{\prime}}{\mathcal H}.
\label{NM3}
\end{equation}
Equation (\ref{NM3}) has been derived in the case of scalar field matter in Refs.  \cite{chibisov,KS,chibisova} (see also \cite{br1,bard2,br1a,bard2a})  but these analyses follow the same logic of Ref. \cite{lukash}. 
Once the curvature perturbations are computed (either from Eq. (\ref{NM2}) or from Eq. (\ref{NM3})) the 
metric fluctuations can be easily derived in a specific gauge. Since ${\mathcal R}$ is gauge-invariant, 
its value is, by definition, the same in any coordinate system even if its expression changes from one gauge 
to the other. For instance, in the synchronous and longitudinal  gauges  
the expression of ${\mathcal R}$ is, respectively
\begin{equation} 
{\mathcal R}^{(S)} = \frac{{\mathcal H}}{{\mathcal H}^2 - {\mathcal H}'} \xi' + \xi, \qquad {\mathcal R}^{(L)} = - \psi_{L} - \frac{{\mathcal H} ( {\mathcal H} \phi_{L} + \psi_{L}^{\prime})}{{\mathcal H}^2 - {\mathcal H}^{\prime}}.
\label{NMR4}
\end{equation}
Even if the expressions ${\mathcal R}^{(L)}$ and ${\mathcal R}^{(S)}$ of Eq. (\ref{NMR4}) are formally different, the invariance 
under infinitesimal coordinate transformations implies that the values of ${\mathcal R}$ 
computed in different gauges must coincide, i.e. ${\mathcal R}^{(S)} = {\mathcal R}^{(L)} = {\mathcal R}$.

\subsection{Supplementary polarizations?} 
The LIGO/Virgo interferometers have different orientations and they can probe the polarizations
arising in general relativity and in its extensions  \cite{SIXPOL1a,SIXPOL1}. Quantitative tests of the same kind have been envisaged in the case of future space-borne detectors. It is actually well known that a massless spin $2$ particle has two polarizations while a massive one has five. Furthermore, general metric theories of gravity (e.g. scalar-tensor theories or scalar-vector-tensor theories) allow up to six polarization states \cite{SIXPOL2,SIXPOL3}. 
The polarization states will not be equally accessible and their potential discovery depends on the 
nature of the signal. If the gravitational wave signal is of {\em short duration}, then even the advanced LIGO
configuration (with two coaligned detectors) will probably be unable to resolve the six polarization states \cite{SIXPOL4}. For this purpose, it will be essential to have more than two detectors; ideally 
six interferometers would be necessary to analyze the general polarization state \cite{SIXPOL4}. 
For short duration signals it is planned to have, in the future, a network involving the advanced LIGO \cite{LV1}
and advanced Virgo detectors \cite{LV2}, the Kagra detector \cite{kagra1,kagra2} 
possibly supplemented by the so-called LIGO-India \cite{LIGOindia}. The story is different in the case of {\em sources of long duration} like in the case of a stochastic background of relic gravitons. In this case the LIGO detectors, in their advanced configuration, could directly assess the polarization of gravitational waves \cite{SIXPOL1a,SIXPOL1}. This is the reason why the search for relic graviton backgrounds will be particularly important not only for conceptual reasons but also for more practical tests of the possible extensions of general relativity. By formally setting $a(\tau)\to 1$ in Eqs. (\ref{COS1})--(\ref{COS2}) we can always select a gauge 
where $B=0$, $\phi=0$ and $Q_{i} =0$. In this case the metric fluctuation becomes, in Fourier space, 
\begin{equation}
\delta^{(1)} g_{ij}(\vec{k}, \tau) = - h_{ij}(\vec{k},\tau) + 2 \biggl[ \delta_{ij} \, \psi(\vec{k},\tau) + k_{i} \, k_{j} \, E(\vec{k},\tau) \biggr]  - i \biggl[k_{i} \,W_{j}(\vec{k},\tau) + k_{j} \,W_{i}(\vec{k},\tau) \biggr];
\label{POLDEF1}
\end{equation}
note that $\delta g_{ij}^{\ast}(\vec{k}, \tau) = \delta g_{ij}(-\vec{k}, \tau)$ and similarly for the individual 
Fourier amplitudes appearing in Eq. (\ref{POLDEF1}). 
Equation (\ref{POLDEF1}) describes up to $6$ polarization states: $2$ for the tensors, $2$ for the vectors and $2$ for the 
scalars. In general relativity the only two propagating degrees of freedom in flat space-time are associated 
with $h_{ij}$ but in more general theories of gravity this is not the case. If we define a triplet of mutually orthogonal 
unit vectors\footnote{As we shall stress later on these unit 
vectors do not have to coincide with the Cartesian directions. However, for pure illustrative purposes,
it is useful to select the coordinate system in this way.}
$\hat{x}^{i}$, $\hat{y}^{i}$ and $\hat{z}^{i}$ we can set the direction 
of propagation of the wave along $\hat{z}^{i}$ (i.e. $\hat{k} = \hat{z}$). 
In this case the two tensor polarizations are:
\begin{equation}
e_{ij}^{\oplus}= \hat{x}_{i} \, \hat{x}_{j} - \hat{y}_{i} \, \hat{y}_{j}, \qquad 
e_{ij}^{\otimes}= \hat{x}_{i} \, \hat{y}_{j} + \hat{y}_{i} \, \hat{x}_{j}.
\label{POLDEF2}
\end{equation}
The two polarizations of the metric corresponding to the vector modes 
can instead be defined as: 
\begin{equation}
e_{ij}^{v_{1}} = \hat{z}_{i} \,\hat{x}_{j} + \hat{z}_{j}\, \hat{x}_{i}, \qquad 
e_{ij}^{v_{2}} = \hat{z}_{i} \,\hat{y}_{j} + \hat{z}_{j} \,\hat{y}_{i}.
\label{POLDEF3}
\end{equation}
Finally the two polarizations of the metric corresponding to the scalar 
modes become:
\begin{equation}
e_{ij}^{s_{1}} = \hat{x}_{i}\, \hat{x}_{j} + \hat{y}_{i} \,\hat{y}_{j}, \qquad 
e_{ij}^{s_{2}} = \hat{z}_{i} \, \hat{z}_{j},
\label{POLDEF4}
\end{equation}
and are sometimes referred to as {\em breathing} and {\em longitudinal} modes respectively. 
Using the decomposition of Eqs. (\ref{POLDEF2}), (\ref{POLDEF3}) and 
(\ref{POLDEF4}) the tensor, vector and scalar modes of the geometry 
can be expanded along the corresponding polarizations. In 2018 the LIGO/Virgo collaboration presented the results from the first directed search for non-tensorial gravitational waves from known pulsars \cite{SIXPOL1a} and the first dedicated search for  tensor, vector, and scalar polarizations in the stochastic gravitational-wave background \cite{SIXPOL1}. The analysis of Ref. \cite{SIXPOL1a} found no evidence of non-tensorial gravitational waves.  The results of Ref. \cite{SIXPOL1} placed the first direct bounds on the contributions of vector and scalar polarizations to the stochastic background at a reference frequency of $25$ Hz:
\begin{equation}
\Omega^{(t)}_{gw} < 5.6\times 10^{-8}, \qquad \Omega^{(v)}_{gw} < 6.4\times 10^{-8}, \qquad \Omega^{(s)}_{gw} < 1.1 \times 10^{-7},
\label{POLDEF4a}
\end{equation}
where $\Omega^{(X)}_{gw}$ (with $X = t\,v,\,s$) denote the spectral energy density of the tensor, vector and scalar polarizations (see also sections \ref{sec3} and \ref{sec7} for 
some related discussions).  There are indeed various gravity theories predicting additional polarizations \cite{SIXPOL4} like the Einstein Aether theories \cite{SIXPOL5,SIXPOL6} positing the existence of a Lorentz symmetry-violating, dynamical, time-like vector field \cite{SIXPOL7}. Another class of examples involves the $f(R)$ 
theories where the action is a generic function of the Ricci scalar \cite{SIXPOL4} and two further scalar polarizations supplement the standard tensor polarizations.
As already suggested above, the supplementary polarizations appearing in Eq. (\ref{POLDEF1}) can be thoroughly studied only with more than three detectors \cite{SIXPOL7a,SIXPOL7b,SIXPOL7c} both on the earth and in space (see also \cite{SIXPOL7d,SIXPOL7e}).  Even if the future measurements will not find  clear evidence for supplementary polarizations, we cannot exclude deviations at earlier time. For instance gravitational waves might acquire an effective index of refraction when they travel in curved space-times \cite{SIXPOL8,SIXPOL9} and this possibility has been recently revisited by studying the parametric amplification of the tensor modes of the geometry during a quasi-de Sitter stage of expansion\footnote{The variation of the refractive index during inflation leads indeed to detectable spectra of cosmic gravitons in the audio band \cite{SIXPOL12} which will be discussed in section \ref{sec6}.} 
\cite{SIXPOL10} (see also \cite{SIXPOL11,SIXPOL11a,SIXPOL12}). 
\subsection{Tensor modes in Fourier space}
The three-dimensional Fourier transform of $h_{i\,j}(\vec{x},\tau)$ will be defined as: 
\begin{equation}
h_{i\, j}(\vec{x},\tau)= \frac{1}{(2\pi)^{3/2}} \int d^{3} k\,\, h_{i\,j}(\vec{k}, \tau) \, e^{- i \vec{k} \cdot\vec{x}}, \qquad 
h_{i\,j}^{\ast}(\vec{k},\tau) = h_{i\, j}( - \vec{k},\tau).
\label{POLDEF5a}
\end{equation}
The Fourier amplitude $h_{i\,j}(\vec{k}, \tau)$ can be also represented as a sum over the linear polarizations
\begin{equation}
h_{i\, j}(\vec{k},\tau) = \sum_{\lambda = \oplus, \,\otimes} e^{(\lambda)}_{i\, j} (\hat{k}) 
\,\, h_{\lambda}(\vec{k},\tau), \qquad h_{\lambda}^{\ast}(\vec{k},\tau) = h_{\lambda}(- \vec{k},\tau).
\label{POLDEF5}
\end{equation}
Even if the unit vectors entering the definition of the tensor polarizations of Eq. (\ref{POLDEF2}) have been 
identified, for the sake of simplicity, with the three Cartesian directions,  $\hat{k}$ might also coincide with the radial direction so that 
\begin{equation}
\hat{k} = \cos{\varphi} \,\sin{\vartheta} \hat{x} + \sin{\varphi} \,\sin{\vartheta} \hat{y} + \cos{\vartheta} \hat{z},
\label{SIXPOL9}
\end{equation}
where $\hat{x}$, $\hat{y}$ and $\hat{z}$ denote, as before, the three Cartesian unit vectors.
In the case of Eq. (\ref{SIXPOL9}) the tensor polarizations will be defined exactly as in Eq. (\ref{POLDEF2})
but the unit vectors will have a different expression in terms of the Cartesian components:
\begin{equation}
e_{i\,j}^{\oplus} = [\hat{a}_{i} \, \hat{a}_{j} - \hat{b}_{i} \hat{b}_{j}], \qquad e_{i\,j}^{\otimes} = [\hat{a}_{i} \, \hat{b}_{j} + \hat{b}_{i} \hat{a}_{j}],
\label{SIXPOL9a}
\end{equation}
where now $\hat{a} = [\sin{\varphi} \,\hat{x} - \cos{\varphi} \, \hat{y}]$ 
and $\hat{b} = [ \cos{\varphi} \, \cos{\vartheta} \, \hat{x} 
+ \sin{\varphi} \,\cos{\vartheta} \,\hat{y} - 
\sin{\vartheta} \, \hat{z}]$ (with $\hat{a} \times \hat{b} = \hat{k}$).
The circular polarizations of the relic gravitons can be introduced as combinations of the linear polarizations: 
\begin{equation} 
 e^{(R)}_{ij} = \frac{1}{\sqrt{2}} \biggl(e_{ij}^{\oplus} + i\,e_{ij}^{\otimes}\biggr), \qquad e^{(L)}_{ij} = \frac{1}{\sqrt{2}} \biggl(e_{ij}^{\oplus} - i\,e_{ij}^{\otimes}\biggr),
 \label{SIXPOL10}
 \end{equation}
where the Fourier amplitudes are now defined as  $ h_{ij}(\vec{k},\tau) = [e^{(R)}_{ij} h_{R}(\vec{k},\tau) + e^{(L)}_{ij} h_{L}(\vec{k},\tau)]$; in this case the left-handed and right-handed components are 
$ h_{L} = (h_{\oplus} + i\, h_{\otimes})/\sqrt{2}$ and $h_{R} =  (h_{\oplus} - i\, h_{\otimes})/\sqrt{2}$.
In spite of the particular basis where they are expressed, the tensor polarizations obey 
the following useful identity:
\begin{equation}
\sum_{\lambda=\oplus,\,\otimes} e^{(\lambda)}_{i\, j} \,\, e^{(\lambda)}_{m\, n} = p_{i\,m}(\hat{k}) \,\,
p_{j\,n}(\hat{k}) + p_{i\,n}(\hat{k}) \,\, p_{j\,m}(\hat{k}) - p_{i\,j}(\hat{k})\,\, p_{m\,n}(\hat{k}),
\label{POLDEF6}
\end{equation} 
where $p_{i\, j}(\hat{k})=( \delta_{i\, j} - \hat{k}_{i} \, \hat{k}_{j})$ and, as usual, $\hat{k}^{i} = k^{i}/|\vec{k}|$.
It is also practical to introduce, for immediate convenience,  the following rank-four tensor in three dimensions:
\begin{equation}
{\mathcal S}_{i\,j\,m\,n}(\hat{k}) = \frac{1}{4} \biggl[p_{i\,m}(\hat{k}) \,\,
p_{j\,n}(\hat{k}) + p_{i\,n}(\hat{k}) \,\, p_{j\,m}(\hat{k}) - p_{i\,j}(\hat{k})\,\, p_{m\,n}(\hat{k})\biggr],
\label{POLDEF7}
\end{equation}
which is symmetric for $i \to j$, $m\to n$ and also for $(i\,j)\to (m,\,n)$. According to Eq. (\ref{POLDEF7}) 
${\mathcal S}_{ijmn}(\hat{k})$ is solenoidal:
\begin{eqnarray}
\hat{k}^{i} \,{\mathcal S}_{i\,j\,m\,n} &=& \hat{k}^{j}\, {\mathcal S}_{i\,j \,m\,n}=
\hat{k}^{m}\, {\mathcal S}_{i\,j \,m\,n} = \hat{k}^{n} \,{\mathcal S}_{i\,j\, m\,n} =0,
\label{POLDEF8}\\
{\mathcal S}_{i\, i\, m \, n}(\hat{k}) &=& {\mathcal S}_{i \,j \, m \,m}(\hat{k}) =0, 
\qquad {\mathcal S}_{i\, j\, i \,j}(\hat{k}) = 1.
\end{eqnarray}
Similarly the following pair of identities easily follows from Eq. (\ref{POLDEF7}):
\begin{equation}
 {\mathcal S}_{i\, j\, m\, n}(\hat{q}) {\mathcal S}_{i\, j\, m\, n}(\hat{p}) = 
 \frac{[ 1 + (\hat{q}\cdot\hat{p})^2] [ 1 + 3  (\hat{q}\cdot\hat{p})^2]}{16},\qquad \sum_{i\, j} e_{i\,j}^{(\lambda)} \, e^{i\, j}_{(\lambda^{\prime})} = 2 \delta^{\,\,\,\lambda}_{\lambda_{\prime}},
\label{ID1}
\end{equation}
where  $\hat{q} = \vec{q}/|\vec{q}|$ and $\hat{p} = \vec{p}/|\vec{p}|$ are two unit vectors defining, for instance,
two different directions of propagation.  In Eq. (\ref{POLDEF5}) the tensor amplitude has been represented in Fourier space by means of a three-dimensional integral over the comoving three-momentum. Denoting with $k = |\vec{k}|$ the modulus of the comoving three-momentum appearing in Eq. (\ref{POLDEF5}), $h_{ij}(\vec{x}, \tau)$ can also be decomposed as 
an integral over the comoving frequency  $\nu= k/(2 \pi)$ 
\begin{equation}
h_{i\, j}(\vec{x},\tau) =  \sum_{\lambda= \oplus,\, \otimes} \,
\int_{-\infty}^{\infty} d \nu \, \int \, d \, \hat{k}  \, e_{i\,j}^{(\lambda)}(\hat{k}) 
\, e^{ 2 i \,\pi\, \nu\, (\tau - \hat{k}\cdot\vec{x})}\, h_{\lambda}(\nu, \, \hat{k}), \qquad
h_{\lambda}^{\ast}(\nu,\hat{k}) = h_{\lambda}(- \nu,\hat{k}),
\label{SIXPOL8}
\end{equation}
where the angular integration is indicated by $d \hat{k} = d \cos{\vartheta} \, d\varphi$.
A shorthand form of Eq. (\ref{SIXPOL8}) is:
\begin{equation}
h_{i\, j}(\vec{x}, \tau) = \int_{-\infty}^{\infty} d \nu \int d\,\hat{k}\, e^{ 2\,i\,\pi\, \nu ( \tau - \hat{k}\cdot\vec{x})} \,\,h_{i\, j}(\nu, \hat{k}) , \qquad  h_{i\, j}(\nu, \hat{k}) = \sum_{\lambda=\oplus,\,\otimes} 
e_{i\, j}^{(\lambda)}(\hat{k}) \, h_{\lambda}(\nu, \, \hat{k}).
\label{OBS13b}
\end{equation}
The tensor amplitude, like the polarization of the electromagnetic field, 
 be expanded in (tensor) spherical harmonics with two slightly different approaches 
illustrated, respectively, in Refs. \cite{EB1,EB2} and in Refs. \cite{EB3a,HBTA}. Postponing 
to section \ref{sec5} the discussion of Refs. \cite{EB1,EB2},  we shall now 
recall some rudiments of the geometrical approach following from the expansion of the tensor 
amplitude in spherical harmonics on the two-sphere \cite{EB3a,HBTA}:
 \begin{equation}
     h_{i\,j}(\nu,\,\hat{k})= 
     \sum_{\ell=2}^\infty\sum_{m=-\ell}^{\ell}
     \left[ a_{\ell m}^{(\mathrm{G})}(\nu)
     Y_{(\ell m)\,i j}^{(\mathrm{G})}(\hat{k}) + a_{\ell m}^{(\mathrm{C})}(\nu) Y_{(\ell m)\, i\,j}^{(\mathrm{C})}
     (\hat{k}) \right],
\label{SIXPOL10a}
\end{equation}
where coefficients
 $a_{\ell m}^{(\mathrm{G})}(\nu)$ and $ a_{\ell m}^{(\mathrm{C})}(\nu)$ represent
the electric and magnetic type components of the polarization,
respectively (note that the sum starts from $\ell=2$,
since relic gravitons generate only perturbations of multipoles
from the quadrupoles up). On the 
 $2$-sphere the metric can be written as $g_{ij}$ where $g_{\vartheta\vartheta}= 1$ and $g_{\varphi\varphi} = \sin^2{\vartheta}$.
For scalar functions defined on the $2$-sphere the
spherical harmonic functions $Y_{\ell m}(\hat{k})$ are the complete
orthonormal basis. For the $2\times 2$ tensors defined on the
$2$-sphere, such as $h_{ij}(\nu,\hat{k})$,  in Eq.(\ref{SIXPOL10a}), the 
complete orthonormal set of tensor spherical harmonics can be written 
as \cite{EB3a}:
\begin{eqnarray}
Y_{(\ell m)i\,j}^{(\rm G)}(\hat{k}) &=& \overline{N}_{\ell}
     \left( \nabla_{i} \nabla_{j} Y_{(\ell m)} - {1\over2} g_{i\,j} \nabla_{c}\nabla^{c}Y_{(\ell m)}\right),
\nonumber\\     
Y_{(\ell m)i\,j}^{(\rm C)} &=& { \overline{N}_{\ell} \over 2}
     \left(\vphantom{1\over 2}
      \nabla_{i}\nabla_{c} Y_{(\ell m)} \epsilon^c{}_j +\nabla_{j} \nabla_{c}Y_{(\ell m)}(\hat{k})\epsilon^c{}_i \right),
\label{SIXPOL10c}
\end{eqnarray}
where $\overline{N}_{\ell}\equiv \sqrt{ {2 (\ell-2)! / (\ell+2)!}}$;  $\nabla_{i}$  and  $\epsilon_{i}^{j}$ denote, respectively, the covariant derivation and the Levi-Civita symbol 
 on the $2$-sphere.  The $ Y_{(\ell m)i\,j}^{(\mathrm{G})}$ and 
$Y_{(\ell m)i\,j}^{(\mathrm{C})}$ can be written in the form of  $2\times 2$ matrices:
\begin{equation}
   Y_{(\ell m)}^{(\mathrm{G}) }(\hat{k})={\overline{N}_{\ell}\over 2} \left( \begin{array}{cc}
   {\mathcal W}_{(\ell m)}(\hat{k}) & {\mathcal X}_{(\ell m)}(\hat{k}) \sin\vartheta\\
   \noalign{\vskip6pt}
   {\mathcal X}_{(\ell m)}(\hat{k})\sin\vartheta & -{\mathcal W}_{(\ell m)}(\hat{k})\sin^2\vartheta \\
   \end{array} \right),
\label{SIXPOL10d}
\end{equation}
and as
\begin{equation}
   Y_{(\ell m)}^{(\mathrm{C})}(\hat{k})={\overline{N}_{\ell}\over 2} \left( \begin{array}{cc}
   - {\mathcal X}_{(\ell m)}(\hat{k}) & {\mathcal W}_{(\ell m)}(\hat{k}) \sin\vartheta \\
   \noalign{\vskip6pt}
   {\mathcal W}_{(\ell m)}(\hat{k})\sin\theta & {\mathcal X}_{(\ell m)}(\hat{k})\sin^2\vartheta \\
   \end{array} \right),
\label{SIXPOL10e}
\end{equation}
where
\begin{equation}
{\mathcal W}_{(\ell m)}(\hat{k}) =  [2\partial^2_{\vartheta} + \ell(\ell+1) ] Y_{(\ell m)}(\hat{k}),\qquad   {\mathcal X}_{(\ell m)}(\hat{k}) = {2\,i\,m \over \sin\vartheta}
     (\partial_{\vartheta} -
     \cot\vartheta ) Y_{(\ell m)}(\hat{k}).
\label{SIXPOL10f}
\end{equation}
In terms of the spin-2 harmonics ${}_{\pm2}Y_{(lm)}(\hat{k})$ discussed in Refs. \cite{EB1,EB2} we have 
\begin{equation}
     {\mathcal W}_{(\ell m)}(\hat k) \pm i {\mathcal X}_{(\ell m)}(\hat k) = \sqrt{{ (\ell+2)!
     \over (\ell-2)!}}\,_{\pm2}Y_{(\ell m)}(\hat{k}).
\label{SIXPOL10g}
\end{equation}
Note that  $a_{\ell m}^{(\mathrm{G})} = a_{\ell m}^{(E)}/\sqrt{2}$ and 
$a_{\ell m}^{(\mathrm{C})} = a_{\ell m}^{(B)}/\sqrt{2}$ where $a_{\ell m}^{(E)}$ and 
$a_{\ell m}^{(B)}$ are the so-called $E$-mode and $B$-mode polarizations of the CMB (see  Eq. (\ref{PM})
and discussion therein). In the tensor case the entries of the polarization matrix 
 ${\mathcal P}^{(t)}_{\lambda\,\lambda^{\,\prime}} = h_{\lambda}(\vec{k}, \tau) \, h_{\lambda^{\,\prime}}^{\ast}(\vec{k},\tau)$
 can be expressed in terms of the appropriate Stokes parameters:
 \begin{equation}
 {\mathcal P}^{(t)}= \left(\matrix{ I^{(t)}+ Q^{(t)}
& U^{(t)} - i \, V^{(t)}&\cr
U^{(t)} + i \,V^{(t)} &I^{(t)} - Q^{(t)}&\cr}\right)=  \frac{1}{2} \left(  I^{(t)}\,{\bf 1} + U^{(t)}\, \sigma_{1} +  V^{(t)}\, \sigma_2 +Q^{(t)} \, \sigma_3 \right),
\label{SIXPOL11}
\end{equation}
where $\sigma_{i}$ (with $i=1,\,\, 2,\,, 3$) are the three Pauli matrices and ${\bf 1}$ is the identity matrix; recalling the definitions of the linear and circular bases given above, the Stokes parameters in the tensor case are:
\begin{eqnarray}
I^{(t)} &=& \bigl| h_{\oplus}(\vec{k},\tau) \bigr|^2 + \bigl| h_{\otimes}(\vec{k},\tau) \bigr|^2 = \bigl| h_{R}(\vec{k},\tau) \bigr|^2 + \bigl| h_{L}(\vec{k},\tau) \bigr|^2, 
\label{SIXPOL12a}\\
Q^{(t)} &=& \bigl| h_{\oplus}(\vec{k},\tau) \bigr|^2 - \bigl| h_{\otimes}(\vec{k},\tau) \bigr|^2 = h_{R}(\vec{k}, \tau) \, h_{L}^{\ast}(\vec{k},\tau) + h_{R}^{*}(\vec{k}, \tau) \, h_{L}(\vec{k},\tau), 
\label{SIXPOL12}\\
U^{(t)} &=& h_{\oplus}(\vec{k},\tau) \, h_{\otimes}^{\ast}(\vec{k},\tau) + h_{\oplus}^{*}(\vec{k},\tau) \, h_{\otimes}(\vec{k},\tau) = i \bigl[h_{R}(\vec{k},\tau)\, h_{L}^{*}(\vec{k},\tau) - h_{R}^{*}(\vec{k},\tau)\, h_{L}(\vec{k},\tau)\bigr],
\label{SIXPOL13a}\\
\qquad V^{(t)} &=& i \bigl[h_{\oplus}(\vec{k},\tau) \, h_{\otimes}^{\ast}(\vec{k},\tau) - h_{\oplus}^{*}(\vec{k},\tau) \, h_{\otimes}(\vec{k},\tau) \bigr] =  \bigl| h_{R}(\vec{k},\tau) \bigr|^2 - \bigl| h_{L}(\vec{k},\tau) \bigr|^2.
\label{SIXPOL13}
\end{eqnarray}
 Under a rotation\footnote{The vectors $\hat{a}$ and $\hat{b}$ can be rotated (in the plane orthogonal to $\hat{k}$) as $ \hat{a}^{\prime}= \cos{\delta} \,\hat{a} + \sin{\delta} \,\hat{b}$ and as $ \hat{b}^{\prime}= - \sin{\delta} \,\hat{a} + \cos{\delta} \,\hat{b}$.} on the plane orthogonal to the direction of propagation of the wave the linear polarizations 
of Eq. (\ref{SIXPOL9a}) are rotated by an angle  $2\delta$ if $\hat{a}$ and $\hat{b}$ are rotated by $\delta$; 
this is a simple consequence of the spin-weight of the function. For the same rotation the circular polarizations 
of Eq. (\ref{SIXPOL10}) transform as $\widetilde{\,e\,}_{i\,j}^{(R)} = e^{- 2 i \delta}\,e_{i\,j}^{(R)}$
and as $\widetilde{\,e\,}_{i\,j}^{(L)} = e^{2 i \delta}\,e_{i\,j}^{(L)}$.
 As in the case of the Stokes parameters associated with the electromagnetic polarizations (see section \ref{sec5}), the intensity and the circular polarizations (i.e. $I^{(t)}$ 
 and $V^{(t)}$) are invariant for a two-dimensional rotation in the plane orthogonal to $\hat{k}$ while the linear polarizations $Q^{(t)}$ and $U^{(t)}$ are not invariant, as it can be explicitly verified from 
 Eqs. (\ref{SIXPOL12a})--(\ref{SIXPOL13}) by recalling that, within the present conventions, $h_{\lambda} = h_{i\, j} e^{i\,j}_{\lambda}/2$.  
 
\renewcommand{\theequation}{3.\arabic{equation}}
\setcounter{equation}{0}
\section{Effective action and energy density of the relic gravitons}
\label{sec3}
The equivalence principle ultimately forbids the localization of the energy-momentum of the gravitational 
field \cite{TENS1,TENS2a} so that the energy density and the pressure of the relic gravitons do not
have a unique gauge-invariant and frame-invariant expression. The equivalence principle in its 
stronger formulation stipulates that the laws of physics are those of special relativity in 
a freely falling (non-rotating) laboratory that occupies a small portion of spacetime. 
But since the coordinates can be transformed to a freely falling frame,  
it is always possible to eliminate the gravitational field locally and this is why it is so 
difficult to propose a unique, local and  gauge-invariant definition of the gravitational energy density.  
Various energy-momentum pseudo-tensors have been proposed through the years. 
{\em The Landau-Lifshitz} \cite{TENS2b} proposal is rooted in the second-order expansion
of the Einstein tensor supplemented by the observation that Bianchi identities 
must be valid to all orders in the perturbative expansion. {\em The Brill-Hartle approach} 
\cite{TENS2c}  aims at separating the terms that evolve faster than the rate of variation 
of the corresponding background and has been used to derive
the Isaacson effective pseudo-tensor \cite{TENS3a,TENS3b} providing a sound description 
of gravitational radiation in the high-frequency limit. {\em The Ford-Parker strategy} 
\cite{HIS7} follows instead from the effective action of the relic gravitons derived by 
perturbing the gravitational action to second order in the tensor amplitude (see also \cite{TENS3d}). The effective energy density and pressure of the relic gravitons are, in this case, both
gauge-invariant and frame-invariant, i.e. their form is the same 
both in the Einstein and in the Jordan frames. Within this approach the various terms of the effective action of the relic gravitons (including the possible parity-violating contributions) can be easily deduced and included.  Even if inside the Hubble radius all the different choices are equivalent,  the only consistent strategy for comoving wavelengths larger than the Hubble radius must be closely related to the original Ford-Parker suggestion \cite{HIS7} where the background metric and the corresponding perturbations are treated as independent fields. 

The other proposals suggested in the current literature, even if superficially different, are related to the ones mentioned in the previous paragraph and will be discussed in the final part of this section. More specifically, Refs. \cite{TENS3cc,TENS3c} employ the Landau-Lifshitz approach appropriately adapted to the case of a cosmological background.  The relations between the Ford-Parker and the Landau-Lifshitz pseudo-tensors have been discussed in Ref. \cite{TENS5} and the relevance of second-order action  has been reinstated in Ref. \cite{TENS5a}. Some elements of comparison between the Brill-Hartle and the Landau-Lifshitz strategies can be found in Refs. \cite{TENS6} and \cite{TENS8}. The Brill-Hartle-Isaacson
scheme has been used in Refs. \cite{TENS5a,TENS9}. 
Through the years the various suggestions have been applied
either for the solution of the concrete problems of backreaction \cite{TENS3c,TENS5,TENS8} or 
in more general terms \cite{TENS5a,TENS6,TENS9}.
 Babak and Grishchuk \cite{TENS4a} came up with a possible definition of a true energy-momentum {\em tensor} of the gravitational field. By treating gravity as a nonlinear tensor field in flat space-time \cite{TENS4aa}, Ref. \cite{TENS4a} claimed a result with all the necessary properties of a true energy-momentum tensor of the gravitational field itself (i.e. symmetry, uniqueness, gauge-invariance and covariant conservation).  The results of Ref.\cite{TENS4a} would imply that 
the problem of localizing the energy and momentum of the gravitational field is 
completely solved.  The perspective of Ref. \cite{TENS4a} has been subsequently questioned by Butcher, Hobson and Lasenby \cite{TENS4b,TENS4c,TENS4d}  who suggested that the proposal of Ref. \cite{TENS4a}  does not have a definite physical significance. In spite of the reasonable concerns of Refs. \cite{TENS4b,TENS4c,TENS4d,TENS4e}, what matters for the present ends is that the geometrical object most closely related to the Babak-Grishchuk suggestion is (again) the Landau-Lifshitz pseudo-tensor \cite{TENS2b} as explicitly recognized by the authors of Ref. \cite{TENS4a}. 

\subsection{Effective action for the relic gravitons}
\subsubsection{Non-covariant derivation}
We start by recalling that  Einstein-Hilbert action can be written in explicit terms by 
isolating the contribution of the total derivatives \cite{TENS2b}. Thanks to this observation 
the sum of the gravity action and of a generic matter action $S_{m}$ becomes:
\begin{equation}
S = \frac{1}{2 \ell_{P}^2} \int d^{4} x \sqrt{-g} \, g^{\alpha\beta}\biggl[ \Gamma_{\alpha\beta}^{\,\,\,\,\,\mu} \Gamma_{\mu\nu}^{\,\,\,\,\,\nu} - \Gamma_{\alpha\nu}^{\,\,\,\,\,\mu} \Gamma_{\mu\beta}^{\,\,\,\,\,\nu} \biggr] + S_{m}
+ \frac{1}{\ell_{P}^2} \int d^{4} x \sqrt{-g}\, \, g^{\alpha\beta} \biggl(\nabla_{\beta} \,\Gamma_{\alpha\lambda}^{\,\,\,\,\,\lambda} - \nabla_{\lambda}  \,\Gamma_{\alpha\beta}^{\,\,\,\,\,\lambda}\biggr),
\label{ACT1}
\end{equation}
where $\ell_{P}^2 = 8 \pi G$ (see also Tab. \ref{SEC1TABLE4} and conventions therein). 
The third term of Eq. (\ref{ACT1}) can be combined in a unique total derivative that does 
not contribute to the second-order variation of the action, i.e.
\begin{equation}
\sqrt{-g}\, \, g^{\alpha\beta}\biggl( \nabla_{\beta} \,\Gamma_{\alpha\lambda}^{\,\,\,\,\,\lambda} - \nabla_{\lambda}  \,\Gamma_{\alpha\beta}^{\,\,\,\,\,\lambda}\biggr) = \nabla_{\gamma}\biggl[ \sqrt{-g}\, \,\biggl( g^{\alpha\gamma} \Gamma_{\alpha\lambda}^{\,\,\,\,\,\lambda}  - g^{\alpha\beta} \, \Gamma_{\alpha\beta}^{\,\,\,\,\,\gamma}\biggr) \biggr].
\label{act1a}
\end{equation}
If we introduce, for convenience, the notation ${\mathcal Z}_{\alpha\beta} = (\Gamma_{\alpha\beta}^{\,\,\,\,\,\mu} \Gamma_{\mu\nu}^{\,\,\,\,\,\nu} - \Gamma_{\alpha\nu}^{\,\,\,\,\,\mu} \Gamma_{\mu\beta}^{\,\,\,\,\,\nu})$, the second-order form of Eq. (\ref{ACT1}) becomes:
\begin{eqnarray}
\delta_{t}^{(2)} S &=& \frac{1}{2 \ell_{P}^2} \int d^{4} x \, 
\biggl\{ \sqrt{- \overline{g}} \biggr[\delta_{t}^{(2)} g^{\alpha\beta} \overline{{\mathcal Z}}_{\alpha\beta}  + \overline{g}^{\alpha\beta}  \delta_{t}^{(2)} {\mathcal Z}_{\alpha\beta} + \delta_{t}^{(1)}\, g^{\alpha\beta} \delta_{t}^{(1)} {\mathcal Z}_{\alpha\beta}\biggr]
\nonumber\\
&+& \delta_{t}^{(2)} \sqrt{-g} \, \overline{g}^{\alpha\beta}\, \overline{{\mathcal Z}}_{\alpha\beta} \biggr\} 
+ \delta_{t}^{(2)} S_{m},
\label{ACT2}
\end{eqnarray}
where $\overline{{\mathcal Z}}_{\alpha\beta}$ denotes the background value\footnote{The components of $\overline{{\mathcal Z}}_{\alpha\beta}$ easily follow and they are $\overline{{\mathcal Z}}_{ij} = 2 \,{\mathcal H}^2 \,\delta_{ij}$ and $\overline{{\mathcal Z}}_{00} =0$.} while 
$\delta_{t}^{(1)}{\mathcal Z}_{\alpha\beta}$ and $\delta_{t}^{(2)} {\mathcal Z}_{\alpha\beta}$ are the first-order and second-order tensor fluctuations of ${\mathcal Z}_{\alpha\beta}$.
 Using the results of Tab. \ref{SEC3TABLE1} 
 the first-order contributions are 
$\delta_{t}^{(1)} \, {\mathcal Z}_{ij} = 2\, {\mathcal H}^2 \, h_{ij}$  and $\delta_{t}^{(1)} \, {\mathcal Z}_{00} = 0$ while the second-order 
contribution can be expressed as:
\begin{eqnarray}
\delta^{(2)} {\mathcal Z}_{00} &=& - \frac{1}{4} h_{k\ell}^{\,\prime} \,  h^{k\ell\,\,\prime} + \frac{{\mathcal H}}{2} h_{k\ell}^{\prime}\, h^{k\ell},
\nonumber\\
\delta^{(2)} {\mathcal Z}_{ij} &=& - \frac{{\mathcal H}}{2} h_{k\ell}^{\prime}\, h^{k\ell} \,\,\delta_{ij} - \frac{1}{4} \biggl[ h_{i}^{\,\,k\,\,\prime}\,\, h_{k\, j}^{\prime} + h_{j}^{\,\,k\,\,\prime}\,\, h_{k\, i}^{\prime}\biggr] 
\nonumber\\
&-& \frac{1}{4} \biggl[ \partial_{\ell} \,h_{i}^{\,\, k} + \partial_{i} h_{\ell}^{\,\, k} - \partial^{k} \,h_{\ell \,i} \biggr] \biggl[ \partial_{k} \,\,h_{j}^{\,\, \ell} + \partial_{j} \,\,h_{k}^{\,\, \ell} - \partial^{\ell} \,\,h_{k \,j} \biggr].
\label{ACT4}
\end{eqnarray}
As usual the prime denotes throughout a derivation with respect to the conformal time 
coordinate $\tau$ (see also Tab. \ref{SEC1TABLE4} and conventions therein). Inserting Eq. (\ref{ACT4}) into Eq. (\ref{ACT2}) the explicit expression
of the second-order action becomes:
\begin{eqnarray}
\delta_{t}^{(2)} S &=& \delta_{t}^{(2)} S_{m} + \frac{1}{ 8 \ell_{P}^2}  \int d^{4} x a^2 \biggl\{ 
h_{ij}^{\prime} \, h^{ij\,\,\prime} - \partial_{k} h_{ij} \partial^{k} h^{\,\,ij} + 
4 {\mathcal H} \biggl[ h_{ij} \, h^{ij \,\,\prime}  +  h_{ij}^{\prime} \, h^{ij} \biggr]  + 6 {\mathcal H}^2 h_{ij}\,h^{ij} 
\nonumber\\
&-&\biggl[ \partial_{k} \biggl( h_{i \ell} \,\,\partial^{\ell} h^{\,\,i k}\biggr)  + \partial_{\ell} \biggl( h_{i k} \,\,\partial^{k} \,h^{\,\,i \ell}\biggr) \biggr] \biggr\},
\label{ACT5}
\end{eqnarray}
so that, neglecting the sources, the effective action of the relic gravitons is:
\begin{equation}
S_{g} = \delta_{t}^{(2)} S  = \frac{1}{8 \ell_{\mathrm{P}}^2} \int d^{4} x \,\,\sqrt{- \overline{g}}\,\,\overline{g}^{\alpha\beta}
\partial_{\alpha} \, h_{i\,j} \, \partial_{\beta} h^{i\, j},
\label{ACT9}
\end{equation}
where we restored the expression of the background metric $\overline{g}_{\alpha\beta}$ and of its determinant. Equation (\ref{ACT9}) coincides 
with the results obtained in Ref. \cite{HIS7} where the second-order tensor action has been obtained for the first time. 
Since the indices of $h_{i\,j}$ are raised and lowered with the help of the three-dimensional Euclidean metric it is 
the same, in practice, to write $\partial_{\alpha} h_{i\,j} \, \partial_{\beta} h^{i\, j}$ and $\partial_{\alpha} h_{i\,j} \, \partial_{\beta} h_{i\, j}$; however it is also useful to keep distinct the two expression 
when defining the corresponding canonically conjugate momenta. 
The possible presence of background sources does not alter the result of Eq. (\ref{ACT9}).
If $\delta^{(2)}_{t} S$ is evaluated by imposing the validity of the background evolution (i.e. on-shell),
the tensor modes decouple from the matter fields provided the anisotropic stress of the sources 
vanishes. For instance, when the sources are parametrized by a perfect relativistic fluid or in terms of a scalar field $\varphi$ we have that the second-order variation of the sources appearing in Eq. (\ref{ACT5}) is:
\begin{eqnarray}
\delta_{t}^{(2)} S_{f} &=& - \frac{1}{4} \int d^{4} x \, \,a^4 \,\,\biggl[ p_{t} \, h_{ij} \,h^{ij} + \Pi_{ij} \,h^{ij} +  \Pi^{ij} \,h_{ij}\biggr],
\label{ACT6}\\
\delta_{t}^{(2)} S_{\varphi} &=& - \frac{1}{4} \int d^{4} x \, a^4 \biggl[ \biggl(\frac{{\varphi}^{\prime\, 2}}{2 a^2} - 
V\biggr) h_{ij} h^{ij} +\Pi_{ij} h^{ij} + \Pi^{ij} h_{ij} \biggr],
\label{ACT7}
\end{eqnarray}
where the contribution of the anisotropic stress $\Pi_{ij}$ has been 
included for the sake of completeness. Inserting Eq. (\ref{ACT6}) into Eq. (\ref{ACT5}) 
the final form of the second-order action becomes:
\begin{eqnarray}
\delta_{t}^{(2)} S &=& \frac{1}{8 \ell_{P}^2} \int d^{4} x \,\, a^2\, \, \biggl[ 
\partial_{\tau} h_{ij} \,\partial_{\tau} h^{\,\,i\,j} - \partial_{k} h_{i\,j} \,\partial^{k} h^{\,\,i\,j} - 
2 a^2 \ell_{P}^2 \Pi_{i\,j} h^{\,i\,j} - 2 a^2 \ell_{P}^2 \Pi^{i\,j} h_{\,i\,j} \biggr] 
\nonumber\\
&-& \frac{1}{2 \ell_{P}^2} \int d^4 x \,\,a^2 h_{ij}\, h^{\,\,ij} \biggl[ \biggl( {\mathcal H}^{\prime} + 
\frac{{\mathcal H}^2}{2} \biggr) + \frac{\ell_{P}^2}{2} a^2 p_{t}\biggr]
\nonumber\\
&+&   \frac{1}{2 \ell_{P}^2} \int d^{4} x \, \biggl\{ \partial_{\tau} \biggl[ a^2 {\mathcal H} h_{ij} \, h^{\,ij} \biggr] - \partial_{k} \biggl[a^2 \,h_{\,i\,\ell} \,\,\partial^{\ell} h^{\,i k} \biggr] \biggr\}.
\label{ACT8}
\end{eqnarray}
Equation (\ref{ACT8}) consists of three distinct contributions that have been formally separated in terms
of three different integrals. The first contribution is just the final form of the effective action while the second contribution  of Eq. (\ref{ACT8}) cancels on-shell\footnote{By combining the background equations (i.e. Eq. (\ref{FL2C}) ) we get  $(2 {\mathcal H}^{\prime} + {\mathcal H}^2 + \ell_{P}^2 a^2 p_{t}) =0$.}. The third contribution of Eq. (\ref{ACT8}) is just a total derivative that does not affect the equations of motion. Neglecting the anisotropic stress, Eq. (\ref{ACT8}) coincides, as anticipated, with Eq. (\ref{ACT9}).  The same steps leading to Eq. (\ref{ACT8}) can be repeated in the case of Eq. (\ref{ACT6}) and  Eq. (\ref{ACT9}) is again recovered.
\begin{table}
\begin{center}
\caption{First- and second-order tensor fluctuations of a conformally flat FRW metric.}
\vskip 0.5truecm
\begin{tabular}{| c | l | c | }
\hline
\quad Implicit form \quad  & 
\quad\quad Explicit expression
      \\ \hline
$ \delta_{t}^{(1)} \Gamma_{i\,j}^{\,\,\,\,\,0}$ & $\frac{1}{2} ( h_{i\,j}^{\prime} + 2 {\mathcal H} h_{i \,j} )$
       \\ \hline 
$  \delta_{t}^{(1)} \Gamma_{i\,0}^{\,\,\,\,\,j}$& $\frac{1}{2} h_{i}^{\,\,\,\,j\,\,\,\prime}$
        \\ \hline
$ \delta_{t}^{(1)} \Gamma_{i\,j}^{\,\,\,\,\,k}$& $\frac{1}{2} ( \partial_{i} \, h_{j}^{\,\,\,k} + \partial_{j} h_{i}^{\,\,\,k} - \partial^{k} h_{i j})$
         \\ \hline
$\delta_{t}^{(2)} \Gamma_{i\, 0}^{\,\,\, j} $ & $- \frac{1}{2} h_{i\, k}^{\,\,\prime}\, h^{j\,k}$
         \\ \hline
$\delta_{t}^{(2)} \Gamma_{i\, j}^{\,\,\, k} $ & $\frac{1}{2} h_{i\, \ell} ( \partial^{\ell} h^{k}_{\,\,\,j} - \partial^{k} \, h^{\ell}_{\,\,\, j} - \partial_{j} \, h^{\, k\ell})$
    \\ \hline
\end{tabular}
\label{SEC3TABLE1}
\end{center}
\end{table}
\subsubsection{Covariant derivation}
From Eqs. (\ref{CUR0b})--(\ref{CUR0c}) and (\ref{CUR0d})--(\ref{CUR0e}) the effective 
action for the relic gravitons can also be derived in a covariant language; for the illustrative 
example of perfect fluid sources the result is:
\begin{equation}
S_{g} = \frac{1}{8 \ell_{P}^2} \int d^{4} x\, \sqrt{-\overline{g}} \biggl[ \overline{\nabla}_{\rho} \overline{h}_{\mu\nu} \, 
\overline{\nabla}^{\rho} \,\overline{h}^{\,\,\mu\nu}
+ 2 \overline{R}_{\rho}^{\,\,\,\,\sigma} \,\, \overline{h}_{\alpha\sigma} \, \overline{h}^{\alpha\rho} + 
2 \, \overline{R}^{\,\mu\,\,\,\,\,\,\,\,\,\,\,\nu}_{\,\,\,\,\,\rho\sigma}\, \overline{h}_{\mu\nu} \, \overline{h}^{\rho \sigma}
+ \ell_{P}^2 ( \rho_{t} - p_{t}) \overline{h}_{\mu\nu} \,\overline{h}^{\mu\nu} \biggr].
\label{ACT9a}
\end{equation}
By extremizing the action (\ref{ACT9a}) with respect to the variation of  $ \overline{h}^{\mu\nu}$
the result of  Eq. (\ref{CUR4}) is recovered. Furthermore, after inserting Eq. (\ref{CUR4a}) into Eq. (\ref{ACT9a}) it follows that the on-shell action decouples from the matter fields:
\begin{equation}
S_{g} = \frac{1}{8 \ell_{P}^2} \int d^{4} x\, \sqrt{-\overline{g}} \biggl[ \overline{\nabla}_{\rho} \overline{h}_{\mu\nu} \,
 \overline{\nabla}^{\rho} \,\overline{h}^{\,\,\mu\nu} + 2 \, \overline{R}^{\mu\,\,\,\,\,\,\,\,\,\,\,\nu}_{\,\,\,\,\,\rho\sigma}\, \overline{h}_{\mu\nu} \, \overline{h}^{\rho \sigma}
\biggr].
\label{ACT9b}
\end{equation}
From Eq. (\ref{CUR7a}) we have that $\overline{h}_{ij} = - a^2 \, h_{ij}$; therefore
 Eq. (\ref{ACT9b}) reproduces exactly (\ref{ACT9}) provided 
$\overline{R}_{k \, \ell\, j\, m} = a^2 \, {\mathcal H}^2 [ \delta_{k \, m} \delta_{\ell\, j} - \delta_{\ell\, m} \delta_{k\, j}]$ which is the relevant component of the background Riemann tensor in the case of a 
conformally flat geometry used to derive Eq. (\ref{ACT9}). The inclusion (or the exclusion) of total derivatives does not alter the equations of motion but the original action (\ref{ACT9}) may take 
different forms. For instance, in terms of the rescaled amplitude $\mu_{ij} = a\, h_{i j}$, Eq. (\ref{ACT9}) can be recast in the  form:
\begin{equation}
S_{g} = \frac{1}{8\ell_{P}^2}\int d^{4} x \biggl[ \partial_{\tau} \mu_{ij}\, \partial_{\tau} \mu^{ij} -  {\mathcal H} \biggl(\mu_{i\,j} \,\partial_{\tau} \mu^{i\,j} + 
\mu^{i\,j}\,\partial_{\tau} \mu_{i\,j} \biggr)
+ {\mathcal H}^2 \mu_{ij} \,\mu^{ij}  - \partial_{k} \mu_{ij}\,\partial^{k}\mu^{ij} \biggr].
\label{ACT9c}
\end{equation}
If the second term inside the square bracket is integrated by parts, Eq. (\ref{ACT9c}) finally reads:
\begin{equation}
S_{g} = \frac{1}{8\ell_{P}^2}\int d^{4} x \biggl[ \partial_{\tau} \mu_{ij} \,\partial_{\tau} \mu^{ij} 
+ ({\mathcal H}^2 + {\mathcal H}^{\prime}) \mu_{ij} \,\mu^{ij}  - \partial_{k} \mu_{ij} \,\partial^{k}\mu^{ij} \biggr].
\label{ACT9d}
\end{equation}
Even if Eqs. (\ref{ACT9}), (\ref{ACT9c}) and (\ref{ACT9d})  are classically equivalent,  they lead to  Hamiltonians \cite{action1,action2} that differ by an appropriate canonical transformations. 

\subsubsection{Effective Hamiltonians}
Denoting by  ${\mathcal L}_{g}(\vec{x}, \tau)$ the Lagrangian density, from  Eq. (\ref{ACT9}) the canonical Hamiltonian is  
\begin{equation}
H_{g}(\tau) = \int d^{3}x\, \biggl[ \pi_{i\,j} \partial_{\tau} h^{i\, j} + \pi^{i\, j} \partial_{\tau} h_{i\,j}  - {\mathcal L}_{g}(\vec{x}, \tau) \biggr], \qquad \pi_{i\,j} = \frac{a^2}{8 \ell_{P}^2} \partial_{\tau} h_{i\, j},
\label{HAM00}
\end{equation}
where  $\pi_{i\,j}$ and $\pi^{i\,j}$ are the canonical momenta (not to be confused with the anisotropic stresses which will be consistently denoted with capital Greek letters). More explicitly Eq. (\ref{HAM00}) becomes:
\begin{equation}
H_{g}(\tau) = \int \, d^{3} x\, \biggl[ \frac{8 \ell_{P}^2}{a^2} \, \pi_{ij} \, \pi^{ij} + \frac{a^2}{8\ell_{P}^2} \partial_{k} h_{i\,j} \, \partial^{k} h^{i\,j}\biggr].
\label{HAM1}
\end{equation}
Since the effective Lagrangian density and the canonical momenta derived from Eq. (\ref{ACT9c}) do not coincide with the ones deduced from Eq. (\ref{ACT9}), the corresponding Hamiltonian will be formally different\footnote{The same remark holds for the Hamiltonian derived from Eq. (\ref{ACT9d}); in this case the corresponding canonical momenta are $\pi_{ij} = \partial_{\tau} \mu_{i\, j} /(8 \ell_{P}^2)$. } from Eq. (\ref{HAM1}):
\begin{eqnarray}
H_{g}(\tau) = \int \, d^{3} x\, \biggl[ 8 \ell_{P}^2 \, \pi_{ij} \, \pi^{ij} + {\mathcal H} \biggl( \mu_{ij} \, \pi^{ij} + \mu^{ij} \, \pi_{ij} \biggr)+ \frac{1}{8\ell_{P}^2} \partial_{k} \mu_{i\,j} \, \partial^{k} \mu^{i\,j}\biggr], 
\,\,\,\,\,\,
\pi_{i\,j} = \frac{\partial_{\tau} \mu_{i\, j} - {\mathcal H} \mu_{i\, j}}{8 \ell_{P}^2}.
\label{HAM2}
\end{eqnarray}
Equations (\ref{HAM1})--(\ref{HAM2}) are related and differ by the (partial) time derivative of the 
functional that generates the corresponding canonical transformation \cite{action1}.  
The state that minimizes a specific Hamiltonian does not coincide, in general, with the state minimizing a canonically related Hamiltonian so that Eqs. (\ref{HAM1}) and (\ref{HAM2}) are associated with different vacua \cite{action3} ultimately responsible for specific corrections to the scalar and tensor power 
spectra; these corrections are often dubbed trans-Planckian and may arise 
from a number of different physical considerations (see, for instance, \cite{action4,action5,action6,action7,action8,action9,action10}) not necessarily related with the physics of the Planck scale.  Indeed the effective description of the tensor modes of the geometry obtained by means of Eq. (\ref{ACT9}) is presumably valid only up to a certain mass scale $M$. If we then demand that physical momenta do not exceed $M$ (i.e. $k/a(\tau)  < M$) there will be a critical conformal time $\tau_{M}$ at which the limit is saturated (i.e. $k/a(\tau_{M}) \simeq M$). 
In the case of an exact de Sitter phase [i.e. $a(\tau) = (-\,H\,\tau)^{-1}$] we will 
have that $k \tau_{M} = - (M/H)$ so that, typically, $|k \tau_{M}| \gg 1$ at least in the context of the conventional inflationary dynamics.
If the mode functions are normalized at $\tau_{M}$ the tensor (and scalar) power spectra inherit computable corrections \cite{action6,action7,action8,action9,action10,action11} that are in principle measurable and that have been observationally scrutinized in the recent past \cite{action12,action13,action14} without any conclusive evidence. These unobserved corrections are given as inverse powers of $| k \tau_{M}|$ and depend on the form of the Hamiltonian since, as explained, canonically related Hamiltonians define different vacua. In the quasi-de Sitter case, the Hamiltonian derived from the action (\ref{ACT9}) leads to a correction ${\mathcal O}( |k\tau_{M}|^{-1})$ whereas 
the Hamiltonians derived from the actions (\ref{ACT9c}) and (\ref{ACT9d}) lead to corrections 
${\mathcal O}( |k\tau_{M}|^{-2})$ and ${\mathcal O}( |k\tau_{M}|^{-3})$ respectively. 
 Since the different vacua obtained from the canonically related Hamiltonians gravitate, they will be responsible for different back-reaction effects \cite{action1} and on this 
basis the largest corrections to the power spectrum can be excluded. A posteriori 
the lack of observations of these corrections \cite{action12,action13,action14}  
seems to naively confirm the validity of the backreaction constraints derived in \cite{action1}. 

\subsubsection{Frame-invariance of the effective action}
The effective action (\ref{ACT9}) {\em is invariant under infinitesimal coordinate transformations and  it is also frame-invariant}. By this we mean that the evolution of the tensor modes 
can be deduced either in the Einstein frame (where the gravitational action 
takes the Einstein-Hilbert form) or in a conformally related frame. In a generalized Jordan frame the scalar-tensor action can always be written as:
\begin{equation}
S_{J} = \int d^4 x\, \sqrt{ - G}\,\biggl[ -  \frac{A(\overline{\varphi})}{ 2 \ell_{P}^2}\, R_{J} + \frac{B(\overline{\varphi})}{2} G^{\alpha\beta} \partial_{\alpha} \,\overline{\varphi} \partial_{\beta} \overline{\varphi} 
- W(\overline{\varphi}) \biggr],
\label{ACT10}
\end{equation}
where $A(\overline{\varphi}) $ and $B(\overline{\varphi})$ are dimensionless and depend on the scalar field $\overline{\varphi}$ (that should not be necessarily identified with the inflaton); note that in the limit $A\to 1$ and $B\to 1$ the standard minimally coupled action is recovered. The conformal rescaling bringing the action 
(\ref{ACT10}) in its Einstein frame form is:
\begin{equation}
A \,\, G_{\alpha\beta} = g_{\alpha\beta}, \qquad d \overline{\varphi} \,\,\sqrt{ \frac{B}{A} + \frac{3}{2 \ell_{P}^2} \biggl(\frac{d \ln{A}}{d \overline{\varphi}}\biggr)^2 } = d \varphi, \qquad V = \frac{W}{A^2}.
\label{EIN11}
\end{equation}
Note that in Eq. (\ref{EIN11}) the scalar field has been redefined to obtain a canonical kinetic term in the new frame.  Thanks to Eq. (\ref{EIN11}), Eq. (\ref{ACT10}) becomes, up to a total derivative, 
\begin{equation}
S_{E} = \int d^4 x\, \sqrt{ - g}\,\biggl[ - \frac{R}{ 2 \ell_{P}^2} \,+ \,\frac{1}{2} g^{\alpha\beta} \partial_{\alpha} \varphi \partial_{\beta} \varphi 
- V(\varphi) \biggr].
\label{EIN10}
\end{equation}
For the background and for the first-order tensor fluctuations the metric rescaling of Eq. (\ref{EIN11}) becomes more explicit and it is given by:
\begin{equation}
a_{J}^2 \,\,A = a^2, \qquad A \,\,a^2_{J} \,\,h^{(J)}_{ij} = a^2\,\, h_{ij}.
\label{ACT11a}
\end{equation}
The first equality of Eq. (\ref{ACT11a}) follows from the conformal rescaling of the 
background (i.e. $A \overline{G}_{\alpha\beta} = \overline{g}_{\alpha\beta}$) 
while the second equality is implied by the relation between the first-order tensor fluctuations in the two frames (i.e. $A \delta_{t}^{(1)} G_{ij} = \delta_{t}^{(1)} g_{ij}$). In the Jordan frame the second-order variation of Eq. (\ref{ACT10}) 
 can be easily obtained by repeating the same steps leading to Eqs. (\ref{ACT9}) and the result reads:
\begin{eqnarray}
 \delta^{(2)}_{t} S_{J} &=&\int d^{4} x  \, \biggl\{ \frac{1}{2 \ell_{P}^2}\,\biggl[ A(\overline{\varphi}) \, \overline{G}^{\alpha\beta} \, \overline{{\mathcal Z}}_{\alpha\beta} \,\,\delta^{(2)}_{t} \sqrt{-G} + A(\overline{\varphi})\, \sqrt{ -\overline{G}} \biggl( \delta^{(2)}_{t} G^{\alpha\beta} \,\overline{{\mathcal Z}}_{\alpha\beta} +
 \delta^{(1)}_{t} G^{\alpha\beta} \,  \delta^{(1)}_{t} {\mathcal Z}_{\alpha\beta} 
 \nonumber\\
 &+& \overline{G}^{\alpha\beta}\,\,\delta^{(2)}_{t} {\mathcal Z}_{\alpha\beta} \biggr)
- \delta^{(2)}_{t}\biggl( \sqrt{-G} \, G^{\alpha\beta} \, \Gamma_{\alpha\lambda}^{\,\,\,\,\,\lambda} \, \partial_{\beta} A(\overline{\varphi})\biggr) 
+  \delta^{(2)}_{t}\biggl( \sqrt{-G} \, G^{\alpha\beta} \, \Gamma_{\alpha\beta}^{\,\,\,\,\,\lambda} \, \partial_{\lambda} A(\overline{\varphi})\biggr) \biggr]
\nonumber\\
&+&  \delta_{t}^{(2)} \sqrt{-G} \,\biggl(\frac{B(\overline{\varphi})}{2} \overline{G}^{\alpha\beta} \partial_{\alpha} \overline{\varphi} \,\partial_{\beta}\overline{\varphi} 
- W(\overline{\varphi})\biggr) + \sqrt{-\overline{G}} \,\frac{B(\overline{\varphi})}{2} \delta_{t}^{(2)} G^{\alpha\beta} \,\partial_{\alpha} \overline{\varphi} \partial_{\beta}\overline{\varphi} \biggr\}.
\label{ACT10a}
\end{eqnarray}
 After some lengthy but straightforward algebra the explicit form of Eq. (\ref{ACT10a}) reads:
\begin{eqnarray}
&& \delta^{(2)} S_{J} = \frac{1}{8 \ell_{P}^2} \int d^{4}x \sqrt{-\overline{G}} \, \overline{G}^{\alpha\beta} \, \, A(\overline{\varphi}) \, \partial_{\alpha} h^{\,\,\,\,(J)}_{i\,j}
\partial_{\beta} h^{(J)\,\,\,i\,j} 
- \frac{1}{8\ell_{P}^2} \int d^{4} x \, a^2_{J} A(\overline{\varphi}) h^{\,\,\,\,(J)}_{k\ell}  h^{(J)\,\,k \ell} \bigg[ 4 {\mathcal H}_{J}^{\prime} 
\nonumber\\
&& + 2 {\mathcal M}^{\prime}+ 2 ({\mathcal H}_{J}^2 + {\mathcal H}_{J} {\mathcal M} + {\mathcal M}^2) 
+ \frac{ 2 \ell_{P}^2}{A} \biggl( \frac{B}{2} \overline{\varphi}^{\prime \, \,2} - W\, a_{J}^2\biggr)\biggr],
\label{ACT10b}
\end{eqnarray}
where ${\mathcal M} = A^{\prime}/A$ and ${\mathcal H}_{J} = a_{J}^{\prime}/a_{J}$. The tensor amplitude $h^{(J)}_{ij}$ entering Eq. (\ref{ACT10b}) is defined directly
 in the Jordan frame, i.e.  $\delta_{t}^{(1)} G_{ij} = -a_{J}^2 \, h^{\,(J)}_{ij}$;  
 $a_{J}$ is the scale factor appearing in  the $J$-frame, i.e. $\overline{G}_{\alpha\beta} = a^2_{J} \,\,\eta_{\alpha\beta}$.The expression inside the squared bracket of Eq. (\ref{ACT10b}) vanishes identically since it 
corresponds to the $(ij)$ component of the  background equations derived from the extremization of the action (\ref{ACT10a}) with 
respect to the variation of the metric. Therefore the final result is\footnote{The value of $A(\overline{\varphi})$ must be evaluated on the background; however the first- and second-order fluctuations 
of  $A(\overline{\varphi})$  only contribute to the effective action of the scalar modes and do not affect the tensor problem discussed here.}
\begin{eqnarray}
&& \delta^{(2)} S_{J} = \frac{1}{8 \ell_{P}^2} \int d^{4}x \sqrt{-\overline{G}} \, \overline{G}^{\alpha\beta} \, \, A(\overline{\varphi}) \, \partial_{\alpha} h^{\,\,\,\,(J)}_{i\,j}
\partial_{\beta} h^{(J)\,\,\,i\,j}
\nonumber\\
&& =  \frac{1}{8 \ell_{P}^2} \int d^{4}x \, a_{J}^2\, A(\overline{\varphi}) \, \eta^{\alpha\beta}\,\partial_{\alpha} h^{\,\,\,\,(J)}_{i\,j}
\partial_{\beta} h^{(J)\,\,\,i\,j} \equiv \frac{1}{8 \ell_{P}^2} \int d^{4}x \, a^2 \biggl[ \partial_{\tau} h_{i\,j}\partial_{\tau} h^{i\,j} -\partial_{k} h_{i\,j}\partial^{k} h^{i\,j} \biggr].
\label{ACT11}
\end{eqnarray}
The first equality in Eq. (\ref{ACT11}) follows from the explicit form of the effective action in the case of a conformally flat background. The third relation in Eq. (\ref{ACT11}) follows instead from the two conditions of Eq. (\ref{ACT11a}) and it ultimately 
implies that Eqs. (\ref{ACT11}) and (\ref{ACT9}) are one and the same equation. 

\subsection{Polarized backgrounds of relic gravitons and effective theories} 
Equation (\ref{EIN10}) can be regarded as the first term of a 
generic effective field theory of inflation \cite{EFPOL1} where the higher derivatives are suppressed by the 
negative powers of a large mass scale $M< M_{P}$ that characterizes the fundamental theory underlying 
the effective description. Even if we shall always consider dimensionful scalar fields, in this particular 
context it is useful to deal with an appropriate dimensionless scalar $\phi = \varphi/M$ in terms of which 
the action (\ref{EIN10}) can be expressed as\footnote{The field $\phi$ introduced in this section has nothing to do with the scalar metric fluctuation introduced in Eq. (\ref{COS2}). Since these two quantities will never appear in the same context, no confusion is possible.}:
\begin{equation}
S_{0} = \int d^{4} x \, \sqrt{-g} \biggl[ - \frac{\overline{M}_{P}^2}{2} R + \frac{M^2}{2} g^{\mu\nu} \partial_{\mu} \phi \partial_{\nu} \phi - M_{P}^2 U(\phi)\biggr], 
\qquad U(\phi) = \frac{V(M \phi)}{M_{P}^2},
\label{EIN10a}
\end{equation} 
where, as usual in our notations, $\overline{M}_{P} = M_{P}/\sqrt{8 \pi}$.
The leading correction to Eq. (\ref{EIN10a}) consists of all possible terms containing four derivatives \cite{EFPOL1}:
\begin{eqnarray}
&& {\mathcal L}_{corr} = \sqrt{-g} \biggl[ c_{1}(\phi) \biggl(g^{\alpha\beta} \partial_{\alpha} \phi \partial_{\beta} \phi\biggr)^2 
+ c_{2}(\phi) g^{\mu\nu} \partial_{\mu} \phi \partial_{\nu} \phi \Box \phi + 
c_{3}(\phi) \bigl( \Box \phi \bigr)^2 
\nonumber\\
&& + c_{4}(\phi) \, R^{\mu\nu} \, \partial_{\mu} \phi \partial_{\nu} \phi
+ c_{5}(\phi) \, R\, g^{\mu\nu} \partial_{\mu} \phi \partial_{\nu} \phi
+ c_{6}(\phi) R \, \Box \phi + c_{7}(\phi) R^2 + c_{8}(\phi) \, R_{\mu\nu} \,R^{\mu\nu} 
\nonumber\\
&& + c_{9}(\phi) R_{\mu\alpha\nu\beta} \, R^{\mu\alpha\nu\beta} + c_{10}(\phi) C_{\mu\alpha\nu\beta} \, C^{\mu\alpha\nu\beta}
+ c_{11}(\phi)  R_{\mu\alpha\nu\beta} \, \widetilde{\,R\,}^{\mu\alpha\nu\beta} + c_{12}(\phi) C_{\mu\alpha\nu\beta} \, \widetilde{\,C\,}^{\mu\alpha\nu\beta}\biggr],
\label{EFFPP1}
\end{eqnarray}
where $R_{\mu\alpha\nu\beta}$ and $C_{\mu\alpha\nu\beta}$ denote the Riemann and  Weyl tensors 
while 
\begin{equation}
\widetilde{\,R\,}^{\mu\alpha\nu\beta} = \frac{1}{2} E^{\mu\alpha\rho\sigma} R_{\rho\sigma}^{\,\,\,\,\,\,\,\,\nu\beta},\qquad  \widetilde{\,C\,}^{\mu\alpha\nu\beta} = \frac{1}{2} E^{\mu\alpha\rho\sigma} C_{\rho\sigma}^{\,\,\,\,\,\,\,\,\nu\beta}, \qquad E^{\mu\alpha\rho\sigma} = \frac{\epsilon^{\mu\alpha\rho\sigma}}{\sqrt{-g}}
\label{dual}
\end{equation}
 are the corresponding dual tensors and $\epsilon^{\mu\alpha\rho\sigma}$ is the Levi-Civita symbol.
Equation (\ref{EFFPP1}) contains $12$ distinct terms but it can be argued that the truly 
independent terms are only $10$ \cite{EFPOL1} since the four  bilinears in $R_{\alpha\mu\beta\nu}$ and $C_{\alpha\mu\beta\nu}$ are in fact equivalent 
 so that we must only include the terms involving either the Riemann or the Weyl tensor; we prefer however to keep 
 distinct the four bilinears in the Weyl and Riemann tensors\footnote{The Weyl tensor vanishes for a spatially flat Friedmann- Robertson Walker metric since it is the traceless part of the Riemann tensor; the derivation of the second-order action describing the tensor modes is comparatively easier in the Weyl rather than in the Riemann case.}. Further possible terms are equivalent 
 to the ones already appearing in Eq. (\ref{EFFPP1}) up to total derivatives \cite{EFPOL1a,EFPOL1b}.
If we took Eq. (\ref{EFFPP1}) literally, we would find more than just the usual two adiabatic modes for single-field inflation.
However, if we follow the logic of the effective theory  \cite{EFPOL1},  all second time derivatives and time derivatives of auxiliary fields must 
be eliminated by using the field equations derived from the leading terms of Eq. (\ref{EIN10a}).
If the coupling terms and the potential are appropriately redefined, the only three terms 
contributing to the effective action besides the leading terms appearing in Eq. (\ref{EIN10a}) are
\begin{equation}
 {\mathcal L}_{corr} = \sqrt{-g} \biggl[ \widetilde{\,c\,}_{1}(\phi) \biggl(g^{\alpha\beta} \partial_{\alpha} \phi \partial_{\beta} \phi\biggr)^2
 + f_{1}(\phi)  R_{\mu\alpha\nu\beta} \, \widetilde{\,R\,}^{\mu\alpha\nu\beta} + f_{2}(\phi) C_{\mu\alpha\nu\beta} \, \widetilde{\,C\,}^{\mu\alpha\nu\beta}\biggr].
 \label{EFFPP3a}
 \end{equation}
While the first term in Eq. (\ref{EFFPP3a}) does not affect the evolution of the tensor modes, the other two 
(containing either $f_{1}(\phi)$ or $f_{2}(\phi)$) may polarize the backgrounds of relic gravitons.
Theories containing either $f_{1}(\phi)$ of $f_{2}(\phi)$ have been discussed in different contexts \cite{EFPOL3,EFPOL4,EFPOL5,EFPOL5a} related to the Chern-Simons modifications 
of the general relativistic dynamics. For practical reasons the terms affecting the tensor effective action 
can be normalized as:
\begin{eqnarray}
S_{1} = - \frac{\beta_{1}}{8} \int d^{4} x \sqrt{-g} f_{1}(\phi) \, R_{\mu\alpha\nu\beta}\,\widetilde{\,R\,}^{\mu\alpha\nu\beta},\qquad S_{2}  =  - \frac{\beta_{2}}{8} \int d^{4} x \sqrt{-g} f_{2}(\phi)\, C_{\mu\alpha\nu\beta}\,\widetilde{\,C\,}^{\mu\alpha\nu\beta}, 
\label{POLGRAV2}
\end{eqnarray}
where  $\beta_{1}$ and $\beta_{2}$ are just two numerical constants; even if $f_{1}(\phi)$ and $f_{2}(\phi)$ 
 have been introduced as dimensionless couplings depending upon the (dimensionless) inflaton $\varphi$, they may also contain the coupling to some other spectator field.  If we switch on the gauge fields, a further class of vertices 
able to polarize the relic gravitons can be obtained;
by contracting the Riemann tensor with the gauge field strength $Y^{\mu\alpha}$ and its dual  $\widetilde{Y}^{\nu\beta}$ the effective vertex becomes \cite{EFPOL6}:
\begin{eqnarray}
S_{3} =  - \frac{1}{2\ell_{P}^2 M^4} \int d^4 x \,f_{3}(\phi) \,R_{\mu\alpha\nu\beta} \, Y^{\mu\alpha} \, \widetilde{\,Y\,}^{\nu\beta},\qquad 
S_{4}  =  - \frac{1}{2\ell_{P}^2 M^4} \int d^4 x \,f_{4}(\phi) \,C_{\mu\alpha\nu\beta} \, Y^{\mu\alpha} \, \widetilde{\,Y\,}^{\nu\beta}.
\label{POLGRAV3}
\end{eqnarray}

\subsubsection{Spinor action for polarized backgrounds of relic gravitons}
The effective action of the relic gravitons can be written in a more compact form by rearranging the two polarizations in a single bispinor $\Psi$ \cite{EFPOL6}. This strategy is reminiscent of the so-called Jones calculus\footnote{In optical applications the Jones calculus stipulates that the electric fields of the waves are organized in a two-dimensional column vector. The Jones approach is customarily contrasted with the Mueller calculus where the polarization is described by a four-dimensional (Mueller) column vector whose components are the four Stokes parameters \cite{EFPOL2}. Both formalisms also play a relevant role in the analysis 
of the polarization anisotropies of the CMB to be discussed in section \ref{sec5}.}. The action describing the dynamics of  $\Psi$ must be invariant under infinitesimal diffeomorphisms; it must contain (at most) two derivatives with respect to the conformal time coordinate $\tau$ and it should reduce to Eq. (\ref{ACT9}) when the interactions between the  polarizations are absent. Putting together these three requirements we are led to the following form of the effective action in Fourier space \cite{EFPOL6}: 
\begin{eqnarray}
S_{pol} &=& \frac{1}{2} \int d^{3} k \int d\tau \biggl\{ a^2(\tau)\biggl[ \partial_{\tau} \Psi^{\dagger} \partial_{\tau} \Psi - 
k^2 \Psi^{\dagger} \Psi  \biggr] +  \Psi^{\dagger} (\vec{v}\cdot\vec{\sigma}) \Psi +   \partial_{\tau} \Psi^{\dagger} (\vec{r}\cdot\vec{\sigma}) \partial_{\tau} \Psi
\nonumber\\
&+&  \Psi^{\dagger} (\vec{p}\cdot\vec{\sigma}) \partial_{\tau} \Psi +  \partial_{\tau} \Psi^{\dagger} (\vec{q}\cdot\vec{\sigma}) 
\Psi  \biggr\},\qquad \Psi = \left(\matrix{h_{\oplus}\cr h_{\otimes}} \right),
\label{POLGRAV4}
\end{eqnarray}
where, as in Eq. (\ref{SIXPOL11}), $\vec{\sigma} =(\sigma_{1}, \, \sigma_{2},\, \sigma_{3})$ and $\sigma_{i}$ (with $i=1,\,\, 2,\,, 3$) are the three Pauli matrices and ${\bf 1}$ is the identity matrix.  The dagger denotes, as usual, the transposed and complex conjugate of the corresponding spinor or matrix. Note that $h_{\oplus}$ and $h_{\otimes}$ are the Fourier 
amplitudes appearing in Eq. (\ref{POLDEF5}). In the parametrization of Eq. (\ref{POLGRAV4}) the vector $\vec{r}(k,\tau)$ is dimensionless while the remaining three vectors $\vec{v}(k,\tau)$, $\vec{p}(k,\tau)$, $\vec{q}(k,\tau)$ are all dimensional and may otherwise contain an arbitrary dependence on $k$. Finally, since the quantum Hamiltonian associated with the action (\ref{POLGRAV4}) must be Hermitian we are led to demand that $\vec{p} = \vec{q}$.  Thanks to these plausible requirements the terms containing a single conformal time derivative can be eliminated by dropping a total time derivative while $\vec{b}$ is redefined as $\vec{v} \to \vec{b}= \vec{v} - \partial_{\tau} \vec{p}$. Therefore the canonical form of Eq. (\ref{POLGRAV4}) is always expressible as:
\begin{equation}
S_{pol} = \frac{1}{2} \int d^{3} k \int d\tau \biggl\{ a^2(\tau)\biggl[\partial_{\tau} \Psi^{\dagger} \partial_{\tau} \Psi - k^2 \Psi^{\dagger} \Psi  \biggr] +  \Psi^{\dagger} (\vec{b}\cdot\vec{\sigma}) \Psi +  \partial_{\tau} \Psi^{\dagger} (\vec{r}\cdot\vec{\sigma}) \partial_{\tau} \Psi  \biggr\}.
\label{POLGRAV5}
\end{equation}
In the non-interacting limit (i.e. when all the vectors are identically vanishing) the actions (\ref{POLGRAV4})--(\ref{POLGRAV5})  coincide with the result obtained from Eq. (\ref{ACT9}) and the two polarizations evolve independently. The physical model and the form of the interaction is specified by the components of the vectors $\vec{b}(k,\tau)$ and $\vec{r}(k,\tau)$. As an example the effective action  (\ref{POLGRAV5})  can be derived in the case where the only contribution besides 
the leading-order result is represented by the interaction term $S_{1}$ given in Eq. (\ref{POLGRAV2}): 
\begin{eqnarray}
 S_{pol} &=& \frac{1}{2} \int d^{3} k \int d\tau \biggl\{ a^2 \biggl[ \partial_{\tau} h_{\oplus} \partial_{\tau} h_{\oplus}^{*} 
+ \partial_{\tau} h_{\otimes} \partial_{\tau} h_{\otimes}^{*} - k^2 \biggl( h_{\oplus} h_{\oplus}^{*} 
+  h_{\otimes} h_{\otimes}^{*} \biggr)\biggr]
\nonumber\\
&+& i k \beta_{1} \,\ell_{P}^2\, \partial_{\tau} f_{1} \biggl[  \partial_{\tau} h_{\oplus} \partial_{\tau} h_{\otimes}^{*} - \partial_{\tau} h_{\otimes} \partial_{\tau} h_{\oplus}^{*} - k^2 \biggl(  h_{\oplus} h_{\otimes}^{*}  -h_{\otimes} h_{\oplus}^{*}\biggr) \biggr] \biggr\}.
\label{POLGRAV6}
\end{eqnarray}
If the two linear polarizations appearing in Eq. (\ref{POLGRAV6}) are arranged into the components of the bispinor $\Psi$ the resulting action is much simpler:
\begin{equation}
S_{pol} = \frac{1}{2} \int d^{3} k \int d\tau \biggl\{ a^2(\tau)\biggl[\partial_{\tau} \Psi^{\dagger} \partial_{\tau} \Psi - k^2 \Psi^{\dagger} \Psi  \biggr] +  b_{2}(k,\tau) \Psi^{\dagger} \sigma_{2} \Psi +  r_{2}(k,\tau) \partial_{\tau} \Psi^{\dagger} \sigma_{2}  \partial_{\tau} \Psi  \biggr\},
\label{POLGRAV7}
\end{equation}
where $b_{2} = - k^3 \beta_{1} \ell_{P}^2 (\partial_{\tau} f_{1})$ and $ r_{2} =  k \beta_{1} \ell_{P}^2 (\partial_{\tau} f_{1})$.
The following unitary transformation (i.e. $U^{\dagger} = U^{-1}$) diagonalizes Eq.  (\ref{POLGRAV7}) 
\begin{equation}
\Psi = U \Phi, \qquad U = \left(\matrix{1/\sqrt{2}
& 1/\sqrt{2}\cr i/\sqrt{2}&- i/\sqrt{2}}\right),  \qquad \Phi = \left(\matrix{h_{R}\cr h_{L}} \right),
\label{sixc}
\end{equation}
where $h_{L}$ and $h_{R}$ denote the left-handed and right-handed combinations (see Eq. (\ref{SIXPOL10}) and discussion thereafter). Since $U^{\dagger} \sigma_{2} U = \sigma_{3}$, Eq. (\ref{POLGRAV7}) becomes:
\begin{equation}
S_{pol} = \frac{1}{2} \int d^{3} k \int d\tau \biggl\{ a^2(\tau)\biggl[\partial_{\tau} \Phi^{\dagger} \partial_{\tau} \Phi - k^2 \Phi^{\dagger} \Phi  \biggr] +  b_{2}(k,\tau) \Phi^{\dagger} \sigma_{3} \Phi +  r_{2}(k,\tau) \partial_{\tau} \Phi^{\dagger} \sigma_{3}  \partial_{\tau} \Phi  \biggr\}.
\label{POLGRAV8}
\end{equation}
Equation (\ref{POLGRAV8}) can be expressed in an even more compact form 
by introducing two appropriate matrices $Z$ and $W$:
\begin{equation}
S_{pol} = \frac{1}{2} \int d^{3}k \,\int d\tau\, \biggl[ \partial_{\tau} \,\Phi^{\dagger} \,\,Z\,\, \partial_{\tau} \,\Phi - \Phi^{\dagger}\,\, W \,\,\Phi \biggr].
\label{POLGRAV10}
\end{equation}
The two matrices appearing in Eq. (\ref{POLGRAV10}) are $Z(k,\tau)= \{[a^2(\tau) +  r_{2}(k,\tau) ] P_{R} +  [a^2(\tau) -  r_{2}(k,\tau) ] P_{L}\}$
and $W(k,\tau) = \{[k^2 a^2(\tau) -  b_{2}(k,\tau)] P_{R} +  [k^2 a^2(\tau) +  b_{2}(k,\tau)] P_{L}\}$
where $P_{L}= (I- \sigma_{3})/2$ and $P_{R} =  (I+ \sigma_{3})/2$ denote the left and right projectors while $I$ is the identity matrix.
In the concordance paradigm the degree of circular polarization induced by the effective terms of Eq. (\ref{POLGRAV2}) 
is rather small; for instance if $f_{1}(\phi) = \phi$ we would have that ${\mathcal O}(10^{-13})$ for typical values of the 
slow-roll parameters. However larger valued may be expected in more exotic scenarios \cite{EFPOL6}. 

\subsection{Classical and quantum fluctuations in cosmology}
During an inflationary stage of expansion the classical and quantum fluctuations 
obey the same evolution equations in the linearized approximation but 
while the classical fluctuations are given once forever (on a given space-like hypersurface),
the quantum fluctuations keep on reappearing all the time during the inflationary phase. 
Assuming a sudden reheating, the kinematical problems of a decelerated 
cosmological scenario are solved by an early phase of accelerated 
expansion lasting (at least) $65$ $e$-folds \cite{HIS10b,HIS10c,HIS10d,HIS10e}.
While the initial classical fluctuations (i.e. spatial gradients) are exponentially suppressed 
during inflation \cite{class1,class2,class3,class4,class5,class6} what ultimately decides 
the potential effects of classical and quantum inhomogeneities 
is the overall duration of the inflationary stage of expansion.  In the framework of the conventional inflationary scenario the role of the classical and of the quantum 
fluctuations depends, in practice, on the total number of $e$-folds $N_{t}$: if $N_{t} \gg {\mathcal O}(65)$ 
the only possible source of inhomogeneity is provided by the quantum mechanical inhomogeneities \cite{book1,book2,book3} and this conclusion holds, in particular, for low-frequency modes of the spectrum of the cosmic gravitons (i.e. for comoving frequencies in the aHz region). In the opposite situation, i.e. when
$N_{t} \leq {\mathcal O}(65)$,  the classical fluctuations undoubtedly present in the initial state may play a 
relevant role especially at low frequencies (see in this respect the subsection \ref{subs56}).  Denoting with $t$ the cosmic time coordinate, the gradient expansion in the proximity of the big-bang singularity is formally defined in the limit $t\to 0^{+}$  where the spatial gradients turn out to be subdominant in comparison with the extrinsic curvature: close to the singularity the geometry may be highly anisotropic but rather homogeneous \cite{HIS1b,HIS1c,HIS1d}. As soon as an inflationary event horizon is formed, a complementary gradient expansion is defined in the limit $t \to \infty$ \cite{class1,class2,class3,class4,class5,class6}.  In this framework it is plausible that any finite portion of the event horizon gradually loses the memory of an initially imposed anisotropy or inhomogeneity so that the metric attains the observed regularity regardless of the initial boundary conditions; this is the logic of the so-called cosmic no-hair conjecture. In recent years 
various authors tried to replace or complement an inflationary stage of expansion with some other less conventional alternatives involving directly or indirectly a contracting phase. Generally speaking, if  all matter components have standard equations of state  then a contracting universe is unstable against perturbative deviations from isotropy so that an initially imposed anisotropy may increase and bring the metric in a Kasner-like form\footnote{This means, in three dimensions, that the scale factors associated with the spatial dimensions (i.e. $a_{i}(t) \propto (- t)^{p_{i}}$ with $i= 1,\,2,\, 3$) satisfy 
$\sum_{i} p_{i}^2 = \sum_{i} p_{i} =1 $.}. In the simplest case all the spatial dimensions except for one shrink away and the metric jumps from one Kasner form to another one: these are the so-called chaotic
Belinsky-Lifshitz- Khalatnikov oscillations \cite{HIS1d,HIS1e}. There are however specific realizations
where this does not happen\footnote{The simplest example along this direction is provided by a minimally 
coupled scalar field; the presence of only one scalar field leads to a monotonic power-law (asymptotic)
 solution near the singularity \cite{HIS1f}. However the addition of vector fields restores the  
 Belinsky-Lifshitz- Khalatnikov oscillations.}: in these situations the quantum treatment of the scalar 
 and tensor modes is essential as in the case of conventional inflationary models and this is the reason 
 why it is useful to discuss the problem in general terms. 

\subsubsection{The quantum theory of parametric amplification}
If we express the normal modes $\mu_{ij}$ and the anisotropic stress $\Pi_{ij}$ in terms of their polarization components 
\begin{equation}
\mu_{ij}(\vec{x},\tau) = a \,h_{ij} =\sqrt{2} \, \ell_{P} \sum_{\alpha= \oplus,\,\otimes} \, \mu_{\alpha} \, e^{(\alpha)}_{ij}, 
\qquad \Pi_{ij}(\vec{x},\tau) = \sqrt{2} \, \ell_{P} \sum_{\alpha= \oplus,\,\otimes} \, \Pi_{\alpha} \, e^{(\alpha)}_{ij},
\label{QTP2}
\end{equation}
the resulting tensor Hamiltonian derived from the action Eq. (\ref{ACT9c}) is
\begin{equation}
H_{g}(\tau) = \frac{1}{2} \sum_{\alpha} \int d^{3} x \biggl[ \pi_{\alpha}^2 + 2 {\mathcal H} \mu_{\alpha} \pi_{\alpha} + \partial_{k} \mu_{\alpha} \partial_{k} \mu_{\alpha} + 
4 \ell_{P}^2 a^3 \mu_{\alpha} \Pi_{\alpha}\biggr],
\label{QTP3}
\end{equation}
where $\pi_{\alpha} = \mu_{\alpha}^{\prime} - {\mathcal H} \mu_{\alpha}$. The normal modes of the field can then be promoted to the status of field operators obeying the canonical commutation relations at equal times:
\begin{equation}
[ \hat{\mu}_{\alpha}(\vec{x}, \tau),\,\, \hat{\pi}_{\beta}(\vec{y}, \tau) ] = i \, \delta^{(3)}(\vec{x} - \vec{y}) \, \delta_{\alpha\beta}.
\label{QTP4}
\end{equation}
The Hermitian operators $\hat{\mu}_{\alpha}(\vec{x}, \tau)$ and $\hat{\pi}_{\beta}(\vec{y}, \tau)$ can be represented 
in Fourier space as:
\begin{equation}
\hat{\mu}_{\alpha}(\vec{x}, \tau) = \frac{1}{(2\pi)^{3/2}} \int d^{3} p \, \hat{\mu}_{\vec{p},\,\alpha}(\tau) 
e^{- i\, \vec{p}\cdot\vec{x}}, \qquad 
\hat{\pi}_{\alpha}(\vec{x}, \tau) = \frac{1}{(2\pi)^{3/2}} \int d^{3} p \, \hat{\pi}_{\vec{p},\,\alpha}(\tau) 
e^{- i\, \vec{p}\cdot\vec{x}}.
\label{QTP5}
\end{equation}
Using Eqs. (\ref{QTP5}) the Hamiltonian (\ref{QTP3}) takes then the form:
\begin{eqnarray}
\hat{H}_{g}(\tau) &=& \frac{1}{2} \int d^{3} p \sum_{\alpha} \biggl\{ \hat{\pi}_{-\vec{p},\,\alpha}\, \hat{\pi}_{\vec{p},\,\alpha} + p^2 \hat{\mu}_{-\vec{p},\,\alpha}\, \hat{\mu}_{\vec{p},\,\alpha} 
+ {\mathcal H} \biggl[ \hat{\pi}_{-\vec{p},\,\alpha}\, \hat{\mu}_{\vec{p},\,\alpha} + 
\hat{\mu}_{-\vec{p},\,\alpha}\, \hat{\pi}_{\vec{p},\,\alpha} \biggr] 
\nonumber\\
&+& 2 \ell_{P}^2 \, a^3 \biggl[ \hat{\Pi}_{-\vec{p},\,\alpha}\, \hat{\mu}_{\vec{p},\,\alpha} + 
\hat{\mu}_{-\vec{p},\,\alpha}\, \hat{\Pi}_{\vec{p},\,\alpha}\biggr]\biggr\}.
\label{QTP7}
\end{eqnarray}
Inserting Eq. (\ref{QTP5}) into Eq. (\ref{QTP4}), the corresponding commutation relation can be written as 
$[ \hat{\mu}_{\vec{k},\,\alpha}(\tau),\,\,\hat{\pi}_{\vec{p},\,\beta}(\tau)] = i \, \delta^{(3)}(\vec{k} + \vec{p}) \, \delta_{\alpha\beta}$ and it implies that $\hat{\mu}_{\vec{p} \, \alpha}$ and $\hat{\pi}_{\vec{p} \,\alpha}$ are expressible in terms of the creation and annihilation operators of 
gravitons with opposite three-momenta:
\begin{equation}
\hat{\mu}_{\vec{p} \, \alpha} = \frac{1}{\sqrt{2 p}} \biggl[ \hat{a}_{\vec{p},\,\alpha} + \hat{a}^{\dagger}_{-\vec{p},\,\alpha} \biggr], \qquad \hat{\pi}_{\vec{p} \,\alpha} = - i\,\sqrt{\frac{p}{2}} 
\biggl[ \hat{a}_{\vec{p},\,\alpha} - \hat{a}^{\dagger}_{-\vec{p},\,\alpha} \biggr],\qquad [\hat{a}_{\vec{k},\,\alpha}, \, \hat{a}_{\vec{p},\,\beta}^{\dagger}] 
= \delta^{(3)}(\vec{k} - \vec{p})\, \delta_{\alpha\beta}.
\label{QTP8}
\end{equation}
The decomposition of Eq. (\ref{QTP8}) can be used into  Eq. (\ref{QTP7}) 
so that the final form of the quantum Hamiltonian becomes:
\begin{eqnarray}
\hat{H}_{g} &=& \frac{1}{2} \int d^{3} p \sum_{\alpha} \biggl\{ p \biggl[ \hat{a}^{\dagger}_{\vec{p},\,\alpha} \hat{a}_{\vec{p},\,\alpha} 
+ \hat{a}_{- \vec{p},\,\alpha} \hat{a}^{\dagger}_{-\vec{p},\,\alpha} \biggr] 
+\lambda \hat{a}^{\dagger}_{-\vec{p},\,\alpha} \hat{a}_{\vec{p},\,\alpha}^{\dagger} 
+ \lambda^{*} \hat{a}_{\vec{p},\,\alpha} \hat{a}_{-\vec{p},\,\alpha}
\nonumber\\
&+& \gamma_{-\vec{p},\, \alpha} \hat{a}_{\vec{p}\,\alpha} + \gamma_{-\vec{p},\, \alpha}^{*} \hat{a}_{\vec{p},\,\alpha}^{\dagger}
+  \gamma_{\vec{p},\, \alpha} \hat{a}_{- \vec{p},\,\alpha} +  \gamma_{\vec{p},\,\alpha}^{*} \hat{a}_{- \vec{p},\,\alpha}^{\dagger} \biggr\},
\label{QTP9}
\end{eqnarray}
where $ \lambda = i {\mathcal H}$; note that $\gamma_{\pm\vec{p}\,\alpha}$ is related to the presence of the anisotropic stress and may lead to a coherent component.  The Hamiltonian for the scalar modes of the geometry has the same form of Eq. (\ref{QTP9}) but a different pump field\footnote{If we want to discuss the quantum mechanical generation of curvature perturbations the Hamiltonian 
describing the production of primordial phonons has the same form of Eq. (\ref{QTP9}) but the pump 
field $\lambda$ is $\lambda_{s} = i z^{\prime}/z$ where $z = a \varphi^{\prime}/{\mathcal H}$ and $\varphi$ denotes the inflaton field.} (i.e. a different $\lambda$).  
The relic gravitons are produced by the pumping action of the gravitational field exactly as photons may be produced by a laser beam impinging on a material with nonlinear susceptibility \cite{QM0}. Equation (\ref{QTP9}) is the continuous mode generalization of the Hamiltonian originally discussed by Mollow and Glauber \cite{QM1,QM1a} in the quantum theory of parametric amplification.  A similar description has been discussed later by Stoler \cite{QM2,QM2a} who generalized the notion of minimum uncertainty wavepacket and derived some of the properties of what we would call today one-mode  squeezed states. The Hamiltonian (\ref{QTP9}) may also describe an interacting Bose gas at zero temperature; in this case the free part of the Hamiltonian corresponds to the kinetic energy while the interaction terms account for the two-body collisions with small momentum transfer  \cite{QM3,QM4}.
Neglecting, for simplicity, the presence of a coherent component, the evolution equations in the Heisenberg description follow from the Hamiltonian  (\ref{QTP9}) and they are: 
\begin{eqnarray}
\frac{d \hat{a}_{\vec{p},\,\alpha}}{d\tau} &=& i \, [ \hat{H}_{g},\hat{a}_{\vec{p},\,\alpha}] =  - i\, p \, \hat{a}_{\vec{p},\,\alpha} -  i \, \lambda \hat{a}_{- \vec{p},\,\alpha}^{\dagger},
\nonumber\\
\frac{d \hat{a}_{-\vec{p},\,\alpha}^{\dagger}}{d\tau} &=& i \, [ \hat{H}_{g},\hat{a}^{\dagger}_{-\vec{p},\,\alpha}]=  i\, p \, \hat{a}_{-\vec{p},\,\alpha}^{\dagger} +  i \, \lambda^{*} \hat{a}_{\vec{p},\,\alpha}.
\label{PAM1}
\end{eqnarray}
The solution of Eq. (\ref{PAM1}) depends on the values of the 
creation and annihilation operators at a certain reference time $\tau_{i}$:
 \begin{eqnarray}
\hat{a}_{\vec{p},\,\alpha}(\tau) &=& u_{p,\,\alpha}(\tau,\tau_{i}) \hat{b}_{\vec{p},\,\alpha}(\tau_{i}) -  
v_{p,\,\alpha}(\tau,\tau_{i}) \hat{b}_{-\vec{p},\,\alpha}^{\dagger}(\tau_{i}),  
\label{PAM2}\\
\hat{a}_{-\vec{p},\,\alpha}^{\dagger}(\tau) &=& u_{p,\,\alpha}^{*}(\tau,\tau_{i}) \hat{b}_{-\vec{p},\,\alpha}^{\dagger}(\tau_{i})  -  v_{p,\,\alpha}^{*}(\tau,\tau_{i}) \hat{b}_{\vec{p},\,\alpha}(\tau_{i}).
\label{PAM3}
\end{eqnarray}
Equations (\ref{PAM2}) and (\ref{PAM3}) describe the parametric amplification in a quantum
mechanical language. Indeed in Eqs. (\ref{PAM1}) the evolution of the operators is determined by the 
relative weight of $p$ and of ${\mathcal H}= a \, H$ which defines the rate of variation of the 
background geometry. The initial conditions in Eq. (\ref{PAM1}) are assigned when all the relevant wavenumbers (or frequencies) exceed the rate of variation of the geometry, i.e. 
$p \gg a_{i} \, H_{i}$. As the evolution proceeds the wavenumbers might get into the regime 
$p \leq a\, H$ where the creation and annihilation operators get mixed.
Inserting Eqs. (\ref{PAM2}) and (\ref{PAM3}) into Eq. (\ref{PAM1}) the 
evolution equations for $u_{p\,\,\alpha}(\tau,\tau_{i})$ and $v_{p\,\,\alpha}(\tau,\tau_{i})$  obey:
\begin{eqnarray}
\frac{d u_{p,\,\alpha}}{d\tau} = - i p\, u_{p,\,\alpha}  + i \lambda  v_{p,\,\alpha}^{\ast}, \qquad 
\frac{d v_{p,\,\alpha}}{d\tau} = - i p\, v_{p,\,\alpha} + i \lambda u_{p,\,\alpha}^{\ast}.
\label{PAM3b}
\end{eqnarray}
 The two complex functions $u_{p,\,\alpha}(\tau,\tau_{i})$ and $v_{p,\,\alpha}(\tau,\tau_{i})$ appearing in Eqs. (\ref{PAM2}) and (\ref{PAM3}) are determined by the dynamical evolution of the pump field $\lambda$ and are also subjected, by unitarity, to the condition $|v_{p,\,\alpha}(\tau,\tau_{i})|^2 - |u_{p,\,\alpha}(\tau,\tau_{i})|^2 =1$. For each of the two tensor polarizations,  $u_{p,\,\alpha}(\tau,\tau_{i})$ and $v_{p,\,\alpha}(\tau,\tau_{i})$ depend upon one amplitude and two phases so that they can always be parametrized as: 
\begin{equation}
u_{k,\,\alpha}(\tau,\tau_{i}) = e^{- i\, \delta_{k,\,\alpha}} \cosh{r_{k,\,\alpha}}, 
\qquad v_{k,\,\alpha}(\tau,\tau_{i}) =  e^{i (\theta_{k,\,\alpha} + \delta_{k,\,\alpha})} \sinh{r_{k,\,\alpha}},
\label{PAM4}
\end{equation}
where $\delta_{k,\,\alpha}$, $\theta_{k,\,\alpha}$ and $r_{k\,\alpha}$ are all real and depend on $\tau$ and $\tau_{i}$. The evolution of the squeezing parameter and of the phases is then given by:
\begin{eqnarray}
r_{p,\alpha}^{\prime} = - {\mathcal H} \cos{\theta_{p,\alpha}}, \qquad \delta_{p,\alpha}^{\prime} = p - {\mathcal H} \sin{\theta_{p,\alpha}} \tanh{r_{p,\alpha}}, \qquad 
\theta_{p,\alpha}^{\prime} = - 2 p  + 2 \frac{{\mathcal H} \sin{\theta_{p,\alpha}}}{\tanh{2 r_{p,\alpha}}},
\label{PAM6}
\end{eqnarray}
and it follows by inserting  Eq. (\ref{PAM4}) into Eq. (\ref{PAM3b}). 
While Eq. (\ref{PAM3b}) is linear and can be solved in various cases, the nonlinear equations 
(\ref{PAM6}) are, in general, more difficult to discuss. 

\subsubsection{Continuous mode squeezed states}
The quantum theory of parametric amplification necessarily leads to squeezed quantum states\footnote{The squeezed quantum states of light are also relevant for the noise reduction in wide-band interferometers, as we shall briefly discuss at the end of this section.}  \cite{yuen,QM6,caves,cavesb,QM7,QM7a}. The concrete experimental realizations of quantum amplifiers produces squeezed states of the radiation field when the interactions with the medium are characterized by nonlinear 
susceptibilities of either second-order or third-order \cite{sqgen1,sqgen2,sqgen3}; however in the experimental situation only few modes of the field are actually taken into account.  The very first applications of the quantum theory of parametric amplification to the case of relic gravitons  have been presented by Grishchuk and Sidorov \cite{HIS15} (see also \cite{QM9,QM10,QM10h}). Different representations of the quantum states of scalar and tensor modes of the geometry have been discussed in the literature \cite{QM10c,QM10d,QM10e,QM10g} however the most convenient ones are based, directly or indirectly,  on the two-mode squeezing operators \cite{QM7}. This is because gravitons are produced in pairs with opposite momenta and this is ultimately the reason for the relevance of the two-mode formalism. It follows from the well known
Baker-Campbell-Hausdorff reduction formula \cite{QM0} that the transformation of Eqs. (\ref{PAM2}) and (\ref{PAM3}) is generated by the action of two unitary operators: the (multimode) squeezing operator ${\mathcal S}(z)$ and by the rotation operator ${\mathcal R}(\delta)$:\begin{equation}
{\mathcal S}(z) = e^{\sigma(z)/2}, \qquad {\mathcal R}(\delta)= e^{- i n(\delta)/2},
\label{PAM5a}
\end{equation}
where $\sigma(z)$ and $n(\delta)$ involve both an integral over the modes and a sum over the two tensor 
polarizations:
\begin{eqnarray}
\sigma(z) &=& \sum_{\lambda} \int d^{3} k \,\, \biggl[ z_{k,\,\lambda}^{*} \, \hat{b}_{\vec{k},\,\lambda} 
\hat{b}_{-\vec{k},\,\lambda} - z_{k,\,\lambda} \, \hat{b}^{\dagger}_{-\vec{k},\,\lambda} 
\hat{b}^{\dagger}_{\vec{k},\,\lambda} \biggr],
\nonumber\\
n(\delta) &=& \sum_{\lambda}  \int d^{3}k\, \delta_{k,\,\lambda}\,\biggl[ \hat{b}_{\vec{k},\,\lambda}^{\dagger} 
 \hat{b}_{\vec{k},\,\lambda} + \hat{b}_{-\vec{k},\,\lambda}  \hat{b}_{-\vec{k},\,\lambda}^{\dagger} \biggr], 
 \label{PAM6a}
 \end{eqnarray}
where $\delta_{k,\,\lambda}$ is real  while $z_{k,\, \lambda} = r_{k,\,\lambda} e^{i \theta_{k,\,\lambda}}$ is complex. Similar operators appear in the derivation of the ground state wavefunction of an interacting Bose gas at zero temperature \cite{QM3,QM4}; the same approach has been used to describe the superfluid ground state \cite{QM11,QM12}. The  parametrization of Eq. (\ref{PAM4}) follows from Eqs. (\ref{PAM5a}) and (\ref{PAM6a}) by appreciating that\footnote{Note, incidentally, that $\delta_{k,\,\lambda}$ denotes the phase $\delta$ with modulus of three-momentum $k$ and polarization $\lambda$; this quantity cannot be confused with the Kronecker $\delta_{\alpha\beta}$ where $\alpha$ and $\beta$ are instead two generic tensor polarizations.}:
 \begin{eqnarray}
{\mathcal R}^{\dagger}(\delta) \, {\mathcal S}^{\dagger}(z) \, \hat{b}_{\vec{k},\,\alpha} 
{\mathcal S}(z) \, {\mathcal R}(\delta) 
= e^{- i\, \delta_{k,\,\alpha}} \cosh{r_{k,\,\alpha}} \, \hat{b}_{\vec{k},\,\alpha} -  e^{i (\theta_{k,\,\alpha} + \delta_{k,\,\alpha})} \sinh{r_{k,\,\alpha}} \, \hat{b}^{\dagger}_{-\vec{k},\,\alpha},
\label{PAM8}\\
{\mathcal R}^{\dagger}(\delta) \, {\mathcal S}^{\dagger}(z) \, \hat{b}_{-\vec{k},\,\alpha}^{\dagger} 
{\mathcal S}(z) \, {\mathcal R}(\delta) 
= e^{i\, \delta_{k,\,\alpha}} \cosh{r_{k,\,\alpha}}\, \hat{b}^{\dagger}_{-\vec{k},\,\alpha} -  e^{- i (\theta_{k,\,\alpha} + \delta_{k,\,\alpha})} \sinh{r_{k,\,\alpha}}\, \hat{b}_{\vec{k},\,\alpha}.
\label{PAM9}
\end{eqnarray}
In the present context,  recalling Eqs. (\ref{PAM8}) and (\ref{PAM9}), the squeezed states of the field are given by:
\begin{equation}
|\{ z \, \delta \}\rangle = {\mathcal S}(z) \, {\mathcal R}(\delta)\, |\mathrm{vac}\rangle, \qquad  |\mathrm{vac}\rangle = 
\prod_{\vec{p}} |0_{\vec{p}} \, \, 0_{-\vec{p}}\rangle.
\label{PAM9a}
\end{equation}
Equation (\ref{PAM9a}) describes the parametric amplification in the Schr\"odinger representation 
from a vacuum state has vanishing three-momentum.  From Eqs. (\ref{PAM8}) and (\ref{PAM9}) we can compute the averaged multiplicity of the state $|\{ z \, \delta \}\rangle$ for gravitons with a given polarization:
\begin{equation}
\langle \{ \delta\, z \}| \biggl[\hat{N}_{\vec{k}, \alpha}  + \hat{N}_{-\vec{k}, \alpha} \biggr] |\{ z \, \delta \}\rangle= 2 \sinh^2{r_{k,\,\alpha}} = 2 \,\overline{n}_{k,\, \alpha},
\label{PAM9b}
\end{equation}
where $\hat{N}_{\vec{k}, \alpha} = \hat{b}^{\dagger}_{\vec{k},\,\alpha} \,\hat{b}_{\vec{k},\,\alpha}$ and 
$\hat{N}_{-\vec{k}, \alpha} = \hat{b}^{\dagger}_{-\vec{k},\,\alpha} \,\hat{b}_{-\vec{k},\,\alpha}$; $\overline{n}_{k,\, \alpha}$ 
denotes the averaged multiplicity of the pairs of gravitons with polarization $\alpha$. If the initial state of the evolution coincides with $ |\mathrm{vac}\rangle$ the corresponding multiplicity distribution is of Bose-Einstein type:
\begin{equation}
P_{n_{k}}(\overline{n}_{k,\,\alpha}) =\bigl| \langle \{ n\, \}  |\{ z \, \delta \}\rangle\bigr|^2=  
\frac{\overline{n}_{k,\, \alpha}^{\,\,n_{k}}}{(\overline{n}_{k,\,\alpha} +1)^{\,\,n_{k}+1}}, \qquad \sum_{n_{k}=0}^{\infty} P_{n_{k}}(\overline{n}_{k,\,\alpha})=1.
\label{PAM9c}
\end{equation}
Equation (\ref{PAM9c}) implies that the squeezed vacuum is a pure state whose associated 
multiplicity distribution has a Bose-Einstein form. This peculiar feature suggested through the years 
various applications directly or indirectly related with the possibility of associating a Shannon 
entropy to squeezed states. Yurke and Potasek \cite{yurke} working with two-mode squeezed states 
noted that by tracing the density matrix $\hat{\rho}$ of the state over one of the two oscillators the reduced density matrix (i.e. $\hat{\rho}_{red}$) leads to a Von-Neumann entropy (i.e. $S= - \mathrm{Tr}[ \hat{\rho}_{red} \ln{ \hat{\rho}_{red}}]$).  The specific approach to the decoherence of the density matrix  depends on the basis where the wavefunctionals are defined. Different bases have been 
explored such as the basis where the quantum fields fluctuate above (or below) the quantum noise \cite{QM10a,QM10aa,QM10bb}, the Fock basis  \cite{QM10b,QM10bbb} and others. When the number 
of produced gravitons is large the entropy density per each Fourier mode goes as $ 2 \, r_{k,\alpha}$ where $r_{k,\alpha}$ is the squeezing parameter introduced in Eq. (\ref{PAM4}). Using the results of Refs. \cite{QM10a,QM10aa,QM10bb} the authors of Ref. \cite{QM10ee} speculated that even in the absence of interactions the density matrix gets naturally reduced by suggesting some 
sort of decoherence without decoherence. However a strongly squeezed 
state is just a strongly correlated multiparticle state that preserves the area of the phase 
space. This conclusion can be reached within the celebrated formalism of the Wigner 
functionals \cite{wigner1,wigner2,wigner3}: the Wigner representation is an extremely  useful tool to express quantum mechanics in a phase space formalism. Note incidentally that the Wigner 
representation has been applied to the case of relic gravitons by Grishchuk and Sidorov 
in one of the first papers on the subject \cite{HIS15}. A more physical notion of decoherence 
cannot avoid a specific mechanism for the reduction of the density matrix \cite{QM10a,QM10aa}
and it is normally associated with a growth of a phase-space area \cite{QM10f}. The 
statistical property expressed by Eq. (\ref{PAM9c}) is typical of the squeezed vacuum state:  when the initial state is different from the vacuum (e.g. a Fock state) the multiplicity distribution does not have the Bose-Einstein form of Eq. 
(\ref{PAM9c}) \cite{QM10gg} and oscillations may appear. In this case the area of the phase
space is not well defined since the corresponding Wigner functions are not positive semi-definite \cite{wigner3}.

Even if the present considerations are not directly related with the technical aspects 
of the actual noise reduction in wide-band interferometers, it may be useful to stress that the squeezed states appear ubiquitously also in experimental strategies for the actual detection of the relic gravitons 
(see \cite{SQL1,SQL2} for two recent reviews). The main observation, in this context, goes back to the 
pioneering analyses of Caves who suggested to inject a squeezed vacuum state inside the interferometer \cite{caves,cavesb} instead of having a coherent state entering the output port.
The quantum noise reduction obtained via the squeezed light is experimentally relevant ever if the typical values of the squeezing parameter $r_{k}$ barely exceed $1$ \cite{SQL1,SQL2,SQL3}. 
For comparison, in the case of the relic gravitons, $r_{k}= {\mathcal O}(100)$ for typical frequencies ${\mathcal O}(\mathrm{aHz})$ assuming a conventional inflationary phase with a sufficient 
number of $e$-folds. An important property of squeezed states is that their 
statistics is, in general, not Poissonian (as it happens in the case of coherent states) \cite{QM0}. In the context of the advanced LIGO \cite{LV1} and advanced Virgo \cite{LV2} projects the use of squeezed light may provide a relevant increase in the sensitivity in the 
high-frequency part of the audio band\footnote{See also some more specific discussions on wide-band interferometers and relic gravitons in section \ref{sec7}.} (i.e. above $200$ Hz). 
The noise reduction at high-frequency is however compensated by a decrease of the sensitivity between few Hz and $100$ Hz: this is the reason why the forthcoming generation of wide-band interferometers will probably have to use a frequency-dependent approach leading to a  high-frequency squeezing and to a low-frequency desqueezing. It is amusing to note that the analogy between with the relic gravitons is even more complete since, in the present case, the squeezing parameter is actually frequency-dependent as a consequence of the original Hamiltonian which is defined for a continuum of modes. 

\subsubsection{Two-point functions and power spectra} 
The functions $u_{k}$ and $v_{k}$ obeying Eq. (\ref{PAM3b}) are a linear combination of the  mode functions of the problem
\begin{equation}
u_{k,\,\lambda} - v_{k,\,\lambda}^{\ast} = \sqrt{2 k} \, f_{k,\,\lambda}, \qquad u_{k,\,\lambda} + v_{k,\,\lambda}^{\ast} = i \sqrt{\frac{2}{k}} g_{k,\,\lambda}.
\label{PAM10}
\end{equation}
From Eq. (\ref{PAM3b}) it follows that the mode functions $f_{k}$ and $g_{k}$ obey the following pair of equations 
\begin{equation}
f_{k,\,\lambda}^{\prime} = g_{k,\,\lambda} + {\mathcal H} f_{k,\,\lambda}, \qquad g_{k,\,\lambda}^{\prime} = - k^2 f_{k,\,\lambda} - {\mathcal H} g_{k,\,\lambda}. 
\label{PAM11}
\end{equation}
The mode functions of  Eq. (\ref{PAM10}) can also be expressed in their rescaled form:
\begin{equation}
F_{k,\,\lambda} = \frac{f_{k,\,\lambda}}{a}, \qquad G_{k,\,\lambda} = F_{k,\,\lambda}^{\prime} \equiv  \biggl(\frac{f_{k,\,\lambda}}{a}\biggr)^{\prime} = \frac{g_{k,\,\lambda}}{a},
\label{PAM12}
\end{equation}
where the chain of equalities defining $G_{k,\,\lambda}$ follows by applying Eq. (\ref{PAM11}). In the absence 
of any source of anisotropic stress we therefore have that the evolution of $F_{k,\,\lambda}$ and $G_{k,\,\lambda}$ is: 
\begin{equation}
F_{k,\,\lambda}^{\prime\prime} + 2 {\mathcal H} F_{k,\,\lambda}^{\prime} + k^2 F_{k,\,\lambda}=0, \qquad G_{k,\,\lambda} = F_{k,\,\lambda}^{\prime}.
\label{PAM12a}
\end{equation}
Recalling the explicit form of Eqs (\ref{QTP5}) and (\ref{QTP8}), the expression of the field operators and of their derivative can be written in terms of the mode functions of Eq. (\ref{PAM12}):
\begin{eqnarray}
\hat{h}_{ij}(\vec{x},\tau) &=& \frac{\sqrt{2} \,\ell_{P}}{(2\pi)^{3/2}}\sum_{\lambda} \int \, d^{3} k \,\,e^{(\lambda)}_{ij}(\hat{k})\, [ F_{k,\,\lambda}(\tau)\, \hat{a}_{\vec{k},\,\lambda } e^{- i \vec{k} \cdot \vec{x}} + F^{*}_{k,\,\lambda}(\tau)\, \hat{a}_{\vec{k},\,\lambda }^{\dagger} e^{ i \vec{k} \cdot \vec{x}} ],
\label{PAM13}\\
\hat{H}_{ij}(\vec{x},\tau) &=& \frac{\sqrt{2} \,\ell_{P}}{(2\pi)^{3/2}}\sum_{\lambda} \int \, d^{3} k \,\,e^{(\lambda)}_{ij}(\hat{k})\, [ G_{k,\,\lambda}(\tau) \,\hat{a}_{\vec{k},\,\lambda } e^{- i \vec{k} \cdot \vec{x}} + G^{*}_{k,\,\lambda}(\tau)\, \hat{a}_{\vec{k},\,\lambda }^{\dagger} e^{ i \vec{k} \cdot \vec{x}} ],
\label{PAM14}
\end{eqnarray}
where the shorthand notation $\hat{H}_{ij}(\vec{x},\tau) = \partial_{\tau} \hat{h}_{ij}(\vec{x},\tau)$ has been used for convenience. In Fourier space the operators $\hat{h}_{ij}(\vec{k},\tau)$ and $\hat{H}_{ij}(\vec{k},\tau)$ are given by:
\begin{eqnarray}
\hat{h}_{ij}(\vec{k},\tau) &=& \sqrt{2} \ell_{P}\sum_{\lambda} \biggl[ e^{(\lambda)}_{ij}(\hat{k})\, F_{k,\,\lambda}(\tau) \hat{a}_{\vec{k},\,\lambda } + e^{(\lambda)}_{ij}(-\hat{k})\, F^{*}_{k,\,\lambda}(\tau) \hat{a}^{\dagger}_{-\vec{k},\,\lambda }\biggr],
\label{PAM15a}\\
\hat{H}_{ij}(\vec{k},\tau) &=& \sqrt{2} \ell_{P} \sum_{\lambda} \biggl[ e^{(\lambda)}_{ij}(\hat{k})\, G_{k,\,\lambda}(\tau) \hat{a}_{\vec{k},\,\lambda } + e^{(\lambda)}_{ij}(-\hat{k})\, G^{*}_{k,\,\lambda}(\tau) \hat{a}^{\dagger}_{-\vec{k},\,\lambda }\biggr].
\label{PAM15}
\end{eqnarray}
The field operators 
(\ref{PAM15a}) and (\ref{PAM15}) consist of a positive and of a negative frequency part, i.e. 
$\hat{h}_{ij}(x) = \hat{h}^{(+)}_{ij}(x) + \hat{h}^{(-)}_{ij}(x)$ with $ \hat{h}^{(+)\,\,\dagger}_{ij}(x) =  \hat{h}^{(-)}_{ij}(x)$.
If $| \mathrm{vac} \rangle$ is the state that minimizes the tensor Hamiltonian when all the frequencies 
are larger than the rate of variation of the background  the operator $\hat{h}^{(+)}_{ij}(x)$ annihilates 
the vacuum [i.e. $ \hat{h}^{(+)}_{ij}(x) \,| \mathrm{vac} \rangle=0$ and $\langle \mathrm{vac} | \,\hat{h}^{(-)}_{ij}(x) =0$]. For instance in the context of the conventional inflationary modes the two-point functions associated with $\hat{h}_{ij}$ and  $\hat{H}_{ij}$ are obtained from the appropriate 
expectation values over the vacuum:
\begin{eqnarray}
\langle \mathrm{vac} | \hat{h}_{ij}(\vec{x}, \tau)\, \hat{h}^{ij}(\vec{x} + \vec{r}, \tau) | \mathrm{vac}\rangle 
&=& \int \,d \ln{k} \,\,P_{T}(k,\tau) \,\,j_{0}(k r), 
\label{fourEa}\\
\langle \mathrm{vac} | \hat{H}_{ij}(\vec{x}, \tau) \,\hat{H}^{ij}(\vec{x} + \vec{r}, \tau) | \mathrm{vac}\rangle 
&=& \int \,d \ln{k} \,\,Q_{T}(k,\tau) \,\,j_{0}(k r),
\label{fourEb}
\end{eqnarray}
where $j_{0}(k r)$ are spherical Bessel functions of zeroth order \cite{abr1,abr2} while $P_{T}(k,\tau)$ and 
$Q_{T}(k,\tau)$ are the corresponding power spectra:
\begin{equation}
P_{T}(k,\tau) = \frac{4 \ell_{P}^2}{\pi^2} \,\,k^3 \,\,\bigl| F_{k}(\tau) \bigr|^2, \qquad Q_{T}(k,\tau) = \frac{4 \ell_{P}^2}{\pi^2} \,\,k^3 \,\,\bigl| G_{k}(\tau) \bigr|^2.
\label{fiveE}
\end{equation}
In Eq. (\ref{fiveE}) the polarization has been suppressed: if the background is not polarized the equations obeyed by the two polarizations coincide. If the polarizations of the relic gravitons interact via parity-violating terms entering the general action of Eq. (\ref{POLGRAV4}), the
mode functions of the different polarizations  do not evolve independently 
as it can be explicitly verified by deriving the evolution equations for instance from Eq. (\ref{POLGRAV7}). The expectation values of Eqs. (\ref{fourEa})--(\ref{fourEb}) can be directly computed in Fourier space by using Eqs. (\ref{PAM15a}) and (\ref{PAM15}) and the result is
\begin{eqnarray}
\langle \hat{h}_{ij}(\vec{k},\tau) \, \hat{h}_{mn}(\vec{k}^{\prime}, \tau) \rangle &=& \frac{2 \pi^2}{k^3}\, P_{T}(k,\tau) \, \delta^{(3)}(\vec{k} + \vec{k}^{\prime}) \, 
{\mathcal S}_{i j m n}(\hat{k}),
\label{PAM16}\\
\langle \hat{H}_{ij}(\vec{k},\tau) \, \hat{H}_{mn}(\vec{k}^{\prime}, \tau) \rangle &=& \frac{2 \pi^2}{k^3}\, Q_{T}(k,\tau) \, \delta^{(3)}(\vec{k} + \vec{k}^{\prime}) \, 
{\mathcal S}_{i j m n}(\hat{k}),
\label{PAM17}
\end{eqnarray}
where ${\mathcal S}_{ijmn}(\hat{k})$ has been already introduced in Eqs. (\ref{POLDEF7}) and (\ref{POLDEF8});
note that we used the shorthand notation $\langle \mathrm{vac}| .\,.\,.\,. | \mathrm{vac} \rangle = \langle \,.\,.\,. \rangle$. Of course the  choice of $|\mathrm{vac} \rangle$  in Eqs. (\ref{fourEa}) and (\ref{fourEb}) follows by assuming 
a sufficiently long inflationary phase. However, if the inflationary phase has duration that is close 
to minimal the initial state can be different and we shall discuss later on some examples 
along this direction. To compute correctly the averaged energy density and pressure 
in some of the prescriptions suggested in the literature it is also relevant to keep track 
of the mixed expectation values, namely 
\begin{equation}
\langle \hat{h}_{ij}(\vec{k},\tau) \, \hat{H}_{mn}(\vec{k}^{\prime}, \tau) \rangle = \frac{2 \pi^2}{k^3}\, S_{T}(k,\tau) \, \delta^{(3)}(\vec{k} + \vec{k}^{\prime}) \, 
{\mathcal S}_{i j m n}(\hat{k}).
\label{PAM18}
\end{equation}
Note that the result of $\langle \hat{H}_{ij}(\vec{k},\tau) \, \hat{h}_{mn}(\vec{k}^{\prime}, \tau) \rangle$ 
is simply the complex conjugate of the expression at the right hand side of Eq. (\ref{PAM18}) while $S_{T}(k,\tau)$ and $S_{T}^{*}(k,\tau)$ are given by
\begin{equation}
S_{T}(k,\tau) =  \frac{4 \ell_{\mathrm{P}}^2\,\, k^3}{\pi^2} F_{k}(\tau) G_{k}^{\ast}(\tau),\qquad S_{T}^{*}(k,\tau) =  \frac{4 \ell_{\mathrm{P}}^2\,\, k^3}{\pi^2} F_{k}^{*}(\tau) G_{k}(\tau).
\label{PAM21}
\end{equation}

\subsection{Effective energy density of the relic gravitons}
The explicit expression of the energy density of the relic gravitons depends on their
frequency. In generic curved backgrounds the only 
sound physical limit would seem the one where the typical frequency of the graviton 
greatly exceeds the rate of variation of the background geometry. This 
limit is often imposed in the case of the  Landau-Lifshitz \cite{TENS2b} approach or 
in the framework of the Brill-Hartle-Isaacson \cite{TENS2c,TENS3a,TENS3b} proposal (see also \cite{TENS5a,TENS6,TENS9}). It is well known that in this 
approximation the high-frequency gravitons behave as a perfect 
relativistic fluid whose barotropic index is $1/3$, exactly as in the case of photons. 
There are however physical situations where the frequencies 
of the waves are much smaller than the rate of variation of the corresponding 
background and this normally happens in cosmological backgrounds. For instance during an accelerated stage of expansion the particle horizon diverges while the event horizon is proportional to the 
Hubble rate. The wavelengths  of the gravitons become  larger than the 
Hubble radius but are still shorter than the typical size of  causally connected 
domains\footnote{By a mode being beyond the horizon we only mean 
that the physical wavenumber $k/a$ is much less than the expansion rate $H$ and
this regime is not forbidden by causality \cite{TENS2aa}.}.
The effective energy density of the relic gravitons obtained from Eq. (\ref{ACT9}) 
is well defined when the frequencies of the gravitons are smaller than the rate 
of variation of the background; the other proposals, in the same limit, give 
contradictory answers.

\subsubsection{The rate of variation of the background}  
\begin{figure}[!ht]
\centering
\includegraphics[width=0.7\textwidth]{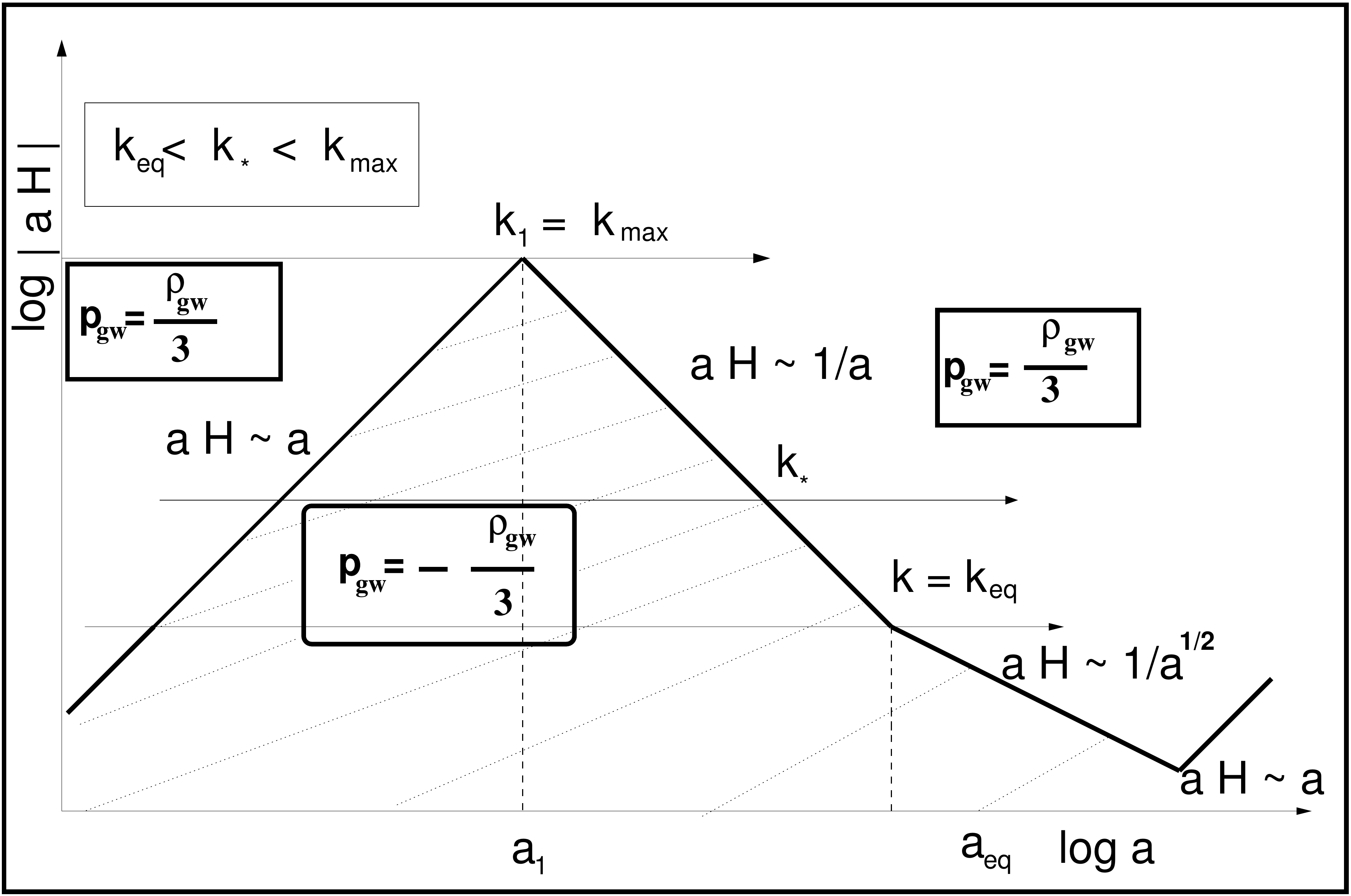}
\caption[a]{The evolution of the effective horizon is schematically illustrated 
in the minimal situation where the radiation background suddenly dominates {\em after }
inflation. In Planck units the maximal height of the thick black curve is around $-6$
since, at the end of inflation, $(H/M_{P}) = {\mathcal O}(10^{-6})$.
While the minimum should be around $-61$ since today 
$H/M_{P}= {\mathcal O}(10^{-61})$. }
\label{SEC3FIG1} 
\end{figure}
In the Heisenberg description the two scales governing the parametric amplification  
 are the wavenumber $k$ and the rate of variation of the geometry ${\mathcal H} = a\, H$ (see e.g. 
 Eqs. (\ref{PAM1}) and (\ref{PAM3b})). From the semi-classical viewpoint the same scales appear in 
 the second-order evolution of the mode function derived from Eq. (\ref{PAM11}):
\begin{equation}
f_{k}'' +\biggl( k^2 - \frac{a''}{a}\biggr) f_{k} =0,\qquad F_{k} = \frac{f_{k}}{a}.
\label{RGC2}
\end{equation}
Once $f_{k}(\tau)$ has been determined from Eq. (\ref{RGC2}), it is immediate to deduce $g_{k}$ and 
$G_{k}$ from Eqs. (\ref{PAM12})--(\ref{PAM12a}).  With simple algebra it is possible to show that the pump field $a^{\prime\prime}/a$ of Eq. (\ref{RGC2}) can be approximated by $a^2 \, H^2$ in various physical situations:
\begin{equation}
\frac{a^{\prime\prime}}{a} =   {\mathcal H}^2 + {\mathcal H}^{\prime} = - \frac{a^2}{6} \, \overline{R} = a^2 H^2 \biggl( 2  + \frac{\dot{H}}{H^2} \biggr),
\label{RGC1}
\end{equation}
where $\overline{R}=- 6 \, a^{\prime\prime}/a^3$ denotes the Ricci scalar of the background in the case of a conformally flat FRW background;  the last equality of Eq. (\ref{RGC1}) follows from the trivial observation that ${\mathcal H} = a \, H$ and from the relation between cosmic and conformal times.
The thick black line in Fig. \ref{SEC3FIG1} illustrates the common logarithm of $a\, H$ as a 
function of the common logarithm of the scale factor in the 
case of the concordance paradigm and for a standard 
thermal history. During a quasi-de Sitter stage of expansion $a H$ increases linearly and the conventional slow-roll 
parameters\footnote{The first and second derivatives of the potential with respect to $\varphi$
are denoted by $V_{,\varphi}$ and $V_{,\varphi\varphi}$ respectively.} will be defined as:
\begin{equation}
\epsilon = - \frac{\dot{H}}{H^2} = \frac{\overline{M}_{\mathrm{P}}^2}{2} \biggl(\frac{V_{,\varphi}}{V}\biggr)^2,\qquad 
 \eta = \frac{\ddot{\varphi}}{H \dot{\varphi}} = \epsilon - \overline{\eta},\qquad \overline{\eta} = 
\overline{M}_{\mathrm{P}}^2 \frac{V_{,\varphi\varphi}}{V},
\label{RGC4}
\end{equation}
see Tab. \ref{SEC1TABLE4} for the conventions on the reduced Planck mass $\overline{M}_{P}$ and its 
relation to $M_{P}$. In the case of Eq. (\ref{RGC4}) the pump field of Eq. (\ref{RGC1}) becomes:
\begin{equation}
\frac{a''}{a} = a^2 H^2 \biggl(2 - \epsilon \biggr) =  \frac{( 2 -\epsilon)}{ (1 - \epsilon)^2 \tau^2},\qquad a H = - \frac{1}{(1 - \epsilon)\tau},
\label{RGC3}
\end{equation}
holding in the situation where $\epsilon \ll 1$ is approximately constant. During the radiation epoch\footnote{While the pump field is ultimately determined by $a^2 H^2$ when the plasma is dominated by radiation 
$\dot{H}/H^2 = -2$ and Eq. (\ref{RGC1}) implies that $a^{\prime\prime}/a =0$.}
$a\, H$ decreases proportionally to $1/a$ while in  the matter dominated phase we have instead that $a\, H \propto 1/\sqrt{a}$. As soon as the dark energy dominates, $a\, H$ increases again and it is proportional to $a$. 
\begin{figure}[!ht]
\centering
\includegraphics[width=0.7\textwidth]{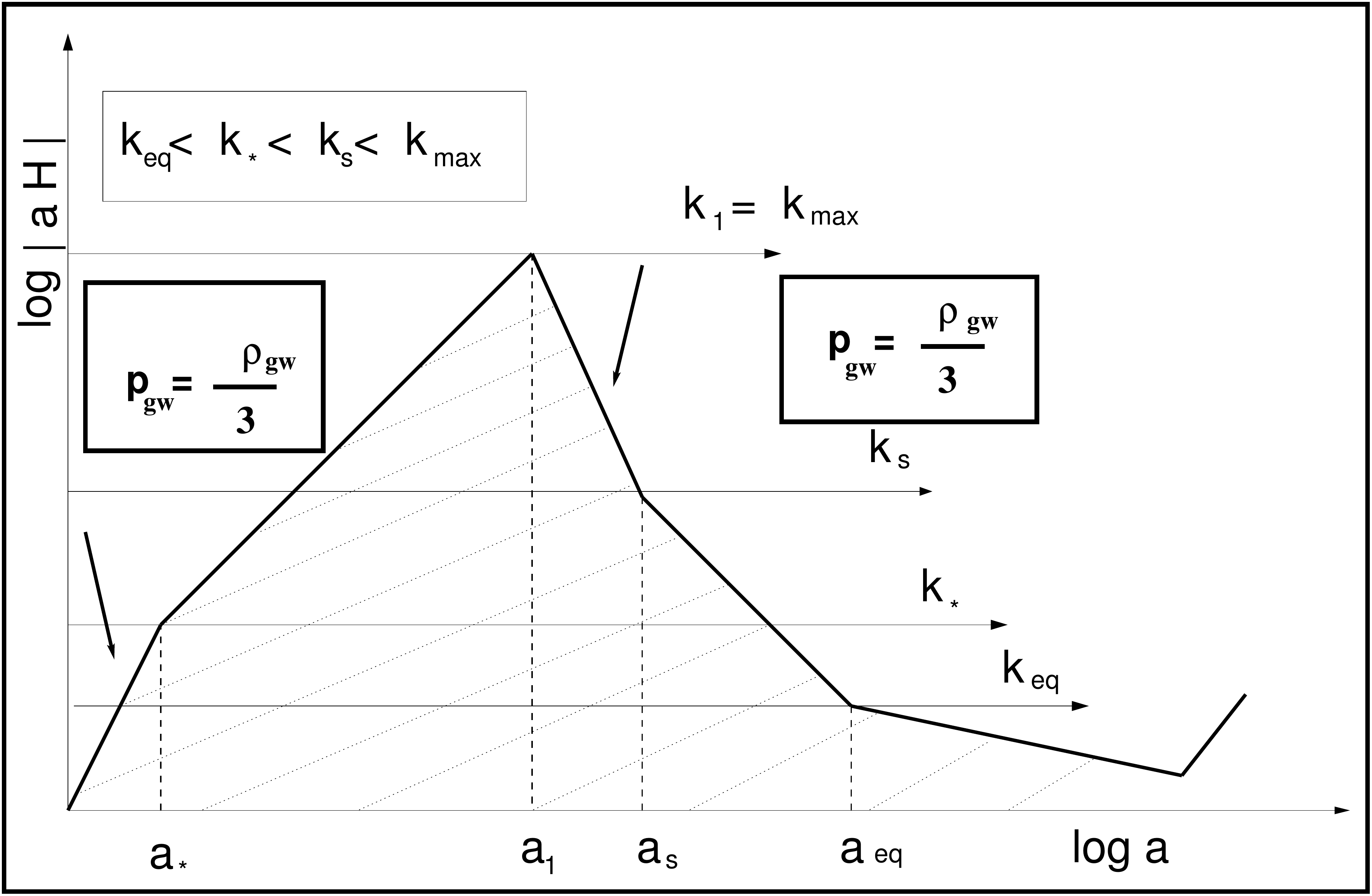}
\caption[a]{The evolution of the effective horizon is schematically illustrated
when the dominance of radiation is delayed by  a stiff phase 
 (indicated by an arrow). Prior to the onset of a conventional stage of inflation a further (pre-inflationary) phase (denoted by a second arrow) has been also added.}
\label{SEC3FIG2} 
\end{figure}
A slightly different scenario of background evolution is illustrated, for comparison, in Fig. \ref{SEC3FIG2} where, for $a < a_{*}$, a conventional inflationary phase is complemented by a pre-inflationary stage illustrated by the left arrow. Later on, when $a > a_{1}$, the dominance of radiation is also delayed by the presence of a stiff phase (illustrated by the right arrow)  where $a H$ decreases faster then in the case of a radiation plasma. More precisely, in the example of Fig. \ref{SEC3FIG2} $a\, H$ scales as $1/a^2$ between $a_{1}$ and $a_{s}$; this behaviour is typical of phases where the equation of state of the plasma is stiffer than radiation. 

\subsubsection{Super-adiabatic amplification}
When the comoving wavenumber exceeds the thick lines in Figs. \ref{SEC3FIG1} and 
\ref{SEC3FIG2} the mode functions $f_{k}$ and $g_{k}$ appearing in Eq. (\ref{PAM11}) 
are both oscillating (i.e. $f_{k} \simeq e^{ \pm i\, k\tau}$). In the same high-frequency regime (i.e. $k > a\, H$) the mode function $F_{k}$ of Eq. (\ref{PAM12a})  is suppressed as $1/a$ when the universe expands: 
this has been originally referred to as the {\em adiabatic suppression} as opposed to the phenomenon 
of {\em super-adiabatic amplification} \cite{HIS5,HIS6} that instead occurs in the 
opposite regime when the rate of variation of the background exceeds the comoving
wavenumber (i.e. $ k < a \, H$, below the thick lines in Figs. \ref{SEC3FIG1} and \ref{SEC3FIG2}). In the super-adiabatic regime $F_{k}$ and $G_{k}$ can be determined by iteration from the following pair of integral equations:
\begin{eqnarray}
F_{k}(\tau) &=& F_{k}(\tau_{ex}) + a_{ex}^2 \, G_{k}(\tau_{ex}) \int_{\tau_{ex}}^{\tau} \, \frac{d\tau_{1}}{a^2(\tau_{1})} 
- k^2 \int_{\tau_{ex}}^{\tau} \frac{d\tau_{2}}{a^2(\tau_{2})} \int_{\tau_{ex}}^{\tau_{2}} a^2(\tau_{1}) \, F_{k}(\tau_{1}) \, d\tau_{1},
\label{PAM30b}\\
G_{k}(\tau) &=& \frac{a_{ex}^2}{a^2(\tau)} \ G_{k}(\tau_{ex}) - \frac{k^2}{a^2(\tau)} \int_{\tau_{ex}}^{\tau} a^2(\tau_{1}) 
F_{k}(\tau_{1}) \, d\tau_{1}.
\label{PAM30c}
\end{eqnarray}
 In Eq. (\ref{PAM30c}) and (\ref{PAM30b}) $\tau_{ex}$ denotes the conformal time coordinate corresponding  $k^2 \simeq a^2_{ex} H^2_{ex} $ (i.e. for $k \tau_{ex} = {\mathcal O}(1)$ the corresponding wavelength is said to exit the Hubble radius $|a\, H|^{-1}$). In Figs. \ref{SEC3FIG1} and 
\ref{SEC3FIG2} the super-adiabatic regime corresponds to the shaded region under the thick lines. 
From Eq. (\ref{fiveE}) the power spectra $P_{T}(k,\tau)$ and $Q_{T}(k,\tau)$  in the limit $k \gg a \, H$ are  related as:
\begin{equation}
 Q_{T}(k, \tau) = k^2 P_{T}(k,\tau) \biggl[ 1 
+ {\mathcal O}\biggl(\frac{a^2 \, H^2}{k^2} \biggr) \biggr], \qquad k > a \, H.
\label{PAM30a}
\end{equation}
In the opposite limit (i.e. $k < a\, H$) the qualitative form of the power spectra depends on
 the evolution of the scale factor;  more specifically from Eqs. (\ref{PAM30b}) and (\ref{PAM30c}) we will have that 
 \begin{eqnarray}
k^2 P_{T}(k,\tau) &\simeq& \frac{4 \ell_{P}^2\, \, k^5}{\pi^2} \bigl| F_{k}(\tau_{ex}) \bigr|^2 \gg Q_{T}(k,\tau), \qquad \dot{a} >0, \qquad k < a\, H,
\label{SUP2}\\
Q_{T}(k,\tau) &\simeq& \frac{4 \ell_{P}^2\, \, k^3}{\pi^2} \bigl| G_{k}(\tau_{ex}) \bigr|^2 \, \biggl(\frac{a_{ex}}{a} \biggr)^4  \gg k^2 P_{T}(k,\tau), \qquad \dot{a} < 0,\qquad k < a\, H.
\label{SUP3}
\end{eqnarray}
If the background expands  (i.e. $\dot{a} >0$), $k^2 P_{T}$ dominates against $Q_{T}$. 
If the background contracts the scale factor shrinks (i.e. $a \ll a_{ex}$) so that, at some 
point, $Q_{T}$ will always exceed $k^2 P_{T}$.

\subsubsection{Effective energy-momentum pseudo-tensor relic gravitons}
Since the tensor fluctuations and the background metric can be regarded as independent variables the variation of Eq. (\ref{ACT9}) with respect to the background metric defines the effective energy-momentum pseudo-tensor of the relic gravitons:
\begin{equation}
\delta S_{g} = \frac{1}{2} \int d^{4} x\,\, \sqrt{-\overline{g}} \,\, T^{(gw)}_{\mu\nu}  \,\,\delta \overline{g}_{\mu\nu}.
\label{fourB}
\end{equation} 
To distinguish the result (\ref{fourB}) from the other approaches the explicit form of $T^{(gw)}_{\mu\nu}$ will be denoted by ${\mathcal F}_{\mu\nu}$: 
 \begin{equation}
 {\mathcal F}_{\mu\nu} = \frac{1}{4 \ell_{\mathrm{P}}^2} \biggl[ \partial_{\mu} h_{i j} \partial_{\nu} h^{i j} 
- \frac{1}{2} \overline{g}_{\mu \nu} \,\biggl(\overline{g}^{\alpha\beta}\, \partial_{\alpha} h_{ij} \partial_{\beta} h^{ij} \biggr)\biggr].
\label{fiveB}
\end{equation}
Since the indices of ${\mathcal F}_{\mu\nu}$ are raised and lowered with the help of the background metric (i.e. ${\mathcal F}_{\mu}^{\nu} = \overline{g}^{\alpha\nu} {\mathcal F}_{\alpha\nu}$), the components of ${\mathcal F}_{\mu}^{\nu}$ are: 
\begin{eqnarray}
{\mathcal F}_{0}^{0} = \rho_{gw}, \qquad 
{\mathcal F}_{i}^{j} = - p_{gw} \,\delta_{i}^{j} + \Pi_{i}^{\,\,j},\qquad {\mathcal F}_{i}^{0} = S_{i}= 
\frac{1}{4 \ell_{\mathrm{P}}^2 a^2} \partial_{\tau} h_{k\ell} \,\partial_{i} h^{k\ell},
\label{sixB}
\end{eqnarray}
where $ S_{i}$ denotes the energy flux while $\rho_{gw}$ and $p_{gw}$ the energy density and the pressure:
\begin{eqnarray}
\rho_{gw} &=& \frac{1}{8 \ell_{\mathrm{P}}^2 a^2} \biggl( \partial_{\tau} h_{k \ell}\, \partial_{\tau}h^{k \ell} + \partial_{m} h_{k\ell} \partial^{m} h^{k\ell}\biggr),
\label{rhoF}\\
p_{gw} &=&   \frac{1}{8 \ell_{\mathrm{P}}^2 a^2} \biggl( \partial_{\tau} h_{k\ell}\partial_{\tau} h^{\,k\ell} - \frac{1}{3} \partial_{m} h_{k \ell} \,\partial^{m} h^{\,k\ell}  \biggr).
\label{pF}
\end{eqnarray}
The anisotropic stress of the relic gravitons is, by definition, traceless (i.e. $\Pi_{i}^{i} =0$) and its 
explicit form is:
\begin{equation}
\Pi_{i}^{\,\,j} = \frac{1}{4 \ell_{\mathrm{P}}^2 a^2} \biggl( - \partial_{i} h_{k\ell} \partial^{j} h^{k\ell} + \frac{1}{3} \delta_{i}^{j} \,\,\partial_{m} h_{k \ell} \,\partial^{m} h^{k\ell} \biggr).
\label{anF}
\end{equation}
The energy density, the pressure and the energy flux are combined in the following identity: 
\begin{eqnarray}
 \partial_{\tau} \rho_{gw} + 3 {\mathcal H} \bigl(\rho_{gw}  + p_{gw} \bigr) = \frac{h_{k\ell}^{\prime}}{4 \ell_{P}^2 a^2}
\biggl( h_{k \ell}^{\prime\prime} + 2 {\mathcal H} h_{k\ell}^{\prime} - \nabla^2 h_{k \ell} \biggr) + \vec{\nabla} \cdot \vec{S},
\label{sevenB}
\end{eqnarray}
and  the first term at the right hand side of Eq. (\ref{sevenB}) vanishes because of the evolution of the 
tensor amplitude. In quantum mechanical language the energy density 
and the pressure are replaced by the corresponding operators $\hat{\rho}_{gw}$ and $\hat{p}_{gw}$ constructed from the bilinear combinations of $\hat{h}_{i\,j}$ and $\hat{H}_{i\,j}$ (see, in this respect, Eqs. (\ref{PAM13}) and (\ref{PAM14})): 
\begin{eqnarray}
\hat{\rho}_{gw} &=& \frac{1}{8 \ell_{\mathrm{P}}^2 a^2} \biggl(  \hat{H}_{k \ell}\, \hat{H}^{k \ell} + \partial_{m} \hat{h}_{k\ell} \partial^{m} \hat{h}^{k\ell}\biggr),
\label{rhoFop}\\
\hat{p}_{gw} &=&   \frac{1}{8 \ell_{\mathrm{P}}^2 a^2} \biggl(\hat{H}_{k\ell}\,\,\hat{H}^{\,k\ell} - \frac{1}{3} \partial_{m} \hat{h}_{k \ell} \,\partial^{m} \hat{h}^{\,k\ell}  \biggr),
\label{pFop}
\end{eqnarray}
 where, as in the classical case, $ \hat{H}_{k\ell} = \partial_{\tau} \hat{h}_{k\ell}$. The mean energy density and pressure are obtained by averaging Eqs. (\ref{rhoFop}) and (\ref{pFop}) over the 
vacuum state $| \mathrm{vac}\rangle$ exactly as in the case of the two-point functions of Eqs. (\ref{PAM15a}) and (\ref{PAM15}): 
\begin{eqnarray}
\overline{\,\rho\,}_{gw} &=&   \langle \mathrm{vac} | \hat{\rho}_{gw} | \mathrm{vac} \rangle = 
\frac{1}{8 \ell_{\mathrm{P}}^2 a^2} \biggl( \langle \hat{H}_{k \ell}\, \hat{H}^{k \ell} \rangle + \langle \partial_{m} \hat{h}_{k\ell} \partial^{m} \hat{h}^{k\ell} \rangle \biggr),
\label{rhoFav}\\
\overline{\,p\,}_{gw} &=&  \langle \mathrm{vac} | \hat{p}_{gw} | \mathrm{vac} \rangle = 
 \frac{1}{8 \ell_{\mathrm{P}}^2 a^2} \biggl( \langle  \hat{H}_{k\ell}\,\,\hat{H}^{\,k\ell} \rangle - \frac{1}{3} \langle \partial_{m} \hat{h}_{k \ell} \,\partial^{m} \hat{h}^{\,k\ell} \rangle \biggr).
\label{pFav}
\end{eqnarray}
Using Eqs. (\ref{PAM16}) and (\ref{PAM17}) inside Eqs. (\ref{rhoFav}) and (\ref{pFav}) the expectation 
values of the energy density and of the pressure can be expressed in terms of the tensor power 
spectra $P_{T}(k,\tau)$ and $Q_{T}(k,\tau)$:
\begin{eqnarray}
\overline{\rho}_{\,gw\,} &=& \frac{1}{8 \ell_{\mathrm{P}}^2 a^2} \int_{0}^{\infty} \frac{d\,k}{k} \biggl[ Q(k,\tau) + k^2\, P_{T}(k,\tau) \biggr],
\label{rhoFav1}\\
\overline{\,p\,}_{gw} &=& \frac{1}{8 \ell_{\mathrm{P}}^2 a^2} \int_{0}^{\infty} \frac{d\,k}{k} \biggl[ Q(k,\tau) - \frac{k^2}{3} \, P_{T}(k,\tau) \biggr].
\label{pFav1}
\end{eqnarray}
The spectral energy density in critical units follows from Eq. (\ref{rhoFav1}):
\begin{equation}
\Omega_{gw}(k,\tau) = \frac{1}{\rho_{crit}} \, \frac{d \,\overline{\,\rho\,}_{gw}}{d \ln{k}} \equiv  \frac{1}{24 \, H^2 \, a^2 }  \biggl[ Q_{T} + k^2 P_{T}\biggr],\qquad \rho_{crit} = 3 \, H^2 \, \overline{M}_{P}^2.
\label{SPAM3}
\end{equation}
 In analogy with Eq. (\ref{SPAM3}), the  spectral pressure in critical units can be defined from Eq. (\ref{pFav1}):
\begin{equation}
\Sigma_{gw}(k,\tau) = \frac{1}{\rho_{crit}} \, \frac{d \,\overline{\,p\,}_{gw}}{d \ln{k}} \equiv \frac{1}{24 \, H^2 \, a^2 }  \biggl[ Q_{T} -  \frac{k^2}{3} P_{T}\biggr].
\label{SPAM4}
\end{equation}
From Eqs. (\ref{rhoFav1})--(\ref{pFav1}) and (\ref{SPAM3})--(\ref{SPAM4}) the approximate form of the barotropic indices can be determined in the high-frequency regime from Eq. (\ref{PAM30a}):
 \begin{equation}
 \lim_{k\gg a H } \,\,\frac{\Sigma_{gw}(k,\tau)}{\Omega_{gw}(k,\tau)} = \frac{1}{3} \biggl( 1 + \frac{a^2\, H^2}{k^2} \biggr).
 \label{tenF}
 \end{equation}
Equation (\ref{tenF}) confirms then the general wisdom: when the wavelengths are inside the Hubble radius (i.e. frequencies larger than the rate of variation of the geometry)  the barotropic index of the relic gravitons coincides approximately with $1/3$ to leading order in $a^2\, H^2/k^2 \ll 1$.  However, in the low-frequency limit  $k \ll |a\, H|$ (i.e.  {\em below} the thick curve in Figs. \ref{SEC3FIG1} and \ref{SEC3FIG2}) Eqs. (\ref{SUP2}) and (\ref{SUP3}) imply instead: 
\begin{eqnarray}
w_{gw} &=& \frac{\Sigma_{gw}(k,\tau) }{\Omega_{gw}(k,\tau)} \to - \frac{1}{3} , \qquad  k^2 \bigl| F_{k}(\tau_{ex})\bigr|^2 \gg   \biggl(\frac{a_{ex}}{a}\biggr)^4 \bigl| G_{k}(\tau_{ex})\bigr|^2,
\label{fifteenF}\\
w_{gw} &=& \frac{\Sigma_{gw}(k,\tau)}{\Omega_{gw}(k,\tau)} \to 1 , \qquad  k^2 \bigl| F_{k}(\tau_{ex})\bigr|^2 \ll   \biggl(\frac{a_{ex}}{a}\biggr)^4 \bigl| G_{k}(\tau_{ex})\bigr|^2.
\label{sixteenF}
\end{eqnarray}
All in all, when the typical frequency of the gravitons exceeds the rate of variation of the geometry 
(i.e. $k \gg a\, H$)  the high-frequency gravitons behave as a perfect relativistic fluid and the barotropic index is 
$1/3$. In the opposite limit  (i.e. $k \ll a\, H$) the effective barotropic index 
becomes $-1/3$ if the background expands while it becomes $1$ if the background contracts. 
For an expanding background the barotropic 
index of relic gravitons interpolates between $1/3$ (corresponding to the case of a perfect fluid of massless particles) and $-1/3$ (corresponding to the dominance of the spatial gradients in the energy-momentum tensor).  The anisotropic stress and the energy flux, once averaged over the quantum state of the relic gravitons, are both vanishing:
\begin{equation}
 \langle \mathrm{vac} | \hat{\Pi}_{i}^{\,\,j} | \mathrm{vac} \rangle = \langle \mathrm{vac} | \hat{S}_{i} | \mathrm{vac} \rangle =0.
 \label{sevenBa}
 \end{equation}
Using the result of Eq. (\ref{sevenBa}),
Eq. (\ref{sevenB}) can also be averaged (term by term) with the purpose of obtaining an evolution equation for the expectation values of the energy density and of the pressure, i.e. 
\begin{equation}
\partial_{\tau} \overline{\,\rho\,}_{gw} + 3 {\mathcal H} \bigl(\overline{\,\rho\,}_{gw}  + \overline{\,p\,}_{gw} \bigr) =0.
\label{sevenBb}
\end{equation}
Equation (\ref{sevenB}) demonstrates that $\overline{\rho}_{gw} \propto a^{-4}$ in the high-frequency limit where $\overline{\,p\,}_{gw}/\overline{\,\rho\,}_{gw} \to 1/3$. 
Inserting then Eqs. (\ref{fifteenF})--(\ref{sixteenF}) into Eq. (\ref{sevenBb}), in the low-frequency limit we will have that the mean energy density of the relic gravitons scales as $\overline{\rho}_{gw} \propto a^{-2}$ (if the background expands) and as $\overline{\rho}_{gw} \propto a^{-6}$ (if the background contracts). Needless to say, the approach of the effective action of Eqs. (\ref{fourB}) and (\ref{fiveB}) can be easily 
extended to other situations; in particular, recalling the results of Eq. (\ref{ACT11}), the  effective energy-momentum tensor of the relic gravitons in the Jordan frame is: 
\begin{equation}
T_{\mu\nu}^{(J)} = \frac{A}{4 \ell_{P}^2} \biggl[ \partial_{\mu} \overline{h}^{\,\,\,\,(J)}_{k\ell} \partial_{\nu} \overline{h}^{(J)\,\,\,\, k\ell} - 
\frac{1}{2} \overline{G}_{\mu\nu} \biggl( \overline{G}^{\alpha\beta} \partial_{\alpha} \overline{h}^{\,\,\,\,(J)}_{k\ell} \partial_{\beta} \overline{h}^{(J)\,\, k\ell}\biggr) \biggr],
\label{threeP}
\end{equation}
in full analogy with the result of Eq. (\ref{fiveB}). The energy density in the $J$-frame becomes
\begin{equation}
\rho_{gw}^{(J)} = \frac{A}{8 \ell_{P}^2 a^2_{J}} \biggl[ \partial_{\tau} \, \overline{h}^{\,\,\,\,(J)}_{k\ell} \partial_{\tau} \overline{h}^{(J)\,\,k\ell}
+ \partial_{m} \overline{h}^{\,\,(J)}_{k\ell}\partial^{m} \overline{h}^{(J)\,\,k\ell} \biggr].
\label{fourP}
\end{equation}
If we now compare Eq. (\ref{fourP}) with Eq. (\ref{rhoF}) and recall Eq. (\ref{ACT11a}) 
that connects the tensor modes of the geometry in the two frames we have that
 the energy densities in the two frames are related as:
\begin{equation}
\rho_{gw}^{(J)} = A^2 \, \rho_{gw}^{(E)} \equiv\, \frac{\sqrt{- \overline{g}}}{\sqrt{ - \overline{G}}} \rho_{gw}^{(E)}. 
\label{sixP}
\end{equation}
The result of Eq. (\ref{sixP}) seems to imply that the energy density is not frame-invariant. However this impression is incorrect since the energy density of the background also scales in a different ways in two 
conformally related frames. So, for instance,
the energy density of a radiation plasma scales as $\rho_{r}^{(J)} = A^2 \, \rho^{(E)}_{r}$; thanks to Eq. (\ref{ACT11a}) we will then have that $\rho_{gw}^{(J)}/\rho_{r}^{(J)}= \rho_{gw}^{(E)}/\rho_{r}^{(E)}$. For the same reason the spectral energy distributions in critical units will be the same in the two frames, i.e.  $\Omega_{gw}^{(J)} = \Omega_{gw}^{(E)}$.

\subsubsection{Drawbacks of other proposals for the energy density of the relic gravitons}
The effective energy-momentum pseudo-tensor of Eqs. (\ref{fourB}) and (\ref{fiveB}) leads to a sound and 
consistent definition of the energy density of the relic gravitons both in the high-frequency regime and in the low-frequency limit. The two complementary approaches associated with the  Landau-Lifshitz pseudo-tensor \cite{TENS2b} (see also \cite{TENS3cc,TENS3c}) and with Brill-Hartle-Isaacson pseudo-tensor \cite{TENS2c,TENS3a,TENS3b} agree with Eqs. (\ref{fourB}) and (\ref{fiveB}) in the high-frequency limit. However, at low-frequencies, the different proposals lead to expressions that are inconsistent 
and also rather bizarre; the main differences have been swiftly summarized in Tab. \ref{SEC3TAB2}.
The Landau-Lifshitz approach is based on the 
analysis of the nonlinear corrections to the Einstein tensor consisting, to lowest order,  
of quadratic combinations of the tensor amplitude $h_{ij}$. In contrast with the derivation 
of Eq. (\ref{fiveB}) the first-order evolution of the tensor amplitude must be heavily used in the second-order expressions.  The Landau-Lifshitz pseudo-tensor ${\mathcal L}_{\mu}^{\nu}$ can be formally expressed as: 
\begin{equation}
\ell_{P}^2 {\mathcal L}_{\mu}^{\nu} = - \delta^{(2)}_{t} {\mathcal G}_{\mu}^{\nu},\qquad {\mathcal G}_{\mu}^{\nu} 
= R_{\mu}^{\nu} - \frac{1}{2} \,\delta_{\mu}^{\nu} \, R.
\label{oneC}
\end{equation}
Since the Bianchi identity $\nabla_{\mu} {\cal G}_{\nu}^{\mu}=0$
 should be valid to all orders, we will also have to demand  $\delta_{\rm t}^{(2)} ( \nabla_{\mu} {\cal G}^{\mu}_{\nu}) =0$ and this condition leads to a relation analog to Eq. (\ref{sevenB}) \cite{TENS5}. Thus, in the Landau-Lifshitz approach the pressure and the energy density of the relic gravitons are \cite{TENS5,TENS3cc,TENS3c}:
\begin{eqnarray}
\rho_{gw}^{(L)} &=&  \frac{1}{a^2 \ell_{P}^2} \biggl[ {\mathcal H} \,
\partial_{\tau}h_{k\ell }\, h^{k\ell} + \frac{1}{8} \biggl( \partial_{m} h_{k\ell}\, \partial^{m} h^{k\ell} + 
\partial_{\tau} h_{k\ell} \partial_{\tau} h^{k\ell}\biggr)\biggr],
\label{rhoL}\\
p_{gw}^{(L)} &=& - \frac{1}{24 a^2 \ell_{P}^2}\biggl( 5 \,\partial_{\tau}h_{k\ell}\,\partial_{\tau}h^{k\ell} - 7
\partial_{m} h_{k\ell}\, \partial^{m} h^{k\ell} \biggr).
\label{pL}
\end{eqnarray}
The associated anisotropic stress is also traceless and will be omitted for the sake of conciseness. Within the covariant approach the energy-momentum tensor following from the Brill-Hartle average is given by: 
\begin{equation}
{\mathcal B}_{\mu\nu} = \frac{1}{4 \ell_{P}^2} \overline{\nabla}_{\mu} \overline{h}_{\alpha\beta} \, \overline{\nabla}_{\nu} \, 
\overline{h}^{\alpha\beta}.
\label{threeD}
\end{equation}
Equation (\ref{threeD}) is the result of an averaging procedure that excludes, by construction, 
the frequencies that are smaller than the rate of variation of the geometry \cite{TENS2c,TENS3a,TENS3b}.  To compare the covariant approach with the other proposals we recall that $ \overline{h}_{ij} = - a^2 h_{ij}$; after inserting this expression into Eq. (\ref{threeD}), the result for the various components of ${\mathcal B}_{\mu}^{\,\,\nu}$ becomes:
\begin{equation}
{\mathcal B}_{0}^{\,\,0} = \rho_{gw}^{(B)}, \qquad {\mathcal B}_{i}^{\,\,j} = - p_{gw}^{(B)} \delta_{i}^{\,\, j} + \Pi_{i}^{(B)\,\, j}, \qquad  {\mathcal B}_{i}^{\,\,0} = S_{i}^{(B)} = \frac{1}{4 \ell_{P}^2 a^2} \, h_{k\ell}^{\prime} \partial_{i} h^{k \ell},
\label{fourD}
\end{equation}
where the superscript reminds that the result has been obtained within the Brill-Hartle average:
\begin{eqnarray}
\rho_{gw}^{(B)} &=& \frac{1}{4 \ell_{P}^2 a^2} \partial_{\tau} h_{k \ell} \, \partial_{\tau} h^{k \ell},\qquad
p_{gw}^{(B)} = \frac{1}{12 \ell_{P}^2 a^2} \partial_{a} h_{k \ell} \, \partial^{a} h^{k \ell},
\label{pB}\\
\Pi_{i}^{(B)\,\,j} &=&  \frac{1}{4 \ell_{P}^2 a^2} \biggl( - \partial_{i} h_{k \ell} \partial^{j} h^{k \ell}  + \frac{1}{3} \delta_{i}^{\,\, j} \partial_{a} h_{k \ell} \partial^{a} h^{k \ell} \biggr).
\label{anB}
\end{eqnarray}
Let us now compare the canonical expression of the energy density of the gravitons $\rho_{gw}$ (see Eq. (\ref{rhoFop}))
with the Landau-Lifshitz result $\rho_{gw}^{(L)}$ (see Eq. (\ref{rhoL})) and with the Brill-Hartle-Isaacson 
paremetrization $\rho_{gw}^{(B)}$ (see first relation in Eq. (\ref{pB})); the result of this comparison can be 
summarized by giving the explicit relations connecting $\rho_{gw}^{(L)}$ and $\rho_{gw}^{(B)}$ to $\rho_{gw}$
\begin{equation}
\rho^{(L)}_{gw} = \rho_{gw} + \frac{{\mathcal H} h_{k\ell}^{\prime} \, h^{k \ell}}{a^2 \ell_{P}^2},
\qquad
\rho_{gw}^{(B)} = 2 \biggl(\rho_{gw}- \frac{1}{8 \ell_{P}^2 a^2} \partial_{m} h_{k\ell} \partial^{m} h^{k\ell}\biggr).
\label{BtoF}
\end{equation}
Since in the high-frequency limit $k^2 P_{T} \to Q_{T}$ for $k \gg |a\, H|$ (see Eq. (\ref{PAM30a})), 
the relation among the different forms of the mean energy density follows immediately from Eq. (\ref{BtoF}) and it is given by:
\begin{equation}
\overline{\,\rho\,}_{gw} = \overline{\rho}^{(L)}_{gw} = \overline{\,\rho\,}^{(B)}_{gw} = \frac{1}{4 \ell_{P}^2 a^2} \int 
\frac{d \, k}{k} \, k^2 \, P_{T}(k,\tau), \qquad k \gg a\, H.
\label{comp1}
\end{equation}
Recalling now the definition given in Eq. (\ref{SPAM3}) it follows from Eq. (\ref{comp1}) 
 that the spectral energy density coincides in all the approaches provided 
$k \gg a\, H$, i.e. 
\begin{equation} 
\Omega_{gw}(k,\tau) = \Omega^{(L)}_{gw}(k,\tau) = \Omega^{(B)}_{gw}(k,\tau) = \frac{k^2 P_{T}(k,\tau)}{12 H^2 a^2}, \qquad k \gg a\, H.
\label{comp2}
\end{equation}
In the opposite limit the comparison between the different proposals is a bit more involved and can be found in Ref. \cite{mgcanonical}; a concise summary of the 
main features of the different choices in given in Tab. \ref{SEC3TAB2}.
\begin{table}[!ht]
\begin{center}
\caption{Summary of the salient properties of the different pseudo-tensors in the case of expanding backgrounds; 
$w_{gw}$ denotes the ratio of the spectral pressure and of the spectral energy density in the various cases.}
\vskip 0.5 cm
\begin{tabular}{||c|c|c|c|c||}
\hline
\hline
\rule{0pt}{4ex} pseudo-tensor& $w_{gw}$ ($k \gg a\, H$)  & $w_{gw}$ ($k \ll  a\, H$) & $\rho_{gw}^{(X)}$ ($k \gg a\, H$) & $\rho_{gw}^{(X)}$($k \ll a\, H$) \\
\hline
\hline
${\mathcal F}_{\mu\nu}$& $1/3$ & $-\,1/3$  & $\overline{\rho}_{gw} \geq 0$ & $\overline{\rho}_{gw} \geq 0$ \\
${\mathcal L}_{\mu\nu}$& $1/3$ & $-1/3$ & $\overline{\rho}_{gw}^{(L)}\geq 0 $ & $\overline{\rho}_{gw}^{(L)}\leq 0 $  \\
${\mathcal B}_{\mu\nu}$ & $1/3$ & $\infty$ & $\overline{\rho}_{gw}^{(B)}  \geq 0$ & $\overline{\rho}_{gw}^{(B)}  \geq 0$  \\
\hline
\hline
\end{tabular}
\label{SEC3TAB2}
\end{center}
\end{table}
According to Tab. \ref{SEC3TAB2} we have that in the high-frequency limit  
the expression of the energy-momentum pseudo-tensor is immaterial since all the proposals converge 
to the same results even if they are formally different. Beyond the Hubble radius (i.e. $k \tau \ll 1$) we have instead 
that in the Landau-Lifshitz parametrization the energy density {\em is typically negative}. In the Brill-Hartle-Isaacson 
approach {\em the barotropic index diverges} when $k\ll a\, H$ and this conclusion simply follows from the fact that $\overline{\rho}_{gw}^{(B)} \to 0$ in the same limit.
The inconsistencies of the Brill-Hartle-Isaacson  approach can be mended by using the 
covariant form of the Isaacson energy-momentum tensor (i.e. without imposing the Brill-Hartle 
average). We can deduce this covariant form of the energy-momentum tensor 
from the variation of the action (\ref{ACT9b}) with respect to $\overline{g}_{\mu\nu}$:
\begin{eqnarray}
T_{\mu\nu}^{(gw)} &=& \frac{1}{4\ell_{P}^2} \biggl[ \overline{\nabla}_{\mu} \overline{h}_{\alpha\beta}  \overline{\nabla}_{\nu} \overline{h}^{\,\,\alpha\beta}  +  \overline{\nabla}_{\alpha} \overline{h}_{\mu\beta}  \overline{\nabla}^{\alpha} \overline{h}^{\,\,\,\,\beta}_{\nu} +  \overline{\nabla}_{\alpha} \overline{h}_{\nu\beta}  \overline{\nabla}^{\alpha} \overline{h}^{\,\,\,\,\beta}_{\mu} 
\nonumber\\
&+& 2 \overline{R}^{\gamma}_{\,\,\,\,\,\,\mu \rho \alpha} \, \overline{h}_{\gamma}^{\,\,\alpha} \, \overline{h}_{\nu}^{\,\,\rho}
+ 2 \overline{R}^{\,\,\gamma}_{\,\,\,\,\,\, \nu \rho \alpha} \, \overline{h}_{\gamma}^{\,\,\,\alpha} \, \overline{h}^{\rho}_{\,\,\,\mu}
- \frac{1}{2} \overline{g}_{\mu\nu} \biggl( \overline{\nabla}_{\rho} \overline{h}_{\alpha\beta} \overline{\nabla}^{\rho} \overline{h}^{\alpha\beta} + 2 \overline{R}^{\,\,\gamma}_{\,\,\,\,\,\alpha\beta\rho} \, \overline{h}_{\gamma}^{\,\,\,\,\rho} \overline{h}^{\,\alpha\beta} \biggr)\biggr].
\label{sevenH}
\end{eqnarray}
If we would now apply the tenets of the Brill-Hartle procedure \cite{TENS2c} the covariant gradients average out to zero. After flipping the covariant derivative from one amplitude to the other with the terms inside the squared bracket  of Eq. (\ref{sevenH}) we can obtain terms like $\overline{h}^{\alpha\beta} \overline{\nabla}_{\rho} \overline{\nabla}^{\rho} \overline{h}_{\alpha\beta}$. All these terms will produce various Riemann tensors 
that will be neglected so that, at the very end, the only term surviving the average will be the first contribution 
of Eq. (\ref{sevenH}), i.e. 
\begin{equation}
T_{\mu\nu}^{(gw)} = {\mathcal B}_{\mu\nu}  = \frac{1}{4 \ell_{P}^2} \langle \overline{\nabla}_{\mu} \overline{h}_{\alpha\beta}  \overline{\nabla}_{\nu} \overline{h}^{\alpha\beta} \rangle_{BH}  
= \frac{1}{4 \ell_{P}^2} \langle \partial_{\mu} h_{ij}  \partial_{\nu} \, h^{i j} \rangle_{BH}. 
\label{eightH}
\end{equation}
Equation (\ref{eightH}) coincides with Eq. (\ref{threeD}) and it shows that the Brill-Hartle average 
{\em effectively neglects all the terms that are relevant beyond the Hubble radius}. 
However, if we want a result applicable beyond the Hubble radius we have therefore to consider 
 Eq. (\ref{sevenH}) {\em without imposing the Brill-Hartle averaging} but rather the quantum averaging
 discussed above. If we then directly use Eq. (\ref{sevenH}) and express it in the conformally flat case 
 (i.e. $\overline{g}_{\mu\nu} = a^2 \eta_{\mu\nu}$ and $ \overline{h}_{ij} = - a^2 h_{ij}$)
 we shall obtain, after a lengthy but straightforward calculation, the same expression 
 of the effective energy-momentum tensor ${\mathcal F}_{\mu\nu}$ given in Eqs. (\ref{rhoF})--(\ref{pF}).

In conclusion, the most sound form of the energy-momentum pseudo-tensor of the relic gravitons follows from the variation of the second-order action with respect to the background metric. The other approaches are fully equivalent in the high-frequency limit (i.e. inside the Hubble radius) but lead to various drawbacks when the wavelengths exceed the Hubble radius (i.e. in the low-frequency regime). Since the rate of variation of the space-time curvature can be both larger and smaller than the typical frequencies 
of the relic gravitons, the super-adiabatic amplification does not imply any violation of causality. It is 
therefore crucial to adopt a definition for the energy-momentum pseudo-tensor that is well defined 
in spite of of the frequency of the gravitons.

\subsection{Relations among the observables}
The backgrounds of relic gravitons are described in terms of four pivotal variables: 
the {\em tensor power spectrum} (see e.g. (\ref{fiveE})), {\em the spectral energy density in critical units} (see e.g. (\ref{SPAM3})),  {\em the chirp amplitude} and {\em the spectral amplitude}. The chirp amplitude (denoted hereunder by $h_{c}(k,\tau)$) 
is related to the tensor power spectrum as $2 \,h_{c}^2(k,\tau) = P_{T}(k,\tau)$, i.e. up to a factor of $2$ the square of the chirp amplitude coincides with our definition of the tensor power spectrum.  The spectral amplitude is customarily introduced in the analysis of stationary random processes where the auto-correlation function only depends on the difference between the times at which the random functions 
are evaluated. In what follows the power spectra, the spectral energy density in critical units, the chirp amplitude and the spectral amplitude will be generically referred to as the {\em observables} 
since they can be determined, at least in principle, from dedicated observations. 
To clarify the connections between the observables it is practical we start from the effective energy density of the relic gravitons deduced in Eq. (\ref{rhoF}):
\begin{equation}
\rho_{gw} = \frac{1}{8 \ell_{\mathrm{P}}^2 a^2} \biggl( \partial_{\tau} h_{k \ell}\, \partial_{\tau}h^{k \ell} + \partial_{m} h_{k\ell} \partial^{m} h^{k\ell}\biggr).
\label{OBS0}
\end{equation}
If the relic gravitons are in a specific quantum state,  the averaged version of Eq. (\ref{OBS0}) follows from the appropriate quantum mechanical expectation values (see Eq. (\ref{PAM9a})
and discussions therein). However, if the tensor amplitudes of Eq. (\ref{OBS1}) are viewed just as isotropic random fields,  the mean energy density can be formally defined as:
\begin{equation}
\overline{\,\rho\,}_{gw} = \frac{1}{8 \ell_{\mathrm{P}}^2 a^2} \biggl(\langle \partial_{\tau} h_{k \ell}\, \partial_{\tau}h^{k \ell} \rangle + \langle \partial_{m} h_{k\ell} \partial^{m} h^{k\ell} \rangle\biggr).
\label{OBS1}
\end{equation}
The assumption that the stochastic background is isotropic and unpolarized implies that the ensemble average of the Fourier amplitudes introduced in Eq. (\ref{POLDEF5}) is given by:
\begin{eqnarray}
\langle h_{ij}(\vec{k},\tau) \, h_{mn}(\vec{k}^{\prime}, \tau) \rangle &=& \frac{2 \pi^2}{k^3}\, P_{T}(k,\tau) \, \delta^{(3)}(\vec{k} + \vec{k}^{\prime}) \, 
{\mathcal S}_{i \,j \,m \,n}(\hat{k})
\label{OBS3}\\
\langle H_{ij}(\vec{k},\tau) \, H_{mn}(\vec{k}^{\prime}, \tau) \rangle &=& \frac{2 \pi^2}{k^3}\, Q_{T}(k,\tau) \, \delta^{(3)}(\vec{k} + \vec{k}^{\prime}) \, 
{\mathcal S}_{i\,j\,m\,n}(\hat{k}),
\label{OBS4}
\end{eqnarray}
where, as before, $H_{ij} = \partial_{\tau} h_{ij}$. Equations (\ref{PAM16})--(\ref{PAM17}) are the quantum mechanical counterpart of Eqs. (\ref{OBS3})--(\ref{OBS4}) where however $h_{ij}$ and $H_{ij}$ are not
operators but simply isotropic random fields.  In terms of $P_{T}(k,\tau)$ and $Q_{T}(k,\tau)$ the mean energy density is then given by: 
\begin{equation}
\overline{\rho}_{gw} = \frac{1}{8\, \ell_{P}^2\, a^2} \int_{0}^{\infty} \, \frac{d k}{k} \biggl[ k^2 P_{T}(k, \tau) + Q_{T}(k,\tau) \biggr],
\label{OBS5}
\end{equation}
and from Eq. (\ref{OBS5})  the spectral energy density in critical units is formally defined in the same 
manner already mentioned in Eq. (\ref{SPAM3}): 
\begin{equation}
\Omega_{gw}(k,\tau) = \frac{1}{\rho_{crit}} \,  \frac{d\, \overline{\,\rho\,}_{gw}}{d \ln{k}} = \frac{1}{24 H^2 a^2 } \biggl[ k^2 P_{T}(k, \tau) + Q_{T}(k,\tau) \biggr],
\label{OBS6}
\end{equation}
where $\rho_{crit} = 3 H^2/\ell_{\mathrm{P}}^2$.
Equations (\ref{OBS3})--(\ref{OBS4}) and (\ref{OBS6}) express the general relation of the power spectra with spectral energy density in critical units. Since $Q_{T}(k,\tau) \to k^2 P_{T}(k,\tau) $ for $k \gg a H$, the spectral energy density  for typical wavelengths shorter that the Hubble radius follows from Eq. (\ref{OBS6}): 
\begin{equation}
\Omega_{gw}(k,\tau) =\frac{k^2}{12 H^2 a^2 } P_{T}(k,\tau), \qquad k \gg a\, H.
\label{OBS7}
\end{equation}
In the opposite limit (i.e.  when the wavelengths are larger than the Hubble radius) and 
for an expanding background geometry (see Eqs. (\ref{SUP2})--(\ref{SUP3}) and discussion therein), $Q_{T}(k,\tau_{ex}) \ll k^2 P_{T}(k,\tau_{ex}) $
where $\tau_{ex}$ is defined by $k \simeq a_{ex} H_{ex}$. Thus  Eq. (\ref{OBS6}) implies: 
\begin{equation}
\Omega_{gw}(k,\tau) =\frac{k^2}{24 H^2 a^2 } P_{T}(k,\tau_{ex}), \qquad k \ll a\, H.
\label{OBS8a}
\end{equation}
Equations (\ref{OBS7}) and (\ref{OBS8a}) provide the wanted relation between the power spectra of the random fields and their spectral energy density in critical units.  If the tensor amplitude is expanded as in  Eqs. (\ref{SIXPOL8})--(\ref{OBS13b})
\begin{equation}
h_{i\, j}(\vec{x}, \tau) = \int_{-\infty}^{\infty} d \nu \int d\,\hat{k}  \, e^{ 2\,i\,\pi \, \nu\,( \tau - \hat{k}\cdot\vec{x})} \,h_{i\, j}(\nu, \hat{k}), \qquad 
\,h^{*}_{i\, j}(\nu, \hat{k}) = \,h_{i\, j}(-\nu, \hat{k}),
\label{OBS11}
\end{equation}
the expectation value of the random  amplitudes will be:
\begin{equation}
\langle h_{\lambda} (\nu, \, \hat{k}) \, h_{\lambda^{\prime}} (\nu^{\prime}, \, \hat{k}^{\prime}) \rangle = 
{\mathcal A}_{c} S_{h}(|\nu|) \, \delta(\nu + \nu^{\prime}) \, \delta^{(2)}(\hat{k} - \hat{k}^{\,\prime}),
\label{OBS13a}
\end{equation}
where ${\mathcal A}_{c}$ is an overall constant that shall be fixed later and it parametrizes the different choices 
currently adopted by different authors; the choices for ${\mathcal A}_{c}$ will be discussed in a moment. In any case, consistently with the notations of Eq. (\ref{SIXPOL8}), the angular delta function is $\delta^{(2)}(\hat{k} - \hat{k}^{\,\prime}) = 
\delta(\varphi -\varphi^{\prime})\, \delta(\cos{\vartheta} - \cos{\vartheta}^{\prime})$.
The expectation value  $\langle h_{i\, j}(\nu, \hat{k}) h_{\ell\, m}(\nu^{\prime}, \hat{k}^{\prime}) \rangle$ 
follows from Eqs. (\ref{OBS13a})--(\ref{OBS13b}) and from the definition of Eq. (\ref{POLDEF7}):
\begin{equation}
\langle h_{i\, j}(\nu, \hat{k}) h_{\ell\, m}(\nu^{\prime}, \hat{k}^{\prime}) \rangle = 4 \, {\mathcal A}_{c}{\mathcal S}_{i\, j\, \ell\, m}(\hat{k}) \, S_{h}(|\nu|) \, \delta^{(2)}(\hat{k} - \hat{k}^{\prime}) \, \delta(\nu + \nu^{\prime}).
\label{OBS13c}
\end{equation}
Using Eq. (\ref{OBS13c}) we can also compute the expectation value of two amplitudes 
at two different times:
\begin{equation}
\Gamma(\vec{x}, \tau- \tau^{\,\prime}) = \langle h_{ij}(\vec{x}, \tau)\, h^{ij}(\vec{x}, \, \tau^{\prime}) \rangle 
= 16 \pi {\mathcal A}_{c} \, \int_{-\infty}^{\infty} \, d \nu\,\,e^{ 2 \,i \, \pi \,\nu(\tau- \tau^{\,\prime})} S_{h}(|\nu|),
 \label{OBS13cc}
 \end{equation}
 According to Eq. (\ref{OBS13cc}) the spectral amplitude and the autocorrelation function of the process 
form a Fourier transform pair; this statement is often referred to as Wiener-Khintchine theorem (see e.g. \cite{QM0}). It should be clear that the possibility of defining a spectral amplitude relies on the stationary nature 
of the underlying random process (see Ref. \cite{STOC} for an introduction to stochastic processes). In the case of relic gravitons, i.e. gravitons coming from the 
early universe, this is not the case since the power spectra are characterized by standing 
waves that are nothing but the so-called Sakharov oscillations mentioned earlier on in this article. There are claims in the current literature \cite{AFP} suggesting that, for all practical purposes, the non-stationary nature of the random processes associated with relic gravitons cannot be observationally determined; however, from a conceptual viewpoint, the relic gravitons do not lead in general to  random processes that are truly stationary. It can be argued that {\em the stochastic process is close to stationary only inside the Hubble radius} where the full expression of the effective energy density given in Eq. (\ref{OBS1}) 
coincides with the Brill-Hartle-Isaacson result (see Eq. (\ref{pB}) and discussion therein); 
the energy density for an approximately stationary process can then be expressed as
\begin{equation}
\overline{\,\rho\,}_{gw} = \frac{1}{4 \ell_{P}^2 a^2} \langle \partial_{\tau} h_{ij} \, \partial_{\tau} h^{ij} \rangle, \qquad k \gg a\, H.
\label{OBS10a}
\end{equation}
Inserting Eq. (\ref{OBS13b}) into Eq. (\ref{OBS10a}) the mean energy density becomes 
\begin{equation}
\overline{\rho}_{gw} = \frac{16 \, \pi^3\, {\mathcal A}_{c}}{ \ell_{P}^2 \, a^2} \int_{-\infty}^{\infty} \, d\nu \, \nu^2 \, S_{h}(|\nu|)
\equiv\frac{32 \, \pi^3\, {\mathcal A}_{c}}{ \ell_{P}^2 \, a^2} \int_{0}^{\infty} \, \frac{d\nu}{\nu} \, \nu^3 \, S_{h}(|\nu|).
\label{OBS13d}
\end{equation}
The relations between the spectral energy density in critical units, the spectral density and the chirp amplitude are then given by:
\begin{equation}
 \Omega_{gw}(\nu) = \frac{32 \, \pi^3\, \nu^{3} {\mathcal A}_{c} }{3 H^2 a^2} \,S_{h}(|\nu|), \qquad P_{T}(\nu) = 32 \,\pi \,\nu\, {\mathcal A}_{c} \,S_{h}(|\nu|),\,
\qquad P_{T}(\nu) = 2\, h_{c}^2(\nu).
\label{OBS13e}
\end{equation}
The LIGO/Virgo collaboration is normally setting ${\mathcal A}_{c} = 1/(16\, \pi)$ so that, at the present time (i.e. $a_{0} =1$),
\begin{equation}
 \Omega_{gw}(\nu) = \frac{2 \, \pi^2\, \nu^{3}  }{3 H_{0}^2} \,S_{h}(\nu),\qquad \langle h_{ij}(\vec{x},\tau) \, h^{ij}(\vec{x},\tau) \rangle = 
 2 \int_{0}^{\infty}  d \,\nu \, S_{h}(|\nu|). 
 \label{OBS13f}
\end{equation}
Equation (\ref{OBS13f}) coincides, incidentally, with the one employed when setting the recent limits on the stochastic backgrounds of relic gravitons in Ref. \cite{SIXPOL1}.  More precisely Eq. (6) of Ref. \cite{SIXPOL1} coincides exactly with Eq. (\ref{OBS13f}) obtained from ${\mathcal A}_{c} = 1/(16\, \pi)$. Another conventional choice is ${\mathcal A}_{c} = 1/(32\, \pi)$ implying, from Eq. (\ref{OBS13e}), that the spectral energy density becomes:
\begin{equation}
 \Omega_{gw}(\nu) = \frac{\pi^2\, \nu^{3}  }{3 H_{0}^2} \,S_{h}(|\nu|),\qquad \langle h_{ij}(\vec{x},\tau) \, h^{ij}(\vec{x},\tau) \rangle = 
 \int_{0}^{\infty}  d \,\nu \, S_{h}(|\nu|).
 \label{OBS13g}
\end{equation}
The choice of Eq. (\ref{OBS13g}) has been used by some authors (see e.g. Eqs. (11) and (12) of Ref. \cite{NC}) but will not be the one preferentially adopted here.
From Eqs. (\ref{OBS13a}) and (\ref{OBS13e}) the relations between the different observables 
related to relic gravitons can also be expressed in more explicit terms and 
for a pivotal frequency of $100$ Hz roughly corresponding to the  region where 
the advanced LIGO/Virgo interferometers will be more sensitive to a stochastic backgrounds:
\begin{equation}
{\mathcal S}_{h}(|\nu|) = \frac{3 a_{0}^2 H_{0}^2}{32 \pi^3 \nu^3} \Omega_{gw}(\nu)=   3.175\times 10^{-44} \,\,\biggl(\frac{100\,\mathrm{Hz}}{\nu}\biggr)^3 \,\, h_{0}^2 \,\Omega_{gw}(\nu,\tau_{0})\,\, \mathrm{Hz}^{-1},
\label{OBS14}
\end{equation}
where we assumed, for simplicity, ${\mathcal A}_{c} =1$.  The relation between the chirp amplitude and the spectral energy density becomes explicitly:
\begin{equation}
h_{c}(\nu) = 1.263 \times 10^{-20} \biggl(\frac{100 \,\, \mathrm{Hz}}{\nu}\biggr) \, \sqrt{h_{0}^2 \,\Omega_{gw}(\nu,\tau_{0})},
\label{OBS15}
\end{equation} 
and does not depend on the value of ${\mathcal A}_{c}$. 
In the case of conventional inflationary models, for $\nu \simeq 100$ Hz, the spectral energy density is  $h_{0}^2 \Omega_{gw} = {\mathcal O}(10^{-16.5})$; thus Eqs. (\ref{OBS15}) imply  $h_{c} = {\mathcal O}(10^{-29})$
which is between $8$ and $9$ orders of magnitude smaller than the signals currently observed by wide-band interferometers. For the same choice of $h_{0}^2 \Omega_{gw}$ the spectral density is $\sqrt{S_{h}} = {\mathcal O}(10^{-30})\, \mathrm{Hz}^{-1/2}$.

\renewcommand{\theequation}{4.\arabic{equation}}
\setcounter{equation}{0}
\section{Relic gravitons from the early universe}
\label{sec4}
\begin{figure}[!ht]
\centering
\includegraphics[width=0.7\textwidth]{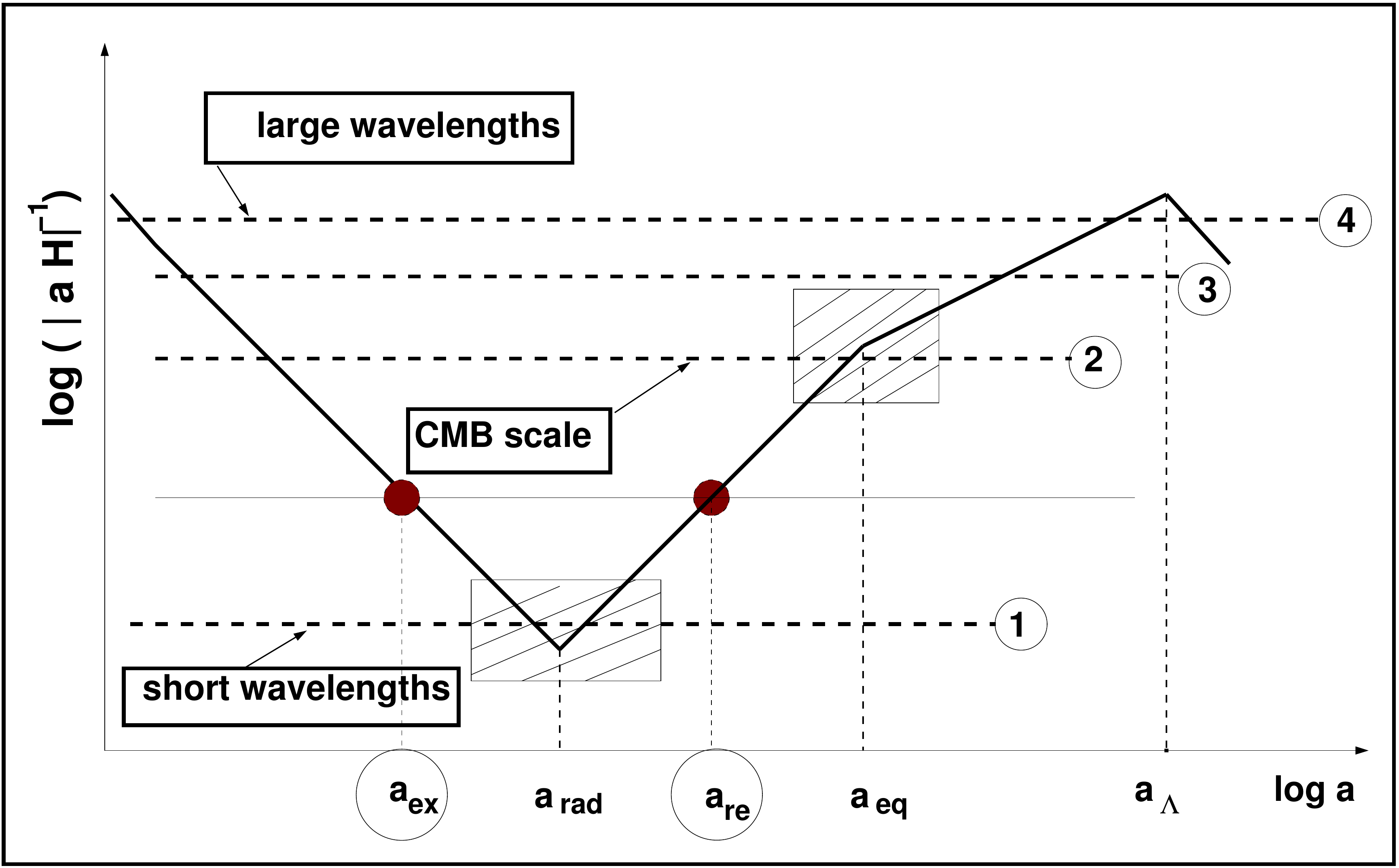}
\caption[a]{The evolution of the Hubble radius is illustrated when the concordance paradigm is complemented 
by an early phase of conventional inflationary evolution. As in the case of Figs. \ref{SEC3FIG1} and \ref{SEC3FIG2}, the vertical axis  in Fig. \ref{SEC4FIG1} has been rescaled for practical reasons. Again the value of the thick curve at the minimum is around $6$ since, at the end of inflation, $(H/M_{P})^{-1} = {\mathcal O}(10^{6})$.
The maximal value of $\bigl| a_{0}\, H_{0} \bigr|^{-1}$  is ${\mathcal O}(61)$ in natural gravitational units and it is reached 
at the present epoch since today $(H_{0}/M_{P})^{-1}= {\mathcal O}(10^{61})$.}
\label{SEC4FIG1} 
\end{figure}

According to the conventional wisdom the concordance paradigm should be complemented at early times by an inflationary stage of expansion. Denoting with $a$ the scale factor, the evolution of the Hubble radius consists schematically of four distinct epochs: ({\em i})  an early inflationary phase (taking place for $a< a_{rad}$);  ({\em ii}) the radiation epoch (when $a_{rad} < a < a_{eq}$);
({\em iii}) the matter phase (i.e.  $a_{eq} <a < a_{\Lambda}$); ({\em iv}) the stage of dark energy for dominance (for  $a > a_{\Lambda}$). In this situation the common logarithm of the Hubble radius is illustrated in Fig. \ref{SEC4FIG1} as a function of the common logarithm of the scale factor. A generic wavelength is said to {\em exit} the Hubble radius 
when it hits (for the first time) the thick black line in Fig. \ref{SEC4FIG1}.  
The same given wavelength will {\em reenter} later on by 
crossing (for the second time) the Hubble radius. In Fig. \ref{SEC4FIG1} the scale factors at the exit and at the reentry are denoted by $a_{ex}$ and $a_{re}$ respectively.  
The full thin line in  Fig. \ref{SEC4FIG1} illustrates the first and second crossings 
 of a generic (comoving) wavelength exiting the Hubble radius during inflation and reentering 
 during the radiation-dominated epoch.  
 
\begin{figure}[!ht]
\centering
\includegraphics[width=0.6\textwidth]{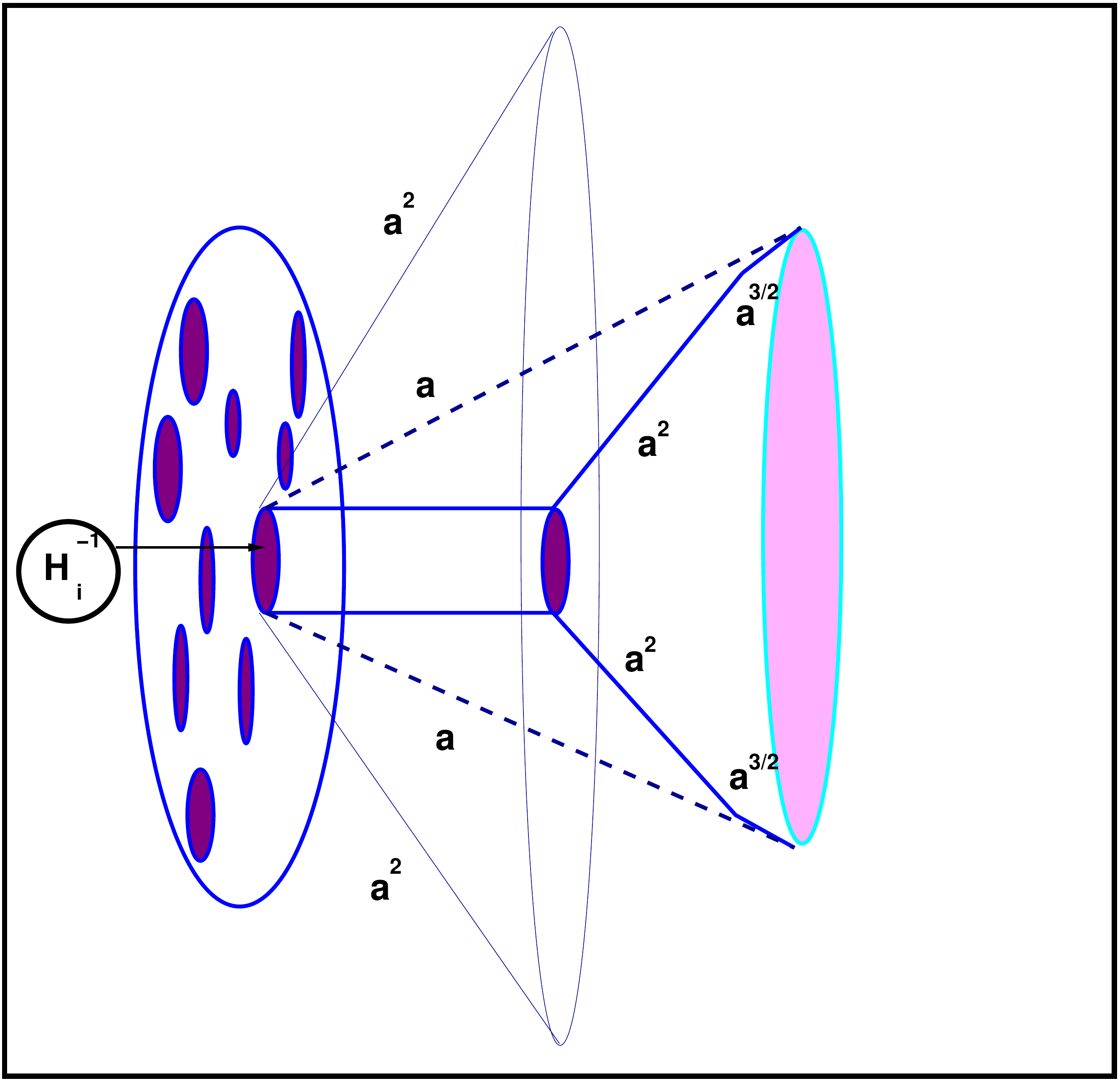}
\caption[a]{From left to right the evolution of the typical size of a causally connected domain is illustrated when the concordance 
paradigm is complemented by a phase of conventional inflationary expansion. }
\label{SEC4FIG2} 
\end{figure}

In the standard decelerated evolution typical of the radiation or matter-dominated epochs  there exists a particle horizon $d_{\mathrm{p}}(t) \simeq H^{-1}(t)$ measuring the typical size of causally connected regions at time $t$. Since the particle horizon evolves {\em faster} than the scale factor, the current extension of the Hubble radius scaled back at the reference time when the initial conditions are set (for instance the Planck time) consisted of a huge number of causally disconnected domains (approximately $10^{30}$).  To cure this drawback (closely related to the features of the standard decelerated expansion) the idea of the inflationary scenario is to
complement the radiation-dominated evolution by an accelerated phase 
where the particle horizon does not exist. In the conventional (slow-roll) 
situation when inflation starts at $t_{i}$ the particle horizon is absent but the {\em event} horizon is of the order of $H_{i}^{-1}$. As we go forward in time $H_{i}^{-1}$ will be approximately constant. In Fig. \ref{SEC4FIG2} the cylinder describes, from left to right, the approximate evolution of $H_{i}^{-1}$; as inflation ends the Hubble radius will evolve 
faster than the scale factor and it will eventually get to its present size.
 However, since the typical size 
of causally connected domains will now scale as the scale factor (i.e. $H_{i}^{-1} (a/a_{i})$) and not 
faster, the whole observable universe will be contained inside a single 
event horizon at $t_{i}$ {\em provided} the inflationary phase is sufficiently long. 
In Fig. \ref{SEC4FIG2} we report (with the dashed lines) the evolution of a causally connected domain 
that can be as large as the current Hubble patch represented by the rightmost filled disc. With 
the thin lines we also show what happens if the inflationary phase disappears: in the case the event horizon is absent so that the particle horizon evolves as $a^2$ (assuming a radiation-dominated evolution), i.e.
faster than the scale factor. In this section we shall assume that the duration of the inflationary phase is sufficiently long. Since the initial classical fluctuations (e.g. spatial gradients) are exponentially (or quasi-exponentially) suppressed during inflation \cite{class1,class2,class3,class4,class5,class6} the quantum fluctuations of the scalar and tensor modes of the geometry will be the only source of 
inhomogeneity\footnote{An exception to this statement is represented by the models where the duration of inflation is minimal or close to minimal. We shall more specifically  discuss in the final part of section \ref{sec5}.} and will then determine the initial conditions.
When the reentry of the wavelengths occurs either during radiation or during matter the 
mode functions associated with the relic gravitons will exhibit the typical patterns of Sakharov 
oscillations \cite{SAK1}. The power spectra and the spectral energy density of the relic gravitons 
will now be scrutinized in the conventional framework of Figs. \ref{SEC4FIG1} and \ref{SEC4FIG2}.  

\subsection{Scales exiting during inflation}
When the concordance paradigm is supplemented by an early inflationary phase (see Fig. \ref{SEC4FIG1}) 
{\em all the wavelengths possibly accessible to present observations exited the Hubble radius during inflation} and reentered at different epochs during the subsequent phases.  If the slow-roll parameters 
of Eqs. (\ref{RGC4}) and (\ref{RGC3}) are approximately 
constant\footnote{When the slow-roll parameters are not constant the evolution 
of the mode functions (see e.g. Eq. (\ref{RGC2}) and discussion therein) can be 
analyzed within the Wentzel-Kramers-Brioullin (WKB) approach, as we shall see in section \ref{sec8}.},
 from the relation between cosmic and conformal times 
[i.e.  $a(\tau) d\tau = d t$] we can deduce (after integration by parts) that $a H = - 1/[\tau ( 1 - \epsilon)] \simeq a$.  The solutions of 
Eqs. (\ref{PAM12}) and  (\ref{RGC2}) can be expressed as:
\begin{eqnarray}
F_{k}(\tau) &=& \frac{f_{k}}{a(\tau)}=  \frac{{\mathcal N}}{\sqrt{2 k}\,\,a(\tau)} \, \sqrt{- k\tau} \, H_{\nu}^{(1)}(- k \tau), \qquad {\mathcal N} 
= \sqrt{\frac{\pi}{2}} e^{i \pi (2 \nu +1)/4},
\label{RGC5a}\\
G_{k}(\tau) &=&  F_{k}^{\,\prime}(\tau) = \frac{{\mathcal N}}{a(\tau)} \sqrt{\frac{k}{2}} \biggl[ -\frac{2 \, \nu}{\sqrt{- k \tau}} H_{\nu}^{(1)}(- k \tau) + \sqrt{- k \tau} \, H_{\nu+1}^{(1)}(- k \tau) \biggr],
\label{RGC5b}
\end{eqnarray}
where $H_{\nu}^{(1)}(-k\tau)$ are the Hankel functions of the first kind \cite{abr1,abr2} and 
the phase appearing in ${\mathcal N}$ has been fixed by requiring that in the limit $\tau \to -\infty$, $f_{k}(\tau) \to e^{- i k \tau}/\sqrt{2 k}$. 
During a quasi-de Sitter stage of expansion we can also express the Bessel index as $\nu = (\beta +1/2)$ since ${\mathcal H} = - \beta/\tau$ and, to first-order  in the slow-roll expansion
\begin{equation}
\beta = \frac{1}{1 - \epsilon} = \biggl(\nu - \frac{1}{2}\biggr), \qquad a_{r} \, H_{r} = - \frac{1}{(1 - \epsilon)\tau_{r}}, \qquad \nu = \frac{(3 - \epsilon)}{2 (1 - \epsilon)}, 
\label{RGC5g}
\end{equation}
where $-\tau_{r}$ denotes the end of the inflationary stage.
Recalling Eqs. (\ref{fiveE}) and the explicit solutions of Eqs. (\ref{RGC5a})--(\ref{RGC5b}) the power spectra $P_{T}(k,\tau)$ and $Q_{T}(k,\tau)$ become
\begin{eqnarray}
P_{T}(k,\tau) &=& \frac{k^2}{\pi \,a^2(\tau) \, \overline{M}_{P}^2} (- k \tau) \bigl| H_{\nu}^{(1)}(- k \tau)\bigr|^2,
\label{RGC5c}\\
Q_{T}(k,\tau) &=&  \frac{k^4}{\pi \,a^2(\tau) \, \overline{M}_{P}^2} \biggl\{ \frac{4 \nu^2}{(- k \tau)} 
\bigl| H_{\nu}^{(1)}(- k \tau)\bigr|^2 + (- k\tau) \bigl| H_{\nu+1}^{(1)}(- k \tau)\bigr|^2
\nonumber\\
&-& 2 \nu \bigg[  H_{\nu}^{(1)}(- k \tau)\,H_{\nu+1}^{(2)}(- k \tau) +  H_{\nu}^{(2)}(- k \tau)\,H_{\nu+1}^{(1)}(- k \tau)\biggr]\biggr\}.
\label{RGC5d}
\end{eqnarray}
The spectral energy density if the relic gravitons can be obtained directly from the power spectra of 
 Eqs. (\ref{RGC5c}) and (\ref{RGC5d}) by recalling the results of Eqs. (\ref{rhoFav1}) and (\ref{SPAM3}): 
\begin{eqnarray}
\Omega_{gw}(k,\tau) &=& \frac{k^{4}}{24 \, \pi\, H^2\, a^4\,\overline{M}_{P}^2} \biggl\{ \biggl[(- k\tau)  + \frac{4 \nu^2}{(- k \tau)}  \biggr] \bigl| H_{\nu}^{(1)}(- k \tau)\bigr|^2 +  (- k \tau) \bigl| H_{\nu + 1}^{(1)}(- k \tau)\bigr|^2 
\nonumber\\
&-& 2 \nu \bigg[  H_{\nu}^{(1)}(- k \tau)\,H_{\nu+1}^{(2)}(- k \tau) +  H_{\nu}^{(2)}(- k \tau)\,H_{\nu+1}^{(1)}(- k \tau)\biggr]\biggr\}.
\label{RGC5e}
\end{eqnarray}
Equation (\ref{RGC5c}) implies that for 
 typical wavenumbers smaller than the rate of variation of the geometry (i.e. $k \ll \bigl| a\, H\bigr|$) 
 during the inflationary  stage of expansion (i.e. $a< a_{r}$) the tensor power spectrum $P_{T}(k,\, \tau)$ 
 is constant in time:
\begin{eqnarray}
\lim_{|k \tau| \ll 1} P_{T}(k,\tau) &\to& \overline{P}_{T}(k,\tau_{r}) = \frac{ 2^{2 \nu} \Gamma^2(\nu)}{\pi^3} \biggl(
\nu - \frac{1}{2} \biggr)^{-2} \, \biggl(\frac{H_{r}}{\overline{M}_{P}}\biggr)^2 \,\, \bigl| k \, \tau_{r} \bigr|^{ 3 - 2 \nu} 
\nonumber\\
&=& \frac{ 2^{2 \nu + 3 } \Gamma^2(\nu)}{\pi^2} \biggl(
\nu - \frac{1}{2} \biggr)^{ 1 - 2 \nu} \, \biggl(\frac{H_{r}}{M_{P}}\biggr)^2 \,\, \biggl( \frac{k}{a_{r} H_{r}}\biggr)^{ 3 - 2 \nu},
 \label{RGC5f}
\end{eqnarray}
as it follows by evaluating $ |F_{k}(\tau)|^2$ (and the corresponding Hankel functions) in the small argument limit  $|k \tau| \ll 1$ \cite{abr1,abr2}.
The two different expressions of the limit reported in Eq. (\ref{RGC5f}) coincide by virtue of Eq. (\ref{RGC5g}).
Using the standard slow-roll relations\footnote{The results of 
Eqs. (\ref{RGC9})--(\ref{RGC10}) as well as the results obtainable for the scalar modes (see the following subsection) follow by recalling the stqandard slow-roll results  $3 \overline{M}_{\mathrm{P}}^2 H^2 \simeq V$ and $3 H \dot{\varphi} \simeq - V_{\,\,,\varphi}$, where $V_{\,\,,\varphi}$ denotes a derivation of the scalar potential $V$ with respect to the inflaton field $\varphi$. It is finally useful to mention that, to lowest order in $\epsilon$, $\Gamma^2(\nu) \simeq \pi/4$.} and rearranging the obtained expression, the general results of Eqs. (\ref{RGC5f})--(\ref{RGC5g}) can be directly phrased in terms 
of the inflationary potential and of its derivatives for $k \simeq H_{r} a_{r}$:
\begin{equation}
\overline{P}_{T}(k,\tau_{r})= \frac{2^{2\nu}}{\pi^3} \Gamma^2(\nu) ( 1 - \epsilon)^{2\nu-1}
\biggl(\frac{H_{r}}{\overline{M}_{\mathrm{P}}}\biggr)^2
\simeq \frac{2}{3 \pi^2} \biggl(\frac{V}{\overline{M}_{\mathrm{P}}^4}\biggr)_{k\simeq a_{r} H_{r}}   = \frac{128}{3 } \biggl(\frac{V}{M_{\mathrm{P}}^4}\biggr).
\label{RGC9}
\end{equation}
Since all the slow-roll parameters are much smaller than one during inflation (and become of order ${\mathcal O}(1)$ as inflation ends) the Bessel indices  can be expanded in powers of the slow-roll parameters so, for instance 
$\nu \simeq 3/2 + \epsilon + {\mathcal O}(\epsilon^2)$. The tensor spectral index $n_{T}= 3 - 2 \nu$ becomes 
$n_{T} = - 2 \epsilon + {\mathcal O}(\epsilon^2)$, or, more precisely:
\begin{equation}
n_{T} = - \biggl(\frac{V_{,\varphi}}{V}\biggr)^2 \frac{\overline{M}_{\mathrm{P}}^2}{1 -\epsilon} \simeq - \frac{2\epsilon}{1 - \epsilon} \simeq - 2\epsilon + {\mathcal O}(\epsilon^2).
\label{RGC10}
\end{equation}
When the rate of variation of the background exceeds the frequencies of the gravitons the spectral  energy density in critical units given in Eq. (\ref{RGC5e}) can be evaluated in the limit $|k \tau | \ll1$; after using the small argument limit of Hankel functions \cite{abr1,abr2} inside Eq. (\ref{RGC5e}) we obtain:
\begin{equation}
\lim_{|k \tau| \ll 1} \Omega_{gw}(k,\tau) \to \frac{k^2}{ 24 H^2 a^2} \overline{P}_{T}(k,\tau_{r}) = \frac{k^2 \tau^2}{24} ( 1- \epsilon)^2 \, \overline{P}_{T}(k,\tau_{r}),
\label{RGC11}
\end{equation}
where $\overline{P}_{T}(k,\tau_{r})$ denotes the constant value of the 
tensor power spectrum deduced in Eq. (\ref{RGC5f}) and valid for typical wavelengths larger than the Hubble radius during the inflationary phase. Equation (\ref{RGC11}) also demonstrates, as expected, that whenever  $k \tau \ll1$ the spectral energy density of the relic gravitons is suppressed for typical wavelengths larger than the Hubble radius and for $a < a_{r}$. Since in the limit  $\epsilon \to 0$ (or $\nu \to 3/2$) the scale factor takes the exact de Sitter form [i.e. $a(\tau) = (- \tau/\tau_{r})^{-1}$], the tensor power spectrum of Eqs. (\ref{RGC5c}) and (\ref{RGC5f})--(\ref{RGC5g}) becomes
\begin{equation}
\lim_{\epsilon \to 0} \overline{P}_{T}(k,\tau_{r}) \to \frac{2}{\pi^2} \biggl(\frac{H_{r}}{\overline{M}_{P}}\biggr)^2 = \frac{16}{\pi} \biggl(\frac{H_{r}}{M_{P}}\biggr)^2.
\label{RGC7a}
\end{equation}
The result of Eq. (\ref{RGC7a}) has no analog in the scalar case:
while the tensor modes of the geometry are amplified during an exact stage of de 
Sitter expansion, the opposite is true in the case of the scalar modes of the geometry. 

\subsubsection{Consistency conditions}
The tensor power spectra are customarily related to the 
scalar power spectra by assuming that the (adiabatic and Gaussian) curvature perturbations arise 
in the context of single-field inflationary models.  To deduce the consistency 
conditions\footnote{This parametrization is rather common even if 
it is not the most general one and could even be misleading since it assumes, to 
some extent, a specific class of scenarios.} we start from the effective action of the curvature perturbations on comoving orthogonal hypersurfaces (see Eqs. (\ref{NM2}) and (\ref{NM3})):
\begin{equation}
S_{s} = \frac{1}{2} \int d^{4} \, \sqrt{-\overline{g}} \biggl(\frac{\varphi^{\prime}}{{\mathcal H}}\biggr)^2 \, 
\overline{g}^{\alpha\beta} \, \partial_{\alpha} {\mathcal R} \,\partial_{\beta} {\mathcal R},
\label{RGC11a}
\end{equation}
where, as usual,  $\overline{g}_{\alpha\beta}$ is the background metric.
By extremizing the action (\ref{RGC11a}) with respect to the variation of ${\mathcal R}$ it is easy to derive the 
evolution equation of the curvature perturbations mentioned in Eq. (\ref{NM3}) so that 
the mode expansion for the corresponding field operator will be given by 
\begin{equation}
\hat{{\mathcal R}}(\vec{x},\tau) = \frac{1}{(2\pi)^{3/2}} \int \, d^{3} k  \biggl[  \hat{a}_{\vec{k} } \, \widetilde{\,F\,}_{k}(\tau)\,e^{- i \vec{k} \cdot \vec{x}} + \hat{a}_{\vec{k} }^{\dagger}\,\widetilde{\,F\,}^{*}_{k}(\tau)   \, e^{ i \vec{k} \cdot \vec{x}} \biggr].
\label{RGC11b}
\end{equation}
In analogy with the tensor case the two-point function and the corresponding power spectrum 
can be computed from Eq. (\ref{RGC11}) by evaluating the expectation value of the field operators over the vacuum state 
of the scalar modes (i.e. $|\widetilde{\mathrm{vac}} \rangle$) which is defined 
in analogy with the vacuum of the tensors introduced in Eqs. (\ref{fourEa})--(\ref{fourEb}) (see also \cite{book1,book2,book3})
\begin{equation}
\langle \widetilde{\mathrm{vac} }| \hat{\mathcal R}(\vec{x},\tau) \hat{\mathcal R}(\vec{x} + \vec{r},\tau) |\widetilde{ \mathrm{vac}} \rangle= 
\int P_{{\mathcal R}}(k,\tau) \frac{\sin{k r}}{kr} d\ln{k},\qquad  P_{{\mathcal R}}(k,\tau) = \frac{k^3}{2\pi^2} |\widetilde{\,F\,}_{k}(\tau)|^2,
\label{RGC12}
\end{equation}
where  $\widetilde{\,F\,}_{k}(\tau)$ is the scalar mode function derived from Eq. (\ref{NM3}) and obeying
\begin{equation}
\widetilde{\,F\,}_{k}^{\prime\prime} + 2 \,\frac{z_{\varphi}^{\prime}}{z_{\varphi}} \,\widetilde{\,F\,}_{k}^{\prime} + k^2\, \widetilde{\,F\,}_{k} =0, \qquad 
\widetilde{\,G\,}_{k}= \widetilde{\,F\,}_{k}^{\prime},
\label{RGC13}
\end{equation}
where $z_{\varphi} = a \varphi^{\prime}/{\mathcal H}$ has been already introduced in Eq. (\ref{NM3}).
After normalizing the scalar 
mode functions it is easy to deduce the explicit form of the scalar analog power spectrum for 
 typical wavenumbers smaller than the rate of variation of the geometry (i.e. $k \ll \bigl| a\, H\bigr|$) 
 during the inflationary  stage of expansion (i.e. $a< a_{r}$)
\begin{equation}
\overline{P}_{{\mathcal R}}(k,\tau) = \frac{2^{2\widetilde{\,\nu\,}-3}}{\pi^3} \Gamma^2(\widetilde{\,\nu\,}) (1 -\epsilon)^{1 -2\widetilde{\,\nu\,}} 
\biggl(\frac{k}{aH}\biggr)^{3-2\widetilde{\,\nu,}} \biggl(\frac{H_{r}^2}{\dot{\varphi}}\biggr)^2 \simeq \frac{1}{4\pi^2} \biggl(\frac{H_{r}^2}{\dot{\varphi}}\biggr)^2_{k\simeq a_{r}\,H_{r}} = \frac{8}{3\, M_{\mathrm{P}}^4} \biggl( \frac{V}{\epsilon}\biggr)_{k\simeq a_{r} H_{r}},
\label{RGC14}
\end{equation}
where we recalled that, to lowest order in the slow-roll expansion, $3 H^2 \, \overline{M}_{P}^2 \simeq V$ 
and $\dot{\varphi}^2 = - 2 \dot{H} \, \overline{M}_{P}^2$; using the definition of $\epsilon = - \dot{H}/H^2$ 
and the relation between $\overline{M}_{P}$ and $M_{P}$ (see Tab. \ref{SEC1TABLE4}) the final result 
of Eq. (\ref{RGC14}) easily follows.  Equation (\ref{RGC14}) is the scalar analog of Eq. (\ref{RGC9}); 
in the scalar case the index of the Hankel function is given by $\widetilde{\,\nu\,}$ and it generally differs from 
the one of the tensor mode function defined in Eq. (\ref{RGC5a}). In the limit $\epsilon \ll 1$ and $\eta\ll 1$ we have
\begin{equation}
\widetilde{\,\nu\,} = \frac{3 +\epsilon + 2 \eta}{2(1-\epsilon)} \simeq \frac{3}{2} + 2\epsilon +\eta 
+ {\mathcal O}(\epsilon^2) + {\mathcal O}(\eta^2) + {\mathcal O}(\eta\,\epsilon).
\label{RGC16}
\end{equation}
The difference between $\nu$ and $\widetilde{\,\nu\,}$ arises as a first-order correction that depends upon (both) $\epsilon$ and $\eta$. As a consequence of Eq. (\ref{RGC16}) the scalar spectral index can be expressed in terms of the slow-roll parameters and it is given by:
\begin{equation}
n_{s} -1 = \biggl(\frac{\dot{\varphi}}{H_{r}}\biggr) \frac{1}{1-\epsilon} \biggl[ \frac{V_{,\varphi}}{V} - \frac{\epsilon_{,\varphi}}{\epsilon}\biggr] \equiv 3 - 2 \widetilde{\,\nu\,}= 1 - 6 \epsilon + 2 \overline{\eta},
\label{RGC17}
\end{equation}
where we included the explicit relation between the scalar spectral index, the Hankel index and 
the slow-roll parameters by recalling that
\begin{equation}
\frac{\epsilon_{,\varphi}}{\epsilon} = 2 \frac{V_{,\varphi\varphi}}{V_{,\varphi}} - 2 \biggl(\frac{V_{,\varphi}}{V}\biggr),\qquad 
\frac{\dot{\varphi}}{H_{r}} = - \frac{V_{,\varphi}}{3 H_{r}^2}.
\label{RGC18}
\end{equation}
The scale-invariant limit (i.e. $n_{s} \to 1$)  cannot be 
dynamically realized when the inflationary phase is driven by 
 a single scalar field while in the tensor case the scale-invariant limit  of Eq. (\ref{RGC7a})
 is well defined and it corresponds to the case of an exact de Sitter expansion supporting the tensor modes of the geometry but not the scalar ones (see also \cite{HIS10}).  All in all, recalling the notation established in Eq. (\ref{RGC9}) the tensor and scalar power spectra can be expressed as:
\begin{equation}
\overline{P}_{T}(k,\tau_{r}) = \frac{128}{3 } \biggl(\frac{V}{M_{\mathrm{P}}^4}\biggr),\qquad 
\overline{P}_{\mathcal R}(k,\tau_{r}) = \frac{8}{3}\biggl(\frac{V}{\epsilon\,\, M_{\mathrm{P}}^4}\biggr).
\label{RGC19}
\end{equation}
Bearing in mind the explicit form of the tensor spectral index derived in Eq. (\ref{RGC10}), the ratio between the tensor and the scalar spectrum is proportional both to the slow-roll parameter $\epsilon$ and to the tensor 
spectral index:
\begin{equation}
r_{T}= \frac{\overline{P}_{T}(k,\tau_{r})}{\overline{P}_{{\mathcal R}}(k,\tau_{r})} = 16 \,\epsilon = - 8 \,n_{T},
\label{RGC21}
\end{equation}
which is often referred to as consistency condition \cite{book1,book2,book3}. Equation (\ref{RGC21}) holds
in the case of a slow-roll dynamics: it would be incorrect to take the limit 
$\epsilon\to 0$ in Eq. (\ref{RGC21}) since, by doing so, we would conclude 
that $\overline{P}_{T}(k,\tau_{r}) \to 0$; on the contrary, as argued in Eq. (\ref{RGC9})
in the limit $\epsilon \to 0$ the tensor power spectrum is constant and finite while 
$\overline{P}_{{\mathcal R}}(k,\tau_{r}) \to 0$. Instead of expressing the scalar and tensor power spectra 
in terms of the parameters of a given inflationary scenario 
it is sometimes more useful to use the following parametrization:
\begin{equation}
\overline{P}_{{\mathcal R}}(k,\tau_{r}) = {\mathcal A}_{{\mathcal R}}   \biggl(\frac{k}{k_{\mathrm{p}}}\biggr)^{n_{s}-1}, \qquad \overline{P}_{T}(k,\tau_{r}) = {\mathcal A}_{T} \biggl(\frac{k}{k_{\mathrm{p}}}\biggr)^{n_{T}},
\label{RGC22}
\end{equation}
where $n_{T} = - 2\epsilon$ and $n_{s} = 1 - 6\epsilon + 2 \overline{\eta}$ and  
\begin{equation} 
{\mathcal A}_{{\mathcal R}} = {\mathcal O}(2.4)\times 10^{-9},
 \qquad  {\mathcal A}_{T} = r_{T} \,{\mathcal A}_{{\mathcal R}}, \qquad k_{\mathrm{p}} = 0.002 \,\, \mathrm{Mpc}^{-1}.
\label{RGC23}
\end{equation}
We stress that {\em the parametrization of Eqs. (\ref{RGC21}), (\ref{RGC22}) and (\ref{RGC23}) is valid within the concordance paradigm and when the corresponding wavelengths are larger than the Hubble radius prior to matter-radiation equality}.

\subsubsection{The maximal comoving frequency and the number of $e$-folds}
The maximal comoving frequency of the spectrum, denoted hereunder by $\nu_{max}$, corresponds to {\em the shortest amplified wavelength} that roughly coincides with the absolute minimum of the Hubble radius illustrated in 
Fig. \ref{SEC4FIG1}. It is relevant to appreciate that the maximal frequency of the graviton spectrum ultimately depends on the {\em whole post-inflationary thermal history};  in the conventional case summarized by Fig. \ref{SEC4FIG1} its final expression is given by:
\begin{eqnarray}
\nu_{max}  &=& \frac{1}{4\pi} \biggl(2 \pi \, r_{T} \, {\mathcal A}_{{\mathcal R}} \, \Omega_{R0} \biggr)^{1/4}\, \sqrt{H_{0} \, M_{P}}
\nonumber\\
&=& 0.27 \,\biggl(\frac{r_{T}}{0.064}\biggr)^{1/4} 
\biggl(\frac{{\mathcal A}_{\mathcal R}}{2.41 \times 10^{-9}}\biggr)^{1/4}
\biggl(\frac{h_{0}^2 \Omega_{R0}}{4.15 \times 10^{-5}}\biggr)^{1/4} \,\mathrm{GHz}.
\label{RGC24}
\end{eqnarray}
 Equation (\ref{RGC24}) coincides with Eq. (\ref{ONEb}) (already mentioned in the preliminary 
discussion of section \ref{sec1}) by using the consistency 
condition of Eq. (\ref{RGC21}), i.e. $r_{T} =16\,\epsilon$. The dependence of $\nu_{max}$ upon 
$r_{T}$ follows from the explicit form of the Hubble rate at the end of inflation 
\begin{equation}
\frac{H}{M_{P}} = \frac{\sqrt{\pi\, {\mathcal A}_{{\mathcal R}} \, r_{T}}}{4}=
2.17\times 10^{-6} \,\sqrt{\frac{r_{T}}{0.01}}\,\, \sqrt{\frac{{\mathcal A}_{{\mathcal R}}}{2.4\times 10^{-9}}}.
\label{RGC25}
\end{equation}
 Depending on the post-inflationary thermal history it can happen that the maximal frequency $\nu_{max}$ is either smaller or slightly larger than the value of Eq. (\ref{RGC24}).  If energy is lost during inflation and in the reheating phase the maximal curvature scale at the end of inflation could be drastically different from the curvature scale during reheating. However, for slow roll inflation it is not a bad approximation to neglect this energy loss as customarily done when estimating $N_{max}$ the maximal number of inflationary $e$-folds 
accessible to large-scale CMB measurements; 
by demanding that the inflationary event horizon, redshifted at the present epoch, coincides 
with the present size of the Hubble radius \cite{HOR} we obtain the explicit value of $N_{max}$ i.e. 
\begin{equation}
e^{N_{max}} = \frac{[2 \pi\, \Omega_{R 0} \,{\mathcal A}_{{\mathcal R}}\,r_{T}]^{1/4}}{4}\, \biggl(\frac{M_{P}}{H_{0}}\biggr)^{1/2}\, \biggl(\frac{H}{H_{r}} \biggr)^{1/2- \gamma},
\label{RGC27}
\end{equation}
which implies that 
 \begin{equation}
 N_{max} = 61.20 + \frac{1}{4} \ln{\biggl(\frac{h_{0}^2 \Omega_{R 0}}{4.15 \times 10^{-5}} \biggr)} - \ln{\biggl(\frac{h_{0}}{0.7}\biggr)}
 + \frac{1}{4} \ln{\biggl(\frac{{\mathcal A}_{{\mathcal R}}}{2.4 \times 10^{-9}}\biggr)} + \frac{1}{4} \ln{\biggl(\frac{r_{T}}{0.064}\biggr)} + \biggl(\frac{1}{2} - \gamma\biggr) 
 \ln{\biggl(\frac{H}{H_{r}}\biggr)}.
 \label{RGC28}
\end{equation} 
In Eqs. (\ref{RGC27}) and (\ref{RGC28})  $\gamma$ accounts for the possibility of a delayed reheating terminating at a putative scale $H_{r}$ smaller (or even much smaller) than the Hubble rate during inflation.  Since the nucleosynthesis scale must always exceed the reheating scale $H_{r}$, we have that 
 $H_{r}> 10^{-44} \overline{M}_{\mathrm{P}}$. Whenever $\gamma > 1/2 $ $N_{max}$ diminishes in comparison with the sudden reheating (i.e. $H=H_{r}$) and it may be reduced to ${\mathcal O}(47)$ when $\gamma = 2/3$ (as in the case 
 of a delayed reheating dominated by dust). Conversely if $\gamma <1/2 $ $N_{max}$ increases
 and it may even become ${\mathcal O}(76)$ (as in the situation when the post-inflationary background is dominated by stiff sources and $\gamma \simeq  1/3$ ). Finally, if $H_{r} = H$ (i.e. $\gamma=1/2$) there is a sudden transition 
between the inflationary regime and the post-inflationary epoch dominated by radiation. Thus, in spite of the 
empirical indeterminations involving ${\mathcal A}_{{\mathcal R}}$ and $r_{T}$, 
$N_{max}$ has a definite theoretical error depending on our ignorance 
of the post-inflationary thermal history; given the maximal excursion of $\gamma$ (say between $2/3$ and $1$) we have that  $N_{max} = 61.49 \pm  14.96$ which agrees with the results of Ref. \cite{HOR}. It is wise to bear in mind 
these figures when discussing possible predictions associated with the primordial backgrounds of relic gravitons.

\subsection{Sakharov oscillations and standing waves}
The relic gravitons are produced in strongly correlated pairs of opposite comoving three-momenta (see Eqs. (\ref{PAM9a})--(\ref{PAM9b}) and discussion therein). From the semi-classical viewpoint this process can be viewed as the conversion of traveling waves into standing waves;  the same phenomenon occurs for the scalar modes of the geometry and it leads to the so-called Sakharov oscillations \cite{SAK1} (see also \cite{SAK2,SAK3}). The general occurrence of the standing waves is  not limited to the context of the concordance paradigm and has been independently suggested by Peebles and Yu \cite{SAK2} (see also \cite{SAK3,SAK4,SAK5}) in their pioneering contribution on the adiabatic initial conditions of the Einstein-Boltzmann hierarchy prior to matter-radiation equality. While the Sakharov oscillations in the pre-decoupling plasma determine the structure of the acoustic oscillations of the temperature anisotropies of the CMB \cite{pav1,pav2,SAK6,SAK7,SAK8}, the attention 
will be concentrated here on the tensor power spectra.

\subsubsection{Standing waves in a radiation-dominated plasma}
A smooth evolution of the extrinsic curvature of the background  
demands the continuity of the inflationary scale factor and of its first derivative across 
the transition to the radiation-dominated phase. The explicit form of of the scale factor
during inflation is  $a_{inf}(\tau) = (- \tau_{r}/\tau)^{\beta}$  (for  $\tau < - \tau_{r}$)
where, as usual, $\beta = 1/(1-\epsilon)$ (see also Eq. (\ref{RGC5g}) and discussion therein). 
Conversely, during the radiation-dominated phase the scale factor is\footnote{It can be explicitly verified from Eq. (\ref{TRAD3}) that $a_{inf}(-\tau_{r}) = a_{r}(-\tau_{r})$ and that $ a_{inf}^{\prime}(-\tau_{r}) = a_{r}^{\prime}(-\tau_{r})$.}:
\begin{equation}
a_{r}(\tau) = \frac{\beta \tau + (\beta+1) \tau_{r}}{\tau_{r}}, \qquad -\tau_{r} < \tau \leq \tau_{eq},
\label{TRAD3}
\end{equation}
where $\tau_{eq}$ denotes the 
time-scale of matter-radiation equality. If the mode functions  $f_{k}(\tau)$ and $g_{k}(\tau)$ are continuous across 
$-\tau_{r}$ the continuity of the scale factor and of its first derivative also implies the continuity of the mode expansions 
of Eqs. (\ref{PAM13}) and (\ref{PAM14}) (and, in particular, of $F_{k}(\tau)$ and of $G_{k}(\tau)$).
Thanks to Eq. (\ref{TRAD3}) the expression of $f_{k}(\tau)$ for $\tau > - \tau_{r}$ is a combination of plane waves:
\begin{equation}
f_{k}(\tau,\tau_{r}) = \frac{1}{\sqrt{2 k}} \biggl[ c_{+}(k,\,k_{r}) \, \,e^{- i \,  x(\tau)} + c_{-}(k,\,k_{r})\,\, e^{i  \,x(\tau)} \biggr], \qquad x(\tau) = k\,\biggl(\tau 
+ \frac{\beta +1}{\beta} \tau_{r} \biggr),
\label{TRAD4}
\end{equation}
where $c_{\pm}(k,\,k_{r})$ depend both on $k$ and $k_{r}$ and are fixed by imposing the continuity of the mode functions and of their first derivatives during inflation (i.e. Eqs. (\ref{RGC5a})--(\ref{RGC5b})) and in the subsequent radiation epoch (i.e. Eq. (\ref{TRAD4})); the result of this strategy is:
\begin{equation}
c_{\pm}(k,\, k_{r}) = \frac{{\mathcal N}}{2} \sqrt{ k\tau_{r}} e^{ \pm\, i\, \frac{k\tau_{r}}{\beta}} \, H_{\nu}^{(1)}(k\,\tau_{r}) \biggl\{ 1 \mp 
i \, \biggl[ \biggl( \nu + \frac{1}{2} \biggr) \frac{1}{k\tau_{r}} - \frac{H_{\nu+1}^{(1)}(k\tau_{r})}{H_{\nu}^{(1)}(k\tau_{r})} \biggr]\biggl\}.
\label{TRAD5}
\end{equation}
where $c_{\pm}(k, \, k_{r})$ obey $|c_{+}(k,\,k_{r})|^2 - |c_{-}(k,\, k_{r})|^2 =1$, as it can be explicitly 
verified from Eq. (\ref{TRAD5}). In the limit $\epsilon\to 0$, $\beta \to 1$ and $\nu \to 3/2$ (see also Eq. (\ref{RGC7a}))  the coefficients $c_{\pm}(k,\, k_{r})$ become: 
\begin{equation}
 c_{+}(k, \,k_{r}) = e^{2 i k \tau_{r}} \biggl[ 1 - \frac{i}{\bigl| k \tau_{r}\bigr|} - \frac{1}{2 \bigr| k \tau_{r} \bigl|^2}\biggr], \qquad c_{-}(k,\,k_{r}) = \frac{1}{2 \bigl| k \tau_{r}\bigr|^2}.
\label{TRAD6}
\end{equation}
The largest amplified wavenumber of the spectrum is  $k_{r} = a_{r} H_{r} = {\mathcal O}(\tau_{r}^{-1})$ and it corresponds to the minimum of the Hubble radius illustrated in Fig. \ref{SEC4FIG1} with a shaded rectangle. The smallest comoving wavelengths hitting the Hubble radius 
are of the order of $\lambda_{r} = 2 \pi (a_{r} \, H_{r})^{-1}$.  Since $\lambda_{r}$ is comparable
with the absolute minimum of the comoving Hubble radius,  all the wavelengths of the spectrum 
will be larger than $\lambda_{r}$ and it will also be generally true that $ k \tau_{r} = x_{r} < 1$. This means that
$c_{\pm}(k,k_{r})$ must be evaluated, for all practical purposes, in the limit $k< k_{r}$ which is equivalent to $k \tau_{r} < 1$.
In this physical limit the mixing coefficients of Eq. (\ref{TRAD5}) are given by:
\begin{equation}
c_{\pm}(k, \,k_{r}) = - i {\mathcal N}\,\frac{ 2^{\nu-1}}{\pi} \Gamma(\nu) \,\, e^{\pm \,i\,\frac{k \tau_{r}}{\beta}} \, \,k \tau_{r}^{-\nu} \,\, \biggl[ \sqrt{k \tau_{r}} \pm i \, \biggl( \frac{1}{2} - \nu\biggr) \frac{1}{\sqrt{k \tau_{r}}} \biggr].
\label{TRAD7}
\end{equation}
Consequently the semi-classical conditions for the production of the relic gravitons demand that 
 the rate of variation of the pump field must exceed the three-momentum of the most energetic 
 graviton of the spectrum. The condition for the amplification corresponds therefore to $k < k_{r}$ and, in this frequency domain, 
Eq. (\ref{TRAD7}) implies exactly that $c_{+}(k,\, k_{r})$ and $c_{-}(k,\, k_{r})$ {\em have the same modulus but opposite 
signs}:
\begin{equation} 
c_{\pm}(k,\, k_{r}) = \pm\,  {\mathcal N}\,\frac{ 2^{\nu-1}}{\pi} \,\Gamma(\nu)\,\biggl( \frac{1}{2} - \nu\biggr) \bigl| k \tau_{r} \bigr|^{- \nu - 1/2} \biggl[ 1 
+ {\mathcal O}( \bigl| k \tau_{r}\bigr|) \biggr].
\label{SOSC1}
\end{equation}
If  Eq. (\ref{SOSC1}) is inserted back into Eq. (\ref{TRAD4}) the obtained expression 
will describe standing waves and not traveling waves as it happens instead 
during inflation when the mode functions have been initially normalized. 

\subsubsection{Standing waves in a dust-dominated plasma}
If the scale factor is continuous with its first derivative across the radiation-matter 
transition occurring for $\tau = \tau_{eq}$, the explicit form of $a_{eq}(\tau)$ is\footnote{From Eqs. (\ref{TRAD3}) and (\ref{SAKm1}) 
it can be readily verified that 
$a_{r}(\tau_{eq}) = a_{eq}(\tau_{eq})$ and $a_{r}^{\prime}(\tau_{eq}) = a_{eq}^{\prime}(\tau_{eq})$.}:
\begin{equation}
a_{eq}(\tau) = \frac{[\beta (\tau + \tau_{eq}) + 2 (\beta+1) \tau_{r}]^2}{4 \, \tau_{r}\, [ \beta \tau_{eq} + (\beta+1) \tau_{r}]},\qquad y(\tau) = k \biggl[ \tau + \tau_{eq} + \frac{2 (\beta +1)}{\beta} \tau_{r} \biggr], \qquad \tau > \tau_{eq},
\label{SAKm1}
\end{equation}
where the dimensionless variable $y(\tau)$ plays the same role of  $x(\tau)$ (see Eq. (\ref{TRAD4})) and it governs the phases of the Sakharov oscillations in the matter-dominated plasma. Thanks to Eq. (\ref{SAKm1})  the mode function can be expressed directly as a function of $y(\tau)$
\begin{equation}
f_{k}^{(m)}(\tau) = \frac{1}{\sqrt{2 k}} \biggl[ d_{-}(k,\,k_{r},\, k_{eq}) \, \biggl( 1 - \frac{i}{y(\tau)} \biggr) \, e^{-\,i \,y(\tau)} +  d_{+}(k,\,k_{r},\, k_{eq}) \, \biggl( 1 + \frac{i}{y(\tau)} \biggr) \, e^{i \,y(\tau)} \biggr].
\label{SAKm3}
\end{equation}
The continuity of the mode function and of its first derivative leads to the explicit expression 
of $d_{\pm}(k,\,k_{r},\, k_{eq})$ whose phase dependence  is essential for the determination 
of the correct structure of the Sakharov oscillations:
\begin{eqnarray}
d_{-}(k,\,k_{r},\, k_{eq})  &=& \frac{ 1}{8 k^2\,[  \tau_{r} + \beta ( \tau_{r} +  \tau_{eq})]^2} 
\biggl\{ - e^{ - 3 \, i\, \alpha} \,\beta^2 \, c_{+}(k,\, k_{r}) 
\nonumber\\
&+& e^{ - i\, \alpha} \biggl[ 8 k^2 \biggl( \tau_{r} + \beta ( \tau_{r} +  \tau_{eq})\biggr)^2 - \beta^2 - 4 \, i\, \beta \, k\,\biggl(  \tau_{r} + \beta(  \tau_{r} +  \tau_{eq})\biggr)\biggr] c_{-}(k,\, k_{r})\biggr\},
\label{SAKm3a}\\
d_{+}(k,\,k_{r},\, k_{eq})  &=& \frac{ 1}{8 k^2 [ \tau_{r} + \beta ( \tau_{r} + \tau_{eq})]^2} 
\biggl\{ - e^{ 3 \, i\, \alpha} \,\beta^2 \, c_{-}(k,\, k_{r})
\nonumber\\
&+& e^{  i\, \alpha} \biggl[ 8 k^2 \biggl( \tau_{r} + \beta ( \tau_{r} +  \tau_{eq})\biggr)^2 - \beta^2 + 4 \, i\, \beta \, k \biggl(  \tau_{r} + 
\beta(  \tau_{r} +  \tau_{eq})\biggr)\biggr] c_{+}(k,\, k_{r})\biggr\},
\label{SAKm3b}
\end{eqnarray}
where $\alpha = [k \tau_{eq} + (\beta+1) k \tau_{r}/\beta]$. 
Equations (\ref{SAKm3a}) and (\ref{SAKm3b}) can now be expanded in two concurrent physical limits, i.e. 
$\bigl| k \tau_{r} \bigr|\ll 1$ and $\bigl| k \tau_{eq} \bigr| \ll 1$ by also demanding that $\tau_{eq} \gg \tau_{r}$.
To take the limit in a consistent manner  Eqs. (\ref{SAKm3a})--(\ref{SAKm3b})  are first expanded 
for $k \tau_{eq} \ll 1$; the obtained results are then expanded  for $k \tau_{r} \ll 1$; the result of this double expansion is 
\begin{equation}
d_{\pm}(k,\,k_{r},\, k_{eq}) = \frac{3 \, i}{ 8 \, k \, \tau_{eq}} \biggl[ c_{+}(k, \, k_{r}) - c_{-}(k,\, k_{r}) \biggr] + {\mathcal O}( \bigl| k \tau_{eq} \bigr|).
\label{SAKm3c}
\end{equation}
The condition (\ref{SAKm3c}), once  inserted back into Eq. (\ref{SAKm3}), leads to the 
correct structure of the Sakharov oscillations in the matter-dominated 
stage of expansion and also determines the tensor power spectra in the low-frequency 
range, as we shall see hereunder.  There is actually a sign difference between Eqs. 
(\ref{SOSC1}) and (\ref{SAKm3c}): while $c_{+}(k,k_{r})$ and $c_{-}(k,k_{r})$ {\em have the same 
modulus but opposite signs}, $d_{\pm}(k,\,k_{r},\, k_{eq})$ {\em have the same modulus and the same signs}. 
The standing waves will thus be sinusoidal in the radiation epoch and (predominantly) cosinusoidal 
after equality.

\subsection{Wavelengths reentering when the radiation dominates}
The comoving wavelengths  $ \lambda_{eq} < \lambda < \lambda_{r}$ left the Hubble radius 
 during inflation and reentered when the universe was dominated by radiation.
The short wavelengths are labeled by $1$ in Fig. \ref{SEC4FIG1} and correspond to 
the frequencies ${\mathcal O}(100)$ MHz already discussed in Eq. (\ref{RGC24}); they exited 
the Hubble radius at the end of inflation and reentered shortly after the onset 
of the radiation epoch. The intermediate wavelengths (determining the graviton spectrum from $100$ aHz up to the  MHz region) fall instead between the dashed lines labeled by $1$ and $2$ in
 Fig. \ref{SEC4FIG1}; they reentered before the beginning of the matter-dominated epoch
 and were still larger than the Hubble radius prior to equality. The intermediate 
 wavelengths affect the initial conditions of the Einstein-Boltzmann hierarchy 
 and indirectly determine the temperature and the polarization anisotropies of the CMB that 
 will be specifically discussed in section \ref{sec5}.

\subsubsection{The power spectra in a plasma dominated by radiation}
Equations (\ref{TRAD4}),  (\ref{TRAD7}) and (\ref{SOSC1}) imply that 
the mode functions of Eqs. (\ref{PAM12}) can be expressed as: 
 \begin{eqnarray}
 F_{k}^{(r)}(\tau) &=& \overline{F}_{k}^{(r)} \, \frac{\sin{x(\tau)}}{x(\tau)}, \qquad \overline{F}_{k}^{(r)} = \frac{i}{\sqrt{\pi\, k}} \, 2^{\nu -1}\, \Gamma(\nu) e^{ i \pi ( 2 \nu +1)/4} \, \bigl|k \tau_{r}\,\bigr|^{1/2 - \nu},
 \label{TRAD7a}\\
G_{k}^{(r)}(\tau) &=& k \,\overline{F}_{k}^{(r)} \biggl[ \frac{\cos{x(\tau)}}{x(\tau)} - 
 \frac{\sin{x(\tau)}}{x^2(\tau)} \biggr], \qquad k_{r} < k < k_{eq}.
\label{TRAD7b}
\end{eqnarray}
The superscripts in Eqs. (\ref{TRAD7a})--(\ref{TRAD7b}) remind that the mode functions apply in the radiation epoch and 
 $\overline{F}^{(r)}_{k}$ denotes the constant amplitude of the mode function\footnote{ By comparing 
 $\bigl| \overline{F}_{k}^{(r)}\bigr|^{2}$ with Eq. (\ref{RGC5f}) we have that $\bigl| \overline{F}_{k}^{(r)}\bigr|^{2} = \pi \, \overline{M}_{P}\sqrt{\overline{P}_{T}(k, \tau_{r})}/( 2 \, k^{3/2})$. }. Using the shorthand notation $x(\tau) = x$ the tensor power spectra during the radiation epoch follow by inserting the results of Eqs. (\ref{TRAD7a}) and (\ref{TRAD7b}) into Eq. (\ref{fiveE}):
\begin{eqnarray}
 P_{T}^{(r)}(k,\tau, \tau_{r}) &=& \frac{4 \ell_{P}^2}{\pi^2} \, k^3 \, \bigl| \overline{F}_{k}^{(r)} \bigr|^2 \, \frac{\sin^2{x}}{x^2}  = \overline{P}_{T}(k, \tau_{r})  \frac{\sin^2{x}}{x^2},  
 \label{TRAD7c}\\
 Q_{T}^{(r)}(k,\tau, \tau_{r}) &=& \frac{4 \ell_{P}^2}{\pi^2} \, k^5 \, \bigl| \overline{F}_{k}^{(r)} \bigr|^2 \, \biggl[\frac{\cos{x}}{x} - \frac{\sin{x}}{x^2}\biggr]^2.
 \label{TRAD7d}
\end{eqnarray}  
Equations (\ref{TRAD7c}) and (\ref{TRAD7d}) must then be inserted into  Eqs. (\ref{OBS5}) and (\ref{OBS6}) so that 
the explicit form of the spectral energy density in critical units becomes:
\begin{equation}
\Omega_{gw}(k,\, \tau,\, \tau_{r}) = \frac{ k^2 \overline{P}_{T}}{24\, H^2 \, a^2 \, x^2} \biggl[ 1 + \frac{\sin{x}}{x^2} - \frac{\sin{2 \, x}}{x} \biggr].
 \label{TRAD7e} 
 \end{equation}
The wavelengths that reentered the Hubble radius after the onset of radiation are characterized 
by an oscillating power spectrum that is suppressed as $| k \tau|^{-2}$ in the limit $|k \tau | \gg 1$; 
however, for the wavelengths that are larger than the Hubble radius prior to matter-radiation equality 
the power spectrum is constant and it coincides exactly with the inflationary result. In these two opposite 
limits the power spectrum is given by:
\begin{eqnarray}
\lim_{k \tau_{r} \ll 1, \,\, x \gg 1} P_{T}^{(r)}(k,\tau, \tau_{r}) &\to& \frac{\overline{P}_{T}(k, \tau_{r})}{ 2 x^2}  =  \frac{\overline{P}_{T}(k, \tau_{r})}{ 2} \biggl[\frac{a_{r}(\tau_{r})}{a_{r}(\tau)}\biggr]^2_{k \tau_{r} \simeq 1},\qquad k \tau \gg 1,
\nonumber\\
\lim_{k \tau_{r} \ll 1, \,\, x \ll 1} P_{T}^{(r)}(k,\tau, \tau_{r}) &\to& \overline{P}_{T}(k, \tau_{r}), \qquad k\tau \ll 1.
\label{TRAD8a}
\end{eqnarray}
Note that in Eq. (\ref{TRAD8a}) the Sakharov oscillations have been smoothed by replacing $\sin^2{x}\to 1/2$. This strategy is grossly correct and it is customarily employed since the leading-order result for 
$P^{(r)}_{T}(k, \tau, \tau_{r})$ and $Q_{T}^{(r)}(k,\tau, \tau_{r})$ implies a strongly oscillating functions in the limit $k\tau \gg 1$.  In the case of $Q^{(r)}_{T}(k,\tau, \tau_{r})$ the same analysis leading to Eq. (\ref{TRAD8a}) implies instead:
\begin{eqnarray}
\lim_{k \tau_{r} \ll 1, \,\, x \gg 1} Q_{T}^{(r)}(k,\tau, \tau_{r}) &\to& \frac{k^2 \overline{P}_{T}}{ 2 x^2}  =  \frac{k^2\,\overline{P}_{T}}{ 2} \biggl[\frac{a_{r}(\tau_{r})}{a_{r}(\tau)}\biggr]^2_{k \tau_{r} \simeq 1},\qquad k \tau \gg 1,
\nonumber\\
\lim_{k \tau_{r} \ll 1, \,\, x \ll 1} Q_{T}^{(r)}(k,\tau, \tau_{r}) &\to& |k\,\tau|^2 \frac{k^2\,\overline{P}_{T}}{9}, \qquad k\tau \ll 1,
\label{TRAD9a}
\end{eqnarray}
where, as in Eq. (\ref{TRAD8a}) the dominant oscillating contribution (going this time as $\cos^2{x}$) 
has been replaced by $1/2$. Unlike $P_{T}^{(r)}(k,\tau, \tau_{r})$, the result for 
 $Q_{T}^{(r)}(k,\tau, \tau_{r})$ is strongly suppressed in the limit $|k \tau | \ll 1$ and it practically does not 
 affect the spectral energy density when the corresponding wavelengths are larger than the Hubble radius. 
 Instead of artificially averaging the Sakharov oscillations in the limit $k \tau > 1$ it is wiser to deal with the spectral energy density of Eqs. (\ref{OBS5}) and  (\ref{TRAD7e}) where the leading order result is not oscillating even if a remnant of the Sakharov  oscillations still  appears in the corrections for $k\tau \gg 1$:
\begin{eqnarray}
\Omega_{gw}^{(r)}(k,\tau, \tau_{r}) &=&  \frac{\overline{P}_{T}(k, \tau_{r})}{24} \biggl[ 1 + {\mathcal O}\biggl(\frac{1}{k^2\tau^2}\biggr) \biggr], \qquad k \tau \gg 1,
\label{TRAD10a}\\
\Omega_{gw}^{(r)}(k,\tau, \tau_{r}) &=& \frac{k^2 \,\tau^2 \overline{P}_{T}(k,\tau_{r})}{24} \, \biggl[ 1 + {\mathcal O}(k^2 \tau^2)\biggr], \qquad 
k \tau \ll 1.
\label{TRAD10b}
\end{eqnarray}
The property expressed by Eqs. (\ref{TRAD10a})--(\ref{TRAD10b}) explains why the 
spectral energy density is often considered as one of the pivotal variables for the analysis 
of the relic graviton backgrounds. 

\subsubsection{The effect of the evolution of the relativistic species}
The quantitative expressions of the energy and of the entropy densities 
of the pre-decoupling plasma depend on the temperature and on the 
composition of the plasma. Denoting by $T$ the temperature of the plasma 
 the total energy density and the total entropy density can be 
expressed as:
\begin{equation}
\rho_{t} = g_{\rho}(T) \frac{\pi^2}{30} T^4,\qquad s_{t} = g_{s}(T) \frac{2 \pi^2}{45} T^3,
\label{EFF1}
\end{equation}
where $g_{\rho}(T)$ and $g_{s}(T)$ are \cite{book1,book2,book3}: 
\begin{equation}
g_{\rho}(T) = \sum_{b=1}^{N_{b}} g_{b} \biggl(\frac{T_{b}}{T}\biggr)^4 + 
\frac{7}{8} \sum_{f=1}^{N_{f}} g_{f} \biggl(\frac{T_{f}}{T}\biggr)^4,\qquad g_{s}(T) = \sum_{b=1}^{N_{b}} g_{b} \biggl(\frac{T_{b}}{T}\biggr)^3 + 
\frac{7}{8} \sum_{f=1}^{N_{f}} g_{f} \biggl(\frac{T_{f}}{T}\biggr)^3.
\label{grhogs}
\end{equation}
In Eq. (\ref{grhogs})  $N_{f}$ and $N_{b}$ are the number of fermionic and bosonic species;  $T_{f}$ and  
$T_{b}$ are instead the  corresponding temperatures. If all the fermionic and bosonic species are in thermal equilibrium at a common temperature $T$, then $T_{b}  = T_{f} = T$ and $g_{\rho} = g_{s}$. If at least one of the various species has a different temperature, then $g_{\rho} \neq g_{s}$. Assuming that the plasma has a temperature $T$ larger than the top mass, all the species of the  $SU_{L}(2)\otimes U_{Y}(1) \otimes SU_{c}(3)$ theory are in thermodynamic equilibrium so that:
\begin{equation}
g_{\rho} = g_{s} = \sum_{b} g_{b} + \frac{7}{8} \sum_{f} g_{f}.
\label{gT}
\end{equation}
The sum now extends over all the fermionic and bosonic 
species; globally the bosons carry $28$  degrees of freedom while the fermions carry $90$ 
degrees of freedom\footnote{ The fermionic species are constituted 
by the six quarks, by the three massive leptons and by the neutrinos 
(that we will take massless for consistency with the customary assumptions of the concordance paradigm). For the quarks the number of relativistic 
degrees of freedom is given by $(6 \times 2\times 2\times 3)= 72$, i.e. 
$6$ particles times a factor $2$ (for the corresponding antiparticles) 
times another factor $2$ (the spin) times a factor $3$ (since each quark 
may come in three different colors). Leptons do not carry color so the 
effective number of relativistic degrees of freedom of $e$, $\mu$ 
and $\tau$ (and of the corresponding neutrinos) is $18$. 
Globally, the fermions carry $90$ degrees of freedom.
The eight massless gluons each of them with two physical polarizations, 
amount to $8\times 2=16$ bosonic degrees of freedom. The $SU_{L}(2)$ 
(massless) gauge bosons and the $U_{Y}(1)$ (massless) 
gauge boson lead to $3\times 2 + 2 =8$
bosonic degrees of freedom. Finally the Higgs field (an $SU_{L}(2)$ complex doublet) carries $4$ degrees of freedom.}. Thus from Eq. (\ref{GT}) $g_{\rho} =g_{s} = 28 + (7/8)\times 90 =106.75$. 
When the temperature drops below $170$ GeV, the top quarks 
start annihilating and $g_{\rho} \to 96.25$. Below $80$ GeV the gauge bosons annihilate and
 $g_{\rho}\to 86.25$; below $4$ GeV, the bottom
quarks start annihilating and $g_{\rho}\to 75.75$ and so on.
While the electroweak phase transition takes place around 
$150$ GeV the quark-hadron phase transition takes place 
around $150$ MeV. 
According to Einstein's equations the temperature of the Universe determines directly the Hubble 
rate and the specific relation is $H = 1.66 \sqrt{g_{\rho}} \,T^2/M_{P}$,
Since the adiabatic evolution implies the conservation of the total entropy, 
also the entropy density defined in Eqs. (\ref{EFF1})-(\ref{grhogs}) will mildly affect the quantitative expression of the scale factor since $ a^3\, s_{t} \, g_{s}$ must be constant throughout the cosmological evolution. 
To estimate the impact of the evolution of the relativistic species let us consider again the spectral energy density 
in critical units for the wavelengths that already reentered the Hubble radius during radiation, i.e. 
Eq. (\ref{TRAD10a}). This equation has been obtained directly from Eq. (\ref{TRAD7e}) in the limit $k \tau \gg 1$.
Let us now reexamine that derivation and recall that according to the parametrization of Eqs. (\ref{TRAD3}) and (\ref{TRAD4}) 
$k^2/x^2(\tau) = H_{r}^2 a_{r}^4/a^2$. Using the latter relation inside the prefactor of Eq. (\ref{TRAD7e}) we can obtain a slightly different expression 
for Eq. (\ref{TRAD10a}), namely 
\begin{equation}
\Omega_{gw}^{(r)}(k,\tau, \tau_{r}) =  \frac{\overline{P}_{T}(k,\tau_{r})}{24} \biggl(\frac{a_{r}^4\, H_{r}^2}{ a^4\, H^2}\biggr)\biggl[ 1 + {\mathcal O}\biggl(\frac{1}{k^2\tau^2}\biggr) \biggr], \qquad k \tau \gg 1.
\label{TRAD10aa}
\end{equation}
Equations (\ref{TRAD10a}) and (\ref{TRAD10aa}) are identical provided 
$H \sim a^2$; however, strictly speaking, this expression 
must also take into account the evolution of the relativistic species. We can therefore immediately 
see that while we took $H\sim a^2$ in Eq. (\ref{TRAD10a}), the evolution 
of the relativistic species would instead imply a slightly different result:
\begin{equation}
\biggl(\frac{a_{r}^4 H_{r}^2}{a^{4} H^2}\biggr) = \biggl(\frac{g_{\rho}(T_{r})}{g_{\rho}(T)}\biggr) \biggl(\frac{g_{s}(T)}{g_{s}(T_{r})}\biggr)^{4/3}.
\label{EFF1a}
\end{equation}
For temperatures much larger than the top quark mass, all the known species of the minimal standard model of particle interactions are in local thermal 
equilibrium, then $g_{\rho} = g_{s} = 106.75$. Below $T \simeq 173$ GeV the various species 
start decoupling and the notion of thermal equilibrium is replaced by the notion of kinetic equilibrium. The 
time evolution of the number of relativistic degrees of freedom effectively changes the evolution of the Hubble rate. 
In principle if a given mode $k$ reenters the Hubble radius at a temperature $T_{k}$ the spectral energy density 
of the relic gravitons is (kinematically) suppressed by a factor which can be estimated as \cite{extrarel1,extrarel2,extrarel3}
 $[g_{\rho}(T_{k})/g_{\rho0}] [g_{s}(T_{k})/g_{s0}]^{-4/3}$; 
at the present time  $g_{\rho0}= 3.36$ and $g_{s0}= 3.90$. In general terms the 
effect parametrized by  Eq. (\ref{EFF1a}) will cause a frequency-dependent suppression, i.e. a further modulation of the spectral energy density $\Omega_{gw}(k,\tau_{0})$.  The maximal suppression one can expect 
can be obtained by inserting into Eq. (\ref{EFF1a}) the largest $g_{s}$ and $g_{\rho}$. 
So, in the case of the minimal standard model this would imply that the suppression (on $\Omega_{gw}(k,\tau_{0})$) will be of the order of ${\mathcal O}(0.38)$. In popular supersymmetric extension of the minimal standard model $g_{\rho}$ and $g_{s}$  can be as large as ${\mathcal O}(230)$ reducing 
 the previous estimate to ${\mathcal O}(0.29)$.

\subsubsection{The effect of neutrino free streaming}
The results of Eqs. (\ref{TRAD7c}), (\ref{TRAD7d}) and (\ref{TRAD7e}) hold 
under the assumption that the evolution of the mode functions is not affected by 
any source of anisotropic stress. Barring for other less conventional contributions that will be 
mentioned later on, already the neutrinos are a relevant source of anisotropic stress in the concordance paradigm. 
The effect of neutrino free streaming on the spectral energy density of the relic gravitons has been first 
pointed out by in Ref. \cite{STRESSNU1} (see also \cite{STRESSNU2}) where the author argued that the correction to the spectrum could be of the order of the $10$ \%. After neutrino decoupling, the neutrinos free stream and the effective energy-momentum tensor acquires, to first-order in the amplitude of the plasma fluctuations, an anisotropic stress. From Eq. (\ref{ACT5}) the evolution 
of the tensor amplitude in the presence of an anisotropic stress is\footnote{The anisotropic stress 
enters the spatial components of total energy-momentum tensor of the fluid 
as $T_{i}^{j} = - p_{t} \delta_{i}^{j} + \Pi_{i}^{\,\,j}$ and it may support scalar, vector or tensor 
inhomogeneities. In what follows we shall assume that $\Pi_{i}^{\,\,j}$ coincides with
 tensor component of the anisotropic stress and postpone to section \ref{sec7} a more specific discussion on this issue.}
\begin{equation}
h_{i\,j}^{\prime\prime} + 2 {\mathcal H} \, h_{i\,j}^{\prime }- \nabla^2 h_{i\,j} = - 2 \ell_{P}^2  a^2 \Pi_{i\,j}.
\label{ANIS2}
\end{equation}
If the neutrinos are the only source of anisotropic stress Eq. (\ref{ANIS2}) can be reduced to an integro-differential equation:
\begin{equation}
h_{\lambda}^{\prime\prime} + 2 {\mathcal H} h_{\lambda}^{\prime} + k^2 h_{\lambda} = - 24 \Omega_{\nu}
{\mathcal H}^2 \int_{0}^{\tau} K[ k (\tau- \tau_{1})] \, h_{\lambda}^{\prime}(\tau_{1}) \, d\tau_{1},
\label{ANIS2a}
\end{equation}
where $\lambda$ denotes alternatively one of the two tensor polarizations, $\Omega_{\nu} = \rho_{\nu}/\rho_{t}$  is the density parameter for the neutrinos and $K(s)$  is given by $K(s) = [- \sin{s}/s^3 - 3\cos{s}/s^4 + 3 \sin{s}/s^5]$.
Equation (\ref{ANIS2a}), has been analyzed for the first time in Ref. \cite{STRESSNU1};
in Ref. \cite{STRESSNU2} analytic solutions in the short wavelength limit 
have been discussed. Equation (\ref{ANIS2a}) can also be presented in the following manner \cite{STRESSNU3}
\begin{equation}
(1+ \alpha)\chi^{\prime\prime}(\alpha) + \left( \frac{2(1+\alpha)}{\alpha} + \frac{1}{2}\right) \chi^{\prime}(\alpha) 
+ q^{2}\chi(\alpha) = - \frac{24 \, R_{\nu}}{\alpha^{2}} \int_{0}^{\alpha} K(\alpha,\alpha_{1}) \frac{d\chi(\alpha_{1})}{d\alpha_{1}} d\alpha_{1},
\label{ANIS2c}
\end{equation}
with initial conditions $\chi(0) =1$ and $\chi^{\prime}(0)=0$ and normalized wavenumber $q= \sqrt{2} k/k_{eq}$.
Deviating for convenience from the notations established by Tab. \ref{SEC1TABLE4}, in  Eq. (\ref{ANIS2c}) the prime denotes a derivation with respect to the normalized scale factor $\alpha=a(t)/a_{eq}$ and not with respect to the conformal time coordinate. For notational convenience, in Eq. (\ref{ANIS2c}), $R_{\nu}$ 
denotes the neutrino fraction in the radiation plasma\footnote{Note that $3$ counts the degrees of freedom associated with the 
massless neutrino families, $(7/8)$ arises because neutrinos follow 
the Fermi-Dirac statistics; the factor  $(4/11)^{4/3}$ stems 
from the relative reduction of the neutrino (kinetic) temperature (in comparison 
with the photon temperature) after weak interactions fall out of thermal 
equilibrium.}
\begin{equation}
R_{\nu} = \frac{\rho_{\nu}}{\rho_{\gamma} + \rho_{\nu}} = \frac{3 \times (7/8)\times (4/11)^{4/3}}{ 1 + 3 \times (7/8)\times (4/11)^{4/3}} = 0.4052.
\label{ANIS2d}
\end{equation}
Explicit solutions of Eq. (\ref{ANIS2c}) have been 
studied in \cite{STRESSNU3} and the authors demonstrated how to handle the solution 
not only in the asymptotic limits $q^{2} \gg 1$ (already analyzed in \cite{STRESSNU1}) 
and $q^2\ll 1$ (already treated in \cite{STRESSNU2}) but also in the intermediate regime. 
A different analytic approach to the solution of Eq. (\ref{ANIS2a}) can be found in Ref. \cite{STRESSNU6}.
The effect of the massless neutrinos on the relic gravitons can be complemented by the effect 
of other free-streaming species possibly contributing to the total anisotropic stress; this 
viewpoint has been pursued in Refs. \cite{STRESSNU6a,STRESSNU6b}. 
Note that also dark matter might lead to a small anisotropic stress even if the estimates of Ref. \cite{STRESSNU7}
suggest a minute result.  Always assuming that the only collisionless 
species in the thermal history of the Universe are the neutrinos and recalling Eq. (\ref{ANIS2d}), the amount 
of suppression can be parametrized by the function
\begin{equation}
{\mathcal F}(R_{\nu}) = 1 - 0.539 R_{\nu} + 0.134 R_{\nu}^2.
\label{ANIS3}
\end{equation}
This suppression is effective for relatively small frequencies which are larger than $\nu_{eq}$ and smaller than the frequency corresponding to the Hubble radius at the time 
of big-bang nucleosynthesis, i.e. 
\begin{equation}
\nu_{bbn} =  \frac{H_{bbn}}{2\pi} \biggl(\frac{a_{bbn}}{a_{0}}\biggr) = \biggl( \frac{g_{\rho} \, \Omega_{R0}}{90 \pi}\biggr)^{1/4} \, 
\sqrt{\frac{H_{0}}{M_{P}}} \, T_{bbn}. 
\label{ANIS5}
\end{equation}
where $g_{\rho}$ denotes the effective number of relativistic species and the nucleosynthesis epoch.
Recalling that $H_{0}= 1.742\times h_{0}\,\,10^{-61} \, M_{P}$, Eq. (\ref{ANIS5}) can also be expressed as 
\begin{equation}
\nu_{bbn} = 8.17 \times 10^{-33} g_{\rho}^{1/4} \, T_{bbn} \biggl(\frac{h_{0}^2 \Omega_{R0}}{4.15\times 10^{-5}} \biggr).
\label{ANIS5a}
\end{equation}
If we take $g_{\rho} =10.75$ and $T_{bbn} = 1$ MeV, $\nu_{bbn} = {\mathcal O}(0.01)$ nHz and the result 
of Eq. (\ref{FF4}) is recovered. The effects of the neutrinos can be sensitive to the higher-order effects coming from the  scalar modes of the geometry or from the fluid sources.
There are various studies trying to include consistently the second-order effects in the anisotropic stress 
of free-streaming species \cite{STRESSNU4a,STRESSNU4b,STRESSNU4c,STRESSNU4d,STRESSNU4e}. The effects of free-streaming neutrinos are customarily addressed in spatially flat cosmologies 
since this is the most relevant phenomenological case but there exist some analyses in the case 
of spatially closed cosmologies \cite{STRESSNU8}.

\subsection{Wavelengths reentering when the dust dominates}
The range of comoving wavelengths  that left the Hubble radius during inflation and 
reentered when the universe was dominated by dust are generically represented by 
the dashed line labeled by $3$ in Fig. \ref{SEC4FIG1}. Because of the corresponding Sakharov 
condition (see Eq. (\ref{SAKm3c})) the dominant oscillation mode of
the tensor power spectrum goes, in this case, like a cosine square. In practice 
these wavelengths affect the spectral energy density of the relic gravitons between few aHz and $100$ 
aHz \cite{HIS11} (see also \cite{HIS12,HIS12a}). 
\subsubsection{The power spectra in a plasma dominated by dust}
Using the dimensionless variable $y(\tau)$ of Eq. (\ref{SAKm1}) the rescaled 
mode functions follow  by inserting Eq. (\ref{SAKm3}) into Eqs. (\ref{PAM10})--(\ref{PAM12})
\begin{eqnarray}
F_{k}^{(m)}(\tau) &=& \overline{F}_{k}^{(m)}  \biggl[ \frac{\cos{y(\tau)}}{y^2(\tau)} - \frac{\sin{y(\tau)}}{y^3(\tau)} \biggr],
\label{TFF2a}\\
G_{k}^{(m)}(\tau) &=& k\,  \overline{F}_{k}^{(m)} \biggl[ 3 \frac{\sin{y(\tau)}}{y^{4}(\tau)} - \frac{\sin{y(\tau)}}{y^2(\tau)} - 3 \frac{\cos{y(\tau)}}{y^3(\tau)} \biggr], 
\label{TFF2b}\\
\overline{F}_{k}^{(m)} &=& - \frac{3 \,i}{\sqrt{\pi\, k}} \, 2^{\nu -1}\, \Gamma(\nu) e^{ i \pi ( 2 \nu +1)/4} \, \bigl|k \tau_{r}\,\bigr|^{1/2 - \nu}.
\label{TFF2c}
\end{eqnarray}
where the superscript reminds that the mode functions apply in the matter-dominated (or dust-dominated) epoch and $\overline{F}^{(m)}_{k}$ denotes the constant amplitude of the mode function.
Inserting Eqs. (\ref{TFF2a}) and (\ref{TFF2b}) into Eq. (\ref{fiveE}) and using the shorthand notation 
$y(\tau) = y$, the general form of the tensor power spectra for $\tau > \tau_{eq}$ becomes:
\begin{eqnarray}
P_{T}(k,\, \tau,\, \tau_{r},\, \tau_{eq}) &=& 9\,\overline{P}_{T}(k,\tau_{r}) \,\biggl[ \frac{\cos{y}}{y^2} - \frac{\sin{y}}{y^3} \biggr]^2, 
\label{TFF2ca}\\
Q_{T}(k,\, \tau,\, \tau_{r},\, \tau_{eq}) &=&9\,k^2\,\overline{P}_{T}(k,\tau_{r}) \,\biggl[ 3 \frac{\sin{y}}{y^{4}} - \frac{\sin{y}}{y^2} - 3 \frac{\cos{y}}{y^3} \biggr]^2.
\label{TFF2cb}
\end{eqnarray}
The wavelengths amplified across the transition to the dust dominance imply that the corresponding 
wavenumbers must satisfy $k \tau_{eq} < 1$. When the wavelengths are larger than the Hubble radius (i.e. $|k \tau| \ll 1$), the tensor power spectrum of  Eq. (\ref{TFF2ca}) tends to a constant value while $Q_{T}(k,\, \tau,\, \tau_{r}\, \tau_{eq})$  is instead suppressed as $| k \tau|^2$ in the same limit:
\begin{equation}
\lim_{k \tau_{eq} \ll 1, \,\, k\tau\ll 1}P_{T}(k,\, \tau,\, \tau_{r},\, \tau_{eq}) = \overline{P}_{T}(k,\tau_{r}),\qquad 
\lim_{k \tau_{eq} \ll 1, \,\, k\tau \ll 1}Q_{T}(k,\, \tau,\, \tau_{r},\, \tau_{eq}) = \frac{9 |k \tau|^2}{225} \overline{P}_{T}(k,\tau_{r}).
\label{TFF2cc}
\end{equation}
From Eqs. (\ref{TFF2ca}) and (\ref{TFF2cb}) the spectra can be approximated  in the simultaneous limits $k \tau_{eq} \ll 1$ and $ k \tau \gg 1$:
\begin{eqnarray}
\lim_{k \tau_{eq} \ll 1, \,\, k \tau \gg 1}P_{T}(k,\, \tau,\, \tau_{r},\, \tau_{eq}) &=& \frac{9 \overline{P}_{T}(k,\tau_{r})}{ 2\, y^4} \biggl[ 1 + {\mathcal O}\biggl(\frac{1}{|k \tau|}\biggr)\biggr],
\nonumber\\
\lim_{k \tau_{eq} \ll 1, \,\, k \tau \gg 1}Q_{T}(k,\, \tau,\, \tau_{r},\, \tau_{eq}) &=& \frac{9  k^2 \overline{P}_{T}(k,\tau_{r})}{2 \,y^4}\biggl[ 1 + {\mathcal O}\biggl(\frac{1}{|k \tau|}\biggr)\biggr],
\label{TFF2cca}
\end{eqnarray}
where as in the case of Eqs. (\ref{TRAD8a}) and (\ref{TRAD9a}) we replaced $ \sin^2y \to 1/2$ and $\cos^2y \to 1/2$.  Again, this strategy is not ideal and it is better to employ the spectral energy density where the Sakharov oscillations disappear from the leading-order result (but not from the corrections)
\begin{equation}
\Omega_{gw}(k, \, \tau,\, \tau_{r},\, \tau_{eq}) = \frac{3 \, k^2 \, \overline{P}_{T}(k,\tau_{r})}{ 8 \, a^2 \, H^2 y^4} \biggl[ 1 + 2 \frac{\sin{ 2 y}}{y} 
+ \frac{9 \cos^2{y} - 5 \sin^2{y}}{y^2} - 9 \,\frac{\sin{2 y}}{y^3}  + 9\,\frac{\sin^2{ y}}{y^4} \biggr].
\label{TFF2d}
\end{equation}
In the limit $y \gg 1$ (i.e. $k \tau \gg 1$) the wavelengths are inside the Hubble radius and the spectral energy density becomes:
\begin{equation}
\Omega_{gw}(k,\tau, \tau_{r}, \tau_{eq}) = \frac{3}{32} \, \frac{\overline{P}_{T}(k,\tau_{r})}{y^2}\biggl[ 1 + {\mathcal O}\biggl( \frac{1}{k \tau} \biggr)\biggr], \qquad y\gg 1.
\label{TFF2e}
\end{equation}
Conversely  the limit $y \ll 1$ (i.e. $k \tau \ll 1$) the wavelengths are all inside the Hubble radius and the spectral energy density is:
\begin{equation}
\Omega_{gw}(k,\tau, \tau_{r}, \tau_{eq}) = \frac{k^2 \,\overline{P}_{T}(k,\tau_{r})}{24\, a^2 \, H^2 y^4} \biggl[ 1 - \frac{36\, y^2}{225} + {\mathcal O} (y^4)\biggr], \qquad y \ll 1.
\label{TFF2f}
\end{equation}

\subsubsection{The concept of transfer function}
The analytic expressions of tensor power spectra of Eqs. (\ref{TFF2ca}) and (\ref{TFF2cb}) 
are a sound approximation in the asymptotic regimes but across the transition regions both in time and in frequency 
(i.e. $ \tau \sim \tau_{eq}$ and $ k \tau \sim 1$) the analytic methods must be complemented by numerical strategies.
In the simplest situation the evolution equations of the mode functions (see e.g. Eqs. (\ref{PAM12})--(\ref{PAM12a}))
must then be integrated across the matter-radiation transition when the sources in the background equations (\ref{FL2C}) are given by $\rho_{t} = (\rho_{M} + \rho_{R} + \rho_{\Lambda})$ and by 
$(\rho_{t} + p_{t}) =(4 \rho_{R}/3 + \rho_{M})$;  $\rho_{\Lambda}$, $\rho_{M}$ and $\rho_{R}$ denote the energy densities of the dark energy, of the dust and of the radiation components respectively. Since the dark energy will only dominate
much later, it is possible to solve analytically Eq. (\ref{FL2C}) across the equality transition:
\begin{equation}
a(\tau) = a_{eq}\biggl[ \biggl(\frac{\tau}{\tau_{1}}\biggr)^2 + 2 \biggl(\frac{\tau}{\tau_{1}}\biggr)
\biggr],\qquad \tau_{1} = \tau_{eq}(\sqrt{2} +1).
\label{TFF4}
\end{equation}
Thanks to Eq. (\ref{TFF4}) $\tau_{eq}$, $k_{eq}$ and $\nu_{eq}$ can be determined analytically with some 
ambiguities that will have ultimately some impact on the normalization of the power spectrum and of the spectral energy density.  For instance, from Eq. (\ref{TFF4}) the value of $\tau_{eq}$ is immediately determined 
by consistency with the equations of motion so that 
\begin{equation}
\tau_{eq} = 
\frac{2 (\sqrt{2} -1)}{H_{0}} \frac{\sqrt{ \Omega_{R0}}}{ \Omega_{M0}} 
= 113.38\,\biggl(\frac{h_{0}^2 \Omega_{M0}}{0.1411}\biggr)^{-1} \biggl(\frac{h_{0}^2 \Omega_{R0}}{4.15 \times 10^{-5}}\biggr)^{1/2}\,\,\mathrm{Mpc}.
\label{TFF5}
\end{equation}
According to Eq. (\ref{TFF5}) the condition of horizon crossing is given by $\widetilde{k}_{eq} = 1/\tau_{eq}$.
It is also possible to determine $k_{eq}$ directly from the condition $k_{eq} = H_{eq} a_{eq}$; in this case recalling 
that $H_{eq}/H_{0} = 2 \Omega_{M0} \, (a_{0}/a_{eq})^{3/2}$ (and that $a_{0}/a_{eq} = \Omega_{M0}/\Omega_{R0}$) 
we have that 
\begin{equation}
k_{eq} = \sqrt{2} \frac{\Omega_{M0}}{\sqrt{\Omega_{R0}}} H_{0} = 
 0.0732\,\, h_{0}^2 \Omega_{\mathrm{M}0} \, \,\mathrm{Mpc}^{-1}.
\label{TFF6}
\end{equation}
The two determinations (i.e. $\widetilde{k}_{eq}$ and $k_{eq}$) are compatible 
since $k_{eq}/\widetilde{k}_{eq} = 2/(\sqrt{2} +1) = 0.82$. However, instead of defining $\widetilde{k}_{eq} = 1/\tau_{eq}$ we could determine the horizon crossing condition from $\widetilde{k}_{eq} = {\mathcal H}_{eq} = 2/\tau_{eq}$.
This estimate follows from the observation that during matter we approximately have $a(\tau) \sim \tau^2$ and 
${\mathcal H} \sim 2/\tau$; with this choice we would have that $k_{eq}/\widetilde{k}_{eq} = 2/(\sqrt{2} +1) = 1.65$.
Even if we shall use Eq. (\ref{TFF6}) as pivotal estimate of $k_{eq}$ it should be borne in mind that 
the  horizon crossing condition is subjected to an indetermination that can even ${\mathcal O}(2)$. 
With this caveat from Eq. (\ref{TFF6}) the typical equality frequency $\nu_{eq}$ becomes 
exactly the one already quoted in Eq. (\ref{FF3}). 

\begin{figure}[!ht]
\centering
\includegraphics[height=6.1cm]{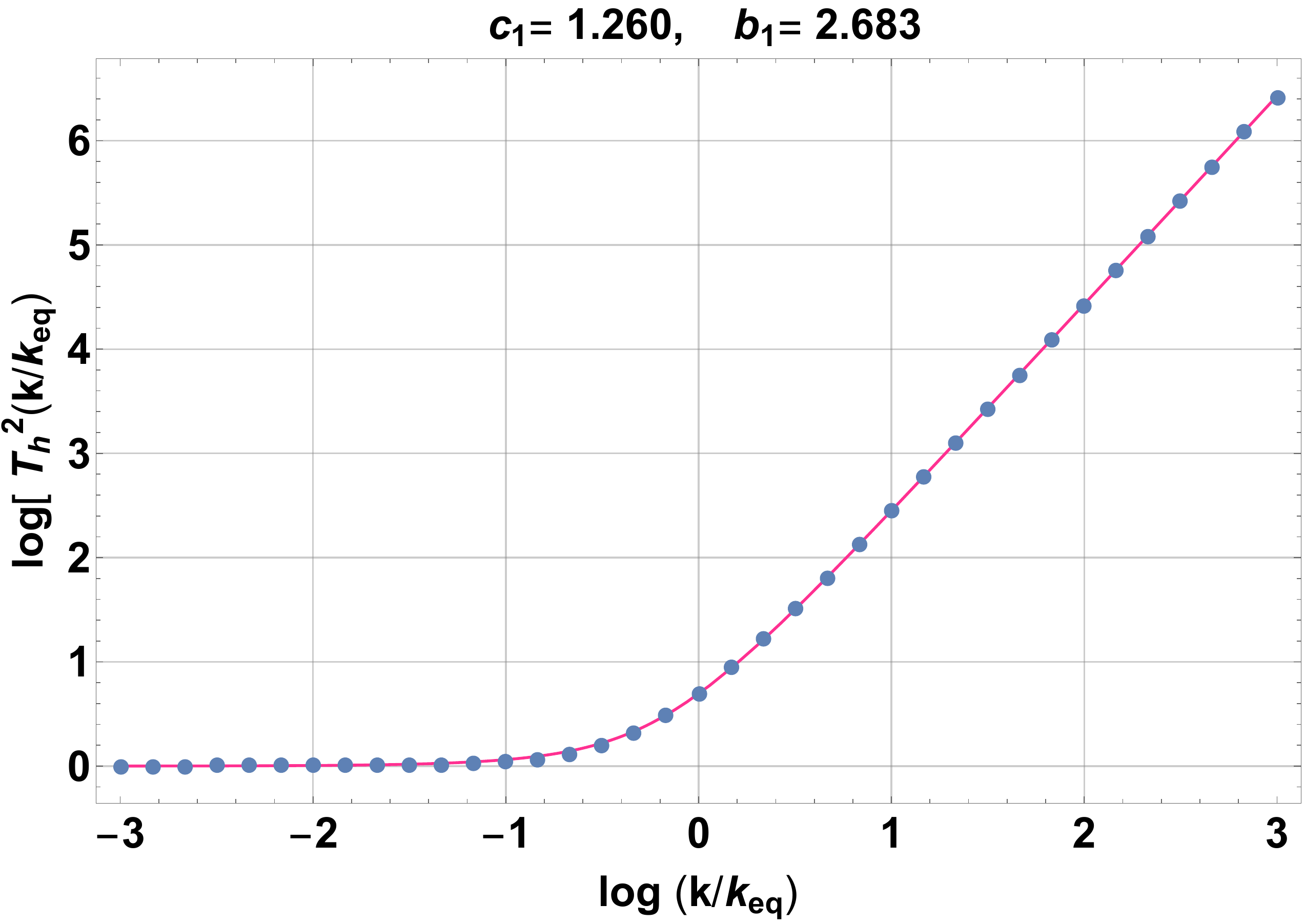}
\includegraphics[height=6.1cm]{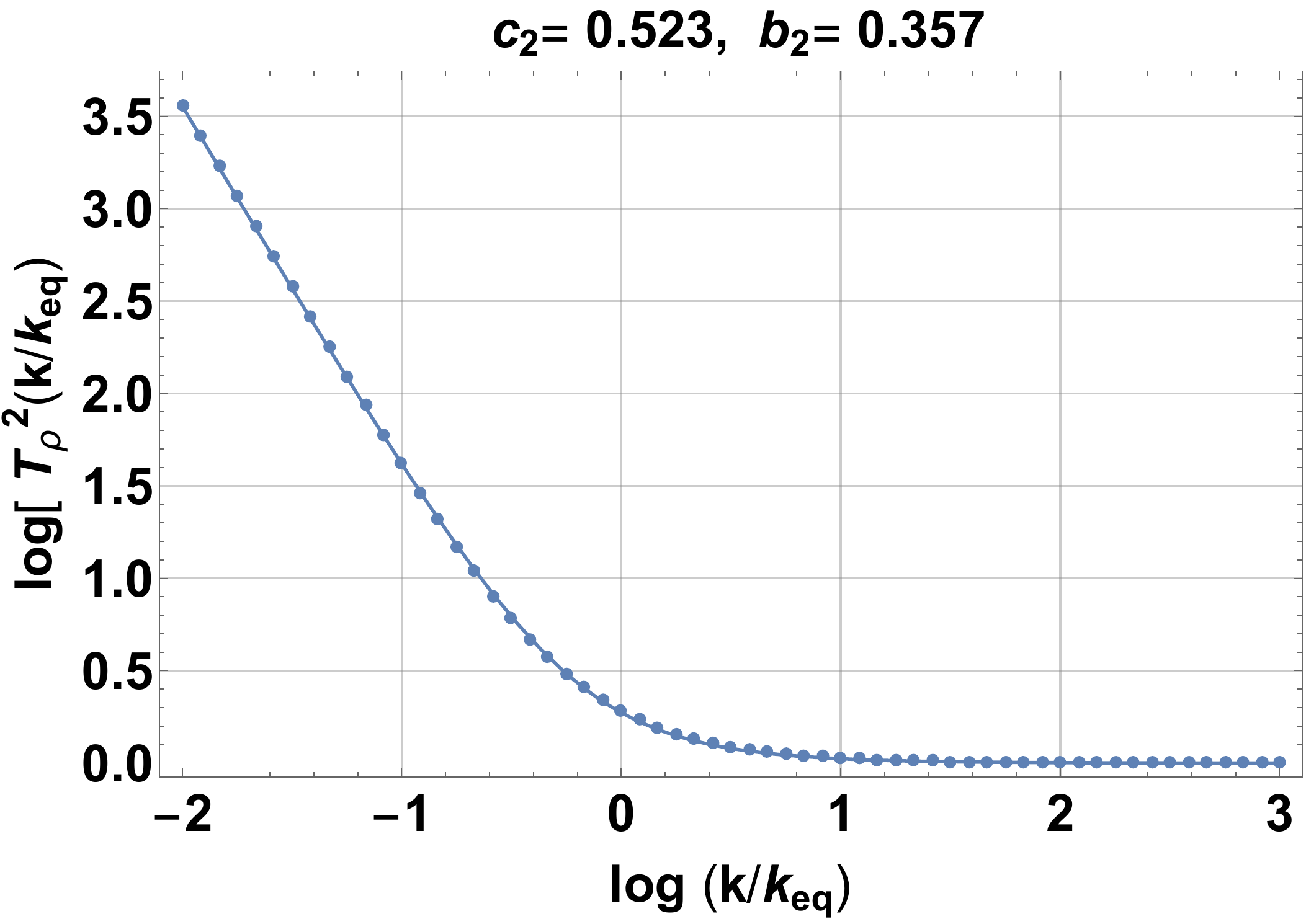}
\caption[a]{The numerical form of the transfer function for the amplitude (left plot) and for the spectral energy density (right plot).}
\label{SEC4FIG3}      
\end{figure}

The transfer function for the amplitude \cite{TRANS1} 
(see also \cite{TRANS1a,TRANS2,TRANS3,TRANS4,TRANS8a})
is determined by integrating Eqs. (\ref{PAM12a}) and (\ref{FL2C})
across a given transition. In a more generic perspective, if the background evolves across a given
transition time $\tau_{*}$, the power spectrum  for $\tau > \tau_{*} \gg \tau_{i}$ 
can be related to $\overline{P}_{T}(k,\tau_{i})$ as:
\begin{equation}
P_{T}(k,\tau) = \biggl| \frac{F^{(app)}_{k}(\tau)}{F_{k}(\tau_{i})}\biggr|^2 \, T^2_{h}(k/k_{*}) \, \overline{P}_{T}(k,\tau_{i}).
\label{TFF3a}
\end{equation}
In Eq. (\ref{TFF3a}) $F^{(app)}_{k}(\tau)$ denotes an 
approximate (but analytic) form of the mode function holding for 
$\tau \gg \tau_{*}$ while $F_{k}(\tau_{i})$ is the value of the mode function for $\tau_{i} < \tau_{*}$. 
Since the mode functions inside the Hubble radius are oscillating 
the numerical determination of the mode function always involves a 
smoothing\footnote{As the wavelengths become shorter than the Hubble radius, $F_{k}(\tau)$ oscillates and the 
explicit form of the transfer function
 $T_{h}(k/k_{*})$ can only be obtained by carefully averaging the oscillations; this is the meaning of the expectation 
values appearing in  Eq. (\ref{TFF3b}).} so that 
$T_{h}(k/k_{*})$ can be formally expressed as:
\begin{equation}
T_{h}(k/k_{*}) = \sqrt{\frac{\langle|F_{k}(\tau)|^2\rangle}{\langle|F^{(app)}_{k}(\tau)|^2\rangle}},
\label{TFF3b}
\end{equation}
where $F_{k}(\tau)$ denotes the full numerical solution. The calculation of $T_{h}(k/k_{*})$ requires a careful matching over the  phases between the numerical and the approximate analytical solution. If we now apply the  logic of Eqs. (\ref{TFF3a}) and (\ref{TFF3b}) across $\tau_{eq}$, the 
approximate solution of Eq. (\ref{PAM12a})  follows from Eq. (\ref{TFF2a}) by requiring
$ y(\tau) \simeq k \tau$ and it is obviously given by $F^{(app)}_{k}(\tau) = 3 j_{1}(k\tau) F_{k}(\tau_{i})/|k\tau|$ 
(where $j_{1}(z)$ is the spherical Bessel function  of first kind \cite{abr1,abr2}). 
In this case the transfer function $T_{h}(k/k_{eq})$ can be determined as:
\begin{equation}
T_{h}(k/k_{eq}) = \sqrt{1 + c_{1} \biggl(\frac{k}{k_{eq}}\biggr) + b_{1} \biggl(\frac{k}{k_{eq}}\biggr)^2}.
\label{TFF9}
\end{equation}
By applying the standard tools of the regression analysis  $c_{1}$ and $b_{1}$ can be determined as $c_{1}= 1.260$ and $b_{1}= 2.683$ \cite{TRANS8}. The latter result agrees with the findings of  \cite{TRANS1} who obtain $ \overline{c}_{1}= 1.34$ and $\overline{b}_{1} = 2.50$.  In Fig. \ref{SEC4FIG3} (left plot) we illustrate 
the transfer function for the amplitude with the technique of Ref. \cite{TRANS8}. With the help of Eq. (\ref{TFF9}) the power spectrum inside 
the Hubble radius at the present time (and neglecting the contribution of the cosmological constant) can also be written as: 
\begin{equation}
 P_{T}(k,\tau_{0}) = {\mathcal T}^2(k/k_{eq}, \,\tau_{0}) \overline{P}_{T}(k,\tau_{r}),\qquad
{\mathcal T}(k/k_{eq}, \,\tau_{0}) = \frac{3\,j_{1}(k\tau_{0})}{|k\tau_{0}|} T_{h}(k/k_{eq}).
 \label{TFF10}
\end{equation}
Note that since ${\mathcal T}(k/k_{eq}, \,\tau)$ is always given in a factorized 
form, it is easy to compute its derivative: 
\begin{equation}
\partial_{\tau} {\mathcal T}(k/k_{eq}, \,\tau) = \frac{ [3 \sin{ k \tau} - 9 j_{1}(k\tau)]}{k \tau^2} T_{h}(k/k_{eq})
\label{TFF10a}
\end{equation}
Thanks to Eq. (\ref{OBS7}) and its descendants the spectral energy density can be always approximately related to the power spectrum; therefore, having determined the tensor power spectrum for wavelengths shorter than the Hubble radius, 
the spectral energy density becomes:
\begin{equation}
\lim_{k \gg k_{eq}} \Omega_{gw}(k,\tau_{0})  \simeq \frac{3 b_{1}}{8 a_{0}^2 H_{0}^2 \tau_{0}^4 k_{eq}^2} {\mathcal A}_{T} \biggl(\frac{k}{k_{\mathrm{p}}}\biggr)^{n_{T}}.
\label{TFF11}
\end{equation}
The leading order result for $\Omega_{gw}(k,\tau_{0})$ obtained in Eq. (\ref{TFF11}) oscillates but this is just an artefact of the parametrization as it can be argued by comparing the result with the analog expression of Eq. (\ref{TFF2e}). 
The final expression given in Eq. (\ref{TFF11}) involves therefore an averaging over the oscillations; in fact for $k \gg k_{eq}$  the cosine will dominate the expression of $j_{1}(k\tau_{0})$ and the second result of Eq. (\ref{TFF11}) arises by replacing $\cos^2({k\tau_{0}}) \to 1/2$. The different determinations of the coefficients in Eq. (\ref{TFF9}) are compatible: if $\overline{b}_{1} = 2.5$ \cite{TRANS1} in the second relation of Eq. (\ref{TFF11}), then $3 b_{1}/8 \equiv 15/16 = 0.9375 $. If 
we take instead the result of Ref. \cite{TRANS8} (i.e. $b_{1} = 2.683$) we will get, for the same quantity, $1.006$. Since the calculation of the transfer function for the amplitude involves a delicate matching on the phases of the tensor mode functions \cite{TRANS1}, the approach of Eq. (\ref{TFF3b}) is practical for the power spectrum but it is not ideal if we ultimately want to assess the spectral energy density. 

If the transfer function is computed directly for the spectral energy density, the oscillatory contributions are suppressed as the wavelengths get shorter than the Hubble radius. This aspect can be understood by considering the spectral energy distribution written in terms of the mode functions $f_{k}$ and $g_{k}$ (see Eqs. (\ref{PAM11})--(\ref{PAM12}) and discussions thereafter); in this case we have that $\Omega_{gw}(k,\tau)$ can be expressed as: 
\begin{equation}
\Omega_{gw}(k,\tau) = \frac{k^3}{ 2 \,\pi^2 \, a^4\, \rho_{crit} } \Delta_{\rho}(k,\tau),\qquad 
\Delta_{\rho}(k,\tau) = |g_{k}(\tau)|^2 + k^2 |f_{k}(\tau)|^2,
\label{TFF12}
\end{equation}
where $\Delta_{\rho}(k,\tau)$ follows from the general expression of the spectral energy density 
(\ref{OBS5})--(\ref{OBS6}). If the numerical integration is performed in terms of the two dimensionless 
variables $\kappa = k/k_{eq}$ and $ k \tau $, the explicit form of Eq. (\ref{TFF12}) implies that $\kappa \,\Delta_{\rho}(\kappa, k \tau)$  reaches a constant value when the relevant modes are
evaluated deep inside the Hubble radius. The {\em transfer function for the energy density} is therefore defined as
\begin{equation}
\lim_{k\tau \gg 1} \Delta_{\rho}(\kappa,\, k\tau) \equiv T^{2}_{\rho}(\kappa) \Delta_{\rho}(\kappa, \, k \tau_{\mathrm{i}}), \qquad 
k\, \tau_{\mathrm{i}} \ll 1.
\label{TFF13}
\end{equation}
The procedure suggested by Eqs. (\ref{TFF12})--(\ref{TFF13}) and discussed in Ref. \cite{TRANS9} is more direct
that the one based on the transfer function of the amplitude: instead of deducing the transfer function 
of the amplitude with the ultimate purpose of computing the spectral energy density, Eqs. (\ref{TFF12}) and 
(\ref{TFF13}) lead to the spectral energy density in one shot and without arbitrary averages of trigonometric functions.
This happens since, inside the Hubble radius, the oscillating contributions appearing in $\Delta_{\rho}(\kappa,\,k\tau)$ 
are suppressed as $(k\tau)^{-1}$ for $k\tau\gg 1$ so in practice they only contribute around $k\tau \simeq 1$. The analytical form of the fit is:
\begin{equation}
T_{\rho}(k/k_{eq}) = \sqrt{1 + c_{2}\biggl(\frac{k_{eq}}{k}\biggr) + b_{2}\biggl(\frac{k_{eq}}{k}\biggr)^2},\qquad c_{2}= 0.5238,\qquad
b_{2}=0.3537.
\label{TFF14}
\end{equation}
In the right plot of Fig. \ref{SEC4FIG3}  we illustrate the explicit form of the transfer function for the 
energy density. The transfer function for the amplitude (see Eq. (\ref{TFF9}) and left plot 
in Fig. \ref{SEC4FIG3}) is dominated by the coefficient $b_{1}$  in the limit $k \gg k_{eq}$ ; in the 
same limit the transfer function for the energy density (see Eq. (\ref{TFF14}) and right plot 
in Fig. \ref{SEC4FIG3}) goes to $1$ by construction. In the concordance 
paradigm, $T_{\rho}(k/k_{eq})$ turns out to be extremely useful for the analytical determination 
of the  quasi-flat plateau of spectral energy density for $k \gg k_{eq}$, as we shall see in a moment.

\subsubsection{Approximate normalization of the spectrum}
Since there are slightly different explicit analytic estimates of spectral energy density 
in the region of the plateau (i.e. for $k \gg k_{eq}$),  Eq. (\ref{TFF11}) shall be first written in the case 
$b_{1} = 2.5$:
\begin{equation}
\Omega_{gw}(k,\tau_{0})  \to \overline{\Omega}_{gw} \biggl(\frac{k}{k_{\mathrm{p}}}\biggr)^{n_{T}},
\qquad \overline{\Omega}_{gw}  = \frac{15}{16\, a_{0}^2 H_{0}^2 \tau_{0}^4 k_{eq}^2} {\mathcal A}_{T},\qquad k \gg k_{eq},
\label{NORR1}
\end{equation}
where $\overline{\Omega}_{gw} $ denotes the spectral energy density in the region of the plateau.
Equation (\ref{NORR1}) is the expression adopted in Refs. \cite{NORM1,NORM1a};
the authors of Ref. \cite{NORM1} estimate $\tau_{0}$ via the (comoving) angular diameter 
distance at decoupling in a spatially flat Universe which is ${\mathcal O}(1.41) \times 10^{4}$ Mpc:
\begin{equation}
d_{\mathrm{A}}(z_{*}) =  \frac{1}{H_{0}} \int_{0}^{z_{*}} \frac{d z}{\sqrt{\Omega_{\mathrm{M}0} ( 1 + z)^3 + \Omega_{\Lambda} + \Omega_{R0} (1 + z)^4}} = \frac{\overline{d}_{0}}{a_{0} H_{0}}= {\mathcal O}(1.41)\,\, \mathrm{Mpc},\qquad z_{*} = 1090,
\label{NORR3}
\end{equation}
where $\overline{d}_{0} = {\mathcal O}(3.4)$. Estimating $\tau_{0} = \overline{d}_{0}/(a_{0} H_{0})$ we can then deduce from Eq. (\ref{NORR1}) that $\overline{\Omega}_{gw} = [15 \,r_{T} \, {\mathcal A}_{{\mathcal R}} \, a_{0}^2 \, H_{0}^2]/[16 \, \overline{d}_{0}^4\,a_{eq}^2 H_{eq}^2]$.
Thus the standard form of the transfer function of the amplitude given in Eq. (\ref{TFF9}) implies:
\begin{equation}
h_{0}^2 \overline{\Omega}_{gw} = 1.23 \times 10^{-15} \biggl(\frac{b_{1}}{2.5}\biggr) 
\biggl(\frac{\overline{d}_{0}}{3.4}\biggr)^{-4} \biggl(\frac{r_{T}}{0.07}\biggr) \biggl(\frac{{\mathcal A}_{{\mathcal R}}}{2.41\times 10^{-9}} \biggr) \biggl(\frac{h_{0}^2 \Omega_{R0}}{4.15\times 10^{-5}} \biggr)
\biggl( \frac{h_{0}^2 \Omega_{M0}}{0.1411}\biggr)^{-2}.
\label{NORR6}
\end{equation}
Note, incidentally, that for a different definition of the spectral energy density the result does not change 
in any way as long as the relevant wavelengths are inside the Hubble radius. 
For instance if the energy density is defined in terms of the Brill-Hartle-Isaacson 
prescription \cite{TENS2c,TENS3a,TENS3b} (see also Eq. (\ref{eightH})) the spectral energy density 
follows from the quadratic expectation value of the time derivative of the tensor amplitude 
\begin{equation}
\Omega_{gw}(k,\tau_{0}) = \frac{1}{12 H_{0}^2 a_{0}^2} \biggl[\partial_{\tau} {\mathcal T}(k/k_{eq},\tau_{0}) \biggr]^2 \overline{P}_{T}(k,\tau_{r}).
\label{NORR7}
\end{equation}
This form of the spectral energy density is often employed in the literature and, in particular 
in Refs. \cite{NORM1a,NORM1b,NORM3,NORM4}. In particular this parametrization 
has been used in the discussion of the reconstruction of the inflaton potential \cite{NORM1b} and by in the analysis of the WMAP three-year data \cite{WMAP2a}. Equation (\ref{NORR7}), when evaluated 
in explicit terms, leads exactly\footnote{The reason is that the limit of Eq. (\ref{TFF10a}) for $k\tau_{0} \gg 1$ leads to $3 [\sin{(k \tau_{0})}/|k \tau_{0}|^2] T_{h}(k/k_{eq})$. The latter expression must then be squared (with the caveat that $\sin^2{(k \tau_{0})}\to 1$) and then inserted into Eq. (\ref{NORR7}). The result of this procedure coincides with the estimate of Eq. 
(\ref{TFF11}).  } to Eqs. (\ref{TFF11}) and hence to Eq. (\ref{NORR1}) (in the case $b_{1} = 2.5$). 
The transfer function of the spectral energy density (defined in Eqs. (\ref{TFF13}) and (\ref{TFF14})) is particularly useful for the determination of $\overline{\Omega}_{gw}$ since in the limit $k \gg k_{eq}$ it does not depend on the fit parameters but it goes to $1$ by construction. In this case we have that 
$\overline{\Omega}_{gw}$ will be given by:
\begin{equation}
\overline{\Omega}_{gw} = \frac{ H_{eq}^2 \, a_{eq}^2 }{24\, H_0^2\, a_0^2} \, 
\biggl(\frac{a_{eq}}{a_{0}} \biggl)^2 {\mathcal A}_{T} \simeq \frac{1}{24 \, z_{eq}}\  {\mathcal A}_{T}.
\label{NORR8}
\end{equation}
The approximation of Eq. (\ref{NORR8}) is rather crude and it comes by using that $a_{eq}/a_{0} = (\tau_{eq}/\tau_{0})^2$ and that ${\mathcal H}_{eq}^2/{\mathcal H}_{0}^2 = (\tau_{0}/\tau_{eq})^2$. A better approximation 
is obtained by using that  $H_{0}^2/H_{eq}^2= (a_{eq}/a_{0})^3/(2 \Omega_{M0})$ and $(a_{0}/a_{eq})= \Omega_{M0}/\Omega_{R0}$; in this case the result will be:
\begin{equation}
h_{0}^2 \overline{\Omega}_{gw} = 2.91\times 10^{-16} \biggl(\frac{r_{T}}{0.07}\biggr) \biggl(\frac{{\mathcal A}_{{\mathcal R}}}{2.41\times 10^{-9}} \biggr) \biggl(\frac{h_{0}^2 \Omega_{R0}}{4.15\times 10^{-5}} \biggr).
\label{NORR9}
\end{equation}
Since Eq. (\ref{NORR9}) has been determined from $T_{\rho}(k/k_{eq})$ in the limit $k \gg k_{eq}$, $h_{0}^2 \overline{\Omega}_{gw}$ does not explicitly depend on the fit parameters of the transfer function.
The evolution of the effective relativistic species and the effect of neutrino free-streaming 
further reduces the height of the plateau and can be easily included by using Eqs. (\ref{ANIS3}) and 
(\ref{EFF1a}).

\subsection{Wavelengths reentering when the dark energy dominates}
The wavelengths labeled by $3$ and $4$ in Fig. \ref{SEC4FIG1} 
 left the Hubble radius at the beginning of inflation and reentered either 
during the matter-dominated epoch or when dark energy was already dominant. 
The evolution of $ \bigl| a\, H \bigr|^{-1}$ is monotonically 
increasing for $a_{rad} < a < a_{\Lambda}$ and this means that all the wavelengths that 
left the  Hubble radius during inflation will remain within the Hubble radius after 
they reenter either during radiation or during matter dominance.
If we also consider the evolution for $a > a_{\Lambda}$,  the behaviour of $\bigl|a\, H\bigr|^{-1}$ 
is, overall, non-monotonic: a bunch wavelengths (labeled by 
$4$ in in Fig. \ref{SEC4FIG1}) that reentered after equality will again exit after dark energy dominates.
 \begin{figure}[!ht]
\centering
\includegraphics[height=6.2cm]{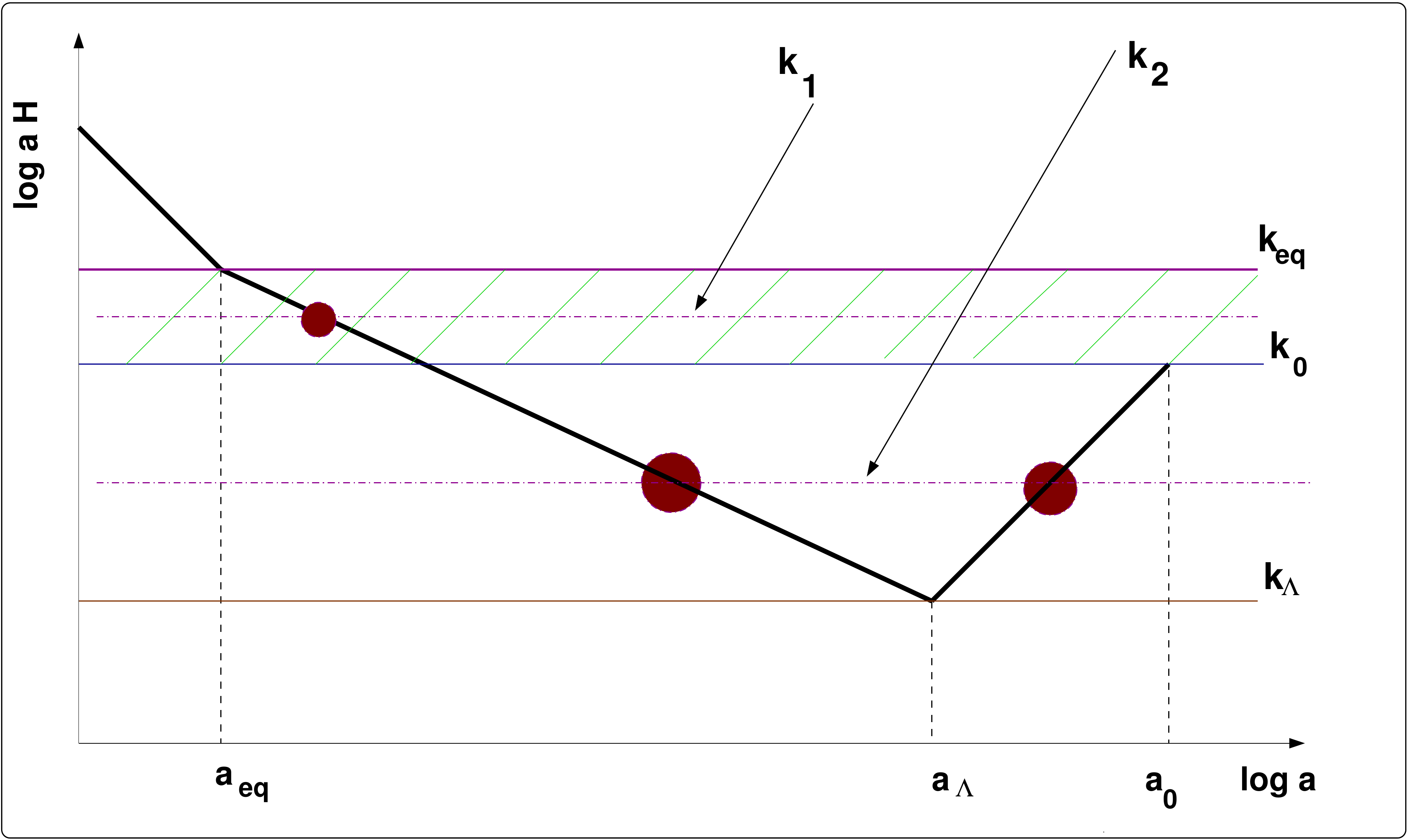}
\caption[a]{The evolution of $|a\, H|$  is illustrated around $a_{\Lambda}$ with particular attention 
to the effect of the dark energy background.}
\label{SEC4FIG4}      
\end{figure}
In Fig. \ref{SEC4FIG4} we illustrate the common logarithm of the 
Hubble rate as function of the scale factor around the time of dark energy 
dominance\footnote{Figure \ref{SEC4FIG4} illustrates, in practice, a detail of the inverse of Fig. \ref{SEC3FIG1};
in both figures, the time range of the dark energy dominance 
is purposely exaggerated. This is done to distinguish  
the different scales of the problem which are however not so 
well separated in practice.}. If the evolution would be monotonically decreasing in Fig. \ref{SEC4FIG4} 
(as it happens for $a< a_{\Lambda}$), two different wavenumbers $k_{1} > k_{2}$ crossing the Hubble rate at two different epochs would imply that $a_{1} H_{1} > a_{2} H_{2}$: this is not true anymore since, as shown by  Fig. \ref{SEC4FIG4}, the evolution of $ a H$ is not monotonic around $a_{\Lambda}$. 
The redshift of $\Lambda$-dominance is simply  defined by 
\begin{equation}
1 + z_{\Lambda} = \biggl(\frac{a_{0}}{a_{\Lambda}}\biggr) = 
 \biggl(\frac{\Omega_{\Lambda}}{\Omega_{\mathrm{M}0}}\biggr)^{1/3}.
\label{LAM1}
\end{equation}
Consider now the mode $k_{\Lambda} = a_{\Lambda} H_{\Lambda}$, i.e. the mode 
approximately reentering the Hubble radius at $\tau_{\Lambda}$. Since $H$ is approximately constant for $a > a_{\Lambda}$ we will have that $H_{\Lambda} \equiv H_{0}$ where $H_{0}$ is the present value of the Hubble rate.
This means that the typical wavenumber which is of Hubble size today 
(i.e. $k_{0} = a_{0} H_{0}$) {\em is larger  than $k_{\Lambda}$} even if, according to a superficial 
intuition, it should be smaller. This aspect  is illustrated in Fig. \ref{SEC4FIG4} 
where it is more clear why $k_{eq} < k_{\Lambda} < k_{0}$; note, however, the the
numerical difference between $k_{0}$ and $k_{\Lambda}$ is rather insignificant since $k_{\Lambda} = (\Omega_{M0}/\Omega_{\Lambda})^{1/3} k_{0}$ 
and $(\Omega_{M0}/\Omega_{\Lambda})^{1/3} \simeq {\mathcal O}(0.7)$. For this reason $k_{0}$, $k_{\Lambda}$ and $k_{p}$ are all coinciding within one order of magnitude (note that $k_{0} \sim 8 k_{p} $).\subsubsection{Wavelengths that are currently inside the Hubble radius}
The wavenumbers falling in the interval $k_{0} < k < k_{eq}$ (and labeled by $k_{1}$ in Fig. \ref{SEC4FIG4}) correspond to wavelengths reentering the Hubble radius 
during the matter-dominated epoch and remaining inside the horizon later on.
In this case the power spectrum at the present epoch is given by 
$P_{T}(k,\, \tau_{re}) (a_{re}/a_{0})^2$
\begin{equation}
P_{T}(k, \, \tau_{0}) =P_{T}(k, \tau_{re})  \biggl(\frac{a_{re}}{a_{\Lambda}}\biggr)^2_{\mathrm{mat}} \, 
\biggl(\frac{a_{\Lambda}}{a_{0}}\biggr)^2_{\Lambda} \equiv P_{T}(k, \tau_{re}) 
\biggl(\frac{k_{\Lambda}}{k}\biggr)^{4} \, \biggl(\frac{\Omega_{M0}}{\Omega_{\Lambda}}\biggr)^{2/3},
 \qquad k_{0} < k < k_{eq}.
\label{LAM2}
\end{equation}
The result of Eq. (\ref{LAM2}) is obtained from the simple observation that when the modes are inside the Hubble radius the power spectrum is suppressed as $a^{-2}$ from the reentry time onwards. 
When computing the suppression we must then take into account that the scale factor behaves 
differently in the different epochs\footnote{This is, however, not strictly necessary since it is easy to find (both analytically and numerically) the appropriate interpolating form of the scale factor like, for instance 
$(a/a_{0}) = (\Omega_{M\,0}/\Omega_{\Lambda})^{1/3}  \bigl\{
\sinh{\bigl[3 \sqrt{\Omega_{\Lambda}} H_{0} (t -t_0)/\bigr]}\bigr\}^{2/3}$.
This solution interpolates between a matter-dominated Universe expanding in a decelerated way 
as $t^{2/3}$ and an exponentially expanding Universe which is also accelerating.}. From Eq. (\ref{LAM2})
we can also derive the spectral energy density when the relevant wavelengths are inside the Hubble radius and 
the result is 
\begin{eqnarray}
\Omega_{gw}(k,\tau_{0}) = \frac{P_{T}(k,\tau_{re})}{12} \biggl(\frac{k}{k_{0}}\biggr)^{-2} \biggl(\frac{k_{\Lambda}}{k_{0}}\biggr)^4 \biggl(\frac{\Omega_{M0}}{\Omega_{\Lambda}}\biggr)^{2/3} =  \frac{P_{T}(k,\tau_{re})}{12}\, \biggl(\frac{k}{k_{0}}\biggr)^{-2} \,  \biggl(\frac{\Omega_{M0}}{\Omega_{\Lambda}}\biggr)^{2}, \,\,\,\,\,k_{0} < k < k_{eq},
\label{LAM4}
\end{eqnarray}
where we used that, by definition, $k_{\Lambda}/k_{0} = (\Omega_{M0}/\Omega_{\Lambda})^{1/3}$. 
If the dominance of dark energy is completely neglected,  Eq. (\ref{LAM4}) can be written in the same 
form where, however, the term $(\Omega_{M0}/\Omega_{\Lambda})^2$ is absent. We conclude that, in this 
branch of the spectrum, the dominance of dark energy suppresses the spectrum by a factor 
$(\Omega_{M0}/\Omega_{\Lambda})^2 = (0.44)^2 = 0.193$. As anticipated this reduction of almost a factor 
of $5$ is more relevant qualitatively than quantitatively.

\subsubsection{Wavelengths that are currently outside the Hubble radius}
The wavenumbers falling instead in the interval $k_{\Lambda} < k < k_{0}$ (and labeled by 
$k_{2}$ in Fig. \ref{SEC4FIG4}) correspond to wavelengths reentering the Hubble radius during the matter-dominated 
epoch and exiting again the Hubble radius when dark energy is already dominant. The
same logic leading  to Eq. (\ref{LAM2}) implies that the power spectrum in the interval
$k_{\Lambda} < k < k_{0}$
is given by $ P_{T}(k,\, \tau_{re})\, (a_{re}/a_{ex})^2$ or, more precisely, 
\begin{equation}
P_{T}(k, \, \tau_{0}) = P_{T}(k, \tau_{re})  \biggl(\frac{a_{re}}{a_{\Lambda}}\biggr)^2_{\mathrm{mat}} \, 
\biggl(\frac{a_{\Lambda}}{a_{ex}}\biggr)^2_{\Lambda} \equiv P_{T}(k, \tau_{re}) 
\biggl(\frac{k_{\Lambda}}{k}\biggr)^{6},
 \qquad k_{\Lambda} < k < k_{0}.
\label{LAM5}
\end{equation}
Equation (\ref{LAM5}) gives the power spectrum for typical wavelengths that are larger 
than the Hubble radius at the present time. The wavelengths belonging to the interval 
$k_{\Lambda} < k < k_{0}$ reenter the Hubble radius but then exit again and while they are 
larger than the Hubble radius their power spectrum remains unchanged. This
is why, in Eq. (\ref{LAM5}), the power spectrum is not suppressed by a further factor 
$(a_{re}/a_{0})^2$: these wavelengths are larger than the Hubble radius at the present time 
and therefore the power spectrum remains constant exactly as in the case of the scales 
exiting the Hubble radius at the onset of inflation. Using the identity $k_{\Lambda}/k_{0} = (\Omega_{M0}/\Omega_{\Lambda})^{1/3}$ the spectrum of Eq. (\ref{LAM5}) can be written as:
\begin{equation}
P_{T}(k, \, \tau_{0}) = P_{T}(k, \tau_{re}) \biggl(\frac{k_{0}}{k}\biggr)^{6} \biggl(\frac{\Omega_{M0}}{\Omega_{\Lambda}}\biggr)^2.
\label{LAM6}
\end{equation}
We can finally compute the spectral energy density in this interval and we obtain 
\begin{eqnarray}
\Omega_{gw}(k,\tau_{0}) = \frac{P_{T}(k,\tau_{re})}{12}\, \biggl(\frac{k}{k_{0}}\biggr)^{-4} \,  \biggl(\frac{\Omega_{M0}}{\Omega_{\Lambda}}\biggr)^{2},  \qquad k_{\Lambda} < k < k_{0}.
\label{LAM7}
\end{eqnarray}
If the presence of dark energy would be neglected, $\Omega_{gw}(k,\tau_{0})$ would 
scale as $(k/k_{0})^{-2}$ and the suppression going as $(\Omega_{M0}/\Omega_{\Lambda})^{2}$ 
would be totally absent. Note finally that when $k = k_{\Lambda}$ Eq. (\ref{LAM6}) implies that 
$\Omega_{gw}(k_{\Lambda},\tau_{0})$ is only suppressed as $(\Omega_{M0}/\Omega_{\Lambda})^{4/3}$.
The analysis of the effects 
of dark energy on the relic graviton background has been discussed in Ref. \cite{COSMC1}
and subsequently improved (always with analytic methods) in Refs. \cite{COSMC3,COSMC4} (see also \cite{COSMC5}). The effect of the cosmological constant can also be assessed with fully numerical methods and this approach has been taken in Refs. \cite{TRANS8,TRANS9,ABSOLUTE}.
All in all the contribution of the dark energy to the graviton spectrum is conceptually 
important even if, from the purely quantitative viewpoint, the 
suppression of the spectral energy density roughly amounts to a 
factor of $2$. 

\subsection{The graviton spectra in the concordance paradigm}
\begin{figure}[!ht]
\centering
\includegraphics[height=6.3cm]{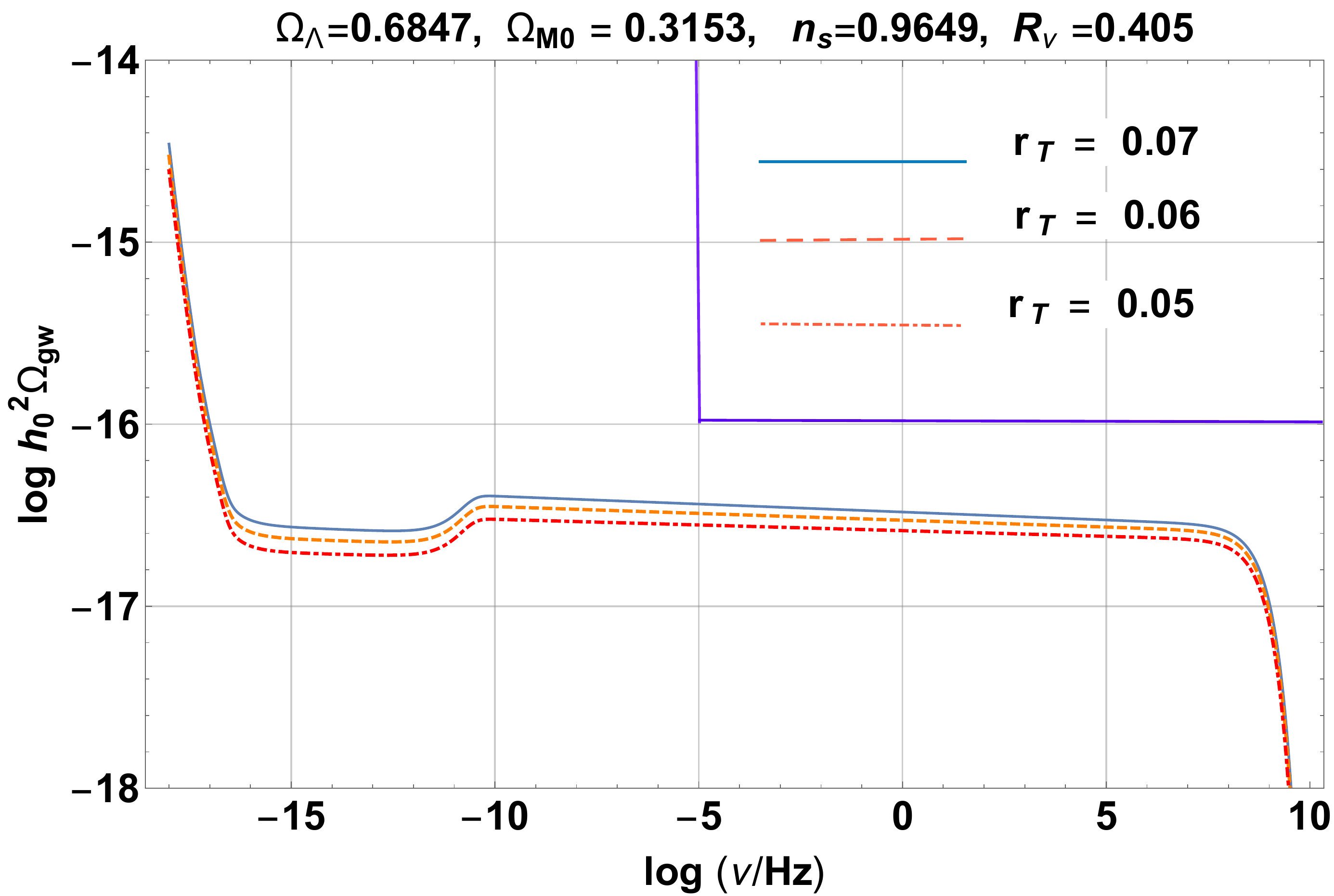}
\caption[a]{The spectral energy density of the relic gravitons is illustrated  in the context 
of the concordance scenario and for different values of $r_{T}$. The combined effect of the various sources 
of damping discussed in the second part of this section and the limits on $r_{T}$ imply that 
in the region of operation of ground-based interferometers (like LIGO and Virgo \cite{LV1,LV2})
the signal of the concordance paradigm should be slightly smaller than ${\mathcal O}(10^{-16.5})$.
The flatness of the spectral slope implies that the same figure 
holds also in the case of space-borne interferometers. A more complete discussion can be found 
in section \ref{sec8}.}
\label{SEC4FIG5}      
\end{figure}

\begin{figure}[!ht]
\centering
\includegraphics[height=6.cm]{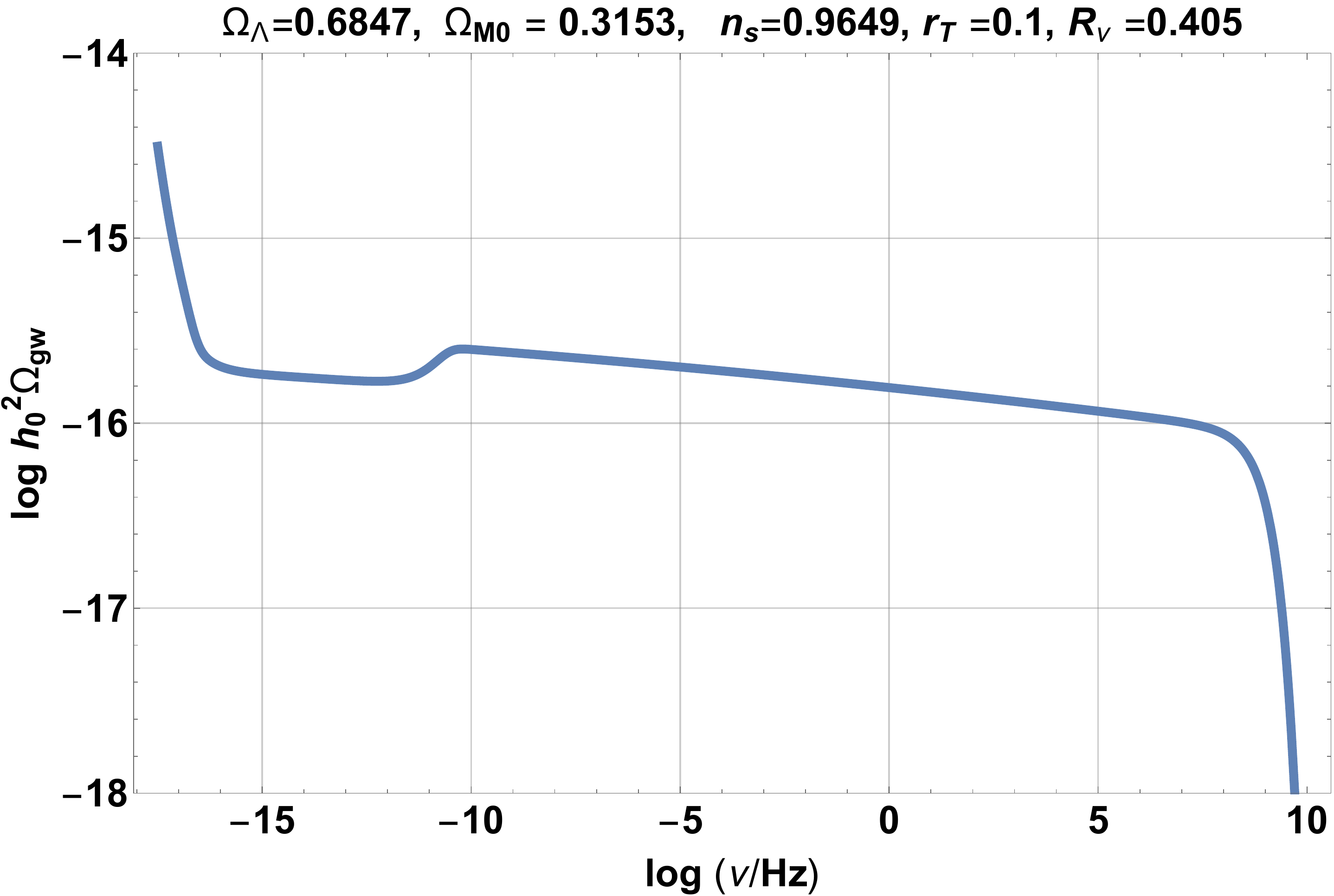}
\caption[a]{We illustrate the possible running of the tensor spectral index 
and purposely assume an excessive value of $r_{T}$ to make the overall effect more dramatic.}
\label{SEC4FIG6}      
\end{figure}

 In Fig. \ref{SEC4FIG5} the spectral energy density of the relic gravitons is illustrated in the context 
 of the concordance paradigm; the cosmological parameters have been selected 
 as in Eq. (\ref{PARSET}) with the difference that the values of $r_{T}$ 
 appearing in Fig. \ref{SEC4FIG5} now vary between $0.05$ and $0.07$.
On the vertical axis the common logarithm of the spectral energy density is illustrated; on the horizontal axis 
we report the common logarithm of the comoving frequency (recall that $\nu = 2 \, \pi\, k$ and that, today, 
comoving and physical frequencies coincide since the scale factor is normalized to $1$ at the present epoch). 
The spectral energy density illustrated in Fig. \ref{SEC4FIG5} encompasses all wavelengths 
that reentered throughout the radiation- and dust-dominated phases; in particular 
the low-frequency branch decreasing as $\nu^{-2}$ and extending approximately between $\nu_{p}$ and $\nu_{eq}$ corresponds to the modes that left the Hubble radius during inflation and reentered during the matter-dominated phase. 
The quasi-flat slope in Fig.  \ref{SEC4FIG5} (between $100$ aHz and $200$ MHz corresponds to the 
modes that exited the Hubble radius during inflation and reentered during the radiation-dominated epoch.
 The decreasing branch of the spectrum (see e.g. Eq. (\ref{TFF2e})) and the quasi-flat 
plateau can be estimated with analytical methods with different levels of accuracy (see e.g. Eqs. (\ref{TRAD10a}), (\ref{ANIS3}) and (\ref{EFF1a})); however the numerical discussion (preferentially based on the transfer function of the spectral energy density, see Eq. (\ref{TFF14}) and discussion thereafter) is essential in the transition regimes and it is 
anyway preferable in comparison with the approximation by straight lines that was 
somehow popular in the past and sometimes used even today. 
 The intermediate suppression in the flat plateau of Fig. \ref{SEC4FIG5} 
is due to the neutrino free-streaming and it approximately occurs for frequencies  $\nu < \nu_{bbn}$ 
since neutrinos decouple for plasma temperatures shorter than the MeV. There are situations where the tensor spectra index does not evolve like $-2\epsilon$ but it has a more complicated dependence and in these instances it said to be ``running'' (see e.g. \cite{COSMC4,COSMC5,COSMC6}). The case 
of running is only relevant as long as $r_{T}$ is sufficiently large.
In Fig. \ref{SEC4FIG6} we also illustrate the case where the tensor spectral index is not only given by 
$ - 2 \epsilon$ (as in the case of Fig. \ref{SEC4FIG5}) but it runs according to the following expression:
$n_{T}(k) =  - 2 \epsilon + (\alpha_{T}/2)\ln{(k/k_{p})}$ where $\alpha_{T} = (r_{T}/8)[(n_{s} -1) + (r_{T}/8)]$ 
now measures the running of the tensor spectral index and it can be estimated by going to second-order in the slow-roll
expansion.
The damping effects estimated in this section  have been included in Figs. \ref{SEC4FIG5} and \ref{SEC4FIG6}.
All the damping effects examined in this sectiondecrease the amplitude of the quasi-flat plateau which is the characteristic feature 
of the spectral energy density computed within the concordance paradigm. Each of the damping effects suppresses $h_{0}^2 \,\Omega_{gw}(\nu,\tau_{0})$ by less than one order of magnitude. The net result of the combined effects is however a suppression of $h_{0}^2 \Omega_{gw}(\nu,\tau_{0})$ which ranges between $10^{-2}$ and $10^{-3}$ (for $100 \,\,\mathrm{aHz} < \nu <\,0.01 \, \mathrm{nHz}$). Since all the phenomena discussed in this section diminish the spectral energy density, it is plausible to conclude that, within the concordance scenario, the spectral energy density above the $mHz$ is at most $h_{0}^2 \Omega_{gw}(\nu, \tau_{0}) = {\mathcal O}(10^{-16.5})$, given the present bounds on $r_{T}$. The various frequency bands of the relic gravitons will be specifically scrutinized in the forthcoming sections by always bearing in mind the reference signal of the concordance paradigm and its indetermination.

\renewcommand{\theequation}{5.\arabic{equation}}
\setcounter{equation}{0}
\section{The low-frequency band}
\label{sec5}
The temperature and the polarization anisotropies of the CMB are directly affected by gravitons with 
typical frequencies ranging between the aHz and the fHz. 
The initial conditions of the Einstein-Boltzmann hierarchy are assigned 
over the wavelengths (labeled by $2$ in Fig. \ref{SEC4FIG1}) that are about to reenter around $a_{eq}$ but are still larger than the Hubble radius prior to matter-radiation equality. Both the scalar and the tensor modes of the geometry  induce temperature and polarization anisotropies with the difference that the tensor 
modes may also generate a $B$-mode polarization that is absent in the scalar 
case\footnote{A $B$-mode polarization is also induced through the lensing of the primary anisotropies; this secondary $B$-mode polarization has been already detected, as we shall briefly discuss hereunder.}. 

\subsection{Temperature and polarization anisotropies}
The temperature and the polarization anisotropies of the CMB 
are caused by the curvature inhomogeneities which are Gaussian and 
(at least predominantly) adiabatic. This evidence has been established first 
by the WMAP collaboration \cite{WMAP1,WMAP1a,WMAP1b}  all the 
subsequent observations confirmed and partially refined the results of Refs. \cite{WMAP1,WMAP1a,WMAP1b}. The initial conditions for the  temperature and polarization anisotropies  
have been originally scrutinized by Peebles \cite{SAK2,pee1,pee2}, Silk \cite{silk},  
Harrison \cite{harrisons}, Novikov and Zeldovich \cite{zeldovichnovikov}.  
Around the same time it was also realized that the repeated Thomson scattering 
of the primeval radiation during the early phases of an anisotropic Universe would modify 
the black-body spectrum and produce linear polarization \cite{rees}. Even if the polarization 
anisotropies have an entirely different angular power spectrum from the temperature fluctuations, their 
initial conditions are common and they come, predominantly, from the large-scale inhomogeneities 
in the spatial curvature.  

The initial data of the Einstein-Boltzmann hierarchy can be either adiabatic or non-adiabatic \cite{book1,book2,book3}. In the concordance paradigm, when the dark energy does not fluctuate, there are, overall one adiabatic \cite{SAK2,SAK3,bertschingerma,ad1a} and four non-adiabatic \cite{nad1,nad1a,nad2,nad2a,nad3,nad3a} Cauchy data\footnote{The field equations for cosmological perturbations in the Newtonian gauge always have an adiabatic solution, for which ${\mathcal R}$ is nonzero and constant in the limit of large wavelength \cite{ad1a}. The four non-adiabatic modes are the CDM radiation mode, the baryon-entropy mode, the neutrino entropy mode and the neutrino isocurvature 
velocity mode.  The solutions of the Einstein-Boltzmann hierarchy must be regular. If they are 
divergent they may be unphysical unless they can be smoothly described in at least 
one gauge.}. It is also possible to consider mixed sets of initial conditions where non-adiabatic modes are correlated (or anticorrelated) with the adiabatic initial condition. The current data suggest that the adiabatic and Gaussian data dominate the space of the initial conditions and the non-adiabatic modes, if present, must be subleading. 
\begin{figure}[!ht]
\centering
\includegraphics[height=6.4cm]{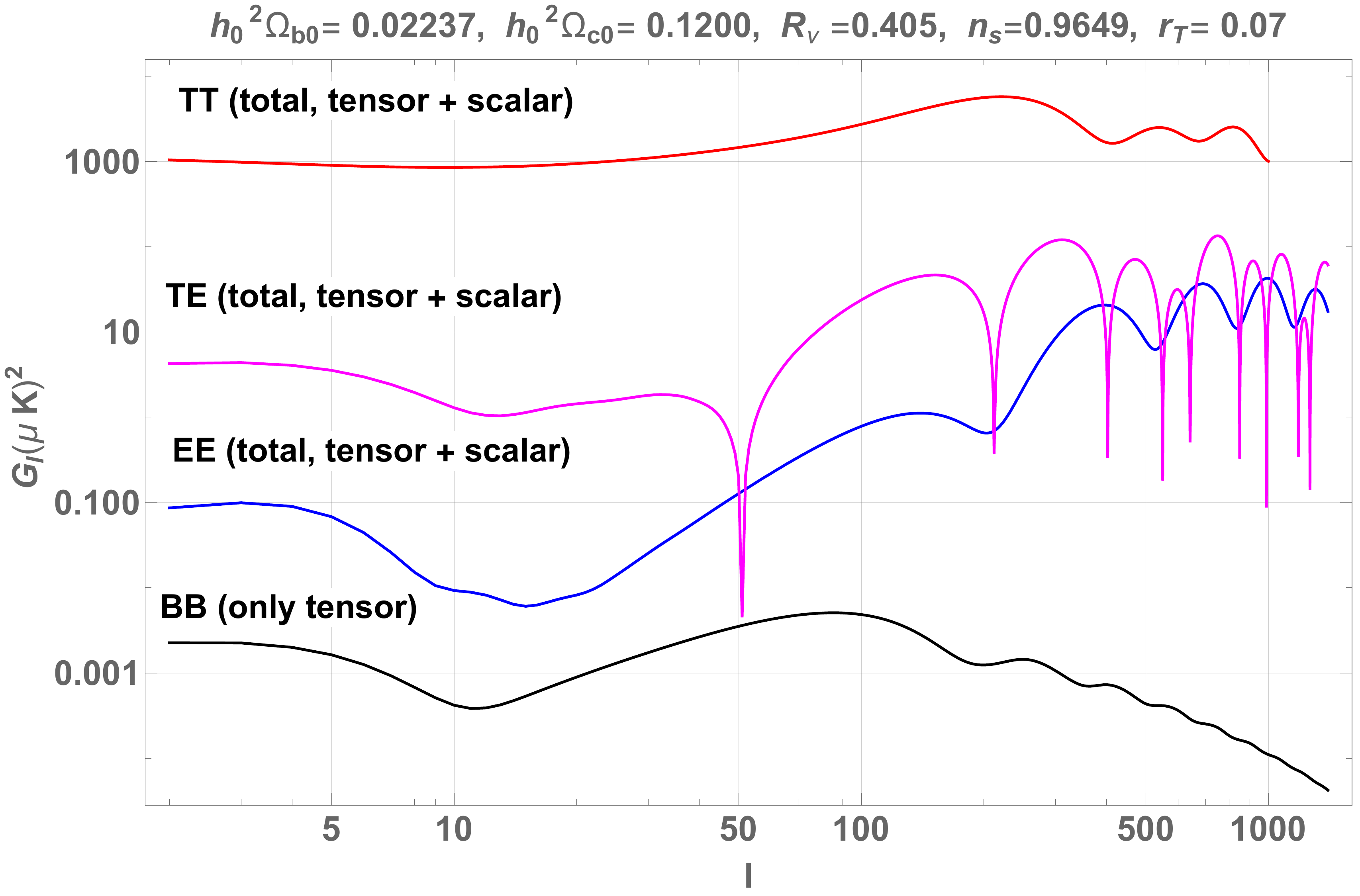}
\caption[a]{We illustrate the $TT$, $EE$ and $TE$ correlations given by the sum of the scalar 
component and of the tensor contribution. For comparison we also illustrate the $BB$ correlations 
coming from the relic gravitons.}
\label{SEC5FIG1}      
\end{figure}
The scalar brightness perturbations $\Delta^{(s)}_{\mathrm{I}}(\hat{n},\tau)$ are customarily expanded in ordinary spherical harmonics so that the angular power spectrum of the temperature autocorrelations is defined as:
\begin{equation}
 C_{\ell}^{(TT)} = \frac{1}{2\ell + 1} \sum_{m} \langle 
 a^{(T)*}_{\ell\,m} a^{(T)}_{\ell\,m}\rangle,\qquad 
 a^{(T)}_{\ell m} 
=  \int d \hat{n}\, Y_{\ell m}^{*}(\hat{n}) \,\Delta_{I}(\hat{n}, \tau),
\label{E1a}
\end{equation}
where $Y_{\ell m}(\hat{n})$ are the (scalar) spherical harmonics. While the intensity fluctuations 
do not change under a rotation in the plane orthogonal to the direction of propagation 
of the photon (i.e. $\hat{n}$),  the two orthogonal combinations of the brightness perturbations related to the 
$U$ and $Q$ Stokes parameters [i.e. $\Delta^{(s)}_{\pm}(\hat{n},\tau) = \Delta^{(s)}_{Q}(\hat{n},\tau) \pm i \Delta^{(s)}_{U}(\hat{n},\tau)$] transform as fluctuations of spin-weight $\pm 2$ respectively.
Owing to this observation,  $\Delta^{(s)}_{\pm}(\hat{n},\tau)$ can be expanded in terms of spin-$\pm2$ spherical harmonics \cite{EB1,EB2} $_{\pm 2}Y_{\ell\,m}(\hat{n})$, i.e. 
\begin{equation}
\Delta^{(s)}_{\pm}(\hat{n},\tau) = \sum_{\ell \, m} a_{\pm2,\,\ell\, m} \, _{\pm 2}Y_{\ell\, m}(\hat{n}),
\label{int2aa}
\end{equation}
where the superscript in Eq. (\ref{int2a}) reminds that we are discussing the dominant 
contribution to the brightness perturbations coming from the scalar curvature inhomogeneities. The 
spin $\pm2$ spherical harmonics already appeared in the discussion of the polarization 
of the gravitons (see e.g. Eq. (\ref{SIXPOL10a}) and discussion thereunder) and will again appear when addressing the brightness perturbations induced by the tensor modes.
The so-called $E$- and $B$-modes coming form the curvature inhomogeneities 
are, up to a sign, the real and the imaginary 
parts of $a_{\pm 2,\ell\,m}$, i.e. 
\begin{equation}
a^{(E)}_{\ell\, m} = - \frac{1}{2}(a_{2,\,\ell m} + a_{-2,\,\ell m}), \qquad  
a^{(B)}_{\ell\, m} =  \frac{i}{2} (a_{2,\,\ell m} - a_{-2,\,\ell m}).
\label{int3aa}
\end{equation}
In real space (as opposed to Fourier space), the fluctuations constructed from 
$a^{(E)}_{\ell\,m}$ and $a^{(B)}_{\ell\,m}$ have the 
property of being {\em invariant under rotations on a plane orthogonal 
to $\hat{n}$}.  They can therefore be expanded in terms of (ordinary) spherical harmonics:
\begin{equation}
\Delta^{(s)}_{E}(\hat{n},\tau) = \sum_{\ell\, m} N_{\ell}^{-1} \,  a^{(E)}_{\ell\, m}  \, Y_{\ell\, m}(\hat{n}),\qquad 
\Delta^{(s)}_{B}(\hat{n},\tau) = \sum_{\ell\, m} N_{\ell}^{-1} \,  a^{(B)}_{\ell\, m}  \, Y_{\ell\, m}(\hat{n}),
\label{int4aa}
\end{equation}
where $N_{\ell} = \sqrt{(\ell - 2)!/(\ell +2)!}$.  The $EE$ and $BB$ and $TE$ angular power spectra are then defined as: 
\begin{equation}
C_{\ell}^{(XX)} = \frac{1}{2\ell + 1} \sum_{m = -\ell}^{\ell} 
\langle a^{(X)*}_{\ell m}\,a^{(X)}_{\ell m}\rangle,\qquad
C_{\ell}^{(TE)} = \frac{1}{2\ell + 1} \sum_{m=-\ell}^{\ell} 
\langle a^{(T)*}_{\ell m}\,a^{(E)}_{\ell m}\rangle,
\label{int5aa}
\end{equation}
where $X=E,\,B$. In the minimal version of the $\Lambda$CDM paradigm where the tensor modes 
are absent, the adiabatic fluctuations of the scalar curvature lead to the condition $a_{2,\,\ell m} = a_{-2,\,\ell m}$ that implies $a_{\ell m}^{(B)} =0$ from Eq. (\ref{int3aa}).  The normalized temperature and polarization anisotropies are customarily measured in 
units of $(\mu \mathrm{K})^2$ and they can be written, with shorthand notation, as: 
\begin{equation}
 G^{(TT)}_{\ell} =  \frac{\ell (\ell  +1)}{2 \pi} C_{\ell}^{(TT)}, \qquad 
 G^{(EE)}_{\ell} =  \frac{\ell (\ell  +1)}{2 \pi} C_{\ell}^{(EE)},\qquad  G^{(TE)}_{\ell} =  \frac{\ell (\ell  +1)}{2 \pi} C_{\ell}^{(TE)}.
\label{int7aa}
\end{equation}
With the prefactor $\ell(\ell+1)/(2\pi)$, the various $G_{\ell}$ measure the power per logarithmic interval of $\ell$ of the corresponding correlation function in $\ell$-space. In Fig. \ref{SEC5FIG1} the observables defined in Eq. (\ref{int7aa}) are illustrated for the fiducial set of cosmological parameters of Eq. (\ref{PARSET}); common logarithms are used both on the vertical and on the horizontal axis.  The $TT$ correlations  are characterized by a quasi-flat 
plateau (for $\ell < 200$) and by a sequence of three peaks: the first (Doppler) peak located for 
$\ell_{d} = {\mathcal O}(220)$ (where approximately $G_{\ell_{d}}^{(TT)} = {\mathcal O}(5700)\, \mu\, \mathrm{K}^2$ and two further acoustic peaks for $\ell_{2} = {\mathcal O}(535)$ and $\ell_{3} = {\mathcal O}(820)$. The polarization autocorrelations (in the jargon $EE$ correlations) approximately reach their absolute maximum  for $\ell_{max} = {\mathcal O}(1000)$ where $G_{\ell_{max}}^{(EE)} = {\mathcal O}(50)\,\mu\, \mathrm{K}^2$.  For adiabatic initial conditions the $G_{\ell}^{(TT)}$ exhibits 
the first acoustic peak for $\ell_{\mathrm{d}}\simeq 220$. The first (anticorrelation) peak of $G_{\ell}^{(TE)}$ occurs instead for $\ell_{\mathrm{ac}} \simeq 3 \,\ell_{\rm d}/4 < \ell_{\mathrm{d}} \simeq 150$.  Since 
 $G_{\ell}^{(TE)}$ is not positive definite, in Fig. \ref{SEC5FIG1}  the common logarithm of $|G_{\ell}^{(TE)}|$ has been illustrated: the cusps correspond then to the zeros of the $TE$ correlations. The first data release of the WMAP experiment convincingly demonstrated that the relative hierarchy between $\ell_{\mathrm{d}}$ and 
$\ell_{\mathrm{ac}}$ is the best evidence of the adiabatic nature of large-scale curvature perturbations\footnote{The large-scale contribution to the $TT$ correlation goes, in Fourier space, as a cosine \cite{WMAP1,WMAP1a,WMAP1b,hus} (see also \cite{book1,book2,book3}). For the same set of adiabatic initial conditions the $TE$ correlation oscillates, in Fourier space, as a sine. The first (compressional) peak of the temperature autocorrelation corresponds to $\ell_{d}$ while  the first peak of the cross-correlation will arise for $\ell_{ac} \simeq 3\, \ell_{d}/4$. These analytic results can be obtained by working to first-order in the tight-coupling expansion \cite{WMAP1,WMAP1a,WMAP1b,SAK2,hus,TC2,TC3}.}.
\begin{figure}[!ht]
\centering
\includegraphics[height=6.4cm]{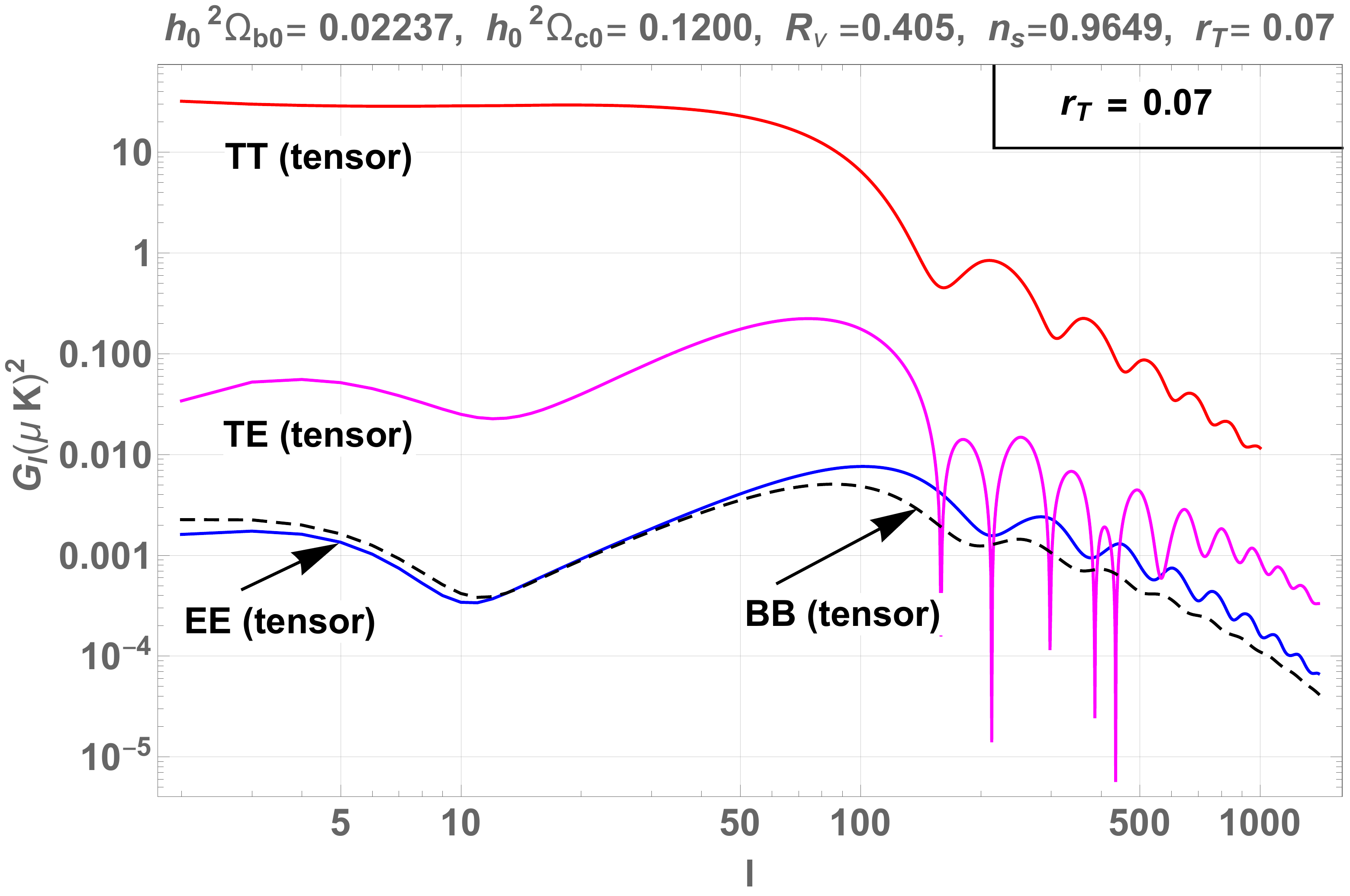}
\caption[a]{We illustrate the $TT$, $EE$, $BB$ and $TE$ correlations coming solely 
from the tensor modes of the geometry for a realistic value of the tensor-to-to scalar 
ratio $r_{T}$.}
\label{SEC5FIG2}      
\end{figure}
\subsection{The effects of the relic gravitons}
\begin{figure}[!ht]
\centering
\includegraphics[height=6.4cm]{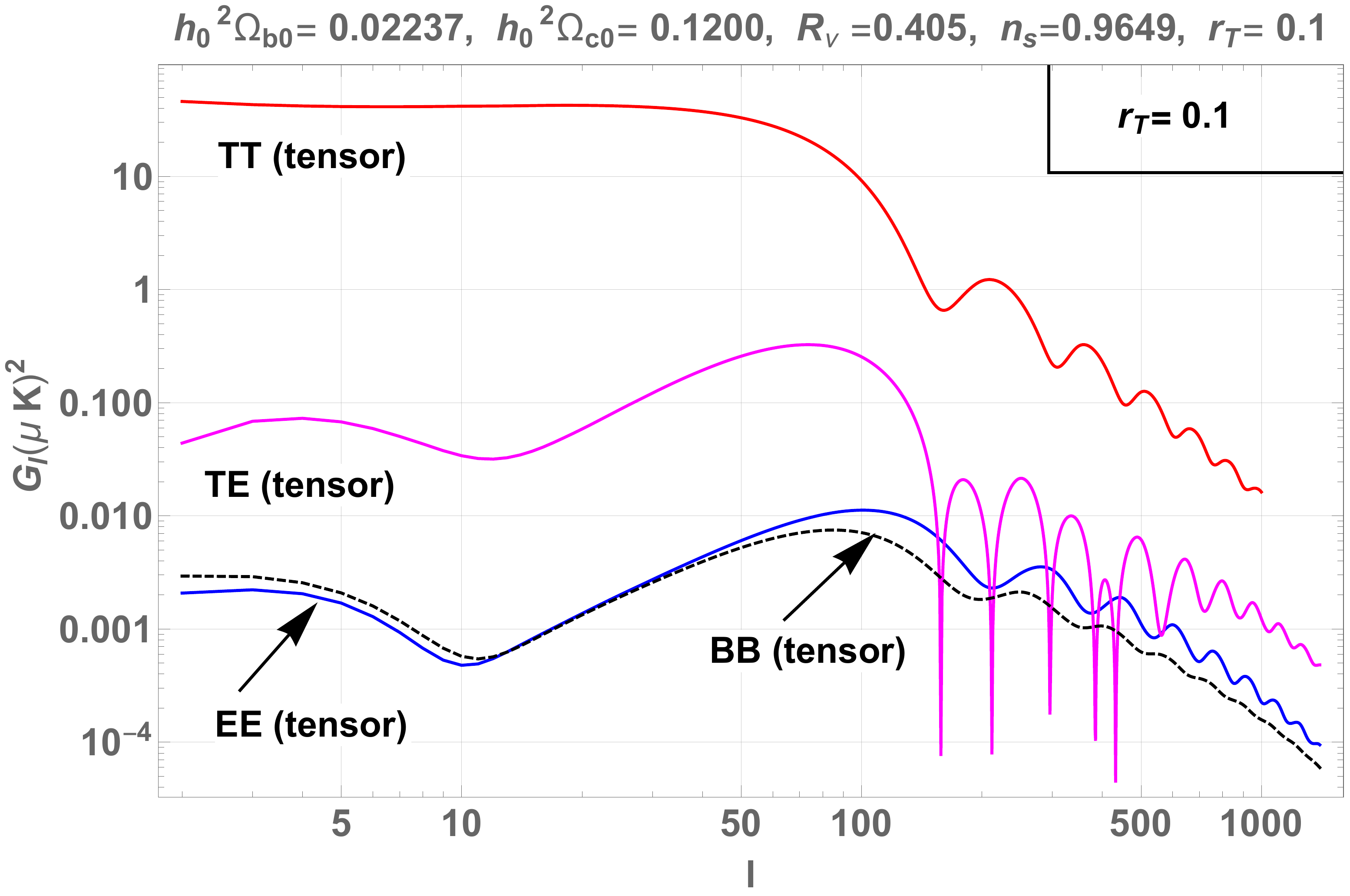}
\caption[a]{The $TT$, $EE$, $BB$ and $TE$ correlations coming 
from the tensor modes of the geometry are reported in the same plot;  a purposely exaggerated value of 
$r_{T}$ has  been used. This value is incidentally {\em not compatible} with the current 
observational data but is anyway quite useful for illustrative purposes.}
\label{SEC5FIG3}      
\end{figure}

In the concordance paradigm the tensor modes a $B$-mode autocorrelation (see last curve at the bottom in Fig. \ref{SEC5FIG1}) but also $TT$, $TE$ and $EE$ correlations  \cite{HIS11,HIS12,HIS12a,HIS12aa}
with signatures that are rather different from the ones of the scalar modes.
The tensor contributions in the case $r_{T} =0.07$ are separately illustrated in Fig. \ref{SEC5FIG2}.
In Fig. \ref{SEC5FIG3} the angular power spectra induced by the tensor mode have been reported
in the (unrealistic) case $r_{T} =0.1$ which is already excluded by the current observational limits. 
The influence of the gravitational waves of the heat transfer equations have been first discussed in Ref. \cite{HIS12a} (see also \cite{HIS12aa,GWHT1}) and subsequently studied by many  authors \cite{GWHT2,GWHT3,GWHT3a,GWHT4,GWHT5,GWHT5a}. The solution of the heat transfer equations in the presence of the tensor modes of the geometry can be obtained by using the integration along the line of sight in full analogy with what is customarily done in the scalar case \cite{GWHT6,GWHT7}. The 
temperature and the polarization anisotropies induced by the tensor modes are estimated in most of the current Boltzmann integrators starting with CMBFAST \cite{bertschingerma,GWHT7a,GWHT7c}. There have been various studies aiming at a more complete analytical understanding of the polarization anisotropies induced by gravitational waves \cite{GWHT8,GWHT9,GWHT10}. These attempts provide a physical insight into the origins of the various features appearing in the angular power spectra especially in the limit of large multipoles \cite{GWHT11} where the asymptotic approximations to the spherical Bessel functions are more accurate \cite{abr1,abr2} (see also, in this respect, some general discussion \cite{book1,book2,book3} on the analytical methods employed in the discussion of CMB observables).
The observables illustrated in Figs. \ref{SEC5FIG2} and \ref{SEC5FIG3} are not affected only by low-frequency gravitons but, at least in principle, by the whole spectrum. However, the typical damping scale of the tensor contribution occurs for $\ell = {\mathcal O}(500)$ and this means that the corresponding angular power spectra 
are practically unaffected by frequencies of the gravitons much larger than $100$ aHz. 
A $B$-mode polarization is also induced through the lensing of the primary anisotropies; this secondary $B$-mode polarization has been already detected by the South Pole Telescope \cite{chone42}. The $B$-mode polarization induced by the lensing of the CMB anisotropies is, however, qualitatively different from the $B$-mode induced by the tensor modes of the geometry.
The Bicep2 experiment \cite{chone43} reported the detection of a primordial $B$-mode component compatible with a tensor to scalar ratio $r_{T}=0.2^{+0.07}_{-0.05}$.  Unfortunately the signal turned out to be affected by  serious contaminations 
of a polarized foreground. In the presence of magnetic random fields there are further sources of $B$-mode polarization that have however a rather different signature from the relic gravitons and are mainly 
induced by the Faraday effect \cite{far1,far2,far3,far4,far5,far6}.

 \subsection{The scalar and tensor Einstein-Boltzmann hierarchies}
The coupled evolution of the background, of its fluctuations and of the various Boltzmann 
equations for the scalar and tensor modes of the geometry is often referred to as the {\em Einstein-Boltzmann 
hierarchy}. The scalar and tensor fluctuations of the geometry both contribute to the total brightness perturbation so that, as already discussed in connection with Fig. \ref{SEC5FIG1},
\begin{equation}
 \Delta_{X}(\vec{x},\tau) = \Delta^{(s)}_{X}(\vec{x},\tau) + \Delta^{(t)}_{X}(\vec{x},\tau),
\label{BRDEC1}
\end{equation}
where the subscript $X=I,\,Q,\,U,\,V$ denotes, generically, one of the four Stokes parameters while the superscripts refer, respectively, to the scalar and tensor contributions. The relation between the fluctuation 
of the intensity (that depends on $q$) and the brightness perturbation (that does not depend on $q$) can be written as:
\begin{equation}
I(\vec{x},\tau, q, \hat{n})= f_{0}(q) \biggl[ 1 - \frac{\partial \ln{f_{0}}}{\partial \ln{q}} \Delta_{\mathrm{I}}(\vec{x},\tau, \hat{n})\biggr],
\label{add1}
\end{equation}
where $f_{0}(q)$ is the (unperturbed) Bose-Einstein distribution, $q$ is the modulus of the comoving three-momentum {\em of the photon} and $\hat{n}$ denotes its direction; the same notations hold for the remaining Stokes parameters so that the four brightness perturbations can either be organized in a $2\times 2$ matrix (as in Jones calculus) or in a column vector with four entries (as in the case of Mueller calculus) \cite{EFPOL2}. 
In analogy with the case of the gravitons of Eq. (\ref{SIXPOL11}), the polarization tensor of the electromagnetic field $ {\mathcal P}_{ij}=  {\mathcal P}_{ji}= E_{i}\, E_{j}^{*}$ corresponds to a $2\times2$ 
matrix whose explicit form is:
\begin{equation}
{\mathcal P}= \frac{1}{2} \left(\matrix{ \Delta_{I} + \Delta_{Q}
& \Delta_{U} - i \Delta_{V} &\cr
\Delta_{U} + i \Delta_{V} & \Delta_{I} - \Delta_{Q} &\cr}\right) = \frac{1}{2} \left( \Delta_{I}\,{\bf 1} + \Delta_{U}\, \sigma_{1} + \Delta_{V}\, \sigma_2 +\Delta_{Q}\, \sigma_3 \right),
\label{PM}
\end{equation}
where  ${\bf 1}$ denotes the identity matrix while $\sigma_{1}$, $\sigma_{2}$ and 
$\sigma_{3}$ are the three Pauli matrices. The matrix (\ref{PM}) depends on $4$ angles: 
{\it i)} $\Omega= (\vartheta,\varphi)$ defines the directions of the scattered photons, 
{\it ii)} $\Omega^{\prime} = (\vartheta',\varphi')$ accounts for the directions of the incident 
photons\footnote{
Hereunder $\Omega$ and $\Omega^{\prime}$ will denote hereunder the angular variables.  They must not be confused with the energy density in critical units. Since the two variables are newer used in the 
same context no confusion is possible. The radial, azimuthal and polar directions of the scattered radiation will be defined as $\hat{r} = (\cos{\varphi} \sin{\vartheta},\, \sin{\varphi} \sin{\vartheta},\, \cos{\vartheta})$, $\hat{\vartheta} = (\cos{\varphi} \cos{\vartheta},\, \sin{\varphi} \cos{\vartheta},\, -\sin{\vartheta})$ and  $\hat{\varphi} = ( - \sin{\varphi},\, \cos{\varphi},\, 0)$ respectively. As it can be checked the orientation of the unit vectors is such that $\hat{r} \times \hat{\vartheta} = \hat{\varphi}$.}.  The evolution of the matrix ${\mathcal P}$ can be formally written as
\begin{equation}
 \frac{d {\mathcal P}}{d\tau} + \epsilon' {\mathcal P} =  \frac{3 \epsilon^{\prime}}{8 \pi} \int\, d\Omega^{\prime}\,  M(\Omega,\Omega')\, {\mathcal P}(\Omega') \,M^{\dagger}(\Omega,\Omega'),
\label{EV1}
\end{equation}
where $d\Omega' = \sin{\vartheta^{\prime}}\, d\vartheta^{\prime} \,d\varphi^{\prime}$  and the dagger in Eq. (\ref{EV1}) defines, as usual, the complex conjugate of the transposed matrix. Defining the rate of electron-photon scattering $\Gamma_{\gamma\mathrm{e}}$ we have that the differential optical depth, 
the Thomson cross section and the classical radius of the electron will be denoted by:
\begin{equation}
\epsilon' =a \,\Gamma_{\gamma\mathrm{e}} = a \widetilde{n}_{\mathrm{0}} x_{\mathrm{e}} \sigma_{\mathrm{e}\gamma}, \qquad \sigma_{\gamma\mathrm{e}} = \frac{8}{3} \pi r_{\mathrm{e}}^2, \qquad r_{\mathrm{e}}= \frac{e^2}{m_{\mathrm{e}}}.
\label{diffop}
\end{equation}
Neglecting the contribution of the ions in the differential optical depth (because 
of the mass difference between electrons and ions) the scattered electric fields are computed by superimposing the contribution of the electrons and of the ions:
\begin{equation} 
\vec{E}^{(\mathrm{out})}_{(\mathrm{e})} = - e \frac{\vec{r} \times [\vec{r} \times \vec{a}_{(\mathrm{e})}]}{r^3},\qquad 
\vec{E}^{\mathrm{out}}_{(\mathrm{i})} =  e \frac{\vec{r} \times[\vec{r} \times \vec{a}_{(\mathrm{i})}]}{r^3},
\label{sc1}
\end{equation}
where $\vec{a}_{(\mathrm{e})}$ and $\vec{a}_{(\mathrm{i})}$ are the corresponding accelerations. 
Thus the outgoing electric fields are:
\begin{eqnarray}
\vec{E}^{(\mathrm{out})} &=& \vec{E}^{(\mathrm{out})}_{(\mathrm{e})}  + \vec{E}^{\mathrm{out}}_{(\mathrm{i})}=- e \frac{\vec{r} \times [\vec{r} \times \vec{A}]}{r^3} \equiv - \frac{e}{r} \biggl[ \hat{r} (\vec{A}\cdot\hat{r}) - \vec{A} \biggr],
\nonumber\\
\vec{A} &=& \vec{a}_{(\mathrm{e})}- \vec{a}_{(\mathrm{i})} = - \frac{e}{m_{e}} \biggl( 1 + \frac{m_{e}}{m_{i}}\biggr) \, \vec{E}^{(\mathrm{in})} \simeq - \frac{e}{m_{e}} \, \vec{E}^{(\mathrm{in})},
\label{sc1a}
\end{eqnarray}
where the masses of the ions have been consistently neglected in  the acceleration three-vector $\vec{A}$.
With these specifications the components of the outgoing electric fields become:
\begin{eqnarray}
&& E^{(\mathrm{out})}_{\vartheta}(\Omega, \Omega^{\prime}) = \frac{r_{\mathrm{e}}}{r} \biggl[ M_{\vartheta\vartheta}(\Omega, \Omega^{\prime},)  E^{(\mathrm{in})}_{\vartheta}(\Omega^{\prime}) + 
M_{\vartheta\varphi}(\Omega, \Omega^{\prime}) E^{(\mathrm{in})}_{\varphi}(\Omega^{\prime})  \biggr],
\nonumber\\
&& E^{(\mathrm{out})}_{\varphi}(\Omega, \Omega^{\prime})  = \frac{r_{\mathrm{e}}}{r} \biggl[ M_{\varphi\vartheta}(\Omega, \Omega^{\prime}) E^{(\mathrm{in})}_{\vartheta}(\Omega^{\prime}) + M_{\varphi\varphi}(\Omega, \Omega^{\prime}) 
E^{(\mathrm{in})}_{\varphi}(\Omega^{\prime})  \biggr],
\label{electric}
\end{eqnarray}
and $M_{ij}(\Omega,\Omega^{\prime})$ denote the components of the $2\times2$ matrix $M(\Omega,\Omega^{\prime})$
\begin{equation}
M(\Omega,\Omega^{\prime}) = \left(\matrix{ M_{\vartheta\vartheta}(\Omega,\Omega^{\prime})
&M_{\vartheta\varphi}(\Omega,\Omega^{\prime}) &\cr
M_{\varphi\vartheta}(\Omega,\Omega^{\prime}) &M_{\varphi\varphi}(\Omega,\Omega^{\prime})&\cr}\right).
\label{LIMM}
\end{equation}
The explicit expressions of the four entries of  $M(\Omega,\Omega^{\prime})$ are:
\begin{eqnarray}
&& M_{\vartheta\vartheta}(\mu,\varphi,\mu^{\prime},\varphi^{\prime}) = - \sqrt{ 1 - \mu^2} \sqrt{1 - \mu^{\,\prime\,2} }- \mu \nu \cos{(\varphi'- \varphi)},
\nonumber\\
&& M_{\varphi\varphi}(\mu,\varphi,\mu^{\prime},\varphi^{\prime})  = - \cos{(\varphi' - \varphi)},\qquad  M_{\varphi\vartheta}(\mu,\varphi,\mu^{\prime},\varphi^{\prime})  = - \mu^{\,\prime} \sin{(\varphi' - \varphi)},
\nonumber\\
&& M_{\vartheta\varphi}(\mu,\varphi,\mu^{\prime},\varphi^{\prime}) = \mu \sin{(\varphi' - \varphi)}, 
\label{LIM2}
\end{eqnarray}
where $\mu = \cos{\vartheta}$ and $\mu^{\,\prime} = \cos{\vartheta'}$. Equation (\ref{LIM2}), modulo the different 
conventions, coincides exactly with the standard results  (see e.g. \cite{chandra,pera}). From Eqs. (\ref{PM}) and (\ref{EV1}), using the explicit form of the matrix elements 
of Eq. (\ref{LIM2}), it is possible to express the evolution in terms of a new column matrix (be it $\Delta$) 
whose components are the four Stokes parameters (i.e. respectively, $\Delta_{I}$, $\Delta_{Q}$, $\Delta_{U}$ and $\Delta_{V}$) \cite{EFPOL2}:
 \begin{equation}
 \frac{d \Delta}{d\tau} + \epsilon' \Delta = \frac{3 \epsilon^{\prime}}{16\pi} \int d\Omega' \,\, {\mathcal N}(\Omega, \Omega') \Delta(\Omega'),
 \label{EV2}
 \end{equation}
where now ${\mathcal N}(\Omega,\Omega')$ is a  $4\times4$ matrix. For a general radiation field, 
of the $16$ entries of the matrix ${\mathcal N}(\Omega,\Omega')$
only $10$ are non-vanishing. Furthermore one component involving the circular polarization is purely 
kinematical and therefore the matrix ${\mathcal N}(\Omega,\Omega')$  is effectively a $3\times3$ matrix
only involving the intensity and the linear polarizations. 

\subsection{The evolution of the temperature and of the polarization anisotropies} 
The collision term relevant to the temperature anisotropies is specified by the first line of the matrix 
${\mathcal N}(\Omega, \Omega^{\prime})$ and it can be explicitly written as:
\begin{equation}
\frac{d \Delta_{I}}{d \tau} + \epsilon^{\, \prime} \Delta_{I} = \frac{3 \epsilon^{\,\prime}}{16 \pi} \int d \Omega^{\, \prime} \biggl[ {\mathcal N}_{I\, I}(\Omega,\Omega^{\prime})  \Delta_{I}(\Omega^{\prime}) + {\mathcal N}_{I\, Q}(\Omega,\Omega^{\prime})  \Delta_{Q}(\Omega^{\prime})  + {\mathcal N}_{I\, U}(\Omega^{\prime}) \Delta_{U}(\Omega^{\prime})\biggl],
\label{EV3}
\end{equation}
where the three components of the matrix ${\mathcal N}(\Omega,\Omega^{\,\prime})$ appearing 
in Eq. (\ref{EV3}) are: 
\begin{eqnarray}
{\mathcal N}_{I\,I} &=& \frac{1}{2} \biggl[ 3 - \mu^2 - \mu^{\prime 2}(1 - 3 \mu^2) + 4 \mu \mu^{\prime} \sqrt{1 - \mu^2} 
\sqrt{1 - \mu^{\prime 2}} \cos{(\varphi' - \varphi)}  
\nonumber\\
&+& (1 - \mu^2) ( 1 - \mu^{\prime 2}) \cos{2 (\varphi' - \varphi)} \biggr],
\nonumber\\
{\mathcal N}_{I\,Q} &=& \frac{1}{2} \biggl[ (3 \mu^2 -1)(\mu^{\prime 2} -1) + 4 \mu \sqrt{1 - \mu^2} \sqrt{1 - \mu^{\prime 2}} \cos{(\varphi' - 
\varphi)} 
\nonumber\\
&+& (\mu^{\prime 2} + 1 ) (\mu^2 -1) \cos{2 (\varphi' - \varphi)}\biggr],
\nonumber\\
{\mathcal N}_{I\,U} &=& - 2\biggl[ \mu \sqrt{1 - \mu^2} \sqrt{1 - \mu^{\prime 2}} + (\mu^2 -1) \mu^{\prime} \cos{(\varphi' - \varphi)} \biggr] \sin{(\varphi' - \varphi)}.
\label{EV4}
\end{eqnarray} 
In the scalar case the collision integral defined by Eqs. (\ref{EV3}) and (\ref{EV4}) can be  simplified since the corresponding brightness perturbations, unlike the tensor contribution, do not depend on $\varphi$ or $\varphi^{\prime}$. Thus, the integral over $d \Omega^{\prime} = 
d \mu^{\,\prime} \, d\varphi^{\,\prime}$ implies:
\begin{equation}
\frac{d \Delta^{(s)}_{I}}{d \tau} + \epsilon^{\, \prime} \Delta^{(s)}_{I} =  \frac{3\, \epsilon^{\prime}}{16} {\mathcal C}^{(s)}_{I}(\mu,k,\tau), 
\label{EV5}
\end{equation}
where ${\mathcal C}^{(s)}_{I}(\mu,k,\tau)$ is explicitly given by:
\begin{equation}
{\mathcal C}^{(s)}_{I}(\mu,\,k,\,\tau) = \int_{-1}^{1}d\mu^{\prime} 
\biggl\{\biggl[3 - \mu^2 - ( 1 - 3 \mu^2) \, \mu^{\prime\, 2}\biggr] \Delta_{I}^{(s)}(\mu^{\prime}, k,\tau)  
+ ( 1 - 3 \mu^2)( 1 - \mu^{\, \prime\, 2}) \Delta_{Q}^{(s)}(\mu^{\prime}, k,\tau) \biggr\}.
\label{EV5a}
\end{equation}
The integration over $\varphi^{\prime}$ eliminates two distinct terms
in each of the matrix elements ${\mathcal N}_{I\,I}$ and ${\mathcal N}_{I\,Q}$ and it also eliminates 
the contribution of all the terms contained in ${\mathcal N}_{I\,U}$.
To facilitate the integration over $\mu^{\prime}$ in Eq. (\ref{EV5}),
$\Delta_{I}^{(s)}$ and $\Delta_{Q}^{(s)}$ are expanded in a series of Legendre polynomials $P_{\ell}(\mu^{\prime})$ as
\begin{equation}
\Delta^{(s)}_{X}(\mu^{\,\prime},k,\tau) = \sum_{\ell=0}^{\infty} (-i)^{\ell} (2\ell + 1) \, P_{\ell}(\mu^{\,\prime})\, \Delta^{(s)}_{X\,\ell}(k,\tau).
\label{EV6}
\end{equation}
Using Eq. (\ref{EV6}) into Eq. (\ref{EV5a}), the well known result for the scalar collision term  is: 
\begin{equation}
{\mathcal C}^{(s)}_{I}(\mu, \,k,\,\tau) = \frac{16}{3} \Delta_{I\,0} + \frac{4}{3} (1 - 3 \mu^2) \biggl(\Delta_{I\,2} + \Delta_{Q\, 2} +\Delta_{Q\, 0}\biggr).
\label{EV7}
\end{equation}
In the conformally Newtonian gauge and in Fourier space the collisionless part of Eq. (\ref{EV5}) 
inherits an explicit dependence on the metric fluctuation and the same will be true for the tensor modes:
\begin{equation}
\frac{d \Delta^{(s)}_{I}}{d \tau} + \epsilon^{\prime} \Delta^{(s)}_{I} = \partial_{\tau} \Delta^{(s)}_{I}  + ( i k \mu + \epsilon^{\prime})\Delta^{(s)}_{I} +  \frac{1}{q} \biggl(\frac{d q}{d\tau}\biggr)_{s} , \qquad  \biggl(\frac{d q}{d\tau}\biggr)_{s} = - q \partial_{\tau} \psi + i k \mu q  \phi.
\label{EV8}
\end{equation}
Inserting Eqs. (\ref{EV7}) and (\ref{EV8}) into Eq. (\ref{EV5}) we therefore obtain the evolution of the intensity 
in its complete form:
\begin{equation}
\partial_{\tau} \Delta^{(s)}_{I}
+ ( i k\mu + \epsilon') \Delta^{(s)}_{I} = \partial_{\tau} \psi - i k \mu \phi + \epsilon'  \,\biggl[\Delta_{I 0} + \mu v_{\mathrm{b}} - \frac{P_{2}(\mu)}{2} S^{(s)}_{P} \biggr],
\label{EV9}
\end{equation}
where $S^{(s)}_{P} =  (\Delta_{I 2} + \Delta_{Q 0} + \Delta_{Q 2})$ 
and the term  $\epsilon^{\prime} \mu v_{b}$ accounts for the scalar component of the Doppler term associated with the baryon velocity field (note that $\vec{v}^{(s)}= \vec{k} v_{\mathrm{b}}$).
Equation (\ref{EV9}) is standard even if it can be written in different gauges 
and with slightly different notations\footnote{The conventions
leading to Eq. (\ref{EV9}) change: there are some who do not include the term $(2 \ell+1)$ in Eq. (\ref{EV6}); there are some others 
who do not include the term $(-\,i)^{\ell}$.} (see e.g. \cite{book1,book2,book3}). 
In the case of the tensor modes the derivation of the brightness perturbations {\em follows exactly the same steps leading to the result of Eq. (\ref{EV9})} with the difference that the evolution is automatically gauge-invariant
and this is ultimately related to the gauge-invariance of $h_{ij}$. In spite of this advantage the contribution of the metric inhomogeneity  to the collisionless part of the Boltzmann equation now contains an explicit dependence on the polarization (and hence on the angle $\varphi$):
\begin{equation}
\frac{d \Delta^{(t)}_{I}}{d \tau} + \epsilon^{\prime} \Delta^{(t)}_{I} = \partial_{\tau} \Delta^{(t)}_{I}  + ( i k \mu + \epsilon^{\prime})\Delta^{(t)}_{I} +  \frac{1}{q} \biggl(\frac{d q}{d\tau}\biggr)_{t}, \qquad 
\biggl(\frac{d q}{d\tau}\biggr)_{t} = - \frac{q}{2} \, \hat{n}^{i}\, \hat{n}^{j} \, \partial_{\tau}h_{ij}.
\label{EV10}
\end{equation}
The tensor polarizations entering Eq. (\ref{EV10}) lead overall to the following term 
\begin{equation}
\hat{n}^{i} \hat{n}^{j} \partial_{\tau} h_{i j}(\vec{k},\tau) =  \biggl[ (\hat{n}\cdot\hat{a})^2 
- (\hat{n}\cdot\hat{b})^2\biggr]\, \partial_{\tau} h_{\oplus}(\vec{k},\tau) + 2 (\hat{n}\cdot\hat{a}) (\hat{n} \cdot \hat{b})\,
\partial_{\tau} h_{\otimes}(\vec{k},\tau),
\label{EV11}
\end{equation}
where $\hat{a}$ and $\hat{b}$ denote two unit vectors mutually orthogonal and orthogonal to $\hat{k}$. In the 
coordinate system where $\hat{a} = \hat{x}$, $\hat{b} = \hat{y}$ (see also Eqs. (\ref{POLDEF2}) and (\ref{SIXPOL9a}))  Eq. (\ref{EV10}) takes the following form:
\begin{equation}
\partial_{\tau} \Delta^{(t)}_{I}  + ( i k \mu + \epsilon^{\prime})\Delta^{(t)}_{I}  - \frac{1}{2} (1 - \mu^2) \cos{2 \varphi} \partial_{\tau} h_{\oplus}  - \frac{1}{2} (1 - \mu^2) \sin{2 \varphi} \partial_{\tau} h_{\otimes}= \frac{3 \epsilon^{\,\prime}}{16 } {\mathcal C}^{(t)}_{I}(\varphi,\,\mu,\,k,\,\tau),
\label{EV11a}
\end{equation}
clearly showing that the brightness perturbations $\Delta^{(t)}_{X}(\varphi,\,\mu,\,k,\,\tau)$  (with $X= I, \, Q,\, U,\, V$) must all depend on $\varphi$. By looking at Eq. (\ref{EV11a}) we can decouple the angular dependence 
 by introducing the following parametrization employed, with a slightly different notation, by Polnarev \cite{HIS12a} (see also \cite{HIS12aa,GWHT1,GWHT5,GWHT9}):
\begin{eqnarray}
\Delta_{I}^{(t)}(\varphi,\mu,k,\tau) &=& (1 - \mu^2) \biggl[ \cos{2 \varphi} \,{\mathcal Z}_{\oplus}(\mu,k,\tau) + \sin{2 \varphi} \,{\mathcal Z}_{\otimes}(\mu, k, \tau)\biggr],
\label{EV12}\\
\Delta_{Q}^{(t)}(\varphi,\mu,k,\tau) &=& (1 + \mu^2) \biggl[ \cos{2 \varphi}\, {\mathcal T}_{\oplus}(\mu,k,\tau) + \sin{2 \varphi}\, {\mathcal T}_{\otimes}(\mu,k,\tau)\biggr],
\label{EV13}\\
\Delta_{U}^{(t)}(\varphi,\mu,k,\tau) &=& 2 \mu \biggl[ -\sin{2 \varphi} \, {\mathcal T}_{\oplus}(\mu,k,\tau) + \cos{2 \varphi}\, {\mathcal T}_{\otimes}(\mu,k,\tau)\biggr],
\label{EV14}\\
\Delta_{V}^{(t)}(\varphi,\mu,k,\tau) &=& 2 \mu \biggl[ \cos{2 \varphi} \, {\mathcal S}_{\oplus}(\mu,k,\tau) + \sin{2 \varphi}\, {\mathcal S}_{\otimes}(\mu,k,\tau)\biggr].
\label{EV15}
\end{eqnarray}
Equations (\ref{EV12})--(\ref{EV13}) and (\ref{EV14})--(\ref{EV15}) can be inserted into Eqs. (\ref{EV3}) and (\ref{EV11a}) and this step allows for a direct integration of the collision term over $\varphi^{\,\prime}$. The result of this straightforward but lengthy manipulation is:
\begin{eqnarray}
{\mathcal C}^{(t)}_{I}(\mu,\varphi, k, \tau) &=& \frac{1}{2}  (\mu ^2+1)  \int_{-1}^{1} \, d\mu^{\prime} 
\biggl\{\sin{2 \varphi} \biggl[\left(\mu^{\prime\,4}+6 \mu^{\prime\,2}+1\right) {\mathcal T}_{\otimes}-\left(\mu^{\prime \,2}-1\right)^2
   {\mathcal Z}_{\otimes}\biggr]
 \nonumber\\  
&+& \cos{2 \varphi} \biggl[\left(\mu^{\prime \,4}+6 \mu^{\prime\, 2}+1\right) {\mathcal T}_{\oplus}-\left(\mu^{\prime \, 2}-1\right)^2
   {\mathcal Z}_{\oplus}\biggr]\biggr\}.
\label{EV16}
\end{eqnarray}
The collision term appearing in Eq. (\ref{EV16}) contains a $\varphi$ dependence that matches exactly 
the angular dependence arising from the tensor fluctuation of the geometry appearing in the collisionless part 
of the Boltzmann equation (see Eq. (\ref{EV11}) and discussion 
thereafter). As a consequence the dependence upon $\varphi$ can be factorized and Eq. (\ref{EV11a})
reduces to a pair of separate equations for ${\mathcal T}_{\lambda}$ and ${\mathcal Z}_{\lambda}$ \begin{eqnarray}
\partial_{\tau} {\mathcal Z}_{\lambda} + (i k\mu + \epsilon') {\mathcal Z}_{\lambda} - \frac{1}{2}\partial_{\tau} h_{\lambda} =  
\epsilon' \Sigma^{(t)},\qquad \partial_{\tau}{\mathcal T}_{\lambda} +  (i k\mu + \epsilon') {\mathcal T}_{\lambda}  = -  \epsilon' \Sigma^{(t)},
\label{EV18}
\end{eqnarray}
where $\lambda= \oplus,\,\otimes$ denotes each of the two independent tensor polarizations.
If the relic gravitons are not polarized (see however Eq. (\ref{POLGRAV7}) and discussion therein) the polarization index 
can be dropped since the two linear polarizations evolve independently and, in this case,  ${\mathcal Z}$, ${\mathcal T}$  shall be expanded with the same logic leading to Eq. (\ref{EV6}): 
 \begin{equation}
 {\mathcal Z}(\nu,k,\tau)  = \sum_{\ell=0}^{\infty} \,(-i)^{\ell} (2\ell + 1) \, P_{\ell}(\nu)\, {\mathcal Z}_{\ell}(k,\tau), \qquad 
{\mathcal T}(\nu,k,\tau)  = \sum_{\ell=0}^{\infty} \, (-i)^{\ell} (2\ell + 1) \, P_{\ell}(\nu)\, {\mathcal T}_{\ell}(k,\tau),
\label{EV20}
\end{equation}
where, as usual, ${\mathcal Z}_{\ell}$ and ${\mathcal T}_{\ell}$ denote the $\ell$-th multipoles of the corresponding functions. After this step the source term $\Sigma^{(t)}$ appearing in Eq. (\ref{EV18}) becomes:
\begin{eqnarray}
\Sigma^{(t)} &=& \frac{3}{32} \int_{-1}^{1} d \nu 
[  (1 - \nu^2)^2 {\mathcal Z}(\nu)
- ( 1 + \nu^2)^2 {\mathcal T}(\nu) - 4 \nu^2 {\mathcal T}(\nu)] 
\nonumber\\
&=& \frac{3}{70}{\mathcal  Z}_{4} + \frac{{\mathcal Z}_{2}}{7} - \frac{{\mathcal Z}_{0}}{10}- \frac{3}{70} {\mathcal T}_{4} + \frac{6}{7} {\mathcal T}_{2} - \frac{3}{5} {\mathcal T}_{0},
\label{EV21}
\end{eqnarray}
In the current literature Eqs. (\ref{EV12}), (\ref{EV13}) and (\ref{EV14}) are sometimes expressed in slightly different ways. For instance a complementary choice is to introduce the $R$ and $L$ combinations (see also Eqs. (\ref{SIXPOL10})--(\ref{sixc}) and discussion therein):
\begin{equation}
{\mathcal T}_{R} = \frac{1}{\sqrt{2}} \biggl[{\mathcal T}_{\oplus} - i\, {\mathcal T}_{\otimes}\biggr], \qquad 
{\mathcal T}_{L}= \frac{1}{\sqrt{2}} \biggl[{\mathcal T}_{\oplus} + i\, {\mathcal T}_{\otimes}\biggr],
\label{EV21a}
\end{equation}
and similarly for ${\mathcal Z}_{R}$ and  $Z_{L}$ by changing ${\mathcal T}_{\oplus} \to {\mathcal Z}_{\oplus}$
and ${\mathcal T}_{\otimes} \to {\mathcal Z}_{\otimes}$.  
Inserting Eqs. (\ref{EV21a}) and its analog for ${\mathcal Z}_{\lambda}$ into Eqs. (\ref{EV12}), (\ref{EV13}) and (\ref{EV14}) we obtain 
\begin{eqnarray}
\Delta_{I}^{(t)}(\varphi,\mu,k,\tau) &=& \frac{(1 - \mu^2)}{\sqrt{2}} \biggl[ e^{2\, i\, \varphi} \,{\mathcal Z}_{R}(\mu,k,\tau) + e^{- 2 \, i\, \varphi}\,{\mathcal Z}_{L}(\mu, k, \tau)\biggr],
\label{EV22}\\
\Delta_{\pm}^{(t)}(\varphi,\mu,k,\tau)
&=& \frac{1}{\sqrt{2}} \biggl[ 
e^{ 2 \,i\, \varphi} ( 1 \mp \mu)^2 {\mathcal T}_{R}(\mu,k,\tau) +  e^{ -2 \,i\, \varphi} ( 1 \pm \mu)^2 {\mathcal T}_{L}(\mu,k,\tau)\biggr],
\label{EV23}
\end{eqnarray}
where $ \Delta_{\pm}^{(t)}(\varphi,\mu,k,\tau) = \Delta_{Q}^{(t)}(\varphi,\mu,k,\tau) \pm i \Delta_{U}^{(t)}(\varphi,\mu,k,\tau)$. When the two polarization evolve independently Eqs. (\ref{EV22}) and (\ref{EV23}) can be expressed in an even simpler form
\begin{eqnarray} 
\Delta_{I}^{(t)}(\varphi,\mu,\vec{k},\tau) &=& (1 - \mu^2) \biggl[ e^{2\, i\, \varphi} \,q_{R}(\vec{k}) + e^{- 2 \, i\, \varphi}\,q_{L}(\vec{k})\biggr] \, \Delta_{T}^{(t)}(k,\mu\,\tau),
\label{EV24}\\
\Delta_{\pm}^{(t)}(\varphi,\mu,\vec{k},\tau) &=& \biggl[ 
e^{ 2 \,i\, \varphi} ( 1 \mp \mu)^2 q_{R}(\vec{k}) +  e^{ -2 \,i\, \varphi} ( 1 \pm \mu)^2 q_{L}(\vec{k})\biggr]\,\Delta_{P}^{(t)}(k,\mu\,\tau),
\label{EV25}
\end{eqnarray}
where  $q_{R}(\vec{k})$ and $q_{L}(\vec{k})$ are random functions of the three-momentum\footnote{In numerical analyses it 
 is often customary to factor the dependence of the initial power spectrum; in this case we will have that $\langle q_{R,\, L}(\vec{k}) \, q_{R,\, L}(\vec{p})\rangle = (2\pi^2 /k^3) \overline{P}_{T}(k)\delta^{(3)}(\vec{k} + \vec{p})$ while $\langle q_{R}(\vec{k}) \, q_{L}^{*}(\vec{p})\rangle  = 
 \langle q_{L}(\vec{k}) \, q_{R}^{*}(\vec{p})\rangle =0$.} while, from Eq. (\ref{EV18})  $\Delta_{T}^{(t)}(k,\mu\,\tau)$ and $\Delta_{P}^{(t)}(k,\mu\,\tau)$ obey: 
\begin{equation}
\Delta_{T}^{(t)\,\,\prime} + ( i k \mu + \epsilon^{\,\prime})\Delta_{T}^{(t)} = \frac{h^{\prime}}{2} + \epsilon^{\prime} \, \Psi, \qquad 
\Delta_{P}^{(t)\,\,\prime} + ( i k \mu + \epsilon^{\,\prime})\Delta_{P}^{(t)} = - \epsilon^{\prime} \, \Psi,
\label{EV27}
\end{equation}
where $\Psi$ has exactly the same form of $\Sigma^{(t)}$ given in Eq. (\ref{EV21}) with the following two replacements:
 ${\mathcal Z}_{\ell} \to \Delta_{T\,\,\ell}^{(t)}$ and ${\mathcal T}_{\ell} \to \Delta_{P\,\,\ell}^{(t)}$. By integrating Eqs. 
 (\ref{EV27}) along the line of sight we have that Eqs. (\ref{EV24}) and (\ref{EV25}) become:
\begin{eqnarray} 
\Delta_{I}^{(t)}(\hat{n},\vec{k},\tau_{0}) &=& (1 - \mu^2) \biggl[ e^{2\, i\, \varphi} \,q_{R}(\vec{k}) + e^{- 2 \, i\, \varphi}\,q_{L}(\vec{k})\biggr] \, 
\int_{0}^{\tau_{0}} \, d\tau\, e^{- i \, \mu \, x} S_{T}^{(t)}(k,\tau),
\label{EV24a}\\
\Delta_{\pm}^{(t)}(\hat{n},\vec{k},\tau) &=& \biggl[ 
e^{ 2 \,i\, \varphi} ( 1 \mp \mu)^2 q_{R}(\vec{k}) +  e^{ -2 \,i\, \varphi} ( 1 \pm \mu)^2 q_{L}(\vec{k})\biggr]\,\int_{0}^{\tau_{0}} \, d\tau\, e^{- i \, \mu \, x} S_{P}^{(t)}(k,\tau),
\label{EV25a}
\end{eqnarray}
where $x = k(\tau_{0} - \tau)$ while the source terms $ S_{T}^{(t)}(k,\tau)$ and $S_{P}^{(t)}(k,\tau)$ have been redefined as
\begin{equation}
S_{T}^{(t)}(k,\tau) = \frac{h^{\prime}}{2} e^{- \epsilon(\tau,\tau_{0})} + {\mathcal K}(\tau) \Psi(k,\tau), \qquad 
S_{P}^{(t)}(k,\tau) = - {\mathcal K}(\tau) \Psi(k,\tau).
\label{EV28}
\end{equation}
The visibility function ${\mathcal K}(\tau)$ gives the probability that a photon is last scattered between $\tau$ and $\tau + d\tau$ and it has a peak around the redshift of recombination. 
In analytic discussions the visibility function is often approximated 
by means of a Gaussian profile (see, e. g., \cite{zeld1,wyse,pav1,pav2}) and it is 
defined as 
\begin{equation}
{\mathcal K}(\tau) = \epsilon' e^{-\epsilon(\tau,\tau_{0})},\qquad \epsilon(\tau,\tau_{0}) 
= \int_{\tau}^{\tau_{0}} a \widetilde{n}_{\mathrm{e}}\, x_{\mathrm{e}} \sigma_{\gamma\, e}\, d\tau, \qquad 
\epsilon' =  a \widetilde{n}_{\mathrm{e}}\, x_{\mathrm{e}} \sigma_{T}.
\label{LS4}
\end{equation}
where $\epsilon(\tau,\tau_{0})$ is the optical depth. In the sudden decoupling approximation 
the visibility function becomes infinitely thin \cite{book1,book2,book3}. The parametrizations 
of the visibility function are essential to derive a number of useful analytic results \cite{HIS12a} 
(see also \cite{GWHT4,GWHT5,GWHT5a}). 

To complete the discussion,  the evolution of the polarization anisotropies follows 
from Eq. (\ref{EV3}):
\begin{eqnarray}
\frac{d \Delta_{Q}}{d \tau} + \epsilon^{\, \prime} \Delta_{Q} &=& \frac{3 \epsilon^{\,\prime}}{16 \pi} \int d \Omega^{\, \prime} \biggl[ {\mathcal N}_{Q\, I}(\Omega,\Omega^{\prime})  \Delta_{I}(\Omega^{\prime}) + {\mathcal N}_{Q\, Q}(\Omega,\Omega^{\prime})  \Delta_{Q}(\Omega^{\prime})  + {\mathcal N}_{Q\, U}(\Omega^{\prime}) \Delta_{U}(\Omega^{\prime})\biggl],
\label{EV29}\\
\frac{d \Delta_{U}}{d \tau} + \epsilon^{\, \prime} \Delta_{U} &=& \frac{3 \epsilon^{\,\prime}}{16 \pi} \int d \Omega^{\, \prime} \biggl[ {\mathcal N}_{U\, I}(\Omega,\Omega^{\prime})  \Delta_{I}(\Omega^{\prime}) + {\mathcal N}_{U\, Q}(\Omega,\Omega^{\prime})  \Delta_{Q}(\Omega^{\prime})  + {\mathcal N}_{U\, U}(\Omega^{\prime}) \Delta_{U}(\Omega^{\prime})\biggl],
\label{EV30}
\end{eqnarray}
and the $6$ further matrix elements appearing in Eqs. (\ref{EV29}) and (\ref{EV30}) are given by 
\begin{eqnarray}
{\mathcal N}_{Q\,I} &=& \frac{1}{4} \biggl[ (\mu^2 -1) (3 \mu^{\prime 2} -1) + 4 \mu \mu^{\prime} \sqrt{1 - \mu^2} \sqrt{1 - \mu^{\prime 2}} \cos{(\varphi' - \varphi)}
\nonumber\\
&+& (\mu^2 +1 ) (\mu^{\prime 2} -1) \cos{2(\varphi' - \varphi)} \biggr],
\nonumber\\
{\mathcal N}_{Q\,Q} &=& \frac{1}{4} \biggr[3 (\mu^2 -1) (\mu^{\prime 2} -1) + 4 \mu \mu^{\prime} \sqrt{1 - \mu^{\prime 2}} \sqrt{1 - \mu^2} \cos{(\varphi' - \varphi)} 
\nonumber\\
&+& ( 1 + \mu^2) (1 + \mu^{\prime 2}) \cos{2 (\varphi' - \varphi)}\biggr],
\nonumber\\
{\mathcal N}_{Q\,U} &=&  - \biggl[ \mu \sqrt{1 - \mu^2} \sqrt{1 - \mu^{\prime 2}} + ( \mu^2 +1) \mu^{\prime} \cos{(\varphi' - \varphi)} \biggr] \sin{(\varphi' - \varphi)},
\nonumber\\
{\mathcal N}_{U\, I} &=&\biggl[ \mu^{\prime} \sqrt{1 - \mu^2} \sqrt{1 - \mu^{\prime 2}} + ( \mu^{\prime 2} -1) \mu \cos{(\varphi' - \varphi)} \biggr] \sin{(\varphi' - \varphi)},
\nonumber\\
{\mathcal N}_{U\, Q} &=& \biggl[ \mu^{\prime} \sqrt{1 - \mu^2} \sqrt{1 - \mu^{\prime 2}} + ( \mu^{\prime 2} +1) \mu \cos{(\varphi' - \varphi)} \biggr] \sin{(\varphi' - \varphi)},
\nonumber\\
{\mathcal N}_{U\, U} &=& \sqrt{1 -\mu^2} \sqrt{1 - \mu^{\prime 2}} \cos{(\varphi' - \varphi)} + \mu \mu^{\prime} \cos{2 (\varphi' - \varphi)}.
\label{EV31}
\end{eqnarray}
In the scalar case it is possible to integrate the collision term over $\varphi^{\prime}$ and, as result, 
the scalar brightness perturbations evolve according to 
\begin{equation}
\partial_{\tau} \Delta_{\pm}^{(s)} + ( i k \mu + \epsilon^{\prime}) \Delta_{\pm}^{(s)} = \frac{3}{4} (1 - \mu^2) \epsilon^{\prime} \, S^{(s)}_{P}.
\label{cor5a}
\end{equation}
where $S^{(s)}_{P}$ has been already introduced in Eq. (\ref{EV9})  and $\Delta^{(s)}_{\pm} = (\Delta_{Q}^{(s)} \pm i \Delta_{U}^{(s)})$ denote the two orthogonal combinations of the brightness perturbations sourced by the scalar modes. In the tensor case the evolution of the polarization anisotropies is even simpler than in the scalar case. In the absence of circular components Eqs. (\ref{EV13}) 
and (\ref{EV14}) already give the evolution of $\Delta_{Q}^{(t)}$ and $\Delta_{U}^{(t)}$. Thus,
once the solutions for ${\mathcal Z}_{\lambda}$ and ${\mathcal T}_{\lambda}$ 
are known, the temperature and the polarization anisotropies follow automatically.  
It can be explicitly checked (but this discussion will be omitted) that Eqs. (\ref{EV29}) and 
(\ref{EV30}) lead {\em exactly} to the same evolution equation for ${\mathcal T}_{\lambda}$ once the brightness perturbations are used in the form given in Eqs. (\ref{EV12}), (\ref{EV13}) and (\ref{EV14}).

\subsection{Angular power spectra for the $E$-mode and $B$-mode polarizations}
The angular power spectra introduced in Eqs. (\ref{E1a}) and (\ref{int5aa}) give the scalar 
contribution to the total temperature and polarization anisotropies. 
In real space the scalars $\Delta^{(s)}_{E}(\hat{n},\tau) $ and $\Delta^{(s)}_{B}(\hat{n},\tau)$ can be expressed in terms of the generalized 
ladder operators \cite{EB1} raising and lowering the spin-weight of a given function \cite{EB1,EB2,EB3}:
\begin{eqnarray}
&& \Delta^{(s)}_{E}(\hat{n},\tau) = - \frac{1}{2} \biggl\{ K_{-}^{(1)}(\hat{n})[K_{-}^{(2)}(\hat{n})
\Delta_{+}(\hat{n},\tau)] +  K_{+}^{(-1)}(\hat{n})[K_{+}^{(-2)}(\hat{n}) \Delta_{-}(\hat{n},\tau)]\biggr\},
\label{BE1}\\
&&  \Delta^{(s)}_{B}(\hat{n},\tau) =  \frac{i}{2} \biggl\{ K_{-}^{(1)}(\hat{n})[K_{-}^{(2)}(\hat{n})
\Delta_{+}(\hat{n},\tau)] -  K_{+}^{(-1)}(\hat{n})[K_{+}^{(-2)}(\hat{n}) \Delta_{-}(\hat{n},\tau)]\biggr\}.
\label{BE2}
\end{eqnarray}
The differential operators appearing in Eqs. (\ref{BE1}) and (\ref{BE2}) 
are  generalized ladder operators \cite{EB1} whose action 
either raises or lowers the spin weight of a given fluctuation. They are defined within the present conventions as 
acting on a fluctuation of spin weight $j$: 
\begin{eqnarray}
 K_{+}^{j}(\hat{n}) = - (\sin{\vartheta})^{j}\biggl[ \partial_{\vartheta} + 
\frac{i}{\sin{\vartheta}} \partial_{\varphi}\biggr] \frac{1}{(\sin{\vartheta})^{j}}, \qquad K_{-}^{j}(\hat{n}) = - \frac{1}{(\sin{\vartheta})^{j}}
\biggl[ \partial_{\vartheta} -
\frac{i}{\sin{\vartheta}} \partial_{\varphi}\biggr] (\sin{\vartheta})^{j}.
\label{Km}
\end{eqnarray}
For instance $K_{-}^{(2)}\Delta_{+}$  transforms as a function of spin-weight 1 while 
$K_{-}^{(1)}[K_{-}^{(2)}\Delta_{+}]$ is, as anticipated, as scalar. Using Eq. (\ref{Km}) 
inside Eqs. (\ref{BE1}) and (\ref{BE2}) the explicit expressions of the E-mode and of the B-mode 
are, in real space:
\begin{eqnarray}
\Delta^{(s)}_{E} (\hat{n},\tau) &=& - \frac{1}{2}\biggl\{( 1 -\mu^2) \partial_{\mu}^2 (\Delta^{(s)}_{+} + \Delta^{(s)}_{-}) - 4 \mu  \partial_{\mu}(\Delta^{(s)}_{+} + \Delta^{(s)}_{-}) - 2  (\Delta^{(s)}_{+} + \Delta^{(s)}_{-}) 
\nonumber\\
&-& 
\frac{\partial_{\varphi}^2 (\Delta^{(s)}_{+} + \Delta^{(s)}_{-})}{1 - \mu^2 } 
+
2 i\biggl[ \partial_{\varphi}  \partial_{\mu}(\Delta^{(s)}_{+} - \Delta^{(s)}_{-}) - \frac{\mu}{1 - \mu^2} \partial_{\varphi}  (\Delta^{(s)}_{+} - \Delta^{(s)}_{-})\biggr] \biggr\},
\label{BE3}\\
 \Delta^{(s)}_{B} (\hat{n},\tau) &=& \frac{i}{2} \biggl\{( 1 -\mu^2) \partial_{\mu}^2(\Delta^{(s)}_{+} - \Delta^{(s)}_{-}) - 4 \mu  \partial_{\mu}(\Delta^{(s)}_{+} - \Delta^{(s)}_{-}) - 2  (\Delta^{(s)}_{+} - \Delta^{(s)}_{-})  
\nonumber\\
&-& 
\frac{\partial_{\varphi}^2  (\Delta^{(s)}_{+} - \Delta^{(s)}_{-})}{1 - \mu^2 }
+ 
2 i\biggl[ \partial_{\varphi}  \partial_{\mu}(\Delta^{(s)}_{+} + \Delta^{(s)}_{-}) - \frac{\mu}{1 - \mu^2} \partial_{\varphi}  (\Delta^{(s)}_{+} + \Delta^{(s)}_{-})\biggr] \biggr\}.
\label{BE4}
\end{eqnarray}
The lengthy expressions reported in Eqs. (\ref{BE3}) and (\ref{BE4}) simplify greatly 
since the scalar modes do not have azimuthal dependence:
\begin{equation}
\Delta^{(s)}_{E} (\hat{n},\tau) =  - \frac{1}{2} \partial_{\mu}^2\, \biggl[ ( 1 -\mu^2)  \biggl(\Delta^{(s)}_{+} + \Delta^{(s)}_{-}\biggr)\biggr], \qquad \Delta^{(s)}_{B} (\hat{n},\tau) = \frac{i}{2} \partial_{\mu}^2\, \biggl[ ( 1 -\mu^2)  \biggl(\Delta^{(s)}_{+} - \Delta^{(s)}_{-}\biggr)\biggr].
 \label{BB}
 \end{equation}
We can finally determine $a_{\ell m}^{(E)}$ and $a_{\ell m}^{(B)}$:
\begin{eqnarray}
a_{\ell m}^{(E)} &=& - \frac{N_{\ell}}{2 (2 \pi)^{3/2}} \int d \hat{n} \, Y_{\ell m}^{*}(\hat{n})\int  d^{3} k \,  \partial_{\mu}^2 \biggl\{ (1 - \mu^2) \biggl[\Delta^{(s)}_{+}(\vec{k},\tau) + \Delta_{-}^{(s)}(\vec{k},\tau)\biggr]\biggr\},
\nonumber\\
a_{\ell m}^{(B)} &=&  \frac{i\, \,N_{\ell}}{2 (2 \pi)^{3/2}} \int d \hat{n} \, Y_{\ell m}^{*}(\hat{n})\int  d^{3} k \, \partial_{\mu}^2 \biggl\{ (1 - \mu^2) \biggl[\Delta_{+}^{(s)}(\vec{k},\tau) - \Delta_{-}^{(s)}(\vec{k},\tau)\biggr]\biggr\}.
\label{cor5}
\end{eqnarray}
But recalling now Eq. (\ref{cor5a}) the two (scalar) orthogonal combinations
$\Delta^{(s)}_{\pm} = (\Delta_{Q}^{(s)} \pm i \Delta_{U}^{(s)})$ actually obey 
the same equation and share the same source term. From Eq. (\ref{cor5}) (or directly from Eq. (\ref{BB})) it then follows that the $B$-mode polarization vanishes in the scalar case. Also the tensor brightness perturbations $\Delta^{(t)}_{\pm}(\hat{n},\tau)$ can be expanded in terms of spin-$\pm2$ spherical harmonics $_{\pm 2}Y_{\ell\,m}(\hat{n})$, i.e. 
\begin{equation}
\Delta^{(t)}_{\pm}(\hat{n},\tau) = \sum_{\ell \, m} \,\, b_{\pm 2,\,\ell\, m} \,\, _{\pm 2}Y_{\ell\, m}(\hat{n}).
\label{int2a}
\end{equation}
In full analogy with the previous case the $E$- and $B$- modes are defined as
\begin{equation}
b^{(E)}_{\ell\, m} = - \frac{1}{2}(b_{2,\,\ell m} + b_{-2,\,\ell m}), \qquad  
b^{(B)}_{\ell\, m} =  \frac{i}{2} (b_{2,\,\ell m} - b_{-2,\,\ell m}).
\label{int3a}
\end{equation}
The fluctuations constructed from 
$b^{(E)}_{\ell\,m}$ and $b^{(B)}_{\ell\,m}$ have the 
property of being invariant under rotations on a plane orthogonal 
to $\hat{n}$ and they can be expanded in terms of (ordinary) spherical harmonics:
\begin{equation}
\Delta^{(t)}_{E}(\hat{n},\tau) = \sum_{\ell\, m} N_{\ell}^{-1} \,  b^{(E)}_{\ell\, m}  \, Y_{\ell\, m}(\hat{n}),\qquad 
\Delta^{(t)}_{B}(\hat{n},\tau) = \sum_{\ell\, m} N_{\ell}^{-1} \,  b^{(B)}_{\ell\, m}  \, Y_{\ell\, m}(\hat{n}),
\label{int4a}
\end{equation}
where $N_{\ell} = \sqrt{(\ell - 2)!/(\ell +2)!}$.  From Eqs. (\ref{BE1}) and (\ref{BE2}) written in the tensor case we can also deduce $b^{(E)}_{\ell m}$ and $b^{(B)}_{\ell m}$ and, in particular, we shall have 
\begin{equation}
b_{\ell m}^{(X)} = \frac{N_{\ell}}{(2 \pi)^{3/2}} \int d \hat{n} \, Y_{\ell m}^{*}(\hat{n})\int  d^{3} k \,  \Delta_{X}^{(t)}(\hat{n},\, k,\,\tau),
\label{cort5}
\end{equation}
where $X= E,\, B$ while $\Delta_{E}^{(t)}(\hat{n},\, k,\,\tau)$ and $\Delta_{B}^{(t)}(\hat{n},\, k,\,\tau)$ are: 
\begin{eqnarray}
\Delta_{E}^{(t)}(\hat{n},\, k,\,\tau) &=& (1 - \mu^2)\biggl[ e^{2\, i\, \varphi} \,q_{R}(\vec{k}) + e^{- 2 \, i\, \varphi}\,q_{L}(\vec{k}) \biggr]
\Lambda_{E}(x) \, \int_{0}^{\tau_0} e^{- i \mu x}\, S_{P}^{(t)}(k,\tau),
\label{EMM}\\
\Delta_{B}^{(t)}(\hat{n},\, k,\,\tau) &=& - (1 - \mu^2)\biggl[ e^{2\, i\, \varphi} \,q_{R}(\vec{k}) - e^{- 2 \, i\, \varphi}\,q_{L}(\vec{k}) \biggr]
\Lambda_{B}(x) \, \int_{0}^{\tau_0} e^{- i \mu x}\, S_{P}^{(t)}(k,\tau), 
\label{BMM}
\end{eqnarray}
where $\Lambda_{E}(x) = [-12+ x^2( 1 - \partial_{x}^2) - 8 x \partial_{x}]$ and $\Lambda_{B}(x) = [8 x + 2 x^2 \partial_{x}]$ are two differential operators; the $EE$ and $BB$ angular power spectra are then defined as: 
\begin{equation}
C_{\ell}^{(EE)} = \frac{1}{2\ell + 1} \sum_{m = -\ell}^{\ell} 
\langle b^{(E)*}_{\ell m}\,b^{(E)}_{\ell m}\rangle,\qquad 
 C_{\ell}^{(BB)} = \frac{1}{2\ell + 1} \sum_{m=-\ell}^{\ell} 
\langle b^{(B)*}_{\ell m}\,b^{(B)}_{\ell m}\rangle.
\label{int5a}
\end{equation}
Recalling Eqs. (\ref{EMM}) and (\ref{BMM}) we finally have 
 \begin{eqnarray}
 &&{\mathcal C}_{\ell}^{(EE)} = 4 \pi \int \frac{d k}{k} \bigl| \Delta_{E\ell}^{(t)}(k,\tau)\bigr|^2, \qquad \Delta_{E\ell}^{(t)}(k,\tau) = \int_{0}^{\tau} d\tau_{1} {\mathcal K}(\tau_{1}) S_{P}^{(t)}(k,\tau_{1}) \, \Lambda_{E}(x) \frac{j_{\ell}(x)}{x^2}, 
 \label{CCT1}\\
&& {\mathcal C}_{\ell}^{(BB)} = 4 \pi \int \frac{d k}{k} \bigl| \Delta_{B\ell}^{(t)}(k,\tau)\bigr|^2, \qquad 
 \Delta_{B\ell}^{(t)}(k,\tau) = \int_{0}^{\tau} d\tau_{1} {\mathcal K}(\tau_{1}) S_{P}^{(t)}(k,\tau_{1}) \, \Lambda_{B}(x) \frac{j_{\ell}(x)}{x^2},
\label{CCT3}
\end{eqnarray}
where $j_{\ell}(x)$ are the spherical Bessel functions \cite{abr1,abr2}.

\subsection{Low-frequency modifications of the tensor power spectra}
\label{subs56}
Even if the number of $e$-folds presently accessible to large-scale observations depends on the post-inflationary thermal history (see e.g. Eq. (\ref{RGC28}) and discussion therein) already in the 
conventional situation the initial state of the tensor (and scalar) inhomogeneities may not 
coincide with the vacuum if the  total number of $e$-folds is ${\mathcal O}(65)$.
As long as the largest wavelengths cross the Hubble radius during the protoinflationary stage of expansion, an initial quantum state different from the vacuum will potentially modify the low-frequency branch of the spectrum. It is difficult to determine the nature of the initial state (possibly different  from the vacuum) just by looking at the properties of the power spectra and this is already evident by considering similar analyses attempted in the scalar case \cite{infrascal1,infrascal2}. If the initial power spectra are not assumed in the standard power-law form of Eq. (\ref{RGC23}) the data-driven analyses are feasible but lead to 
ambiguous results\footnote{See, in this respect, section \ref{sec7} and the remarks on the Hanbury Brown-Twiss interferometry applied to relic gravitons.}  \cite{infrascal1,infrascal2}. 
Different sets of initial states of relic gravitons and relic phonons have been explored through the years.
A coherent state of relic gravitons can be constructed from the generalization of the Glauber displacement operator to a continuum of modes \cite{kibble}:
\begin{equation}
{\mathcal D}(\alpha) = e^{d(\alpha)}, \qquad d(\alpha) = \sum_{\lambda} \int d^{3} k \biggl[ \alpha_{\vec{k}\,\lambda} \hat{a}^{\dagger}_{\vec{k},\lambda} - \alpha^{*}_{\vec{k}\,\lambda} \hat{a}_{\vec{k},\lambda}\biggr].
\label{coh1}
\end{equation}
The multimode Glauber operator is analogous to the multimode squeezing and rotation operators in Eqs. (\ref{PAM8}) and (\ref{PAM9}). If the initial state has a coherent component the final state of the 
relic gravitons will be a multimode squeezed-coherent state. The squeezed coherent states of relic gravitons can be introduced in two complementary perspectives mirroring their quantum optical analogs
originally discussed by  Yuen \cite{yuen} (the so-called two-photon coherent states) and by 
Caves \cite{caves} (sometimes also referred to as ideal squeezed state).
In the Caves representation the initial state is rotated,  squeezed and finally displaced\footnote{We recall that, by definition, $| \{z\, \delta \}\rangle = {\mathcal S}(z) \,{\mathcal R}(\delta) |\mathrm{vac}\rangle$ and $| \{ \beta \}\rangle = {\mathcal D}(\beta) |\mathrm{vac}\rangle$. } i.e. $| \{\alpha\, z\, \delta \}\rangle = {\mathcal D}(\alpha) | \{z\, \delta \}\rangle$. In the Yuen representation \cite{yuen} the squeezed-coherent states of relic gravitons are instead defined as $|\{  z\, \delta\, \beta \}\rangle = {\mathcal S}(z) {\mathcal R}(\delta) |\{ \beta\}\rangle$.  Provided the eigenvalues of the coherent states 
in the two representations are connected as 
\begin{equation}
\alpha_{\vec{q},\lambda} = e^{- i\, \delta_{q,\lambda}} \cosh{r_{q,\lambda}} \beta_{\vec{q},\lambda} -  e^{i (\theta_{q,\lambda} 
+ \delta_{q,\lambda}}) \sinh{r_{q,\lambda}} \beta^{*}_{-\vec{q},\lambda},
\label{BB10}
\end{equation}
the strategies of Refs. \cite{yuen} and \cite{caves} are fully equivalent. Initial states dominated 
by a coherent component have been discussed in the past \cite{QM10c,QM10d}  and will be further 
mentioned in section \ref{sec7} in connection with the possible applications of Hanbury-Brown-Twiss 
interferometry to the case of the relic gravitons. 

The initial state of the relic gravitons does not need to be 
pure since the protoinflationary phase might be dominated 
by a plasma in thermal or in kinetic equilibrium.  Assuming 
either thermal or kinetic equilibrium, the initial state can then be
described density matrix with Bose-Einstein distribution i.e. for each species 
\begin{equation}
\hat{\rho} = \sum_{\{n\}} P_{\{n \}} |\{n \}\rangle \langle \{n \}|,
\qquad P_{\{n\}} = \prod_{\vec{k}} \frac{\overline{n}_{k}^{n_{k}}}{( 1 + \overline{n}_{k})^{n_{k} + 1 }},
\label{twelve}
\end{equation}
where $\overline{n}_{k}$ is the average multiplicity of each Fourier mode and $ |\{n \}\rangle = |n_{\vec{k}_{1}} \rangle \otimes  |n_{\vec{k}_{2}} \rangle \otimes  |n_{\vec{k}_{3}} \rangle\,.\,.\,.$ and the ellipses standing for all the occupied modes of the field. If the species are in local thermal equilibrium $\overline{n}_{k}$ will be given by the standard Bose-Einstein distribution.
In terms of Eq. (\ref{twelve}) the average multiplicities for the gravitons and for the phonons are given by:
\begin{equation}
\overline{n}^{\mathrm{ph}}_{k} = \frac{1}{e^{\omega^{\mathrm{ph}}/k_{T}}-1}, \qquad \overline{n}^{\mathrm{gr}}_{k} = \frac{1}{e^{\omega^{\mathrm{gr}}/k_{T}} -1},
\label{occup}
\end{equation}
having denoted with $k_{T} = T$ the effective temperature of the phonons and of the gravitons\footnote{We shall posit that phonons and gravitons have the same temperature but this assumption can be dropped if only approximate (kinetic) equilibrium holds between 
the different species.}. Note that $\omega^{\mathrm{ph}}= k c_{s}$ while $\omega^{\mathrm{gr}}= k$; here $c_{s}$ denotes the sound speed of the primordial phonons and it does not coincide necessarily with the $c_{st}$ of Eq. (\ref{NM2}) which is the sound speed of the whole plasma. Multiplying the occupation numbers of Eq. (\ref{occup})
by the corresponding energy quanta and integrating the result over the appropriate phase space, the energy density of phonons and gravitons 
becomes $g_{\mathrm{eff}}\pi^2 T_{*}^4/30$ where $g_{\mathrm{eff}}$ denotes the effective number of relativistic degrees of freedom. If only gravitons and phonons are considered $g_{\mathrm{eff}} =3$ but
$g_{\mathrm{eff}}$ maybe larger than $3$ because of other species in local thermal equilibrium.
If  $T_{max} = [ 45/(4\pi ^3 g_{\mathrm{eff}})]^{1/4} 
\sqrt{H_{*} M_{P}}$ the energy density associated with the initial state is ${\mathcal O}( H_{*}^2 M_{P}^2)$. For the forthcoming applications it will be practical to define $T_{*} = Q\, T_{max}$ where $Q\leq 1$ measures the protoinflationary temperature in units of $T_{max}$. 
The initial mixed quantum state of the system is the direct product of the quantum state of the phonons and of the gravitons, i.e. $|\Phi \rangle = |\psi_{\mathrm{phonons}}\rangle\, \otimes\, |\phi_{\mathrm{gravitons}}\rangle$ and the corresponding power spectra are obtained from the following two expectation values: 
\begin{eqnarray}
&& \langle \Phi | \hat{{\mathcal R}}(\vec{x}, \tau) \hat{{\mathcal R}}(\vec{x} + \vec{r}, \tau) | \Phi \rangle = 
\int \frac{d k}{k} \,P_{{\mathcal R}}(k,\tau) j_{0}(k r),\,\,\,\, P_{{\mathcal R}}(k,\tau) = \frac{k^3}{2\pi^2} |\widetilde{\,F\,}_{k}(\tau)|^2 ( 2 \overline{n}^{\mathrm{ph}}_{k} +1),
\label{SP1}\\
&& \langle \Phi | \hat{h}_{ij}(\vec{x}, \tau) \hat{h}^{ij}(\vec{x} + \vec{r}, \tau) | \Phi \rangle = 
 \int \frac{d k}{k} \, P_{{T}}(k,\tau) j_{0}(k r),\,\, P_{T}(k,\tau) = \frac{4 \ell_{\mathrm{P}}^2\, k^3}{\pi^2} \,  |F_{k}(\tau)|^2 (2 \overline{n}^{\mathrm{gr}}_{k} +1).
\label{SP2}
\end{eqnarray}
The spectra (\ref{SP1}) and (\ref{SP2}) must be evaluated in the limits $k/(a_{*} H_{*}) \gg 1$ (corresponding to wavelengths smaller then the Hubble radius at the protoinflationary transition) and $k/(a H) \ll 1$  (since the phenomenologically relevant wavelengths are still larger than the Hubble radius at the epoch of the matter radiation equality). In these concurrent limits Eqs. (\ref{SP1})--(\ref{SP2}) imply
\begin{equation}
r_{T}(k,\epsilon,c_{s}, k_{T}) = 16 \epsilon \biggl(\frac{k}{k_{\mathrm{p}}}\biggr)^{1 - n_{s} - n_{T}} 
 \frac{\coth{[k /(2 k_{T})]}}{\coth{[k\, c_{s}/(2 k_{T})]}}.
 \label{RATIO}
 \end{equation}
In the concordance paradigm  radiation takes places almost suddenly after inflation and
long phases (different from radiation) are excluded down to the curvature scale of matter-radiation equality. 
Under this approximation the ratio $(k/k_{T})$ is given by
\begin{equation}
\frac{k}{k_{T}} = \frac{\pi}{Q} g_{\mathrm{eff}}^{1/4}\, \biggl(\frac{4 {\mathcal A}_{{\mathcal R}}\, \epsilon}{45} \biggr)^{1/4} \biggl(\frac{k}{H_{0}}\biggr) \, e^{ - (N_{max} - N_{t})}, 
\label{num1}
\end{equation}
where $N_{max}$ has been already introduced in Eq. (\ref{RGC28}) and $N_{t}$ denotes, as usual, the total 
number of inflationary $e$-folds. If $N \gg N_{max}$ Eqs. (\ref{RATIO}) and (\ref{num1})  imply that 
 $r_{T}(k_{\mathrm{p}}) \to 16 \epsilon$. Similarly, for $c_{s} \to 1$
(i.e. when the sound speed coincides with the speed of light) $r_{T}(k_{\mathrm{p}}) = 16 \epsilon$ in spite of the total number of $e$-folds. This is the so-called consistency regime stipulating that, at the pivot scale, $r_{T} \simeq 16\epsilon \simeq - 8 n_{T}$. Whenever the speed of sound does not equal the speed of light the consistency relation is violated. According to Eqs. (\ref{num1}), if  $N_{t} \simeq {\mathcal O}(N_{max})$ then $k/k_{T} \simeq {\mathcal O}(1)$ as long as $60\leq N_{t} \leq64$,  $0.01\leq Q\leq 1$ and 
 $0.26 g_{\mathrm{eff}}^{1/4} \leq (k_{\mathrm{p}}/k_{T})\leq 0.48 g_{\mathrm{eff}}^{1/4}$. Thus
$r_{T}(k_{\mathrm{p}},\epsilon,c_{s}, k_{T})$  can be expanded for $ k_{\mathrm{p}}/(2k_{T}) < 1$ so that, approximately, $r_{T} \simeq 16 \,\epsilon \, c_{s}(T_{\mathrm{gr}}/T_{\mathrm{ph}})$ where the case of different (kinetic) temperatures has been included for completeness (in the thermal equilibrium case $T_{\mathrm{ph}}\simeq T_{\mathrm{gr}}$).  This means that the consistency relation is violated and the sound speed  of the primordial phonons can be potentially measured from independent estimates of the tensor to scalar 
ratio and of the tensor spectral index since it is still true that $n_{T} = - 2 \epsilon$.

The parametrization of the initial state based on  Eqs. (\ref{twelve}) and (\ref{occup}) 
has been firstly discussed in the case of relic gravitons in \cite{infra1} and later generalized to the case 
scalar phonons \cite{infra2,infra3,infra3a,infra3b}. Given the explicit form of the scalar and tensor 
power spectra [i.e.  Eqs. (\ref{SP1})--(\ref{SP2}) and (\ref{RATIO})] the modifications of the CMB anisotropies 
are only expected over large angular scales \cite{infra1,infra2} while over smaller angular scales (i.e. $\ell \gg 50$) the corresponding effects are negligible both in the case of gravitons and for curvature phonons; the CMB observables 
have been explicitly computed  and compared with the available data with the aim of constraining $k_{T}$ 
\cite{infra3a,infra3b,infra4} especially from the analysis of the $B$-mode polarization induced by thermal 
gravitons. Needless to say, the presence of curvature phonons in the initial state can be viewed 
as the stimulated production of inflaton quanta \cite{infra5,infra6}. The generic initial conditions 
for the scalar and tensor modes of the geometry suggested in this framework may also lead to 
computable non-Gaussianities that are particularly difficult to detect \cite{infra5,infra6}.
Equation (\ref{num1}) implies that a protoinflationary phase dominated by fluid phonons 
invalidates the consistency paradigm by impacting on the tensor to scalar ratio; the spectral dependence 
of $r_{T}$ is determined by the sound speed of the phonons and by the potentially different temperatures of the various species in kinetic equilibrium \cite{infra8,infra9}. Reversing the argument the consistency relations, even if heuristically assumed in nearly all experimental analyses, are not always tenable and bear the mark of an essential completion of the early inflationary dynamics, i.e. the protoinflationary transition. Independent measurements of $r_{T}$ and $n_{T}$ from $B$-mode polarization may offer a potentially novel diagnostic for the role played by primordial phonons in setting the initial conditions of large-scale gravitational perturbations \cite{infra10}. The violation 
of the consistency relations leads to non-Gaussianities that are too small to be observed in the 
future \cite{infra11} but that may affect the $B$-mode signal \cite{infra8,infra9,infra10,infra12}. The primordial phonons may also produce secondary spectra of relic gravitons \cite{infra10}. The 
difficulty of building metastable vacua in string theory, has led some to conjecture that in the string theory landscape the inflationary potentials must satisfy peculiar criteria (the so-called swampland criteria) which go against the paradigm provided by single-field inflationary scenarios. Initial mixed states of the relic gravitons 
 relax the tension between the swampland criteria and single-field inflationary models \cite{infra13} even if it is unclear why the possibility of this fine-tuning is excluded by more general initial states \cite{infra14}.
 
\subsection{Direct and indirect bounds on the tensor to scalar ratio}
Direct upper limits on the $B$-mode autocorrelation exist in the literature but, besides the measurements 
of the contributions coming from the lensing of the primary anisotropies, 
the empirical evidences are still lacking. The limits on the $B$-mode autocorrelations depend on the frequency band and on the specific range of harmonics. The pivot frequencies of the microwave background polarization experiments can be conventionally divided into two ranges denoted hereunder by $\nu_{low}$ and $\nu_{high}$:
\begin{equation}
26 \,\, \mathrm{GHz} \leq \nu_{low} \leq 36\,\, \mathrm{GHz}, \qquad
100 \,\, \mathrm{GHz} \leq \nu_{high} \leq 150\,\, \mathrm{GHz}.
\label{nulowhigh}
\end{equation}
To avoid possible confusions we stress that the frequencies of Eq. (\ref{nulowhigh}) 
refer to the CMB photons and {\em not} to the gravitons. The DASI (Degree Angular Scale Interferometer) \cite{LCDM1,LCDM1a,LCDM1b} and the CBI (Cosmic Background Imager) \cite{cbi1,cbi2} experiments were both working in a range coinciding exactly with $\nu_{low}$. 
Four other experiments have been conducted around $\nu_{high}$ and they are:
{\it (i)} Boomerang (Balloon Observations of Millimetric Extragalactic Radiation and Geophysics) working at $145$ GHz with 
four pairs of polarization sensitive bolometers \cite{boom}; {\it (ii)}
Maxipol (Millimeter Anisotropy experiment Imaging array) working at $140$ GHz with 12 polarimeters
 \cite{maxipol,maxipola}; {\it( iii)} Quad\footnote{An acronym or a contraction between the Quest (Q and U extragalactic sub-mm telescope) and the DASI experiments.}
working with 31 pairs of polarization sensitive bolometers: 12 at $100$ GHz and $19$ at $150$ GHz \cite{quad1,quad1a,quad2,quad2a}; 
{\it (iv)} Bicep2 \cite{BICEP2,bicep2new} and its precursor Bicep1 \cite{bicep1,bicep1a}  working, respectively, at $150$ GHz and $100$ GHz. There have been finally four polarization sensitive experiments working in mixed or intermediate frequency ranges. They include:
{\it (a)} the WMAP experiment  (see also \cite{WMAP9,WMAP9a}) spanning five frequencies from $23$ 
to $94$ GHz; {\it (b)} the Capmap experiment (Cosmic Anisotropy Polarization Mapper) \cite{capmap,capmapa}, with $12$ receivers operating between $84$ and $100$
GHz and four receivers operating between $35$ and $46$ GHz; {\it (c)} the Quiet (Q/U imager experiment) \cite{quiet,quieta} operating at $43$ GHz (during the first season of the experiment) and at $95$ GHz (during the second season of the experiment). Finally we have the Planck experiment: the three low frequency channels (i.e. $30,\,44,\,70$ GHz) belonged to the low frequency instrument (LFI);  six channels (i.e. $100,\,143,\,217,\,353,\,545,\,857$ GHz) belonged to the high 
frequency instrument (HFI). The DASI collaboration provided upper limits for the $B$-mode polarization implying $G_{\ell}^{(BB)}(\nu_{low}) < 2.12 \,\mu\mathrm{K}^2$ (for $28\leq \ell < 245$) and 
$G_{\ell}^{(BB)}(\nu_{low}) < 6.45 \,\mu\mathrm{K}^2$ (for $246\leq \ell < 420$)
at $95\%$ confidence level. As a representative of the experiments working in a mixed frequency range, it is useful to mention the WMAP bound stipulating  $G_{\ell}^{(BB)}(\nu_{V})< 0.25 \, \mu\mathrm{K}^2$ for $50 \leq\ell \leq100$ and for 
$\nu_{V} = 41$ GHz (the so called V band) \cite{WMAP9,WMAP9a}. 
The WMAP collaboration also produced a stronger bound on $G_{\ell}^{(BB)}$ for $2 \leq \ell \leq 7$. Interesting bounds on the $B$-mode power spectrum on sub-degree angular scales (i.e. $500 < \ell <2100$) have been reported also by the Polarbear experiment \cite{polar1,polar2}.

Even if direct bounds on $r_{T}$ can only be obtained from the  $B$-mode polarization,
in the context of specific inflationary models indirect bounds on $r_{T}$ may follow 
from the theoretical consistency of the inflationary models. For slow-roll inflation from the definition of the excursion of the scalar field and by using 
the consistency relations we have $\Delta\, \varphi \simeq (M_{P}/8)  \sqrt{r_{T}/\pi} \Delta\, N$.
Since the background inflates by $\Delta N \simeq 4$ during the period that wavelengths corresponding to CMB multipoles 
$2 < \ell \leq 100$ cross the Hubble radius the previous equation leads to the bound 
$\Delta \varphi > M_{P} \sqrt{ r_{T}/(4 \pi)}$ \cite{lbound}. According to this 
limit, high values of $r_{T}$ require changes in $\Delta\varphi = {\mathcal O}(M_{P})$.  A more stringent bound has been found in Ref. \cite{efs} by a statistical analysis of $2 \times 10^{6}$ slow-roll inflationary models, i.e. $\Delta \varphi \simeq 6 \, r_{T}^{1/4} \, M_{P}$ \cite{efs} (see also \cite{east,krause,baum,ads,antusch}). These bounds are not model-independent; for instance they have been claimed to be violated in the context 
of quintessential inflationary models \cite{hoss}.

\subsection{Plateau-like potentials and a theory for the scalar spectral index}
In spite of the absence of sensitive measurements of the $B$-mode autocorrelations, 
the bottom line of Tab. \ref{SEC1TABLE3} demonstrates that the limits on the 
tensor-to-scalar ratio $r_{T}$ are becoming progressively more constraining just 
by gauging the combined effects of the low-frequency gravitons on the $TT$, $EE$ and $TE$ correlations. 
Even if the bound $r_{T} < 0.1$ at a scale $k_{p}=0.002\, \mathrm{Mpc}^{-1}$ seems still reasonable, in Ref. \cite{RT3} the combination of different data sets implies $r_{T} <0.07$ while in Ref. \cite{RT4} the limit $r_{T} < 0.064$  has been claimed. The single-field inflationary scenarios where $r_{T} \ll 1$ (as opposed to the ones where 
$r_{T} \leq 1$) generally fall in the class of plateau-like potentials (see e.g. \cite{pl1,pl3,pl4,pl5} and \cite{pl6} for a review). A prototypical example along this direction is given by \cite{pl1}
\begin{equation}
W(\varphi) = \frac{3}{4}\, M^2 \, \overline{M}_{P}^2 \biggl[ 1 - e^{- \sqrt{2/3} (\varphi/\overline{M}_{P})}\biggr]^2.
\label{STT8}
\end{equation}
For the class of potentials exemplified by Eq. (\ref{STT8}), inflation occurs in the limit $\varphi \gg \overline{M}_{P}$ where $W \to 3 \, M^2 \, \overline{M}_{P}^2/3$. While in the Einstein frame the inflaton is minimally coupled and it is characterized by the potential $W(\varphi)$, the original higher-derivative action reads \cite{pl1}
\begin{equation}
S = - \frac{1}{2 \ell_{P}^2} \int \, d^4 x \, \sqrt{- g} \, f(R), \qquad f(R) = R - \frac{R^2}{6 M^2},
\label{STT8a}
\end{equation}
where $f(R)$ is a function of the Ricci scalar; for the sake of conciseness we shall stick to the explicit expression reported in Eq. (\ref{STT8a}). Indeed  by adding an appropriate Lagrange multiplier the action (\ref{STT8a}) can be 
conformally transformed to the Einstein frame \cite{pl7,pl8} and the mapping between the two 
descriptions is:
\begin{equation}
 \Omega = \frac{\partial f}{\partial R}, \qquad \biggl(\frac{\varphi}{ \overline{M}_{P}} \biggr)= \sqrt{\frac{3}{2}}\, \, \ln{\Omega},\qquad  W(R) = \frac{\overline{M}_{P}^2}{2}\frac{[f(R) - R (\partial f/\partial R)]}{(\partial f/\partial R)^2}.
\label{STT7}
\end{equation}
From Eq. (\ref{STT8}) the expression for the total number of $e$-folds $N_{t}$: 
\begin{equation}
N_{t}= \int_{\varphi_{f}}^{\varphi_{i}}\frac{d \varphi}{\overline{M}_{P}  \sqrt{2 \epsilon}}, \qquad \epsilon= \frac{4}{3 \,(e^{\sqrt{2/3} \varphi/\overline{M}_{P}} -1)^2},
\label{STT10}
\end{equation}
implies that the scalar spectral index and the consistency relations can all be expressed in terms of $N_{*}$ thanks to the definitions of the slow-roll parameters; more precisely we have\footnote{ Since $N_{t}$ denotes the total number of $e$-folds the two expressions in Eq. (\ref{STT11}) can be approximated as $n_{s} \simeq 1 - (2/N_{t}) + {\mathcal O}(N_{t}^{-2}) $ and as $ r_{T} = 12/N_{t}^2 + {\mathcal O}(N_{t}^{-3})$. Note that, according to the consistency relation of Eq. (\ref{RGC21}), the tensor spectral index only depends on $r_{T}$ since $r_{T} = - 8 n_{T}$.}
\begin{equation}
n_{s} -1 = \frac{8(4 N_{t} +9)}{(4 N_{t} +3)^2} , \qquad r_{T} = \frac{192}{(4 N_{t} +3)^2}.
\label{STT11}
\end{equation}
It follows from Eq. (\ref{STT11}) that  $r_{T}$  is suppressed as $N_{t}^{-2}$ and this occurrence is ultimately caused by the explicit form of the potential (\ref{STT8}) which is characterized by a new scale $M$ measuring the energy density of the inflaton in Planck units. According to Eqs. (\ref{STT8}) and (\ref{STT11}) the energy density of the inflaton at the onset of inflation must be 
constrained as $\rho_{inf}(t_{i}) < 10^{-12} \overline{M}_{P}^4$ (see also Fig. \ref{SEC5FIG1}). 
While different plateau-like potentials will lead to slightly different figures, the overall 
logic remains substantially the same and a plausible viewpoint suggests that this requirement 
is at odds with the logic of inflationary dynamics \cite{pl9,pl10}. The standard inflationary paradigm was originally conceived as a theory of the initial conditions of the whole universe with the aim of explaining (or relaxing) 
the drawbacks the standard cosmological scenario by starting from a rather generic 
set of Cauchy data \cite{pl11}. In the context of single-field inflationary models this requirement implies an effective
equipartition between the kinetic energy of the inflaton, its potential and the spatial gradients (or the spatial curvature)
\begin{equation}
\frac{\dot{\varphi}^2}{2} \simeq W(\varphi) \simeq \frac{1}{a^2} \partial_{k} \varphi \,\partial^{k} \varphi \simeq \frac{k}{a^2} \leq {\mathcal O} (M_{P}^4).
\label{EQP1}
\end{equation}
Since the kinetic energy, the spatial curvature (as well as other potential fluid sources) are all diluted faster than $W(\varphi)$,  Eq. (\ref{EQP1}) demands, in the conventional lore, that the Universe inflates before becoming inhomogeneous and this conclusion holds provided the background was expanding prior to $t_{i}$. Equation (\ref{STT8}) together with the suppression of the tensor to scalar ratio suggests instead that $W(\varphi) \ll M_{P}^4$. To avoid that the universe becomes inhomogeneous before inflating,  the spatial curvature 
and the gradients of the inflaton (as well as its kinetic energy) should not exceed the initial values of the potential (i.e. $k/a^2 \simeq \dot{\varphi}^2 \simeq (\partial_{k}\varphi)^2/a^2 \ll M_{P}^4)$.  This occurrence can be quantitatively scrutinized by following the evolution of the spatial gradients during the preinflationary phase \cite{pl12}.  While a complete theory of the initial conditions should account for the emergence of the observable universe from a sufficiently generic set of initial data, the logic emerging from the plateau-like potential suggests instead that inflation should be primarily regarded as a model of the spectral indices rather than a theory of the initial data. 

The suppression of $r_{T}$ may also arise for other reasons and  it has been recently suggested that,
within the Palatini approach,  the slow-roll parameters and the tensor-to-scalar ratio could be reduced in comparison with the conventional situation \cite{anto1,enqv,anto2,tenka,maeda,gialamas}. If the action 
\begin{equation}
S = - \frac{1}{2 \ell_{P}^2} \int d^{4} x\,\,\sqrt{-g} \,f(R)  + \int d^{4} x\,\ \sqrt{-g}\,\, \biggl[ \frac{1}{2} g^{\alpha\beta} \partial_{\alpha} \varphi \partial_{\beta} \varphi - V(\varphi) \biggr],
\label{oneP} 
\end{equation}
is treated within the Palatini formulation $r_{T}$ is suppressed without changing the amplitude 
of the adiabatic mode. This conclusion can be explicitly analyzed in the case where $f(R) = R - \overline{\alpha} R^2$ where 
$\overline{\alpha} = \alpha/M^2$ where $M$ is not related to the scale of the potential (\ref{STT8}). In this case the tensor to scalar ratio is suppressed as 
\begin{equation}
r_{T}^{(\alpha)} = \frac{r_{T}^{(0)}}{ \sqrt{ 1 + 8 \overline{\alpha} \ell_{P}^2 V}}, \qquad {\mathcal A}_{R}^{(0)}= {\mathcal A}_{R}^{(\alpha)}, 
\qquad 
n_{s}^{(0)} = n_{s}^{(\alpha)},
\label{twoP}
\end{equation}
where ${\mathcal A}_{R}^{(0)}$ denotes the amplitude of the scalar power spectrum for $\overline{\alpha} =0$ while ${\mathcal A}_{R}^{(\alpha)}$ denotes the scalar amplitude in the case $\alpha\neq 0$; the same notation has been employed for the scalar spectral index. The results of Eqs. (\ref{oneP}) and (\ref{twoP}) suggest that while the scalar power spectrum and the spectral index remain unaffected,  the tensor $r_{T}^{(\alpha)}$ is suppressed so that, in this framework, monomial potential currently excluded by large-scale data (e. g. $V(\varphi) \simeq \lambda \varphi^4$ ) can become again viable. Note that the potential $V$ is suppressed for $\overline{\alpha} \neq 0$ as $V\to W = V/(1 + 8 \overline{\alpha} \ell_{P}^2 V)$. In the limit $\alpha \gg 1$, 
the rescaled potential $W \to \overline{M}_{P}^2 M^2/(8 \alpha)$  
is reminiscent of the plateau of Eq. (\ref{STT8}).

\renewcommand{\theequation}{6.\arabic{equation}}
\setcounter{equation}{0}
\section{The intermediate frequency band}
\label{sec6}
The intermediate frequency band ranges approximately from the pHz to a fraction of the Hz. 
In this region the spectral energy density of the relic gravitons is bounded by the millisecond 
pulsar pulses and also constrained by the success of big-bang nucleosynthesis. 
While these limits are practically immaterial in the context of the concordance 
paradigm (where the spectral energy density decreases in frequency and exhibits 
therefore a red slope), they are essential in the case of blue or violet spectra. 
In the intermediate frequency domain large anisotropic stresses 
may come from different late-time phenomena (e.g. hypercharge fields, strongly first-order 
phase transitions, topological defects) as well as from many unresolved sources of 
gravitational radiation at the present time.  All these presumed signals have characteristic spectral slopes
and could act as diffuse foregrounds for the primordial spectra of cosmic gravitons.

\subsection{Early deviations from the concordance scenario}
The evolution of the primeval plasma may involve a contracting 
stage (i.e. $\dot{a} <\,0$  and $H<0$), as historically suggested, along slightly different perspectives, by 
Tolman and Lema\^itre \cite{orig1,orig1a}. Various recent suggestions 
can be traced back, directly or indirectly, to the original ideas of Refs. \cite{orig1,orig1a} 
even if the context and the underlying motivations are unrelated. Some rudiments of bouncing cosmologies will now be swiftly discussed with the only purpose of reviewing the ideas that are germane to our theme. 

\begin{figure}[!ht]
\centering
\includegraphics[height=5.9cm]{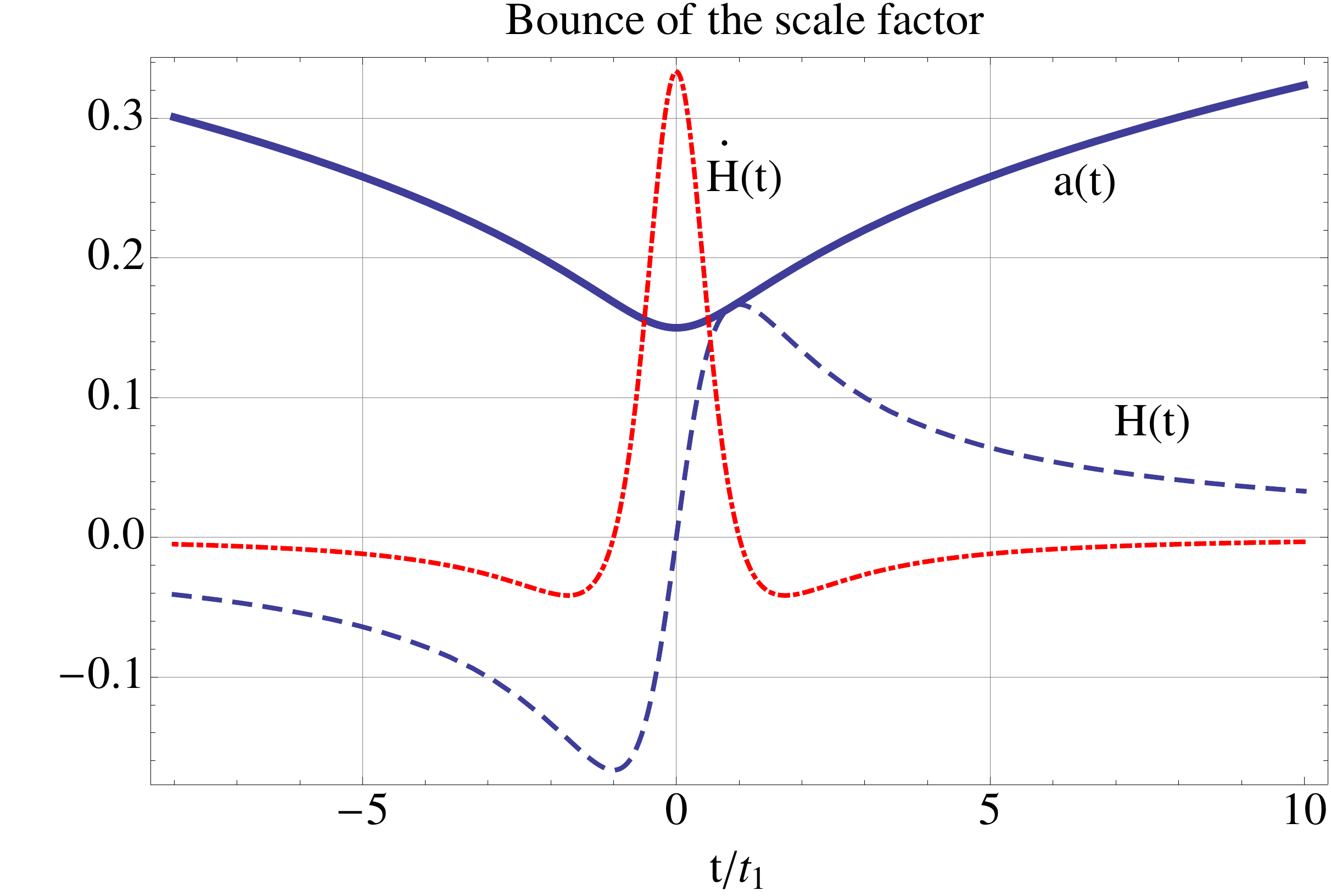}
\includegraphics[height=5.9cm]{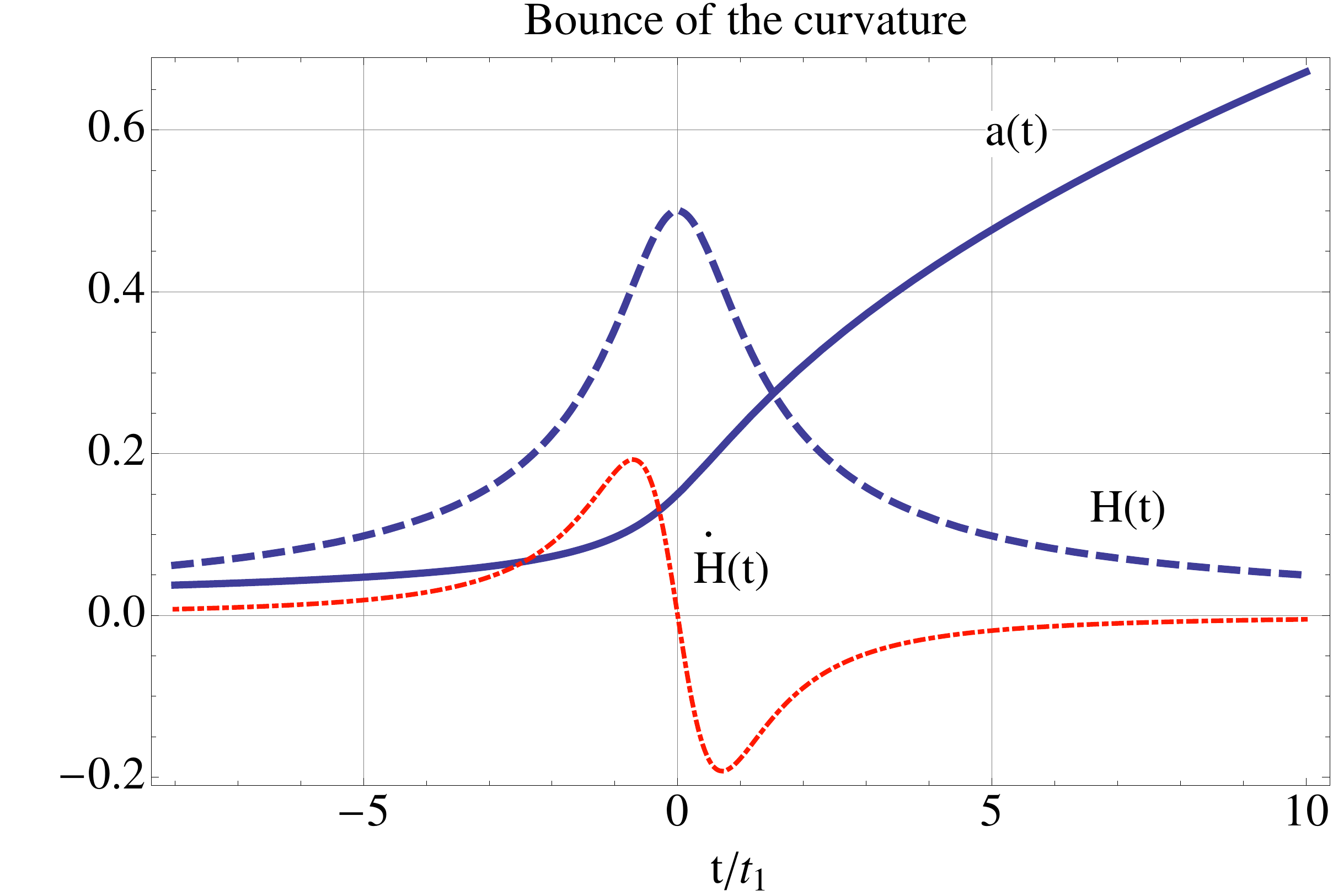}
\caption[a]{A bounce of the scale factor (plot at the left) and a curvature bounce
(plot at the right) are illustrated. The rescaled cosmic time coordinate 
is plotted on the horizontal axis while on the vertical axis we illustrate, in both plots, three complementary functions: 
the scale factor (full thick line), the Hubble rate (dashed line) and the first derivative of the Hubble rate (dot-dashed line). While the scale factors are dimensionless, the Hubble rates and their derivatives have been reported in natural gravitational units (i.e. $\ell_{P}^2 =1$).}
\label{SEC6FIG1}      
\end{figure}

The decelerated expansion (i.e. $\dot{a} > 0$ and $ \ddot{a} < 0$) of the standard cosmology
could be the late counterpart of an earlier accelerated contraction (i.e. $\dot{a} < 0$ and $ \ddot{a} < 0$).
Prior to the formulation of conventional inflationary models various dynamical realizations of this viewpoint 
have been suggested and they range from higher-order gravitational actions \cite{orig2} to the evolution of the quantum corrections 
of scalar particles minimally coupled to a classical  gravitational field \cite{orig3} (see also \cite{orig3a,orig4}).
The solutions connecting a contracting phase to an expanding regime are commonly referred to as 
{\em bounces} and a very economical example of this dynamics follows from Ref. \cite{orig3} whose numerical solutions can be parametrized as:
\begin{equation}
a(t) = a_{\ast}\biggl[(t/t_{\ast})^2 + 1\biggr]^{\alpha/2}, \qquad a_{-}(t) = a_{+}(-t),\qquad 0 < \alpha < 1,
\label{bounce1} 
\end{equation}
where, in the terminology of Ref. \cite{orig3}, $a_{-}(t)$ and $a_{+}(t)$ denote  the scale factors in the two 
asymptotic regimes $t\ll - \,t_{\ast}$ and $ t\gg t_{\ast}$. The asymptotic form of the scale factor suggested by Eq. (\ref{bounce1}) for $t \ll - t_{*}$ implies 
$a_{-}(t) = (- t/t_{\ast})^{\alpha}$ and it potentially describes three complementary sets of initial 
data for the evolution of the background; these possibilities are concisely summarized in Tab. \ref{SEC6TAB0}.
\begin{table}[!ht]
\begin{center}
\caption{Summary of the kinematic evolution of the scale factor $a_{-}(t)$ for different values of $\alpha$. }
\vskip 0.4 cm
\begin{tabular}{||c|c|c|c|c|c||}
\hline
\hline
\rule{0pt}{4ex} $\alpha$ & $\dot{a}$ & $\ddot{a}$ & $\dot{H}$ & $\epsilon= - \dot{H}/H^2$ & model \\
\hline
\hline
$\alpha<0 $& $ > 0 $ & $>0$  & $>0$ & $< 0 $ & accelerated expansion\\
$0 < \alpha < 1$& $<0$ & $ < 0 $ & $< 0 $ & $>0 $  & accelerated contraction\\
$\alpha> 1$ & $<0$ & $> 0 $ & $< 0$ & $>0 $ & decelerated contraction \\
\hline
\hline
\end{tabular}
\label{SEC6TAB0}
\end{center}
\end{table}
According to Eq. (\ref{bounce1})  the bounce of the scale factor is realized, strictly speaking, 
only when $0 < \alpha < 1$ and this situation is illustrated in the left plot of Fig. \ref{SEC6FIG1} 
(in the particular case $\alpha =1/3$). For a bounce of the scale factor the Hubble rate changes sign 
once and $\dot{H}$ changes sign twice.  While in Ref. \cite{orig3} the form of Eq. (\ref{bounce1}) 
was dictated by the concrete expression of the pressure, in the intervening time different bouncing solutions
have been critically analyzed and discussed: bounces must not be necessarily  symmetric and may even involve a finite series of contractions and expansions before the final expanding stage is reached. 
When $\alpha <0$ the asymptotic Cauchy data correspond (see Tab. \ref{SEC6TAB0}) to 
an accelerated expansion so that the scale factor, in a technical sense, does not bounce; 
nonetheless this set of Cauchy data can describe a {\em curvature bounce}, i.e. a solution 
where $H$ is always positive but $\dot{H}$  changes sign at least once.  One of the simplest 
examples of this kind of evolution follows by flipping the symmetry of Eq. (\ref{bounce1}) 
by requiring, for instance, that $a_{-}(t)/a_{\ast} = a_{\ast}/a_{+}(-t)$; a scale factor satisfying 
this condition can be written as\footnote{It can be explicitly verified that Eq. (\ref{ap40}) satisfies also $a(-t)/a_{\ast} = a_{\ast}/a(t)$.} :
\begin{equation}
a(t) = a_{\ast}\,\biggl[ (t/t_{\ast}) + \sqrt{(t/t_{\ast})^2 +1}\,\, \biggr]^{\gamma}, \qquad \gamma > 0.
\label{ap40}
\end{equation}
The asymptotic form of Eq.~(\ref{ap40}) for $ t \ll -t_{\ast}$ is given by 
$a_{-}(t) = (- t/t_{\ast})^{-\gamma}$ and this possibility is also covered by Tab. \ref{SEC6TAB0} but 
with $\alpha = -\gamma$; an explicit example of this evolution is illustrated 
in the right plot of Fig. \ref{SEC6FIG1} for $\gamma = 1/2$.
By changing the conformal frame a curvature bounce may turn into a bounce of the scale factor (and vice-versa).
This means that the bouncing dynamics is not invariant under conformal rescaling. 
However, even if bounces are {\em not} frame-invariant, the spectral energy density of the relic gravitons {\em is ultimately invariant under conformal rescaling}, as argued 
in Eq. (\ref{ACT10}) by working directly in terms of the effective action. Therefore, provided we work at sufficiently low curvatures 
the slopes of the spectral energy density of the relic gravitons have an invariant meaning and can be described 
in any frame. For these reasons we shall often employ the collective terminology {\em bouncing dynamics} to denote 
simultaneously the curvature bounces and the bounces of the scale factor.

After the first formulations of conventional inflationary models the bouncing solutions have been temporarily superseded 
but it was subsequently realized that an ever expanding de Sitter space-time is not geodesically complete in past  time-like directions \cite{DISC1,DISC2,DISC3,DISC4} and this property implies that causal geodesics cannot be extended indefinitely in the past as a function of their affine parameter. In some cases a geodesically incomplete space-time could be completed. For instance a sound geodesic completion of a conventional inflationary stage could be the so-called de Sitter bounce where the scale factor evolves as $a(t) = \cosh{H_{1} \, t}$. In this case the space-time exhibits a  decelerated contraction for $t< 0$ and an accelerated expansion for $t > 0$. The de Sitter bounce can be obtained, incidentally, also in the presence of a minimally coupled scalar \cite{DISC5}.

While it is not our purpose to describe in detail the various possibilities explored in the past, there 
is no doubt that the recent years witnessed a proliferation of bouncing scenarios which have been constructed with the aim of either complementing or even challenging the conventional inflationary ideas. Bouncing models appear in a variety of frameworks motivated by physical premises that are very different or even opposite and various reviews are currently available (see e.g. \cite{bouncing1}). In the Einstein frame both the pre-big bang \cite{bouncing2c} and the ekpyrotic \cite{bouncing3} scenarios are based on an early phase of accelerated contraction whose rate depends on the respective models. Even more recently non-singular (or quantum) bounces have been claimed to be a potential prediction of loop (quantum) cosmology \cite{bouncing3a}, of modified theories of gravity \cite{bouncing3b} and even of the so-called genesis models \cite{bouncing0,san1}. Often the solutions obtained in a certain theoretical framework are recycled, mutatis mutandis, in an entirely different situation: the kinematical features, however remain always the same even if the various authors do not seem to be aware of it. An indirect motivation for bouncing universes may also come from recent considerations stipulating that string theory provides a large landscape of vacua encompassing a number of consistent low-energy effective theories. The landscape must be however surrounded by the swampland \cite{swamp1,swamp2,swamp3,swamp4,swamp5,swamp6}, a region populated by inconsistent semi-classical effective theories.  In other words, the string landscape is surrounded by an even larger swampland of consistent-looking semi-classical effective field theories, which are actually inconsistent. A particular issue, in this context, concerns the maximally symmetric vacua and, in particular, the Minkowski, the anti-de Sitter and the de Sitter spacetimes. Since in string theory it is rather difficult to obtain consistent de Sitter vacua leading to more or less conventional inflationary scenarios, it has been proposed that de Sitter space-time is not part of the landscape, but rather lives in the swampland. The authors of Ref. \cite{swamp6} concluded that conventional inflationary models are generically in tension with the swampland criteria.

Bouncing universes are occasionally associated with various coordinate-invariant restrictions on the total energy-momentum tensor $T_{\mu\nu}$. These restrictions are conventionally referred to as energy conditions \cite{encon1,encon2} (see also \cite{encon3,encon4}) and play a central role, together with the notion of geodesic completeness, in various singularity theorems. The energy conditions are practically expressed as restrictions on certain scalar functions coming from the contraction of arbitrary time-like or null vectors with the energy-momentum tensor $T_{\mu\nu}$ whose general form will be taken to coincide, for simplicity, with 
Eq. (\ref{CUR1}).  For instance, the {\em null energy condition} demands 
that $T_{\mu\nu} k^{\mu} k^{\nu} \geq 0$ where $k^{\mu}$ is a null vector (i.e. $g_{\mu\nu} k^{\mu} k^{\nu} =0$).
For a perfect relativistic fluid o Eq. (\ref{CUR1}) the null energy condition requires that $(p_{t} + \rho_{t}) \geq 0$;
this means that the energy density is allowed to be negative as long as it is compensated by a positive 
pressure. Besides the null energy condition, the weak energy condition stipulates that 
$T_{\mu\nu} q^{\mu} q^{\nu} \geq 0$ where $q^{\mu}$ is a 
time-like vector (i.e. $q^{\mu} q^{\nu} g_{\mu\nu} =1$).  Conversely, the strong energy condition 
implies that $T_{\mu\nu} p^{\mu} p^{\nu} \geq T^{\lambda}_{\lambda} p^{\sigma} p_{\sigma}/2$
where $p^{\mu}$ is now a non-spacelike vector (i.e. either timelike or null). For spatially flat 
bouncing scenarios the most relevant energy conditions
are probably the null and the dominant energy condition; the dominant energy condition stipulates 
that $\rho_{t} \geq |p_{t} |$ \cite{encon3,encon4} and it is somehow more restrictive than the 
null energy condition. In general terms the bouncing
 solutions may violate the strong, the dominant, the null or the dominant-null energy conditions.
Although there are regions where the null energy condition can be locally violated, it may perhaps be satisfied in some averaged sense. The notion of averaged energy conditions \cite{encon2,encon5}  
aims at preventing the violation of the second law of thermodynamics and plays also a relevant role in the derivation of the so-called quantum interest conjecture \cite{encon6,encon7,encon8}. 
A direct way of introducing the averaged null energy condition is to consider 
the following integral bound along the null geodesic $\gamma$, i.e. 
$ \int_{\gamma} T_{\mu\nu} k^{\mu} k^{\nu} \, d\lambda \geq 0$ where $k^{\mu} = d x^{\mu}/d\lambda$, 
 is the tangent vector to the null geodesic and $\lambda$ 
 is an affine parameter with respect to which the tangent vector to the geodesic is defined. In general relativity the preceding integral condition can also be phrased by substituting $T_{\mu\nu}$ with the Ricci tensor $R_{\mu\nu}$: the two conditions are in fact equivalent (up to numerical factors) by definition of null vector and thanks to the field equations. It is interesting 
to consider the possibility that the null (or dominant) energy conditions 
are locally violated {\em but not in average} \cite{encon9}. 

\subsection{WKB approach and the blue spectral slopes}
A {\em sufficient} condition for a blue spectrum of relic gravitons at intermediate frequencies is the existence of a bouncing solution (either in the curvature or in the scale factor) preceding (or even replacing) a conventional stage of inflationary expansion.
This means that in the absence of bouncing solutions (and in the absence of a phase where 
the curvature scale increases in absolute value) it is still possible, as we shall see later on, 
to have blue and violet spectra at intermediate frequencies. This is why the condition is only sufficient. 
To prove this statement in general terms, the Wentzel-Kramers-Brioullin (WKB) strategy is very effective especially in the cases when the evolution of the Hubble radius is only approximately known and deviates from the conventional evolution\footnote{Note incidentally that also the conventional evolution discussed in section \ref{sec4} can be investigated within the WKB approach. By comparing the 
approximate treatment with the exact one it is possible to assess the accuracy of the WKB treatment; some elements for this 
comparison will be briefly outlined hereunder.}.  The WKB approach to the relic graviton spectra 
has been firstly discussed in the context of the conventional scenario based on an early inflationary evolution \cite{HIS16aa}. While during a stage of conventional inflationary expansion $\bigl| a H \bigr|\, \simeq a^{-1}$, in both situations illustrated in Fig. \ref{SEC6FIG1} the 
Hubble radius evolves as $\bigl |a \, H\bigr|^{-1} \propto a^{- \delta}$ with $\delta > 1$ for $t \ll - t_{*}$.
\begin{figure}[!ht]
\centering
\includegraphics[height=6.5cm]{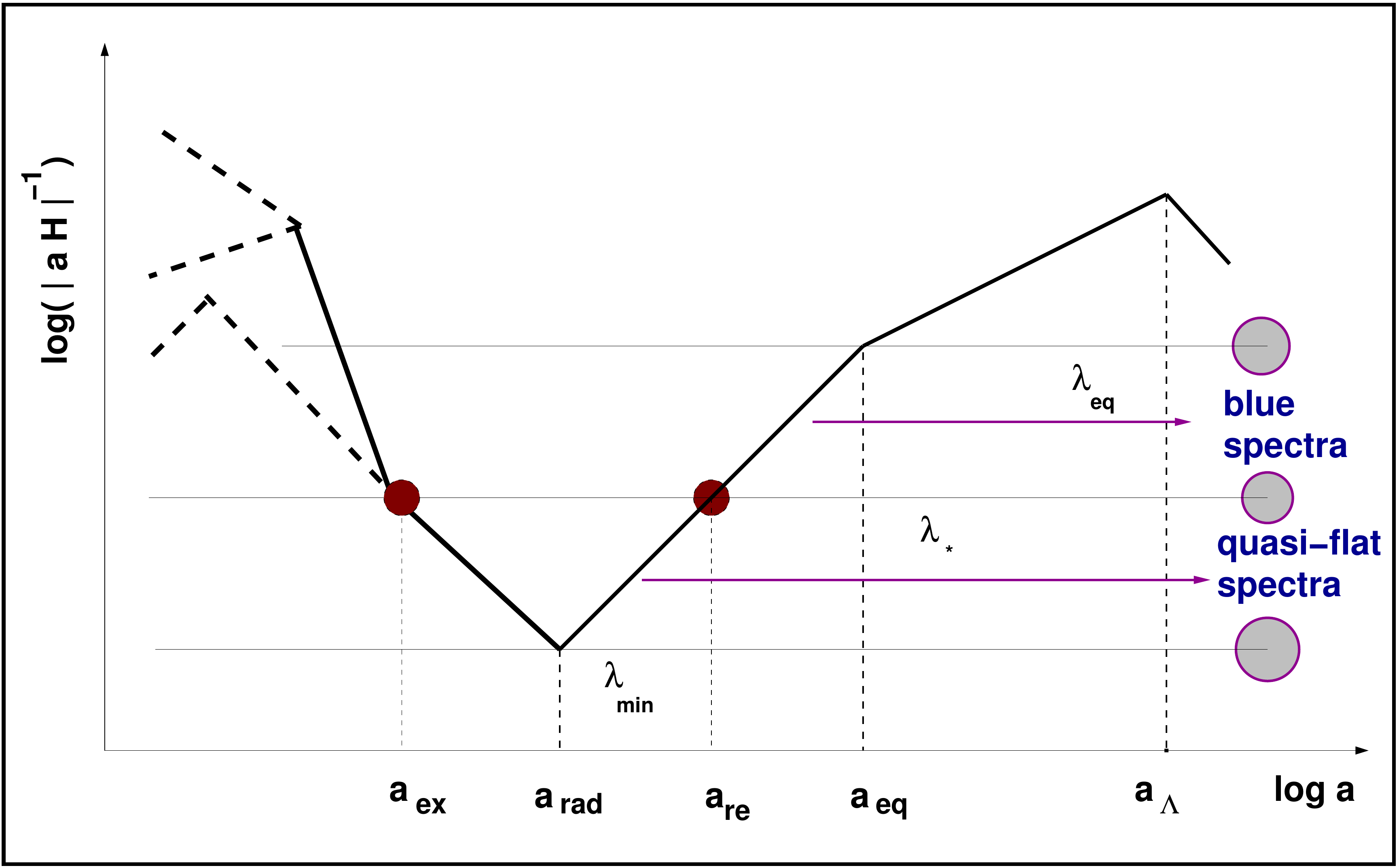}
\caption[a]{The approximate evolution of the Hubble radius is illustrated when the conventional inflationary phase does not indefinitely extend in the past and it is preceded by an epoch where the Hubble radius decreases faster than $a^{-1}$.}
\label{SEC6FIG3}      
\end{figure}
In Fig. \ref{SEC6FIG3}, with the full line we illustrate the evolution of $| a\, H \,|^{-1}$ when the conventional inflationary phase is preceded by a stage where $\delta >1$; the dashed line shows instead the conventional inflationary evolution already discussed\footnote{The first raise in the Hubble radius in Fig. \ref{SEC6FIG3} corresponds to the situation where inflation is preceded by a decelerated phase (i.e. $\ddot{a} < 0$ and $\dot{a}>0$). This possibility has been already considered at the end of section \ref{sec5} and it leads to a suppression of $r_{T}$ in the low-frequency regime whenever the total number of $e$-folds is close to critical.}
in Fig. \ref{SEC4FIG1}.  Within the WKB approximation  the evolution of the pump field can be generically expressed as 
\begin{equation}
\frac{a^{\prime\prime}}{a} = a^2 H^2 \biggl[ 2 - \epsilon(a)\biggr], \qquad \epsilon(a) = - \frac{\dot{H}}{H^2},
\end{equation}
where $\epsilon(a)$ is now a function of the scale factor (or of the conformal time coordinate) and it is not necessarily 
constant.
Recalling Eqs. (\ref{PAM30b})--(\ref{PAM30c}), the evolution of the mode functions between the turning points 
 $a_{ex}$ and $a_{re}$ (i.e.  where the solutions change their analytical form) can be approximately written as: 
\begin{equation}
F_{k}(\tau) = \frac{e^{- i\, k \, \tau_{ex}}}{a_{ex}\, \sqrt{2 k}} \biggl[ 1 - \biggl( i\, k + {\mathcal H}_{ex}\biggr) \int_{\tau_{ex}}^{\tau} \biggl(\frac{a_{ex}}{a} \biggr)^2 \, d\tau\biggr], \qquad \tau_{ex} < \tau < \tau_{re}.
\end{equation}
After the corresponding wavelengths reentered the Hubble radius the explicit form of the mode function becomes 
\begin{equation}
F_{k}(\tau) = \frac{1}{a \sqrt{2 k}} \biggl[ C_{+}(k,\tau_{ex},\tau_{re}) e^{ - i k \tau} +  C_{-}(k,\tau_{ex},\tau_{re}) e^{ i k \tau}\biggr], \qquad \tau > \tau_{re},
\label{WKB2}
\end{equation}
where $C_{\pm}(k,\tau_{ex},\tau_{re})$ follows from the continuity of $F_{k}$ and of $F_{k}^{\,\,\prime}$ in $\tau_{re}$; recalling that ${\mathcal H}_{ex} = a_{ex} \, H_{ex}$ and ${\mathcal H}_{re} = a_{re} \, H_{re}$ we have that the coefficients of Eq. (\ref{WKB2}) are:
\begin{eqnarray}
C_{\pm}(k, \tau_{ex},\tau_{re})  &=& \frac{e^{- ik (\tau_{ex} \mp \tau_{re}) }}{2 i k} \biggl[\pm \frac{a_{ex}}{a_{re}} ({\mathcal H}_{ex} + i k) \mp \frac{a_{re}}{a_{ex}} ({\mathcal H}_{re} - i k) 
\nonumber\\
&\pm&  a_{re} a_{ex} ({\mathcal H}_{ex} + i k) ({\mathcal H}_{re} \mp i k) {\mathcal J}(\tau_{ex},\tau_{re})\biggr],
\label{ap11}\\
 {\mathcal J}(\tau_{ex}, \tau_{re}) &=&  \int_{\tau_{ex}}^{\tau_{re}} \frac{d\tau}{a^2(\tau)} \equiv \int_{a_{ex}}^{a_{re}} \frac{d \,a}{H\, a^4}.
\label{ap13}
\end{eqnarray}
Since 
${\mathcal J}(\tau_{ex}, \tau_{re})$ appearing in Eq. (\ref{ap13}) can be integrated by parts, its ultimate expression becomes:
\begin{equation}
{\mathcal J}(\tau_{ex}, \tau_{re}) = \frac{1}{3 - \overline{\,\epsilon\,}}\biggl(\frac{1}{{\mathcal H}_{ex} a_{ex}^2} - \frac{1}{{\mathcal H}_{re} a_{re}^2}\biggr), \qquad \overline{\,\epsilon\,} = \frac{\int_{\tau_{ex}}^{\tau_{re}} \epsilon(\tau) a^{-2} \, d\tau}{\int_{\tau_{ex}}^{\tau_{re}} a^{-2} \, d\tau},
\label{ap18}
\end{equation}
where $\overline{\,\epsilon\,}$ is mean value of $\epsilon$ calculated using the weighting factor $a^{-2}$
while each given wavelength (exiting in $a_{ex}$ and reentering in $a_{re}$) is still larger 
than the Hubble radius. Inserting Eq. (\ref{ap18}) into Eq. (\ref{ap11}) 
the general expressions of $C_{\pm}(k,\tau_{ex}, \tau_{re})$ become:
\begin{eqnarray}
C_{\pm}(k, \tau_{ex}, \tau_{re})  &=& \frac{e^{- ik (\tau_{ex} \mp \tau_{re}) }}{2 i k} \biggl\{i k \biggl[ \biggl(\frac{a_{re}}{a_{ex}}\biggr)\pm \biggl(\frac{a_{ex}}{a_{re}}\biggr)\biggr]
\nonumber\\
&+& \biggl[\pm {\mathcal H}_{ex} \biggl(\frac{a_{ex}}{a_{re}}\biggr) \mp  {\mathcal H}_{re} \biggl(\frac{a_{re}}{a_{ex}}\biggr)\biggr]
\biggl[1 - \frac{1}{3 - \overline{\,\epsilon\,}}\biggl(1 + \frac{i k}{{\mathcal H}_{ex}}\biggr) \biggl(1 \mp \frac{i k}{{\mathcal H}_{re}}\biggr) \biggr] \biggr\}.
\label{ap19}
\end{eqnarray}
The mixing coefficients appearing in  Eq. (\ref{ap19}) assume a much simpler form 
when the different classes of turning points are specified. The two turning points follow from the solutions of the equation  $k^2 = [ 2 - \epsilon(a)] \, a^2 H^2$ at the exit and at the reentry of a given wavelength (see also Fig. \ref{SEC6FIG3}):
\begin{equation}
k^2  = ( 2 - \epsilon_{ex}) \, a_{ex}^2 \, H_{ex}^2 , \qquad k^2 = ( 2 - \epsilon_{re})\,a_{re}^2 \, H_{re}^2, \qquad \epsilon_{ex} = \epsilon(a_{ex}), \qquad \epsilon_{re} = \epsilon(a_{re}).
\label{ap21}
\end{equation}
If the wavelengths reenter during a decelerated stage of expansion 
$\epsilon_{re} = {\mathcal O}(1)$ (e.g. $\epsilon_{re} = 3(1 +  w)/2$ where
 $w$ denotes the barotropic index of the dominant component of the matter sources).
The case of radiation (i.e. $w \to 1/3$) implies $\epsilon_{re}\to 2$ so that the pump field 
{\em exactly vanishes at the turning point}. It is therefore important to appreciate that if the
 reentry (or the exit) of a given scale takes place  when $| 2 - \epsilon|\to 0$, 
then, from Eq. (\ref{ap21}), the condition $k/|a \, H| = \sqrt{|2 - \epsilon|} \ll 1$ will be verified
in the vicinity of the corresponding turning point. Conversely, whenever $\epsilon\neq 2$, the turning points are 
simply determined as $k \simeq |a_{ex}\,H_{ex}|$ and $k \simeq | a_{re}\, H_{re}|$ and this 
 conclusion applies, in particular, when the wavelengths exit (or enter) sufficiently far from the 
radiation dominated epoch where $ \epsilon \to 2$. 

In summary we can say that  if one of the conditions $ |\epsilon_{ex} -2|\ll1$, $ |\epsilon_{re} -2|\ll1$ (or both of them)
is satisfied,  {\em then at least one of the turning points (or both) will satisfy the condition
$|a \, H | \gg k$}; so for instance if a given wavelength reenters during the radiation epoch we will have that, at reentry, $ k \ll a_{re}\, H_{re}$. In spite of the value of $\epsilon$ at the turning points, Eq. (\ref{ap19}) implies  that $q_{\pm}$ are two complex numbers with $|q_{\pm}| = {\mathcal O}(1)$
where $q_{\pm}$ are defined as\footnote{This conclusion follows immediately when $\epsilon_{re} \neq 2$ and $\epsilon_{ex} \neq 2$; in this case, $k \simeq {\mathcal H}_{re}$ and $k \simeq {\mathcal H}_{ex}$.  In the complementary case (i.e. $ |\epsilon_{ex} -2|\ll1$ and $ |\epsilon_{re} -2|\ll1$) the conditions $a_{ex}\,H_{ex} \gg k$ and $a_{re} H_{re} \gg k$, once inserted in Eq. (\ref{ap22}), imply that $q_{\pm}$ are complex numbers ${\mathcal O}(1)$. The same result holds when the pump field vanishes in the vicinity of {\em only one} of the turning points (e.g.   $ |\epsilon_{ex} -2|\ll1$ and $\epsilon_{re} \neq 0$). For these arguments we are assuming that, in average, $\overline{\,\epsilon\,} \neq 3$. The case  $\overline{\,\epsilon\,} \to 3$ would correspond to a single phase where $w \to 1$ and.
For this reason when $w\to 1$ and $\overline{\,\epsilon\,} \to 3$ the integration by parts of Eq. (\ref{ap18}) should be simply avoided by going back to the original expression of Eq. (\ref{ap13}). }:
\begin{equation}
q_{\pm} = 1 - \frac{1}{3 - \overline{\,\epsilon\,}}\biggl(1 + \frac{i k}{{\mathcal H}_{ex}}\biggr) \biggl(1 \mp \frac{i k}{{\mathcal H}_{re}}\biggr).
\label{ap22}
\end{equation}
Taking into account that $|q_{\pm}| = {\mathcal O}(1)$, Eq. (\ref{ap19}) can be further simplified as:
\begin{equation}
C_{\pm}(k, \tau_{ex}, \tau_{re})  = \frac{e^{- ik (\tau_{ex} \mp \tau_{re}) }}{2 i k} \biggl\{\biggl[\pm {\mathcal H}_{ex} \biggl(\frac{a_{ex}}{a_{re}}\biggr) \mp  {\mathcal H}_{re} \biggl(\frac{a_{re}}{a_{ex}}\biggr)\biggr] q_{\pm} + i k \biggl[ \biggl(\frac{a_{re}}{a_{ex}}\biggr)\pm \biggl(\frac{a_{ex}}{a_{re}}\biggr)\biggr]\biggr\}.
\label{ap19a}
\end{equation}
We stress that Eq. (\ref{ap19a}) does not assume any specific dynamics of the 
scale factors so that we shall simply demand that $(a_{re}/a_{ex}) \gg 1$. If the 
background expands the validity of this condition is obvious. If the background contracts 
(and then expands again) the condition $(a_{re}/a_{ex}) \gg 1$ may be still verified 
provided the contracting phase is shorter than the subsequent expanding stage (i.e. $a_{re}/a_{1} \gg a_{ex}/a_{1} > 1$); this 
is, after all, the physically interesting situation so that, in the limit $a_{ex} H_{ex} \gg k$ and $ a_{re} \,H_{re} \gg k$, Eq. (\ref{ap19a}) 
becomes:
\begin{equation}
C_{\pm}(k, \tau_{ex}, \tau_{re}) =  \frac{e^{- ik (\tau_{ex} \mp \tau_{re}) }}{2 i k} \biggl[ \pm {\mathcal H}_{ex} \biggl(\frac{a_{ex}}{a_{re}}\biggr) \mp  {\mathcal H}_{re} \biggl(\frac{a_{re}}{a_{ex}}\biggr)\biggr] q_{\pm}, \,\,\,\, a_{ex} H_{ex} \gg k, \,\,\,\, a_{re} \,H_{re} \gg k.
\label{ap23}
\end{equation}
The result of Eq. (\ref{ap23}) applies, for instance, to those wavelengths exiting the Hubble radius 
during a conventional inflationary phase (i.e. $\epsilon_{ex} \ll 1$) and reentering when the 
background is dominated by dusty matter (i.e. $\epsilon_{re} \simeq 3/2$). 
Conversely, when $a_{ex}\, H_{ex} \simeq k$ and $a_{re} \,H_{re} \simeq k$ the explicit 
form of Eq. (\ref{ap19a}) is yet different:
\begin{equation}
C_{\pm}(k, \tau_{ex}, \tau_{re}) =  \frac{e^{- ik (\tau_{ex} \mp \tau_{re}) }}{2 i k}  \biggl[ i k  \mp  q_{\pm} {\mathcal H}_{re} \biggr] \biggl(\frac{a_{re}}{a_{ex}}\biggr) , \qquad a_{ex}\, H_{ex} \simeq k, \qquad a_{re}\, H_{re} \simeq k.
\label{ap24}
\end{equation}
Finally,  when $a_{ex} \, H_{ex} \simeq k$ while 
$a_{re}\, H_{re} \gg k$ Eq. (\ref{ap19a}) 
demands\footnote{Note that, in the case $a_{re}\, H_{re} \gg k$ and $a_{ex} \, H_{ex} \simeq k$ Eq. (\ref{ap22}) implies 
that $q_{+} = q_{-} = q = {\mathcal O}(1)$ and this is why $q_{\pm}$ disappeared from Eq. (\ref{ap26}).}:
\begin{equation}
C_{\pm}(k,\tau_{ex}, \tau_{re})  = \mp \frac{e^{- ik (\tau_{ex} \mp \tau_{re}) }}{2 i k} {\mathcal H}_{re} \biggl(\frac{a_{re}}{a_{ex}}\biggr), \qquad a_{ex} \, H_{ex} \simeq k, \qquad a_{re}\, H_{re} \gg k.
\label{ap26}
\end{equation}
Equation (\ref{ap26}) applies for the scales leaving the Hubble radius during a conventional stage of inflationary expansion (i.e. $\epsilon_{ex} < 1$) and reentering during radiation (i.e. $\epsilon_{re} \simeq 2$). From the purely algebraic viewpoint there is a fourth case 
where  $a_{ex}\, H_{ex} \gg k$ and $a_{re} \,H_{re} \simeq k$ which is however easily deduced 
from Eq. (\ref{ap19a}) by following the same steps leading to Eq. (\ref{ap26}). Equations (\ref{ap23}), (\ref{ap24}) 
and (\ref{ap26}) cover in practice all physically relevant 
cases and shall now be used to estimate the power spectra and the spectral energy density.

Before plunging into the actual estimates, it seems appropriate to compare the WKB results 
for the wavelengths leaving the Hubble radius during 
inflation and reentering in the radiation-dominated phase (i.e.  
$a_{ex} \, H_{ex} \simeq k$ while $a_{re}\, H_{re} \gg k$). In this case Eq. (\ref{ap26}) leads to the 
standard form of the Sakharov oscillations and, in particular to the power 
spectrum $P_{T}(k,\tau) = \overline{P}_{T}(k) \sin^2{k \, \tau}/|k\tau|^2$ where 
$\overline{P}_{T}(k) = (2/\pi^2) (H_{ex}/\overline{M}_{P})^2$.
If the exit occurs during a conventional inflationary phase 
where the scale factor goes as $a(\tau) = (-\tau/\tau_{r})^{-\beta}$ the result for $\overline{P}_{T}(k,\tau_{r})$ has been reported
in Eq. (\ref{RGC5g}) and the comparison of the two results implies:
\begin{equation}
\overline{P}_{T}(k,\,\tau_{r}) = A^2(\beta) \overline{P}_{T}, \qquad A(\beta) = \frac{2^{\beta}}{\sqrt{\pi}} \Gamma\biggl(\beta + \frac{1}{2}\biggr) \, \beta^{- \beta},
\label{ap28}
\end{equation}
where $H_{ex}$ has been estimated by recalling that $H_{ex} = k/a_{ex} = a_{r}\, H_{r} \, \beta^{\beta -1} \, |\,k\tau_{r}\, |^{1- \beta}$.  Equations (\ref{RGC5g})--(\ref{RGC5f})  and the corresponding WKB estimate
differ by a factor $A(\beta) = {\mathcal O}(1)$ which goes exactly to $1$ when $\beta \to 1$ 
and $\nu\to 3/2$. This means that the WKB results exact for the spectral slopes and 
lead to ${\mathcal O}(1)$ corrections for the corresponding amplitudes.

\subsection{Sufficient conditions for blue and violet spectra}
For wavelengths shorter than the Hubble radius {\em after reentry} (i.e. $k\tau \gg 1$ and  $\tau \geq  \tau_{re}$) the general 
 expression of the spectral energy density is:
 \begin{equation}
\Omega_{gw}(k,\tau) = \frac{k^4}{3 H^2 \overline{M}_{P}^2 \pi^2 a^4} \bigl| C_{-}(k,\tau_{ex}, \tau_{re})\bigr|^2 \biggl[ 1 + {\mathcal O} \biggl(\frac{1}{k^2 \tau^2}\biggr) \biggr],\qquad k\tau > 1, \qquad \tau > \tau_{re}.
\label{ap30}
\end{equation}
If the wavelength exited the Hubble radius when $a_{ex} H_{ex} \simeq k$ and reentered
in the vicinity of $\epsilon_{re} \to 2$ (i.e. during the radiation phase) the spectral 
energy density is obtained from the squared modulus of Eq. (\ref{ap26}) 
\begin{equation}
|C_{+}(k)|^2 \simeq |C_{-}(k)|^2 \simeq \frac{{\mathcal H}_{re}^2}{4 k^2} \biggl(\frac{a_{re}}{a_{ex}}\biggr)^2 \simeq \frac{  H^2_{re} H^2_{ex} a_{re}^4}{4 k^4}.
\label{ap32}
\end{equation}
In the second equality of Eq. (\ref{ap32}) the condition
${\mathcal H}_{ex} = a_{ex} H_{ex} \simeq k$ has been used; if 
instead  $a_{re} H_{re} \simeq k$ (i.e. $\epsilon_{re} \neq 2$), Eq. (\ref{ap24}) implies 
\begin{equation}
| C_{+}(k)|^2 \simeq |C_{-}(k)|^2 \simeq \biggl(\frac{a_{re}}{a_{ex}}\biggr)^2 \simeq \biggl(\frac{H_{ex}}{H_{re}}\biggr)^2.
\label{ap31}
\end{equation}
If $a \, H $ is roughly constant (like in the case of a Milne universe with $\ddot{a} =0$) the 
energy density of Eq. (\ref{ap30}) scales like $k^4$. Equation (\ref{ap31}) also applies 
when the wavelengths reenter during the matter-dominated epoch (where 
$\epsilon_{re} \neq 0$ and hence, $k\simeq a_{re}\, H_{re}$).

\begin{figure}[!ht]
\centering
\includegraphics[height=7cm]{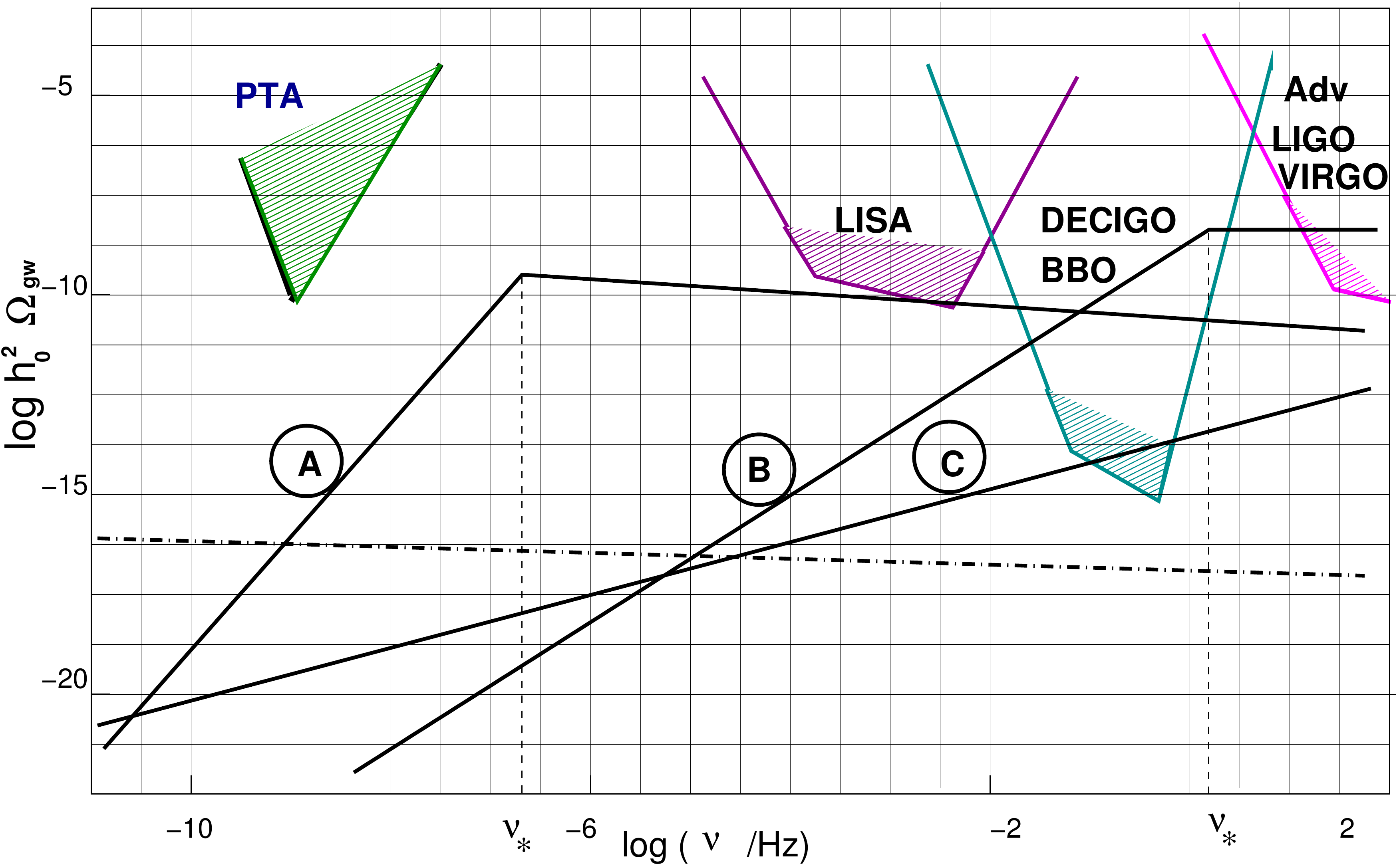}
\caption[a]{We illustrate the general features of the blue-tilted spectra at intermediate frequencies as they follow from 
the evolution of the Hubble radius of Fig. \ref{SEC6FIG3} within the WKB approximation. The shaded region at the left 
labeled by PTA denotes the approximate bounds from the pulsar timing array (see Eqs. (\ref{PUL0a}) and the specific discussion thereafter). With similar logic the shaded regions at the right illustrate the hoped 
sensitivities of the space-borne interferometers (i.e. LISA, BBO and DECIGO) and of the advanced LIGO/Virgo detectors. A thorough discussion of the interplay between the relic gravitons and the wide-band detectors of gravitational radiation can be found in section \ref{sec8}. }
\label{SEC6FIG4}      
\end{figure}
The results of Eqs. (\ref{ap30}), (\ref{ap32}) and (\ref{ap31}) allow for an accurate 
estimate of the spectral energy density in a variety of scenarios and, in particular, 
in the situations illustrated in Fig. \ref{SEC6FIG3}. For the wavelengths 
that exited the Hubble radius when $a > a_{*}$ (i.e. during a conventional 
inflationary phase) the spectral energy density following from Eqs. (\ref{ap30}) 
and (\ref{ap32}) is:
\begin{eqnarray}
\Omega_{gw}(k,\tau_{0}) &=& \frac{1}{12 \, \pi^2} \biggl(\frac{H_{r}}{\overline{M}_{P}}\biggr)^2 \, 
\biggl(\frac{a_{re}^4 \, H_{re}^2}{a_{0}^4 \, H_{0}^2}\biggr) \biggl| A\biggl( \frac{1}{1- \epsilon}\biggr)\biggr|^2 
\nonumber\\
&=& \frac{2\, \Omega_{R\,0}}{3 \pi} \biggl(\frac{H_{r}}{M_{P}}\biggr)^2 \,\biggl| A\biggl( \frac{1}{1- \epsilon}\biggr)\biggr|^2\, \biggl| \frac{k}{a_{r} \, H_{r}} \biggr|^{ - 2\epsilon/(1 - \epsilon)}
\to \frac{2\, \Omega_{R\,0}}{3 \pi}\, \biggl(\frac{H_{r}}{M_{P}}\biggr)^2,
\label{ap33}
\end{eqnarray}
where as in Eq. (\ref{ap28}) we used that $H_{ex} = k/a_{ex} = a_{r}\, H_{r} \, \beta^{\beta -1} \, |\,k\tau_{r}\, |^{1- \beta}$. 
Note that, when $a< a_{*}$,  the approximate evolution of the scale factor can be parametrized
 as $a(t) \simeq (- t/t_{*})^{\alpha}$ so that, in the conformal time coordinate, the corresponding expression of the scale factor is given by $a(\tau) =(- \tau/\tau_{*})^{-\beta}$ where now $\beta = \alpha/(\alpha -1)$.  From Eqs. (\ref{ap30}) and (\ref{ap32}) 
the spectral energy density corresponding to the wavelengths that exited the Hubble radius for $a <a_{*}$ and reentered for $a< a_{eq}$ (i.e. $\lambda_{*} < \lambda< \lambda_{eq}$)  is given by: 
\begin{equation}
\Omega_{gw}(k,\tau_{0}) = \frac{2\, \Omega_{R\, 0}}{3 \pi} \, \biggl(\frac{H_{r}}{M_{P}} \biggr)^2 
\biggl(\frac{k}{a_{*}\, H_{*}}\biggr)^{n_{T}}, \qquad n_{T} = \frac{2}{1 -\alpha},
\label{ap34}
\end{equation}
where $n_{T}$ will denote hereunder the spectral slope in the intermediate frequency range.
Finally, inserting Eq. (\ref{ap31}) inside Eq. (\ref{ap30}) and recalling Fig. \ref{SEC6FIG3}, the spectral energy density 
in the larger-wavelength limit (i.e.  $\lambda > \lambda_{eq}$) is :
\begin{equation}
\Omega_{gw}(k,\tau_{0}) = \frac{2 \Omega_{R0}}{3 \pi} \,\biggl(\frac{H_{r}}{M_{P}}\biggr)^2\, 
\biggl(\frac{k}{a_{*}\, H_{*}}\biggr)^{n_{T}}\, \biggl(\frac{k}{a_{eq}\, H_{eq}}\biggr)^{-2}.
\label{ap35}
\end{equation}
By putting together all the results of  Eqs. (\ref{ap33}), (\ref{ap34}) and (\ref{ap35}) we can derive the following {\em physical template 
family} for the spectral energy distribution of the relic gravitons:
\begin{eqnarray}
\Omega_{gw}(\nu,\tau_{0}) &=& \frac{2 \Omega_{R0}}{ 3\pi} \biggl(\frac{H_{r}}{M_{P}}\biggr)^2, \qquad 
\nu_{*} < \, \nu <\, \nu_{r}, 
\label{ap36}\\
\Omega_{gw}(\nu,\tau_{0}) &=& \frac{2 \Omega_{R0}}{ 3\pi} \biggl(\frac{H_{r}}{M_{P}}\biggr)^2 \biggl(\frac{\nu}{\nu_{*}}\biggr)^{n_{T}}, \qquad 
\nu_{eq} < \, \nu <\, \nu_{*}, 
\label{ap37}\\
\Omega_{gw}(\nu,\tau_{0}) &=& \frac{2 \Omega_{R0}}{ 3\pi} \biggl(\frac{H_{r}}{M_{P}}\biggr)^2 \biggl(\frac{\nu}{\nu_{*}}\biggr)^{n_{T}} \biggl(\frac{\nu}{\nu_{eq}}\biggr)^{-2}, \qquad 
 \nu <\, \nu_{eq}.
\label{ap38}
\end{eqnarray}
The template given in Eqs. (\ref{ap36}), (\ref{ap37}) and (\ref{ap38}) is illustrated 
in Fig. \ref{SEC6FIG4} for three typical situations where the spectral energy density 
{\em increases over the intermediate frequency range}.  According to Eqs.  (\ref{ap34}) 
and (\ref{ap37}) {\em  the spectral energy density grows for $\nu_{eq} < \nu < \nu_{*}$  provided $n_{T} > 0$ (i.e. 
$\alpha < 1$)}. Since the condition $\alpha < 1$ is realized both in the case of a 
negative $\alpha$ (i.e. $\alpha < 0$) and for positive $\alpha$ (i.e. $0< \alpha < 1$)
the spectral energy density increases both in the case of 
accelerated contraction (i.e. $\dot{a} <0$ and $\ddot{a} <0$) 
and for accelerated expansion with growing curvature 
(i.e. $\dot{a} >0$, $\ddot{a} >0$ and $\dot{H}>0$). Bouncing solutions 
of the type discussed in  in Tab. \ref{SEC6TAB1} and Fig. \ref{SEC6FIG1}
{\em are therefore a sufficient condition for a blue or violet spectral index over intermediate 
frequencies}.

The observation that a phase of accelerated expansion with $\dot{H} >0$ may lead to a 
growing spectrum of relic gravitons has been suggested in Ref. \cite{HIS16a}. In a different conformally 
related frame the evolution of the scale factor may change but the spectral slope is frame-invariant 
\cite{HIS18} (see also Eq. (\ref{ACT11}) and discussion therein).  In the case $\alpha < 0$ we can always set, 
for convenience, $\alpha = -\gamma$ so that the relevant 
scale factor will be, asymptotically, $a(t) \simeq (- t/t_{\ast})^{-\gamma}$ and it coincides with the 
 limit of Eq. (\ref{ap40}) for $t \ll - t_{*}$. According to Eq. (\ref{ap34}) the spectral slope is given, in this case, by 
$n_{T} = \gamma/(\gamma +1)$. If $\gamma = 1/2$ the spectrum is mildly violet 
(i.e. $n_{T} = 4/3$) while Eq. (\ref{ap40}) implies that the asymptotic evolution for $t >0$ 
corresponds to a radiation-dominated background. This spectrum and the related scale factor (\ref{ap40}) for $\gamma =1/2$ arose in a simple model of curvature bounce driven by bulk viscous stresses in the Einstein frame \cite{BS1,BS2}. Explicit examples with a similar analytic form may occur in a conformally related frame\footnote{In the  string frame four-dimensional 
solutions of the type (\ref{ap40}) with $\gamma = 1/\sqrt{3}$ have been discussed in \cite{BS4,BS5}; these solutions describe 
bounces of the scale factor in the Einstein frame. In Ref. \cite{BS3} more general solutions  (with a generic exponent) 
have been studied in the conformal time parametrization. For the reheating mechanisms in bouncing universes 
see e.g. \cite{BS3a}.} 
 \cite{BS4,BS5,BS3} or in the context of double field theory \cite{BS6,BS7} which was proposed to realize $T$-duality explicitly at the level of component fields of closed string field theory: since the nonlocal potentials of \cite{BS4,BS5,BS3} depend on a $T$-duality invariant combination, the corresponding solutions can also be interpreted in a double field theory context \cite{BS8,BS9,BS10}.  In the Einstein frame curvature bounces {\em exactly} in the form (\ref{ap40}) have been also dubbed genesis scenarios  (see e.g. \cite{bouncing0,bouncing0a,bouncing0b} and references therein). Even if the same kind of solutions keep on reappearing in different constructions (but with a different nomenclature) what matters here is that a phase of accelerated expansion with growing Hubble rate (see Tab. \ref{SEC6TAB0} and Eq. (\ref{ap40})) leads to a growing spectral energy density. 

In the case $0< \alpha < 1$,  Eq. (\ref{ap34}) implies that the spectral energy density 
is still increasing even if the slopes may be even larger. For instance when  $\alpha = 1/3$ we would have  $n_{T} \to 3$ from Eq. (\ref{ap34}): we recover the standard case of dilaton-driven evolution typical of the pre-big bang scenario \cite{HIS18,HIS20b,bouncing2c}. In the case of the ekpyrotic scenario \cite{HIS20c,HIS20d} (see also 
\cite{bouncing3}) the contraction is much slower than in the case $\alpha = 1/3$; as in Eq. (\ref{ap34})  
the spectral index is then given by $n_{T} = 2/( 1 - \alpha)$ with $\alpha\to 0$, i.e. $n_{T} \simeq {\mathcal O}(2)$. The case $\alpha =1/3$ has been recently claimed to be natural or anyway possible in the context of bouncing models coming from effective theories \cite{bouncing0}. Since some of the current approaches even renounce general covariance and often impose ad hoc symmetries (like the shift symmetry), it is possible that the propagating speed of the scalar modes becomes negative \cite{bouncing0c,bouncing0d,bouncing0e}. This is the so-called 
gradient instability (see also \cite{BS3} and discussion therein) that may add further constraints 
on the practical realization of a given scenario. 

The template discussed in Eqs. (\ref{ap36}), (\ref{ap37}) and (\ref{ap38}) contains in a nutshell 
three parameters:  the maximal Hubble rate in Planck units (i.e.  $(H_{r}/M_{P})$), the spectral index $n_{T}$ and the frequency range of the blue (or violet) spectrum, i.e. $(\nu_{*}/\nu_{r})$. The maximal Hubble rate in Planck units ultimately determines $\nu_{r}$ (i.e. the  maximal frequency of the spectrum) 
\begin{equation}
\nu_{r} = 1.17 \times 10^{2} \, \biggl(\frac{h_{0}^2 \Omega_{R\, 0}}{4.15 \times 10^{-5}} \biggr) 
\, \biggl(\frac{H_{r}}{M_{P}}\biggr)^{1/2} \,\, \mathrm{GHz}.
\label{ap39}
\end{equation}
Equation (\ref{ap39}) has the same content of Eqs. (\ref{ONEb}) and (\ref{RGC24}) with the 
difference that $H_{r}$ is not fixed as it the case of conventional inflationary models. 
The results of Eqs. (\ref{ap36}), (\ref{ap37}) and (\ref{ap38}) 
are schematically illustrated in Fig. \ref{SEC6FIG4} where the cases labeled by $A$, $B$ and $C$ 
denote the three relevant phenomenological situations that may arise when $n_{T} >0$. In the case $A$ the spectral energy density increases for $\nu < \nu_{*}$ while it is either flat or decreasing for $\nu > \nu_{*}$. When $\mathrm{nHz} < \nu_{*} < \mathrm{mHz}$
(see the $A$-curve in Fig. \ref{SEC6FIG4})  the spectral energy density is constrained 
by the pulsar timing arrays, by the BBN and by the direct bounds obtained by wide-band interferometers (see section \ref{sec7}). If $\mathrm{mHz} < \nu_{*} < \mathrm{kHz}$ (see the $B$-curve in Fig. \ref{SEC6FIG4}) the pulsar bounds are practically immaterial.  Finally if $\nu_{*} > \mathrm{kHz}$ (see the $C$-curve in Fig. \ref{SEC6FIG4}) the only relevant constraint is provided by the BBN limits.

In Fig. \ref{SEC6FIG4} we also illustrated the approximate sensitivities of the advanced LIGO/Virgo detectors \cite{LV1,LV2} and of the futuristic space-borne detectors such as LISA
 (Laser Interferometer Space Antenna) \cite{LISA,LISAa}, BBO (Big Bang Observer) \cite{BBO}, and DECIGO (Deci-hertz Interferometer Gravitational Wave Observatory) \cite{DECIGO1}. As we shall see in section \ref{sec8} 
 the current limits on the relic graviton backgrounds from the operating interferometers are quite constraining 
 in the range between few Hz and $10$ kHz; since the signal to noise ratio depends on the slope 
 of the spectral energy density also the bounds will depend on the frequency behaviour of $\Omega_{gw}$. 
 In general terms the bounds {\em are comparatively more constraining for growing spectral slopes than for 
 flat slopes}. The bounds coming from the detectors 
 now operating have been taken into account in Fig. \ref{SEC6FIG4}. If the bounds from big-bang nucleosynthesis 
 are combined with the pulsar timing constraints\footnote{For the big-bang nucleosynthesis bound see Eq. (\ref{BBN1}) and discussions therein. For the pulsar timing constraints see Eq. (\ref{PUL0a}) and discussion 
 thereafter.} and with the current bounds from wide-band detectors  the most promising options for the 
 advanced LIGO/Virgo detectors seem the models falling in the class $B$ and $C$.
 The models falling in the class $A$ may lead to signals that are very similar to the ones 
 obtainable from the cosmic defects. It is finally interesting that 
 LISA, BBO/DECIGO and the advanced LIGO/Virgo detectors will together cover a rather 
 broad frequency interval (of about $10$ orders of magnitude) from the $\mu$Hz region up to the $kHz$ band.

\subsection{Early evolution of the refractive index of the relic gravitons} 
In the preceding subsection we stressed that the bouncing dynamics  is a condition 
for the growth of the spectral energy density  over intermediate frequencies.
The derived condition is {\em only sufficient } since the spectral energy distribution can increase over intermediate and high frequencies 
 also because of more mundane possibilities. For instance the relic gravitons may develop an effective index of refraction when they travel in curved space-times \cite{SIXPOL8,SIXPOL9} and their spectral energy distribution becomes comparatively larger than in the conventional situation \cite{SIXPOL10}. If the refractive index
{\em increases during a quasi-de Sitter stage of expansion}, the propagating speed 
diminishes and $\Omega_{gw}(\nu,\tau_{0}) \propto 
\nu^{n_{T}}$ (with $n_{T}>0$) for $\nu$ ranging, approximately, between the nHz and $100$ MHz.
The evolution of the tensor modes of the geometry in the presence of a dynamical 
refractive index $n(\tau)$ can be parametrized by the following effective action:
\begin{equation}
 S = \frac{1}{8 \ell_{P}^2}  \int d^{3} x \int d\tau a^2 \biggl[\partial_{\tau} h_{ij} \partial_{\tau} h_{ij} - 
 \frac{1}{n^2} \partial_{k} h_{ij} \partial_{k} h_{ij} \biggr].
\label{3firstone}
\end{equation}
When $n\to 1$ Eq. (\ref{3firstone}) reproduces the standard effective 
action of Eq. (\ref{ACT9}). After the suggestion of Ref. \cite{SIXPOL10}, other authors 
concocted slightly different parametrizations \cite{SIXPOL11,SIXPOL11a} that are related to 
Eq. (\ref{3firstone}) by trivial conformal rescalings \cite{SIXPOL12}.
In both situations the modes of the gravitational field are amplified because of the presence of an effective horizon whose specific evolution is affected by the dynamics of the refractive index. It is 
sufficient to introduce in Eq. (\ref{3firstone}) a generalized time coordinate, conventionally denoted by $\eta$
and obeying  $n(\eta) d\eta = d\tau$. In terms of the new time coordinate the action 
of Eq. (\ref{3firstone}) becomes 
\begin{equation}
 S = \frac{1}{8 \ell_{P}^2}  \int d^{3} x \int d\eta\, b^2(\eta) \biggl[\partial_{\eta} h_{ij} \partial_{\eta} h_{ij} - 
 \partial_{k} h_{ij} \partial_{k} h_{ij} \biggr], \qquad b(\eta) = \frac{a}{\sqrt{n}}.
\label{3first}
\end{equation}
The function $b(\eta)$ in Eq. (\ref{3first}) plays the role of an effective scale factor since, in the limit $n\to 1$,  $\eta$ coincides with $\tau$, i.e. $b(\eta) \to a(\tau)$. When $n\neq 1$ the evolution of $b(\eta)$ defines an effective horizon ${\mathcal F} = \partial_{\eta} b/b$. 
\begin{figure}[!ht]
\centering
\includegraphics[height=5.7cm]{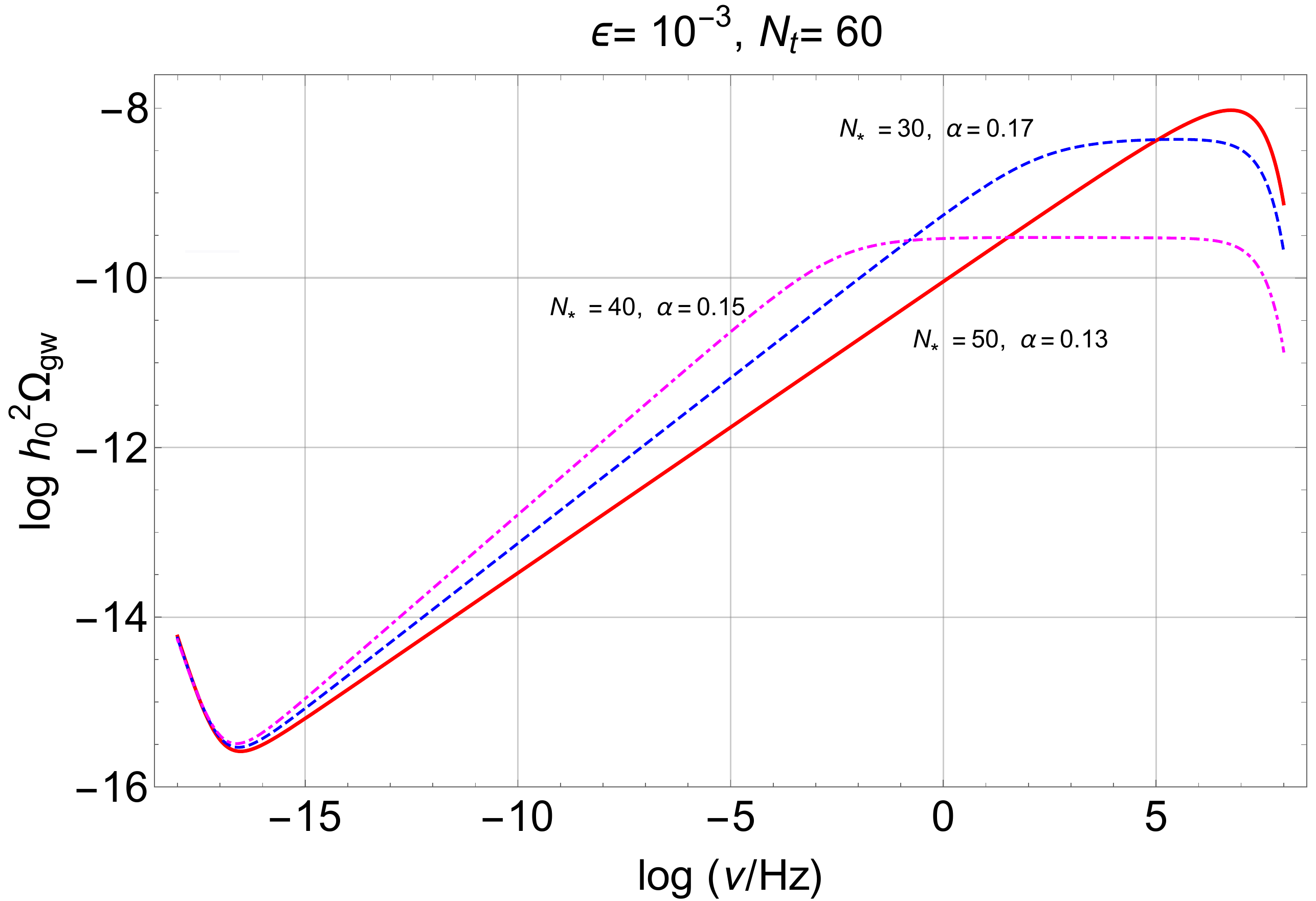}
\includegraphics[height=5.7cm]{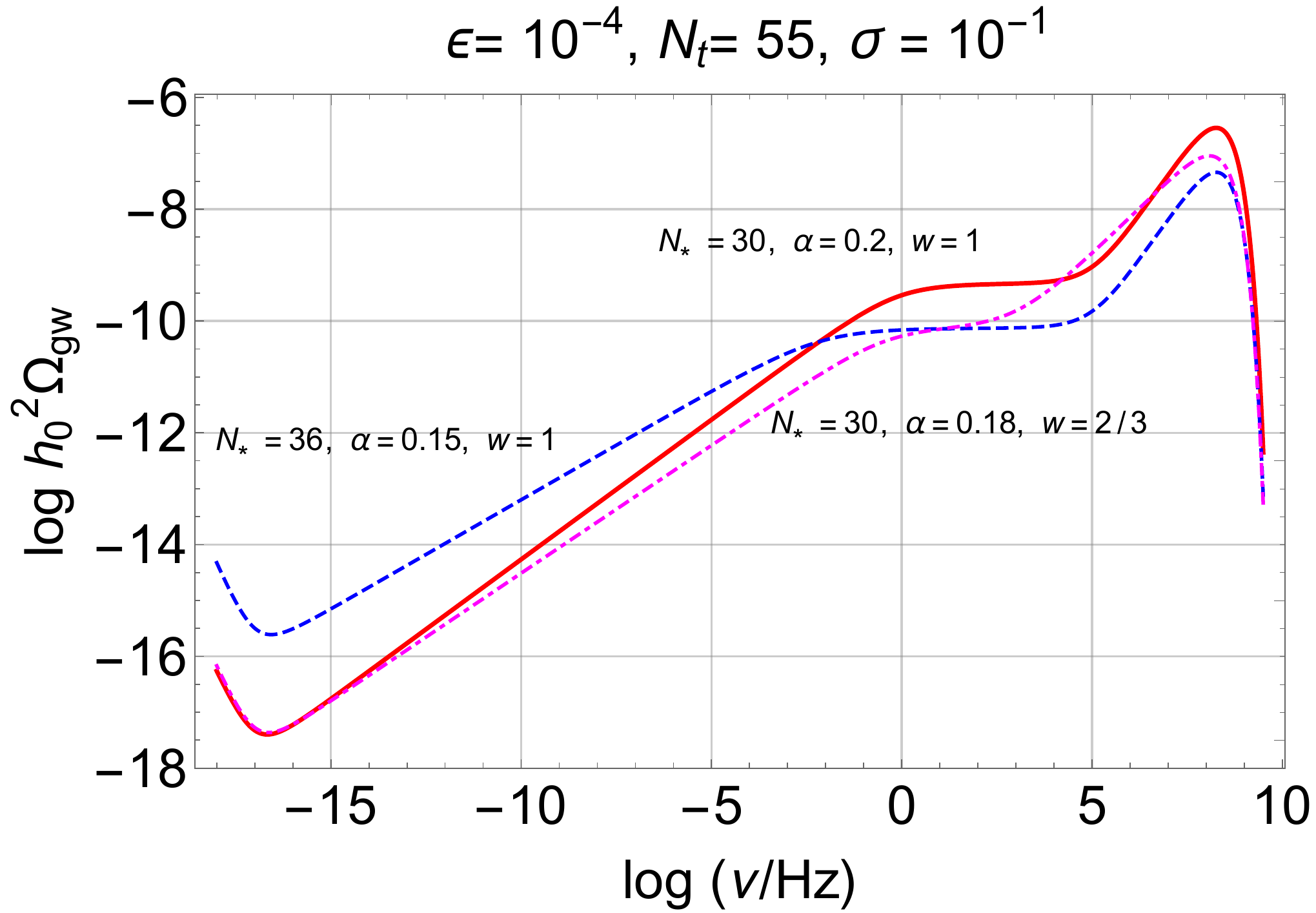}
\caption[a]{The spectral energy distribution of the relic gravitons produced 
by the variation of the refractive index during inflation is illustrated as a function of the comoving 
frequency for two broad classes of post-inflationary evolutions. The 
energy density is measured in critical units and common logarithms are employed on both axes. }
\label{SEC6FIG5}      
\end{figure}
The spectral energy distribution of the relic gravitons produced by the 
variation of the refractive index is illustrated in Fig. \ref{SEC6FIG5} where, in both plots, $N_{t}$ and 
$N_{*}= a_{*}/a_{i}\leq N_{t}$ denote, respectively, the total and the critical number of $e$-folds beyond 
which the refractive index goes back to $1$. While the explicit evolution may vary \cite{SIXPOL8,SIXPOL9,SIXPOL10}, the results of Fig. \ref{SEC6FIG5} refer to the situation where 
the refractive index evolves as a power of the scale factor $a$ during inflation\footnote{The rate of variation of the refractive index during the inflationary stage of expansion is given by $\alpha$ in units of the Hubble rate;
note also that, by definition, $n_{*} = n_{i} (a_{*}/a_{i})^{\alpha}$ with 
$n_{i} = 1$. } i.e. $n(a) = n_{*} (a/a_{*})^{\alpha}$ for $a< a_{*}$, while  $n(a) \to 1$ when $a > a_{*}$.  An explicit profile with this 
property is $n(x) = (n_{*} x^{\alpha} e^{- \xi x} +1)$ where $x = a/a_{*}$; $\xi$ controls 
the sharpness of the transition and we shall bound the attention to the case $\xi>1$ 
(in practice $\xi  = 2$ even if larger values of $\xi$ do not change the conclusions reported here). According to Fig. \ref{SEC6FIG5}, if $N_{*}$ is just slightly smaller than $N_{t}$ 
the spectral energy density is increasing in the whole range 
of comoving frequencies between $\nu_{eq}= {\mathcal O}(100) \, \mathrm{aHz}$ and $\nu_{max}= {\mathcal O}(200)\, \mathrm{MHz}$. As  soon as $N_{*}$ diminishes substantially, $\Omega_{gw}(\nu,\tau_{0})$
develops a quasi-flat plateau whose slope is controlled by the slow-roll parameter $\epsilon$ (see dashed and dot-dashed curves in the left plot of Fig. \ref{SEC6FIG5}).  The knee and the end point of the spectral energy distribution are fixed by the typical frequencies $\nu_{*}$ and $\nu_{max}$:
\begin{equation}
\nu_{*} = p(\alpha, \epsilon, N_{*}, N_{t}) \, \nu_{max}, \qquad  
p(\alpha, \epsilon, N_{*}, N_{t})= \biggl| 1 + \frac{\alpha}{1- \epsilon}\biggr| e^{N_{*}(\alpha+ 1) - N_{t}}.
\label{ONEaa}
\end{equation}
 In the right plot of Fig. \ref{SEC6FIG5} the spike at the end of the quasi-flat plateau
extends into the GHz band. Figure \ref{SEC6FIG5} demonstrates that the spectral energy density may consist of two, three or even 
four different branches: the infrared branch (between the aHz and $100$ aHz) is supplemented by a second mildly increasing branch extending between $100$ aHz and $200$ MHz. 
When the variation of the refractive index terminates before the end of inflation 
a third branch develops between $\nu_{*}$ and $\nu_{max}$ and corresponds to a quasi-flat plateau. 
Finally if the post-inflationary evolution is dominated, for some time, 
by a stiff barotropic fluid, then a fourth branch arises between few kHz 
and the GHz region. For a variety of post-inflationary histories, including the conventional one, 
a dynamical index of refraction leads to a potentially detectable spectral energy density in the kHz and in the 
mHz regions while all the relevant phenomenological constraints are concurrently satisfied \cite{refr19}.
The comparison of Figs. \ref{SEC6FIG4} and \ref{SEC6FIG5} shows a substantial analogy 
in the two cases and supports the sufficiency of the conditions that produce an increasing spectral energy distribution\footnote{
To further support the sufficiency of the requirements deduced in the previous subsection, in section \ref{sec7} a further class of examples will be examined and it will be shown that blue spectral 
slopes at high-frequencies may also arise because of the presence of stiff post-inflationary phases. }.

\subsection{Big-bang nucleosynthesis}
Approximately one quarter of the baryonic matter in the Universe is in the 
form of $^{4}\mathrm{He}$. The remaining part is made, predominantly, by Hydrogen
in its different combinations (atomic, molecular and ionized). During BBN the protons and neutrons 
(formed during the quark-hadron phase transition) combine to form nuclei. The quark-hadron phase transition takes place when the Universe 
is approximately old\footnote{Since during the radiation epoch $H^{-1} = 2 t$, 
$t_{H} = 23\,\,(106.75/g_{\rho})^{1/2} (100\, \mathrm{GeV}/T) \,\,{\mathrm{psec}}$ 
which shows that the Hubble time is of the order of $20$ psec at the electroweak time while it is $t\simeq 0.73 \,\mathrm{sec}$ right before electron positron annihilation and for $T\simeq\mathrm{MeV}$.}  of $20 \mu\mathrm{sec}$ to be compared with $t_{\mathrm{BBN}}\simeq \mathrm{sec}$.
During BBN only light nuclei are formed and, more specifically, $^{4}\mathrm{He}$, $^{3}\mathrm{He}$,
$D$, $^{7}\mathrm{Li}$.  The $^{4}\mathrm{He}$ has the largest binding energy for nuclei with 
atomic number $A<12$ (corresponding to carbon) and this is why it is the most abundant element
among the light nuclei that are supposed to provide stars with the 
initial set of reactions necessary to turn on the synthesis of heavier elements (iron, cobalt and so on). In short, the logic of BBN can be easily summarized by the following three relevant stages.  After the quark-hadron phase transition the antinucleons annihilate with the nucleons, thus 
the total baryon concentration will be given by $n_{\mathrm{B}} = n_{\mathrm{n}} + n_{\mathrm{p}}$.
If the baryon number is conserved (as it is rather plausible) the baryon concentration 
will stay constant and, in particular, it will be $n_{\mathrm{B}} \simeq 10^{-10} n_{\gamma}$ where 
$n_{\gamma}$ denotes, as usual, the photon concentration \cite{book1,book2,book3}.
For temperatures lower than  $T\simeq {\mathcal O}(\mathrm{MeV})$ weak interactions 
fall out of equilibrium; at this stage the concentrations of neutrons and protons are determined 
from their equilibrium values, i.e. approximately
\begin{equation}
\frac{n_{\mathrm{n}}}{n_{\mathrm{p}}} = e^{- \frac{Q}{T}}\simeq \frac{1}{6},\qquad T\simeq 0.73\,\, \mathrm{MeV},\qquad Q = m_{\mathrm{n}} -m_{\mathrm{p}}.
\label{1/6}
\end{equation}
If nothing else would happen, the neutron concentration would be progressively 
depleted by free neutron decay, i.e. $n \to p + e^{-} + \overline{\nu}_{e}$.
However, when $T\simeq 0.1$ MeV the reactions for the formation of the deuterium (D) are 
in equilibrium, i.e. $p + n \to D+ \gamma$ and $D + \gamma \to p + n$. The reactions for the formation of deuterium fall out of thermal equilibrium only at a much 
lower temperature (i.e. $T_{D} \simeq 0.06$ MeV).
As soon as deuterium is formed, $^{4}\mathrm{He}$ and $^{3}\mathrm{He}$ can arise
according to the following chain of reactions:
\begin{eqnarray}
&& D + n \to T +\gamma,\qquad T + p \to ^{4}\mathrm{He} + \gamma,
\label{He1}\\
&& D + p \to ^{3}\mathrm{He} + \gamma,\qquad ^{3}\mathrm{He}+ n \to ^{4}\mathrm{He} +\gamma.
\label{He2}
\end{eqnarray}
As soon as the helium is formed there is some sort of very lucky coincidence: since the temperature of equilibration 
of the helium is rather large (i.e. $T\simeq 0.3$ MeV), the helium is not in equilibrium at the moment when it is formed. In fact the helium can only be formed when deuterium is already present. As a consequence the reactions (\ref{He1}) and (\ref{He2}) only take place from right to left.
When the deuterium starts being formed, free neutron decay has already depleted a bit the 
neutron to proton ratio which is equal to $1/7$. 
Since each $^{4}\mathrm{He}$ has two neutrons, we have that $n_{\mathrm{n}}/2$ nuclei of 
 $^{4}\mathrm{He}$ can be formed per unit volume. Therefore, the  $^{4}\mathrm{He}$ mass fraction will be 
 \begin{equation}
 Y_{p} = \frac{4(n_{\mathrm{n}}/2)}{n_{\mathrm{n}} + n_{\mathrm{p}}} \simeq \frac{2(1/7)}{1 + 1/7} \simeq 0.25.
\label{heab}
\end{equation}
The abundances of the other light elements are comparatively smaller than the one of  $^{4}\mathrm{He}$ and, in
particular:
\begin{equation}
D/H \simeq 10^{-5},\qquad ^{3}\mathrm{He}/H \simeq 10^{-3},\qquad ^{7}\mathrm{Li}/H \simeq 10^{-10}.
\end{equation}
The abundances of the light elements computed from BBN calculations agree with the observations.
The simplest BBN scenario implies that the only two free parameters are the temperature and 
\begin{equation}
\eta_{b0} = \frac{n_{b0}}{n_{\gamma0}} = 6.27 \times 10^{-10} 
\biggl(\frac{h_{0}^2 \Omega_{\mathrm{b}0}}{0.023}\biggr),
\label{etab}
\end{equation}
i.e. the baryon to photon ratio which must be of the order of $10^{-10}$ to agree with experimental data. BBN represents, therefore, one of the most notable successes 
of the standard cosmological model.

\subsection{The big-bang nucleosynthesis bound}
Any additional energy density speeds up the expansion and the cooling of the universe. 
For this reason the typical time scale of BBN is reduced in comparison with the standard case. The additional radiation-like energy density may be attributed to some extra relativistic species whose statistics may be either bosonic or fermionic. Since the supplementary species may be fermionic, they have been customarily parametrized, for historic reasons, in terms of the effective number of neutrino species:
$N_{\nu} = 3 + \Delta N_{\nu}$.
The standard BBN  results are in agreement with the observed abundances for $\Delta N_{\nu} \leq 1$ \cite{Dn1,Dn2,Dn3,Dn4}. The most constraining bound for the intermediate and  high-frequency branches of the relic graviton spectrum is represented by big-bang nucleosynthesis as argued long ago by Schwartzman \cite{bbn1} who was 
probably the first one to realize that potential bounds on $\Delta N_{\nu}$ also apply to the case of relic gravitons. Indeed, relic gravitons, being relativistic, can potentially increase the expansion rate at the BBN epoch. The increase in the expansion rate will affect, in particular, 
the synthesis of $^{4}\mathrm{He}$. To avoid the overproduction 
of  $^{4}\mathrm{He}$ the expansion rate the number of relativistic species 
must be bounded from above. An (ultra)relativistic fermion species with
two internal degrees of freedom
and in thermal equilibrium contributes $2\cdot7/8 = 7/4 = 1.75$ to $g_{\rho}$.  Before
neutrino decoupling the contributing relativistic particles are photons,
electrons, positrons, and $N_\nu = 3$ species of neutrinos, giving
\begin{equation}
    g_{\rho} = \frac{11}{2} + \frac{7}{4}N_{\nu} = 10.75.
\end{equation}
The neutrinos have decoupled before electron-positron annihilation so that they
do not contribute to the entropy released in the annihilation.
While they are relativistic, the neutrinos still retain an equilibrium energy
distribution, but after the annihilation
their (kinetic) temperature is lower, $T_\nu = (4/11)^{1/3}T$.  Thus
\begin{equation}
    g_{\rho} = 2 + \frac{7}{4}N_\nu\left(\frac{T_\nu}{T}\right)^4 = 2 + 0.454N_\nu = 3.36,
\end{equation}
after electron-positron annihilation.
By now assuming that there are some additional relativistic degrees of
freedom, which also have decoupled by the time of electron-positron
annihilation,  or just some additional component $\rho_X$ to the energy
density with a radiation-like equation of state (i.e. $p_{X} = \rho_{X}/3$), the effect on the 
expansion rate will be the same as that of having
some (perhaps a fractional number of)
additional neutrino species.  Thus its contribution can be represented by
replacing $N_\nu$ with $ N_\nu + \Delta N_{\nu}$ in the above.  Before
electron-positron annihilation we have $\rho_X = (7/8)\Delta N_{\nu} \rho_\gamma$
and after electron-positron annihilation we have
$\rho_X = (7/8) (4/11)^{4/3} \,\Delta N_{\nu} \,\rho_{\gamma} \simeq 0.227\,\Delta N_{\nu} \,\rho_\gamma$.
The critical fraction of CMB photons can be directly computed from the 
value of the CMB temperature and it is notoriously given by
$h_{0}^2 \Omega_\gamma \equiv \rho_\gamma/\rho_{\mathrm{crit}} = 2.47\times10^{-5}$.
If the extra energy density component has stayed radiation-like until today,
its ratio to the critical density, $\Omega_X$, is given by
\begin{equation}
h_{0}^2   \Omega_X \equiv h_{0}^2\frac{\rho_X}{\rho_{\mathrm{c}}} = 5.61\times10^{-6}\Delta N_{\nu} 
\biggl(\frac{h_{0}^2 \Omega_{\gamma0}}{2.47 \times 10^{-5}}\biggr).
\end{equation}
If the additional species are relic gravitons, then  \cite{bbn2,bbn3}: 
\begin{equation}
h_{0}^2  \int_{\nu_{bbn}}^{\nu_{max}}
  \Omega_{gw}(\nu,\tau_{0}) d\ln{\nu} = 5.61 \times 10^{-6} \Delta N_{\nu} 
  \biggl(\frac{h_{0}^2 \Omega_{\gamma0}}{2.47 \times 10^{-5}}\biggr),
\label{BBN1}
\end{equation}
where $\nu_{bbn}$ is given by Eqs. (\ref{ANIS5})--(\ref{ANIS5a}) (see also Eq. (\ref{FF4})).
The frequency $\nu_{max}$ corresponds to the maximal frequency of the spectrum. In the case 
of the relic gravitons produced within the concordance scenario $\nu_{max}$ is given by Eq. (\ref{RGC24}) 
(see also Eq. (\ref{ONEb}) and discussion therein). For different signals $\nu_{max}$ can take different values. 
Thus the constraint of Eq. (\ref{BBN1}) arises from the simple consideration 
that new massless particles could eventually increase the expansion rate 
at the epoch of BBN. The extra-relativistic species do not have to be, however, fermionic \cite{bbn1,bbn2} 
and therefore the bounds on $\Delta N_{\nu}$ can be translated  into bounds 
on the energy density of the relic gravitons. 

Depending on the combined data sets (i.e. various light elements abundances and different combinations of CMB observations), the standard BBN scenario implies that the bounds on $\Delta N_{\nu}$ range from $\Delta N_{\nu} \leq 0.2$ 
to $\Delta N_{\nu} \leq 1$.  All the relativistic species present inside the 
Hubble radius at the BBN contribute to the potential increase in the expansion rate and this  explains why the integral in Eq. (\ref{BBN1}) must be performed from $\nu_{bbn}$ to $\nu_{max}$ (see also 
 where this point was stressed in the framework of a specific model).  
The existence of the exponential suppression for $\nu>\nu_{max}$  
guarantees the convergence of the integral also in the case when the integration 
is performed up to $\nu \to \infty$. The constraint of Eq. (\ref{BBN1})  can be relaxed in some 
non-standard nucleosynthesis scenarios \cite{bbn2}, but, in what follows, the 
validity of Eq. (\ref{BBN1}) will be enforced by adopting 
$\Delta N_{\nu} \simeq 1$  which implies, effectively 
\begin{equation}
h_{0}^2  \int_{\nu_{bbn}}^{\nu_{max}}
  \Omega_{gw}(\nu,\tau_{0}) d\ln{\nu} < 5.61\times 10^{-6} \biggl(\frac{h_{0}^2 \Omega_{\gamma0}}{2.47 \times 10^{-5}}\biggr). 
\label{BBN2}
\end{equation}
The spectral energy density of arising in the concordance paradigm and illustrated in Figs. \ref{SEC4FIG5} and \ref{SEC4FIG6} 
are compatible with the BBN bound. The BBN constraint, however, does not forbid a potentially detectable signal in the 
high-frequency branch of the relic graviton spectrum, as we shall see in section \ref{sec7}.

The basic considerations discussed here can be complemented by other bounds which are, however,  less constraining 
than the ones mentioned above. The same logic employed for the derivation of Eq. (\ref{BBN1}) can be applied 
at the decoupling of matter and radiation \cite{cmbconst1,cmbconst2}. While the typical frequency of BBN is ${\mathcal O}(0.1)$ nHz the frequency
of matter--radiation equality is ${\mathcal O}(100)$ aHz. Since the decoupling between matter and radiation occurs after equality we have that 
\begin{equation}
h_{0}^2  \int_{\nu_{dec}}^{\nu_{max}}
  \Omega_{gw}(\nu,\tau_{0}) d\ln{\nu} \leq 8.7 \times 10^{-6}.
 \label{CMBconst} 
  \end{equation}
While the bound itself is numerically similar to the one of Eq. (\ref{BBN1}) the lower extremum of integration is 
smaller since $\nu_{dec} \ll \nu_{bbn}$. The bound (\ref{CMBconst}) (discussed in Ref. with slightly 
different notations) has been also taken into account in the present analysis. However, since we are dealing here with growing spectral energy distributions, Eq. (\ref{CMBconst}) is less constraining: for the same (increasing) slope the lower extremum of integration of Eq. (\ref{CMBconst}) gives a smaller contribution than the one of Eq. (\ref{BBN1}).

The BBN limits examined so far can be relaxed in nonstandard BBN scenarios. For example, the $^{4}\mathrm{He}$ yield can be affected by changing the electron neutrino spectrum, like in the model with active-sterile neutrino mixing, or by $\nu_{e}$ degeneracy, which changes the amount of thermal $\nu_{e}$. It may also happen that the BBN takes place in the presence of matter--antimatter 
domains \cite{bbn2}. In the standard BBN scenario $\eta_{b0}$ is homogeneous but it can also be inhomogeneous and it can change sign:
this is the scenario of BBN with matter--antimatter domains. Since the neutron diffusion distance is larger than the proton diffusion distance, matter and antimatter are mixed by (anti)neutron diffusion. More neutrons than protons are annihilated and thus the $n/p$ ratio is reduced compared to standard BBN. The reduction in the abundance of $^{4}\mathrm{He}$ can be compensated by a larger expansion rate due to a larger number of extra-relativistic species \cite{bbn2}: in this case the bound of Eq. (\ref{BBN1}) can be relaxed also 
by one order of magnitude. 

\subsection{The bounds from the pulsar timing array}
Gravitational waves of relatively long wavelengths  can influence 
the propagation of electromagnetic signals. Sazhin \cite{PUL0a} suggested long ago 
that the arrival times of pulsar's pulses could be 
used for direct searches gravitational radiation.
Shortly after Detweiler \cite{PUL0b} made a similar point and derived 
what could be interpreted today as the first upper limit on the relic gravitons 
from pulsars. This aspect was subsequently reinstated in more 
accurate terms in Ref. \cite{PUL0c}. Direct bounds on relic gravitons from pulsars 
have been obtained in Refs. \cite{PUL1,PUL2} and the current limits can be summarized as
\begin{equation}
h_{0}^2 \Omega_{gw}(\nu_{pulsar},\tau_{0}) < {\mathcal O}(10^{-10}),\qquad 
\nu_{pulsar} =  {\mathcal O}(\mathrm{nHz}),
\label{PUL0a}
\end{equation}
where $\nu_{\mathrm{pulsar}}$ roughly corresponds to the inverse 
of the observation time during which the pulsars timing has been monitored \cite{PUL3,PUL4,PUL5, PUL6}. Equation (\ref{PUL0a}) implies a bound on the corresponding chirp amplitude, i.e.$ h_{c} (\nu_{pulsar}, \tau_{0}) < {\mathcal O}(4) \times 10^{-14}$ where $h_{c}(\nu, \tau_{0})$ has been introduced in Eq. (\ref{OBS15}). As time goes by the typical value of $\nu_{pulsar}$ {\em decreases} since the typical observation time {\em increases}. The development of the observations suggested the idea of pulsar timing array (PTA) collaborations that are now exploiting the timing precision of millisecond pulsars over decades of observations to search for correlated timing deviations induced by gravitational waves. The typical frequencies fall in the nHz band but they can also arrive up to the mHz, at least in principle. 
The gross bound illustrated by Eq. (\ref{PUL0a})  can become slightly more specific by considering the figures provided by the different collaborations. The quoted $95$ \% confidence limits of the 
 European Pulsar Timing Array (EPTA collaboration) are \cite{PUL7}: 
 \begin{equation} 
 h_{0}^2 \Omega_{gw}(\nu_{pulsar},\tau_{0}) < 1.2 \times 10^{-9},\qquad h_{c}(\nu_{pulsar}, \tau_{0}) < 8.75\times 10^{-15}, \qquad 
\nu_{pulsar} =  5 \,\mathrm{nHz}.
\label{PUL0b}
\end{equation}
In the case of North American Nanohertz Observatory for Gravitational Waves (NANOGrav collaboration) the analog limit \cite{PUL8,PUL9,PUL10} reads: 
 \begin{equation} 
 h_{0}^2 \Omega_{gw}(\nu_{pulsar},\tau_{0}) < 4.2 \times 10^{-10},\qquad h_{c}(\nu_{pulsar}, \tau_{0}) < 7.8\times 10^{-15}, \qquad 
\nu_{pulsar} =  3.3\, \mathrm{nHz}.
\label{PUL0c}
\end{equation}
Finally, in the case of the Parkes Pulsar Timing Array (PPTA collaboration) \cite{PUL6,PUL11,PUL12} we have:
 \begin{equation} 
 h_{0}^2 \Omega_{gw}(\nu_{pulsar},\tau_{0}) <  10^{-10},\qquad h_{c}(\nu_{pulsar}, \tau_{0}) < 4.5\times 10^{-15}, \qquad 
\nu_{pulsar} =  2.8\, \mathrm{nHz}.
\label{PUL0d}
\end{equation}
In the nHz band the relic gravitons can be observed by the influence they have on the arrival time of pulsar signals at the Earth. The measured versus the predicted arrival times of the pulses, as a function of time, are referred to as the timing residuals.  A propagating graviton will pass both the pulsar and the Earth, affecting their local space-time at different moments. The analysis of the pulsar timing can detect the relic gravitons through an individual pulsar (the so-called {\it pulsar term}), and can also determine the passage of a gravitational wave through the Earth (the so-called {\it Earth term}) as a signal correlated between pulsars. While we can detect the Earth or pulsar terms in one pulsar, a gravitational wave can only be confidently detected by observing the correlated influence of the potential signal on multiple pulsars. This is why 
the consortia mentioned above (i.e.  the Parkes, the North American, and European groups) are now converging towards the  International Pulsar Timing Array (IPTA) \cite{PUL13a}.
In this respect the Square Kilometer Array (SKA) project together with the PTA 
could provide a further opportunity \cite{PUL13}. 

\subsection{Potential signals from cosmic strings}
The phase transitions in the early universe and
the symmetry breaking  associated with grand unified theories may produce networks 
of cosmic strings and other topological defects that have been repeatedly studied through the years starting from the seminal contribution of T. W. Kibble \cite{CS1,CS2,CS3}. String theory itself also suggests, according to some, the presence of networks of fundamental strings, especially through the mechanism of brane inflation \cite{CS4,CS5}. These networks may have various effects such as the imprints (via lensing) on the CMB and possibly through the presence of a diffuse background of gravitational waves. Cosmic strings and other topological defects have been proposed in the past as a  sound alternative paradigm for structure formation. In particular topological defects could induce temperature and polarization anisotropies \cite{CS2}.
\begin{figure}[!ht]
\centering
\includegraphics[height=7.cm]{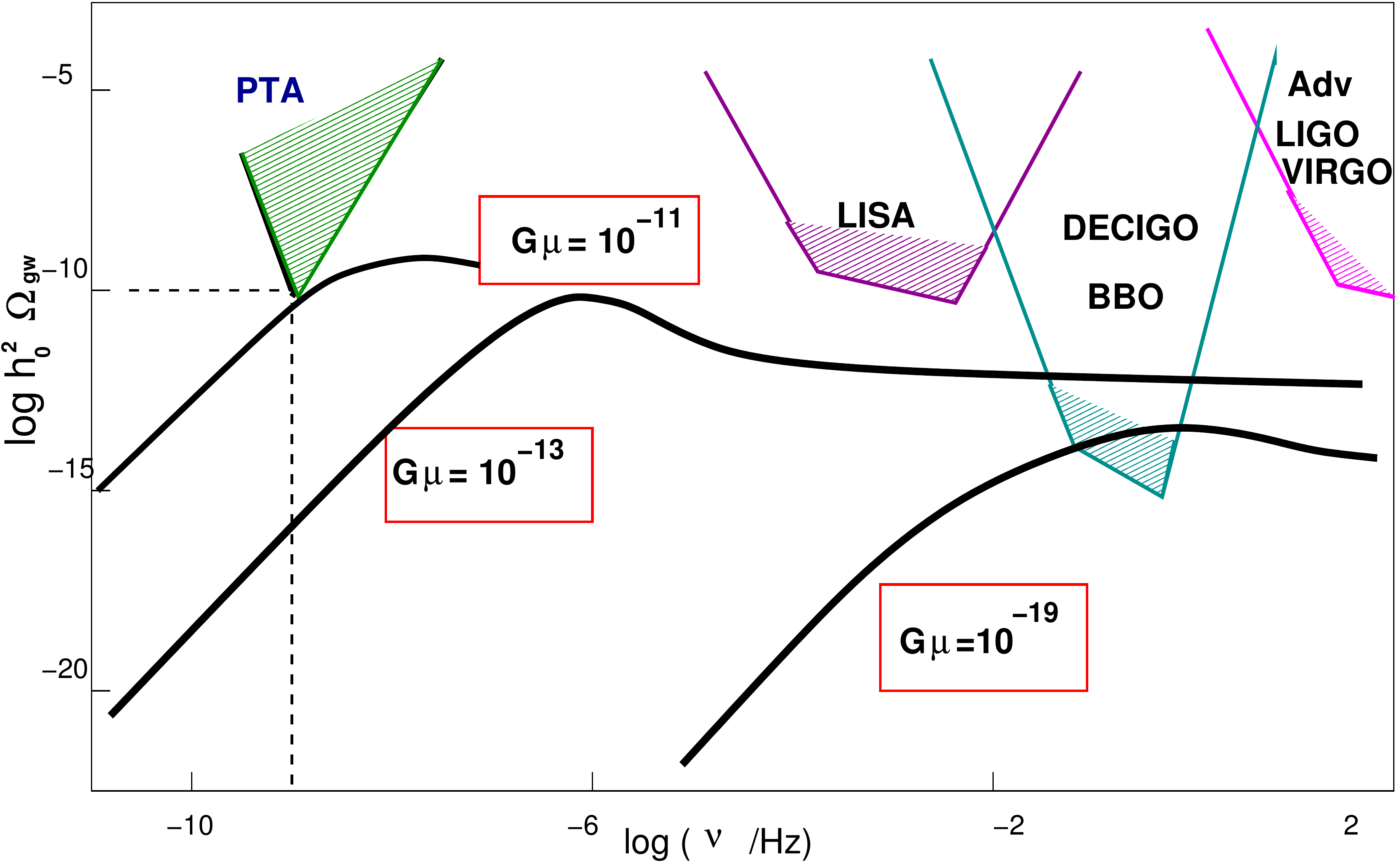}
\caption[a]{The stochastic backgrounds induced by cosmic strings at intermediate frequencies. }
\label{SEC6FIG6}      
\end{figure}
This perspective has been partially abandoned already after the COBE observations \cite{CMB3} and even more convincingly after first release of the WMAP data \cite{WMAP1,WMAP1a,WMAP1b}. At the 
moment it is clear that cosmic defects cannot lead to the adiabatic initial conditions discussed 
in section \ref{sec5} since they are unable to mimic, at once, the patterns of the acoustic peaks of the $TT$ 
power spectrum and the first anticorrelation peak in the corresponding $TE$ power spectrum.
It has been subsequently argued that cosmic strings could be produced at the end of the reheating phase.
The post-reheating cosmic strings must therefore have a concentration that is sufficiently small 
not to affect the observed CMB anisotropies but sufficiently large to emit bursts of gravitational radiation from various non-linear structures arising through their evolution (e.g. cusps, kinks, oscillating loops). 
If the concentration is appropriately tuned the signals from these non-linear structures may hopefully lead to a diffuse background of gravitational radiation \cite{CS2}. 

The gravitational waves emitted by oscillating loops at different epochs have been argued to produce a stochastic background  \cite{CS6,CS7,CS8} with quasi-flat spectral energy density which is typically larger than the inflationary signal (see e.g. Fig. \ref{SEC6FIG6} and compare it with Fig. \ref{SEC4FIG6}). These predictions are derived through various numerical steps and some disagreements exist on the slopes of the signals. Cosmic strings are usually modelled by solving the field theory equations in flat or expanding space-time, or through an effective Nambu-Goto prescription with weak coupling to gravity \cite{CS3}.  From the stress-energy of a network of strings, one can then compute (in the weak gravity limit) the induced backgrounds of relic gravitons. The collisions of traveling kinks and cusps along the strings are expected to produce bursts of gravitational radiation. The bursts events have been computed using the Nambu-Goto approximation, again in the weak field limit \cite{CS6,CS7,CS8}. The proposed methods do not seem to agree on the primary energy loss mechanism of the cosmic strings \cite{CS9,CS10,CS11,CS12,CS13,CS14}. There have been recently attempts to go beyond the weak field limit \cite{CS15}.  In the last thirty years various detailed analyses have been presented with different degrees of accuracy by exploiting some of the best numerical facilities available at the corresponding times (see e.g. \cite{CS16,CS17,CS18,CS19,CS20,CS21,CS22,CS23}). For the present purposes it is fair to say that the nature of the signal changes depending on three basic parameters: the string tension in Planck units (i.e. $G \mu$); the typical size of the loops normalized at the formation time; the emission efficiency of the loop.
The quoted values of $G\mu$ may range between $10^{-8}$ and $10^{-23}$ while the typical size of the loop may vary between $10^{-10}$ and $10^{-1}$. The large interval of variation of the parameters makes it obvious that different signals can be expected. From symmetry breaking in the grand unified context the typical values  of $G \mu$ could be as large as ${\mathcal O}(10^{-6})$. These values 
would cause however measurable temperature and polarization anisotropies of the CMB and have been ruled out \cite{WMAP1,WMAP1a,WMAP1b,WMAP2,WMAP2a}, as already mentioned. Current limits from CMB observations demand $G\mu < {\mathcal O} (10^{-8})$.

The pulsar timing constraints of Eqs. (\ref{PUL0b}), (\ref{PUL0c}) and (\ref{PUL0d}) are even more stringent and they seem to imply, quite generically, that $ G \mu = {\mathcal O}(10^{-11})$. In the cartoon of Fig. \ref{SEC6FIG1} we illustrate the spectral energy density of gravitational waves due to cosmic strings for three different values of $G \mu$ as they appear in the current literature. We also report, for comparison, the hoped sensitivities of various instruments. 
The pulsar timing limit sets a bound for typical frequencies in the nHz range and it is clearly more constraining than the limits at lower frequencies. The leftmost curve in Fig. \ref{SEC6FIG1} (corresponding to $G \mu = 10^{-11}$) roughly illustrates the class of models that saturate the pulsar limits. We also illustrate, for comparison, the cases $G \mu =10^{-13}$ and $G \mu =10^{-19}$. To be compatible with the pulsar limits and 
to get a larger signal  below the Hz (and possibly also in the audio band) the spectral energy density must 
first increase (even below the nHz) and then flatten out. The typical slopes in the increasing 
branch are violet, i.e.  $h_{0}^2 \Omega_{gw} \propto \nu^{\alpha}$ with $1< \alpha \leq 2$. In
the flat branch  $\alpha$ can be slightly negative (i.e. $-0.2 < \alpha < 0$).
The spectral energy density  depends, in general, on the distribution of loop sizes, the cosmological evolution, and the power emitted by each loop at each frequency.  The high frequency background today consists of gravitational waves emitted in the radiation era. At lower frequencies, the spectrum is increasingly dominated by waves emitted in the matter era.  At still lower frequencies, the background spectrum falls rapidly, because 
there are few loops large enough to emit such long waves. The signals due to topological defects, if present, will act as  {\em diffuse foregrounds} for the relic gravitons produced at intermediate frequencies. In this respect Figs. \ref{SEC6FIG4} and \ref{SEC6FIG5} could be usefully compared with Fig. \ref{SEC6FIG6}. It seems plausible to argue that a distinction among the various signals over different ranges of frequency is possible even if it is probably too early to discuss this topic given the current uncertainties on the planned instruments and on their sensitivities. 

In Fig. \ref{SEC6FIG5} we also illustrated the minimal spectral energy density potentially 
detectable by the advanced LIGO/Virgo detectors \cite{LV1,LV2} together with the 
hoped sensitivities of the space-borne detectors \cite{LISA,LISAa,BBO,DECIGO1}.
While a more specific discussion will be presented in section \ref{sec8}, 
the current bounds on the diffuse backgrounds of relic gravitons are already
quite constraining( see e.g. Tab. \ref{SEC7TABLE1} of section \ref{sec8}). Even though 
the constraints depend on the spectral slope of $\Omega_{gw}$,
in the case of the cosmic defects the limits derived in the case 
of an exactly scale-invariant spectrum can be approximately applied and they 
imply $\Omega_{gw} \leq 10^{-8}$. If the signal from cosmic defects is the one reported in Fig. \ref{SEC6FIG5} 
we have that the current bounds from wide-band interferometers 
are always less constraining than the limits coming from the pulsar timing arrays;
this conclusion follows from the red slope of the spectrum after the hump. 
We finally note that the cases $A$, $B$ and $C$ of Fig. \ref{SEC6FIG4} as well as the examples 
of Fig. \ref{SEC6FIG5} lead to a set of potential templates very similar 
to the ones obtainable in the case of cosmic defects.

\subsection{Potential signals from the electroweak epoch}
Strongly first-order phase transitions may also lead to bursts of gravitational radiation 
 \cite{PTT1,PTT2,PTT3}. Provided the phase transition proceeds thanks 
to the collision of bubbles of the new phase, the lower frequency scale of the burst 
is at most comparable with the Hubble radius at the corresponding epoch. Denoting by $\nu_{b}$ the 
frequency of the burst, in the case of the electroweak phase transition we must have $\nu_{b} \geq  {\mathcal O}(\nu_{ew})$ [see Eq. (\ref{FF5})].  Various scenarios for the physics beyond the standard electroweak theory seem to predict, under various assumptions, a first-order electroweak phase transition. Recently these models have enjoyed renewed attention exactly because a strong phase transition would also be a potential source of observable gravitational wave (see e.g. \cite{PTT4,PTT5} and references therein). 
In general terms the signal of a strongly first-order phase transition will act as a foreground 
from the primordial signal due to relic gravitons; since the burst of gravitational radiation from phase transitions is expected to be localized in frequency, it could be easily disentangled from a competing signal 
with broader spectral energy distribution.  Since the mass of the Higgs boson is much larger than the mass of the $W^{\pm}$ and $Z^{0}$ bosons the phase diagram of the standard electroweak theory can only be scrutinized with non-perturbative methods  \cite{PT0a,PT0b}. Lattice simulations \cite{PT1,PT2a,PT2b} suggest the presence of a cross-over regime which is conceptually similar to the behaviour experienced by ordinary chemical compounds above their triple point. In this regime the collision of bubbles of the new phase cannot happen but gravitational waves are still emitted, even provided the electroweak plasma hosts maximally gyrotropic configurations of the hypermagnetic field. This idea has been suggested in Refs. \cite{PT3a,PT3b}. 

Inside the electroweak particle horizon, hypermagnetic knots (HK in what follows) are characterized by a non-vanishing gyrotropy, i.e. $\vec{B} \cdot \vec{\nabla} \times \vec{B}$ where $\vec{B}$ denotes the comoving hypermagnetic field. The dynamical production of HK and Chern-Simons waves may also be related to an unconventional approach for the generation of the baryon asymmetry of the universe \cite{PT3c,PT3d,PT3e}. The minimal comoving frequency of the gravitational waves potentially emitted by the HK is: 
\begin{equation}
\nu_{ew}  \,\,= \,\, \frac{H_{ew}}{2\pi} \biggl(\frac{a_{ew}}{a_{0}}\biggr) = \biggl( \frac{g_{\rho} \, \Omega_{R0}}{90 \pi}\biggr)^{1/4} \, 
\sqrt{\frac{H_{0}}{M_{P}}} \, T_{ew},
\label{HK1}
\end{equation}
which coincides with the numerical value of Eq. (\ref{FF5}) for $T_{ew} = 173\, \mathrm{GeV}$ and $g_{\rho} = 106.75$. The non-screened vector modes of the electroweak plasma correspond to the hypercharge field which has a chiral coupling to fermions. The axial currents may arise either as a finite density effect (implying a non-trivial evolution of the chemical potential), or they can be associated with the presence of an axion field. Anomalous magnetohydrodynamics (AMHD) aims at describing the dynamical evolution of the gauge fields in a plasma containing both vector and axial-vector currents: while the axial currents are not conserved because of the triangle anomaly, the vector currents are responsible for the Ohmic dissipation. For  wavelengths shorter than  the Hubble radius at the electroweak time the relevant subset of the AMHD equations for the hypermagnetic field and for the vorticity $\vec{\omega}$  \cite{AMHD1,AMHD2} can be simplified 
as follows:
\begin{eqnarray}
\partial_{\tau} \vec{B} &=& \vec{\nabla}\times(\vec{v} \times \vec{B}) + \frac{\nabla^2 \vec{B}}{ 4 \pi \sigma} + \frac{\vec{\nabla} \times (g_{\omega} \vec{\omega})}{4 \pi \sigma} - 
\frac{\vec{\nabla} \times(g_{B} \vec{B})}{4 \pi \sigma},
\label{seca}\\
\partial_{\tau} \vec{\omega} &=& \vec{\nabla}\times(\vec{v} \times \vec{\omega})  + \frac{\vec{\nabla} \times(\vec{J} \times\vec{B})}{a^4 (\rho + p)} +  \frac{\eta_{s}}{a^4(\rho + p)}\nabla^2 \vec{\omega}, 
\label{secb}
\end{eqnarray}
where  $\eta_{s}$ is the shear viscosity and  $\sigma$ is the comoving conductivity of the electroweak plasma. Equations (\ref{seca}) and (\ref{secb}) hold when the two-fluid effects can be neglected in the slow branch of the AMHD spectrum  \cite{AMHD1}.  In Eq. (\ref{seca}) $g_{B}$ and $g_{\omega}$ are the coefficients of the magnetic and the vortical currents. There are specific situations where the chemical potential and the axion field can be simultaneously present  and both contribute to $g_{B}$. In spite of the  richness of the spectrum of AMHD, the perfectly conducting limit strongly suppresses the anomalous contributions. According to Eq. (\ref{seca}) when the conductivity is very large  the magnetic and the vortical currents are damped in comparison with the remaining term (which is the electroweak analog of the standard dynamo contribution). In a shorthand notation we then have that $\partial_{\tau} \vec{B} = \vec{\nabla}\times(\vec{v} \times \vec{B}) + {\mathcal O}(g_{B}/\sigma)$. Defining the vector potential in the Coulomb gauge, Eq. (\ref{seca}) becomes, up to small corrections due to the conductivity, $\partial_{\tau} \vec{A}= \vec{v} \times (\vec{\nabla}\times\vec{A})$. In the highly conducting limit, thanks to classic analyses of  Chandrasekhar Kendall and Woltjer \cite{AMHD2a,AMHD2b},  the magnetic energy density shall then be minimized in a fiducial volume $V$ under the assumption of constant magnetic helicity by introducing the Lagrange multiplier $\zeta$. The configurations minimizing the functional 
\begin{equation}
{\mathcal G} =\int_{V} d^{3} x\{ |\vec{\nabla} \times \vec{A}|^2 - \zeta \vec{A} \cdot (\vec{\nabla}\times\vec{A}) \},
\label{secd}
\end{equation}
are the Beltrami fields satisfying $\vec{\nabla} \times \vec{B} = \zeta \vec{B}$ where $1/\zeta = \lambda_{B}$ has dimensions of a length characterizing the spatial extension of the solution. The constant-$\zeta$ solutions represent the lowest state of hypermagnetic energy which a closed system may attain also in the case where anomalous currents are present, provided the ambient plasma is perfectly conducting \cite{AMHD1,AMHD1a,AMHD2,AMHD2aa,AMHD2bb}. The configurations minimizing the energy density with the constraint that the magnetic helicity be conserved coincide then with the ones obtainable in ideal magnetohydrodynamics (i.e. without anomalous currents). This is why it is possible to derive hypermagnetic knot solutions in a hot plasma from their magnetic counterpart \cite{AMHD3,AMHD3a}.  The gyrotropic configurations minimizing Eq. (\ref{secd}) exclude the backreaction on the flow. Indeed, because of the smallness of the hyperelectric fields 
the Ohmic current is given by $\vec{J} = \vec{\nabla}\times \vec{B}/(4\pi)$;  at the same time, $\vec{\nabla} \times \vec{B} = \zeta \vec{B}$.
Putting together the two previous observations the second term at the right hand side of Eq. (\ref{secb}) cancels exactly. Consequently HK have the unique property of allowing resistive decay of the field without introducing stresses on the evolution of the bulk velocity of the plasma. Thus, the maximal frequency of the spectrum is only be determined by the conductivity and it is given by:
\begin{equation}
\nu_{\sigma} = 58.28 \biggl(\frac{T_{ew}}{10^{2}\, \mathrm{GeV}} \biggr)^{1/2} \biggl(\frac{g^{\prime}}{0.3}\biggr)^{-1}\biggl(\frac{N_{eff}}{106.75}\biggr)^{1/4}\, \mathrm{kHz},
\label{condfreq}
\end{equation}
where $g^{\prime}$ is the hypercharge coupling constant. Equation (\ref{condfreq}) is obtained by computing the comoving wavenumber $k_{\sigma}$ corresponding to the hypermagnetic diffusivity. In general terms we have that $k_{\sigma}^{-2} = \int^{\tau} d\tau^{\prime}/(4\pi\sigma)$ however when the 
comoving conductivity is constant  the previous expression simplifies as $k_{\sigma} \simeq \sqrt{\sigma {\mathcal H}_{ew} }$. 

The frequencies ranging between $\nu_{ew}$ and $\nu_{\sigma}$  encompass the operating window of space-borne interferometers and of the terrestrial wide-band detectors. To determine the template family for the emission of gravitational radiation it is possible to analyze first the case of a single knot configuration and then move to the case of stochastic collection of knots. The evolution equations can be expressed as:
\begin{equation}
\mu_{ij}^{\prime\prime} - \nabla^2 \mu_{ij} - \frac{a^{\prime\prime}}{a} \mu_{ij} = - 2 \ell_{P}^2 a^3(\tau) \Pi_{ij}, \qquad \mu_{ij} = a h_{ij}.
\label{gw2}
\end{equation}
Since the backreaction on the flow vanishes because the Ohmic current and the hypermagnetic field are parallel,  the solution for HK can be factorized as 
\begin{equation}
\vec{B}(\vec{x},\tau) = \vec{b}(\vec{x}) f(\zeta,\tau), \qquad f(\zeta,\tau)= \exp{[ - \zeta^2 \tau/(4 \pi\sigma) ]}
\label{gw2a}
\end{equation}
where $\vec{b}(\vec{x})$ minimizes Eq. (\ref{secd}) and $f(\zeta,\tau)$ accounts for the contribution of the Ohmic dissipation in the case of a single configuration with constant $\zeta$. According to the Chandrasekhar-Kendall representation \cite{AMHD2a},  $\vec{b}(\vec{x})$ can always be expressed as:
\begin{equation}
\vec{b}(\vec{x}) = \lambda_{B} \vec{\nabla}\times [ \vec{\nabla} \times ( \hat{u} \Psi)] + \vec{\nabla} \times  ( \hat{u} \Psi),
\label{third}
\end{equation}
where $\hat{u}$ is a unit vector denoting the direction of the knot and $\Psi$ obeys $\nabla^2 \Psi+ \zeta^2 \Psi=0$. If $\zeta$ is constant,  the hypermagnetic knot configuration can be represented through the Chandrasekhar-Kendall solution \cite{AMHD2a}. The direction of the HK introduces a difference in the evolution of the two polarizations of the gravitational wave. The anisotropic stress associated with the 
HK [i.e. $\Pi_{ij}= (B_{i} B_{j} - B^2 \delta_{ij}/3)/(4 \pi a^4(\tau))$ ]  can always be projected along the two tensor polarizations and since the direction of the knot does not need to coincide with the direction of propagation of the gravitational wave the emission due to hypermagnetic knots is generically polarized \cite{PT3a,PT3b}. A stochastic collection of HK is characterised, in Fourier space,  by the two-point function:
\begin{equation}
\langle B_{i}(\vec{k},\tau) \, B_{j}(\vec{p},\tau^{\prime}) \rangle = \frac{2\pi^2}{k^3} \epsilon_{ijk} \hat{k}^{k} P_{hk}(k,\tau,\tau^{\prime}) \, \delta^{(3)}(\vec{k} + \vec{p}),
\label{gw12}
\end{equation}
where the power spectrum $P_{hk}(k,\tau)$ has the dimensions of an energy density. As in the case of a single knot the resistive decay of the field does not introduce stresses on the evolution of the bulk velocity.
As in the case of Eq. (\ref{third}) the evolution can be factorized and also the power spectrum can be written as  $P_{hk}(k,\tau,\tau^{\prime}) = A_{hk} (k/k_{ew})^{\beta} e^{- k^2(\tau+\tau^{\prime})/(4 \pi \sigma)}$. The power-law is the simplest parametrization for the power spectrum and it appears in the context of different models (see e.g. \cite{PT3a,PT3b,PT3d,PT3e}). 
If we then consider a collection, the template for the emission of gravitational waves can be determined by solving Eq. (\ref{gw2}) 
and by reabsorbing all the parameters in a single amplitude $\Omega_{B}$ and in a spectral slope $\alpha$. 
The template family for the emission of gravitational waves from maximally gyrotropic configurations of the hypermagnetic field 
between the $\mu$Hz and the kHz can therefore be expressed as \cite{AMHD4}
\begin{equation}
\Omega_{gw}(\nu,\tau_0)=  \frac{\Omega_{B}^2}{(1 + z_{eq}) (1 + z_{\Lambda})^3}  \biggl(\frac{\nu}{\nu_{ew}}\biggr)^{\alpha} e^{ - 2 (\nu/\nu_{\sigma})^2},\qquad \nu\geq \nu_{ew}.
\label{gw18}
\end{equation}
The spectral slope $\alpha$ $\Omega_{B}^2$ which is chiefly determined by the dimensionless ratio $A_{hk}^2/(H_{ew}^4 \overline{M}_{P}^2)$ and $\alpha = 2 \beta$. For typical values of the parameters (i.e. $ 0 \geq \alpha \leq 1$ and $ 10^{-8} \leq \Omega_{B} < 10^{-2}$) there are regions in the parameter space where $h_{0}^2 \Omega_{gw}(\nu_{\mathrm{mHz}},\tau_0) > 10^{-12}$ and $h_{0}^2 \Omega_{gw}(\nu_{\mathrm{audio}},\tau_0) > 10^{-10}$ where we took $\nu_{\mathrm{mHz}} = 1\, \mathrm{mHz}$ and $\nu_{\mathrm{audio}}= 100\, \mathrm{Hz}$. While the considerations illustrated here hold in the symmetric phase of the electroweak theory, there are also different approaches where ordinary magnetic fields 
are extrapolated in different regimes of the life of the universe with the purpose 
of computing the emitted gravitational radiation (see e.g. \cite{AMHD5} for a recent review also 
treating some of these issues).  Equation (\ref{gw18}) shows that the localized 
signal customarily attributed to a strongly first-order phase transition can also be mimicked 
by the evolution of hypermagnetic fields inside the electroweak horizon. Again, the frequency range 
of the signal and its feature can be used to disentangle potentially similar spectral energy distributions. 

 Because of the seismic noise ground-based interferometers cannot measure relic gravitons 
with frequencies smaller than the Hz. This limitation disappears in outer space where, however, 
the common sense suggests that different sources of noise will appear. Bearing in mind this general caveat, the operating window of space-borne interferometers falls in the mHz range and so far three complementary projects have been proposed: 
 LISA (Laser Interferometer Space Antenna) \cite{LISA,LISAa} (see also \cite{LISA2,LISA3,LISA4,LISA5}),  BBO (Big Bang Observer) \cite{BBO}, and DECIGO (Deci-hertz Interferometer Gravitational Wave Observatory) \cite{DECIGO1} (see also \cite{DECIGO2,DECIGO3}).  These futuristic detectors will be swiftly examined in section \ref{sec8} only in connection 
 with their potential relevance for the primordial backgrounds of relic gravitons.

\renewcommand{\theequation}{7.\arabic{equation}}
\setcounter{equation}{0}
\section{The high-frequency band}
\label{sec7}
The high-frequency domain  (i.e. between few Hz and the $100$ GHz) 
encompasses the operating window of wide-band interferometers (i.e.  between few Hz and $10$ kHz). The potential signal of the concordance paradigm (supplemented by an early inflationary phase and by the standard thermal history) is negligible while in the GHz region the spectral energy density is dominated by the thermal gravitons (see Eqs. (\ref{MAGN2})--(\ref{MAGN3}) and discussion therein). Modified post-inflationary histories and reheating dynamics also provide high-frequency signals that will be hopefully scrutinized in the future with a new generation of electro-mechanical detectors eventually operating in the MHz range. In this range the statistical properties of relic gravitons might be deduced by looking at the intensity correlations 
(see section \ref{sec7} and discussion therein).
\subsection{High-frequency gravitons and post-inflationary thermal history}
If the concordance paradigm is supplemented by an early inflationary phase and by a standard thermal history, the spectral energy density at high frequencies is, at most, ${\mathcal O}(10^{-16.5})$ 
and various sorts of corrections, specifically discussed in section \ref{sec4}, tend to reduce this 
estimate even further. However, if the inflationary evolution is followed by a stiff phase the spectral energy density may increase at high-frequencies \cite{HIS23} by leaving unmodified the 
signal at lower frequencies. This possibility has been later suggested also in connection with quintessential inflationary models \cite{HIS24a} where the inflaton and the 
quintessence field are unified in a single physical entity (see also \cite{HIS24b}).
In the presence of a stiff post-inflationary phase the spectral energy density in the MHz  
region is between five and six orders of magnitude larger than in the conventional case \cite{HIS24,TRANS8,TRANS9}. The signal is characterized by a broad spike 
with a maximum in the GHz region where the spectral energy density can even be eight 
orders of magnitude larger than the standard concordance signal \cite{HIS24,STIFF0,STIFF0a,STIFF1}. 
\begin{figure}[!ht]
\centering
\includegraphics[height=6.3cm]{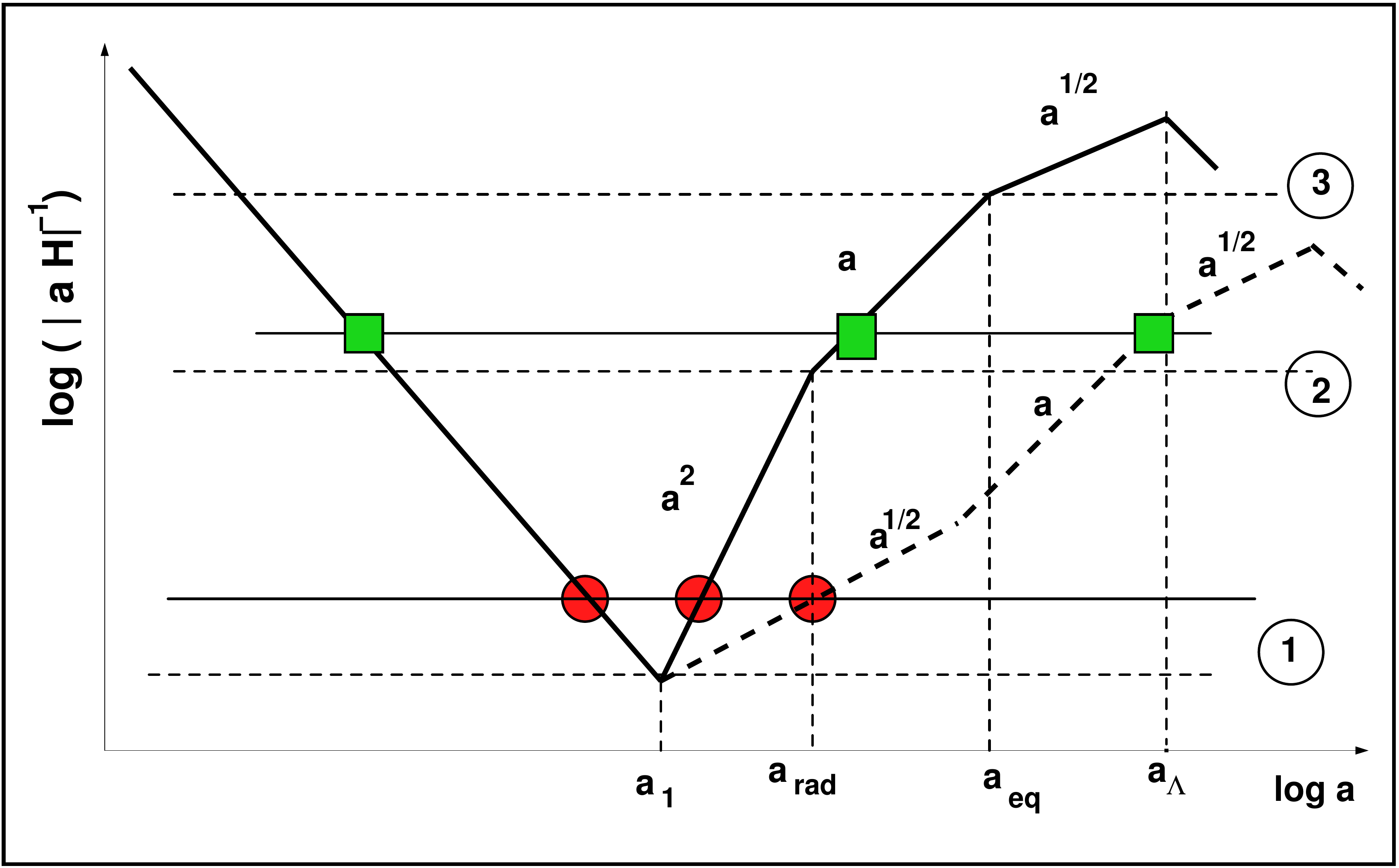}
\caption[a]{The evolution of the Hubble radius is reported when the transition to the radiation 
phase is delayed by a stiff epoch that follows conventional inflationary stage of expansion. For comparison the dashed line illustrates the case of a delayed reheating dominated by a dusty contribution.}
\label{SEC7FIG1}      
\end{figure}
Unlike the case of bouncing scenarios the presence of a stiff post-inflationary phase 
does not crucially affect the intermediate frequency region and the potential signal is simultaneously constrained at low and higher frequencies \cite{HIS24a,STIFF1a,STIFF2,STIFF3,STIFF4,STIFF6}.  
In the framework of the concordance paradigm the large-scale observations constraining the 
tensor to scalar ratio at low-frequencies and the observations of interferometers in the audio band are unrelated. However if the post-inflationary evolution is not immediately dominated by radiation,
the low-frequency observations and the data from the wide-band interferometers must be analyzed together \cite{TRANS8,TRANS9}. In this way the post-inflationary thermal history can be directly constrained  \cite{STIFF8,STIFF9,STIFF10,STIFF11} rather than assumed or imposed by uncertain (and sometimes arbitrary) theoretical premises. Various extensions of these suggestions have been discussed from specific reheating mechanisms \cite{STIFF13,STIFF13a,STIFF17,STIFF18} to modified gravity theories \cite{STIFF14,STIFF14a}. In this extension of the conventional cosmological paradigm, (sometimes dubbed tensor-$\Lambda$CDM  \cite{TRANS8}) gravitons are 
copiously produced at high-frequencies without conflicting with the bounds on the tensor to scalar ratio stemming from the combined analysis of the standard cosmological data sets (i.e. cosmic microwave background anisotropies, large-scale structure and supernovae).

There are neither direct nor indirect tests constraining the post-inflationary thermal 
history prior to the formation of light nuclei. This means, in particular, that 
the sound speed of the plasma does not need to coincide with $1/\sqrt{3}$ throughout 
the whole thermal history of the universe down to the epoch of matter-radiation equality.
Conversely the sound speed could be larger than $1/\sqrt{3}$ at most equaling the speed of light (i.e. $1$ in the present units). To illustrate the possibility of growing spectra as a result of a modified post-inflationary thermal history, it is simpler to consider first the wavelengths leaving the Hubble radius during inflation and reentering 
during an unknown post-inflationary phase;  the exit and the reentry of this class of wavelengths 
(labeled by $1$ in Fig. \ref{SEC7FIG1}) have been highlighted by two filled circles.
Since the exit takes place during a conventional inflationary phase (i.e. $\epsilon_{ex} \ll 1$) and the reentry occurs far from the radiation epoch (i.e. $\epsilon_{re} \neq 2$) the spectral energy density in the WKB approximation is determined, in general terms, from Eqs. (\ref{ap30}) and (\ref{ap31}):
\begin{equation}
\Omega_{gw}(k,\tau) = \frac{4}{3\, \pi} \biggl(\frac{H_{1}}{M_{P}}\biggr)^2 \, \biggl(\frac{a}{a_{1}}\biggr)^{3 w -1 } \biggl| 
\frac{k}{a_{1}\, H_{1}}\biggr|^{\overline{n}_{T}}, \qquad  
\overline{n}_{T} = \frac{12\, w \, (1- \epsilon) - 2 ( 3 w +1 )}{(3 w+1 ) ( 1 - \epsilon)},
\label{SSTT1}
\end{equation}
where $\epsilon$ coincides with $\epsilon_{ex}$ and $w$ denotes the 
barotropic index of the unknown post-inflationary phase. In the limit $\epsilon \to 0$ and $w\to 1$ 
the spectral index in the high-frequency region is blue, i.e. $\overline{n}_{T} \to 1$. Different values of $w$ slightly reduce the slope (e.g. for $ w \to 2/3$ we have $\overline{n}_{T} \to 2/3$) so that  $0< \overline{n}_{T} \leq 1 $ as long as  $1/3 < w \leq 1$.  Equation (\ref{SSTT1}) 
follows from Eqs. (\ref{ap24}), (\ref{ap30}) and (\ref{ap31}) since for the wavelengths exiting during inflation and reentering in the stiff epoch the pump fields do not vanish in the vicinity of the turning points. For  $\tau < - \tau_{1}$ the inflationary scale factor is $a_{i}(\tau) = (- \tau/\tau_{1})^{-\beta}$
with $\beta = 1/(1-\epsilon)$ while for  $\tau> - \tau_{1}$ (i.e. $a > a_{1}$ in Fig. \ref{SEC7FIG1}) 
the scale factor can be generically parametrized as:
\begin{equation}
a_{p}(\tau) = \biggl[ \frac{\beta}{\delta} \biggl( \frac{\tau}{\tau_{1}} +1 \biggr) +1\biggr]^{\delta}, \qquad \tau \geq -\tau_{1},\qquad \delta = \frac{2}{3 w+1}.
\label{SSTT2}
\end{equation}
The results expressed by Eq. (\ref{SSTT1}) hold provided $w \neq 1$ since the condition $\overline{\,\epsilon\,} \neq 3$ has been assumed in  Eq. (\ref{ap22}). However for $w\to 1$ Eq. (\ref{SSTT1}) is only logarithmically corrected \cite{HIS23,HIS24a,HIS24} since in the case 
$w \to 1$ the scale factor (\ref{SSTT2}) becomes $a_{p} \propto \sqrt{\tau/\tau_{1}}$ so that 
the integral appearing in ${\mathcal J}(\tau_{ex}, \tau_{re})$  (see Eq. (\ref{ap13}))
will ultimately contain a logarithmic contribution going as $\ln{|k \tau_{1}|}$.  The results of Eq. (\ref{SSTT1}) can also be obtained without appealing to the WKB approximation by 
simply requiring that the mode functions and their first derivatives are continuously matched across $- \tau_{1}$. In this case the generic form of the mode functions for $\tau< -\tau_{1}$ and for $\tau> - \tau_{1}$ is 
 \begin{eqnarray}
 f_{k}^{(i)}(\tau) &=& \frac{{\mathcal N}_{\mu}}{\sqrt{2 k}} \, \sqrt{- k \tau} \, H_{\mu}^{(1)}( -k\tau), \qquad \tau < - \tau_{1},
 \label{SSTT3}\\
  f_{k}^{(p)}(\tau) &=&\frac{1}{\sqrt{ 2 k}} \sqrt{k y}\biggl[ A_{-}(k) {\mathcal N}_{\nu} H_{\nu}^{(1)}(ky) + A_{+}(k) {\mathcal N}^{*}_{\nu} H_{\nu}^{(2)}(ky)\biggr],\qquad \tau> - \tau_{1}, 
  \label{SSTT4}
  \end{eqnarray}
 where, the Bessel indices can be expressed in terms of the expansion rates and the results following from Eq. (\ref{SSTT2}) are $\mu = (\beta + 1/2)$ and  $\nu =\bigl|\delta -1/2 \bigr|$.
 Following usual notations, ${\mathcal N}_{\mu}$ and ${\mathcal N}_{\nu}$ are two constant phases and $y$ 
 is the reduced time coordinate:
 \begin{equation}
 {\mathcal N}_{\mu} = \sqrt{\frac{\pi}{2}} e^{i \pi( 2 \mu+1)/4}, \qquad  {\mathcal N}_{\nu} = \sqrt{\frac{\pi}{2}} e^{i \pi( 2 \nu+1)/4},\qquad y= \tau + \tau_{1} \biggl( 1 + \frac{\delta}{\beta}\biggr). 
 \label{SSTT5}
 \end{equation}
Imposing the continuity of the mode functions (and of their first derivatives) across $-\tau_{1}$, the coefficients $A_{\pm}(k,\tau_{1})$ are
 \begin{eqnarray}
 A_{-}(k,\tau_{1}) &=& \frac{i \, \pi}{4} e^{i \pi(\mu-\nu)/2} \sqrt{ q} \biggl\{H_{\mu}^{(1)}(k \tau_1) H_{\nu}^{(2)}( q \,k \tau_{1}) \biggl(\mu + \frac{1}{2} + \frac{2\nu +1}{2 q}\biggr)
\nonumber\\
 &-& k\,\tau_{1} \biggl[ H_{\mu}^{(1)}(k\,\tau_{1}) H_{\nu+1}^{(2)}(q \,k\,\tau_{1}) + H_{\mu+1}^{(1)}(k\,\tau_{1}) H_{\nu}^{(2)}(q \,k\,\tau_{1})\biggr],
 \nonumber\\
 A_{+}(k,\tau_{1}) &=& -\frac{i \, \pi}{4} e^{i \pi(\mu+\nu)/2} \sqrt{ q} \biggl\{H_{\mu}^{(1)}(k \tau_1) H_{\nu}^{(1)}( q \,k \tau_{1}) \biggl(\mu + \frac{1}{2} + \frac{2\nu +1}{2 q}\biggr)
\nonumber\\
 &-& k\,\tau_{1} \biggl[ H_{\mu}^{(1)}(k\,\tau_{1}) H_{\nu+1}^{(1)}(q \, k\,\tau_{1}) + H_{\mu+1}^{(1)}(k\,\tau_{1}) H_{\nu}^{(1)}(q \,k\,\tau_{1})\biggr].
\label{SSTT7}
\end{eqnarray}
where $q = \delta/\beta$ and $|A_{+}(k,\tau_{1})|^2 - |A_{-}(k,\tau_{1})|^2 = 1$, as it can be explicitly verified. From Eq. (\ref{SSTT7}) the slope of the spectral energy density is  $\overline{n}_{T}= 2(2 - \mu-\nu)$ and it coincides exactly with the result of Eq. (\ref{SSTT1}). As expected for  $\delta \to 1/2$ (i.e. $w\to 1$)
there is a logarithmic enhancement since, in the concurrent limit $\nu\to 0$ and $| k \tau_{1} |\ll 1$,  
the Hankel function approximates a logarithm of its argument (i.e. $H_{0}^{(1)}(k\,\tau_{1}) \to (2\, i/\pi) \ln{|k\,\tau_{1}|}$ )\cite{abr1,abr2}. 
\begin{figure}[!ht]
\centering
\includegraphics[height=6.3cm]{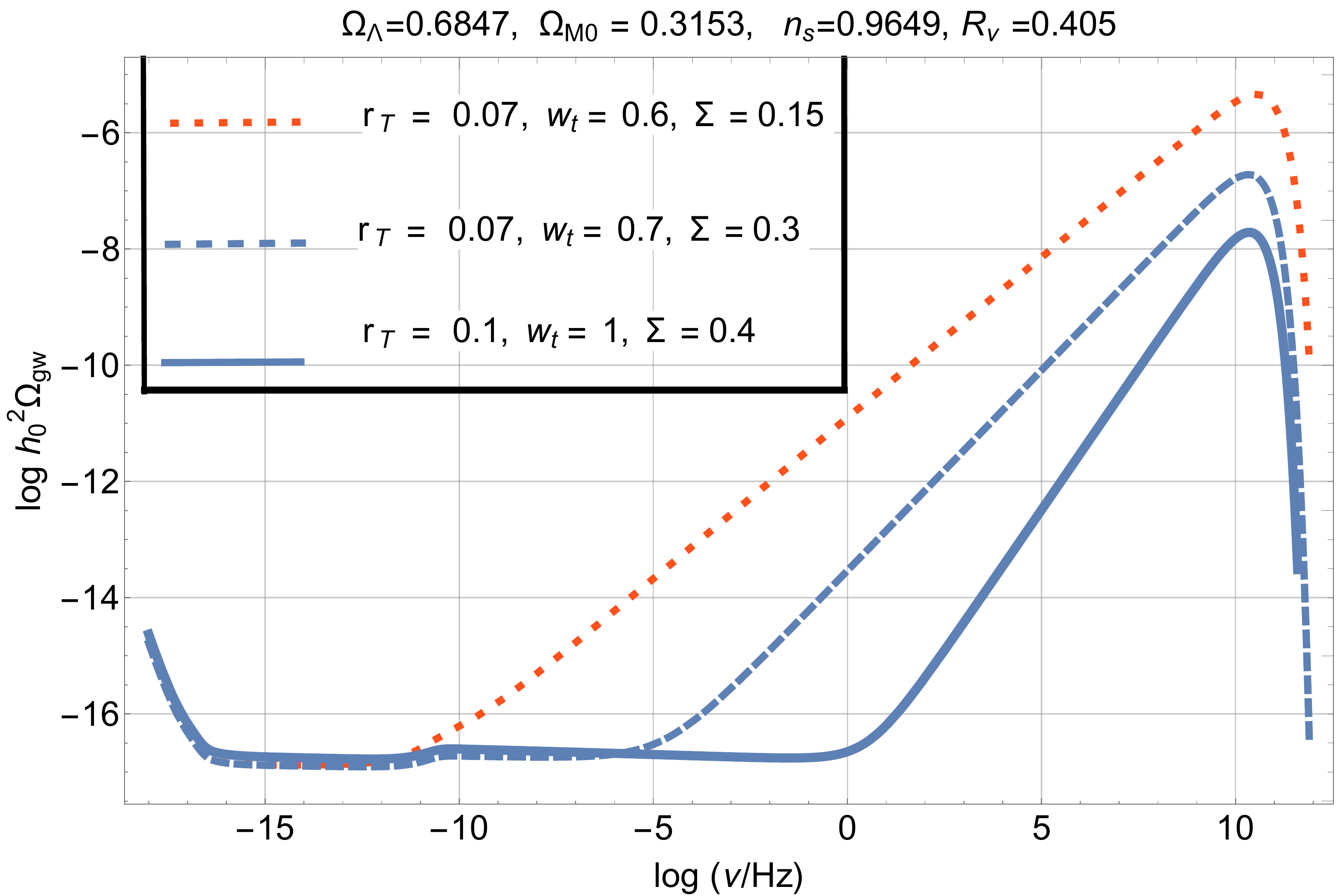}
\caption[a]{We illustrate the spectral energy density of the relic gravitons when the inflationary phase 
is not immediately followed by the radiation-dominated epoch; the barotropic index in the intermediate 
phase is always stiffer than radiation (i.e. $w_{t} > 1/3$).}
\label{SEC7FIG2}      
\end{figure}
Equations (\ref{SSTT1})--(\ref{SSTT2}) and (\ref{SSTT7}) lead to the correct of the spectral energy density but 
to determine more accurately the overall normalization, more specific numerical approaches have been  
employed \cite{TRANS8,TRANS9}. In the case of a stiff equation of state with $w \to 1$ the transfer function 
for the energy density (see Eqs. (\ref{TFF12})--(\ref{TFF13}) and discussion therein) can be computed  and the result is:
\begin{equation}
T_{\rho}^2(k/k_{r}) = 1.0  + 0.204\,\biggl(\frac{k}{k_{r}}\biggr)^{1/4} - 0.980 \,\biggl(\frac{k}{k_{r}}\biggr)^{1/2}  + 3.389 \biggl(\frac{k}{k_{r}}\biggr) -0.067\,\biggl(\frac{k}{k_{r}}\biggr)\ln^2{(k/k_{r})},
\label{ST10}
\end{equation}
where $k_{r} = \tau_{r}^{-1}$. In what follows we shall stress that the value of $k_{r}$ can either be 
computed in an explicit model or left as a free parameter to be estimated from the appropriate 
experimental data.

\subsection{Spikes and scaling violations at high frequencies}
Even if a sufficiently long stiff phase is compatible with all the high-frequency and intermediate frequency constraints (e.g. the pulsar limits and the nucleosynthesis bound), Eq. (\ref{SSTT1})  
implies that the spectral energy density 
evaluated at $k_{1} = {\mathcal O}(a_{1} H_{1})$ becomes critical when $(a_{1}/a_{*}) = (H_{1}/M_{P})^{2/(3 w -1)}$ and this observation suggests the relevance of the backreaction considerations 
associated with the presence of stiff epochs \cite{HIS23,HIS24a,HIS24b,STIFF19} (see also \cite{STIFF9,STIFF10,STIFF11}). The broad spike appearing in Fig. \ref{SEC7FIG2} is characterized by the maximal frequency:
\begin{equation}
\nu_{max} = \frac{M_{\mathrm{P}}}{2\pi} \Sigma^{-1}
 \biggl(\frac{H_{eq}}{M_{\mathrm{P}}}\biggr)^{1/2} \biggl(\frac{a_{eq}}{a_{0}}\biggr)
  = 1.17 \times 10^{11} \Sigma^{-1} 
\biggl(\frac{h_{0}^2 \Omega_{R0}}{4.15 \times 10^{-5}}\biggr)^{1/4}\,\,\mathrm{Hz},
 \label{STFR1}
 \end{equation}
 where, $\Sigma$ ultimately determines the position of $\nu_{max}$. Recalling the definition 
 of $H_{1}$ and $H_{r}$ associated with the Hubble rates at $a_{1}$ and $a_{r}$ (see Fig, \ref{SEC7FIG1}) we have that $\Sigma =  (H_{1}/M_{p})^{\alpha-1} \, (H_{r}/M_{P})^{1/2 - \alpha}$ where $\alpha = 2/[3(w+1)]$ for $w > 1/3$.   After the formulation of conventional inflationary models, Ford \cite{STIFF19} 
noted that gravitational particle production at the end of inflation could account for the entropy of the present universe and observed that the backreaction effects of the created quanta constrain the length of a stiff post-inflationary phase by making the expansion dominated by radiation. It has been later argued by  Spokoiny \cite{HIS24b} that various classes of scalar field potentials exhibit a transition from inflation to a stiff phase dominated by the kinetic energy of the inflaton; the same author also analyzed the conditions under which the expansion is again dominated by the inflaton asymptotically in the future. More recently it became increasingly plausible to have a single scalar field acting as inflaton in the early universe and as quintessence field in the late universe \cite{HIS24a,STIFF20,STIFF21} so that, in the case of quintessential inflation $\Sigma = {\mathcal O}(1)$, $\alpha \simeq 1/3$  and  $\nu_{max} \simeq {\mathcal O}(100)\,\,\mathrm{GHz}$. The wavelengths exiting the Hubble radius during inflation and reentering {\em after} the end of the stiff phase correspond to the straight lines denoted by $2$ in Fig. \ref{SEC7FIG1} and crossing the Hubble radius where there are the filled squares. In this frequency range the spectral energy density is quasi-flat exactly 
as in the case of the concordance paradigm; in Fig. \ref{SEC7FIG2}  the spectral energy density of the relic gravitons has been illustrated in the presence of a stiff post-inflationary phase with different spectral indices \cite{TRANS8,TRANS9,ABSOLUTE}. 
 
 In the simplest case the spectral energy density of the relic gravitons illustrated in  Fig. \ref{SEC7FIG2} depends on three parameters: the tensor to scalar ratio $r_{T}$ (defining the normalization and the spectral slope at low frequencies),   $\nu_{r}$ (denoting the frequency at which the scaling violations start
being relevant) and the spectral slope during the stiff phase\footnote{ Since $\nu_{r}$, $\Sigma$ and $w$ are related, different explicit parametrizations are possible.}.
The frequency $\nu_{r}$ can be explicitly expressed as:
\begin{equation}
\nu_{r} =  1.173 \times 10^{11} \Sigma^{ 1/(1-2\delta)} \,\,(\pi \epsilon {\mathcal A}_{{\mathcal R}})^{\frac{\alpha -1}{2(2\alpha -1)}}\,\,\biggl(\frac{h_{0}^2 \Omega_{R 0}}{4.15 \times 10^{-5}}\biggr)^{1/4}\,\,\mathrm{Hz}.
\label{STFR6}
\end{equation}
To preserve the light nuclei we need $\nu_{r} > \mathrm{nHz}$ and supposing  that the 
baryons were formed at the electroweak epoch we might also demand $\nu > \mathrm{mHz}$
but this second constraint can be evaded if the baryons are formed during the stiff phase as 
suggested long ago \cite{STIFF16a}. In this sense the requirement of $\nu_{r}$ in the mHz range 
is not so compelling and it would be wiser to leave $\nu_{r}$ as a free parameter instead of limiting its value 
to the mHz region, as recently suggested \cite{AMHD5}. We note, in this 
respect, that the possible variation of the refractive index of the relic gravitons during a quasi-de Sitter 
stage of expansion can be combined with the presence of a stiff post-inflationary phase: in this case the spike gets thinner and it is preceded by a quasi-flat plateau at high-frequency \cite{STIFF16}. 
If all the radiation present at the end of inflation comes from amplified quantum fluctuations \cite{STIFF19}
the energy density of the massive quanta is approximately $\rho_{r} \simeq H^4$ as it would follow 
from the Gibbons-Hawking radiation in de Sitter space, i.e. more precisely $\rho_{r}  = \pi^2 N_{eff} T_{H}^4/(30) = N_{eff} H^4/[480 \pi^2]$ where $N_{eff}$ is the number of species contributing 
to the quantum fluctuations during the (quasi)-de Sitter stage of expansion. 
It has been argued that this quantity could be evaluated using a perturbative expansion valid in the limit of quasi-conformal coupling \cite{STIFF19,STIFF22}.  
Given $H$ and $N_{\mathrm{eff}}$ the length of the stiff phase is fixed, in this case,  by 
\begin{equation}
{\mathcal N} H^4 \biggl(\frac{a_{i}}{a_{r}}\biggr)^{4} \simeq  H^2 M_{\mathrm{P}}^2  \biggl(\frac{a_{i}}{a_{r}}\biggr)^{3(w +1)} 
\simeq H^2 M_{P}^2  \biggl(\frac{a_{i}}{a_{r}}\biggr)^{2/\alpha}, 
\label{STFR14}
\end{equation}
where ${\mathcal N}= N_{eff}/(480\pi^2)$. Thanks to Eq. (\ref{STFR14}) the length of the stiff phase can be determined more explicitly by recalling that $H\simeq H_{1}$:
\begin{equation}
\biggl(\frac{a_{i}}{a_{r}}\biggr) \simeq {\mathcal N}^{\frac{\alpha}{2 - 4\alpha}} \biggl(\frac{H_{1}}{M_{P}}\biggr)^{\frac{\alpha}{1 - 2 \alpha}}, \qquad 
\biggl(\frac{H_{r}}{M_{P}}\biggr) \simeq {\mathcal N}^{\frac{1}{2 ( 1 - 2 \alpha)}} \biggl(\frac{H_1}{M_{P}}\biggr)^{\frac{2 ( 1 - \alpha)}{(1 -2 \alpha)}}.
\label{STFR15}
\end{equation}
For comparison, in the context of quintessential inflationary models $\Sigma=0.37$ so that $\nu_{r} = {\mathcal O}(4.3)$ Hz and this situation corresponds to the full thick curve of Fig. \ref{SEC7FIG2}. The backreaction of non-conformally coupled species can also be associated with other reheating 
mechanisms (see e.g. \cite{STIFF13,STIFF17,STIFF18}). In spite of the different possible scenarios 
explored in the current literature the results of Fig. \ref{SEC7FIG1} are compatible 
both with the bounds coming from the pulsar timing arrays of Eqs. (\ref{PUL0b})--(\ref{PUL0d}) and with 
the BBN constraints of Eqs. (\ref{BBN2}).
 
\subsection{Secondary spectra of high-frequency gravitons from reheating}
In the high-frequency band the primary contribution to the spectral energy density of the relic gravitons 
comes from the direct variation of the space-time curvature but there can also be secondary contributions associated with the anisotropic stress of some other species possibly produced inside the Hubble radius after the end inflation.  
Whenever the evolution of the inhomogeneities inside the Hubble radius 
generates a sizable anisotropic stress after the end of inflation, a secondary spectrum 
of relic gravitons corrects the primary contribution that follows in the framework 
of the concordance scenario. Since the secondary effects are generated 
by the evolution of the anisotropic stress, their contribution to the spectral energy 
density involves the analysis of specific convolutions. Secondary spectra 
of relic graviton may generically arise during reheating  (see \cite{RHn} for a review article on reheating mechanisms)  preheating \cite{RH1} (see also \cite{RHnn1,RHnn2} and references therein), in the presence of dynamical waterfall fields \cite{RH5,RWAT1,RWAT2,RWAT3}, when particles are produced at the end  of inflation \cite{RHnn,RHnn3} or even by 
the evolution of scalar perturbations \cite{RHff1,RHff2,RHff3}. 
\begin{figure}[!ht]
\centering
\includegraphics[height=6.3cm]{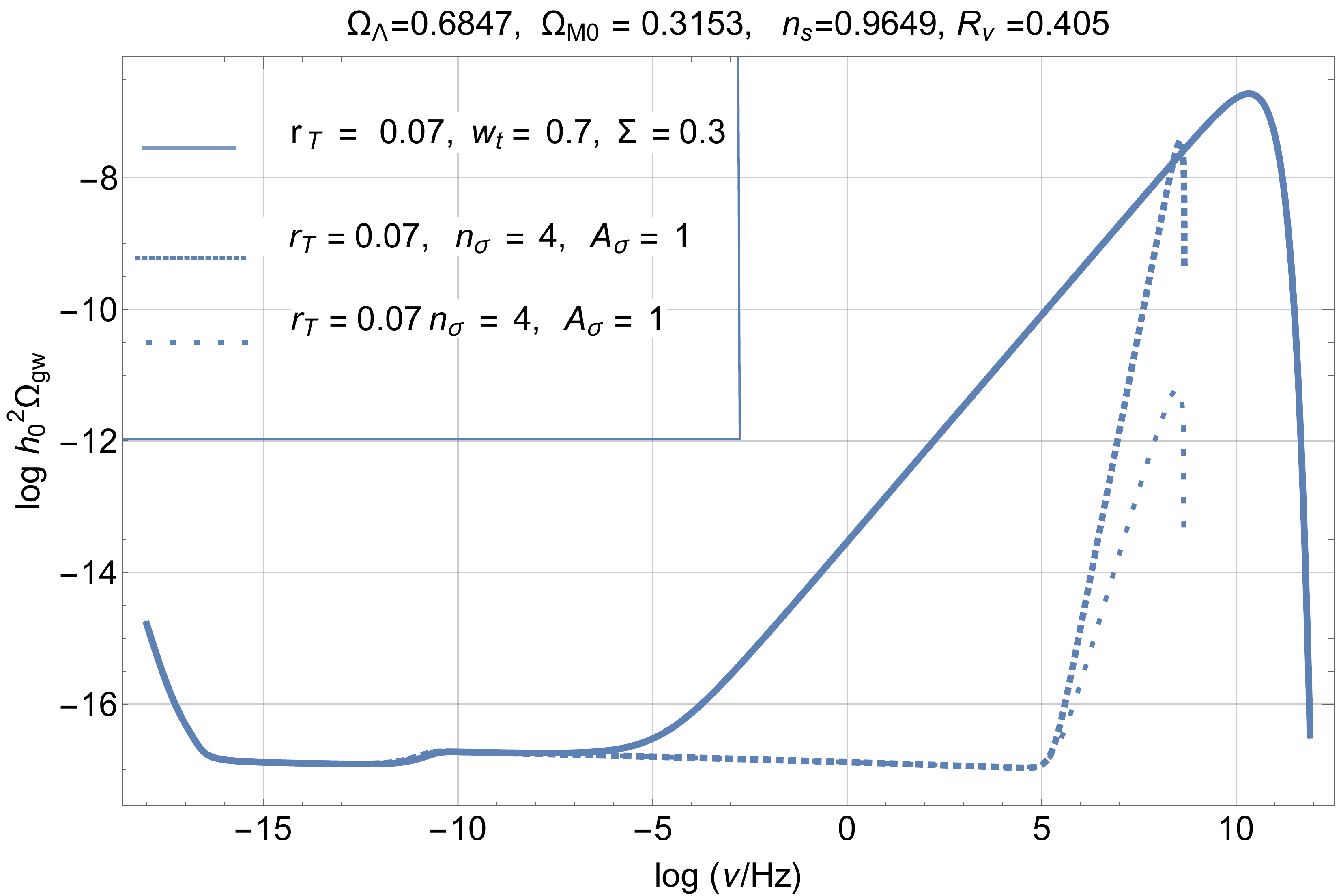}
\caption[a]{The secondary contribution to the spectral energy density is illustrated in the case of waterfall fields and compared 
with the primary contribution arising from the presence of stiff phases. }
\label{SEC7FIG3}      
\end{figure}
For instance in the case of a scalar field $\sigma(\vec{x},\tau)$ the associated anisotropic stress is given by
\begin{equation}
\Pi_{i j}(\vec{x},\tau) = - \frac{1}{a^2} \biggl[\partial_{i} \sigma \,\partial_{j} \sigma - \frac{1}{3} (\partial_{k} \sigma \partial^{k} \sigma) \delta_{i j} \biggr],
\label{WAT1}
\end{equation}
and the  evolution equation for each polarization can  be written, Fourier space, as $ h_{\lambda}'' + 2 {\mathcal H} h_{\lambda}' + q^2 h_{\lambda} = - 2 \,a^2 \,\ell_{\mathrm{P}}^2 \,\Pi_{\lambda}$ where $\Pi_{\lambda}(\vec{q},\tau)$ is:
\begin{equation}
\Pi_{\lambda}(\vec{q},\tau) = - \frac{1}{2 a^2 (2\pi)^{3/2}} \int d^{3} p \, e^{m n}_{\lambda}(\hat{q}) \, p_{m} \, p_{n} \, \sigma(\vec{q} - \vec{p}, \tau)\, \sigma(\vec{p}, \tau). 
\label{WAT2}
\end{equation}
By now recalling the sum over the polarizations of Eq. (\ref{POLDEF7}), Eq. (\ref{WAT2}) can also be written
by introducing the projector ${\mathcal S}_{i\,j\,m\,n}$ defined in Eqs. (\ref{POLDEF7})--(\ref{POLDEF8}) and appearing ubiquitously in the expressions of the two-point functions in Fourier space: 
\begin{equation}
h_{ij}^{\prime\prime} + 2 {\mathcal H} \, h_{ij}^{\prime} + q^2 h_{ij} = -4 \ell_{P}^2 \, a^2 \, {\mathcal S}_{i\, j\, m\, n}(\hat{q}) \Pi^{m\,n}(\vec{q},\tau),
\label{WAT3}
\end{equation}
where $\Pi^{m\,n}(\vec{q},\tau)$ denotes the Fourier transform of Eq. (\ref{WAT1}).
The nature of the scalar, pseudo-scalar or even vector fields contributing to the anisotropic stress depends on the specific model. For instance gravitational radiation can be  produced quite efficiently in interactions of classical waves created by the resonant decay of a coherently oscillating field \cite{RH1} in the presence 
of a quartic inflaton potential (e.g. $\lambda \varphi^4/4$). For $\lambda = 10^{-13}$ the spectral energy density was $\Omega_{gw} = {\mathcal O}(10^{-11})$ in the audio band (i.e. in the kHz region). Through the years 
gravitational waves have been claimed to be a generic consequence of reheating. For instance if inflation is 
described by plateau-like potential, reheating generically features oscillons, i.e. 
 long lived and spatially localized regions where the inflaton oscillates with large amplitude \cite{RHnn1,RHnn2}.
In these cases the spectral energy density in the audio band can be as large as $10^{-9}$ \cite{RHnn1}. 
Another possibility to obtain anisotropic stresses and secondary contributions 
peaked at high-frequencies is represented by waterfall-like fields 
that arise in hybrid inflationary models. If the spectrum of waterfall field is either blue or violet 
\cite{RWAT1,RWAT2,RWAT3} the secondary contribution to the spectral energy density 
is illustrated in Fig. \ref{SEC7FIG3} where $n_{\sigma}$ denotes the spectral slope of the waterfall 
spectrum and $A_{\sigma}$ is the normalized spectral amplitude in Planck units (see also \cite{RH5}).
In general terms the secondary contributions to the spectral energy density are typically
narrower in frequency as shown in Fig. \ref{SEC7FIG3} by comparison with the stiff 
models discussed above. There is finally the possibility that the anisotropic stress 
is induced directly by the scalar modes of the geometry and, in this situation, secondary contributions 
are expected not only in the audio band but also in the mHz band \cite{RHff2,RHff3}. 
Before concluding this section it is useful to remind that the typical signals 
expected from the evaporation of primordial black-holes may even reach beyond the THz region \cite{ST1}
even if the corresponding spectral energy density is quite small in comparison with the one 
analyzed here. 

\subsection{Hanbury-Brown Twiss effect and the quantumness of relic gravitons}
Even if relic gravitons and curvature perturbations are believed to share 
a common quantum mechanical origin, it is often difficult to envisage specific observational strategies 
supporting this conclusion. The vague problem of the quantumness of the large-scale inhomogeneities 
has been framed in a more empirical perspective by applying 
the tenets of the Hanbury Brown-Twiss (HBT) interferometry \cite{HBT1,HBT2,HBT2a} 
 to relic gravitons and relic phonons \cite{HBTA,HBTB}. 
Fields of various origin can well be {\em first-order coherent}
but what matters for an empirical distinction between the statistical properties of the classical and of the quantum fields is the degree of {\em second-order coherence}. While the degree of first-order 
coherence is measured by interfering amplitudes in conventional Young-type experiments,  the degree of second-order coherence follows from the analysis of the correlations between the field intensities. The formulation of the quantum theory of optical coherence, originally developed 
by Glauber and Sudarshan \cite{QM0,QO2,QO3} rests on the  complementarity between 
Young and HBT interferometry whose applications range from stellar astronomy \cite{HBT1,HBT2}  to subatomic physics \cite{revs,revsa}: the interference of the intensities has been also used to determine the hadron fireball dimensions \cite{cocconi1,cocconi1a,cocconi2} corresponding to the linear size of the 
interaction region in proton-proton collisions.  All the current and planned experiments are sensitive to the average multiplicity of the gravitons and hence to the degree of first-order coherence. However the statistical properties of the quantum fields in general (and of relic gravitons in particular) can only be disambiguated by examining the higher degrees of quantum coherence. The large-scale curvature perturbations have been shown to be always first-order coherent \cite{HBTBa,HBTBb} and the same is true in the case of the relic gravitons \cite{HBTA,HBTB} (see also \cite{HBTC,HBTCa}).  The degree of second-order coherence of the relic gravitons is instead always {\em super-Poissonian} \cite{HBTC,HBTCa} (see also \cite{loudon} for a detailed discussion of this 
property in a quantum optical context). Without referring to earlier results, the authors of Refs. \cite{HBTD,HBTDa} claimed instead that in the case of squeezed-coherent states and for an arbitrary choice of phases the degree of second-order coherence can become {\em sub-Poissonian}. Unfortunately the phases of squeezed-coherent states of relic gravitons are dictated by the process of parametric amplification \cite{HBTC,HBTCa}, they can be computed with reasonable accuracy and cannot be arbitrarily selected \cite{HBTB}.  While the canonical theory of optical coherence is formulated in terms of vector fields, it is also common to consider scalars corresponding to a single optical polarization \cite{QM0,QO2,QO3}. In the {\em tensor} case the general Glauber correlation function is given by \cite{HBTBb,HBTC}: 
\begin{eqnarray}
&& {\mathcal T}^{(n,m)}_{(i_{1}\,\,j_{1}), \,.\,.\,.\,(i_{n}\,\,j_{n}), \, (i_{n+1}\,\,j_{n+1}),\, .\,.\,.\,, (i_{n +m}\,\,j_{n+m}) }(x_{1}, \,.\,.\,.\,x_{n}, \, x_{n+1},\, .\,.\,.\,, x_{n +m}) 
\nonumber\\
&& = \mathrm{Tr}\biggl[ \hat{\rho} \, \hat{\mu}_{i_{1}\,\,j_{1}}^{(-)}(x_{1})\,.\,.\,.\, \hat{\mu}_{i_{n}\,\,j_{n}}^{(-)}(x_{n})
\, \hat{\mu}_{(i_{n+1}\,\,j_{n+1})}^{(+)}(x_{n+1})\,.\,.\,.\,\hat{\mu}_{(i_{n+m}\,\,j_{n +m})}^{(+)}(x_{n+m})\biggr],
\label{corrT1}
\end{eqnarray}
where $\hat{\rho}$ denotes here the density operator representing the (generally mixed) state of the field $\hat{\mu}_{ij}$; the notations are the same of the ones already employed in Eqs. (\ref{QTP2})--(\ref{QTP3}). Furthermore, as in Eqs. (\ref{PAM15a}) and (\ref{PAM15}) the field operator has been separated 
into a positive and a negative frequency part, i.e. $\hat{\mu}_{ij}(x) = \hat{\mu}^{(+)}_{ij}(x) + \hat{\mu}^{(-)}_{ij}(x)$ with $ \hat{\mu}^{(+)\,\,\dagger}_{ij}(x) =  \hat{\mu}^{(-)}_{ij}(x)$.
  Equation (\ref{corrT1}) generalizes the Glauber correlator (normally written in the case of photons) to the case of gravitons: instead of $(n+m)$ {\em vector} indices \cite{QO2}, in Eq. (\ref{corrT1}) we have $(n+m)$ pairs of {\em tensor} indices (i.e. $(i_{1}\, j_{1})\,...\,(i_{n}\, j_{n})\,...\, (i_{n+m}\, j_{n+m})$). In the single-polarization approximation Eq. (\ref{corrT1}) becomes:
\begin{eqnarray}
&& {\mathcal S}^{(n,m)}(x_{1}, \,.\,.\,.\,x_{n}, \, x_{n+1},\, .\,.\,.\,, x_{n +m}) 
\nonumber\\
&& = \mathrm{Tr}\biggl[ \hat{\rho} \, \hat{\mu}^{(-)}(x_{1})\,.\,.\,.\, \hat{\mu}^{(-)}(x_{n})
\, \hat{\mu}^{(+)}(x_{n+1})\,.\,.\,.\,\hat{\mu}^{(+)}(x_{n+m})\biggr],
\label{corrT2}
\end{eqnarray}
Equations (\ref{corrT1}) and (\ref{corrT2})  arise when considering the $n$-fold delayed  coincidence measurement of the tensor field at the space-time points $(x_{1}, \,.\,.\,.\,x_{n}, \, x_{n})$. From Eqs. (\ref{corrT1})--(\ref{corrT2}) the first-order Glauber correlators are \cite{HBTC}:
\begin{equation}
{\mathcal T}^{(1)}(x_{1},\, x_{2}) =  \langle  \hat{\mu}^{(-)}_{i\,j}(x_{1}) \, \hat{\mu}^{(+)}_{i\,j}(x_{2}) \rangle, 
\qquad 
{\mathcal S}^{(1)}(x_{1},\, x_{2}) = \langle  \hat{\mu}^{(-)}(x_{1}) \, \hat{\mu}^{(+)}(x_{2}) \rangle.
\label{corrT4}
\end{equation}
The second-order correlation function which is relevant when discussing the HBT
interferometry in the tensor case becomes instead:
\begin{eqnarray}
&& {\mathcal T}^{(2)}(x_{1}, x_{2}) = 
 \langle  \hat{\mu}^{(-)}_{i\,\,j}(x_{1}) \, \hat{\mu}^{(-)}_{k\,\,\ell}(x_{2})\hat{\mu}^{(+)}_{k\,\ell}(x_{2}) \hat{\mu}^{(+)}_{i\,j}(x_{1}) \rangle, 
\label{corrT5}\\
&& {\mathcal S}^{(2)}(x_{1}, x_{2}) = \langle  \hat{\mu}^{(-)}(x_{1}) \, \hat{\mu}^{(-)}(x_{2})\hat{\mu}^{(+)}(x_{2}) \hat{\mu}^{(+)}(x_{1}) \rangle.
\label{corrT6}
\end{eqnarray}
Note that the first relation in Eq. (\ref{corrT4}) as well as Eq. (\ref{corrT5}) both depend on the polarizations through the sum over the tensor indices; the sum {\em does not appear in the second correlator of  Eq. (\ref{corrT4}) and in Eq. (\ref{corrT6})}: the latter correlation functions are derived in  {\em single-polarization approximation}.  From Eq. (\ref{corrT4}) the normalized degrees of first-order coherence are:
\begin{eqnarray}
g^{(1)}(x_{1},\, x_{2}) = \frac{{\mathcal T}^{(1)}(x_1,\, x_{2})}{\sqrt{{\mathcal T}^{(1)}(x_1)} \,\, \sqrt{{\mathcal T}^{(1)}(x_2)}}, \qquad 
\overline{g}^{(1)}(x_{1},\, x_{2}) = \frac{{\mathcal S}^{(1)}(x_1,\, x_{2})}{\sqrt{{\mathcal S}^{(1)}(x_1)} \,\, \sqrt{{\mathcal S}^{(1)}(x_2)}}.
\label{corrT7}
\end{eqnarray}
The  normalized degrees of second-order coherence for the relic gravitons are finally given by given by:
\begin{eqnarray}
g^{(2)}(x_{1},\, x_{2}) = \frac{{\mathcal T}^{(2)}(x_1,\, x_{2})}{{\mathcal T}^{(1)}(x_1) \,\, {\mathcal T}^{(1)}(x_2)},
\qquad 
\overline{g}^{(2)}(x_{1},\, x_{2}) = \frac{{\mathcal S}^{(2)}(x_1,\, x_{2})}{{\mathcal S}^{(1)}(x_1) \,\, {\mathcal S}^{(1)}(x_2)}.
\label{corrT8}
\end{eqnarray}
where  the notation ${\mathcal T}^{(1)}(x_1)={\mathcal T}^{(1)}(x_1,\, x_{1})$ and ${\mathcal S}^{(1)}(x_1)={\mathcal S}^{(1)}(x_1,\, x_{1})$ have been used; $g^{(2)}(x_{1},\, x_{2})$ denotes
the degree of second-order coherence in its general form while $\overline{g}^{(2)}(x_{1},\, x_{2})$ is the degree 
of second-order coherence in the single polarization approximation. Finally, in the single-mode limit $g^{(2)}_{s}$
will be defined as\footnote{The average multiplicity of the gravitons shall be denoted by 
$\langle \hat{n} \rangle = {\mathrm{Tr}}[ \hat{\rho} \, \hat{a}^{\dagger}\,\hat{a}]$ while  
$\sigma^2 = \langle \hat{n}^2 \rangle - \langle \hat{n} \rangle^2$ is the dispersion. With these 
notations the second equality in Eq. (\ref{one}) immediately follows.}:
\begin{equation}
g^{(2)}_{s} = \frac{{\mathrm{Tr}}[ \hat{\rho} \, \hat{a}^{\dagger} \, \hat{a}^{\dagger} \, \hat{a} \,\hat{a} ]}{ \{{\mathrm{Tr}}[ \hat{\rho} \, \hat{a}^{\dagger}\,\hat{a}] \}^2} = 1 + \frac{\sigma^2 - \langle \hat{n}\rangle }{ \langle \hat{n}\rangle^2},
\label{one}
\end{equation}
where the subscript {\em s}  refers to the single-mode approximation while, as usual, $\hat{\rho}$ 
denotes the density operator in the same approximation.
\begin{figure}[!ht]
\centering
\includegraphics[height=6.3cm]{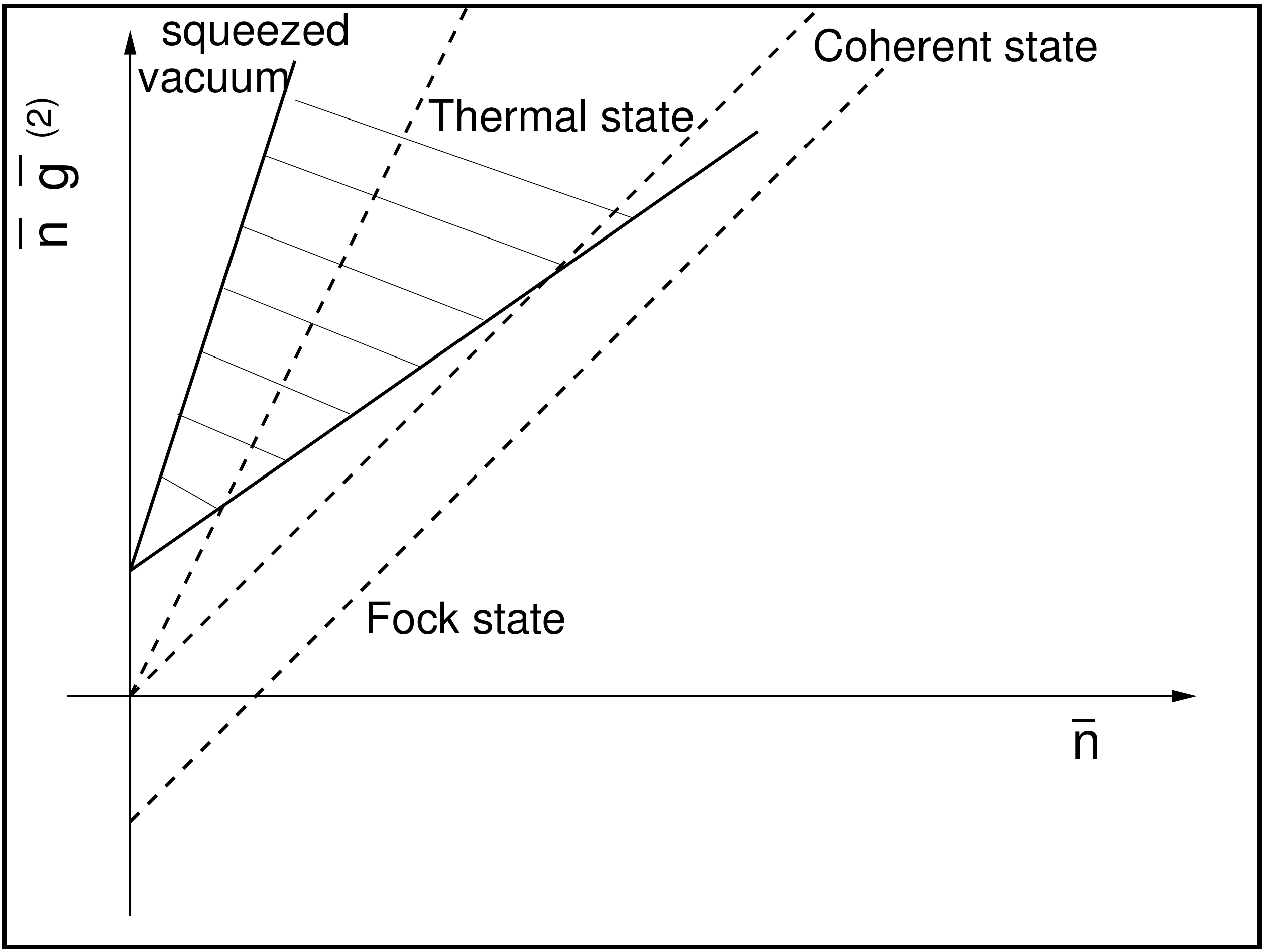}
\caption[a]{The degree of second-order coherence of the relic gravitons is illustrated and it follows 
by the shaded region. The dashed line passing through the origin corresponds to the 
case of a coherent state.}
\label{SEC7FIG4}      
\end{figure}
When the dispersion coincides with the mean value (i.e. $\sigma^2 = \langle \hat{n} \rangle$) as in the case of a Poisson distribution, then $g_{s}^{(2)} \to 1$ and this limit is reached by a single-mode coherent state (and it is commonly referred to as the Poisson limit \cite{QM0}). In Fig. \ref{SEC7FIG4} on the vertical axis 
we illustrate  $(\overline{n} \, \overline{g}^{(2)})$ while on the horizontal axis we report the average multiplicity $\overline{n}$; for simplicity we shall discuss the case of a single-mode approximation but Fig. \ref{SEC7FIG4} holds also in the single-polarization approximation. In Fig. \ref{SEC7FIG4} the Poissonian limit coincides with the dashed straight line passing through the origin. The statistical properties of the relic 
gravitons can be be summarized, in short, as follows. The relic gravitons are always first-order coherent in the zero-delay limit i.e. 
\begin{equation}
\lim_{\tau_{1}\to \tau_{2}} g^{(1)}(\vec{x}_{1}, \, \vec{x}_{1}; \tau_{1},\, \tau_{2}) = \lim_{\tau_{1}\to \tau_{2}}\overline{g}^{(1)}(\vec{x}_{2}, \, \vec{x}_{2}; \tau_{1},\, \tau_{2}) = g_{s}{(1)} =1;
\label{two}
\end{equation}
and this result is true in any approximation both for wavelengths shorter  (i.e. $k\tau > 1$) and larger (i.e. $k\tau< 1$) than the Hubble radius \cite{HBTB,HBTC}. If the initial state of the relic gravitons is the vacuum, the degrees of second-order coherence are \cite{HBTC,HBTCa}
\begin{equation}
g^{(2)}(x_{1}, x_{2}) \to \frac{41}{30}, \qquad \overline{g}^{(2)}(x_{1}, x_{2}) \to 3;
\label{three}
\end{equation}
where the first result refers to the general Glauber correlator (see Eqs. (\ref{corrT1}), first correlator in Eq. (\ref{corrT4}) and Eq. (\ref{corrT5})) while the second result refers to the single-polarization approximation (see Eqs. (\ref{corrT2}), second correlator in Eq. (\ref{corrT4}) and Eq. (\ref{corrT6})). The result of the single-polarization approximation matches 
the analog result obtainable for a single oscillator (see Eq. (\ref{one})) i.e. $g_{s}^{(2)} = 3 + 1/\overline{n}_{sq}$ where $\overline{n}_{sq} = \sinh^2{r}$ is the average multiplicity of the squeezed vacuum state.  Equation (\ref{three}) holds up to corrections $1/\overline{n}_{k}$ where $\overline{n}_{k}$ is the average multiplicity in $k$-mode of the field. The statistics of the relic gravitons is always super-Poissonian and larger than in the case of Bose-Einstein correlations (i.e. $g_{s}^{(2)} = 2$) and this region has been illustrated in the cartoon of Fig. \ref{SEC7FIG4}.
The degree of second-order coherence depends on the dynamical evolution and on the final state of the relic gravitons; the dependence on the initial state is milder: in spite of statistics of the initial state (thermal state, Fock state, coherent state) the relic gravitons is always super-Poissonian. On a general ground, it can be shown that in the single-polarization approximation the degree of second-order coherence of the relic gravitons 
obeys the following general bound \cite{HBTC,HBTCa}:
\begin{equation}
\frac{3}{2} \leq \overline{g}^{(2)} \leq 3.
\label{FS7}
\end{equation}
In Eq. (\ref{FS7}) $3/2$ corresponds to the minimal value associated with an initial Fock state while $3$ is the value of the squeezed vacuum state. The range of Eq. (\ref{FS7}) always exceeds $1$, which is the result valid for a coherent state (or for a mixed state with Poissonian weights). Since the results obtained so far in the current literature demonstrate that the HBT interferometry could indeed be used to disambiguate, in principle, the origin of the relic gravitons, it would be interesting to tailor specific observational strategies along this direction. In the case of curvature inhomogeneities  the new generations of CMB detectors and the HBT effect in the THz region seem to be a reasonable hope in this respect \cite{HBTBa}. Similar techniques might be extended to the case 
of relic gravitons as argued in the past \cite{HBTB}.

We conclude this section by noting that the high-frequency band is also particularly important for the direct detection 
of relic gravitons. In the kHz region the wide-band interferometers are now operating and provide direct bounds 
on spectral energy density of the relic gravitons. At even higher frequencies (e.g. in the MHz range) 
the electro-mechanical where the electromagnetic field and the field of elastic deformations are simultaneously present and are both affected by the high-frequency gravitational wave. Both topics will be addressed in the following section.

\newpage

\renewcommand{\theequation}{8.\arabic{equation}}
\setcounter{equation}{0}
\section{Stochastic backgrounds of relic gravitons: observations}
\label{sec8}
Since the spectrum of relic gravitons extends between 
the aHz and the GHz regions the detectors operating within each of the corresponding bands 
exhibit features that are markedly different. While the CMB experiments (discussed in section 
\ref{sec5}) are sensitive to the lowest branch of the spectrum [i.e. ${\mathcal O}(\mathrm{aHz})$], the space-borne
and the ground-based interferometers seem the most promising options in the intermediate 
and high-frequency regions discussed in sections \ref{sec6} and \ref{sec7} respectively.
Resonant mass detectors have been  the first instruments proposed for a direct 
detection of potential bursts of gravitational radiation. After a swift historic account 
of the pioneering experimental endeavours, the attention will be focussed on the futuristic projects of wide-band interferometers in space possibly operating in the mHz band. The current limits on the diffuse backgrounds of relic gravitons in the audio band will then be discussed in the light of the  LIGO/Virgo results. 
In the last part of the present section we shall briefly account for the main features of the electro-mechanical detectors that could be particularly promising for the gravitons in the MHz and GHz regions. 

\subsection{The early history of the detection of gravitational waves} 
In the late fifties of the past century questions have been raised on the true physical 
nature of  gravitational radiation. A school of thought suggested that gravitational radiation could not  
exist at least in the same empirical sense used to assess the existence of electromagnetic radiation. 
This early debate has been summarized in a classic paper by J. Weber and J. A. Wheeler \cite{WWEB0a}. 
Today we know that gravitational waves indeed modify the space-time curvature and have therefore 
a gauge-invariant meaning very much like the electromagnetic waves. An essential step along this direction has been the observation, made by F. Pirani \cite{WWEB0b} that the space-time curvature induced by gravitational waves modifies the invariant distance between test particles that are infinitesimally close (see e.g.\cite{TENS1,TENS2a,TENS2b}). The propagation of a gravitational wave through the space-time induces 
 a relative deviation of two nearby geodesics and this deviation can be measured though the vector 
 $\eta^{\mu} = \delta x^{\mu}$ connecting them. If the equation for the geodesics 
is perturbed to first-order  the connections must be consistently estimated to the same order in $\delta x^{\mu}$ 
so that the final result for the geodesic deviation leads to the well known equation for the geodesic deviation:
 \begin{equation}
 \frac{D^2 \eta^{\mu}}{ D\, s^2} = R^{\mu}_{\,\,\,\,\,\alpha\nu\beta} \, u^{\alpha} \, u^{\beta} \, \eta^{\nu}, \qquad u^{\mu} \eta_{\mu}=0, \qquad u^{\mu} = \frac{d x^{\mu}}{ds}.
 \label{DT1}
 \end{equation}
Equation (\ref{DT1}) involves the Riemann tensor and, in particular, its electric components 
 \cite{WWEB0b} (i.e. $R_{i\,0\, j\, 0}$); in Eq. (\ref{DT1}) $D$ denotes the covariant derivative along the  geodesic $x^{\mu}(s)$ and $u^{\mu}$ is the tangent velocity four-vector \cite{WWEB0a,WWEB0b}. According to Eq. (\ref{DT1}) 
 gravitational waves can then be measured as long as all the components of the Riemann tensor are experimentally inferred from the relative deviation of two nearby geodesics. 
 
 \subsubsection{Resonant mass detectors}
The experimental quest for gravitational waves  begun shortly after  the controversies described in Refs. \cite{WWEB0a,WWEB0b}:  between 1966 and 1970 J. Weber build the first resonant-mass detectors of gravitational radiation and conducted a series of pioneering observations \cite{WEB0c,WEB1,WEB2,WEB3,WEB4,WEB5} that will remain in the annals of experimental physics. The first gravitational wave antenna consisted of a large Aluminum cylinder (two meters long, half a meter in diameter) held at room temperature and isolated from the possible vibrations in a vacuum chamber. The detector weighted about $1.5\times 10^{3}$ kg and operated at an approximate frequency of $1.6$ kHz. As implied by Eq. (\ref{DT1}), a gravitational wave traveling perpendicular to the  axis of the cylinder produces tidal forces that  stretch and contract the length of the cylinder. The change in the length can be detected provided the frequency of the gravitational wave is sufficiently close to the resonant frequency of the antenna [i.e. ${\mathcal O}(1.6)$ kHz]. In the late sixties and early seventies of the past century the piezoelectric crystals were used to transform the changes in the length into electric currents; the successive generations of resonant detectors (in the eighties and throughout the nineties) used very low noise parametric amplifiers based on the Josephson effect.
 After observing the thermal fluctuations of the instruments \cite{WEB0c} the observations of Weber \cite{WEB1,WEB2} culminated with the analysis of the coincidences over a baseline 
 of ${\mathcal O}(10^{3})$ km at the Argonne National Laboratory and at the University of Maryland \cite{WEB3}.
 These data represented, according to Weber, {\em the first evidence of the existence of gravitational radiation}. 
 Few years later two independent groups by Garwin and Levine \cite{WWC1,WWC2,WWC3} and by Tyson \cite{WWC4,WWC5} severely criticised the results obtained by Weber. While a portion of the community 
 concluded that the Weber's results were not reproducible (and became hence skeptical about the 
 potential success of resonant mass detectors) some other groups tried to improve on the 
 original Weber's results by using cryogenic techniques to beat the thermal noises and increase 
 the sensitivities of the detectors.
 
 Improvements to the detectors occurred over the next two decades so that in the mid nineties of the past century 
 there have been up to five cryogenic resonant mass detectors operating in the kHz band: Niobe \cite{niobe} (Perth, Australia), Allegro \cite{allegro} (Baton Rouge, Lousiana, USA), Auriga 
\cite{auriga} (Legnaro, Italy), Explorer \cite{explorer}
(Geneva, Switzerland) and Nautilus \cite{nautilus} 
(Frascati, Italy). They all had cylindrical shape and for this reason they were commonly referred to as {\em bars}.
Except Niobe (made of Niobium and with approximate mass  of $1.5\times10^{3}$ kg),
all the remaining bars were made of Aluminium and their masses were slightly larger than
$2\times10^{3}$ kg. The mode frequencies corresponding to the various instruments ranged from the $694$ Hz  of Niobe to the $912$ Hz of Auriga. The shape of a resonant mass detector does not need to be cylindrical
even if the bars have been the predominant empirical realization of this type of instruments. 
Proposals of spherical resonant detectors 
have been actively pursued \cite{sph1,sph2,sph3} and an interesting review of the experimental situation until the mid nineties of the pas century can be found in Ref. \cite{pizzella}. The bars have been operating over long periods (see, for instance 
\cite{bars10}) and cooled down to $0.1$ K \cite{bars11}. Resonant mass detectors provided the first direct limits on stochastic backgrounds of gravitational radiation \cite{bars12}. It is useful to remind, for the sake of historical accuracy, 
that the data recorded by the Rome antennas during the celebrated Supernova event SN 1987a in the large 
Magellanic cloud were analyzed in coincidence with the data of the Maryland antenna (by Weber) and with the 
neutrino data of the Nusex experiment located in the Mont Blanc tunnel \cite{SN1,SN2}. In spite of striking coincidences, no discovery could be claimed since the antennas showing the largest ``signal''  were not cooled and effectively operated at room temperature, i.e. with low sensitivity. Resonant-mass detectors demonstrated through the years excellent sensitivities especially in their cryogenic 
versions. The weak point of these instruments was probably the rather narrow bandwidth: bars 
could only detect frequencies around the resonant frequency and this feature was particularly suitable 
for impulsive events like the supernova explosion. For this reason the correlation 
between two bars was also disfavoured in comparison with the hopes and promises 
of interferometers. 

\subsubsection{Interferometric detection of gravitational waves}
 Gertsenshtein and Pustovoit suggested in Ref.  \cite{INT1} the first key element for the interferometric detection of gravitational waves: in a gravitational field the optical length of the interferometer arm is changed and the 
relative difference is proportional to the amplitude of the gravitational wave. Thus ``the gravitational wave produces a periodic displacement of the interference bands'' \cite{INT1}. The main result of Ref. \cite{INT1} is actually one of the standard shorthand expressions stipulating that $\Delta L/L \simeq (h/2)$ where $h$ denotes the amplitude of the gravitational wave. From this 
approximate result we can estimate that for $L = {\mathcal O}(3)$ km we can assess an amplitude 
$h= {\mathcal O}(10^{-21})$ provided $\Delta L= 10^{-16}$ cm. Given that the Compton wavelength of an electron
is roughly $2\times 10^{10}$ cm, the displacement to be assessed corresponds 
to the  Compton wavelength of a particle whose mass is ${\mathcal O}(\mathrm{TeV})$. This figure 
gives the feeling of the orders of magnitude involved in the actual empirical 
detection of gravitational radiation through wide-band interferometers.  The first complete work on the noise competing with the gravitational wave signal in an interferometric antenna is due to the seminal contribution of Weiss \cite{INT2}.  An excellent account of the most important contributions and key papers can be found in Ref. \cite{INT3} by A. Giazotto.
This reference covers most of the work done until the late eighties of the past century.

 In the mid nineties of the past century four interferometers for the detection of gravitational waves have 
 been independently proposed and eventually evolved (sometimes with a tortuous path)  in the interferometers 
 that are operating today or will be operating in the near future. The proposed instruments were: 
 the Tama interferometer \cite{INT4} 
 (see also Ref. \cite{TAMA} already mentioned in the introduction), the Geo interferometer \cite{INT5} (see also 
 Refs. \cite{GEO1,GEO2} for some more recent references), the Virgo project \cite{INT6} and the original 
 LIGO ( Laser Interferometer Gravitational-wave Observatory) proposal \cite{INT7}. 
  The TAMA experiment (sometimes referred to as Tama-300) is a $300$ m interferometer operating near Tokyo. 
 After several years of operation the Tama team is now engaged in the design and construction 
 of a $3$ km interferometer called Kagra (Kamioka Gravitational Wave Detector) \cite{kagra1,kagra2}.
 The Geo detector (sometimes dubbed Geo-600) is a $600$ m interferometer build by a German-English 
 team near Hanover, Germany. Geo-600 achieved sensitivities comparable to the one of larger 
 instruments and has been used in the past as an essential playground for various technologies used 
 in larger instruments \cite{GEO1,GEO2}. This is particularly true for the use of squeezed light \cite{SQL1,SQL2,SQL3};
 this theme has been swiftly accounted for in section \ref{sec3} where the main features of the squeezed 
 states are described.
 
The LIGO/Virgo detectors have been often mentioned in the preceding sections and rather accurate 
descriptions of these instruments are available in the current literature (see e.g. \cite{LV1,LV2}). 
The LIGO interferometer consists in practice of three  Michelson interferometers: a single $4$ km interferometer in Livingston, Louisiana, as well as a pair of interferometers ($4$ km and $2$ km) in the LIGO facility at Hanford, Washington. The sites are separated by roughly 3000 km and are conceived to support coincidence analyses of the events. Virgo is a $3$ km French-Italian-Dutch Michelson interferometer near Pisa, Italy. In most respects, Virgo is quite similar to LIGO but Virgo employed a very sophisticated seismic isolation system that provides extremely good low-frequency sensitivity.  The advanced LIGO/Virgo detectors will probably be complemented in the future by the Indian LIGO \cite{LIGOindia} (sometimes called INDIGO) and, hopefully, by the futuristic Einstein telescope \cite{ET1}. As already discussed in section \ref{sec2} the possibility of developing a network of ground-based interferometers seems to be rather crucial to resolve the six polarization states of the gravitational waves \cite{SIXPOL1a,SIXPOL1,SIXPOL2,SIXPOL3} especially in the case of a short duration signal coming from distant sources (see also Eqs. (\ref{POLDEF2})--(\ref{POLDEF3}) and the discussions thereafter).

During a typical interferometric observation of LIGO the mirrors actually move by about ${\mathcal O}(10^{-16})$ cm.
Recalling that the spectral amplitude $S_{h}(\nu)$ estimates the mean amplitude of the gravitational wave 
(see Eqs. (\ref{OBS13f})--(\ref{OBS13g}) and (\ref{OBS14})--(\ref{OBS15}) we can say that in the typical situation provided by the the GW150914 event \cite{firstevent} (i.e. the first gravitational wave observation of the LIGO/Virgo collaboration) 
\begin{equation}
\Delta L = L\, \sqrt{\langle h_{ij} h^{ij}\rangle} = {\mathcal O}(4) \times 10^{-17} \mathrm{cm}, \qquad \langle h_{ij} h^{ij}\rangle
= \int_{\nu_{1}}^{\nu_{2}} \, S_{h}(\nu) \, d \nu,
\label{DT2}
\end{equation}
where $\sqrt{S_{h}(\nu)} = {\mathcal O}(10^{-23})/\sqrt{\mathrm{Hz}}$ and the integral is performed over the most 
sensitive region for $\Delta \nu = {\mathcal O}(100)$ Hz (i.e. $\nu_{1} = 35$ Hz and $\nu_{2} =250$ Hz); 
note that Eq. (\ref{DT2}) holds within the conventions of Eq. (\ref{OBS13g}).

The observations of gravitational waves by the advanced LIGO/Virgo detectors have been so-far focussed on the coalescence of binary black holes and binary neutron star systems but qualitatively different signals can be 
expected and, among them, the intermediate and high-frequency backgrounds 
of relic gravitons discussed in sections \ref{sec6} and \ref{sec7} respectively. With some optimism we could hope that LISA, BBO/DECIGO and the advanced Ligo/Virgo detectors might cover, on day, a rather  broad frequency interval (of about $10$ orders of magnitude) from the $\mu$Hz region up to the kHz band. The potential detection of a diffuse background in this range will be extremely important 
both per se as well as in connection with the general relativistic tests on the polarizations of the gravitational waves: in the case of a stochastic backgrounds these tests are more feasible (and direct) than in the case 
of short-duration signals. It could also happen that the stochastic backgrounds created by all binary black hole and binary neutron star mergers will act as foregrounds for the true primordial signals: the best antidote for this 
potential drawback will be the possibility of exploring different frequency bands that could resolve 
the slopes of the spectral energy density of the relic gravitons through a whole frequency range, as already pointed out 
in sections \ref{sec6} and \ref{sec7}. 
 
\subsection{Wide-band interferometers in space}
As already mentioned at the end of section \ref{sec6}, in outer space the limitations related to the seismic 
noise disappear so that signals below the Hz may become observationally accessible. In spite of the 
absence of seismic noise it is not excluded that different sources of disturbance, unexpected on the ground, might arise 
and threaten the possibility of sensitive measurements. With these caveats we can say that the operating window of space-borne interferometers falls in the mHz range where three complementary projects have been proposed so far: the 
 LISA (Laser Interferometer Space Antenna) project \cite{LISA,LISAa} (see also \cite{LISA2,LISA3,LISA4}),  the BBO (Big Bang Observer) project \cite{BBO}, and the DECIGO project (Deci-hertz Interferometer Gravitational Wave Observatory) \cite{DECIGO1} (see also \cite{DECIGO2,DECIGO3}).  The LISA pathfinder mission \cite{LISA4} (the precursor of the true LISA mission) already  tested some of the key technologies so that LISA should be operational for about four years hopefully starting in 2034 (or later).  LISA consists of a triplet of satellites located at the vertices of an equilateral triangle; the mutual distance between the satellites is $2.5\times10^{11}$ cm. Each of the three satellites will host one interferometer even if the technical features of these instruments are different from their ground-based counterpart. The measurements of  the relative acceleration noise of two test masses have been successfully tested by the LISA Pathfinder mission \cite{LISA4,LISA5}. The instrument in its final configuration will be probably sensitive to both polarizations of gravitational waves from any direction and its operating band will extend from frequencies smaller than $0.1$ mHz to a fraction of the Hz. 
\begin{table}[!ht]
\begin{center}
\caption{The hoped strain sensitivities of wide-band interferometers in space are roughly compared with their terrestrial counterparts (i.e. the advanced LIGO/Virgo detectors). Three different ranges of frequency are reported for each of the instruments. As far as the space-borne interferometers the reported figures 
are purely illustrative and follow from the optimistic suggestions of the current literature.}
\vskip 0.3 cm
\begin{tabular}{||c|c|c||}
\hline
\hline
\rule{0pt}{4ex} Experiment & strain sensitivity   & frequency range \\
\hline
\hline
LISA& $ \sqrt{S_{h}(\nu)} ={\mathcal O}(10^{-16}) \mathrm{Hz}^{-1/2}$ &  $\nu_{1} = {\mathcal O}(10)\, \mu\mathrm{Hz}$  \\
LISA& $  \sqrt{S_{h}(\nu)} = {\mathcal O}(10^{-20}) \mathrm{Hz}^{-1/2}$ & $ \mathrm{mHz}< \nu_{2} < 10 \,\mathrm{mHz}$\\
LISA&  $  \sqrt{S_{h}(\nu)} \geq {\mathcal O}(10^{-19}) \mathrm{Hz}^{-1/2}$ & $\nu_{3}\geq 100\, \mathrm{mHz}$\\
\hline
\hline
DECIGO & $\sqrt{S_{h}(\nu)} ={\mathcal O}(10^{-21}) \mathrm{Hz}^{-1/2}$ & $\nu_{1} ={\mathcal O}(10)\, \mathrm{mHz}$\\
DECIGO & $\sqrt{S_{h}(\nu)} ={\mathcal O}(10^{-24}) \mathrm{Hz}^{-1/2}$ & $100\,\mathrm{mHz}< \nu_{2} < 10\,\mathrm{Hz}$\\
DECIGO & $\sqrt{S_{h}(\nu)} \geq {\mathcal O}(10^{-22}) \mathrm{Hz}^{-1/2}$ & $\nu_{3} \geq 100 \,\mathrm{Hz}$\\
\hline
\hline
Advanced LIGO/Virgo & $\sqrt{S_{h}(\nu)} = {\mathcal O}(10^{-22}) \mathrm{Hz}^{-1/2}$ &  $\nu = {\mathcal O}(10) \,\mathrm{Hz}$\\
Advanced LIGO/Virgo & $\sqrt{S_{h}(\nu)} = {\mathcal O}(10^{-24}) \mathrm{Hz}^{-1/2}$ &  $50\,\mathrm{Hz}< \nu \leq \,\mathrm{kHz}$\\
Advanced LIGO/Virgo & $\sqrt{S_{h}(\nu)} > {\mathcal O}(10^{-23}) \mathrm{Hz}^{-1/2}$ &  $\nu > \mathrm{kHz}$\\
\hline
\hline
\end{tabular}
\label{SEC6TAB1}
\end{center}
\end{table}
While the frequency range of  LISA does not exceed the Hz, at higher frequencies the DECIGO project \cite{DECIGO1,DECIGO2,DECIGO3} should operate about at the same time between $0.1$ Hz and $10$ Hz.  DECIGO should also consist of 3 satellites, in an equilateral triangle configuration, but with a distance separation of $10^{8}$ cm. Four DECIGO clusters have been proposed so far and 
each cluster should consist of  3 satellites in an equilateral triangle configuration. Two of the DECIGO clusters will be overlapping  and this possibility makes DECIGO especially sensitive to stochastic backgrounds: the ideal situation for the detection of stochastic backgrounds is when the interferometers are colocated since, in this case, the signal to noise ratio will be maximized, as we shall briefly discuss section \ref{sec7} in connection with the attempts of colocated measurements performed by the LIGO experiment. The precursor of the DECIGO interferometer (called B-DECIGO) will soon fly to demonstrate the technological feasibility of the instrument and it will consist of three satellites (separated by $10^{7}$ cm) 
and orbiting around the earth (while the true DECIGO, as LISA, will fly in an heliocentric orbit).
Last but not least, the Big Bang Observer (BBO) \cite{BBO} will always use a triangular configuration, but with arm lengths of $5 \times 10^{9}$ cm.  BBO is designed to look for a cosmologically produced stochastic background, with a sensitivity of $\Omega_{GW} \sim 10^{-17}$ in the range 0.03 Hz to 3 Hz.
The hoped strain amplitudes of the LISA and DECIGO/BBO detectors are summarized in Tab. \ref{SEC6TAB1} and compared with the approximate strain sensitivities 
of the advanced LIGO/Virgo detectors \cite{LV1,LV2}.

The frequency range of DECIGO seems particularly suitable for the searches of stochastic backgrounds 
of relic gravitons since the potential contamination from white dwarf binaries should be 
extremely small in this range. Because of this reduced white dwarf binary foreground, and the sensitivity of DECIGO, it could be possible, in the most optimistic perspective,  to achieve a detection limit for a stochastic background search of $\Omega_{gw} \simeq 2 \times 10^{-16}$ with three years of observations. This hopeful 
figure, however, might not be sufficient to detect the stochastic background of relic gravitons 
in the concordance paradigm (see Fig. \ref{SEC4FIG5} and discussion therein). In the case 
of blue and violet spectra of relic gravitons discussed in sections \ref{sec6} and \ref{sec7} 
the DECIGO sensitivity will be more than sufficient to set direct limits and, hopefully, to 
detect some of the potential signals.

The Big Bang Observer (BBO) \cite{BBO} is similar to DECIGO: like DECIGO, it would have a triangular configuration, but with arm lengths of $5 \times 10^{4}$ km. With two overlapping triangular clusters a cross-correlation can be made between independent detector data sets. BBO is designed to look for a cosmologically produced stochastic background, with a sensitivity of $\Omega_{gw} \sim 10^{-17}$ in the band ranging from $0.03$ Hz to $3$ Hz. This frequency band should hopefully be free of astrophysical contamination.
\begin{figure}[!ht]
\centering
\includegraphics[height=6.3cm]{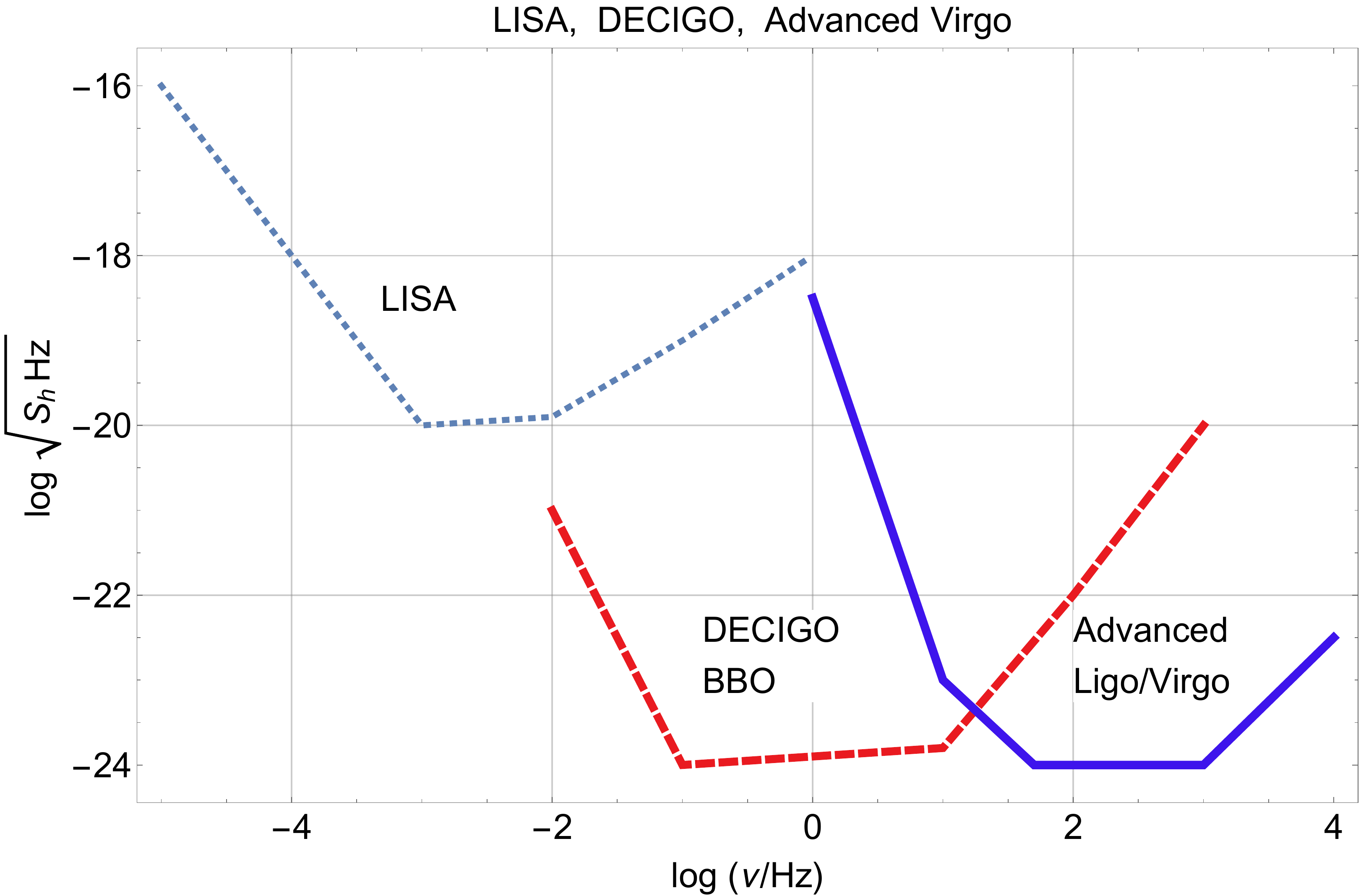}
\caption[a]{Using the figures of Tab. \ref{SEC6TAB1} the approximate strain amplitudes of the LISA and DECIGO/BBO projects are illustrated together with the ones of the advanced LIGO/Virgo detectors.}
\label{SEC6FIG7}      
\end{figure}
It is clear from the cartoon of Fig. \ref{SEC6FIG7} (constructed from Tab. \ref{SEC6TAB1} )
that the BBO/DECIGO range overlaps at low-frequencies with the LISA frequencies and at high-frequency with the audio band 
where ground-based detectors are already operational\footnote{There are already relevant limits for the stochastic 
backgrounds obtained by the ground-based detectors currently operating. These observations will be briefly summarized in section \ref{sec7} since they are related to the audio band.}. It is interesting to go back to Eqs. (\ref{OBS14}) and (\ref{OBS15}) 
and to derive from Tab. \ref{SEC6TAB1} the minimal detectable $h_{0}^2 \Omega_{gw}(\nu,\tau_{0})$. 
According to Tab. \ref{SEC6TAB1} the LISA project will be able to probe strain amplitudes ${\mathcal O}(10^{-20})/\sqrt{\mathrm{Hz}}$;
 from Eqs. (\ref{OBS14}) and (\ref{OBS15}) the minimal detectable spectral energy density and chirp amplitude
will be given by 
\begin{equation}
h_{0}^2 \Omega_{gw}(\nu_{L}, \tau_{0}) = 4 \times 10^{-11}, \qquad h_{c}(\nu_{L},\tau_{0}) = 1.6 \times 10^{-21}, \qquad \nu_{L} = 5\, \mathrm{mHz},
\label{SNL10}
\end{equation}
where we conventionally selected $\nu_{L}$ within the range where the LISA detectors will be hopefully more sensitive. Let us then do the same estimate in the case of DECIGO; in this case choosing $\nu_{D} =  0.1\,\mathrm{Hz}$ (see Tab. \ref{SEC6TAB1}) and recalling 
 Eqs. (\ref{OBS14}) and (\ref{OBS15}) we shall obtain 
\begin{equation}
h_{0}^2 \Omega_{gw}(\nu_{D}, \tau_{0}) = 3 \times 10^{-15}, \qquad h_{c}(\nu_{D},\tau_{0}) = 6.9 \times 10^{-25}, \qquad \nu_{D} = 0.1\,\mathrm{Hz}.
\label{SNL11}
\end{equation}
Finally assuming $\nu_{LV} = 100$ Hz we can compute the same quantities in the LIGO/Virgo case and obtain
\begin{equation}
h_{0}^2 \Omega_{gw}(\nu_{LV}, \tau_{0}) = 3 \times 10^{-6}, \qquad h_{c}(\nu_{LV}, \tau_{0}) = 2.1 \times 10^{-23}, \qquad \nu_{LV} = 100\,\mathrm{Hz}.
\label{SNL12}
\end{equation}
The estimates of Eqs. (\ref{SNL10}), (\ref{SNL11}) and (\ref{SNL12}) illustrate an interesting aspect:
 if we bluntly assume that the minimal detectable chirp (or strain) amplitudes are the same in the 
mHz and in the audio band, the sensitivity in $h_{0}^2 \Omega_{gw}(\nu,\tau_{0})$ {\em increases} as we {\em decrease} the frequency. This conclusion ultimately follows from the relation between 
the spectral energy density and the chirp (or strain) amplitudes discussed in Eqs. (\ref{OBS14}) and (\ref{OBS15}).
This aspect is particularly evident by looking at Fig. \ref{SEC6FIG7}: the approximate strain amplitudes 
of the DECIGO/BBO interferometers are comparable with the ones of the advanced LIGO/Virgo 
but nonetheless the minimal detectable $h_{0}^2 \Omega_{gw}$ of Eq. (\ref{SNL11}) is much smaller than 
the one of Eq. (\ref{SNL12}). To appreciate this difference let us therefore go back to 
Eq. (\ref{OBS14}) which we can write as:
\begin{equation}
h_{0}^2 \Omega_{gw}(\nu, \tau_{0}) = 3.15 \times 10^{-6}  \, \biggl(\frac{\nu}{100\mathrm{Hz}}\biggr)^3 \biggl[ \frac{ {\mathcal S}_{h}(|\nu|)\, \,\mathrm{Hz}}{10^{-49}}\biggr],
\label{SNL12a}
\end{equation}
where we assumed, for the sake of concreteness,  a strain sensitivity $\sqrt{{\mathcal S}_{h}(|\nu|) \, \,\mathrm{Hz}} = 10^{-24.5}$. Now, when $\nu = {\mathcal O}(100)$ Hz, Eq. (\ref{SNL12a}) implies a minimal detectable 
spectral energy density ${\mathcal O}(3) \times 10^{-6}$. Conversely if the operating 
frequency of the instrument is much lower, {\em the minimal detectable $h_{0}^2 \Omega_{gw}(\nu)$ 
increases provided $\sqrt{{\mathcal S}_{h}(|\nu|) \, \,\mathrm{Hz}}$ remains the same}. For instance, when $\nu = 0.1$ Hz we will have instead $h_{0}^2 \Omega_{gw}(\nu_{D}, \tau_{0}) = 3.15 \times 10^{-15}$. With the same logic we can determine the minimal detectable  chirp amplitude 
$h_{c}(\nu)$ (see Eq. (\ref{OBS15}) and definitions therein). To detect the stochastic background of the conventional lore 
discussed in section \ref{sec4} (see, for instance, Fig. \ref{SEC4FIG5}) we should have, for example, 
\begin{equation}
h_{c}(\nu_{D},\tau_{0}) = {\mathcal O}(4) \times 10^{-26}\, \sqrt{\frac{h_{0}^2 \Omega_{gw}}{10^{-17}}}.
\label{SNL12b}
\end{equation}
If is finally useful to remark that, in Fig. \ref{SEC1FIG1}, the minimal detectable spectral energy density by the advanced LIGO/Virgo detectors has been 
quoted as ${\mathcal O}(10^{-10})$. There is no contradiction between this estimate and Eqs. (\ref{SNL12})--(\ref{SNL12a}): as we shall see in the following subsection the sensitivity to a stochastic background of relic gravitons increases by correlating two (separated or colocated) detectors and the minimal detectable spectral energy density ultimately depends on the bandwidth and of the total integration time.

\subsection{Bounds on relic gravitons from operating ground-based detectors}
The LIGO and Virgo interferometers operate in a frequency window approximately 
extending between few Hz and $10$ kHz and this range is commonly referred to as {\em the audio band}. The possible detection of relic gravitons by two or more correlated interferometers relies on the methods developed in the past by various authors \cite{wide1,wide2,wide3,wide4,wide5,wide6}. So far only upper limits on the spectral energy density 
of relic graviton backgrounds have been reported from $20$ Hz to the 
kHz \cite{STone,STonea,STtwo,STthree,STthreea,STfour,STfive}.  These upper limits are
summarized in Tab. \ref{SEC7TABLE1} and generally depend {\em on the spectral slope of the signal}.
\begin{table}[!ht]
\begin{center}
\caption{List of the direct limits on the relic gravitons obtained by wide-band interferometers.}
\vskip 0.4 cm
\begin{tabular}{||c|c|c|c||}
\hline
\hline
\rule{0pt}{4ex}  Year & frequency range [Hz] & Bound & Reference \\
\hline
\hline
& & &\\
$2004$ & $40-314$ & $ \overline{\Omega}_{0} < 23$ & Ref. \cite{STone} \\
$2005$ & $69-156$ & $  \overline{\Omega}_{0} < 8.4 \times 10^{-4}$  & Ref. \cite{STonea} \\
$2012$ & $600-1000$ & $\overline{\Omega}_{3} < 0.32$ & Ref. \cite{STtwo} \\
$2014$ & $41.5-169.25$ & $\overline{\Omega}_{0} < 5.6 \times 10^{-6}$ & Ref. \cite{STthree}\\
$2014$ & $600-1000$ & $\overline{\Omega}_{3} < 0.14$ & Ref. \cite{STthree}\\
$2014$ & $170-600$ & $\overline{\Omega}_{0} < 1.8\times 10^{-4}$ & Ref. \cite{STthree}\\
$2014$ & $1000-1726$ & $\overline{\Omega}_{3} < 1$ & Ref. \cite{STthree}\\
$2015$ & $460-1000$  & $\overline{\Omega}_{3} < 7.7 \times 10^{-4}$ & Ref. \cite{STthreea}\\
$2017$ & $20-86$ &    $\overline{\Omega}_{0} < 1.7 \times 10^{-7}$ & Ref. \cite{STfour}\\
$2017$ & $20-300$ & $\overline{\Omega}_{3} < 1.7 \times 10^{-8}$ & Ref. \cite{STfour}\\
$2019$  & $20-81.9$ & $\overline{\Omega}_{0} < 6 \times 10^{-8}$  & Ref. \cite{STfive}\\
$2019$  & $20-95.2$ & $\overline{\Omega}_{2/3} < 4.8 \times 10^{-8}$ &Ref. \cite{STfive}\\
$2019$ &  $20-301$  & $\overline{\Omega}_{3} < 7.9 \times 10^{-9}$ & Ref. \cite{STfive}\\
\hline
\hline
\end{tabular}
\label{SEC7TABLE1}
\end{center}
\end{table}
In the context of optimal processing  required for the detection of stochastic backgrounds,
the expression of the signal-to-noise ratio (SNR) depends on the spectral energy density 
as \cite{wide1,wide2,wide3,wide4,wide5}: 
\begin{equation}
{\rm SNR} \,=\,\frac{3 H_0^2}{2 \sqrt{2}\,\pi^2}\,F\,\sqrt{T}\,
\left\{\,\int_0^{\infty}\,{\rm d} \nu\,\frac{\gamma^2 (\nu)\,\Omega^2_{gw}(\nu,\tau_{0})}
{\nu^6\,S_n^{\,(1)} (\nu)\,S_n^{\,(2)} (\nu)}\,\right\}^{1/2}\;.
\label{SNR1}
\end{equation}
Even if slightly different definitions of the SNR are possible \cite{wide6} the key elements entering 
Eq. (\ref{SNR1}) are, in short, the following. The term $F$ depends upon the geometry of the two detectors\footnote{In the past resonant-mass detectors were operating (see e.g. the beginning of this section). It was therefore useful to analyze the correlation between detectors with different geometries (be it a bar, a sphere or an interferometer).}  and in the case of the correlation between 
two interferometers $F\to2/5$.  The observation time is denoted by $T$ 
and $S_n^{\,(k)} (f)$ (with $k = 1,\, 2$) are the  (one-sided) noise power 
spectra $k$-th detector. The noise power spectra contain the informations concerning the 
noise source. Generally speaking, the seismic noise is the main source 
of disturbance in the low-frequency region (i.e. between few Hz and say $10$ Hz); between 
$20$ and $100$ Hz the thermal noise dominates while for larger frequencies the shot noise 
(i.e. the photon noise) gives the most important contribution. 
An important quantity appearing in Eq. (\ref{SNR1}) is the so-called overlap reduction function 
$\gamma(\nu)$ which is determined by the relative locations and orientations 
of the two detectors. If the two detectors are {\em colocated and coaligned} $|\gamma(\nu)| = 1$. 
If the two detectors are not colocated at intermediate frequencies the integral of Eq. (\ref{SNR1}) is  sensitive to the form of the overlap reduction function which can be given as a combination of spherical Bessel functions \cite{wide7,wide8}. This observation implies that $\gamma(\nu)$ is in general an oscillating function and the approximate position of its first zero (i.e.  $\nu_{c} = 1/(2\, d)$ where 
$d$ denotes the separation between the two detectors) will determine an effective cut-off for the integrand appearing in Eq. (\ref{SNR1}). According to this rough argument the most sensitive band for the LIGO/Virgo detectors must be for $\nu< \nu_{c} < 100$ Hz. As expected the bounds reported in Tab. \ref{SEC7TABLE1} suggest that $\nu_{c}$ falls between $20$ and $90$ Hz.

In the analyses directly relevant for the relic gravitons \cite{STone,STonea,STtwo,STthree,STthreea,STfour,STfive} the LIGO/Virgo collaboration parametrized the spectral energy density with a power-law slope of the type: 
\begin{equation}
\Omega_{gw}(\nu,\tau_{0}) = \overline{\Omega}_{\delta} \, \biggl(\frac{\nu}{\nu_{ref}} \biggr)^{\delta}, 
\qquad \delta \geq 0,
\label{SNR2}
\end{equation}
where $\nu_{ref}$ is a (conventional) frequency while $\overline{\Omega}_{\delta}$ is a constant amplitude 
that is different depending on the frequency slope $\delta$. Different slopes are characterized by a corresponding overall amplitude so, for instance, $\overline{\Omega}_{0}$ is the amplitude of the scale-invariant spectral energy density while $\overline{\Omega}_{3}$ is the amplitude of a spectral energy density with cubic slope. 
In Refs.\cite{STone,STonea,STtwo,STthree,STthreea,STfour} (see alsoTab. \ref{SEC7TABLE1}) the LIGO/Virgo collaboration often presents the obtained upper limits in the two aforementioned  cases (i.e. $\Omega_{0}$ and $\Omega_{3}$). The scale-invariant case represents the simplest signal grossly compatible with the concordance paradigm; conversely,  if $\delta= 3$ in Eq. (\ref{SNR2}),  the factor $\Omega_{gw}^2/\nu^{6}$ appearing inside the integrand of Eq. (\ref{SNR1}) does not depend on the frequency and the evaluation of the SNR gets somehow simpler. 

In the case of an exactly scale-invariant spectrum the most stringent constraints obtained in Refs. \cite{STfour,STfive} imply that $\overline{\Omega}_{0} < 6\times 10^{-8}$. If this figure is compared with the analog result obtained in Ref. \cite{STthree}, the upper limit is ${\mathcal O}(100)$ times more constraining for the same slope and for the same frequency band (see also Tab. \ref{SEC7TABLE1}).
For $\delta = 2/3$ the constraints of Ref. \cite{STfour} imply $\overline{\Omega}_{2/3} < 4.8 \times 10^{-8}$ with 95\% confidence within the $20$--$95$ Hz frequency band with $\nu_{ref} = 25$ Hz.  The slope $\delta= 2/3$ may actually parametrize a potentially interesting foreground for the 
relic gravitons. In fact depending on the estimates, the rates of black hole  mergers range from ${\mathcal O}(50) \, \mathrm{Gpc}^{-3}\, \mathrm{yr}^{-1}$ to ${\mathcal O}(300) \, \mathrm{Gpc}^{-3}\, \mathrm{yr}^{-1}$. If this is the case we can not only expect to have many more signals 
but also a stochastic foreground coming from  unresolved sources of gravitational radiation. 
In the near future the advanced LIGO/Virgo detectors may offer promising avenues for a further scrutiny of relic gravitons \cite{LV1,LV2}. It is commonly quoted that the advanced LIGO/Virgo instruments should probably cut through the interesting region of the parameter space by reaching sensitivities comparable to  $\Omega_{gw} = {\mathcal O}(10^{-10})$ 
[or, more accurately, $\overline{\Omega}_{0} = {\mathcal O}(10^{-10})$]. In this respect we remind that the wideness of the band is very important when cross-correlating two detectors: typically the minimal detectable $h_{0}^2\Omega_{gw}$ will become smaller (i.e. the sensitivity will increase) by a factor $1/\sqrt{T\,\Delta \nu }$ where $\Delta \nu$ is the bandwidth and $T$, as already mentioned, is the observation time. Naively, if the minimal detectable signal (by one detector) is $h_{0}^2\Omega_{gw} \simeq 10^{-5}$, then the cross-correlation of two identical detector with overlap reduction $\gamma(\nu) =1$ will detect  
 $h_{0}^2\Omega_{gw} \simeq 10^{-10}$ provided $\Delta \nu \simeq 100$ Hz and 
 $T\simeq {\mathcal O}(1\mathrm{yr})$ (recall that $1\mathrm{yr} = 3.15 \times 10^{7} \mathrm{Hz}^{-1}$). 
 
Concerning the limit $|\gamma(\nu)| \to 1$ in Eq. (\ref{SNR1}) a final caveat is in order.
In Ref. \cite{STthreea} a search for stochastic gravitational waves using data from
the two colocated LIGO Hanford detectors has been reported.  At low frequencies (i.e. $40$--$460$ Hz)
the collaboration was unable to mitigate the correlated noise to an acceptable level.  However, at high frequencies, $460$--$1000$ Hz (see also Tab. \ref{SEC7TABLE1}) these techniques were sufficient to set a 95\% confidence level upper limit on the gravitational-wave energy density of $\overline{\Omega}_{3} < 7.7\times 10^{-4}$ for $\nu_{ref} = 900$ Hz. The coherence of the common noises in both detectors (especially for  $\nu < 460$ Hz) seems to suggest that, in the future, the high-frequency bands may offer the optimal sensitivity in the case of colocated and coaligned detectors.

In summary the observations in the audio band by advanced LIGO/Virgo did not detect 
so far the diffuse backgrounds of gravitational radiation. The stochastic signals created 
by all binary black hole and binary neutron star mergers throughout the history of the universe
might be detected in the future but, if this is the case, also the backgrounds induced 
by relic gravitons with growing spectral energy density should have a similar chance 
of being detected. A similar comment holds for the potential signals coming from cosmic defects. 
An important cross-check, in this respect would be the pulsar timing observations and their potential successes (see section \ref{sec6} and discussion therein).

\subsection{Electro-mechanical detectors and high-frequency gravitons}
If the spectral energy density increases either in the intermediate range (see e.g. Fig. \ref{SEC6FIG5}) or at high-frequencies (see e.g. Figs. \ref{SEC7FIG2} and \ref{SEC7FIG3}), large signals are expected in the MHz range. This is why electro-mechanical  detectors seem particularly suitable for the detection of high-frequency gravitons that may interact both with the electromagnetic field and with the field of elastic deformations of the detector \cite{brag1,cav1}.  These instruments can be concisely described as electro-mechanical detectors of gravitational radiation and one of the first concrete proposals along this direction goes back to the seminal work of Braginsky and Menskii \cite{brag1} (see also \cite{brag2}). The Braginsky-Menskii detector consisted of a toroidal waveguide where a wave-packet of electromagnetic radiation propagates: the effect of a circularly polarized gravitational wave translates into a variation of the averaged frequency of the wave-packet. 

Relic gravitons may induce transitions between the different levels of an electro-mechanical apparatus provided the system satisfies some resonance conditions in a quadrupole geometry. While the simplest detector of this kind would consist of two free-falling mirrors oriented on orthogonal axes (with an electromagnetic field propagating inside the cavity), it turns out to be more practical to couple together two electromagnetic resonant cavities \cite{cav1,cav1a,cav1b,cav2,cav2a}. This strategy uses 
static electromagnetic fields and demands an accurate analysis of the interaction of the gravitational wave with the electromagnetic field \cite{cav3,cav3a}.
The analysis of electromagnetic cavities has been complemented by the use of dynamical electromagnetic fields, such as for instance, waveguides \cite{cav4,cav4a}. A prototype for the observation 
of relic gravitons at $100$ MHz has been constructed by Cruise and Ingley \cite{cav5,cav5a}. Other interesting  detectors have been described and partially built \cite{cav6,cav6a,cav7} with frequency of operation of the order of 100 MHz. The possibility of detecting stochastic backgrounds of relic gravitons in the MHz region has also been seriously considered by using small (i.e. 75 cm) interferometers \cite{cav8}. 
 
 A tunable resonant sensor, using optically trapped and cooled dielectric microspheres or microdisks, has been proposed to detect gravitational waves in the frequency range of $50$--$300$ kHz \cite{SENS1}. The  frequency range of these detectors would be smaller than the MHz but much larger than the audio band and would therefore be particularly suitable to investigate the transition region between the ground-based detectors (i.e. LIGO/Virgo) and the MHz band.  Another project potentially relevant for high-frequency gravitons is the
Fermilab Holometer \cite{holo1,holo2} consisting of two Michelson interferometers that are nearly overlapping (a separation of $0.635$ m), with arm lengths of $39.2$ m. The strain sensitivity  is claimed to be better than $10^{-21}$ Hz$^{-1/2}$ in the range between few MHz to 13 MHz. Some limits have been also derived in this range (i.e.  $\Omega_{gw} <  {\mathcal O}(10^{12})$). Even if the obtained limits are not constraining, they are important to demonstrate the feasibility of the measurements. Since we are discussing small detectors of relic gravitons it is appropriate to include at this point some discussion on the atomic gravitational wave interferometric sensor \cite{SENS2,SENS3,SENS4}. These devices operate at frequencies much smaller than the MHz but they have a relatively small size. Using the core technology of the Stanford $10$ m atom interferometer two kinds of atom interferometers have been proposed a terrestrial instrument and a space-borne detector \cite{SENS2}. The terrestrial experiment consists of two $10$ m atom interferometers separated by a $1$ km baseline; they can operate with strain sensitivity $10^{-19}\sqrt{\mathrm{Hz}}$ in the band between few Hz and $10$ Hz. Needless to say, this band is not accessible to the LIGO/Virgo detectors due to seismic noise. The space-borne experiment consists of two atom interferometers separated by a ${\mathcal O}(10^{3})$ km baseline and a potential strain sensitivity ${\mathcal O}(10^{-20})/\sqrt{\mathrm{Hz}}$. 

\newpage
\renewcommand{\theequation}{9.\arabic{equation}}
\setcounter{equation}{0}
\section{Concluding perspectives}
\label{sec9}
The primordial backgrounds of relic gravitons are the true pathfinders for the new physics beyond the standard lore 
of fundamental interactions. Their energy distribution bears the mark of the early history of the 
Hubble rate, of the underlying gravity theory and of the energy budget of the universe throughout all the essential moments 
characterizing the evolution of the primeval plasma. 
If ever observed the $B$-mode polarization will represent a direct probe of the spectral energy density 
of the relic gravitons in the aHz range. Conversely the absence of a confirmed 
$B$-mode detection at low frequencies will presumably cast serious doubts on the validity of some of the 
premises of the so-called concordance scenario. The pulsar timing arrays will be sensitive, in the intermediate nHz domain, to the late-time backgrounds of gravitational radiation, like the ones produced by 
inspiral mergers and supermassive black holes.  From the practical viewpoint the direct observation of a long-duration signal will improve the direct bounds on the supplementary polarizations (e.g. scalar or vector-like) possibly arising in various alternative theories of gravity. Depending on the specific scenario the pulsar observations will still represent a relevant constraint for the relic gravitons 
whose energy density will probably have to compete with various foregrounds, not only in the nHz domain.
In the audio band (i.e. between few Hz and $10$ kHz) the current limits on the cosmic gravitons come from the direct searches of the LIGO/Virgo collaboration. The third generation of ground-based gravitational wave detectors will be one order of magnitude more sensitive than the LIGO/Virgo detectors in their advanced configuration. These sensitivities will hopefully allow for a direct observation of various astrophysical foregrounds. In the most optimistic case the minimal detectable spectral energy density is still too large for the potential signal of the concordance paradigm.  In the mHz region the space-borne interferometers  may achieve sensitivities that could be equally insufficient to probe the signal of the concordance scenario. However the validity of an improved concordance paradigm in the aHz region does not exclude a much larger signal in the audio band. Diverse physical considerations lead to spectral energy densities that increase over intermediate and high frequencies and are potentially detectable by the advanced LIGO/Virgo interferometers: bouncing scenarios, modified post-inflationary histories, refractive indices of relic gravitons, reheating dynamics all lead to a variety of signals that can be separated either by looking at the frequency range or at the spectral slopes. In the intermediate frequency domain hypercharge fields, strongly first order phase transitions and topological defects may also lead to potentially large  anisotropic stresses ultimately producing secondary foregrounds.

The mutual relationships between the low-frequency domain and the high-frequency channels of the cosmic graviton spectrum will  shape a new version of the concordance paradigm: since the relic gravitons are the sole opportunity of scrutinizing the thermodynamic history of the plasma and the underlying gravity theory at small distances, they might also represent the ultimate test for inflation and for all the competing scenarios aiming at a sensible description of the early universe. The considerations developed so far demonstrate that the primordial backgrounds cannot be scrutinized by looking along single line of sight and that their analysis inclusively encompasses various concurrent approaches. For these reasons there are little doubts that the relic gravitons will foster, in the years to come, an interesting cross-disciplinary trialogue involving cosmology, astrophysics and high-energy physics. 

\section*{Acknowledgements} 
I am very grateful to T. Basaglia, A. Gentil-Beccot, S. Rohr and J. Vigen of the CERN Scientific Information Service for their kind assistance. I also thank D. Pedrini for his friendly support.


\newpage

\begin{thebibliography}{99}

\itemsep -2pt
 
\bibitem{LIGO1} B.~P.~Abbott {\it et al.} [LIGO/Virgo Collaboration],  Phys.\ Rev.\ Lett.\  {\bf 119}, 141101 (2017).

\bibitem{LIGO2} B.~P.~Abbott {\it et al.} [LIGO/Virgo Collaboration],  Phys.\ Rev.\ Lett.\  {\bf 119}, 161101 (2017).
  
\bibitem{LIGO3} B.~P.~Abbott {\it et al.} [LIGO/Virgo Collaboration],   Phys.\ Rev.\ Lett.\  {\bf 118}, 221101 (2017).

\bibitem{RT1} G.~Hinshaw {\it et al.} [WMAP Collaboration],   Astrophys.\ J.\ Suppl.  {\bf 208}, 19 (2013).

\bibitem{RT2} C. L. Bennett, {\it et.al.}  [WMAP Collaboration],  Astrophys.\ J.\ Suppl.  {\bf 208}, 20B (2013).

\bibitem{RT3} P.~A.~R.~Ade {\it et al.} [BICEP2 and Keck Array Collaborations],  Phys.\ Rev.\ Lett.\  {\bf 116}, 031302 (2016).
  
\bibitem{RT4}  Y.~Akrami {\it et al.} [Planck Collaboration], arXiv:1807.06211 [astro-ph.CO].

\bibitem{CMB1} A.~A.~Penzias and R.~W.~Wilson,  Astrophys.\ J.\  {\bf 142}, 419 (1965).

\bibitem{CMB2} J.~C.~Mather {\it et al.}  [COBE Collaboration], Astrophys.\ J.\  {\bf 354}, L37 (1990).

\bibitem{CMB4} J.~C.~Mather {\it et al.} [COBE Collaboration], Astrophys.\ J.\  {\bf 420}, 439 (1994).

\bibitem{CMB5} D.~J.~Fixsen,  Astrophys.\ J.\  {\bf 707}, 916 (2009).

\bibitem{CMB3} G.~F.~Smoot {\it et al.} [COBE Collaboration],  Astrophys.\ J.\  {\bf 396}, L1 (1992).

\bibitem{WMAP1} D.~N.~Spergel {\it et al.} [WMAP Collaboration],  Astrophys.\ J.\ Suppl. {\bf 148}, 175 (2003).

\bibitem{WMAP1a} H.~V.~Peiris {\it et al.}  [WMAP Collaboration],  Astrophys.\ J.\ Suppl.\  {\bf 148}, 213 (2003); 

\bibitem{WMAP1b} C.~L.~Bennett {\it et al.}  [WMAP Collaboration],  Astrophys.\ J.\ Suppl.\  {\bf 148}, 1 (2003).

\bibitem{WMAP2} D.~N.~Spergel {\it et al.} [WMAP Collaboration],  Astrophys.\ J.\ Suppl.  {\bf 170}, 377 (2007).

\bibitem{WMAP2a} L.~Page {\it et al.}  [WMAP Collaboration],  Astrophys.\ J.\ Suppl.\  {\bf 170}, 335 (2007).

\bibitem{B1} P.~A.~R.~Ade {\it et al.}  [BICEP2 Collaboration],  Phys.\ Rev.\ Lett.\  {\bf 112}, 241101 (2014).
  
\bibitem{PUL1} V.~M.~Kaspi, J.~H.~Taylor, and M.~F.~Ryba,   Astrophys.\ J.\ {\bf 428}, 713 (1994). 
  
\bibitem{PUL2}  F.~A.~Jenet {\it et al.}, Astrophys.\ J.\ {\bf 653}, 1571 (2006). 

\bibitem{LISA} P.~Amaro-Seoane {\it et al.},  GW Notes {\bf 6}, 4 (2013).

\bibitem{LISAa} S. Vitale, Gen. Relativ. Gravit. {\bf 46}, 1730 (2014).

\bibitem{BBO}  G.~M.~Harry, P.~Fritschel, D.~A.~Shaddock, W.~Folkner and E.~S.~Phinney, Class.\ Quant.\ Grav.\  {\bf 23}, 4887 (2006)  Erratum: [Class.\ Quant.\ Grav.\  {\bf 23}, 7361 (2006)].

\bibitem{DECIGO1} S.~Kawamura {\it et al.}, J.\ Phys.\ Conf.\ Ser.\  {\bf 120}, 032004 (2008).

\bibitem{PT0a} G. Aad  {\it et al.}  Phys. Lett. B {\bf 716}, 1 (2016).

\bibitem{PT0b} S. Chatrchyan {\it et al.}  Phys. Lett. B {\bf 716}, 3 (2016).
 
\bibitem{LV1} J.~Aasi {\it et al.} [LIGO Collaboration],  Class.\ Quant.\ Grav.\  {\bf 32}, 074001 (2015) .

\bibitem{LV2} F.~Acernese {\it et al.} [VIRGO Collaboration],  Class.\ Quant.\ Grav.\  {\bf 32}, 024001 (2015).

\bibitem{kagra1} Y.~Aso {\it et al.} [KAGRA Collaboration],  Phys.\ Rev.\ D {\bf 88}, 043007 (2013).

\bibitem{kagra2} K.~Somiya [KAGRA Collaboration],  Class.\ Quant.\ Grav.\  {\bf 29}, 124007 (2012).

\bibitem{TAMA}  M. Ando et al., Phys. Rev. Lett. {\bf 86}, 3950 (2001). 

\bibitem{LIGOindia} C. S. Unnikrishnan, Int. J. Mod. Phys. D {\bf 22}, 1341010 (2013).

\bibitem{GEO1}  B.~Willke {\it et al.}, Class.\ Quant.\ Grav.\  {\bf 19}, 1377 (2002).

\bibitem{GEO2} H.~Grote [LIGO Collaboration], Class.\ Quant.\ Grav.\  {\bf 27}, 084003 (2010).

\bibitem{ET1} B.~Sathyaprakash {\it et al.},  Class.\ Quant.\ Grav.\  {\bf 29}, 124013 (2012)
  Erratum: [Class.\ Quant.\ Grav.\  {\bf 30}, 079501 (2013)].

\bibitem{brag1} V. B. Braginsky and M. B. Menskii,  Pis'ma Zh. Eksp. Teor. Fiz. {\bf 13}, 585 (1971) [JETP Lett. {\bf 13}, 417 (1971)].

\bibitem{cav1} F. Pegoraro, L.  Radicati, Ph. Bernard, and E. Picasso, Phys. Lett. A {\bf 68}, 165 (1978).
  
\bibitem{LCDM1} J. M. Kovac {\it et al.}, Nature {\bf 420} 772 (2002).

\bibitem{LCDM4a}  G.~Hinshaw {\it et al.} [WMAP Collaboration], Astrophys.\ J.\ Suppl.\  {\bf 180}, 225 (2009).

\bibitem{LCDM4b} M.~R.~Nolta {\it et al.} [WMAP Collaboration], Astrophys.\ J.\ Suppl.\  {\bf 180}, 296 (2009).

\bibitem{LCDM4c} E.~Komatsu {\it et al.} [WMAP Collaboration],   Astrophys.\ J.\ Suppl.  {\bf 180}, 330 (2009).

\bibitem{LCDM5} E.~Komatsu {\it et al.} [WMAP Collaboration],   Astrophys.\ J.\ Suppl.  {\bf 192}, 18 (2011).

\bibitem{LCDM6} P.~A.~R.~Ade {\it et al.} [Planck Collaboration],  Astron.\ Astrophys. {\bf 571}, A16 (2014).

\bibitem{LCDM7} P.~A.~R.~Ade {\it et al.} [Planck Collaboration],  Astron.\ Astrophys.\  {\bf 571}, A22 (2014).

\bibitem{LCDM8} P.~A.~R.~Ade {\it et al.} [Planck Collaboration],  Astron.\ Astrophys.\  {\bf 594}, A13 (2016). 

\bibitem{LCDM9} W.~J.~Percival, B.~A.~Reid, D.~J.~Eisenstein {\it et al.},  Mon.\ Not.\ Roy.\ Astron.\ Soc.\  {\bf 401}, 2148 (2010).

\bibitem{LCDM10} B.~A.~Reid, W.~J.~Percival, D.~J.~Eisenstein {\it et al.}, Mon.\ Not.\ Roy.\ Astron.\ Soc.\  {\bf 404}, 60 (2010).

\bibitem{LCDM11} R.~Kessler, A.~Becker, D.~Cinabro {\it et al.},  Astrophys.\ J.\ Suppl.\  {\bf 185}, 32 (2009).

\bibitem{LCDM12}  M.~Hicken, W.~M.~Wood-Vasey, S.~Blondin {\it et al.}, Astrophys.\ J.\  {\bf 700}, 1097 (2009).

\bibitem{LCDM13} C.~L.~Reichardt, P.~A.~R.~Ade, J.~J.~Bock {\it et al.}, Astrophys.\ J.\  {\bf 694}, 1200 (2009).

\bibitem{LCDM14} M.~Zemcov {\it et al.}  [QUaD collaboration], Astrophys.\ J.\  {\bf 710}, 1541 (2010).

\bibitem{LCDM15} C.~Bischoff {\it et al.} [QUIET Collaboration],  Astrophys.\ J.\  {\bf 741}, 111 (2011).

\bibitem{LCDM17} J.~L.~Sievers {\it et al.} [Atacama Cosmology Telescope Collaboration],  JCAP {\bf 1310}, 060 (2013).

\bibitem{LCDM17a} A.~van Engelen {\it et al.} [Atacama Cosmology Telescope Collaboration ],   Astrophys.\ J.\  {\bf 808}, 7 (2015).

\bibitem{LCDM19} P.~A.~R.~Ade {\it et al.} [POLARBEAR Collaboration],  Phys.\ Rev.\ D {\bf 92}, 123509 (2015).
  
\bibitem{RT5}  N. Aghanim {\it et. al.}  [Planck Collaboration], arXiv:1807.06209 [astro-ph.CO].  
  
\bibitem{book1} P. D. Naselsky, D. I. Novikov, and I. D. Novikov 
{\it The physics of the cosmic microwave background}, (Cambridge University press, Cambridge UK, 2006).  
  
\bibitem{book2} S. Weinberg, {\it Cosmology} (Oxford University Press, Oxford, UK, 2008).

\bibitem{book3} M. Giovannini, {\it A primer on the Physics of the Cosmic Microwave Background}, (World Scientific, Singapore, 2008).  

\bibitem{SAK1}  A. D. Sakharov,  Sov. Phys. JETP {\bf 22}, 241 (1966) [Zh. Eksp. Teor. Fiz. {\bf 49}, 345 (1965)].

\bibitem{SAK2} P.~J.~E.~Peebles and J.~T.~Yu,  Astrophys.\ J.\  {\bf 162} 815 (1970).

\bibitem{SAK3} R.~A.~Sunyaev and Y.~B.~Zeldovich,  Astrophys.\ Space Sci.\  {\bf 7}, 3 (1970).

\bibitem{PAR1} L.~Parker,  Phys.\ Rev.\ Lett.\  {\bf 21}, 562 (1968).

\bibitem{PAR2}  L.~Parker,  Phys.\ Rev.\  {\bf 183}, 1057 (1969).
  
\bibitem{PAR3} L.~Parker, Phys.\ Rev.\ D {\bf 3}, 346 (1971) Erratum: [Phys.\ Rev.\ D {\bf 3}, 2546 (1971)].

\bibitem{PAR4}   L. Parker, Phys. Rev. Lett. {\bf 28}, 705 (1972).

\bibitem{HIS1} E. M. Lifshitz, {\it J. Phys. USSR} {\bf 10}, 116 (1946) [ Zh. Eksp. Teor. Fiz.  {\bf 16} 587 (1946)].

\bibitem{ryan} M.P. Ryan, and L.C. Shepley, {\it ``Homogeneous Relativistic Cosmologies''},  (Princeton University Press, Princeton 1975).

\bibitem{HIS1a} E.M. Lifshitz,  I.M. Khalatnikov, Sov. Phys. JETP {\bf 12}, 108 (1961) [Zh. Eksp. Teor. Fiz. {\bf 39}, 345 (1960)].

\bibitem{HIS1aa} E.M. Lifshitz, I.M. Khalatnikov, Phys. Rev. Lett. {\bf 24}, 76 (1970).

\bibitem{HIS1b} V.A. Belinskii, I.M. Khalatnikov, Sov. Phys. JETP {\bf 30}, 1174 (1970) [Zh. Eksp. Teor. Fiz. {\bf 49}, 2163 (1969)].

\bibitem{HIS1bb} V.A. Belinskii, I.M. Khalatnikov, Sov. Phys. JETP {\bf 36}, 591 (1973) [Zh. Eksp. Teor. Fiz. {\bf 63}, 1121 (1972)].  

\bibitem{HIS1c} I. M. Khalatnikov and E. M. Lifshitz, Phys. Rev. Lett. {\bf 24}, 76 (1969).    
    
\bibitem{HIS1d} V. A. Belinskii, E. M. Lifshitz, and I. M. Khalatnikov,  Advan. Phys. {\bf 19}, 525 (1970).

\bibitem{HIS1e}  C. W. Misner, Phys. Rev. Lett. {\bf 22}, 1071 (1969).

\bibitem{HIS1f} V. A.  Belinskii and I.M. Khalatnikov, Sov. Phys. JETP {\bf 36}, 591 (1973)  [Zh.\ Eksp.\ Teor.\ Fiz.\  {\bf 63}, 1121 (1972)].

\bibitem{HIS5} L.~P.~Grishchuk,   Sov.\ Phys.\ JETP {\bf 40}, 409 (1975)   [Zh.\ Eksp.\ Teor.\ Fiz.\  {\bf 67}, 825 (1974)].

\bibitem{HIS5a} L.~P.~Grishchuk,  Nuovo Cimento Lett. {\bf 12}(2), 60 (1975).

\bibitem{HIS5b} B. V. Vayner and P. D. Naselsky, Sov. Phys. JETP Lett. {\bf 23} 123 (1976) 
[Pis'ma Zh. Eksp. Teor. Fiz. {\bf 23} 141, (1976)].

\bibitem{HIS6} L.~P.~Grishchuk,  Annals N.\ Y.\ Acad.\ Sci.\  {\bf 302}, 439 (1977).

\bibitem{HIS7} L. H. Ford and L. Parker, Phys. Rev. {\bf D16}, 1601 (1977).

\bibitem{HIS8} L. H. Ford and L. Parker, Phys.\ Rev.\ D {\bf 16}, 245 (1977).

\bibitem{HIS9} B. L. Hu and L. Parker, Phys. Lett. A {\bf 63}, 217 (1977).

\bibitem{HIS10} A. A. Starobinsky, JETP Lett. {\bf 30}, 682 (1979) [Pis'ma Zh. Eksp. Teor. Fiz. {\bf 30}, 719 (1979)].

\bibitem{HIS10a} V.~N.~Lukash, Sov.\ Phys.\ JETP {\bf 52}, 807 (1980) [Zh.\ Eksp.\ Teor.\ Fiz.\  {\bf 79} (1980) 1601].

\bibitem{HIS10b}  A.~A.~Starobinsky,  Phys.\ Lett.\ B {\bf 91}, 99 (1980)  [Adv.\ Ser.\ Astrophys.\ Cosmol.\  {\bf 3}, 130 (1987)].

\bibitem{HIS10c}  A.~H.~Guth,  Phys.\ Rev.\ D {\bf 23}, 347 (1981) [Adv.\ Ser.\ Astrophys.\ Cosmol.\  {\bf 3}, 139 (1987)].
  
\bibitem{HIS10d} A.~D.~Linde,  Phys.\ Lett.\  {\bf 108B}, 389 (1982) [Adv.\ Ser.\ Astrophys.\ Cosmol.\  {\bf 3}, 149 (1987)].  
  
\bibitem{HIS10e} A.~Albrecht and P.~J.~Steinhardt,  Phys.\ Rev.\ Lett.\  {\bf 48}, 1220 (1982)  
[Adv.\ Ser.\ Astrophys.\ Cosmol.\  {\bf 3}, 158 (1987)]. 

\bibitem{HIS11} V. A. Rubakov, M. V. Sazhin, and A. V. Veryaskin, Phys. Lett. B {\bf 115}, 189 (1982).

\bibitem{HIS12} R.~Fabbri and M.~d.~Pollock, Phys.\ Lett.\  {\bf 125B}, 445 (1983).
  
 \bibitem{HIS12a}  A. G. Polnarev, Sov. Astron. {\bf 29}, 607 (1985) [Astron. Zh. {\bf 62}, 1041 (1985)].

\bibitem{HIS12aa} A.~A.~Starobinsky, Sov. Astron. Lett. {\bf 11}, 133 (1985) [Astron. zh. {\bf 11}, 323 (1985)].

\bibitem{HIS13} L.~F.~Abbott and M.~B.~Wise, Nucl.\ Phys.\ B {\bf 244}, 541 (1984).
  
\bibitem{HIS14}  S.~W.~Hawking,  Phys.\ Lett.\  {\bf 150B}, 339 (1985).

\bibitem{HIS15} L.~P.~Grishchuk and Y.~V.~Sidorov,  Phys.\ Rev.\ D {\bf 42}, 3413 (1990).

\bibitem{HIS15b} L.~H.~Ford,  Phys.\ Rev.\ D {\bf 51}, 1692 (1995).

\bibitem{HIS16} V.~Sahni, Phys.\ Rev.\ D {\bf 42}, 453 (1990).

\bibitem{HIS16aa} L.~P.~Grishchuk and M.~Solokhin,  Phys.\ Rev.\ D {\bf 43}, 2566 (1991).

 \bibitem{HIS16a} M.~Gasperini and M.~Giovannini, Phys.\ Lett.\ B {\bf 282}, 36 (1992).

\bibitem{HIS18} M.~Gasperini and M.~Giovannini,  Phys.\ Rev.\ D {\bf 47}, 1519 (1993).

\bibitem{HIS19} M.~R.~de Garcia Maia, Phys.\ Rev.\ D {\bf 48}, 647 (1993).

\bibitem{HIS20} J.~D.~Barrow, J.~P.~Mimoso and M.~R.~de Garcia Maia,  Phys.\ Rev.\ D {\bf 48}, 3630 (1993)
  Erratum: [Phys.\ Rev.\ D {\bf 51}, 5967 (1995)].

\bibitem{HIS20a} J.~c.~Hwang,  Class.\ Quant.\ Grav.\  {\bf 15}, 1401 (1998).

 \bibitem{HIS20b} R.~Brustein, M.~Gasperini, M.~Giovannini and G.~Veneziano,  Phys.\ Lett.\ B {\bf 361}, 45 (1995).

\bibitem{HIS20c} J.~Khoury, B.~A.~Ovrut, P.~J.~Steinhardt and N.~Turok,  Phys.\ Rev.\ D {\bf 64}, 123522 (2001).
  
\bibitem{HIS20d} L.~A.~Boyle, P.~J.~Steinhardt and N.~Turok,  Phys.\ Rev.\ D {\bf 69}, 127302 (2004).  

\bibitem{HIS23} M.~Giovannini,  Phys.\ Rev.\ D {\bf 58}, 083504 (1998).
  
\bibitem{HIS24a}   P.~J.~E.~Peebles and A.~Vilenkin,  Phys.\ Rev.\ D {\bf 59}, 063505 (1999).

\bibitem{HIS24b}  B. Spokoiny, Phys. Lett. B {\bf 315}, 40 (1993).

\bibitem{HIS24} M.~Giovannini, Phys.\ Rev.\ D {\bf 60}, 123511 (1999).
  
\bibitem{bard1} J. Bardeen, Phys. Rev. {\bf D22}, 1882 (1980).

\bibitem{tolman} R. Tolman, {\it Relativity, Thermodynamics and Cosmology} (Dover press), p. 364.

\bibitem{hobson} M. P. Hobson, G. Efsthathiou and A. N. Lasenby, {\it General Relativity}, (Cambridge University Press, Cambridge, UK, 2006).

\bibitem{open0} M.~R.~de Garcia Maia and J.~A.~S.~Lima,  Phys.\ Rev.\ D {\bf 54}, 6111 (1996).

\bibitem{open1} S.~W.~Hawking and N.~Turok, Phys.\ Lett.\ B {\bf 425}, 25 (1998).

\bibitem{open2} S.~W.~Hawking, T.~Hertog and N.~Turok,  Phys.\ Rev.\ D {\bf 62}, 063502 (2000).
  
\bibitem{closed1} A.~Lasenby and C.~Doran,  Phys.\ Rev.\ D {\bf 71}, 063502 (2005).  

\bibitem{bertschingerma} C.~P.~Ma and E.~Bertschinger,  Astrophys.\ J.\  {\bf 455}, 7 (1995).

\bibitem{syn1} W. Press and E. Vishniac, Astrophys. J. {\bf 239}, 1 (1980).

\bibitem{syn2} W. Press and E. Vishniac, Astrophys. J. {\bf 236}, 323 (1980).

\bibitem{hw1} J~-c.~Hwang, Astrophys. J.  {\bf 375}, 443 (1990); Class.\ Quant.\ Grav.\  {\bf 11}, 2305 (1994).

\bibitem{hw2} J.~-c.~Hwang and H.~Noh, Phys.\ Rev.\ D {\bf 65}, 124010 (2002).

\bibitem{hw2a}  J.~-c.~Hwang and H.~Noh, Phys.\ Rev.\ D {\bf 73}, 044021 (2006).
 
\bibitem{NoM1} M.~Giovannini,  Phys.\ Rev.\ D {\bf 87}, 083004 (2013).

 \bibitem{lythpert} D.~H.~Lyth,  Phys.\ Rev.\ D {\bf 31}, 1792 (1985).

\bibitem{KSa} M. Sasaki, Prog. Teor. Phys. {\bf 76}, 1036 (1986).

\bibitem{lukash} V.~N.~Lukash,  Sov.\ Phys.\ JETP {\bf 52}, 807 (1980) [Zh. Eksp. Teor. Fiz. {\bf 79}, 1601 (1980)].

\bibitem{strokov}  V.~Strokov,  Astron.\ Rep.\  {\bf 51}, 431 (2007).

\bibitem{luk2} V. N. Lukash and I. D. Novikov, {\it Lectures on the very early universe} in {\it Observational and Physical Cosmology}, 
II Canary Islands Winter School of Astrophysics, eds. F. Sanchez, M. Collados and R. Rebolo (Cambridge University Press, Cambridge UK, 1992), p. 3.

\bibitem{chibisov} G.~V.~Chibisov, V.~F.~Mukhanov,  Mon.\ Not.\ Roy.\ Astron.\ Soc.\  {\bf 200}, 535 (1982).  
 
\bibitem{chibisova} V.~F.~Mukhanov,  Sov.\ Phys.\ JETP {\bf 67}, 1297 (1988)  [Zh. Eksp. Teor. Fiz. {\bf 94}, 1 (1988)].

\bibitem{KS}  H.~Kodama, M.~Sasaki,  Prog.\ Theor.\ Phys.\ Suppl.\  {\bf 78}, 1 (1984).
  
\bibitem{br1} R.~H.~Brandenberger, R.~Kahn and W.~H.~Press,  Phys.\ Rev.\ D {\bf 28}, 1809 (1983).  
  
\bibitem{bard2} J. Bardeen, P. Steinhardt, and M. Turner, Phys. Rev. {\bf D28}, 679 (1983).  
  
\bibitem{br1a} R.~H.~Brandenberger and R.~Kahn,  Phys.\ Rev.\ D {\bf 29}, 2172 (1984).

\bibitem{bard2a} J.~A.~Frieman and M.~S.~Turner, Phys.\ Rev.\ D {\bf 30}, 265 (1984).

\bibitem{SIXPOL1a} B.~P.~Abbott {\it et al.} [LIGO/Virgo Collaboration],  Phys.\ Rev.\ Lett.\  {\bf 120}, 031104  (2018).

\bibitem{SIXPOL1}  B.~P.~Abbott {\it et al.} [LIGO/Virgo Collaboration],  Phys.\ Rev.\ Lett.\  {\bf 120}, 201102 (2018).

\bibitem{SIXPOL2} D. M. Eardley, D. L. Lee, A. P. Lightman, R. V. Wagoner, and C. M. Will, Phys. Rev. Lett. {\bf 30}, 884 (1973).

\bibitem{SIXPOL3}  D. M. Eardley, D. L. Lee, and A. P. Lightman, Phys. Rev. D {\bf 8}, 3308 (1973).

\bibitem{SIXPOL4} C. M. Will, Living Rev. Relativity {\bf 17}, 4 (2014).

\bibitem{SIXPOL5}  T. Jacobson and D.  Mattingly,  Phys. Rev. D {\bf 64}, 024028 (2001). 

\bibitem{SIXPOL6}  T. Jacobson and D.  Mattingly, Phys. Rev. D {\bf 70}, 024003 (2004).

\bibitem{SIXPOL7} C. M. Will and K. Nordtvedt K, Astrophys. J.  {\bf 177}, 757 (1972).

\bibitem{SIXPOL7a} A.~Nishizawa and K.~Hayama, Phys.\ Rev.\ D {\bf 88}, 064005 (2013).
   
\bibitem{SIXPOL7b}  Y.~Hagihara, N.~Era, D.~Iikawa and H.~Asada,  Phys.\ Rev.\ D {\bf 98}, 064035 (2018).
 
\bibitem{SIXPOL7c}  L.~Philippoz, A.~Bo\^itier and P.~Jetzer,  Phys.\ Rev.\ D {\bf 98},  044025 (2018).
 
\bibitem{SIXPOL7d}   A.~Nishizawa, A.~Taruya, K.~Hayama, S.~Kawamura and M.~a.~Sakagami,  
Phys.\ Rev.\ D {\bf 79}, 082002 (2009)

\bibitem{SIXPOL7e} D.~Liang, Y.~Gong, A.~J.~Weinstein, C.~Zhang and C.~Zhang,  Phys.\ Rev.\ D {\bf 99},  104027 (2019).

\bibitem{SIXPOL8} P. Szekeres, Annals Phys. {\bf 64}, 599 (1971). 

\bibitem{SIXPOL9} P. C. Peters, Phys. Rev. D {\bf 9}, 2207 (1974).

\bibitem{SIXPOL10} M.~Giovannini,  Class.\ Quant.\ Grav.\  {\bf 33},  125002 (2016). 

\bibitem{SIXPOL11} Y. Cai, Y. T. Wang and Y. S. Piao Phys. Rev. D {\bf 94}, 043002 (2016).

\bibitem{SIXPOL11a}  Y.~Cai, Y.~T.~Wang and Y.~S.~Piao, Phys.\ Rev.\ D {\bf 93}, 063005 (2016).

\bibitem{SIXPOL12}  M.~Giovannini,  Eur.\ Phys.\ J.\ C {\bf 78}, 442 (2018).

\bibitem{EB1} J. Goldberg,  A. Macfarlane, E. Newman, F. Rohrlich, and E. Sudarshan J.\ Math.\ Phys.\  {\bf 8}, 2155 (1967).

\bibitem{EB2} K.~S.~Thorne, Rev.\ Mod.\ Phys.\  {\bf 52}, 299 (1980).

\bibitem{EB3} M.~Zaldarriaga and U.~Seljak,  Phys.\ Rev.\  D {\bf 55}, 1830 (1997).

\bibitem{EB3a} M. Kamionkowski, A. Kosowsky and A. Stebbins, Phys. Rev. D {\bf 55}, 7368 (1997).

\bibitem{HBTA} M.~Giovannini,  PMC Phys.\ A {\bf 4}, 1 (2010).

\bibitem{TENS1} C. W. Misner, K. S. Thorne, and J. A. Wheeler, {\it Gravitation} (Freeman, New York, 1973), p. 467.

\bibitem{TENS2a} S. Weinberg, {\it Gravitation and Cosmology}, (Wiley, New York, 1972), p.166.

\bibitem{TENS2b} L. D. Landau and E. M. Lifshitz, {\it The Classical Theory of Fields}, (Pergamon Press, New York, 1971).

 \bibitem{TENS2c} D.~R.~Brill and J.~B.~Hartle,  Phys.\ Rev.\  {\bf 135}, B271 (1964).

\bibitem{TENS3a}  R.~A.~Isaacson,  Phys.\ Rev.\  {\bf 166}, 1263 (1968).

\bibitem{TENS3b}  R.~A.~Isaacson,  Phys.\ Rev.\  {\bf 166}, 1272 (1968).

\bibitem{TENS3d} M. A. H. MacCallum and A. H. Taub, Commun. Math. Phys. {\bf 30}, 153 (1973).

\bibitem{TENS3cc}  L.~R.~W.~Abramo, R.~H.~Brandenberger and V.~F.~Mukhanov,  Phys.\ Rev.\ D {\bf 56}, 3248 (1997).

\bibitem{TENS3c} L. R. Abramo, Phys Rev. D {\bf 60}, 064004 (1999).

\bibitem{TENS5} M. Giovannini, Phys.\ Rev.\ D {\bf 73} 083505 (2006).

\bibitem{TENS5a}  L.~C.~Stein and N.~Yunes,  Phys.\ Rev.\ D {\bf 83}, 064038 (2011). 

\bibitem{TENS6} D. Su and Y. Zhang, Phys.\ Rev.\ D {\bf 85}, 104012 (2012).

\bibitem{TENS8} M.~Giovannini,  Phys.\ Rev.\ D {\bf 91},  023521 (2015). 
  
\bibitem{TENS9} M.~Isi and L.~C.~Stein,  Phys.\ Rev.\ D {\bf 98}, 104025 (2018).
 
\bibitem{TENS4a} S.~V.~Babak and L.~P.~Grishchuk, Phys.\ Rev.\ D {\bf 61}, 024038 (2000).

\bibitem{TENS4aa}  L. P. Grishchuk, A. N. Petrov, and A. D. Popova, Commun. Math. Phys. {\bf 94}, 379 (1984).

\bibitem{TENS4b} L.~M.~Butcher, A.~Lasenby and M.~Hobson,  Phys.\ Rev.\ D {\bf 78}, 064034 (2008).
  
\bibitem{TENS4c}   L.~M.~Butcher, M.~Hobson and A.~Lasenby,  Phys.\ Rev.\ D {\bf 80}, 084014 (2009).
  
\bibitem{TENS4d} L.~M.~Butcher, M.~Hobson and A.~Lasenby,  Phys.\ Rev.\ D {\bf 82}, 104040 (2010). 

\bibitem{TENS4e} L.~M.~Butcher, M.~Hobson and A.~Lasenby, Phys.\ Rev.\ D {\bf 86}, 084012 (2012).  
 
\bibitem{action1} M.~Giovannini,  Class.\ Quant.\ Grav.\  {\bf 20}, 5455 (2003).

\bibitem{action2}  V.~Bozza, M.~Giovannini and G.~Veneziano,  JCAP {\bf 0305}, 001 (2003).

\bibitem{action3} C.~Armendariz-Picon and E.~A.~Lim,  JCAP {\bf 0312}, 006 (2003).

\bibitem{action4} M.~Porrati,  Phys.\ Lett.\ B {\bf 596}, 306 (2004).

\bibitem{action5} K.~Schalm, G.~Shiu and J.~P.~van der Schaar,  AIP Conf.\ Proc.\  {\bf 743}, 362 (2004).

\bibitem{action6}   U.~H.~Danielsson,  Phys.\ Rev.\ D {\bf 71}, 023516 (2005).

\bibitem{action7} U.~H.~Danielsson,  Class.\ Quant.\ Grav.\  {\bf 22}, S1 (2005).

\bibitem{action8}  U.~H.~Danielsson,  JCAP {\bf 0603}, 014 (2006).

\bibitem{action9} B.~Greene, M.~Parikh and J.~P.~van der Schaar, JHEP {\bf 0604}, 057 (2006).

\bibitem{action10} M.~G.~Jackson and K.~Schalm,  Phys.\ Rev.\ Lett.\  {\bf 108}, 111301 (2012).

\bibitem{action11} R.~H.~Brandenberger and J.~Martin,  Class.\ Quant.\ Grav.\  {\bf 30}, 113001 (2013).

\bibitem{action12} S.~Bahrami and E.~E.~Flanagan,  JCAP {\bf 1601}, 027 (2016).
  
\bibitem{action13} C.~Zeng, E.~D.~Kovetz, X.~Chen, Y.~Gong, J.~B.~Munoz and M.~Kamionkowski,  Phys.\ Rev.\ D {\bf 99},  043517 (2019).

 \bibitem{action14} P.~A.~R.~Ade {\it et al.} [Planck Collaboration],  Astron.\ Astrophys.\  {\bf 594}, A20 (2016).

\bibitem{EFPOL1} S.~Weinberg, Phys.\ Rev.\ D {\bf 77}, 123541 (2008).

\bibitem{EFPOL1a}  E. Elizalde, A. Jacksenaev, S.  Odintsov, and I.  Shapiro, Phys. Lett. B {\bf 328}, 297 (1994).

\bibitem{EFPOL1b}  E. Elizalde, A. Jacksenaev, S.  Odintsov, and I.  Shapiro, Class. Quant. Grav. {\bf 12}, 1385 (1995).

\bibitem{EFPOL3}  S-Y. Pi and R. Jackiw, Phys. Rev. D {\bf 68}, 104012 (2003).

\bibitem{EFPOL4} A. Lue, L. Wang, and M. Kamionkowski, Phys. Rev. Lett. {\bf 83}, 156 (1999).

\bibitem{EFPOL5} K.~Choi, J.-c.~Hwang and K.~W.~Hwang, Phys.\ Rev.\ D {\bf 61}, 084026 (2000).

\bibitem{EFPOL5a} S.~H.~S.~Alexander,  Phys.\ Lett.\ B {\bf 660}, 444 (2008).

\bibitem{EFPOL6}  M.~Giovannini,  Phys.\ Rev.\ D {\bf 99},  083501 (2019).

\bibitem{EFPOL2} B. A. Robson, {\it The Theory of Polarization Phenomena}, (Clarendon Press, Oxford, 1974).

\bibitem{class1} A. A. Starobinsky, JETP Lett. {\bf 30}, 682 (1979) [Pis'ma Zh. Eksp. Teor. Fiz. {\bf 30}, 719 (1979)]. 

\bibitem{class2} A. A. Starobinsky, JETP Lett. {\bf 37}, 66 (1983) Pis'ma Zh. Eksp. Teor. Fiz. {\bf 37}, 55 (1983)].

\bibitem{class3} R.M. Wald, Phys. Rev. D {\bf 28}, 2118 (1983).

\bibitem{class4}  D.S. Salopek, J.M. Stewart, Class. Quantum Gravity {\bf 9}, 1943 (1992)

\bibitem{class5} J. Parry, D.S. Salopek, J.M. Stewart, Phys. Rev. D {\bf 49}, 2872 (1994).

\bibitem{class6} K. Tomita, Phys. Rev. D {\bf 48},  5634 (1993).

\bibitem{QM0}  L. Mandel and E. Wolf, {\it ``Optical coherence and quantum optics''}, (Cambridge University Press, Cambridge UK, 1995).

\bibitem{QM1}  B. L. Mollow and R. J. Glauber, Phys. Rev. {\bf 160}, 1076 (1967).

\bibitem{QM1a}  B. L. Mollow and R. J. Glauber, Phys. Rev. {\bf 160}, 1097 (1967).

\bibitem{QM2} D.~Stoler,  Phys.\ Rev.\  D {\bf 1}, 3217 (1970).

\bibitem{QM2a} D.~Stoler, Phys.\ Rev.\  D {\bf 4}, 1925 (1971).

\bibitem{QM3}  A.L.~Fetter and J.D.~Walecka, {\it Quantum Theory of Many-Particle Systems} (McGraw-Hill, New York, 1971).

\bibitem{QM4} A. I. Solomon J.\ Math.\ Phys.\  {\bf 12}, 390 (1971).

\bibitem{yuen} H.~P.~Yuen, Phys.\ Rev.\  {\bf A13}, 2226 (1976).

\bibitem{QM6}  J. N. Hollenhorst, Phys. Rev. D. {\bf 19}, 1669 (1979). 

\bibitem{caves} C. M. Caves,  Phys. Rev. D {\bf 23},1693 (1981).

\bibitem{cavesb} C. M. Caves,  Phys. Rev. Lett. {\bf 54},  2465 (1985). 

\bibitem{QM7} C. Caves and B. L. Schumaker,  Phys.  Rev.  A {\bf 31}, 3068 (1985). 

\bibitem{QM7a} C. Caves and B. L. Schumaker,  Phys.  Rev.  A {\bf 31}, 3093 (1985).

\bibitem{sqgen1} R.  Slusher, L. W.  Hollberg, B. Yurke, J. C. Mertz, and J. F. Valley,  Phys . Rev. Lett. {\bf 55}, 2409 (1985).

\bibitem{sqgen2} R. M. Shelby, M. D.  Levenson,  S. H. Perlmutter, R. G. Devoe, and D. F. Walls,  Phys . Rev. Lett. {\bf  57}, 691 (1986).

\bibitem{sqgen3}  L.-A. Wu, H. J. Kimble, J. L. Hall,  and H. Wu,  Phys . Rev . Lett. {\bf 57},  2520 (1986).
  
\bibitem{QM9}  L. P. Grishchuk, H. A.  Haus, and K. Bergman Phys. Rev. D {\bf 46} 1440 (1992).

\bibitem{QM10} L. P. Grishchuk, Class. Quantum Grav. {\bf 10},  2449 (1993).

\bibitem{QM10h} L.~P.~Grishchuk, Lect.\ Notes Phys.\  {\bf 562}, 167 (2001).

\bibitem{QM10c} A.~L.~Matacz,  Phys.\ Rev.\ D {\bf 49}, 788 (1994).

\bibitem{QM10d} A.~L.~Matacz, Phys.Rev. D {\bf 55} 1860 (1997).

\bibitem{QM10e} A.~Albrecht, P.~Ferreira, M.~Joyce and T.~Prokopec,  Phys.\ Rev.\ D {\bf 50}, 4807 (1994).
  
\bibitem{QM10g} L.~P.~Grishchuk, Phys.\ Rev.\ D {\bf 53}, 6784 (1996).

\bibitem{QM11}  J. G. Valatin and D. Butler, Nuovo Cimento {\bf 10}, 37 (1958).

\bibitem{QM12} F. W. Cummings and J. R. Johnston, Phys. Rev. {\bf 151}, 105 (1966).

\bibitem{yurke} B. Yurke and M. Potasek, Phys. Rev. A {\bf 36}, 3464 (1987).

\bibitem{QM10a} M.~Gasperini and M.~Giovannini, Phys.\ Lett.\ B {\bf 301}, 334 (1993).

\bibitem{QM10aa}  M.~Gasperini and M.~Giovannini,   Class.\ Quant.\ Grav.\  {\bf 10}, L133 (1993).

\bibitem{QM10bb} C. G. Bollini and L. E. Oxman, Phys. Rev. A {\bf 47}, 2339 (1993).

\bibitem{QM10b} R.~H.~Brandenberger, T.~Prokopec and V.~F.~Mukhanov, Phys.\ Rev.\ D {\bf 48}, 2443 (1993).

\bibitem{QM10bbb} M.~Kruczenski, L.~E.~Oxman and M.~Zaldarriaga,  Class.\ Quant.\ Grav.\  {\bf 11}, 2317 (1994).

\bibitem{QM10ee}  D.~Polarski and A.~A.~Starobinsky,  Class.\ Quant.\ Grav.\  {\bf 13}, 377 (1996).

\bibitem{wigner1} E.P. Wigner, Phys. Rev. {\bf 40}, 749 (1932).

\bibitem{wigner2}  V.I. Tatarskii, Usp. Fiz. Nauk {\bf 139}, 587 (1983) [Sov. Phys. Usp. {\bf 26}, 311 (1983)].

\bibitem{wigner3} M. Hillery, R.F. O'Connell, M.O. Scully, and E.P. Wigner, Phys. Rep. {\bf 106}, 121 (1984).

\bibitem{QM10f} W. H. Zurek,  Rev. Mod. Phys. {\bf 75}, 715 (2003).

\bibitem{QM10gg} M.~Giovannini,  Phys.\ Lett.\ B {\bf 691}, 274 (2010).

\bibitem{SQL1} R. Schnabel, Phys. Rep. {\bf 684}, 1 (2017).

\bibitem{SQL2}  S. S. Y. Chua, B. J. J. Slagmolen, D. A.  Shaddock,  and D. E.  McClelland, Class. Quantum Grav. {\bf 31}, 183001 (2014).

\bibitem{SQL3}  S.~Dwyer {\it et al.},  Opt.\ Express {\bf 21}, 16, 19047 (2013).

\bibitem{abr1} M. Abramowitz and I. A. Stegun, {\it Handbook of Mathematical Functions} (Dover, New York, 1972).

\bibitem{abr2}  I. S. Gradshteyn and I. M. Ryzhik,  {\it Tables of Integrals, Series and Products (fifth edition)},  (Academic Press, New York, 1994).

\bibitem{TENS2aa} S. Weinberg, Phys. Rev. D {\bf 67},123504 (2003).

\bibitem{mgcanonical}  M.~Giovannini,  Phys.\ Rev.\ D {\bf 100},  083531 (2019).

\bibitem{STOC} S. Karlin and H. M. Taylor, {\it A first course in stochastic processes} (Academic Press, New York, 1975 ).

\bibitem{AFP}  B.~Allen, E.~E.~Flanagan and M.~A.~Papa,  Phys.\ Rev.\ D {\bf 61}, 024024 (2000).
 
\bibitem{NC}  N.  Christensen, Rep. Prog. Phys. {\bf 82}, 016903 (2019).

\bibitem{HOR} A.R. Liddle, S.M. Leach, Phys. Rev. D {\bf 68}, 103503 (2008).

\bibitem{SAK4} P. J. E. Peebles, Astrophys. J. {\bf 248}, 885 (1981).

\bibitem{SAK5} Ya. B. Zeldovich and l. D. Novikov, {\it The Structure and Evolution of the Universe}  (University of Chicago Press, Chicago, 1983).

\bibitem{pav1} P. Naselsky and I. Novikov, Astrophys. J. {\bf 413}, 14 (1993).

\bibitem{pav2} H. Jorgensen, E. Kotok, P. Naselsky, and I Novikov, Astron. Astrophys. {\bf 294}, 639 (1995).

\bibitem{SAK6}  L.~P.~Grishchuk,  Phys.\ Rev.\ D {\bf 50}, 7154 (1994).

\bibitem{SAK7}  L.~P.~Grishchuk,  Phys.\ Rev.\ D {\bf 48}, 5581 (1993).

\bibitem{SAK8}  L.~P.~Grishchuk,  Usp.\ Fiz.\ Nauk {\bf 182}, 222 (2012)  [Phys.\ Usp.\  {\bf 55}, 210 (2012)].

\bibitem{extrarel1}  Y.~Watanabe and E.~Komatsu,  Phys.\ Rev.\  D {\bf 73}, 123515 (2006).

\bibitem{extrarel2} W.~Zhao and Y.~Zhang,  Phys.\ Rev.\  D {\bf 74}, 043503 (2006).

\bibitem{extrarel3} K.~Saikawa and S.~Shirai,  JCAP {\bf 1805}, 05, 035 (2018).

\bibitem{STRESSNU1}   S.~Weinberg,  Phys.\ Rev.\  D {\bf 69}, 023503 (2004).

 \bibitem{STRESSNU2} D.~A.~Dicus and W.~W.~Repko,  Phys.\ Rev.\  D {\bf 72}, 088302 (2005).
 
\bibitem{STRESSNU3}  B.~A.~Stefanek and W.~W.~Repko, Phys.\ Rev.\ D {\bf 88},  083536 (2013).

 \bibitem{STRESSNU6} H.~X.~Miao and Y.~Zhang, Phys.\ Rev.\ D {\bf 75}, 104009 (2007).
  
\bibitem{STRESSNU6a} W. Zhao, Y. Zang, and T. Xia, Phys. Lett. B {\bf 677}, 235 (2009).
 
\bibitem{STRESSNU6b}  K.~W.~Ng, Phys.\ Rev.\ D {\bf 86}, 103510 (2012) 
 
\bibitem{STRESSNU7}  R.~Flauger and S.~Weinberg, Phys.\ Rev.\ D {\bf 97}, 123506 (2018).
 
\bibitem{STRESSNU4a} S.~Saga, K.~Ichiki and N.~Sugiyama, Phys.\ Rev.\ D {\bf 91}, 024030 (2015).

\bibitem{STRESSNU4b} B.~Wang and Y.~Zhang,  Phys.\ Rev.\ D {\bf 98}, 123019 (2018)

\bibitem{STRESSNU4c} B.~Wang and Y.~Zhang,  Phys.\ Rev.\ D {\bf 99},  123008 (2019).

\bibitem{STRESSNU4d}  Y.~Zhang, F.~Qin and B.~Wang,  Phys.\ Rev.\ D {\bf 96}, 103523 (2017).

\bibitem{STRESSNU4e} B.~Wang and Y.~Zhang,  Phys.\ Rev.\ D {\bf 96}, 103522 (2017).

\bibitem{STRESSNU8} J.~Khodagholizadeh, A.~H.~Abbassi and A.~A.~Asgari, Phys.\ Rev.\ D {\bf 90},  063520 (2014).

\bibitem{TRANS1}  M.~S.~Turner, M.~J.~White and J.~E.~Lidsey,  Phys.\ Rev.\  D {\bf 48}, 4613 (1993).

\bibitem{TRANS1a} L.~M.~Krauss and M.~J.~White,  Phys.\ Rev.\ Lett.\  {\bf 69}, 869 (1992).

\bibitem{TRANS2} B.~Allen and S.~Koranda,  Phys.\ Rev.\ D {\bf 50}, 3713 (1994).

\bibitem{TRANS3} K.~w.~Ng and A.~D.~Speliotopoulos,  Phys.\ Rev.\ D {\bf 52}, 2112 (1995).
  
\bibitem{TRANS4}  L.~Knox,  Phys.\ Rev.\ D {\bf 52}, 4307 (1995). 
  
\bibitem{TRANS8a}  L.~A.~Boyle and P.~J.~Steinhardt,  Phys.\ Rev.\ D {\bf 77}, 063504 (2008).

\bibitem{TRANS8}  M.~Giovannini,  Phys.\ Lett.\ B {\bf 668}, 44 (2008).
  
\bibitem{TRANS9} M.~Giovannini,  Class.\ Quant.\ Grav.\  {\bf 26}, 045004 (2009).

\bibitem{NORM1} S.~Chongchitnan and G.~Efstathiou, Phys.\ Rev.\  D {\bf 73}, 083511 (2006).

\bibitem{NORM1a}  S.~Chongchitnan and G.~Efstathiou, Prog.\ Theor.\ Phys.\ Suppl.\  {\bf 163}, 204 (2006).

\bibitem{NORM1b}  J.~E.~Lidsey, A.~R.~Liddle, E.~W.~Kolb, E.~J.~Copeland, T.~Barreiro and M.~Abney,
  Rev.\ Mod.\ Phys.\  {\bf 69}, 373 (1997).

\bibitem{NORM3}   P.~D.~Meerburg, R.~Hlozek, B.~Hadzhiyska and J.~Meyers, Phys.\ Rev.\ D {\bf 91},  103505 (2015).

\bibitem{NORM4} G.~Cabass, L.~Pagano, L.~Salvati, M.~Gerbino, E.~Giusarma and A.~Melchiorri,  Phys.\ Rev.\ D {\bf 93},  063508 (2016).

\bibitem{COSMC1} Y.~Zhang, Y.~Yuan, W.~Zhao and Y.~T.~Chen,  Class.\ Quant.\ Grav.\  {\bf 22}, 1383 (2005).

\bibitem{COSMC3} Y.~Zhang, X.~Z.~Er, T.~Y.~Xia, W.~Zhao and H.~X.~Miao,  Class.\ Quant.\ Grav.\  {\bf 23}, 3783 (2006).
  
 \bibitem{COSMC4}  W.~Zhao,  Chin.\ Phys.\  {\bf 16}, 2894 (2007).

\bibitem{COSMC5} W.~Zhao and Y.~Zhang,  Phys.\ Rev.\ D {\bf 74}, 043503 (2006).

\bibitem{ABSOLUTE} M.~Giovannini,  Class.\ Quant.\ Grav.\  {\bf 31}, 225002 (2014).

\bibitem{COSMC6}  M.~L.~Tong and Y.~Zhang, Phys.\ Rev.\ D {\bf 80}, 084022 (2009).

\bibitem{pee1} P. J. E. Peebles, Astrophys. J. {\bf 142}, 1317 (1965).

\bibitem{pee2} P. J. E. Peebles, Astrophys. J. {\bf 147}, 859 (1967).

\bibitem{silk} J.~Silk,  Astrophys.\ J.\  {\bf 151}, 459 (1968).

\bibitem{harrisons} E. R. Harrison, Phys. Rev. D {\bf 1}, 2726 (1970).

\bibitem{zeldovichnovikov} Ya. B. Zeldovich and I. D. Novikov, Sov. Astron. {\bf 13}, 754 (1970).

\bibitem{rees} M. Rees, Astrophys. J. {\bf 153}, L1 (1968).

\bibitem{ad1a} S.~Weinberg, Phys.\ Rev.\ D {\bf 67}, 123504 (2003).

\bibitem{nad1} K.~Enqvist, H.~Kurki-Suonio and J.~Valiviita, Phys.\ Rev.\  D {\bf 62}, 103003 (2000).

\bibitem{nad1a}  J.~Valiviita and V.~Muhonen,  Phys.\ Rev.\ Lett.\  {\bf 91}, 131302 (2003).

\bibitem{nad2} H.~Kurki-Suonio, V.~Muhonen and J.~Valiviita, Phys.\ Rev.\  D {\bf 71}, 063005 (2005).

\bibitem{nad2a} R.~Keskitalo, H.~Kurki-Suonio, V.~Muhonen and J.~Valiviita, JCAP {\bf 0709}, 008 (2007).

\bibitem{nad3} M.~Giovannini,  Class.\ Quant.\ Grav.\  {\bf 23}, 4991 (2006).

\bibitem{nad3a}  M.~Giovannini and K.~E.~Kunze, Phys.\ Rev.\ D {\bf 77}, 123001 (2008).

\bibitem{hus}  W.~Hu and N.~Sugiyama,  Astrophys.\ J.\  {\bf 471}, 542 (1996);  Astrophys.\ J.\  {\bf 444}, 489 (1995).

\bibitem{TC2}  A. G. Doroshkevich, Ya. B. Zeldovich, and R. A. Sunyaev,  Sov. Astron. {\bf 22}, 523 (1978).

\bibitem{TC3} M.~Zaldarriaga and D.~D.~Harari,  Phys.\ Rev.\ D {\bf 52}  3276 (1995).

\bibitem{GWHT1} M. Basko and A. Polnarev, Sov. Astron. {\bf 24}, 3 (1979).

\bibitem{GWHT2} E. Linder, Astrophys. J. {\bf 326}, 517 (1988).

\bibitem{GWHT3} R. Crittenden, J. Bond, R. L. Davis, G. Efstathiou, and P. J. Steinhardt, Phys. Rev. Lett. {\bf 71}, 324  (1993).
 
 \bibitem{GWHT3a} R.~Crittenden, R.~L.~Davis and P.~J.~Steinhardt,  Astrophys.\ J.\  {\bf 417}, L13 (1993). 
 
\bibitem{GWHT4}  D.~D.~Harari and M.~Zaldarriaga, Phys.\ Lett.\ B {\bf 319}, 96 (1993).
 
 \bibitem{GWHT5} R. A. Frewin, A. G. Polnarev and P. Coles, Mon. Not. R. Astron. Soc. {\bf 266}, L21 (1994).
 
 \bibitem{GWHT5a} K.~L.~Ng and K.~W.~Ng,  Phys.\ Rev.\ D {\bf 51}, 364 (1995).

\bibitem{GWHT6}  M.~Zaldarriaga and D.~D.~Harari,  Phys.\ Rev.\ D {\bf 52}, 3276 (1995).
  
\bibitem{GWHT7} U.~Seljak and M.~Zaldarriaga,  Astrophys.\ J.\  {\bf 469}, 437 (1996). 
  
\bibitem{GWHT7a} E. Bertschinger, arXiv:astro-ph/9506070.

\bibitem{GWHT7c} M. Zaldarriaga, D. N. Spergel, and U. Seljak, Astrophys. J. {\bf 488}, 1 (1997).  
  
\bibitem{GWHT8} J.~R.~Pritchard and M.~Kamionkowski,  Annals Phys.\  {\bf 318}, 2 (2005).
 
\bibitem{GWHT9} D.~Baskaran, L.~P.~Grishchuk and A.~G.~Polnarev,  Phys.\ Rev.\ D {\bf 74}, 083008 (2006). 
  
  \bibitem{GWHT10} Y.~Zhang, W.~Zhao, T.~Xia and Y.~Yuan,  Phys.\ Rev.\ D {\bf 74}, 083006 (2006).
 
\bibitem{GWHT11} R.~Flauger and S.~Weinberg,  Phys.\ Rev.\ D {\bf 75}, 123505 (2007).
 
\bibitem{chone42} D.~Hanson {\it et al.}  [SPTpol Collaboration],  Phys.\ Rev.\ Lett.\  {\bf 111}, 141301 (2013).

\bibitem{chone43} P.~A.~R.~Ade {\it et al.}  [BICEP2 Collaboration],  Phys.\ Rev.\ Lett.\  {\bf 112}, 241101 (2014).

\bibitem{far1} A.~Kosowsky and A.~Loeb, Astrophys.\ J.\  {\bf 469}, 1 (1996).
  
\bibitem{far2}  D.~D.~Harari, J.~D.~Hayward and M.~Zaldarriaga,  Phys.\ Rev.\ D {\bf 55}, 1841 (1997)
  
\bibitem{far3} M.~Giovannini,  Phys.\ Rev.\ D {\bf 56}, 3198 (1997).

\bibitem{far4} C.~Scoccola, D.~Harari and S.~Mollerach,  Phys.\ Rev.\ D {\bf 70}, 063003 (2004).  

\bibitem{far5} M.~Giovannini,  Phys.\ Rev.\ D {\bf 71}, 021301 (2005).

\bibitem{far6} M.~Giovannini and K.~E.~Kunze,  Phys.\ Rev.\ D {\bf 79}, 087301 (2009).

\bibitem{chandra} S.  Chandrasekhar, {\it Radiative Transfer}, (Dover, New York, US, 1966).

\bibitem{pera} A. Peraiah, {\it An Introduction to Radiative Transfer}, (Cambridge University Press, Cambridge UK, 2001).

\bibitem{zeld1} R.~A.~Sunyaev and Y.~B.~Zeldovich,  Astrophys.\ Space Sci.\  {\bf 7}, 3 (1970).

\bibitem{wyse}  B. Jones and R. Wyse, Astron. Astrophys. {\bf 149}, 144 (1985).

\bibitem{infrascal1} D.~K.~Hazra, A.~Shafieloo and T.~Souradeep, JCAP {\bf 1411},  011 (2014).

\bibitem{infrascal2} D.~K.~Hazra, A.~Shafieloo and T.~Souradeep,  Phys.\ Rev.\ D {\bf 87}, 123528 (2013).

\bibitem{kibble}  T.~W.~B.~Kibble, {\it ``Some applications of coherent states''}, Cargese Lect.\ Phys.\  {\bf 2}, 209 (1968).

\bibitem{infra1}  M.~Gasperini, M.~Giovannini and G.~Veneziano,  Phys.\ Rev.\ D {\bf 48}, R439 (1993).

\bibitem{infra2} K.~Bhattacharya, S.~Mohanty and R.~Rangarajan,  Phys.\ Rev.\ Lett.\  {\bf 96}, 121302 (2006).
  
\bibitem{infra3} K.~Bhattacharya, S.~Mohanty and A.~Nautiyal,  Phys.\ Rev.\ Lett.\  {\bf 97}, 251301 (2006).
  
\bibitem{infra3a} E.~Masso, S.~Mohanty, A.~Nautiyal and G.~Zsembinszki,  Phys.\ Rev.\ D {\bf 78}, 043534 (2008).
  
\bibitem{infra3b} S.~Das and S.~Mohanty,  Phys.\ Rev.\ D {\bf 80}, 123537 (2009).
  
\bibitem{infra4} W.~Zhao, D.~Baskaran and P.~Coles,  Phys.\ Lett.\ B {\bf 680}, 411 (2009).
  
\bibitem{infra5} I.~Agullo and L.~Parker,  Phys.\ Rev.\ D {\bf 83}, 063526 (2011).
  
\bibitem{infra6}  S.~Kundu,  JCAP {\bf 1202}, 005 (2012).
  
\bibitem{infra8} M.~Giovannini,  Class.\ Quant.\ Grav.\  {\bf 29}, 155003 (2012).

\bibitem{infra9} M.~Giovannini,  Class.\ Quant.\ Grav.\  {\bf 30}, 015009 (2013).
  
\bibitem{infra10} M.~Giovannini,  Phys.\ Rev.\ D {\bf 88},  021301 (2013).

\bibitem{infra11} S.~Kundu, JCAP {\bf 1404}, 016 (2014).

\bibitem{infra12}  K.~Wang, L.~Santos, J.~Q.~Xia and W.~Zhao,  JCAP {\bf 1701}, 053 (2017).

\bibitem{infra13} A.~Ashoorioon,  Phys.\ Lett.\ B {\bf 790}, 568 (2019).

\bibitem{infra14}  S.~Brahma and M.~Wali Hossain,  JHEP {\bf 1903}, 006 (2019).

\bibitem{LCDM1a} E.~M.~Leitch,  {\it et al.}, Nature {\bf 420}, 763 (2002).

\bibitem{LCDM1b} E. M. Leitch {\it et al.}, Astrophys. J. {\bf 624}, 10 (2005).
  
\bibitem{cbi1} A.~C.~S.~Readhead {\it et al.}, Science {\bf 306}, 836 (2004).

\bibitem{cbi2} J. L. Sievers {\it et al.} , Astrophys. J. {\bf 660},  976 (2007).

\bibitem{boom} T. E. Montroy {\it et al.}, Astrophys. J. {\bf 647}, 813 (2006).

\bibitem{maxipol} B.R. Johnson {\it et al.}, Astrophys. J. {\bf 665}, 42  (2007). 

\bibitem{maxipola}  J.H.P. Wu {\it et al.} , Astrophys. J. {\bf 665}, 55 (2007).

\bibitem{quiet} C.~Bischoff {\it et al.} [QUIET Collaboration],  Astrophys.\ J.\  {\bf 741}, 111 (2011).

\bibitem{quieta}  D.~Araujo {\it et al.}  [QUIET Collaboration], Astrophys.\ J.\  {\bf 760}, 145 (2012).

\bibitem{quad1} P.~A.~R.~Ade {\it et al.}  [QUaD collaboration], Astrophys. J. {\bf 674}, 22 (2008).

\bibitem{quad1a}   C.~Pryke {\it et al.}  [QUaD collaboration],  Astrophys.\ J.\  {\bf 692}, 1247 (2009).
 
\bibitem{quad2}  E.~Y.~S.~Wu {\it et al.}  [QUaD Collaboration],  Phys.\ Rev.\ Lett.\  {\bf 102}, 161302 (2009).

\bibitem{quad2a} M.~L.~Brown {\it et al.}  [QUaD Collaboration],  Astrophys.\ J.\  {\bf 705}, 978 (2009).

\bibitem{BICEP2}  P.~A.~R.~Ade {\it et al.} [BICEP2 Collaboration],  Astrophys.\ J.\  {\bf 792}, 62 (2014).

\bibitem{bicep2new}  P.~A.~R.~Ade {\it et al.} [BICEP2 and Keck Arrary Collaborations],  Phys.\ Rev.\ D {\bf 96}, 102003 (2017).

\bibitem{bicep1}  J.~P.~Kaufman {\it et al.}  [BICEP1 Collaboration],  Phys.\ Rev.\ D {\bf 89}, 062006 (2014).
  
 \bibitem{bicep1a}  D.~Barkats {\it et al.} [BICEP1 Collaboration], Astrophys.\ J.\  {\bf 783}, 67 (2014).  

\bibitem{WMAP9} C.~L.~Bennett {\it et al.} [WMAP Collaboration],  Astrophys.\ J.\ Suppl.\ {\bf 192}, 17 (2011).

\bibitem{WMAP9a}   B.~Gold {\it et al.} [WMAP Collaboration],   Astrophys.\ J.\ Suppl.\ \ {\bf 192}, 15 (2011).

\bibitem{capmap} D.~Barkats {\it et al.}, Astrophys.\ J.\  {\bf 619}, L127 (2005).
  
\bibitem{capmapa}   C. Bischoff {\it et al.} [The CAPMAP Collaboration], Astrophys. J. {\bf 684}, 771 (2008).

\bibitem{polar1} P.~A.~R.~Ade {\it et al.} [POLARBEAR Collaboration],  Phys.\ Rev.\ Lett.\  {\bf 113}, 021301 (2014).
 
\bibitem{polar2} P.~A.~R.~Ade {\it et al.} [POLARBEAR Collaboration],  Phys.\ Rev.\ Lett.\  {\bf 112}, 131302 (2014).   
  
\bibitem{lbound} D.H. Lyth, Phys. Rev. Lett. {\bf 78}, 1861 (1997).

\bibitem{efs} G. Efstathiou, K.J. Mack, J. Cosmol. Astropart. Phys. {\bf 0505}, 008 (2005).

\bibitem{east} R. Easther, W.H. Kinney, B.A. Powell, J. Cosmol. Astropart. Phys. {\bf 0608}, 004 (2006).

\bibitem{krause} A. Krause, J. Cosmol. Astropart. Phys. {\bf 0807}, 001 (2008).

\bibitem{baum} D. Baumann, D. Green, J. Cosmol. Astropart. Phys. {\bf 1205}, 017 (2012).

\bibitem{ads} P. Adshead, E. Martinec, M. Wyman, Phys. Rev. D {\bf 88}, 021302 (2013).

\bibitem{antusch} S. Antusch, D. Nolde, J. Cosmol. Astropart. Phys. {\bf 1405}, 035 (2014).

\bibitem{hoss} M. W. Hossain, R. Myrzakulov, M. Sami, and E. N. Saridakis, Phys. Lett. B {\bf 737}, 191 (2014).

\bibitem{pl1} A. A. Starobinsky, Phys.Lett. B  {\bf 91}, 99 (1980).

\bibitem{pl3} P. Steinhardt and M. S. Turner, Phys. Rev. D {\bf 29}, 2162 (1984).

\bibitem{pl4} D. Salopek, J. Bond, and J. M. Bardeen, Phys. Rev. D {\bf 40}, 1753 (1989).

\bibitem{pl5} R. Fakir and W. Unruh, Phys.Rev. D {\bf 41}, 1783 (1990).

\bibitem{pl6} K. A. Olive, Phys.Rept. {\bf 190}, 307 (1990).

\bibitem{pl7}  B.~Whitt,  Phys.\ Lett.\  {\bf 145B}, 176 (1984).

\bibitem{pl8} T.~P.~Sotiriou and V.~Faraoni,  Rev.\ Mod.\ Phys.\  {\bf 82}, 451 (2010).

\bibitem{pl9} A.~Ijjas, P.~J.~Steinhardt and A.~Loeb,  Phys.\ Lett.\ B {\bf 723}, 261 (2013).

\bibitem{pl10} A.~Ijjas, P.~J.~Steinhardt and A.~Loeb,  Phys.\ Lett.\ B {\bf 736}, 142 (2014).

\bibitem{pl11} A. D. Linde, Phys. Lett. B {\bf 129}, 177 (1983).

\bibitem{pl12} M.~Giovannini,  Phys.\ Lett.\ B {\bf 746}, 159 (2015).

\bibitem{anto1} I.~Antoniadis, A.~Karam, A.~Lykkas and K.~Tamvakis,  JCAP {\bf 1811}, 028 (2018).

\bibitem{enqv} V.~M.~Enckell, K.~Enqvist, S.~Rasanen and L.~P.~Wahlman,  JCAP {\bf 1902}, 022 (2019).

\bibitem{anto2}  I.~Antoniadis, A.~Karam, A.~Lykkas, T.~Pappas and K.~Tamvakis,   JCAP {\bf 1903},  005 (2019).

\bibitem{tenka} T.~Tenkanen,  Phys.\ Rev.\ D {\bf 99},  063528 (2019).

\bibitem{maeda} K.~Shimada, K.~Aoki and K.~i.~Maeda,  Phys.\ Rev.\ D {\bf 99}, 104020 (2019).

\bibitem{gialamas} I.~D.~Gialamas and A.~B.~Lahanas,  arXiv:1911.11513 [gr-qc].

\bibitem{orig1} C. Tolman, Phys. Rev. {\bf 38}, 1758 (1931).

\bibitem{orig1a} G. Lema\^itre, Ann. Soc. Sci. Bruxelles A {\bf 53}, 51 (1933).

\bibitem{orig2}  H. Nariai and K. Tomita, Progr. Theoret. Phys. {\bf 46}, 433 (1971). 

\bibitem{orig3} L.~Parker and S.~A.~Fulling,  Phys.\ Rev.\ D {\bf 7}, 2357 (1973).

\bibitem{orig3a}  L. Parker, Nature {\bf 261}, 20 (1976).

\bibitem{orig4}  A. A. Starobinsky, Sov.\ Astron.\ Lett. {\bf 4}, 82 (1978) [Pis'ma Astron. Zh. {\bf 4}, 155 (1978)].

\bibitem{DISC1} A. Borde and A. Vilenkin, Phys. Rev. Lett. {\bf 72}, 3305 (1994).

\bibitem{DISC2} A. Borde and A. Vilenkin, Int. J. Mod. Phys. D {\bf 5}, 813 (1996).

\bibitem{DISC3} A. Borde and A. Vilenkin, Phys. Rev. D {\bf 56}, 717 (1997).

\bibitem{DISC4} A. Borde, A. H. Guth and A. Vilenkin, Phys. Rev. Lett. {\bf 90}, 151301 (2003).

\bibitem{DISC5}  G.~F.~R.~Ellis and M.~S.~Madsen,  Class.\ Quant.\ Grav.\  {\bf 8}, 667 (1991).

\bibitem{bouncing1} M. Novello and S. E. P. Bergliaffa, Phys. Rept. {\bf 463}, 127 (2008).

\bibitem{bouncing2c} M.~Gasperini and G.~Veneziano,  Phys.\ Rept.\  {\bf 373}, 1 (2003).

\bibitem{bouncing3}  J. L. Lehners,  Phys. Rept. {\bf 465}, 223 (2008). 

\bibitem{bouncing3a} A. Ashtekar and P. Singh, Class. Quant. Grav. {\bf 28}, 213001 (2011).

\bibitem{bouncing3b} S. Nojiri, S. D. Odintsov, and V. K. Oikonomou, Phys. Rep. {\bf 692}, 1 (2017).

\bibitem{bouncing0} S.~Mironov, V.~Rubakov and V.~Volkova,  arXiv:1906.12139 [hep-th].

\bibitem{san1} D.~Pirtskhalava, L.~Santoni, E.~Trincherini and P.~Uttayarat,  JHEP {\bf 1412}, 151 (2014).

\bibitem{swamp1} C. Vafa, {\it ``The String landscape and the Swampland,''} hep-th/0509212.

\bibitem{swamp2} H. Ooguri and C. Vafa, {\it ``Non-supersymmetric AdS and the Swampland,''} Adv. Theor. Math. Phys. {\bf 21}, 1787 (2017).

\bibitem{swamp3} G. Obied, H. Ooguri, L. Spodyneiko and C. Vafa, {\it ``De Sitter Space and the Swampland,''} [hep-th/1806.08362].

\bibitem{swamp4}  P.~Agrawal, G.~Obied and C.~Vafa,  arXiv:1906.08261 [astro-ph.CO].

\bibitem{swamp5} U. H. Danielsson and T. Van Riet, Int.\ J.\ Mod.\ Phys.\ D {\bf 27},  1830007 (2018).

\bibitem{swamp6} P.~Agrawal, G.~Obied, P.~J.~Steinhardt and C.~Vafa,  Phys.\ Lett.\ B {\bf 784}, 271 (2018).

\bibitem{encon1} S. W Hawking and G. F. B. Ellis, {\it The Large Scale Structure of Space-time} (Cambridge University Press, Cambridge UK, 1973).

\bibitem{encon2}  F.~J.~Tipler,  Phys.\ Rev.\ D {\bf 17}, 2521 (1978).

\bibitem{encon3} J.~D.~Barrow,  Phys.\ Lett.\ B {\bf 180}, 335 (1986).

\bibitem{encon4}  J.~D.~Barrow,  Nucl.\ Phys.\ B {\bf 310}, 743 (1988).

\bibitem{encon5} L. H. Ford, Proc. R. Soc. A {\bf 364}, 227 (1978).

\bibitem{encon6} C. I. Kuo and L. H. Ford, Phys. Rev. D {\bf 47}, 4510 (1993).

\bibitem{encon7} L. H. Ford and T. A. Roman, Phys. Rev. D {\bf 51}, 4277 (1995).

\bibitem{encon8}  L. H. Ford and T. A. Roman, Phys. Rev. D {\bf 60}, 104018 (1999).

\bibitem{encon9} M. Giovannini, Phys. Rev. D {\bf 96}, 101302 (2017).

\bibitem{BS1} M. Giovannini, Phys.\ Rev.\ D {\bf 59}, 121301 (1999).

\bibitem{BS2} M.~Cataldo and P.~Mella,  Phys.\ Lett.\ B {\bf 642}, 5 (2006).

\bibitem{BS4}  M.~Gasperini, M.~Giovannini and G.~Veneziano,  Phys.\ Lett.\ B {\bf 569}, 113 (2003).
  
\bibitem{BS5}  M.~Gasperini, M.~Giovannini and G.~Veneziano, Nucl.\ Phys.\ B {\bf 694}, 206 (2004).

\bibitem{BS3} M.~Giovannini, Phys.\ Rev.\ D {\bf 95}, 083506 (2017).

\bibitem{BS3a}  M.~Giovannini,  Class.\ Quant.\ Grav.\  {\bf 21}, 4209 (2004).

\bibitem{BS6} C. Hull and B. Zwiebach, J. High Energy Phys. {\bf 09} 099 (2009).

\bibitem{BS7} O. Hohm, C. Hull, and B. Zwiebach, J. High Energy Phys. {\bf 07}, 016 (2010).

\bibitem{BS8} H. Wu and H. Yang, J. Cosmol. Astropart. Phys. {\bf 07}, 024 (2014).

\bibitem{BS9} C. T. Ma and C. M. Shen, Fortschr. Phys. {\bf 62}, 921 (2014).

\bibitem{BS10} K. Lee and J.H. Park, Nucl. Phys. B880, 134 (2014).

\bibitem{bouncing0a} V. A. Rubakov, Phys. Usp. 57, 128 (2014) [Usp. Fiz. Nauk 184, 137 (2014)].

\bibitem{bouncing0b} M. Libanov, S. Mironov, and V. Rubakov, J. Cosmol. Astropart. Phys. {\bf 08} 037 (2016).

\bibitem{bouncing0c} T. Kobayashi, Phys. Rev. D {\bf 94}, 043511 (2016).

\bibitem{bouncing0d}  Y. Cai, Y. Wan, H. G. Li, T. Qiu, and Y. S. Piao, J. High Energy Phys. {\bf 01}, 090 (2017).

\bibitem{bouncing0e}  Y. Cai, H. G. Li, T. Qiu, and Y. S. Piao, Eur. Phys. J. C {\bf 77}, 369 (2017).

\bibitem{refr19} M.~Giovannini,  Phys.\ Lett.\ B {\bf 789}, 502 (2019).

\bibitem{Dn1} C.J. Copi, D.N. Schramm, and M.S. Turner, Phys. Rev. D {\bf 55}, 3389 (1997).

\bibitem{Dn2}  S. Burles, K.M. Nollett, J.W. Truran, and M.S. Turner, Phys. Rev. Lett. {\bf 82}, 4176 (1999).

\bibitem{Dn3} R. Cyburt, B. Fields, and K. Olive, Astropart. Phys. {\bf 17}, 87 (2002).

\bibitem{Dn4} R. Cyburt, B.~D.~Fields, K.~A.~Olive and T.~H.~Yeh,  Rev.\ Mod.\ Phys.\  {\bf 88}, 015004 (2016).

\bibitem{bbn1}  V.F. Schwartzman, Pis'ma Zh. Eksp. Teor. Fiz. {\bf 9}, 315 (1969) [JETP Lett. {\bf 9}, 184 (1969)].
  
\bibitem{bbn2} M.~Giovannini, H.~Kurki-Suonio and E.~Sihvola, Phys.\ Rev.\  D {\bf 66}, 043504 (2002).

\bibitem{bbn3} R. Cyburt, B.~D.~Fields, K.~A.~Olive, and E.~Skillman, Astropart.\ Phys.\ {\bf 23}, 313 (2005).

\bibitem{cmbconst1} T.~L.~Smith, E.~Pierpaoli and M.~Kamionkowski, Phys.\ Rev.\ Lett.\ {\bf 97}, 021301 (2006). 

\bibitem{cmbconst2} I.~Sendra and T.~L.~Smith,  Phys.\ Rev.\ D {\bf 85}, 123002 (2012).

\bibitem{PUL0a} M. V. Sazhin,  Sov. Astron.  {\bf 22},  36 (1978) [Astron. Zh. {\bf 55}, 65 (1979)].

\bibitem{PUL0b} S. Detweiler, Astrophys. J. {\bf 234},  1100 (1979).

\bibitem{PUL0c} R. W. Hellings and G. S. Downs, Astrophys. J. Lett. {\bf  265} L39 (1983).

\bibitem{PUL3}  P.~B.~Demorest {\it et al.}, Astrophys.\ J.\  {\bf 762}, 94 (2013).

\bibitem{PUL4} W. Zhao, Phys. Rev. D {\bf 83}, 104021 (2011).

\bibitem{PUL5} W. Zhao, Y. Zhang, X.-P. You, Z.-H. Zhu, Phys. Rev. D {\bf 87}, 124012 (2013). 

\bibitem{PUL6}  R.~M.~Shannon {\it et al.},  Science {\bf 349}, 6255, 1522 (2015).

\bibitem{PUL7}   L. Lentati, {\it et al.}, [EPTA collaboration] , Mon. Not. R. Astron. Soc. {\bf 453}, 2756  (2015).

\bibitem{PUL8}   Z. Arzoumanian, {\it et al.}  [NANOGrav collaboration] , Astrophys. J. {\bf 821}, 13 (2016).

\bibitem{PUL9} M. L Jones {\it et. al}  [NANOGrav collaboration], Astrophys. J. {\bf 841}, 125 (2017).

\bibitem{PUL10}   Z. Arzoumanian, {\it et al.}  [NANOGrav collaboration], Astrophys. J. {\bf 859}, 47 (2018).

\bibitem{PUL11} P.D. Lasky, {\it et al.}  [PPTA collaboration], Phys. Rev. X {\bf 6},  011035 (2016). 

\bibitem{PUL12} N. K. Porayko {\it et al.} [PPTA collaboration], Phys Rev D {\bf 98}, 102002 (2018).

\bibitem{PUL13a} R. N. Manchester ,  Class. Quantum Grav. {\bf 30}, 224010 (2013). 

\bibitem{PUL13}  T. J. W. Lazio, Class. Quantum Grav. {\bf 30},  224011 (2013).

\bibitem{CS1}  T. W. B. Kibble, J. Phys. A {\bf 9}, 1387 (1976).

\bibitem{CS2} A. Vilenkin and E. P. S. Shellard, {\it ``Cosmic Strings and Other Topological Defects''} 
(Cambridge University Press, Cambridge UK, 2000).

\bibitem{CS3} E. J. Copeland and T. W. B. Kibble, Proc. Roy. Soc. Lond. A {\bf 466}, 623 (2010).

\bibitem{CS4} L. Pogosian, S. H. H. Tye, I. Wasserman, and M. Wyman, Phys. Rev. D {\bf 68}, 023506 (2003), 
[Erratum: Phys. Rev.D {\bf 73}, 089904 (2006)].

\bibitem{CS5} N. T. Jones, H. Stoica, and S. H. H. Tye, Phys. Lett. B {\bf 563}, 6 (2003).

\bibitem{CS6} A.~Vilenkin,  Phys.\ Lett.\  B {\bf 107}, 47 (1981).

\bibitem{CS7}  C.~J.~Hogan and M.~J.~Rees, Nature {\bf 311}, 109 (1984).

\bibitem{CS8} T.~Damour and A.~Vilenkin, Phys.\ Rev.\ D {\bf 71}, 063510 (2005).
  
\bibitem{CS9}   M. Hindmarsh, J. Lizarraga, J. Urrestilla, D. Daverio, and M. Kunz, Phys. Rev. D {\bf 96}, 023525 (2017).

\bibitem{CS10}  G. Vincent, N. D. Antunes, and M. Hindmarsh, Phys. Rev. Lett. {\bf 80}, 2277 (1998). 

\bibitem{CS11}  K. D. Olum and J. J. Blanco-Pillado, Phys. Rev. D {\bf 60}, 023503 (1999).

\bibitem{CS12}  K. D. Olum and J. J. Blanco-Pillado, Phys. Rev. D {\bf 60}, 023503 (1999).

\bibitem{CS13} J. N. Moore, E. P. S. Shellard, and C. J. A. P. Martins, Phys. Rev. D65, 023503 (2002).

\bibitem{CS14} J. J. Blanco-Pillado, K. D. Olum, and B. Shlaer, Phys. Rev. D {\bf 83}, 083514 (2011).

\bibitem{CS15} T.~Helfer, J.~C.~Aurrekoetxea and E.~A.~Lim,  Phys.\ Rev.\ D {\bf 99}, 104028 (2019).

\bibitem{CS16} R. Caldwell and B. Allen, Phys. Rev. D {\bf 45}, 3447 (1992).

\bibitem{CS17} R. Caldwell, R. Battye and E. Shellard,  Phys. Rev. D {\bf 54} 7146 (1996).

\bibitem{CS18} C.J. Hogan,  Phys. Rev. D {\bf 74}, 043526 (2006).

\bibitem{CS19} M.R. DePies and C.J. Hogan,  Phys. Rev. D {\bf 75}  125006 (2007).

\bibitem{CS20} X. Siemens, V. Mandic and J. Creighton,  Phys. Rev. Lett. {\bf 98}, 111101(2007).

\bibitem{CS21}  S. Olmez, V. Mandic and X. Siemens,  Phys. Rev. D {\bf 81}, 104028 (2010).

\bibitem{CS22} S.~Kuroyanagi, K.~Miyamoto, T.~Sekiguchi, K.~Takahashi and J.~Silk,  Phys.\ Rev.\ D {\bf 86}, 023503 (2012).

\bibitem{CS23} J. J. Blanco-Pillado,  K. D. Olum and X. Siemens, Phys. Lett. B {\bf 778},  392 (2018).

\bibitem{PTT1} A. Kosowsky, M. S. Turner and R. Watkins, Phys. Rev. D {\bf 45}, 4514 (1992).

\bibitem{PTT2} A. Kosowsky and M. S. Turner, Phys. Rev. D {\bf 47}, 4372 (1993).

\bibitem{PTT3} M. Kamionkowski, A. Kosowsky and M. S. Turner, Phys. Rev. D {\bf 49}, 2837 (1994).

\bibitem{PTT4} R.~Jinno, S.~Lee, H.~Seong and M.~Takimoto,  JCAP {\bf 1711}, 050 (2017).

\bibitem{PTT5} J.~Ellis, M.~Lewicki and J.~M.~No,  JCAP {\bf 1904}, 003 (2019).

\bibitem{PT1} M. D'Onofrio and K. Rummukainen  Phys. Rev. D {\bf 93}, 025003 (2016).

\bibitem{PT2a} K. Kajantie, M. Laine, J. Peisa, K. Rummukainen and M. Shaposhnikov  Nucl. Phys. B {\bf 544}, 357 (1999).

\bibitem{PT2b} K. Kajantie, M. Laine, K. Rummukainen and M. Shaposhnikov Phys. Rev. Lett. {\bf 77} 2887 (1996).

\bibitem{PT3a} M. Giovannini, Phys. Rev. D {\bf 61}, 063004 (2000).

\bibitem{PT3b} M. Giovannini, Phys. Rev. D {\bf 61} 063502 (2000). 

\bibitem{PT3c} M.~Giovannini,  Phys.\ Rev.\ D {\bf 92} 121301 (2015).

\bibitem{PT3d}  T.~Fujita and K.~Kamada, Phys.\ Rev.\ D {\bf 93},  083520 (2016).

\bibitem{PT3e}  A.~J.~Long and E.~Sabancilar,  JCAP {\bf 1605}, 05, 029 (2016).

\bibitem{AMHD1} M.~Giovannini,  Phys.\ Rev.\ D {\bf 88}, 063536 (2013).

\bibitem{AMHD1a}  M.~Giovannini,  Phys.\ Rev.\ D {\bf 93}, 103518 (2016).

\bibitem{AMHD2}   N.~Yamamoto, Phys.\ Rev.\ D {\bf 93}, 125016 (2016).

\bibitem{AMHD2aa}  K.~Hattori and Y.~Yin, Phys.\ Rev.\ Lett.\  {\bf 117},  152002 (2016).
  
\bibitem{AMHD2bb} M.~Giovannini,  Phys.\ Rev.\ D {\bf 94}, 081301 (2016).
  
\bibitem{AMHD2a}  S. Chandrasekhar and P. C. Kendall, Astrophys. J. {\bf 126}, 457 (1957).

\bibitem{AMHD2b}  S. Chandrasekhar and L. Woltjer, Proc. Natl. Acad. Sci. U.S.A. {\bf 44}, 285 (1958).
  
\bibitem{AMHD3}  M.~Giovannini and M.~E.~Shaposhnikov,  Phys.\ Rev.\ D {\bf 57}, 2186 (1998).

\bibitem{AMHD3a}  M.~Giovannini and M.~E.~Shaposhnikov, Phys.\ Rev.\ Lett.\  {\bf 80}, 22 (1998).  

\bibitem{AMHD4} M. Giovannini, Class. Quantum Grav. {\bf 34} 135010 (2017).

\bibitem{AMHD5} S.~Kuroyanagi, T.~Chiba and T.~Takahashi,  JCAP {\bf 1811}, 11, 038 (2018).

\bibitem{LISA2}   P.~Amaro-Seoane {\it et al.},  Class.\ Quant.\ Grav.\  {\bf 29}, 124016 (2012).

\bibitem{LISA3} H.~Audley {\it et al.} [LISA Collaboration],  arXiv:1702.00786 [astro-ph.IM].
    
\bibitem{LISA4}  M. Armano {\it et al.},  Phys. Rev. Lett. {\bf 116}, 231101 (2016).

\bibitem{LISA5} M. Armano {\it et al.},  Phys. Rev. Lett. {\bf 120}, 061101 (2018).
   
\bibitem{DECIGO2}  S.~Kawamura {\it et al.} [DECIGO Collaboration],  Class.\ Quant.\ Grav.\  {\bf 28}, 094011 (2011).
    
\bibitem{DECIGO3}  S. Sato {\it et al.} [DECIGO Collaboration] , J. Phys. Conf. Ser. {\bf 840} 012010 (2017).

\bibitem{STIFF0}  M.~Giovannini, Class.\ Quant.\ Grav.\  {\bf 16}, 2905 (1999).

\bibitem{STIFF0a} A.~Riazuelo and J.~P.~Uzan,  Phys.\ Rev.\ D {\bf 62}, 083506 (2000).

\bibitem{STIFF1}   V.~Sahni, M.~Sami and T.~Souradeep, Phys.\ Rev.\  D {\bf 65}, 023518 (2002).

\bibitem{STIFF1a}  M.~Yahiro, G.~J.~Mathews, K.~Ichiki, T.~Kajino and M.~Orito,  Phys.\ Rev.\ D {\bf 65}, 063502 (2002)

\bibitem{STIFF2} M.~Giovannini,  Phys.\ Rev.\ D {\bf 67}, 123512 (2003).

\bibitem{STIFF3}   H.~Tashiro, T.~Chiba and M.~Sasaki, Class.\ Quant.\ Grav.\  {\bf 21}, 1761 (2004).

\bibitem{STIFF4}  T.~J.~Battefeld and D.~A.~Easson,  Phys.\ Rev.\  D {\bf 70}, 103516 (2004).

\bibitem{STIFF6} L.~A.~Boyle and A.~Buonanno,  Phys.\ Rev.\ D {\bf 78}, 043531 (2008).
  
 \bibitem{STIFF8}  M.~W.~Hossain, R.~Myrzakulov, M.~Sami and E.~N.~Saridakis,
  Phys.\ Rev.\ D {\bf 89}, 123513 (2014).
  
 \bibitem{STIFF9}  M.~Wali Hossain, R.~Myrzakulov, M.~Sami and E.~N.~Saridakis,  Int.\ J.\ Mod.\ Phys.\ D {\bf 24}, 1530014 (2015).
  
 \bibitem{STIFF10} J.~de Haro, J.~Amoros and S.~Pan,  Phys.\ Rev.\ D {\bf 93}, 084018 (2016).
  
\bibitem{STIFF11}   M.~Giovannini,  Phys.\ Lett.\ B {\bf 759}, 528 (2016).
  
\bibitem{STIFF13}  L.~Arest\'e Sal\'o and J.~de Haro,  Eur.\ Phys.\ J.\ C {\bf 77},  798 (2017).

\bibitem{STIFF13a} K.~Dimopoulos and T.~Markkanen,  JCAP {\bf 1806}, 021 (2018).
  
\bibitem{STIFF17}  J.~Haro,  Phys.\ Rev.\ D {\bf 99},  043510 (2019).
  
\bibitem{STIFF18}   J.~Haro, W.~Yang and S.~Pan,  JCAP {\bf 1901}, 023 (2019).
  
\bibitem{STIFF14}   S.~D.~Odintsov and V.~K.~Oikonomou,  Phys.\ Rev.\ D {\bf 96}, 104059 (2017).
 
\bibitem{STIFF14a}  M.~Giovannini,  Class. Quantum Grav. {\bf 36}, 235017 (2019).
  
\bibitem{STIFF19}    L. H. Ford, Phys. Rev. D {\bf 35}, 2955  (1987).

\bibitem{STIFF20} P. J. E. Peebles and B. Ratra, Astrophys. J. {\bf 325}, L17 (1988).

\bibitem{STIFF21} R. R. Caldwell, R. Dave, and P. J. Steinhardt, Phys. Rev. Lett. {\bf 80}, 1582 (1998).

\bibitem{STIFF16a}   M.~Joyce,  Phys.\ Rev.\ D {\bf 55}, 1875 (1997).
  
\bibitem{STIFF16}  M.~Giovannini,  Phys.\ Rev.\ D {\bf 98},  103509 (2018). 

\bibitem{STIFF22}  E.~D.~Schiappacasse and L.~H.~Ford,  Phys. Rev. D {\bf 94},  084030 (2016).
  
\bibitem{RH1} S. Y. Khlebnikov and I. I. Tkachev, Phys. Rev. D {\bf 56}, 653 (1997).

 \bibitem{RHnn1} M.~A.~Amin, J.~Braden, E.~J.~Copeland, J.~T.~Giblin, C.~Solorio, Z.~J.~Weiner and S.~Y.~Zhou,  Phys.\ Rev.\ D {\bf 98}, 024040 (2018).

 \bibitem{RHnn2} S.~Antusch, F.~Cefala and S.~Orani,  JCAP {\bf 1803}, 032 (2018).
 
\bibitem{RH5} M.~Giovannini, Phys.\ Rev.\ D {\bf 82}, 083523 (2010).

\bibitem{RWAT1} J.~Fonseca, M.~Sasaki and D.~Wands,  JCAP {\bf 1009}, 012 (2010).

\bibitem{RWAT2} D.~H.~Lyth,  Prog.\ Theor.\ Phys.\ Suppl.\  {\bf 190}, 107 (2011).

\bibitem{RWAT3} S.~Clesse,  Phys.\ Rev.\ D {\bf 83}, 063518 (2011).

 \bibitem{RHnn} J.~L.~Cook and L.~Sorbo,  Phys.\ Rev.\ D {\bf 85}, 023534 (2012) Erratum: [Phys.\ Rev.\ D {\bf 86}, 069901 (2012)].
 
 \bibitem{RHnn3}  S.~Kuroyanagi, C.~Lin, M.~Sasaki and S.~Tsujikawa, Phys.\ Rev.\ D {\bf 97}, 023516 (2018).
  
\bibitem{RHn}   B.~A.~Bassett, S.~Tsujikawa and D.~Wands,  Rev.\ Mod.\ Phys.\  {\bf 78}, 537 (2006).

 \bibitem{RHff1} S.~Pi, M.~Sasaki and Y.~l.~Zhang,  JCAP {\bf 1906}, 049 (2019). 
  
\bibitem{RHff2}  R.~g.~Cai, S.~Pi and M.~Sasaki,  Phys.\ Rev.\ Lett.\  {\bf 122}, 201101 (2019).  

\bibitem{RHff3} K.~Inomata and T.~Nakama,  Phys.\ Rev.\ D {\bf 99}, 043511 (2019).

\bibitem{ST1} R. Dong, W. H. Kinney, D. Stojkovic, JCAP 1610  {\bf 034} (2016). 

\bibitem{HBT1} R. Hanbury Brown and R. Q. Twiss, Nature {\bf 178}, 1046 (1956).

\bibitem{HBT2} R. Hanbury Brown and R. Q. Twiss, Proc. R. Soc. A {\bf 242}, 300 (1957).

\bibitem{HBT2a} R. Hanbury Brown and R. Q. Twiss, Proc. R. Soc. A  {\bf 243}, 291 (1958).

\bibitem{HBTB} M.~Giovannini,  Phys.\ Rev.\ D {\bf 83}, 023515 (2011).

\bibitem{QO2} R.~J.~Glauber,  Phys.\ Rev.\ Lett.\  {\bf 10}, 84 (1963).

\bibitem{QO3} E.~C.~C.~Sudarshan, Phys.\ Rev.\ Lett.\  {\bf 10}, 277 (1963).

\bibitem{revs} D.~H.~Boal, C.~K.~Gelbke, B.~K.~Jennings, Rev.\ Mod.\ Phys.\  {\bf 62}, 553 (1990).

\bibitem{revsa} G. Baym, Acta Phys.\ Polon.\  B {\bf 29}, 1839 (1998).

\bibitem{cocconi1} G.~I.~Kopylov, M.~I.~Podgoretsky,  Sov.\ J.\ Nucl.\ Phys.\  {\bf 15}, 219-223 (1972) [Yad. Fiz. {\bf 15}, 392 (1972)].

\bibitem{cocconi1a} G.~I.~Kopylov, M.~I.~Podgoretsky, Sov.\ J.\ Nucl.\ Phys.\  {\bf 18},  336 (1973) [Yad. Fiz. {\bf 18}, 656 (1973)].
 
\bibitem{cocconi2} G.~Cocconi,  Phys.\ Lett.\  {\bf B49}, 459 (1974).

\bibitem{HBTBa}  M.~Giovannini,  Class.\ Quant.\ Grav.\  {\bf 34}, 035019 (2017).

\bibitem{HBTBb}  M.~Giovannini, Mod.\ Phys.\ Lett.\ A {\bf 32}, 1750191 (2017).

 \bibitem{HBTC} M.~Giovannini, Phys.\ Rev.\ D {\bf 99}, 123507 (2019).
 
\bibitem{HBTCa} M.~Giovannini,   Mod.\ Phys.\ Lett.\ A {\bf 34}, 1950185 (2019).

\bibitem{loudon}  R. Loudon, {\it ``The quantum theory of light''} (Clarendon Press, Oxford, 1983).

\bibitem{HBTD}  S.~Kanno and J.~Soda,  Phys.\ Rev.\ D {\bf 99}, 084010 (2019).

\bibitem{HBTDa} S.~Kanno,  arXiv:1905.06800 [hep-th].

\bibitem{WWEB0a}  J.~Weber and J.~A.~Wheeler,  Rev.\ Mod.\ Phys.\  {\bf 29},  509 (1957).

\bibitem{WWEB0b} F.~A.~E.~Pirani, Acta Phys.\ Polon.\  {\bf 15}, 389 (1956)
  [Gen.\ Rel.\ Grav.\  {\bf 41}, 1215 (2009)].

\bibitem{WEB0c} J.~Weber, Phys. Rev. Lett. {\bf 17}, 1228 (1966).

\bibitem{WEB1} J.~Weber, Phys. Rev. Lett. {\bf 18}, 498 (1967).

\bibitem{WEB2}  J.~Weber, Phys. Rev. Lett. {\bf 20}, 1307 (1968).

\bibitem{WEB3} J.~Weber,  Phys.\ Rev.\ Lett.\  {\bf 22}, 1320 (1969).

\bibitem{WEB4}  J.~Weber,  Phys.\ Rev.\ Lett.\  {\bf 25}, 180 (1970).
  
 \bibitem{WEB5} J.~Weber,  Phys.\ Rev.\ Lett.\  {\bf 24}, 276 (1970).

\bibitem{WWC1}  R.~L.~Garwin and J.~L.~Levine,  Phys.\ Rev.\ Lett.\  {\bf 31}, 176 (1973).

\bibitem{WWC2}   J.~L.~Levine and R.~L.~Garwin,  Phys.\ Rev.\ Lett.\  {\bf 31}, 173 (1973).

\bibitem{WWC3} J.~L.~Levine and R.~L.~Garwin,  Phys.\ Rev.\ Lett.\  {\bf 33}, 794 (1974).

\bibitem{WWC4} J.~A.~Tyson,  Phys.\ Rev.\ Lett.\  {\bf 31}, 326 (1973).
  
\bibitem{WWC5}  J.~A.~Tyson, C.~G.~Maclennan and L.~J.~Lanzerotti,  Phys.\ Rev.\ Lett.\  {\bf 30}, 1006 (1973).  
  
 \bibitem{niobe} D. G.  Blair {\it et al.}, Phys. Rev. Lett. {\bf 74}, 1908 (1995).

\bibitem{allegro} E. Manuceli, {\it et al.}, Phys. Rev. D {\bf 54}, 1264 (1996).

\bibitem{auriga} M. Cerdonio, {\it et al.},  Class. Quantum Grav.  {\bf 14}, 1491 (1997).

\bibitem{explorer} P. Astone {\it et al.} Phys. Rev. D {\bf 47}, 362 (1993).

\bibitem{nautilus} P.  Astone, {\it et al.} Astroparticle Physics, {\bf 7}, 231 (1997).

\bibitem{sph1} J. A. Lobo,  Phys. Rev. D {\bf 52}, 591 (1995).

\bibitem{sph2} E. Coccia,{\it et al.}   Phys. Rev. D {\bf 52}, 3735 (1998).

\bibitem{sph3} E. Coccia, {\it et al.}  Phys. Rev. D {\bf 57}, 2051 (1998).

\bibitem{pizzella} G.~Pizzella,  Nuovo Cim.\ C {\bf 18}, 285 (1995).

\bibitem{bars10}  P.~Astone {\it et al.},  Phys.\ Rev.\ D {\bf 47}, 362 (1993). 

\bibitem{bars11}  P.~Astone {\it et al.}, Astropart.\ Phys.\  {\bf 7}, 231 (1997).  
  
\bibitem{bars12} P.~Astone, G.~V.~Pallottino and G.~Pizzella, Class.\ Quant.\ Grav.\  {\bf 14}, 2019 (1997).
  
\bibitem{SN1}  E.~Amaldi {\it et al.},  Europhys.\ Lett.\  {\bf 3}, 1325 (1987).  
 
\bibitem{SN2} E.~Amaldi {\it et al.},  Annals N.\ Y.\ Acad.\ Sci.\  {\bf 571}, 561 (1989). 
  
\bibitem{INT1}  M.~E.~Gertsenshtein and V.~I.~Pustovoit, Sov.\ Phys.\ JETP {\bf 16}, 433 (1962)
 [ Zh. Eksp. Teor. Fiz. {\bf 43}, 605 (1962)].

\bibitem{firstevent} B.~P.~Abbott {\it et al.} [LIGO/Virgo Collaboration],  Phys.\ Rev.\ Lett.\  {\bf 116}, 061102 (2016).

\bibitem{INT2} R. Weiss, Quart. Progr. Rep. Res. Lab. Electron. MIT {\bf 105}, 54 (1972).

\bibitem{INT3} A.~Giazotto,  Phys.\ Rept.\  {\bf 182}, 365 (1989).

\bibitem{INT4} K.~Kawabe [TAMA Collaboration],  Class.\ Quant.\ Grav.\  {\bf 14}, 1477 (1997).

\bibitem{INT5} K. Danzmann, {\it et al.}   Class. Quantum Grav. {\bf 14}, 1471 (1997).

\bibitem{INT6} B. Caron, {\it et al.},  Class. Quantum Grav. {\bf 14} 1461 (1997).

\bibitem{INT7} A. Abramovici, {\it et al.}, {\it Science} {\bf 256}, 325 (1992).

\bibitem{wide1} P. Michelson, Mon. Not. R. Astron. Soc. {\bf 227}, 933 (1987).

\bibitem{wide2} N. Christensen, Phys. Rev. D {\bf 46}, 5250 (1992).

\bibitem{wide3} E. Flanagan, Phys. Rev. D {\bf 48}, 2389 (1993).

\bibitem{wide4} N. Christensen, Phys. Rev. D {\bf 55}, 448 (1997).

\bibitem{wide5} B. Allen and J. Romano, Phys. Rev. D {\bf 59}, 102001 (1999).

\bibitem{wide6} D.~Babusci and M.~Giovannini,  Phys.\ Rev.\ D {\bf 60}, 083511 (1999).

\bibitem{STone}   B.~Abbott {\it et al.} [LIGO Collaboration], Phys.\ Rev.\ D {\bf 69}, 122004 (2004).
  
\bibitem{STonea} B.~Abbott {\it et al.} [LIGO Collaboration], Phys.\ Rev.\ Lett.\  {\bf 95}, 221101 (2005).  
  
 \bibitem{STtwo}  J.~Abadie {\it et al.} [LIGO/Virgo Collaboration], Phys.\ Rev.\ D {\bf 85}, 122001 (2012).

\bibitem{STthree}  J.~Aasi {\it et al.} [LIGO/Virgo Collaboration], Phys.\ Rev.\ Lett.\  {\bf 113}, 231101 (2014).

\bibitem{STthreea} J.~Aasi {\it et al.} [LIGO/Virgo Collaboration],  Phys.\ Rev.\ D {\bf 91},  022003 (2015).
  
\bibitem{STfour}  B.~P.~Abbott {\it et al.} [LIGO/Virgo Collaboration],
   Phys.\ Rev.\ Lett.\  {\bf 118}, 121101 (2017)  Erratum: [Phys.\ Rev.\ Lett.\  {\bf 119}, 029901 (2017)].

\bibitem{STfive}  B.~P.~Abbott {\it et al.} [LIGO/Virgo Collaboration],Phys.\ Rev.\ D {\bf 100}, 061101(R) (2019).

\bibitem{wide7} D.~Babusci and M.~Giovannini,  Class.\ Quant.\ Grav.\  {\bf 17}, 2621 (2000).

\bibitem{wide8} D.~Babusci and M.~Giovannini,  Int.\ J.\ Mod.\ Phys.\ D {\bf 10}, 477 (2001).

\bibitem{brag2} V. B. Braginsky, L.P. Grishchuk, A. G. Doroshkevich, Ya. B. Zeldovich, I. D. Novikov and M. Sazhin, 
Sov. Phys. JETP {\bf 38}, 865 (1974) [Zh. Eksp. Teor. Fiz. {\bf 65}, 1729 (1973)].

\bibitem{cav1a} F. Pegoraro, E. Picasso, and L. Radicati, J. Phys. A {\bf 11}, 1949 (1978).

\bibitem{cav1b}  E. Jacopini, F. Pegoraro, E. Picasso, and L. Radicati, Phy. Lett. {\bf 73}, 140 (1979).

\bibitem{cav2}  C.  Reece, P. Reiner, and A.  Melissinos, Phys. Lett. A {\bf 104}, 341 (1984).

\bibitem{cav2a} C.  Reece, P. Reiner, and A.  Melissinos, Nucl. Inst. and Methods, A {\bf 245}, 299 (1986).

\bibitem{cav3} Ph. Bernard, G. Gemme, R. Parodi and E. Picasso, Rev. Sci. Instrum. {\bf 72}, 2428 (2001).

\bibitem{cav3a} R. Ballantini, P. Bernard, A. Chincarini, G. Gemme, R. Parodi and E. Picasso, Class. Quant. Grav. {\bf 21}, S1241 (2004).

\bibitem{cav4} A. M. Cruise, Class. Quantum Grav. {\bf 17} , 2525 (2000).

\bibitem{cav4a} A. M. Cruise, Mon. Not. R. Astron. Soc.  {\bf 204}, 485 (1983).

\bibitem{cav5} A. M. Cruise and R.  Ingley, Class. Quantum Grav. {\bf 23}, 6185 (2006).

\bibitem{cav5a} A. M. Cruise and R. M. Ingley,  Class. Quantum Grav. {\bf 22}, S479 (2005).

\bibitem{cav6} F.  Li,  M.  Tang and D.  Shi, Phys. Rev. D {\bf 67}, 104008 (2003).

\bibitem{cav6a} F. Li, Z. Wu and Y.~Zhang, Chin.\ Phys.\ Lett.\  {\bf 20}, 1917 (2003).

\bibitem{cav7} A. Nishizawa {\it et al.}, Phys. Rev. D {\bf 77}, 022002 (2008).

\bibitem{cav8} A. T. Akutsu {\it et al.},  Phys. Rev. Lett. {\bf 101}, 101101 (2008).

\bibitem{SENS1}  A.~Arvanitaki and A.~A.~Geraci,  Phys.\ Rev.\ Lett.\  {\bf 110},  071105 (2013).

\bibitem{holo1} A.~S.~Chou {\it et al.} [Holometer Collaboration],  Phys.\ Rev.\ Lett.\  {\bf 117}, 111102 (2016).

\bibitem{holo2} A.~Chou {\it et al.} [Holometer Collaboration] Class.\ Quant.\ Grav.\  {\bf 34},  065005 (2017).
  
\bibitem{SENS2} S.~Dimopoulos, P.~W.~Graham, J.~M.~Hogan, M.~A.~Kasevich and S.~Rajendran,
    Phys.\ Rev.\ D {\bf 78}, 122002 (2008).
  
\bibitem{SENS3}   S.~Dimopoulos, P.~W.~Graham, J.~M.~Hogan, M.~A.~Kasevich and S.~Rajendran,
   Phys.\ Lett.\ B {\bf 678}, 37 (2009).
  
\bibitem{SENS4} S.~Dimopoulos, P.~W.~Graham, J.~M.~Hogan, M.~A.~Kasevich and S.~Rajendran,
   Phys.\ Rev.\ D {\bf 84}, 028102 (2011).

\end{thebibliography}
\end{document}